\documentclass{cernrep}
\usepackage{epsfig} 
\usepackage{graphicx}
\usepackage{ifpdf}
\usepackage{cite}
\usepackage{eurosym}

\usepackage{axodraw}

\def\ga{\mathrel{\rlap{\raise.6ex\hbox{$>$}}{\lower.6ex\hbox{$\sim$}}}}
\def\la{\mathrel{\rlap{\raise.6ex\hbox{$<$}}{\lower.6ex\hbox{$\sim$}}}}

\providecommand{\GEANTthree}{{\sc{geant3}}\xspace}
\providecommand{\GEANTfour}{{\sc{geant4}}\xspace}

\newcommand{\mum}{\ensuremath{\mathrm{\;\mu m}}\xspace}
\newcommand{\micron}{\ensuremath{\mathrm{\;\mu m}}\xspace}
\newcommand{\cm}{\ensuremath{\mathrm{\; cm}}\xspace}
\newcommand{\mm}{\ensuremath{\mathrm{\; mm}}\xspace}

\newcommand{\keV}{\ensuremath{\mathrm{\; keV}}\xspace}
\newcommand{\MeV}{\ensuremath{\mathrm{\; MeV}}\xspace}
\newcommand{\GeV}{\ensuremath{\mathrm{\; GeV}}\xspace}
\newcommand{\TeV}{\ensuremath{\mathrm{\; TeV}}\xspace}

\newcommand{\GeVc}{\ensuremath{\mathrm{\; GeV}/c}\xspace}

\newcommand{\MeVcc}{\ensuremath{\mathrm{\; MeV}/c^2}\xspace}

\newcommand{\fbinv} {\mbox{\ensuremath{\mathrm{\; fb}^{-1}}}\xspace}

\newcommand{\EE}{{e^+e^-}\xspace}
\newcommand{\MM}{{\mu^+\mu^-}\xspace}

\newcommand{\PT}{\ensuremath{p_{\mathrm{T}}}\xspace}

\newcommand{\MET}{\ensuremath{{E\!\!\!/}_{\mathrm{T}}}\xspace}

\newcommand{\Hz}{\ensuremath{H^{0}}\xspace}

\newcommand{\rpv}{\ensuremath{\rlap{\kern.2em/}R}\xspace}

 \newcommand{\cA}{\mathcal{A}}
\newcommand{\cO}{\mathcal{O}}
\newcommand{\no}{\nonumber}
\newcommand{\ba}{\begin{array}}
\newcommand{\ea}{\end{array}}
\newcommand{\cL}{{\cal L}}
\newcommand{\yuk}{{Y}}
\newcommand{\diagyuk}{{\lambda}}
\newcommand{\identity}{1 \hspace{-.085cm}{\rm l}}
\newcommand{\sw}{s^2_{\rm W}}

\newcommand{\bequ}{\begin{equation}}
\newcommand{\eequ}{\end{equation}}
\newcommand{\bea}{\begin{eqnarray}}
\newcommand{\eea}{\end{eqnarray}}
\newcommand{\nn}{\nonumber}
\newcommand{\bi}{\begin{itemize}}
\newcommand{\ei}{\end{itemize}}

\newcommand{\gev}{\, {\rm GeV}}
\newcommand{\mev}{\, {\rm MeV}}

\def\kpn{K^+\rightarrow\pi^+\nu\bar\nu}

\def\klpn{K_{L}\rightarrow\pi^0\nu\bar\nu}

\newcounter{number}
\newcommand{\num}[1]{\stepcounter{number}\thenumber .$\;$}
\def\ul#1{\underline{\vphantom{g}#1}}
\newcommand{\br}{\mbox{BR}}

\newcommand{\til}{\tilde}

\newcommand{\Mpi}{M_\pi}

\newcommand{\al}{\alpha}

\newcommand{\rh}{\rho}

\newcommand{\ps}{\psi}

\newcommand{\bdm}{\begin{displaymath}}
\newcommand{\edm}{\end{displaymath}}
\newcommand{\beq}{\begin{equation}}
\newcommand{\eeq}{\end{equation}}

\newcommand{\mr}{\mathrm}
\newcommand{\Nf}{{N_{\!f}}}

\newcommand{\fm}{\,\mr{fm}}
\newcommand{\MSbar}{\overline{\mr{MS}}}

 \newcommand{\bsg}{\bar B\to X_{s} \gamma}
\newcommand{\Bqll}{{\bar{B}^0_{q} \to \ell^+ \ell^-}}

\newcommand{\BKll}{{\bar{B} \to K \ell^+ \ell^-}}
\newcommand{\BKastll}{{\bar{B} \to K^\ast \ell^+ \ell^-}}

\newcommand{\BKastgamma}{{\bar{B} \to K^\ast \gamma}}
\newcommand{\BKastKpill}{{\bar{B}^0 \to K^{\ast 0}(\to K^- \pi^+) \ell^+ \ell^-}}

 \newcommand{\psubt}{\mbox{$p_{T}\ $}}
\newcommand{\lt}{\left}
\newcommand{\rt}{\right}
\newcommand{\ov}[1]{\overline{#1}}
\newcommand{\eq}[1]{Eq.~(\ref{#1})}

\newcommand{\eqsto}[2]{Eqs.~(\ref{#1}--\ref{#2})}

\newcommand{\be}{\begin{equation}}
\newcommand{\ee}{\end{equation}}
\newcommand{\DB}{\Delta B}

\providecommand{\dg}{\ensuremath{\Delta \Gamma_{s}}\xspace}
\providecommand{\dgg}{\ensuremath{\Delta \Gamma_{s}/\Gamma_{s}}\xspace}
\providecommand{\gm}{\ensuremath{\bar \Gamma_s}\xspace}
\providecommand{\cp}{\ensuremath{CP}}

\def \bspsiphi {\ensuremath{B_{s}\rightarrow J/\psi\, \phi}\xspace}

\def \bsmunux {\ensuremath{B_{s}\rightarrow D_{s} \mu \nu X} \xspace}
\def \dsphipi {\ensuremath{D_{s}\rightarrow \phi \pi} \xspace}
\def \bsdspi {\ensuremath{B_{s}\rightarrow D_{s} \pi} \xspace}

\newcommand{\phis}{\ensuremath{\phi_s}\xspace}
\newcommand{\dms}{\ensuremath{\Delta m_s}\xspace}
\newcommand{\dmd}{\ensuremath{\Delta m_d}\xspace}
\newcommand{\asl}{\ensuremath{A_{SL}}\xspace}
\newcommand{\apm}[2]{\ensuremath{^{+#1}_{-#2}}}

\def\Bs    {\ensuremath{B_s}\xspace}
\def\Bd    {\ensuremath{B_d}\xspace}
\def\Dsm   {\ensuremath{D^-_s}\xspace}
\def\ellp  {\ensuremath{\ell^+}\xspace}
\def\fs    {\ensuremath{\mbox{\,fs}}\xspace}

\newcommand{\myunit}[1]{\ifmmode{\mathrm{#1}}\else{#1}\fi\xspace}
\newcommand{\MeVctwo}{\myunit{\text{MeV}/\ensuremath{\text{c}^2}}}
\newcommand{\Ps}{\ensuremath{\text{s}}\xspace}

\renewcommand{\Pb}{\ensuremath{\text{b}}\xspace}

\newcommand{\Pab}{\ensuremath{\bar{\text{b}}}\xspace}
\newcommand{\PsB}{\ensuremath{\text{B}_{\text{s}}}\xspace}

\newcommand{\Pphi}{\ensuremath{\phi}\xspace}
\newcommand{\jpsi}{\ensuremath{J/\psi}\xspace}

\newcommand{\bccs}{\ensuremath {\bar b\to\bar cc\bar s}
\xspace}

\newcommand{\Bsetacphi}{\ensuremath {B_s\to\eta_c\phi}
\xspace}
\newcommand{\BsDsDs}{\ensuremath{B_s\to D_s D_s}\xspace}
\newcommand{\Bsjpphi}{\ensuremath{B_s\to J/\psi\phi}
\xspace}
\newcommand{\BsDspi}{\ensuremath{B_s\to D_s\pi}
\xspace}
\newcommand{\Bsjpetappp}{\ensuremath{
B_s\to J/\psi\eta(\pi^+\pi^-\pi^0)}
\xspace}
\newcommand{\Bsjpetagg}{\ensuremath{
B_s\to J/\psi\eta(\gamma\gamma)}\xspace}
\newcommand{\Bsjpetaprimetapp}{\ensuremath{
B_s\to J/\psi\eta'(\eta\pi^+\pi^-)}
\xspace}
\newcommand{\Bsjpetaprimrhog}{\ensuremath{
B_s\to J/\psi\eta'(\rho\gamma)}\xspace}

\newcommand{\bsmix}{\ensuremath{B_s -\bar B_s}\xspace}

\def\bz{{B^0}}
\def\bzb{{\overline{B}{}^0}}

\def\bm{{B^-}}

\newcommand{\sinbb}{{\sin2\phi_1}}

\def\taubz{{\tau_\bz}}

\def\ks{{K_S^0}}
\def\KS{\ks}

\def\kz{{K^0}}

\def\piz{{\pi^0}}

\def\etap{{\eta'}}

\def\cala{{\cal A}}
\def\cals{{\cal S}}
\def\dm{\Delta m_d}
\def\dmd{\dm}
\def\taubz{\tau_\bz}
\def\ks{{K_S^0}}

\def\btosss{b \to s\overline{s}s}

\def\fq{\ensuremath{q}}

\def\btoccs{b \to c\bar{c}s}

\usepackage{relsize}
\def\babar{\mbox{\slshape B\kern-0.1em{\smaller A}\kern-0.1em
    B\kern-0.1em{\smaller A\kern-0.2em R}}}

\def\jpsi     {\ensuremath{{J\mskip -3mu/\mskip -2mu\psi\mskip 2mu}}\xspace}
\newcommand{\etapr}{\ensuremath{\eta^{\prime}}\xspace}

\def\CP                {\ensuremath{C\!P}\xspace}
\def\B       {\ensuremath{B}\xspace}

\newcommand{\abs}[1]{\lvert#1\rvert}
\newcommand{\mt}{m_{\rm t}}
\newcommand{\mc}{m_{\rm c}}
\newcommand{\beqa}{\begin{eqnarray}}
\newcommand{\eeqa}{\end{eqnarray}}
\newcommand{\ord}{\cal O}
\def\kpn{K^+\to\pi^+\nu\bar\nu}
\def\klpn{K_L\to\pi^0\nu\bar\nu}

\newcommand{\Eqsand}[2]{Eqs.~(\ref{#1}) and~(\ref{#2})}
\newcommand{\Sec}[1]{Section~\ref{#1}}
\newcommand{\Secsand}[2]{Sections~\ref{#1}~and~\ref{#2}}
\newcommand{\Fig}[1]{Fig.~\ref{#1}}
\newcommand{\f}{\frac}
\newcommand{\BR}{{\cal B}}
\newcommand{\kpns}{K\to\pi\nu\bar\nu} 
\newcommand{\re}{{\rm Re}}
\newcommand{\im}{{\rm Im}}
\newcommand{\dPcu}{\delta P_{c,u}}

\def \bo{B^0}
\def \ko{K^0}
\def \ob{\overline{B}^0}
\def \ok{\overline{K}^0}
\def \ot{\overline{t}}

\def\op{{P}}
\def\ot{{T}}
\def\cpt{{CPT}}

\def\mtiny{\vrule width 0pt}
\def\mrm#1{\mathrm{#1}}
\def\DZ{\relax\ifmmode{{D}^0}\else{$\mrm{D}^{\mrm{0}}$}\fi}
\def\BZ{\relax\ifmmode{B^0}\else{$\mrm{B}^{\mrm{0}}$}\fi}
\def\BZS{\relax\ifmmode{B_s^0}\else{$\mrm{B_s}^{\mrm{0}}$}\fi}
\def\DZB{\relax\ifmmode{\overline{D}\mtiny^{0}}
        \else{$\overline{\mrm{D}}\mtiny^{\mrm{0}}$}\fi}
\def\BZB{\relax\ifmmode{\overline{B}\mtiny^{0}}
       \else{$\overline{\mrm{B}}\mtiny^{\mrm{0}}$}\fi}
\def\BZBS{\relax\ifmmode{\overline{B_s}\mtiny^{0}}
        \else{$\overline{\mrm{B_s}}\mtiny^{\mrm{0}}$}\fi}

 \def\FourS {$\Upsilon(\rm 4S)$}
\renewcommand{\Re}{{\rm Re}}
\renewcommand{\Im}{{\rm Im}}
\def\rhobar {\ensuremath{\overline \rho}\xspace}
\def\etabar {\ensuremath{\overline \eta}\xspace}

\newcommand{\degrees}{\mbox{${^\circ}$}}
\newcommand{\sphis}{\mbox{$\sigma({\phi_s})$}}
\newcommand{\phisphiphi}{\mbox{$S(\phi\phi)$}}
\newcommand{\sphisphiphi}{\mbox{$\sigma(S(\phi\phi))$}}
\newcommand{\Bstojpsiphi}{\mbox{$B^0_s \to J/\psi\phi$}} 
\newcommand{\Bstophiphi}{\mbox{$B^0_s \to \phi\phi$}} 
\newcommand{\Bdtojpsiks}{\mbox{$B^0 \to J/\psi K^0_S$}} 
\newcommand{\Bdtophiks}{\mbox{$B^0 \to \phi K^0_S$}}

\newcommand{\Bstodsk}{\mbox{$B^0_s \to D_s^{\mp} K^{\pm}$}}

 \newcommand{\MW}{M_W}

\newcommand{\mA}{m_A}
\newcommand{\MA}{M_A}
\newcommand{\mh}{m_h}

\newcommand{\Mh}{M_h}

\def\tu{{\tilde u}}

\def\tg{{\tilde g}}
\def\tq{{\tilde q}}
\def\ttop{{\tilde t}}
\def\sbot{{\tilde b}}
\def\ttau{{\tilde\tau}}

\def\tchi{{\tilde\chi}}
\def\tl{{\tilde\ell}}
\def\lsp{{\tilde\chi_1^0}}

\newcommand{\mste}{m_{\tilde{t}_1}}

\newcommand{\msbe}{m_{\tilde{b}_1}}
\newcommand{\msbz}{m_{\tilde{b}_2}}

\newcommand{\At}{A_t}

\newcommand{\sq}{\tilde{q}}
\newcommand{\gl}{\tilde{g}}
\newcommand{\Mgl}{m_{\tilde{g}}}
\newcommand{\mgl}{m_{\tilde{g}}}

\newcommand{\mcha}[1]{m_{\tilde \chi^\pm_{#1}}}

\newcommand{\tst}{\theta_{\tilde{t}}}
\newcommand{\tsb}{\theta_{\tilde{b}}}

\newcommand{\tb}{\tan \beta}

\def\dofig#1#2{\epsfxsize=#1\centerline{\epsfbox{#2}}}
\def\dofigs#1#2#3{\centerline{\epsfxsize=#1\epsfbox{#2}%
   \hfil\epsfxsize=#1\epsfbox{#3}}}

\def\bmp#1{\begin{minipage}{#1\textwidth}}
\def\emp{\end{minipage}}

\def\simge{
    \mathrel{\rlap{\raise 0.511ex
        \hbox{$>$}}{\lower 0.511ex \hbox{$\sim$}}}}
\def\simle{
    \mathrel{\rlap{\raise 0.511ex
        \hbox{$<$}}{\lower 0.511ex \hbox{$\sim$}}}}

\newcommand{\SLASH}[2]{\makebox[#2ex][l]{$#1$}/}
\newcommand{\mMET}{$\SLASH{E_T}{.5}$\hspace{1.0ex}}

\newcommand{\lsim}
{\;\raisebox{-.3em}{$\stackrel{\displaystyle <}{\sim}$}\;}
\newcommand{\gsim}
{\;\raisebox{-.3em}{$\stackrel{\displaystyle >}{\sim}$}\;}

\begin{document}



\title{$B$, $D$ and $K$ decays
\footnote{Report of Working Group 2 of the CERN Workshop ``Flavour in
  the era of the LHC'', Geneva, Switzerland, November 2005 -- March 2007. }}


\author{%
{\bf Conveners}: G.~Buchalla, T.K.~Komatsubara, F.~Muheim, L.~Silvestrini\\
{\bf Section coordinators}:
New physics scenarios: A.J.~Buras, S.~Heinemeyer, G.~Isidori,
Y.~Okada, F.~Parodi, L.~Silvestrini\\
Weak decays of hadrons and QCD: G.~Buchalla\\
Radiative penguin decays: P.~Gambino, A.~Golutvin\\
Electroweak penguin decays: J.~Berryhill, Th.~Feldmann\\
Neutrino modes: Y.~Grossman, T.~Iijima\\
Very rare decays: U.~Nierste, M.~Smizanska\\
UT angles from tree decays:   M.~Bona, A.~Soni, K.~Trabelsi, G.~Wilkinson\\
B-meson mixing: V.~Lubicz, J.~van~Hunen\\
Hadronic $b\to s$ and $b \to d$ transitions: 
M.~Ciuchini, F.~Muheim\\
Kaon decays: A.J.~Buras, T.K.~Komatsubara\\
Charm physics: D.M.~Asner, S.~Fajfer\\
Prospects for future facilities: T.~Hurth\\
Assessments:  S.~Heinemeyer, F.~Parodi, L.~Silvestrini\\
{\bf Contributing authors:}\\
M.~Artuso$^1$,
D.M.~Asner$^2$,
P.~Ball$^3$,
E.~Baracchini$^4$,
G.~Bell$^5$,
M.~Beneke$^6$,
J.~Berryhill$^7$,
A.~Bevan$^8$,
I.I.~Bigi$^9$,
M.~Blanke$^{10}$,
Ch.~Bobeth$^{11}$,
M.~Bona$^{12}$,
F.~Borzumati$^{13}$,
T.~Browder$^{14}$,
T.~Buanes$^{15}$,
G.~Buchalla$^{16}$,
O.~Buchm\"uller$^{17}$,
A.J.~Buras$^{18}$,
S.~Burdin$^{19}$,
D.G.~Cassel$^{20}$,
R.~Cavanaugh$^{21}$,
M.~Ciuchini$^{22}$,
P.~Colangelo$^{23}$,
G.~Crosetti$^{24}$,
A.~Dedes$^3$,
F.~De~Fazio$^{23}$,
S.~Descotes-Genon$^{25}$,
J.~Dickens$^{26}$,
Z.~Dole\v zal$^{27}$,
S.~D\"urr$^{28}$,
U.~Egede$^{29}$,
C.~Eggel$^{30}$,
G.~Eigen$^{15}$,
S.~Fajfer$^{31}$,
Th.~Feldmann$^{32}$,
R.~Ferrandes$^{23}$,
P.~Gambino$^{33}$,
T.~Gershon$^{34}$,
V.~Gibson$^{26}$,
M.~Giorgi$^{35}$,
V.V.~Gligorov$^{36}$,
B.~Golob$^{37}$,
A.~Golutvin$^{38}$,
Y.~Grossman$^{39}$,
D.~Guadagnoli$^{18}$,
U.~Haisch$^{40}$,
M.~Hazumi$^{41}$,
S.~Heinemeyer$^{42}$,
G.~Hiller$^{11}$,
D.~Hitlin$^{43}$,
T.~Huber$^6$,
T.~Hurth$^{44}$,
T.~Iijima$^{45}$,
A.~Ishikawa$^{46}$,
G.~Isidori$^{47}$,
S.~J\"ager$^{17}$,
A.~Khodjamirian$^{32}$,
T.K.~Komatsubara$^{41}$,
P.~Koppenburg$^{29}$,
T.~Lagouri$^{27}$,
U.~Langenegger$^{48}$,
C.~Lazzeroni$^{26}$,
A.~Lenz$^{49}$,
V.~Lubicz$^{22}$,
W.~Lucha$^{50}$,
H.~Mahlke$^{20}$,
D.~Melikhov$^{51}$,
F.~Mescia$^{52}$,
M.~Misiak$^{53}$,
F.~Muheim$^{54}$,
M.~Nakao$^{41}$,
J.~Napolitano$^{55}$,
N.~Nikitin$^{56}$
U.~Nierste$^5$,
K.~Oide$^{41}$,
Y.~Okada$^{41}$,
P.~Paradisi$^{18}$,
F.~Parodi$^{57}$,
M.~Patel$^{17}$,
A.A.~Petrov$^{58}$,
T.N.~Pham$^{59}$,
M.~Pierini$^{17}$,
S.~Playfer$^{54}$,
G.~Polesello$^{60}$,
A.~Policicchio$^{24}$
A.~Poschenrieder$^{18}$,
P.~Raimondi$^{52}$,
S.~Recksiegel$^{18}$,
P.~\v Rezn\'i\v cek$^{27}$,
A.~Robert$^{61}$,
S.~Robertson$^{62}$,
J.L.~Rosner$^{63}$,
G.~Ruggiero$^{17}$,
A.~Sarti$^{52}$,
O.~Schneider$^{64}$,
F.~Schwab$^{65}$,
L.~Silvestrini$^4$,
S.~Simula$^{22}$,
S.~Sivoklokov$^{56}$,
P.~Slavich$^{66}$,
C.~Smith$^{67}$,
M.~Smizanska$^{68}$,
A.~Soni$^{69}$,
T.~Speer$^{40}$,
P.~Spradlin$^{36}$,
M.~Spranger$^{18}$,
A.~Starodumov$^{48}$,
B.~Stech$^{70}$,
A.~Stocchi$^{71}$,
S.~Stone$^{1}$,
C.~Tarantino$^{22}$,
F.~Teubert$^{17}$,
S.~T'Jampens$^{12}$,
K.~Toms$^{56}$,
K.~Trabelsi$^{41}$,
S.~Trine$^5$,
S.~Uhlig$^{18}$,
V.~Vagnoni$^{72}$,
J.J.~van~Hunen$^{64}$,
G.~Weiglein$^3$,
A.~Weiler$^{20}$,
G.~Wilkinson$^{36}$,
Y.~Xie$^{54}$,
M.~Yamauchi$^{41}$,
G.~Zhu$^{73}$,
J.~Zupan$^{31}$,
R.~Zwicky$^3$.\\
$^1$ Syracuse University, Syracuse, NY, USA\\
$^2$ Carleton University, Ottawa, Canada\\
$^3$ Durham University, IPPP, Durham, UK\\
$^4$ Universit\`a di Roma La Sapienza and INFN, Rome, Italy\\
$^5$ Universit\"at Karlsruhe, Germany\\
$^6$ RWTH Aachen, Aachen, Germany\\
$^7$ Fermi National Accelerator Laboratory, Batavia, IL, USA\\
$^8$ Queen Mary, University of London, United Kingdom\\
$^9$ University of Notre Dame, Notre Dame, IN, USA\\
$^{10}$ Technische Universit\"at M\"unchen, Garching and
        Max-Planck-Institut f\"ur Physik, M\"unchen, Germany\\
$^{11}$ Institut f\"ur Physik, Universit\"at Dortmund,  Germany\\
$^{12}$ LAPP, Universit\'e de Savoie, IN2P3-CNRS, 
        Annecy-le-Vieux, France \\
$^{13}$ ICTP, Trieste, Italy and National Central University, Taiwan\\
$^{14}$ University of Hawaii at Manoa, Honolulu, HI, USA\\
$^{15}$ University of Bergen, Norway\\
$^{16}$ Ludwig-Maximilians-Universit\"at M\"unchen, M\"unchen, Germany\\
$^{17}$ CERN, Geneva, Switzerland \\
$^{18}$ Technische Universit\"at M\"unchen, Garching, Germany\\
$^{19}$ The University of Liverpool, Liverpool, United Kingdom\\
$^{20}$ Cornell University, Ithaca, NY, USA\\
$^{21}$ University of Florida, Gainesville, FL, USA\\
$^{22}$ Universit\`a di Roma Tre and INFN, Rome, Italy\\
$^{23}$ INFN Bari, Italy\\
$^{24}$ Universit\`a di Calabria and INFN Cosenza, Italy\\
$^{25}$ LPT, CNRS/Universit\'e de Paris-Sud 11, Orsay, France\\
$^{26}$ University of Cambridge, Cambridge, United Kingdom\\
$^{27}$ IPNP, Charles University in Prague, Czech Republic\\
$^{28}$ NIC, FZ J\"ulich and DESY Zeuthen, J\"ulich, Germany\\
$^{29}$ Imperial College, London, United Kingdom\\
$^{30}$ ETH, Z\"urich and PSI, Villigen, Switzerland \\
$^{31}$ Ljubljana University and Jozef Stefan Institute, Ljubljana, Slovenia\\
$^{32}$ Universit\"at Siegen, Siegen, Germany\\
$^{33}$ Universit\`a di Torino and INFN, Torino, Italy\\
$^{34}$ University of Warwick, Coventry, United Kingdom\\
$^{35}$ Universit\`a di Pisa, SNS and INFN, Pisa, Italy\\
$^{36}$ University of Oxford, Oxford, United Kingdom\\
$^{37}$ University of Ljubljana, Slovenia\\
$^{38}$ CERN, Geneva, Switzerland and ITEP, Moscow, Russia\\
$^{39}$ Technion, Haifa, Israel\\
$^{40}$ Universit\"at Z\"urich, Z\"urich, Switzerland \\
$^{41}$ KEK and Graduate University for Advanced Studies (Sokendai), 
        Tsukuba, Japan\\
$^{42}$ IFCA, Santander, Spain\\
$^{43}$ CalTech, Pasadena, CA, USA\\
$^{44}$ CERN, Geneva, Switzerland and SLAC, Stanford, CA, USA\\
$^{45}$ Nagoya University, Nagoya, Japan\\
$^{46}$ Saga University, Saga, Japan\\
$^{47}$ SNS and INFN, Pisa and INFN, LNF, Frascati, Italy\\
$^{48}$ ETH, Z\"urich, Switzerland\\
$^{49}$ Universit\"at Regensburg, Regensburg, Germany\\
$^{50}$ Institut f\"ur Hochenergiephysik, \"Osterreichische
        Akademie der Wissenschaften, Wien, Austria\\
$^{51}$ Institut f\"ur Hochenergiephysik, \"Osterreichische
        Akademie der Wissenschaften, Wien, Austria and 
        Nuclear Physics Institute, Moscow State University, Moscow, Russia\\
$^{52}$ INFN, LNF, Frascati, Italy\\
$^{53}$ CERN, Geneva, Switzerland and Warsaw University, Warsaw, Poland\\
$^{54}$ University of Edinburgh, Edinburgh, United Kingdom\\
$^{55}$ Rensselaer Polytechnic Institute, Troy, NY, USA\\
$^{56}$ Skobeltsin Institute of Nuclear Physics,
        Lomonosov Moscow State University, Russia\\
$^{57}$ Universit\`a di Genova and INFN, Genova, Italy\\
$^{58}$ Wayne State University, Detroit, MI, USA\\
$^{59}$ Ecole Polytechnique, CNRS, Palaiseau, France\\
$^{60}$ Universit\`a di Pavia and INFN, Pavia, Italy\\
$^{61}$ Universit\'e de Clermont-Ferrand, Clermont-Ferrand, France\\
$^{62}$ McGill University and IPP, Canada \\
$^{63}$ Enrico Fermi Institute, University of Chicago,
        Chicago, IL, USA\\
$^{64}$ Ecole Polytechnique F\'ed\'erale de Lausanne (EPFL), 
        Lausanne, Switzerland\\
$^{65}$ Universitat Autonoma de Barcelona, IFAE, Barcelona, Spain\\
$^{66}$ CERN, Geneva, Switzerland and LAPTH, Annecy-le-Vieux, France\\
$^{67}$ Universit\"at Bern, Bern, Switzerland\\
$^{68}$ Lancaster University, Lancaster, United Kingdom\\
$^{69}$ Brookhaven National Laboratory, Upton, NY, USA\\
$^{70}$ Universit\"at Heidelberg, Heidelberg, Germany\\
$^{71}$ LAL, IN2P3-CNRS and Universit\'e de Paris-Sud, Orsay, France\\
$^{72}$ Universit\`a di Bologna and INFN, Bologna, Italy\\
$^{73}$ Universit{\"a}t Hamburg, Hamburg, Germany}
%
%
\maketitle





\newpage 
\begin{abstract}
The present report documents the results of
Working Group 2: $B$, $D$ and $K$ decays, of the workshop
on Flavour in the Era of the LHC, held at CERN from 
November 2005 through March 2007.  

With the advent of the LHC, we will be able to probe New Physics (NP)
up to energy scales almost one order of magnitude larger than it has
been possible with present accelerator facilities. While direct
detection of new particles will be the main avenue to establish the
presence of NP at the LHC, indirect searches will provide precious
complementary information, since most probably it will not be possible
to measure the full spectrum of new particles and their couplings
through direct production. In particular, precision measurements and
computations in the realm of flavour physics are expected to play a
key role in constraining the unknown parameters of the Lagrangian of
any NP model emerging from direct searches at the LHC.

The aim of Working Group 2 was twofold: on
one hand, to provide a coherent, up-to-date picture of the status of
flavour physics before the start of the LHC; on the other hand, to
initiate activities on the 
path towards integrating information on NP from high-$p_T$ and flavour
data. 

This report is organized as follows. In Sec.~\ref{sec:nps}, we give an
overview of NP models, focusing on a few examples that have been
discussed in some detail during the workshop, with a short description
of the available computational tools for flavour observables in NP
models. Sec.~\ref{sec:hu} contains a concise discussion of the main
theoretical problem in flavour physics: the evaluation of the relevant
hadronic matrix elements for weak decays. Sec.~\ref{sec:npbc} contains
a detailed discussion of NP effects in a set of flavour observables
that we identified as ``benchmark channels'' for NP searches. The
experimental prospects for flavour physics at future facilities are
discussed in Sec.~\ref{sec:superb}.
Finally, Sec.~\ref{sec:asm} contains some assessments on the work done
at the workshop and the prospects for future developments.

\end{abstract}

\tableofcontents




\newpage \section{New physics scenarios}
\label{sec:nps}

\subsection{Overview}

The Standard Model (SM) of electroweak and strong interactions
describes with an impressive accuracy all experimental data on
particle physics up to energies of the order of the electroweak
scale. On the other hand, we know that the SM should be viewed as an
effective theory valid up to a scale $\Lambda \sim M_W$, since,
among many other things, the SM does not contain a suitable candidate
of dark matter and it does not account for gravitational
interactions. Viewing the SM as an effective theory, however, poses a
series of theoretical questions. First of all, the quadratic
sensitivity of the electroweak scale on the cutoff calls for a low
value of $\Lambda$, in order to avoid excessive fine tuning. Second,
several of the higher dimensional operators which appear in the SM effective
Lagrangian violate the accidental symmetries of the SM. Therefore,
their coefficients must be highly suppressed in order not to clash
with the experimental data, in particular in the flavour
sector. Unless additional suppression mechanisms are present in the
fundamental theory, a cutoff around the electroweak scale is thus
phenomenologically not acceptable since it generates higher
dimensional operators with large coefficients. 

We are facing a formidable task: formulating a natural extension of
the SM with a cutoff close to the electroweak scale and with a very
strong suppression of additional sources of flavour and CP violation.
While the simplest supersymmetric extensions of the SM with minimal
flavour and CP violation, such as Minimal Supergravity (MSUGRA)
models, seem to be the phenomenologically most viable NP options, it
is fair to say that a fully consistent model of SUSY breaking has not
been put forward yet. On the other hand, alternative solutions of the
hierarchy problem based on extra dimensions have recently become very
popular, although they have not yet been tested at the same level of
accuracy as the Minimal Supersymmetric Standard Model (MSSM). Waiting
for the LHC to discover new particles and shed some light on these
fundamental problems, we should consider a range of NP models as wide
as possible, in order to be ready to interpret the NP signals that
will show up in the near future.

In the following paragraphs, we discuss how flavour and CP violation
beyond the SM can be analyzed on general grounds in a
model-independent way. We then specialize to a few popular extensions
of the SM, such as SUSY and little Higgs models, and present their
most relevant aspects in view of our subsequent discussion of 
NP effects in flavour physics.

\newpage \subsection{Model-independent approaches}

\subsubsection{General considerations}
  
In most extensions of the Standard Model (SM), 
the new degrees of freedom that modify the ultraviolet 
behavior of the theory appear only around or above
the electroweak scale ($v \approx 174$~GeV). As long as we are interested 
in processes occurring below this scale (such as $B$, $D$ and $K$ 
decays), we can integrate out the new degrees of freedom 
and describe the new-physics effects --in full 
generality-- by means of an Effective Field Theory (EFT) approach.
The SM Lagrangian becomes the renormalizable part of a more general 
local Lagrangian which includes an infinite tower of higher-dimensional 
operators, constructed in terms of SM fields and 
suppressed by inverse powers of a scale 
$\Lambda_{\rm NP} > v$. 
 
This general bottom-up approach allows us to analyse all realistic 
extensions of the SM in terms of a limited number of parameters 
(the coefficients of the higher-dimensional operators). 
The disadvantage of this strategy is that  
it does not allow us to establish correlations 
of New Physics (NP) effects at low and high energies (the scale
$\Lambda_{\rm NP}$ defines the cut-off of the EFT). 
The number of correlations among different low-energy 
observables is also very limited, unless some restrictive 
assumptions about the structure of the EFT are employed. 

The generic EFT approach is somehow the 
opposite of the standard top-down strategy towards NP, 
where a given theory --and a specific set of parameters-- 
are employed to evaluate possible deviations from the SM.  
The top-down approach usually allows us to establish several correlations,
both at low energies and between low- and high-energy observables. 
However, the price to pay is the loss of generality. This is 
quite a high price given our limited knowledge about the physics 
above the electroweak scale. 

An interesting compromise between these two extreme strategies 
is obtained by implementing specific symmetry restrictions 
on the EFT. The extra constraints increase the number of 
correlations in low-energy observables. The experimental 
tests of such correlations allow us to test/establish 
general features of the NP model (possibly valid both 
at low and high energies). In particular, 
$B$, $D$ and $K$ decays are extremely useful in determining 
the flavour-symmetry breaking pattern of the NP model. 
The EFT approaches based on the Minimal Flavour Violation (MFV) 
hypothesis and its variations 
(MFV at large $\tan\beta$, n-MFV,~\ldots)
have exactly this goal. 

In Sect.~\ref{sect:flav_prob} we illustrate 
some of the main conclusions about NP effects 
in the flavour sector derived so far
within general EFT approaches. 
In Sect.~\ref{sect:MFV} we analyse in more
detail the MFV hypothesis, discussing:
i) the general formulation and the general consequences
of this hypothesis; 
ii) the possible strategies 
to verify or falsify the MFV assumption from low-energy data;
iii) the implementation of the MFV hypothesis in 
more explicit beyond-the-SM frameworks, such as the Minimal 
Supersymmetric SM (MSSM) or Grand Unified Theories (GUTs).

\subsubsection{Generic EFT approaches and the flavour problem}
\label{sect:flav_prob}

The NP contributions to the higher-dimensional operators of the 
EFT  should naturally induce large effects 
in processes which are not mediated by tree-level SM amplitudes, 
such as meson-antimeson mixing ($\Delta F=2$ amplitudes)
or flavour-changing neutral-current (FCNC) rare decays.
Up to now there is no evidence of deviations from the SM in these
processes and this 
implies severe bounds on the effective scale of various dimension-six
operators. For instance, the good agreement between SM 
expectations and experimental determinations of $K^0$--${\bar K}^0$ 
mixing leads to bounds above $10^4$~TeV for the effective scale 
of $\Delta S=2$ operators, i.e.~well above the few TeV 
range suggested by a natural stabilization of the  
electroweak-symmetry breaking mechanism. 
Similar bounds are obtained for the scale of operators 
contributing to lepton-flavour violating (LFV) 
transitions in the lepton sector, such as $\mu\to e\gamma$.

The apparent contradiction between these 
two determinations of  $\Lambda$ is a manifestation of what in 
many specific frameworks (supersymmetry, technicolour, etc.)
goes under the name of {\em flavour problem}:
if we insist on the theoretical prejudice that new physics has to 
emerge in the TeV region, we have to conclude that the new theory 
possesses a highly non-generic flavour structure. 
Interestingly enough, this structure has not been clearly identified yet,
mainly because the SM (the low-energy 
limit of the new theory), doesn't possess an exact flavour symmetry.
Within a model-independent approach, we should try to deduce this structure from data, 
using the experimental information on FCNC
transitions to constrain its form. 

\paragraph{Bounds on $\Delta F=2$ operators}

\begin{figure}[t]
\begin{center}
\includegraphics[width=70mm]{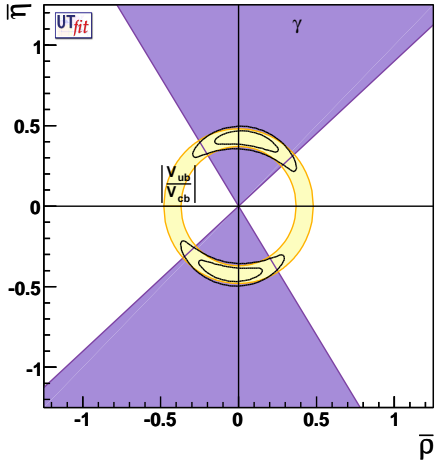}
\caption{Constraints on the $\bar \rho$--$\bar \eta$ plane 
using tree-level observables only, from Ref.~\cite{Bona:2006sa}
(see also Ref.~\cite{Charles:2004jd}). }
\label{fig:UT_tree}
\end{center}
\end{figure}

In most realistic NP models we can safely neglect NP effects in all
cases where the corresponding effective operator is generated at the
tree-level within the SM. This general assumption implies that the
experimental determination of $\gamma$ and $|V_{ub}|$ via tree-level
processes (see Fig.~{\ref{fig:UT_tree}})  is free from the
contamination of NP contributions.  The comparison of the experimental
data on meson-antimeson mixing amplitudes (both magnitudes and phases)
with the theoretical SM expectations (obtained by means of the
tree-level determination of the CKM matrix) allows to derive some of
the most stringent constraints on NP models.

\begin{figure}[t]
\begin{center}
\includegraphics[width=70mm]{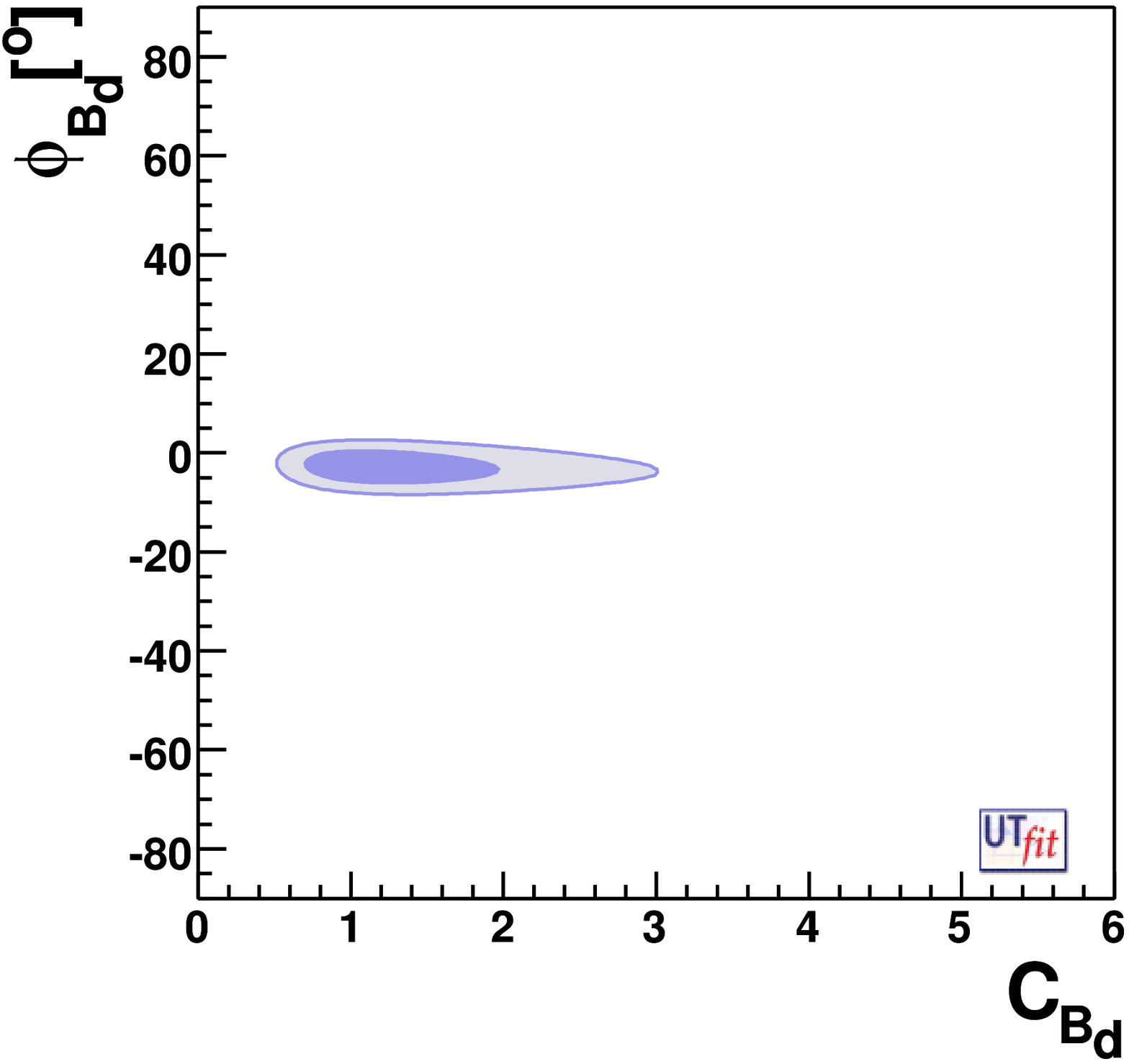}
\includegraphics[width=66mm]{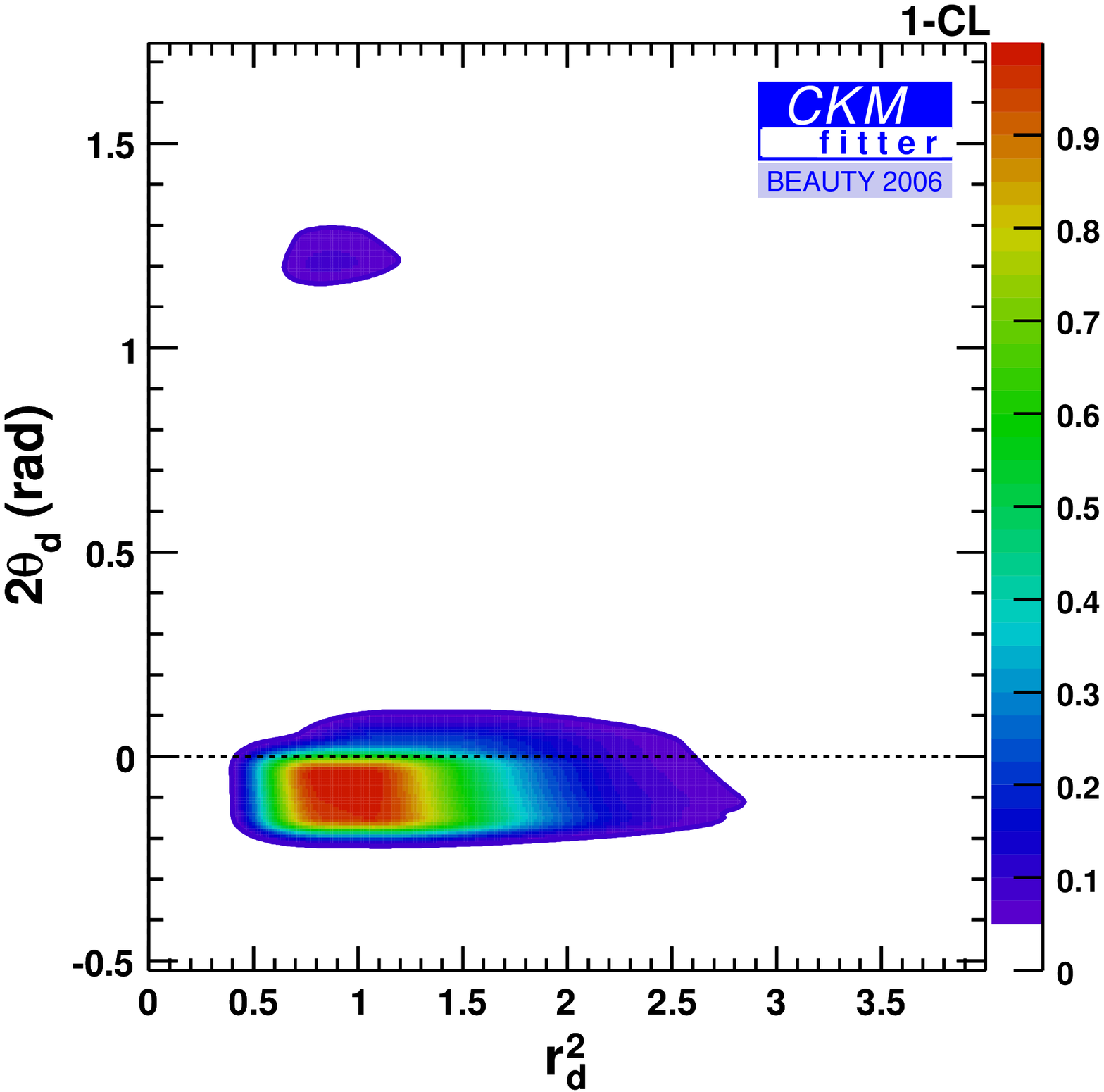}
\caption{%
  Constraints on the effective parameters encoding NP effects in the
  $B_d$--$\bar B_d$ mixing amplitude (magnitude and phase) obtained by
  the UTfit \cite{Bona:2006sa} (left) and CKMfitter
  \cite{Charles:2004jd} (right) collaborations.}
\label{fig:npCphi}
\end{center}
\end{figure}

In a wide class of beyond-the-SM scenarios we expect 
sizable and uncorrelated deviations from the SM in the 
various $\Delta F=2$ amplitudes.\footnote{~As discussed for instance
in Ref.~\cite{Fleischer:2003xx}, there is a rather general limit 
where NP effects in $\Delta F=2$ amplitudes are expected to 
be the dominant deviations from the SM in the flavour sector. 
This happens under the following two general 
assumptions: i) the effective scale 
of NP is substantially higher than the electroweak scale;
ii) the dimensionless effective couplings
ruling $\Delta F=2$ transitions can be expressed
as the square of the corresponding $\Delta F=1$ coupling,
without extra suppression factors.}
As discussed by several 
authors~\cite{Soares:1992xi,Deshpande:1996yt,Silva:1996ih,Cohen:1996sq,Grossman:1997dd},
in this case NP effects can be parameterized in terms of the shift induced
in the $B_q$--$\bar B_q$ mixing frequencies ($q=d,s$) and in the corresponding 
CPV phases,
\begin{equation} 
\frac{\langle
    B_q|H_\mathrm{eff}^\mathrm{full}|\bar{B}_q\rangle} {\langle
    B_q|H_\mathrm{eff}^\mathrm{SM}|\bar{B}_q\rangle}=
    C_{B_q}  e^{2 i \phi_{B_q}} = r_q^2 e^{2 i \theta_{q}}~,
    \label{eq:paranp}
\end{equation}
and similarly for the neutral kaon system.
The two equivalent parameterizations [$(C_{B_q},\phi_{B_q})$ or
$(r_q,\theta_q)$] have been shown to facilitate the 
interpretation of the results of the UTfit~\cite{Bona:2006sa}
and CKMfitter~\cite{Charles:2004jd} collaborations
for the $B_d$ case, shown in Fig.~\ref{fig:npCphi}.

\begin{figure}[t]
\begin{center}
\includegraphics[width=70mm]{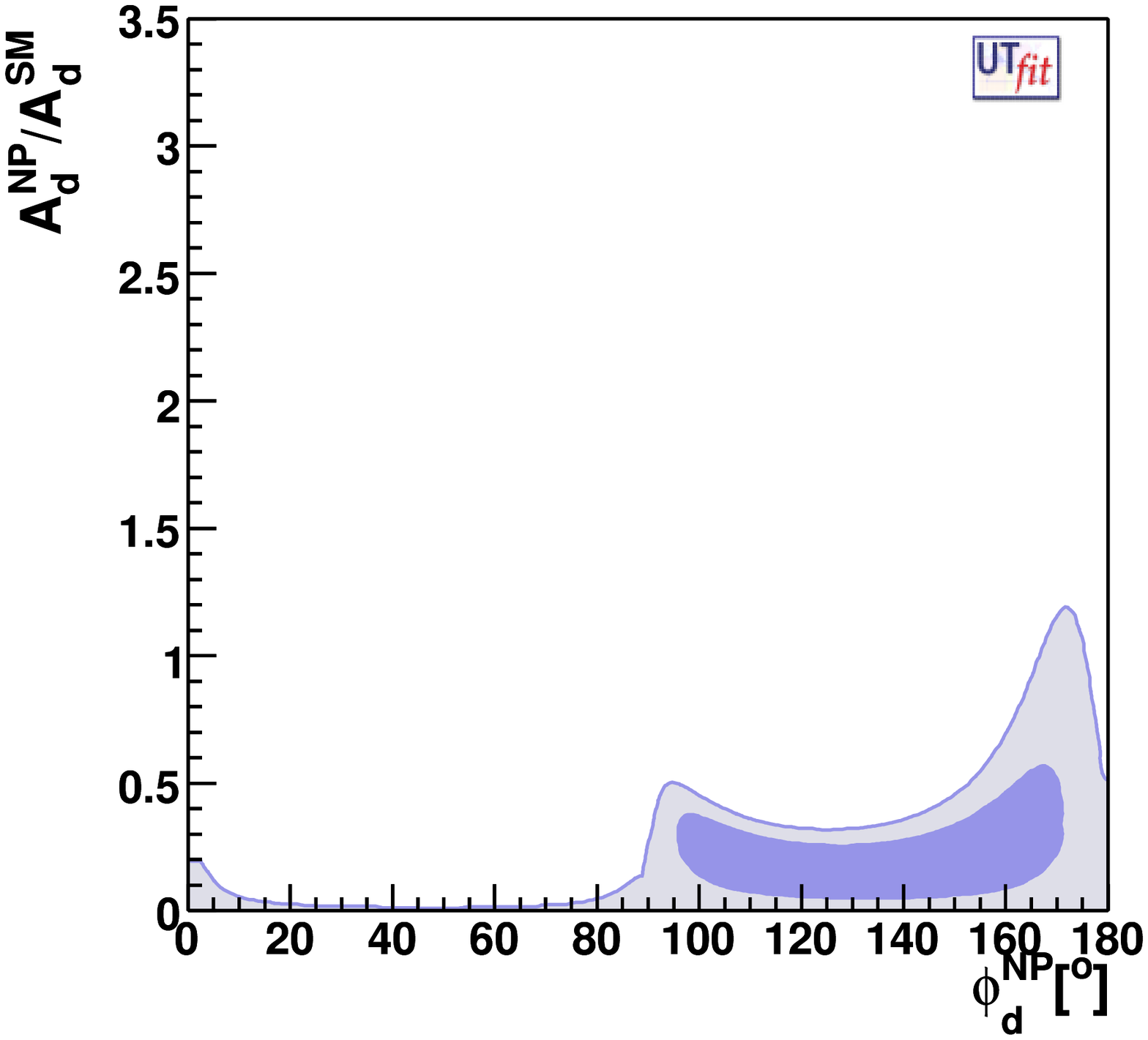}
\includegraphics[width=70mm]{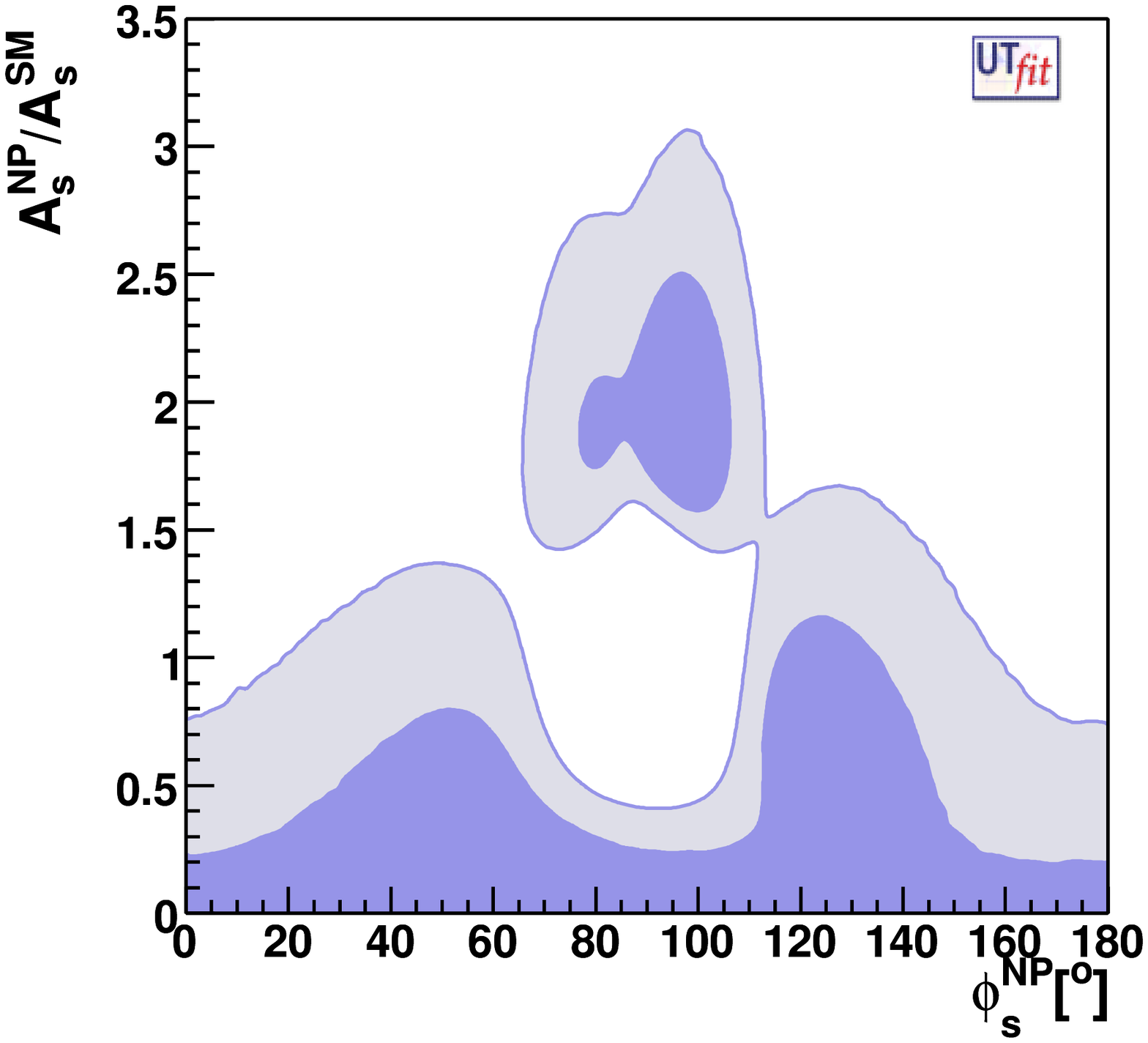}
\caption{%
  Constraints on the absolute value and phase (normalized to the SM)
  of the new physics amplitude in $B_d$--$\bar B_d$ and $B_s$--$\bar
  B_s$  mixing from ref.~\cite{Bona:2007vi}.}
\label{fig:ANP}
\end{center}
\end{figure}

The main conclusions that can be drawn form the 
present analyses of new-physics effects in $\Delta F=2$ 
amplitudes can be summarized as follows: 
\begin{itemize}
\item 
In all the three accessible short-distance amplitudes       
($K^0$--$\bar K^0$, $B_d$--$\bar B_d$, and $B_s$--$\bar B_s$)
the magnitude of the new-physics amplitude cannot exceed, in size, 
the SM short-distance contribution. The latter is suppressed both 
by the GIM mechanism and by the hierarchical structure of the 
CKM matrix ($V$):
\beq
\cA^{\Delta F=2}_{\rm SM} \sim  \frac{ G_F^2 M_W^2 }{2 \pi^2} ~ \left(V_{ti}^* V_{tj} \right)^2  
   \langle \bar  M |  (\bar Q_L^i \gamma^\mu Q_L^j )^2  | M \rangle 
\eeq
Therefore, new-physics models with TeV-scale flavoured 
degrees of freedom and $\cO(1)$ flavour-mixing couplings are 
essentially ruled out. To quantify this statement, we report here the
results of the recent analysis of ref.~\cite{Bona:2007vi}. Writing 
\beqa
&& \cA^{\Delta F=2}_{\rm NP} \sim  \frac{C^k_{ij}}{\Lambda^2} 
   \langle \bar  M | (\bar Q^i \Gamma^k Q^j )^2  | M \rangle\,,
   \no \\
\eeqa
where $\Gamma^k$ is a generic Dirac and colour structure (see
ref.~\cite{Bona:2007vi} for details), one has\footnote{The choice $\Gamma^4=P_L
  \otimes P_R$ gives the most stringent constraints. Constraints from
  other operators are up to one order of magnitude weaker.}
\beqa
&&  
\Lambda
>  \left\{ \ba{l}  
2\times 10^5~{\rm TeV} \times |C^4_{12}|^{1/2}  \\ 
2\times 10^3~{\rm TeV} \times |C^4_{13}|^{1/2}  \\
3\times 10^2~{\rm TeV} \times |C^4_{23}|^{1/2}  \ea \right. 
\no
\eeqa
\item 
As clearly shown in Fig.~\ref{fig:ANP}, in the 
$B_d$--$\bar B_d$ case there is still room 
for a new-physics contribution up to
the SM one. However, this is possible only 
if the new-physics contribution is aligned in phase 
with respect to the SM amplitude ($\phi^\mathrm{NP}_{d}$
close to zero). 
Similar, but thighter, constraints hold also for the new physics
contribution to the $K^0$--$\bar K^0$
amplitude.
\item
Contrary to $B_d$--$\bar B_d$ and $K^0$--$\bar K^0$
amplitudes, at present there is only a very  loose bound on the 
CPV phase of the  $B_s$--$\bar B_s$ mixing amplitude. 
This leaves open the possibility of observing a large 
$\cA_{\rm CP}(B_s \to J/\Psi \phi)$ at LHCb, which 
would be a clear signal of physics beyond the SM.
\end{itemize}
As we will discuss in the following, the 
first two items listed above find a natural
explanation within the so-called hypothesis of 
Minimal Flavour Violation. 

\subsubsection{Minimal Flavour Violation}
\label{sect:MFV}

A very reasonable, although quite pessimistic, solution
to the flavour problem is the so-called 
Minimal Flavour Violation (MFV) hypothesis.
Under this assumption, which will be formalized  
in detail below, flavour-violating 
interactions are linked to the
known structure of Yukawa couplings also beyond the SM. 
As a result, non-standard contributions in FCNC 
transitions turn out to be suppressed to a level consistent 
with experiments even for $\Lambda \sim$~few TeV.
One of the most interesting aspects of the MFV hypothesis 
is that it can naturally be implemented within the 
EFT approach to NP~\cite{D'Ambrosio:2002ex}.
The effective theories based on this symmetry principle
allow us to establish unambiguous correlations 
among NP effects in various rare decays.
These falsifiable predictions are the key ingredients   
to identify in a model-independent way which are the 
irreducible sources of flavour symmetry breaking.

\paragraph{The MFV hypothesis}
\label{sect:qMFV}
The pure gauge sector of the SM is invariant under
a large symmetry group of flavour transformations: 
${\mathcal G}_{\rm SM} = {\mathcal G}_{q} \otimes 
{\mathcal G}_{\ell} \otimes U(1)^5$,
where 
\beq
{\mathcal G}_{q}
= {\rm SU}(3)_{Q_L}\otimes {\rm SU}(3)_{U_R} \otimes {\rm SU}(3)_{D_R},
\qquad 
{\mathcal G}_{\ell}
=  {\rm SU}(3)_{L_L} \otimes {\rm SU}(3)_{E_R}
\eeq
and three of the five $U(1)$ charges can be identified with 
baryon number, lepton number and hypercharge  \cite{Chivukula:1987py}.
This large group and, particularly the ${\rm SU}(3)$ 
subgroups controlling flavour-changing transitions, is 
explicitly broken by the Yukawa interaction
\beq
\cL_Y  =   {\bar Q}_L \yuk_D D_R  H
+ {\bar Q}_L {\yuk_U} U_R  H_c
+ {\bar L}_L {\yuk_E} E_R  H {\rm ~+~h.c.}
\label{eq:LYY}
\eeq
Since ${\mathcal G}_{\rm SM}$ is already broken within the SM, 
it would not be consistent to impose it as an exact symmetry 
beyond the SM: even if absent a the tree-level,
the breaking of ${\mathcal G}_{\rm SM}$ would reappear at the quantum level 
because of the Yukawa interaction.  
The most restrictive hypothesis 
we can make to {\em protect} in a consistent way flavour mixing 
in the quark sector, is to assume that $\yuk_D$ and $\yuk_U$ are the only 
sources of  ${\mathcal G}_{q}$ breaking also beyond the SM.
To implement and interpret this hypothesis in a consistent way, 
we can assume that ${\mathcal G}_{q}$ is indeed a good symmetry, promoting 
$\yuk_{U,D}$ to be non-dynamical fields (spurions) with 
non-trivial transformation properties under this symmetry
\beq
\yuk_U \sim (3, \bar 3,1)_{{\mathcal G}_{q}}~,\qquad
\yuk_D \sim (3, 1, \bar 3)_{{\mathcal G}_{q}}~.\qquad
\eeq
If the breaking of the symmetry occurs at very high energy scales 
 --well above the TeV region where the new degrees of freedom 
necessary to stabilize the Higgs sector should appear--  
at low-energies we would only be sensitive to the background values of 
the $\yuk$, i.e. to the ordinary SM Yukawa couplings. 
Employing the effective-theory language, 
we then define that an effective theory satisfies the criterion of
Minimal Flavour Violation in the quark sector if all higher-dimensional operators,
constructed from SM and $\yuk$ fields, are invariant under CP and (formally)
under the flavour group ${\mathcal G}_{q}$ \cite{D'Ambrosio:2002ex}.

According to this criterion one should in principle 
consider operators with arbitrary powers of the (dimensionless) 
Yukawa fields. However, a strong simplification arises by the 
observation that all the eigenvalues of the Yukawa matrices 
are small, but for the top one, and that the off-diagonal 
elements of the CKM matrix ($V_{ij}$) are very suppressed. 
Using the  ${\mathcal G}_{q}$
symmetry, we can rotate the background values of the auxiliary fields $\yuk$
such that
\beq
\yuk_D = \diagyuk_d~, \qquad
\yuk_U =  V^\dagger \diagyuk_u~,
\label{eq:d-basis}
\eeq
where $\diagyuk$ are diagonal matrices
and $V$ is the CKM matrix.
It is then easy to realize that, similarly to the pure SM case, 
the leading coupling ruling all FCNC transitions 
with external down-type quarks is:
\beq
\label{eq:FC}
(\lambda_{\rm FC})_{ij} = \left\{ \ba{ll} \left( \yuk_U \yuk_U^\dagger \right)_{ij}
\approx \lambda_t^2  V^*_{3i} V_{3j}~ &\qquad i \not= j~, \\
0 &\qquad i = j~. \ea \right.
\eeq
The number of relevant dimension-6 effective 
operators is then strongly reduced (representative examples
are reported in Table~\ref{tab:MFV}, while the complete list can 
be found in Ref.~\cite{D'Ambrosio:2002ex}).

\renewcommand\arraystretch{1.2}
\begin{table}
$$
\begin{array}{lc|ccc}
\multicolumn{1}{c}{\hbox{MFV dim-6 operator}} &\hbox{Main observables}
&\multicolumn{2}{c}{\Lambda\hbox{ [TeV]}}&\\
\hline
 \frac{1}{2} (\bar Q_L  \yuk_U \yuk_U^\dagger \gamma_{\mu} Q_L)^2 
\phantom{X^{X^X}}
&\epsilon_K,~\Delta m_{B_d},~\Delta m_{B_s}       &\ 5.9\  [+]\ &\ 8.8\  [-]  \\
    e H^\dagger \left( {\bar D}_R \yuk_D^\dagger \yuk_U \yuk_U^\dagger \sigma_{\mu\nu}
Q_L \right) F_{\mu\nu}   &
B\to X_s \gamma     &\ 5.0\ [+]\ &\ 9.0\ [-] \\
  (\bar Q_L  \yuk_U \yuk_U^\dagger \gamma_{\mu}   Q_L)(\bar L_L \gamma_\mu L_L )  
&B\to (X) \ell\bar{\ell},\quad K\to \pi \nu\bar{\nu},(\pi) \ell \bar{\ell} \quad
  &\ 3.7\ [+]\ &\  3.2\ [-]  & \\
  (\bar Q_L  \yuk_U \yuk_U^\dagger \gamma_{\mu} Q_L)(H^\dagger i D_\mu H)\qquad  
&B\to(X) \ell\bar{\ell},\quad K\to \pi \nu\bar{\nu},(\pi) \ell \bar{\ell} \quad
 &\  2.0\  [+]\ &\  2.0\ [-] &  \\
\end{array}$$
\caption[X]{\label{tab:MFV}  $95\%$ {\rm CL} bounds on the scale
of representative dimension-six operators in the MFV scenario.
The constraints are obtained on the single operator, with coefficient $\pm1/\Lambda^2$
($+$ or $-$ denote constructive or destructive interference with the SM amplitude). }
\end{table}
\renewcommand\arraystretch{1}

\paragraph{Universal UT and MFV bounds on the effective operators}

\begin{figure}[t]
\begin{center}
\includegraphics[width=70mm]{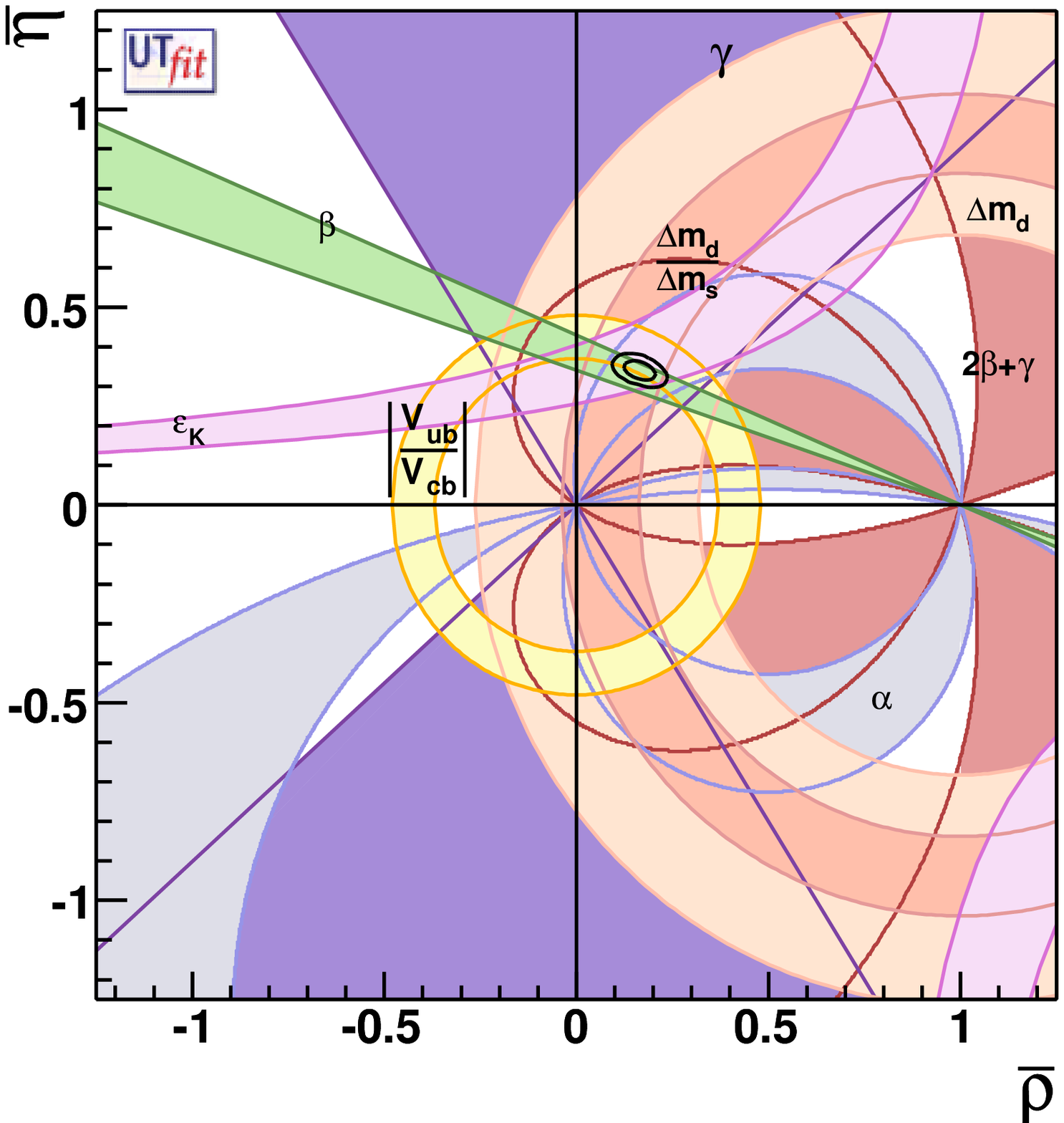}
\includegraphics[width=70mm]{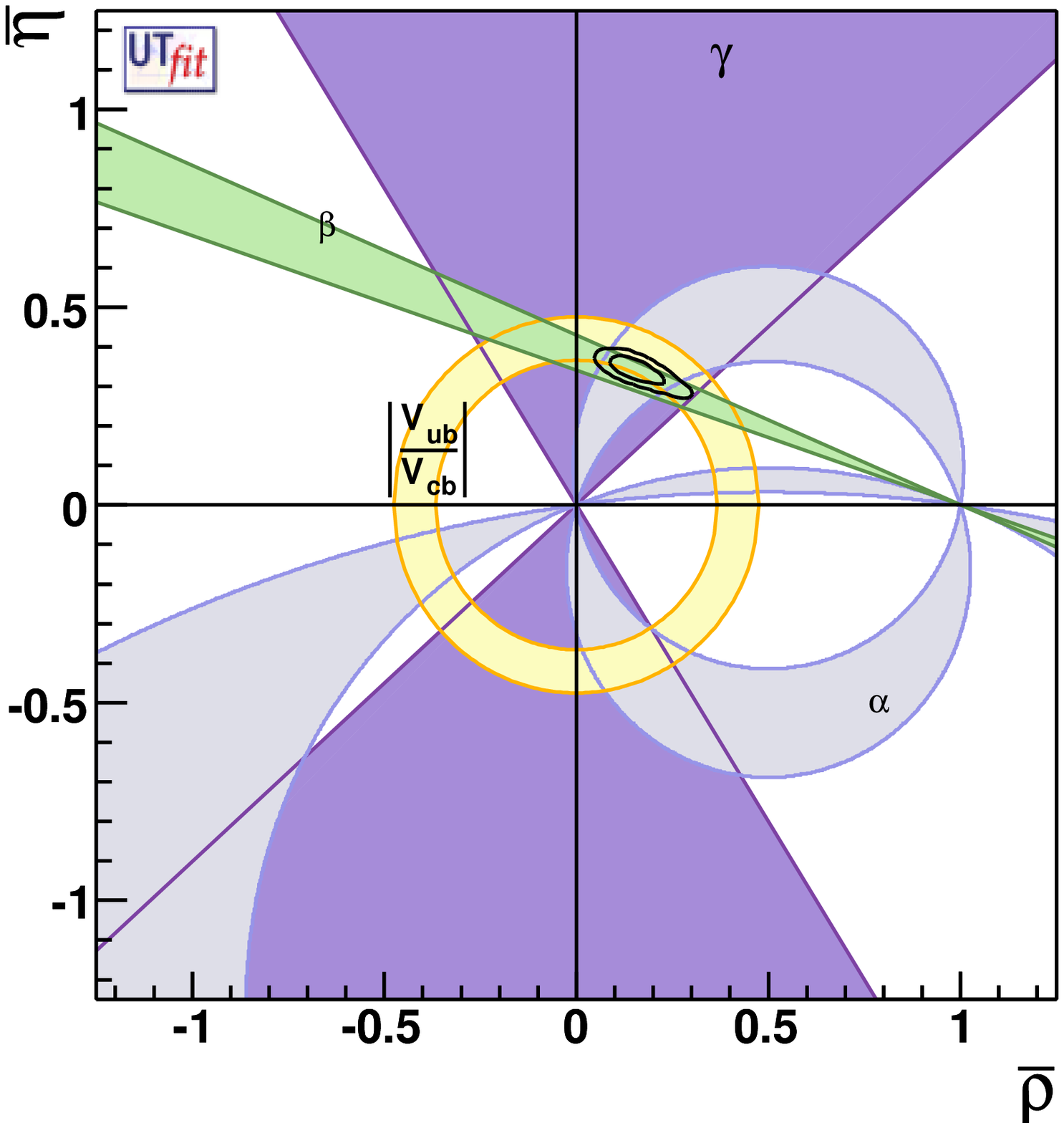}
\caption{\label{fig:UTfits} Fit of the CKM unitarity triangle within the SM (left) and 
in generic extensions of the SM satisfying the MFV hypothesis (right)~\cite{Bona:2006sa}. }
\end{center}
\end{figure}

As originally pointed out in Ref.~\cite{Buras:2000dm}, within the MFV
framework several of the constraints used to determine the CKM matrix
(and in particular the unitarity triangle) are not affected by NP.
In this framework, NP effects are negligible not only in tree-level
processes but also in a few clean observables sensitive to loop
effects, such as the time-dependent CPV asymmetry in $B_d \to J/\Psi
K_{L,S}$. Indeed the structure of the basic flavour-changing coupling
in Eq.~(\ref{eq:FC}) implies that the weak CPV phase of $B_d$--$\bar
B_d$ mixing is arg[$(V_{td}V_{tb}^*)^2$], exactly as in the SM.  The
determination of the unitarity triangle using only these clean
observables (denoted Universal Unitarity Triangle) is shown in
Fig.~\ref{fig:UTfits}.\footnote{The UUT as originally proposed in
  Ref.~\cite{Buras:2000dm} includes $\Delta M_{B_d}/\Delta M_{B_s}$
  and is therefore valid only in models of CMFV (see
  Sec.~\ref{sec:CMFV}). On the other hand, removing $\Delta
  M_{B_d}/\Delta M_{B_s}$ from the analysis gives a UUT that is valid
  in any MFV scenario.}  This construction provides a natural (a
posteriori) justification of why no NP effects have been observed in
the quark sector: by construction, most of the clean observables
measured at $B$ factories are insensitive to NP effects in this
framework.

In  Table~\ref{tab:MFV} we report a few representative 
examples of the bounds on the higher-dimen\-sio\-nal operators
in the MFV framework.
As can be noted, the built-in CKM suppression
leads to bounds on the effective scale of new physics 
not far from the TeV region. These bounds are very similar to the 
bounds on flavour-conserving operators derived by precision electroweak tests. 
This observation reinforces the conclusion that a deeper study of 
rare decays is definitely needed in order to clarify 
the flavour problem: the experimental precision on the clean 
FCNC observables required to obtain bounds more stringent 
than those derived from precision electroweak tests
(and possibly discover new physics) is typically
in the $1\%-10\%$ range.

Although the MFV seems to be a natural solution to the flavour problem, 
it should be stressed that we are still
very far from having proved the validity of this hypothesis from data.
A proof of the MFV hypothesis 
can be achieved only with a positive evidence of physics beyond 
the SM exhibiting the flavour pattern (link between $s\to d$, $b\to d$, and  
$b\to s$ transitions) predicted by the MFV assumption.

\paragraph{Comparison with other approaches (CMFV \& n-MFV)}
\label{sec:CMFV}

The idea that the CKM matrix rules the strength of FCNC 
transitions also beyond the SM has become a very popular 
concept in the recent literature and has been implemented 
and discussed in several works 
(see e.g.~Refs.~\cite{Ali:1999we,Buras:2000dm,Bartl:2001wc,Laplace:2002ik,Buras:2003jf}).

It is worth stressing that the CKM matrix 
represents only one part of the problem: a key role in
determining the structure of FCNCs  is also played  by quark masses, 
or by the Yukawa eigenvalues. In this respect, the MFV 
criterion illustrated above provides the maximal protection 
of FCNCs (or the minimal violation of flavour symmetry), 
since the full structure of Yukawa matrices is preserved. 
At the same time, this criterion is based on a renormalization-group-invariant 
symmetry argument. Therefore, it can be implemented 
independently of any specific hypothesis about the dynamics 
of the new-physics framework. The only two assumptions are:
i) the flavour symmetry and the sources of its breaking; 
ii) the number of light degrees of freedom of the theory 
(identified with the SM fields in the minimal case).
 
This model-independent structure does not hold in 
most of the alternative definitions of MFV models 
that can be found in the literature. For instance, 
the definition of Ref.~\cite{Buras:2003jf} 
(denoted constrained MFV, or CMFV)
contains the additional requirement that the 
effective FCNC operators playing a significant 
role within the SM are the only relevant ones 
also beyond the SM. 
This condition is realized within weakly coupled 
theories at the TeV scale with only one light Higgs doublet, 
such as the model with universal extra dimensions 
analysed in Ref.~\cite{Buras:2003mk}, or the MSSM 
with small $\tan\beta$ and small $\mu$ term.
However, it does not hold in other frameworks, such as 
technicolour models, or the MSSM with large 
$\tan\beta$ and/or large $\mu$ term 
(see Sect.~\ref{sect:MSSM_MFV}),
whose low-energy phenomenology could still be described 
using the general MFV criterion discussed 
in Sect.~\ref{sect:qMFV}.

Since we are still
far from having proved the validity of the MFV hypothesis from data, 
specific less restrictive symmetry assumptions about the flavour-structure 
of NP can also be considered. Next-to-minimal MFV frameworks have
recently been discussed in Ref.~ \cite{Agashe:2005hk,Feldmann:2006jk}.
As shown in Ref.~\cite{Feldmann:2006jk}, a convenient way to systematically 
analyse the possible deviations from the MFV ansatz is to introduce 
additional spurions of the ${\mathcal G}_{\rm SM}$ group.

\paragraph{MFV at large $\tan\beta$}
\label{eq:largetanb}
If the Yukawa Lagrangian contains only one Higgs field,
as in Eq.~(\ref{eq:LYY}), it
necessarily breaks both ${\mathcal G}_{q}$ and two of the 
$U(1)$ subgroups of ${\mathcal G}_{\rm SM}$.
In models with more than one Higgs doublet,
the breaking mechanisms of ${\mathcal G}_{q}$ and the 
$U(1)$ symmetries can be decoupled, allowing a different overall 
normalization of the $Y_{U,D}$ spurions with respect to the 
SM case.

A particularly interesting scenario
is the two-Higgs-doublet model where 
the two Higgses are coupled separately to up-
and down-type quarks:
\beq
\cL_{Y_0}  =   {\bar Q}_L \yuk_D D_R  H_D
+ {\bar Q}_L \yuk_U U_R  H_U
+ {\bar L}_L \yuk_E E_R  H_D {\rm ~+~h.c.}
\label{eq:LY2}
\eeq
This Lagrangian is invariant under a  ${\rm U}(1)$
symmetry, denoted ${\rm U}(1)_{\rm PQ}$, whose
only charged fields are $D_R$ and $E_R$ 
(charge $+1$)  and  $H_D$ (charge $-1$).
The ${\rm U}_{\rm PQ}$ symmetry prevents tree-level
FCNCs and implies that $Y_{U,D}$ are the only 
sources of ${\mathcal G}_{q}$ breaking appearing in the Yukawa
interaction (similar to the one-Higgs-doublet
scenario). Coherently with the MFV hypothesis, 
in order to protect the good agreement between data and SM 
in FCNCs and $\Delta F=2$ amplitudes, we assume that
$Y_{U,D}$ are the only relevant 
sources of ${\mathcal G}_{q}$ breaking appearing 
in all the low-energy effective operators. 
This is sufficient to ensure that flavour-mixing 
is still governed by the CKM matrix, and naturally guarantees
a good agreement with present data in the $\Delta F =2$
sector. However, 
the extra symmetry of the Yukawa interaction allows
us to change the overall normalization of 
$Y_{U,D}$ with interesting phenomenological consequences
in specific rare modes. 

The normalization of the  Yukawa couplings is controlled
by  $\tan\beta = \langle H_U\rangle/\langle H_D\rangle$.
For $\tan\beta \gg 1 $ the smallness of the $b$ quark 
and $\tau$ lepton masses can be attributed to the smallness 
of $1/\tan\beta$ rather than to the corresponding Yukawa couplings.
As a result, for $\tan\beta \gg 1$ we cannot anymore neglect 
the down-type  Yukawa coupling. 
In this scenario the determination of the effective
low-energy Hamiltonian relevant to FCNC processes
involves the following three steps:
\begin{itemize}
\item{} construction of the gauge-invariant basis of
dimension-six operators (suppressed by $\Lambda^{-2}$)
in terms of SM fields and two Higgs doublets;
\item{} breaking of ${\rm SU}(2)\times {\rm U}(1)_Y$ and
integration of the  $\cO(M_H^2)$ heavy Higgs fields;
\item{} integration of the $\cO(M_W^2)$ SM degrees
of freedom (top quark and electroweak gauge bosons).
\end{itemize}
These steps are well separated if we assume the
scale hierarchy $\Lambda \gg M_H \gg M_W$.
On the other hand, if $\Lambda \sim M_H$, the first
two steps can be joined, resembling the
one-Higgs-doublet scenario discussed before.
The only difference is that now, at large $\tan\beta$,
$\yuk_D$ is not negligible and this requires to enlarge
the basis of effective dimension-six operators.
From the phenomenological 
point of view, this implies the breaking of the strong MFV 
link between $K$- and $B$-physics FCNC amplitudes 
occurring in the  one-Higgs-doublet case \cite{D'Ambrosio:2002ex}. 

\medskip 

A more substantial modification of the one-Higgs-doublet
case occurs if we allow sizable sources of 
 ${\rm U}(1)_{\rm PQ}$ breaking. It should be pointed out 
that the ${\rm U}(1)_{\rm PQ}$ symmetry cannot be exact:
it has to be broken at least in the scalar potential
in order to avoid the presence of a massless pseudoscalar Higgs.
Even if the breaking of  ${\rm U}(1)_{\rm PQ}$  and 
 ${\mathcal G}_{q}$ are decoupled, the presence of 
${\rm U}(1)_{\rm PQ}$ breaking sources can have important 
implications on the  structure of the Yukawa interaction.
We can indeed consider new dimension-four operators such as 
\beq
 \epsilon {\bar Q}_L  \yuk_D D_R  (H_U)^c
\qquad {\rm or} \qquad 
 \epsilon {\bar Q}_L  \yuk_U\yuk_U^\dagger \yuk_D D_R  (H_U)^c~,
\label{eq:O_PCU}
\eeq
where $\epsilon$ denotes a generic ${\mathcal G}_q$-invariant
${\rm U}(1)_{\rm PQ}$-breaking source. Even if $\epsilon \ll 1 $, the product
$\epsilon \times  \tan\beta$ can be $\cO(1)$, 
inducing $\cO(1)$ non-decoupling corrections to $\cL_{Y_0}$. 
As discussed in specific supersymmetric scenarios,
for $\epsilon \tan\beta = \cO(1)$ the ${\rm U}(1)_{\rm PQ}$-breaking
terms induce $\cO(1)$ corrections to the down-type Yukawa
couplings~\cite{Hall:1993gn}, the 
CKM matrix elements~\cite{Blazek:1995nv},
and the charged-Higgs 
couplings~\cite{Carena:1999py,Degrassi:2000qf,Carena:2000uj}.
Moreover, sizable FCNC couplings of the down-type quarks to the
heavy neutral Higgs fields 
are
allowed~\cite{Hamzaoui:1998nu,Choudhury:1998ze,Babu:1999hn,Isidori:2001fv,Buras:2002vd,Buras:2002wq}.  
All these effects can be taken into account to all orders
with a proper re-diagonalization of the effective Yukawa 
interaction~\cite{D'Ambrosio:2002ex}.

\medskip 

Since the $b$-quark Yukawa coupling becomes $\cO(1)$, 
the large-$\tan\beta$ regime is particularly interesting 
for helicity-suppressed observables in $B$ physics. 
One of the clearest phenomenological 
consequences is a suppression (typically in the $10-50\%$ range)
of the $B \to \ell \nu$ decay
rate with respect to its SM expectation \cite{Hou:1992sy}.
Potentially measurable effects in the $10-30\%$ range
are expected also in $B\to X_s \gamma$ and $\Delta M_{B_s}$. 
The most striking signature could arise from the 
rare decays $B_{s,d}\to \ell^+\ell^-$, 
whose rates could be enhanced over the SM expectations 
by more than one order of magnitude. 
An enhancement of both $B_{s}\to \ell^+\ell^-$ and 
$B_{d}\to \ell^+\ell^-$ respecting the MFV relation
$\Gamma(B_{s}\to \ell^+\ell^-)/\Gamma(B_{d}\to \ell^+\ell^-)
\approx |V_{ts}/V_{td}|^2$ would be an unambiguous signature 
of MFV at large $\tan\beta$. 

Within the EFT approach where all the 
heavy degrees of freedom except the Higgs fields 
are integrated out, we cannot establish many other 
correlations among the helicity-suppressed $B$-physics 
observables. However, the scenario
becomes quite predictive within a more ambitious EFT: the MSSM
with MFV (see Sect.~\ref{sect:MSSM_MFV}). As recently discussed in 
Ref.~\cite{Isidori:2006pk,Lunghi:2006uf,Isidori:2007jw}, in the MFV-MSSM
with large $\tan\beta$ and heavy squarks, interesting correlations
can be established among all the $B$-physics observables mentioned above 
and several flavour-conserving observables (both at low and high energies).
In particular, while compatible with present $B$-physics 
constraints, this scenario can naturally resolve 
the long-standing $(g-2)_\mu$ anomaly and explain in a natural way,
why the lightest Higgs boson has not been observed yet.
The predictivity, the high-sensitivity to various 
$B$-physics observables, and the natural compatibility with existing
data, make this scenario a very interesting benchmark for correlated 
studies of low- and high-energy data (see Sect.~\ref{sec:asm}).

\paragraph{MFV in Grand Unified Theories}

Once we accept the idea that flavour dynamics obeys a MFV
principle, at least in the quark sector, it is
interesting to ask if and how this is compatible with
Grand Unified Theories (GUTs), where quarks and leptons sit in the same
representations of a unified gauge group. This question has 
recently been addressed in Ref.~\cite{Grinstein:2006cg}, 
considering the exemplifying case of ${\rm SU}(5)_{\rm gauge}$.

Within ${\rm SU}(5)_{\rm gauge}$, the down-type singlet 
quarks ($D^{i}_{R}$) and the lepton doublets 
($L^i_{L}$) belong to the $\bar {\bf 5}$ representation; the quark doublet
($Q^i_{L}$), the up-type ($U^{i}_{R}$) and lepton singlets ($E^{i}_{R}$) 
belong to the ${\bf 10}$ representation, and finally 
the right-handed neutrinos ($\nu^i_{R}$) are singlets.
In this framework the largest 
group of flavour transformation commuting with 
the gauge group is
${\mathcal G}_{\rm GUT} = 
{\rm SU}(3)_{\bar 5} \times {\rm SU}(3)_{10}\times {\rm SU}(3)_1$, 
which is smaller than the direct product 
of the quark and lepton flavour groups compatible 
with the SM gauge sector: ${\mathcal G}_q \times {\mathcal G}_l$.  
We should therefore expect some violations 
of the MFV predictions, either in the quark sector, 
or in the lepton sector, or in both
(a review of the MFV predictions for the lepton sector~\cite{Cirigliano:2005ck}
can be found in the WG3 section of this report).

A phenomenologically acceptable description of the low-energy fermion
mass matrices requires the introduction of at least four irreducible
sources of ${\mathcal G}_{\rm GUT}$ breaking. From this point of view
the situation is apparently similar to the non-unified case: the four
${\mathcal G}_{\rm GUT}$ spurions can be put in one-to-one
correspondence with the low-energy spurions $\yuk_{U,D,E}$ plus the
neutrino Yukawa coupling $\yuk_\nu$ (which is the only low-energy
spurion in the neutrino sector assuming an approximately degenerate
heavy $\nu_R$ spectrum).  However, the smaller flavour group does not
allow the diagonalization of $\yuk_D$ and $\yuk_E$ (which transform in
the same way under ${\mathcal G}_{\rm GUT}$) in the same basis. As a
result, two additional mixing matrices can appear in the expressions
for flavour changing rates \cite{Grinstein:2006cg}.  The hierarchical
texture of the new mixing matrices is known since they reduce to the
identity matrix in the limit $\yuk_E^T = \yuk_D$. Taking into account
this fact, and analysing the structure of the allowed
higher-dimensional operators, a number of reasonably firm
phenomenological consequences can be deduced~\cite{Grinstein:2006cg}:
\begin{itemize}
\item There is a well defined limit in which the standard MFV scenario
  for the quark sector is fully recovered: $|Y_\nu|\ll 1$ and small
  $\tan \beta$.  The upper bound on the neutrino Yukawa couplings
  implies an upper bound on the heavy neutrino masses ($M_\nu$).  In
  the limit of a degenerate heavy neutrino spectrum, this bound is
  about $10^{12}$ GeV.  For $M_\nu \sim 10^{12}$ GeV and small $\tan
  \beta$, deviations from the standard MFV pattern can be expected in
  rare $K$ decays but not in $B$ physics.\footnote{~The conclusion
    that $K$ decays are the most sensitive probes of possible
    deviations from the strict MFV ansatz follows from the strong
    suppression of the $s \to d$ short-distance amplitude in the SM
    [$V_{td}V_{ts}^* =\cO(10^{-4})$], and goes beyond the hypothesis
    of an underlying GUT.  This is the reason why $K \to \pi
    \nu\bar\nu$ decays, which are the best probes of $s \to d$ $\Delta
    F=1$ short-distance amplitudes, play a key role in any extension
    of the SM containing non-minimal sources of flavour symmetry
    breaking.} Ignoring fine-tuned scenarios, $M_\nu \gg 10^{12}$~GeV
  is excluded by the present constraints on quark FCNC transitions.
  Independently from the value of $M_\nu$, deviations from the
  standard MFV pattern can appear both in $K$ and in $B$ physics for
  $\tan\beta \gsim m_t/m_b$.
\item Contrary to the non-GUT MFV framework for the lepton sector, the
  rate for $\mu \to e \gamma$ and other LFV decays cannot be
  arbitrarily suppressed by lowering the mass of the heavy $\nu_R$.
  This fact can easily be understood by noting that the GUT group
  allows also $M_\nu$-independent contributions to LFV decays
  proportional to the quark Yukawa couplings. The latter become
  competitive for $M_\nu \lsim 10^{12}$ GeV and their contribution is
  such that for $\Lambda \lsim 10$ TeV the $\mu \to e \gamma$ rate is
  above $10^{-13}$ (i.e.~within the reach of
  MEG~\cite{Grassi:2005ac}).
\item Within this framework improved experimental searches on $\tau
  \to \mu \gamma$ and $\tau \to e \gamma$ are a key tool: they are the
  best observables to discriminate the relative size of the non-GUT
  MFV contributions with respect to the GUT ones. In particular, if
  the quark-induced terms turn out to be dominant, the
  $\mathcal{B}(\tau\to\mu\gamma)/\mathcal{B}(\mu\to e\gamma)$ ratio
  could reach values of $\cO(10^{-4})$, allowing $\tau\to\mu\gamma$ to
  be just below the present exclusion bounds.
\end{itemize}

\paragraph{The MFV hypothesis in the MSSM}
\label{sect:MSSM_MFV}

A detailed discussion of the so-called Minimal Supersymmetric
extension of the SM will be presented in Sect.~\ref{sec:susy}. Here we
limit ourself to analyse how the MFV hypothesis can be implemented in
this framework, and to briefly summarise its main implications.

It is first worth to recall that the adjective {\em minimal}
in the MSSM acronyms refers to the particle content of the model 
and not to its flavour structure. In general, the MSSM contains a 
huge number of free parameters and most of them are related 
to the flavour structure of the model (sfermion masses
and trilinear couplings). 
Since the new degrees of freedom (in particular the squark fields) 
have well-defined transformation
properties under the quark-flavour group ${\mathcal G}_q$,
the MFV hypothesis can easily be implemented in this 
framework following the general rules outlined in 
Sect.~\ref{sect:qMFV}: 
we need to consider all possible interactions compatible 
with i) softly-broken supersymmetry; ii) the breaking of 
${\mathcal G}_q$ via the spurion fields $Y_{U,D}$. 
This allows to express the squark mass terms and 
the trilinear quark-squark-Higgs couplings 
as follows~\cite{Hall:1990ac,D'Ambrosio:2002ex}:
\bea
{\tilde m}_{Q_L}^2 &=& {\tilde m}^2 \left( a_1 \identity 
+b_1 Y_U Y_U^\dagger +b_2 Y_D Y_D^\dagger 
+b_3 Y_D Y_D^\dagger Y_U Y_U^\dagger
+b_4 Y_U Y_U^\dagger Y_D Y_D^\dagger +\ldots
 \right)~,
\label{prima}\\
{\tilde m}_{U_R}^2 &=& {\tilde m}^2 \left( a_2 \identity 
+b_5 Y_U^\dagger Y_U +\ldots \right)~,\\
{\tilde m}_{D_R}^2 &=& {\tilde m}^2 \left( a_3 \identity 
+b_6 Y_D^\dagger Y_D +\ldots \right)~,\\
A_U &=& A\left( a_4 \identity 
+b_7 Y_D Y_D^\dagger +\ldots \right) Y_U~,\\
 A_D &=& A\left( a_5 \identity 
+b_8 Y_U Y_U^\dagger +\ldots \right) Y_D~,
\label{ultima}
\eea
where the dimensionful parameters $\tilde m$ and $A$ 
set the overall scale of the soft-breaking terms.
In Eqs.~(\ref{prima})--(\ref{ultima}) we have explicitly shown  
all independent flavour structures which cannot be absorbed into 
a redefinition of the leading terms (up to tiny contributions 
quadratic in the Yukawas of the first two families). 
When $\tan\beta$ is not too large and the bottom Yukawa coupling 
is small, the terms quadratic in $Y_D$ can be dropped.

In a bottom-up approach, the dimensionless coefficients 
$a_i$ and $b_i$ in Eqs.~(\ref{prima})--(\ref{ultima})
should be considered as free parameters of the model.
Note that  this structure is 
renormalization-group invariant: the values of 
$a_i$ and $b_i$ change according to the Renormalization Group (RG) 
flow, but the general structure of Eqs.~(\ref{prima})--(\ref{ultima})
is unchanged. This is not the case if the $b_i$ are set to zero
(corresponding to the so-called hypothesis of flavour universality).
If this hypothesis is set as initial condition 
at some high-energy scale $M$, then non vanishing 
$b_i \sim (1/4\pi)^2 \ln M^2/ {\tilde m}^2$ are 
generated by the RG evolution. 
This is for instance what happens in  models with gauge-mediated 
supersymmetry breaking~\cite{Dine:1993yw,Dine:1994vc,Giudice:1998bp}, 
where the scale $M$ is identified
with the mass of the hypothetical messenger particles. 

Using the soft terms in Eqs.~(\ref{prima})--(\ref{ultima}), the physical
$6\times 6$ squark-mass matrices, after electroweak symmetry breaking, 
are given by
\beqa
&&   \!\!\!\!\!\! 
{\tilde M}_U^2 = 
\left(
\ba{cc}
{\tilde m}_{Q_L}^2+Y_UY_U^\dagger v_U^2+\left( \frac{1}{2} -\frac{2}{3}\sw
\right) M_Z^2\cos 2\beta & \left( A_U -\mu Y_U \cot \beta \right) v_U \\
 \left( A_U -\mu Y_U \cot \beta \right)^\dagger v_U & \!\!\!\!\!\!
{\tilde m}_{U_R}^2+Y_U^\dagger Y_U v_U^2+\frac{2}{3}\sw
 M_Z^2\cos 2\beta 
\ea
\right)~, \no \\
&&  \!\!\!\!\!\! 
 {\tilde M}_D^2  =
\left(
\ba{cc}
{\tilde m}_{Q_L}^2+Y_DY_D^\dagger v_D^2-\left( \frac{1}{2} -\frac{1}{3}\sw
\right) M_Z^2\cos 2\beta & \left( A_D -\mu Y_D \tan \beta \right) v_D \\
 \left( A_D -\mu Y_D \tan \beta \right)^\dagger v_D & \!\!\!\!\!\!
{\tilde m}_{D_R}^2+Y_D^\dagger Y_D v_D^2-\frac{1}{3}\sw
 M_Z^2\cos 2\beta 
\ea
\right)~. \no \\
&& \label{mmdo}
\eeqa
where $\mu$ is the higgsino mass parameter and 
$v_{U,D}= \langle H_{U,D}\rangle$~ ($\tan\beta =v_U/v_D$).
The eigenvalues of these mass matrices 
are not degenerate; however, the mass splittings are tightly 
constrained by the specific (Yukawa-type) symmetry-breaking pattern. 

If we are interested only in low-energy processes we can integrate 
out the supersymmetric particles at one loop and project this 
theory into the general EFT discussed in the previous sections. 
In this case the coefficients of the dimension-six effective operators 
written in terms of SM and Higgs fields (see Table~\ref{tab:MFV}) 
are computable in terms of the supersymmetric soft-breaking parameters.
We stress that if $\tan\beta \gg 1$ (see Sect.~\ref{eq:largetanb})
and/or if $\mu$ is large enough~\cite{Altmannshofer:2007cs}, 
the relevant operators thus obtained go beyond the restricted 
basis of the CMFV scenario~\cite{Buras:2003jf}. 
The typical effective scale suppressing these operators 
(assuming an overall coefficient $1/\Lambda^2$) is
\beq
\Lambda \sim ~ 4 \pi ~ \tilde m~.
\eeq
Looking at the bounds in Table~\ref{tab:MFV}, we then conclude that  
if MFV holds, the present bounds on FCNCs do not exclude squarks in 
the few hundred GeV mass range, i.e.~well within the LHC reach.

It is finally worth recalling that  the integration of the 
supersymmetric degrees of freedom may also lead to sizable modifications 
of the renormalizable operators
and, in particular, of the effective Yukawa interactions. 
As a result, in an effective field theory 
with supersymmetric degrees of freedom, the relations between $Y_{U,D}$ 
and the physical quark masses and CKM angles are potentially modified.
As already pointed out in Sect.~\ref{eq:largetanb}, this effect 
is particularly relevant in the large $\tan\beta$ regime.

\newpage \subsection{SUSY models}
\label{sec:susy}


\subsubsection{FCNC and SUSY}
\label{sec:FCNC}

The generation of fermion masses and mixings (``flavour problem'') gives 
rise to a first and important distinction among theories of new physics 
beyond the electroweak standard model. 

One may conceive a kind of new physics that is completely ``flavour
blind'', i.e. new interactions that have nothing to do with the
flavour structure. To provide an example of such a situation, consider
a scheme where flavour arises at a very large scale (for instance the
Planck mass) while new physics is represented by a supersymmetric
extension of the SM with supersymmetry broken at a much lower scale
and with the SUSY breaking transmitted to the observable sector by
flavour-blind gauge interactions. In this case one may think that the
new physics does not cause any major change to the original flavour
structure of the SM, namely that the pattern of fermion masses and
mixings is compatible with the numerous and demanding tests of flavour
changing neutral currents.

Alternatively, one can conceive a new physics that is entangled 
with the flavour problem. As an example consider a technicolour scheme 
where fermion masses and mixings arise through the exchange of new gauge 
bosons which mix together ordinary and technifermions. Here we expect 
(correctly enough) new physics to have potential problems in 
accommodating the usual fermion spectrum with the adequate suppression 
of FCNC. As another example of new physics that is not flavour blind, 
take a more conventional SUSY model which is derived from a 
spontaneously broken N=1 supergravity and where the SUSY breaking 
information is conveyed to the ordinary sector of the theory through 
gravitational interactions. In this case we may expect that the scale at 
which flavour arises and the scale of SUSY breaking are not so different 
and possibly the mechanism of SUSY breaking and transmission itself is 
flavour-dependent. Under these circumstances we may expect 
a potential flavour problem to arise, namely that SUSY contributions to 
FCNC processes are too large.   

The potentiality of probing SUSY in FCNC phenomena was readily
realized when the era of SUSY phenomenology started in the early 80's
\cite{Ellis:1981ts,Barbieri:1981gn}. In particular, the major
implication that the scalar partners of quarks of the same electric
charge but belonging to different generations had to share a
remarkably high mass degeneracy was emphasized.

Throughout the large amount of work in the past decades it became
clearer and clearer that generically talking of the implications of
low-energy SUSY on FCNC may be rather misleading. We have a minimal
SUSY extension of the SM, the so-called Constrained Minimal
Supersymmetric Standard Model (CMSSM), where the FCNC
contributions can be computed in terms of a very limited set of
unknown new SUSY parameters. Remarkably enough, this minimal model
succeeds to pass all FCNC tests unscathed. To be sure, it
is possible to severely constrain the SUSY parameter space, for
instance using $b \to s \gamma$, in a way that is complementary to
what is achieved by direct SUSY searches at colliders.

However, the CMSSM is by no means equivalent to low-energy SUSY. A
first sharp distinction concerns the mechanism of SUSY breaking and
transmission to the observable sector that is chosen. As we mentioned
above, in models with gauge-mediated SUSY breaking (GMSB models
\cite{Dine:1981za,Dimopoulos:1981au,Dine:1981gu,Dine:1982qj,Dine:1982zb,AlvarezGaume:1981wy,Nappi:1982hm,Dimopoulos:1982gm,Dine:1993yw,Dine:1994vc,Dine:1995ag,Poppitz:1996fw,ArkaniHamed:1997jv,Murayama:1997pb,Dimopoulos:1997ww,Dimopoulos:1997je,Luty:1997ny,Hotta:1996ag,Randall:1996zi,Shadmi:1997md,Haba:1997ad,Csaki:1997if,Shirman:1997rm,Langacker:1999hs,Babu:2001gp,Delgado:2007rz})
it may be possible to avoid the FCNC threat ``ab initio'' (notice that
this is not an automatic feature of this class of models, but it
depends on the specific choice of the sector that transmits the SUSY
breaking information, the so-called messenger sector). The other more
``canonical'' class of SUSY theories that was mentioned above has
gravitational messengers and a very large scale at which SUSY breaking
occurs. In this brief discussion we focus only on this class of
gravity-mediated SUSY breaking models. Even sticking to this more
limited choice we have a variety of options with very different
implications for the flavour problem.

First, there exists an interesting large class of SUSY realizations
where the customary R-parity (which is invoked to suppress proton
decay) is replaced by other discrete symmetries which allow either
baryon or lepton violating terms in the superpotential. But, even
sticking to the more orthodox view of imposing R-parity, we are still
left with a large variety of extensions of the MSSM at low energy. The
point is that low-energy SUSY ``feels'' the new physics at the
superlarge scale at which supergravity (i.e., local supersymmetry)
broke down. In the past years we have witnessed an
increasing interest in supergravity realizations without the so-called
flavour universality of the terms which break SUSY explicitly. Another
class of low-energy SUSY realizations, which differ from the MSSM in
the FCNC sector, is obtained from SUSY-GUT's. The interactions
involving superheavy particles in the energy range between the GUT and
the Planck scale bear important implications for the amount and kind
of FCNC that we expect at low energy
\cite{Ciuchini:2003rg,Ciuchini:2007ha,Albrecht:2007ii}. 

\subsubsection{FCNC in SUSY without R-parity}
\label{sec:Rbroken}

It is well known that in the SM case the imposition of gauge symmetry and the 
usual gauge assignment of the 15 elementary fermions of each family lead to 
the automatic conservation of baryon and lepton numbers (this is true
at any order in perturbation theory).

On the contrary, imposing in addition to the usual $SU(3)\otimes SU(2)
\otimes U(1)$ gauge symmetry an N=1 global SUSY does not prevent the
appearance of terms which explicitly break B or L
\cite{Weinberg:1981wj,Sakai:1981pk}. 
Indeed, the superpotential reads: 
\begin{eqnarray}
W&=&h^U Q H_{U}u^c + h^D Q H_{D} d^c + h^L L H_D e^c + \mu H_U H_D \nonumber \\
&+& \mu^\prime H_{U} L + \lambda^{\prime
  \prime}_{ijk}u^c_{i}d^c_{j}d_{k}^c +
\lambda^{\prime}_{ijk}Q_{i}L_{j}d_{k}^c +
\lambda_{ijk}L_{i}L_{j}e_{k}^c \, ,
\label{superp}
\end{eqnarray}
where the chiral matter superfields $Q$, $u^c$, $d^c$, $L$, $e^c$, $H_{U}$ and 
$H_{D}$ transform under the above gauge symmetry as:
\begin{eqnarray}
&\,&Q\equiv (3,2,1/6); \qquad u^c\equiv (\bar{3},1,-2/3);\qquad d^c\equiv
(\bar{3},1,1/3);\\
&\,& L\equiv (1,2,-1/2); \; \; e^c \equiv (1,1,1); \;\; H_{U}\equiv 
(1,2,1/2); \;\; H_{D}\equiv (1,2,-1/2). \nonumber
\label{qnumbers}
\end{eqnarray}
The couplings $h^U$, $h^D$, $h^L$ are $3\times 3$ matrices in the generation 
space; $i$, $j$ and $k$ are generation indices. Using the product of 
$\lambda^\prime$ and $\lambda^{\prime \prime}$ couplings it is immediate to 
construct four-fermion operators leading to proton decay through the exchange 
of a squark. Even if one allows for the existence of $\lambda^\prime$ and 
$\lambda^{\prime \prime}$ couplings only involving the heaviest generation, 
one can show that the bound on the product $\lambda^\prime \times 
\lambda^{\prime \prime}$ of these couplings is very severe (of $O(10^{-7})$)
\cite{Smirnov:1996bg}.

A solution is that there exists a discrete symmetry, B-parity
\cite{Aulakh:1982yn,Hall:1983id,Lee:1984tn,Ellis:1984gi,Ibanez:1991pr},
which forbids the B violating terms 
proportional to $\lambda^{\prime \prime}$ in eq.~(\ref{superp}). 
In that case it is still
possible to produce sizable effects in FC processes. Two general
features of these R-parity violating contributions are:
\begin{enumerate}
\item  complete loss of any correlation to the CKM elements. For instance, in 
the above example, the couplings $\lambda^\prime$ and $\lambda$ have nothing 
to do with the usual angles $V_{tb}$ and $V_{ts}$ which appear in $b \to s l^+ 
l^-$ in the SM;
\item  loss of correlation among different FCNC processes, which are 
tightly correlated in the SM. For instance, in our example $b \to d l^+ l^-$ 
would depend on $\lambda^\prime$ and $\lambda$ parameters which are different 
from those appearing in $B_{d}-\bar{B}_{d}$ mixing.
\end{enumerate}

In this context it is difficult to make predictions given the
arbitrariness of the large number of $\lambda$ and $\lambda^\prime$
parameters. There exist bounds on each individual coupling (i.e.
assuming all the other L violating couplings are zero)
\cite{Barger:1989rk,Enqvist:1992ef}.

Obviously, the most practical way of avoiding any threat of B and L
violating operators is to forbid \underline{all} such terms in
eq.~(\ref{superp}). This is achieved by imposing the usual R matter
parity. This quantum number is $+1$ for every ordinary particle
and $-1$ for SUSY partners. We now turn to FCNC in the framework of
low-energy SUSY with R parity.

\subsubsection{FCNC in SUSY with R-parity - CMSSM framework}
\label{sec:CMSSM}

Even when R parity is imposed the FCNC challenge is not over. It is true
that in this case, analogously to what happens in the SM, no tree
level FCNC contributions arise. However, it is well-known that this is a
necessary but not sufficient condition to consider the FCNC problem
overcome. The loop contributions to FCNC in the SM exhibit the presence of
the GIM mechanism and we have to make sure that in the SUSY case with R
parity some analog of the GIM mechanism is active. 

To give a qualitative idea of what we mean by an effective super-GIM
mechanism, let us consider the following simplified situation where
the main features emerge clearly. Consider the SM box diagram
responsible for $K^0 - \bar{K}^0$ mixing and take only two
generations, i.e. only the up and charm quarks run in the loop. In
this case the GIM mechanism yields a suppression factor of $O((m_c^2 -
m_u^2)/M_W^2)$. If we replace the W boson and the up quarks in the
loop with their SUSY partners and we take, for simplicity, all SUSY
masses of the same order, we obtain a super-GIM factor which looks
like the GIM one with the masses of the superparticles instead of
those of the corresponding particles. The problem is that the up and
charm squarks have masses which are much larger than those of the
corresponding quarks. Hence the super-GIM factor tends to be of $O(1)$
instead of being $O(10^{-3})$ as it is in the SM case. To obtain this
small number we would need a high degeneracy between the mass of the
charm and up squarks. It is difficult to think that such a degeneracy
may be accidental. After all, since we invoked SUSY for a naturalness
problem (the gauge hierarchy issue), we should avoid invoking a
fine-tuning to solve its problems! Then one can turn to some symmetry
reason. For instance, just sticking to this simple example that we are
considering, one may think that the main bulk of the charm and up
squark masses is the same, i.e. the mechanism of SUSY breaking should
have some universality in providing the mass to these two squarks with
the same electric charge. Another possibility one may envisage is that
the masses of the squarks are quite high, say above few TeV's. Then
even if they are not so degenerate in mass, the overall factor in
front of the four-fermion operator responsible for the kaon mixing
becomes smaller and smaller (it decreases quadratically with the mass
of the squarks) and, consequently, one can respect the experimental
result. We see from this simple example that the issue of FCNC may be
closely linked to the crucial problem of how we break SUSY.

We now turn to some more quantitative considerations. We start by
discussing the different degrees of concern that FCNC raise 
according to the specific low-energy SUSY realization one has in mind.
In this section we will consider FCNC in the CMSSM realizations. In
Sect. \ref{sec:MSSMCP} we will deal with CP-violating FCNC phenomena
in the same context.  After discussing these aspects in the CMSSM we
will provide bounds from FCNC and CP violation in a generic SUSY
extension of the SM (Sect. \ref{sec:genFCNC}).

Obviously the reference frame for any discussion in a specific SUSY
scheme is the MSSM. Although the name seems to indicate a well-defined
particle model, we can identify at least two quite different classes
of low-energy SUSY models. First, we have the CMSSM, the minimal SUSY
extension of the SM (i.e. with the smallest needed number of
superfields) with R-parity, radiative breaking of the electroweak
symmetry, universality of the soft breaking terms and simplifying
relations at the GUT scale among SUSY parameters. In this
\emph{constrained} version, the MSSM exhibits only four free
parameters in addition to those of the SM, and is an example of a SUSY
model with MFV. Moreover, some authors impose specific relations
between the two parameters $A$ and $B$ that appear in the trilinear
and bilinear scalar terms of the soft breaking sector, further reducing
the number of SUSY free parameters to three.  Then, all SUSY masses
are just functions of these few independent parameters and, hence, many
relations among them exist.  

In SUSY there are five classes of one-loop diagrams that contribute
to FCNC and CP violating processes. They are distinguished according
to the virtual particles running in the loop: W and up-quarks, charged
Higgs and up-quarks, charginos and up-squarks, neutralinos and
down-squarks, gluinos and down-squarks. It turns out that, in this
\emph{constrained} version of the MSSM, at low or moderate $\tan
\beta$ the charged Higgs and chargino exchanges yield the dominant
SUSY contributions, while at large $\tan \beta$ Higgs-mediated effects
become dominant. 

Obviously this very minimal version of the MSSM can be very
predictive. The most powerful constraint on this minimal model in the
FCNC context comes from $b \to s
\gamma$~\cite{Misiak:1997ei,Ciuchini:1997xe,Ciuchini:1998xy,Degrassi:2000qf}.
For large values of $\tan \beta$, strong constraints are also obtained
from the upper bound on $B_s \to \mu^+ \mu^-$, from $\Delta M_s$ and from 
$B(B \to \tau \nu)$~\cite{Babu:1999hn,Isidori:2001fv,Buras:2002vd,Buras:2002wq,Isidori:2006pk,Freitas:2007dp}. No observable deviations
from the SM predictions in other FCNC processes are expected, given
the present experimental and theoretical uncertainties.

It should be kept in mind that the above stringent results strictly
depend not only on the minimality of the model in terms of the
superfields that are introduced, but also on the ``boundary''
conditions that are chosen.  All the low-energy SUSY masses are
computed in terms of the four SUSY parameters at the Planck scale $M_{Pl}$ 
through the RG evolution. If one relaxes this tight constraint on the 
relation of the
low-energy quantities and treats the masses of the SUSY particles as
independent parameters, then much more freedom is gained. This holds
true even in the MSSM with MFV at small or moderate $\tan \beta$:
sizable SUSY effects can be present both in meson-antimeson mixing and in
rare decays \cite{Buras:2000qz}, in particular for light stop and charginos.

Moreover, flavour universality is by no means a prediction of low-energy SUSY.
The absence of flavour universality of soft-breaking terms may result from 
radiative effects at the GUT scale or from effective supergravities derived
from string theory. For instance, even starting with an exact universality
of the soft breaking terms at the Planck scale, in a SUSY GUT scheme one
has to consider the running from this latter scale to the GUT scale. Due
to the large value of the top Yukawa coupling and to the fact that quarks
and lepton superfields are in common GUT multiplets, we may expect the tau
slepton mass to be conspicuously different from that of the first two
generation sleptons at the end of this RG running. This lack of
universality at the GUT scale may lead to large violations of lepton flavour
number yielding, for instance, $\mu \to e  \gamma$ at a rate in the
ball park of observability \cite{Barbieri:1995rs}. In the non-universal case,
most FCNC processes receive sizable SUSY corrections, and indeed
flavour physics poses strong constraints on the parameter space of
SUSY models without MFV. 

\subsubsection{CP violation in the CMSSM}
\label{sec:MSSMCP}

CP violation has a major potential to exhibit manifestations of new
physics beyond the standard model.  Indeed, it is quite a general
feature that new physics possesses new CP violating phases in addition
to the Cabibbo-Kobayashi-Maskawa (CKM) phase $\left(\delta_{\rm
    CKM}\right)$ or, even in those cases where this does not occur,
$\delta_{\rm CKM}$ shows up in interactions of the new particles,
hence with potential departures from the SM expectations. Moreover,
although the SM is able to account for the observed CP violation, the
possibility of large NP contributions to CP violation in $b \to s$
transitions is still open (see sec. \ref{sec:b2sandb2d} and
ref. \cite{Silvestrini:2007yf} for recent reviews). 
The detection of CP violation in $B_s$
mixing and the improvement of the measurements of CP asymmetries in $b
\to s$ penguin decays will constitute a crucial test of the 
CKM picture within the SM. Again, on general grounds, we expect new
physics to provide departures from the SM CKM scenario. A final remark
on reasons that make us optimistic in having new physics playing a
major role in CP violation concerns the matter-antimatter asymmetry in
the universe. Starting from a baryon-antibaryon symmetric universe,
the SM is unable to account for the observed baryon asymmetry. The
presence of new CP-violating contributions when one goes beyond the SM
looks crucial to produce an efficient mechanism for the generation of
a satisfactory $\Delta$B asymmetry.

The above considerations apply well to the new physics represented by
low-energy supersymmetric extensions of the SM. Indeed, as we will see
below, supersymmetry introduces CP violating phases in addition to
$\delta_{\rm CKM}$ and, even if one envisages particular situations
where such extra-phases vanish, the phase $\delta_{\rm CKM}$ itself
leads to new CP-violating contributions in processes where SUSY
particles are exchanged. CP violation in $b \to s$ transitions has a good
potential to exhibit departures from the SM CKM picture in
low-energy SUSY extensions, although, as we will discuss, the
detectability of such deviations strongly depends on the regions of
the SUSY parameter space under consideration.

In this section we will deal with CP violation in the context of the CMSSM.
In Sec. \ref{sec:genFCNC} we will discuss the CP issue in a
model-independent approach. 

In the CMSSM two new ``genuine'' SUSY CP-violating phases are
present. They originate from the SUSY parameters $\mu$, $M$, $A$ and
$B$. The first of these parameters is the dimensionful coefficient of
the $H_u H_d$ term of the superpotential. The remaining three
parameters are present in the sector that softly breaks the N=1 global
SUSY. $M$ denotes the common value of the gaugino masses, $A$ is the
trilinear scalar coupling, while $B$ denotes the bilinear scalar
coupling. In our notation all these three parameters are
dimensionful. The simplest way to see which combinations of the phases
of these four parameters are physical \cite{Dugan:1984qf} is to notice that
for vanishing values of $\mu$, $M$, $A$ and $B$ the theory possesses
two additional symmetries \cite{Dimopoulos:1995kn}. Indeed, letting $B$ and
$\mu$ vanish, a $U(1)$ Peccei-Quinn symmetry arises, which in
particular rotates $H_u$ and $H_d$.  If $M$, $A$ and $B$ are set to
zero, the Lagrangian acquires a continuous $U(1)$ $R$ symmetry. Then
we can consider $\mu$, $M$, $A$ and $B$ as spurions which break the
$U(1)_{PQ}$ and $U(1)_R$ symmetries. In this way the question
concerning the number and nature of the meaningful phases translates
into the problem of finding the independent combinations of the four
parameters which are invariant under $U(1)_{PQ}$ and $U(1)_R$ and
determining their independent phases. There are three such independent
combinations, but only two of their phases are independent. We use
here the commonly adopted choice:
\begin{equation}
  \label{MSSMphases}
  \Phi_A = {\rm arg}\left( A^* M\right), \qquad
  \Phi_B = {\rm arg}\left( B^* M\right).
\end{equation}
The main constraints on $\Phi_A$ and $\Phi_B$ come from their contribution to
the electric dipole moments of the neutron and of the electron. For instance,
the effect of $\Phi_A$ and $\Phi_B$ on the electric and chromoelectric dipole
moments of the light quarks ($u$, $d$, $s$) lead to a contribution to
$d^e_N$ of 
order
\begin{equation}
  \label{EDMNMSSM}
  d^e_N \sim 2 \left( \frac{100 {\rm GeV}}{\tilde{m}}\right)^2 \sin \Phi_{A,B}
  \times 10^{-23} {\rm e\, cm},
\end{equation}
where $\tilde{m}$ here denotes a common mass for squarks and
gluinos. We refer the reader to the results of Working Group III for a
detailed discussion of the present status of constraints on SUSY from
electric dipole moments. We just remark that the present experimental
bounds imply that $\Phi_{A,B}$ should be at most of
$\mathcal{O}(10^{-2})$, unless one pushes SUSY masses up to 
$\mathcal{O}(1 {\rm TeV})$.

In view of the previous considerations most authors dealing with the
CMSSM prefer to simply put $\Phi_A$ and $\Phi_B$ equal to
zero. Actually, one may argue in favour of this choice by considering
the soft breaking sector of the MSSM as resulting from SUSY breaking
mechanisms which force $\Phi_A$ and $\Phi_B$ to vanish. For instance,
it is conceivable that both $A$ and $M$ originate from the same source
of $U(1)_R$ breaking. Since $\Phi_A$ ``measures'' the relative phase
of $A$ and $M$, in this case it would ``naturally''vanish. In some
specific models it has been shown \cite{Dine:1994vc} that through an
analogous mechanism also $\Phi_B$ may vanish.

If $\Phi_A=\Phi_B=0$, then the novelty of the CMSSM in CP violating
contributions merely arises from the presence of the CKM phase in
loops with SUSY particles 
\cite{Duncan:1983wz,Franco:1983xm,Gerard:1984bg,Gerard:1984pc,Gerard:1984vb,Langacker:1984ak,Dugan:1984qf}. The
crucial point is that the usual GIM suppression, which plays a major
role in evaluating $\varepsilon$ and $\varepsilon^\prime$ in the SM,
is replaced in the MSSM case by a super-GIM cancellation, which has the
same ``power'' of suppression as the original GIM (see previous
section). Again also in the MSSM, as it is the case in the SM, the
smallness of $\varepsilon$ and $\varepsilon^\prime$ is guaranteed not
by the smallness of $\delta_{\rm CKM}$, but rather by the small CKM
angles and/or small Yukawa couplings. By the same token, we do not
expect any significant departure of the MSSM from the SM predictions
also concerning CP violation in $B$ physics. As a matter of fact,
given the large lower bounds on squark and gluino masses, one expects
relatively tiny contributions of the SUSY loops in $\varepsilon$ or
$\varepsilon^\prime$ in comparison with the normal $W$ loops of the
SM. Let us be more detailed on this point.  In the MSSM the gluino
exchange contribution to FCNC is subleading with respect to chargino
($\chi^\pm$) and charged Higgs ($H^\pm$) exchanges. Hence when dealing
with CP violating FCNC processes in the MSSM with $\Phi_A=\Phi_B=0$
one can confine the analysis to $\chi^\pm$ and $H^\pm$ loops. If one
takes all squarks to be degenerate in mass and heavier than $\sim 200$
GeV, then $\chi^\pm-\tilde q$ loops are obviously severely penalized
with respect to the SM $W-q$ loops (remember that at the vertices the
same CKM angles occur in both cases). The only chance to generate
sizable contributions to CP violating phenomena is for a light stop
and chargino: in this case, sizable departures from the SM predictions
are possible \cite{Buras:2000qz}.

In conclusion, the situation concerning CP violation in the MSSM case
with $\Phi_A=\Phi_B=0$ and exact universality in the soft-breaking
sector can be summarized in the following way: the MSSM does not lead
to any significant deviation from the SM expectation for CP-violating
phenomena as $d_N^e$, $\varepsilon$, $\varepsilon^\prime$ and CP
violation in $B$ physics; the only exception to this statement
concerns a small portion of the MSSM parameter space where a very
light $\tilde t$ and $\chi^+$ are present. 

\subsubsection{Model-independent analysis of FCNC and CP violating processes
  in SUSY}
\label{sec:genFCNC}

Given a specific SUSY model it is in principle possible to make a full
computation of all the FCNC phenomena in that context. However, given
the variety of options for low-energy SUSY which was mentioned in the
Introduction (even confining ourselves here to models with R matter
parity), it is important to have a way to extract from the whole host
of FCNC processes a set of upper limits on quantities that can be
readily computed in any chosen SUSY frame.

A useful model-independent parameterization of FCNC effects is the
so-called mass insertion (MI) approximation \cite{Hall:1985dx}.  It
concerns the most peculiar source of FCNC SUSY contributions that do
not arise from the mere supersymmetrization of the FCNC in the
SM. They originate from the FC couplings of gluinos and neutralinos to
fermions and
sfermions~\cite{Duncan:1983iq,Donoghue:1983mx,Bouquet:1984pp}. One
chooses a basis for the fermion and sfermion states where all the
couplings of these particles to neutral gauginos are flavour diagonal,
while the FC is exhibited by the non-diagonality of the sfermion
propagators. Denoting by $\Delta$ the off-diagonal terms in the
sfermion mass matrices (i.e. the mass terms relating sfermions of the
same electric charge, but different flavour), the sfermion propagators
can be expanded as a series in terms of $\delta = \Delta/ \tilde{m}^2$
where $\tilde{m}$ is the average sfermion mass.  As long as $\Delta$
is significantly smaller than $\tilde{m}^2$, we can just take the
first term of this expansion and, then, the experimental information
concerning FCNC and CP violating phenomena translates into upper
bounds on these $\delta$'s
\cite{Gabbiani:1988rb,Hagelin:1992tc,Gabrielli:1995bd,Gabbiani:1996hi}.

Obviously the above mass insertion method presents the major advantage
that one does not need the full diagonalization of the sfermion mass
matrices to perform a test of the SUSY model under consideration in
the FCNC sector. It is enough to compute ratios of the off-diagonal
over the diagonal entries of the sfermion mass matrices and compare
the results with the general bounds on the $\delta$'s that we provide
here from all available experimental information.

There exist four different $\Delta$ mass insertions connecting
flavours $i$ and $j$ along a sfermion propagator:
$\left(\Delta_{ij}\right)_{LL}$, $\left(\Delta_{ij}\right)_{RR}$,
$\left(\Delta_{ij}\right)_{LR}$ and
$\left(\Delta_{ij}\right)_{RL}$. The indices $L$ and $R$ refer to the
helicity of the fermion partners.  Instead of the dimensionful
quantities $\Delta$ it is more useful to provide bounds making use of
dimensionless quantities, $\delta$, that are obtained dividing the
mass insertions by an average sfermion mass.

The comparison of several flavour-changing processes to their
experimental values can be used to bound the $\delta$s in the
different
sectors~\cite{Gabbiani:1996hi,Becirevic:2001jj,Ciuchini:1998ix,Ciuchini:2002uv,Besmer:2001cj,Kane:2002sp,Harnik:2002vs,Okumura:2003hy,Foster:2004vp,Foster:2005wb,Foster:2005kb,Foster:2006ze,Ciuchini:2006dx}. In
these analyses it is customary to consider only the dominant
contributions due to gluino exchange, which give a good approximation
of the full amplitude, barring accidental cancellations. In the same
spirit, the bounds are usually obtained taking only one non-vanishing
MI at a time, neglecting the interference among MIs. This procedure is
justified \emph{a posteriori} by observing that the MI bounds have
typically a strong hierarchy, making the destructive interference
among different MIs very unlikely.

\begingroup
\begin{table}
\begin{center}
\begin{tabular}{|c|c|c|}
 \hline\hline
Observable & Measurement/Bound & Ref.\\[0.2pt] 
 \hline
\multicolumn{3}{|c|}{Sector 1--2}\\
$\Delta M_K$ & $(0.0$ -- $5.3) \times 10^{-3}$ GeV & \cite{Yao:2006px}\\
$\varepsilon$ & $(2.232\pm 0.007) \times 10^{-3}$ & \cite{Yao:2006px} \\
$\vert(\varepsilon^\prime/\varepsilon)_{\mathrm SUSY}\vert$ & $< 2 \times 10^{-2}$ &
--\\
\hline
\multicolumn{3}{|c|}{Sector 1--3}\\
$\Delta M_{B_d}$ & $(0.507\pm 0.005)$ ps$^{-1}$ & \cite{Barberio:2007cr}\\
$\sin 2\beta$ & $0.675\pm 0.026$ & \cite{Barberio:2007cr}\\
$\cos 2\beta$ & $>-0.4$ & \cite{Bona:2006ah}\\
\hline
\multicolumn{3}{|c|}{Sector 2--3}\\
BR$(b\to (s+d)\gamma)(E_\gamma > 2.0~\mathrm{GeV})$ & $(3.06\pm 0.49) \times 10^{-4
}$ & \cite{Chen:2001fj}\\
BR$(b\to (s+d)\gamma)(E_\gamma > 1.8~\mathrm{GeV})$ & $(3.51\pm 0.43) \times 10^{-4
}$ & \cite{Koppenburg:2004fz}\\
BR$(b\to s\gamma)(E_\gamma > 1.9~{\mathrm GeV})$ & $(3.34\pm 0.18\pm 0.48) \times
10^{-4}$ & \cite{Aubert:2005cu}\\
$A_{CP}(b\to s \gamma)$ & $0.004\pm 0.036$ & \\
BR$(b\to s l^+l^-) (0.04~\mathrm{GeV} < q^2 < 1~\mathrm{GeV})$ &
$(11.34\pm 5.96)\times 10^{-7}$ & \cite{Aubert:2004it,Iwasaki:2005sy}\\
BR$(b\to s l^+l^-) (1~\mathrm{GeV} < q^2 < 6~\mathrm{GeV})$ &
$(15.9\pm 4.9)\times 10^{-7}$ & \cite{Aubert:2004it,Iwasaki:2005sy}\\
BR$(b\to s l^+l^-) (14.4~\mathrm{GeV} < q^2 < 25~\mathrm{GeV})$ &
$(4.34\pm 1.15)\times 10^{-7}$ & \cite{Aubert:2004it,Iwasaki:2005sy}\\
$A_{CP}(b\to s l^+l^-)$ & $-0.22\pm 0.26$ & \cite{Yao:2006px}\\
$\Delta M_{B_s}$ & $(17.77\pm 0.12)$ ps$^{-1}$ & \cite{Abulencia:2006ze} \\
\hline\hline
\end{tabular}
\end{center}
\caption{Measurements and bounds used to constrain the hadronic $\delta^d$'s.}
\label{tab:hexp}
\end{table}
\endgroup

The effective Hamiltonians for $\Delta F=1$ and $\Delta F=2$
transitions including gluino contributions computed in the MI
approximation can be found in the literature together with the
formulae of several observables~\cite{Gabbiani:1996hi}. Even the full NLO
calculation is available for the $\Delta F=2$ effective
Hamiltonian~\cite{Ciuchini:1997bw,Ciuchini:2006dw}. See
Refs.~\cite{Okumura:2003hy,Foster:2004vp,Foster:2005wb} for the
calculation of $\tan \beta$-enhanced subleading terms for several $B$
decays in the case of general flavour violation. 

In our study we use the phenomenological constraints collected in
Table~\ref{tab:hexp}.  In particular:
\begin{list}{}
\item{\em Sector 1--2}~ The measurements of $\Delta M_K$,
  $\varepsilon$ and $\varepsilon^\prime/\varepsilon$ are used to
  constrain the $\left(\delta^d_{12} \right)_{AB}$ with $(A,B)=(L,R)$.
  The first two measurements, $\Delta M_K$ and $\varepsilon$
  respectively bound the real and imaginary part of the product
  $\left(\delta^d_{12}\right) \left(\delta^d_{12}\right)$. In the case
  of $\Delta M_K$, given the uncertainty coming from the long-distance
  contribution, we use the conservative range in Table~\ref{tab:hexp}.
  The measurement of $\varepsilon^\prime/\varepsilon$, on the other
  hand, puts a bound on Im($\delta^d_{12}$). This bound, however, is
  effective in the case of the LR MI only. Notice that, given the
  large hadronic uncertainties in the SM calculation of
  $\varepsilon^\prime/\varepsilon$, we use the very loose bound on the
  SUSY contribution shown in Table~\ref{tab:hexp}. The bounds coming
  from the combined constraints are shown in
  Table~\ref{tab:MIquarks}. Notice that, here and in the other
  sectors, the bound on the RR MI is obtained in the presence of the
  radiatively-induced LL MI (see eq.~(\ref{prima}). The product
  $\left(\delta^d_{12}\right)_{LL} \left(\delta^d_{12}\right)_{RR}$
  generates left-right operators that are enhanced both by the QCD
  evolution and by the matrix element (for kaons only). Therefore, the
  bounds on RR MIs are more stringent than the ones on LL MIs.  
\item{\em Sector 1--3}~ The measurements of $\Delta M_{B_d}$ and
  $2\beta$ respectively constrain the modulus and the phase of the
  mixing amplitude bounding the products $\left(\delta^d_{13}\right)
  \left(\delta^d_{13}\right)$. For the sake of simplicity, in
  Table~\ref{tab:MIquarks} we show the bounds on the modulus of
  $\left(\delta^d_{13}\right)$ only.
\item{\em Sector 2--3}~ This sector enjoys the largest number of
  constraints. The recent measurement of $\Delta M_{B_s}$ constrains
  the modulus of the mixing amplitude, thus bounding the products
  $\vert \left(\delta^d_{23}\right) \left(\delta^d_{23}\right)\vert$.
  Additional strong constraints come from $\Delta B=1$ branching
  ratios, such as $b \to s\gamma$ and $b\to s l^+l^-$. Also for this
  sector, we present the bounds on the modulus of
  $\left(\delta^d_{23}\right)$ in Table~\ref{tab:MIquarks}.
\end{list}

All the bounds in Table~\ref{tab:MIquarks} have been obtained using
the NLO expressions for SM contributions and for SUSY where available.
Hadronic matrix elements of $\Delta F=2$ operators are taken from
lattice
calculations~\cite{Becirevic:2001xt,Allton:1998sm,Babich:2006bh,Nakamura:2006eq}.  
The values of the CKM parameters $\bar\rho$ and $\bar\eta$ are taken
from the {\bf UT}{\it fit} analysis in the presence of arbitrary
loop-mediated NP contributions~\cite{Bona:2006sa}. This conservative
choice allows us to decouple the determination of SUSY parameters from
the CKM matrix.  For $b\to s\gamma$ we use NLO expressions with the
value of the charm quark mass suggested by the recent NNLO
calculation~\cite{Misiak:2006zs}. For the chromomagnetic contribution
to $\varepsilon^\prime/\varepsilon$ we have used the matrix element as
estimated in Ref.~\cite{Buras:1999da}. The $95\%$ probability bounds
are computed using the statistical method described in
Refs.~\cite{Ciuchini:2000de,Ciuchini:2002uv}.

Concerning the dependence on the SUSY parameters, the bounds mainly
depend on the gluino mass and on the ``average squark mass''. A mild
dependence on $\tan\beta$ is introduced by the presence of double MIs
$\left(\delta^d_{ij}\right)_{LL} \left(\delta^d_{jj}\right)_{LR}$ in
chromomagnetic operators. This dependence however becomes sizable only
for very large values of $\tan\beta$. Approximately, all bounds scale
as squark and gluino masses. 

\begin{table}
  \centering
  \begin{tabular}{|c|c|c|c|}
    \hline
    $
      \left\vert\left(
        \delta^d_{12}
      \right)_{LL,RR}
      \right\vert$ 
    &
    $
      \left\vert\left(
        \delta^d_{12}
      \right)_{LL=RR}
      \right\vert$
    &
    $\left\vert\left(
        \delta^d_{12}
      \right)_{LR}\right\vert$
    &
    $\left\vert\left(
        \delta^d_{12}
      \right)_{RL}\right\vert$ \\ \hline
    $1\cdot 10^{-2}$ & $2 \cdot 10^{-4}$ & $5\cdot 10^{-4}$ & 
    $5 \cdot 10^{-4}$\\ \hline\hline
    $
      \left\vert\left(
        \delta^u_{12}
      \right)_{LL,RR}
      \right\vert$ 
    &
    $
      \left\vert\left(
        \delta^u_{12}
      \right)_{LL=RR}
      \right\vert$
    &
    $\left\vert\left(
        \delta^u_{12}
      \right)_{LR}\right\vert$
    &
    $\left\vert\left(
        \delta^u_{12}
      \right)_{RL}\right\vert$ \\ \hline
    $3\cdot 10^{-2}$ & $2 \cdot 10^{-3}$ & $6\cdot 10^{-3}$ & 
    $6 \cdot 10^{-3}$\\ \hline\hline
    $
      \left\vert\left(
        \delta^d_{13}
      \right)_{LL,RR}
      \right\vert$ 
    &
    $
      \left\vert\left(
        \delta^d_{13}
      \right)_{LL=RR}
      \right\vert$
    &
    $\left\vert\left(
        \delta^d_{13}
      \right)_{LR}\right\vert$
    &
    $\left\vert\left(
        \delta^d_{13}
      \right)_{RL}\right\vert$ \\ \hline
    $7\cdot 10^{-2}$ & $5 \cdot 10^{-3}$ & $1\cdot 10^{-2}$ & 
    $1 \cdot 10^{-2}$\\ \hline\hline
    $
      \left\vert\left(
        \delta^d_{23}
      \right)_{LL}
      \right\vert$ 
    &
    $
      \left\vert\left(
        \delta^d_{23}
      \right)_{RR}
      \right\vert$ 
    &
    $
      \left\vert\left(
        \delta^d_{23}
      \right)_{LL=RR}
      \right\vert$
    &
    $\left\vert\left(
        \delta^d_{23}
      \right)_{LR,RL}\right\vert$ \\ \hline
    $2\cdot 10^{-1}$ & $7 \cdot 10^{-1}$ & $5\cdot 10^{-2}$ & 
    $5 \cdot 10^{-3}$\\ \hline\hline
  \end{tabular}
\caption{95\% probability bounds on $\vert \left(\delta^{q}_{ij}
  \right)_{AB}\vert$  
  obtained for squark and gluino masses of 350 GeV. See the 
  text for details.} 
\label{tab:MIquarks}
\end{table}

\newpage \subsection{Non-supersymmetric extensions of the Standard Model}


In this Section we briefly describe two most popular non-supersymmetric
    extensions of the Standard Model (SM), paying particular attention to the
    flavour structure of these models. These are Little Higgs models and
    a model with one universal extra dimension.

\subsubsection{Little Higgs models}

\subsubsubsection{Little hierarchy problem and Little Higgs models}

The SM is in excellent agreement with the results of particle physics
experiments, in particular with the electroweak (ew) precision
measurements, thus suggesting that the SM cutoff scale is at least as
large as $10 \,\text{TeV}$.  Having such a relatively high cutoff,
however, the SM requires an unsatisfactory fine-tuning to yield a
correct ($\approx 10^2\,\text{GeV}$) scale for the squared Higgs mass,
whose corrections are quadratic and therefore highly sensitive to the
cutoff.  This ``little hierarchy problem'' has been one of the main
motivations to elaborate models of physics beyond the SM.  While
Supersymmetry is at present the leading candidate, different proposals
have been formulated more recently.  Among them, Little Higgs models
play an important role, being perturbatively computable up to about
$10 \,\text{TeV}$ and with a rather small number of parameters,
although their predictivity can be weakened by a certain sensitivity
to the unknown UV-completion of these models (see below).

In Little Higgs models\cite{Arkani-Hamed:2001nc} the Higgs is
naturally light as it is identified with a Nambu-Goldstone boson (NGB)
of a spontaneously broken global symmetry.  An exact NGB, however,
would have only derivative interactions.  Gauge and Yukawa
interactions of the Higgs have to be incorporated. This can be done
without generating quadratically divergent one-loop contributions to
the Higgs mass, through the so-called {\it collective symmetry
  breaking}.

In the following we restrict ourselves to product-group Little Higgs
models in order not to complicate the presentation. The idea of
collective symmetry breaking has also been applied to simple-group
models \cite{Kaplan:2003uc,Schmaltz:2004de}, however the
implementation is somewhat different there.  (Product-group) Little
Higgs models are based on a global symmetry group $G$, like $G=G'^N$
in the case of moose-type models
\cite{Arkani-Hamed:2001nc,Arkani-Hamed:2002qx} or $G=SU(5)$ in the
case of the Littlest Higgs, that is spontaneously broken to a subgroup
$H\subset G$ by the vacuum condensate of a non-linear sigma model
field $\Sigma$. A subgroup of $G$ is gauged, which contains at least
two $SU(2)\times U(1)$ factors, or larger groups containing such
factors. The gauge group is then broken to the SM gauge group
$SU(2)_L\times U(1)_Y$ by the vacuum expectation value (vev) of
$\Sigma$. The potential for the Higgs field is generated radiatively,
making thus the scale of the ew symmetry breaking $v\simeq
246\,\text{GeV}$ a loop factor smaller than the scale $f$, where the
breaking $G\to H$ takes place.

In order to allow for a Higgs potential being generated radiatively,
interaction terms explicitly breaking the global symmetry group $G$
have to be included as well. However, these interactions have to
preserve enough of the global symmetry to prevent the Higgs potential
from quadratically divergent radiative contributions. Only when two or
more of the corresponding coupling constants are non-vanishing,
radiative corrections are allowed. In particular, only at two or higher loop 
level, quadratically divergent
contributions appear, but these are safely small due to the loop
factor in front. 
This mechanism is referred to as the collective symmetry breaking.

\subsubsubsection{The Littlest Higgs}

The most economical, in matter content, Little Higgs model is the Littlest
Higgs (LH)\cite{Arkani-Hamed:2002qy}, where the global group $SU(5)$ is spontaneously broken
into $SO(5)$ at the scale $f \approx \mathcal{O}(1 \,\text{TeV})$ and
the ew sector of the SM is embedded in an $SU(5)/SO(5)$ non-linear
sigma model. 
Gauge and Yukawa Higgs interactions are introduced by gauging the subgroup of
$SU(5)$: $[SU(2) \times U(1)]_1 \times [SU(2) \times U(1)]_2$, with gauge 
couplings respectively equal to $g_1, g_1^\prime, g_2, g_2^\prime$. 
The key feature for the realization of collective SB is that
the two gauge factors commute with a different $SU(3)$ global symmetry
subgroup of $SU(5)$, that prevents the Higgs from becoming massive when the
couplings of one of the two gauge factors vanish. 
Consequently, quadratic corrections to the squared Higgs mass involve two
couplings and cannot appear at one-loop.
In the LH model, the new particles appearing at the TeV scale are the heavy
gauge bosons ($W^\pm_H, Z_H, A_H$), the heavy top ($T$) and the scalar triplet 
$\Phi$.

In the LH model, significant corrections to ew observables come from
tree-level heavy gauge boson contributions and the triplet vev which
breaks the custodial $SU(2)$ symmetry.  Consequently, ew precision
tests are satisfied only for quite large values of the NP scale $f \ge
2-3 \,\text{TeV}$\cite{Han:2003wu,Csaki:2002qg}, unable to solve the
little hierarchy problem. Since the LH model belongs to the class of
models with Constrained Minimal Flavour Violation
(CMFV)~\cite{Buras:2000dm}, the contributions of the
new particles to FCNC processes turn out to be at most
$10-20\%$~\cite{Buras:2004kq,Buras:2005iv,Buras:2006wk,Choudhury:2004bh,Lee:2004me,Fajfer:2005ke,Huo:2003vd}.

\subsubsubsection{T-parity}

Motivated by reconciling the LH model with ew precision tests, Cheng and 
Low\cite{Cheng:2003ju,Cheng:2004yc} proposed to enlarge the symmetry structure of the theory by
introducing a discrete symmetry called T-parity.
T-parity acts as an automorphism which exchanges the $[SU(2) \times U(1)]_1$ 
and $[SU(2) \times U(1)]_2$ gauge factors. The invariance of the theory under
this automorphism implies $g_1=g_2$ and $g_1^\prime = g_2^\prime$.
Furthermore, T-parity explicitly forbids the tree-level contributions of  heavy gauge bosons and the
interactions that induced the triplet vev.
The custodial $SU(2)$ symmetry is restored and the compatibility with ew
precision data is obtained already for smaller values of the NP scale, $f \ge
500 \,\text{GeV}$\cite{Hubisz:2005tx}.
Another important consequence is that particle fields are T-even or T-odd
under T-parity. The SM particles and the heavy top
$T_+$ are T-even, while the heavy gauge bosons $W_H^\pm,Z_H,A_H$ and the
scalar triplet $\Phi$ are T-odd.
Additional T-odd particles are required by T-parity: 
the odd heavy top $T_-$ and the so-called mirror fermions, i.e.,
fermions corresponding to the SM ones but with opposite T-parity and
$\mathcal{O}(1\,\text{TeV})$ mass~\cite{Low:2004xc}.

\subsubsubsection{New flavour interactions in LHT}

Mirror fermions are characterized by new flavour interactions with SM fermions
and heavy gauge bosons, which involve two new unitary 
mixing
matrices, in the quark sector, analogous to the Cabibbo-Kobayashi-Maskawa (CKM) matrix $V_\text{CKM}$~\cite{Cabibbo:1963yz,Kobayashi:1973fv}.
They are $V_{Hd}$ and
$V_{Hu}$, respectively involved when the SM quark is of down- or up-type,
and satisfying $V_{Hu}^\dagger V_{Hd}=V_\text{CKM}$\cite{Hubisz:2005bd}.
Similarly, two new mixing matrices $V_{H\ell}$ and $V_{H\nu}$, appear in the
lepton sector and are respectively involved when the SM lepton is charged or a
neutrino and related to the PMNS matrix~\cite{Pontecorvo:1957cp,Pontecorvo:1957qd,Maki:1962mu} through $V_{H\nu}^\dagger V_{H\ell}=V_\text{PMNS}^\dagger$. 
Both $V_{Hd}$ and $V_{H\ell}$
  contain $3$ angles, like $V_\text{CKM}$ and $V_\text{PMNS}$, but $3$
  (non-Majorana) phases \cite{Blanke:2006xr}, i.e. 
  two more phases than the SM matrices, that cannot be rotated
  away in this case.

Therefore, $V_{Hd}$ can be parameterized as
\bequ\label{2.12a}
\addtolength{\arraycolsep}{3pt}
V_{Hd}= \begin{pmatrix}
c_{12}^d c_{13}^d & s_{12}^d c_{13}^d e^{-i\delta^d_{12}}& s_{13}^d e^{-i\delta^d_{13}}\\
-s_{12}^d c_{23}^d e^{i\delta^d_{12}}-c_{12}^d s_{23}^ds_{13}^d e^{i(\delta^d_{13}-\delta^d_{23})} &
c_{12}^d c_{23}^d-s_{12}^d s_{23}^ds_{13}^d e^{i(\delta^d_{13}-\delta^d_{12}-\delta^d_{23})} &
s_{23}^dc_{13}^d e^{-i\delta^d_{23}}\\
s_{12}^d s_{23}^d e^{i(\delta^d_{12}+\delta^d_{23})}-c_{12}^d c_{23}^ds_{13}^d e^{i\delta^d_{13}} &
-c_{12}^d s_{23}^d e^{i\delta^d_{23}}-s_{12}^d c_{23}^d s_{13}^d
e^{i(\delta^d_{13}-\delta^d_{12})} & c_{23}^d c_{13}^d\\
\end{pmatrix}
\eequ
and a similar parameterization applies to $V_{H \ell}$.

The new flavour violating interactions involving mirror fermions contain the 
following combinations of elements of the mixing matrices
\begin{equation}\label{2.12}
\xi_i^{(K)}=V^{*is}_{Hd}V^{id}_{Hd}\,,\qquad
\xi_i^{(d)}=V^{*ib}_{Hd}V^{id}_{Hd}\,,\qquad
\xi_i^{(s)}=V^{*ib}_{Hd}V^{is}_{Hd}\, \qquad(i=1,2,3)\,,
\end{equation}
in the quark sector, respectively for $K$, $B_d$ and $B_s$ systems, and
\begin{equation}\label{eq:chi}
\chi_i^{(\mu e)}=V^{*ie}_{H\ell}V^{i\mu}_{H\ell}\,,\qquad
\chi_i^{(\tau e)}=V^{*ie}_{H\ell}V^{i\tau}_{H\ell}\,,\qquad
\chi_i^{(\tau\mu)}=V^{*i\mu}_{H\ell}V^{i\tau}_{H\ell}\,,
\end{equation}
that enter the leptonic transitions $\mu\to e$, $\tau\to e$ and $\tau\to\mu$, respectively.

As the LHT model, in contrast to the LH model without T-parity does not belong to the Minimal
Flavour Violation (MFV) class of models, significant effects in flavour 
violating observables both in the quark and in the lepton sector are possible.
This becomes evident if one looks at the contributions of mirror fermions to
the short distance functions $X$, $Y$ and $Z$ that govern rare and
CP-violating $K$ and $B$ decays.
For example, the mirror fermion contribution to be added to the SM one in the
$X$ function has the following structure~\cite{Blanke:2006eb}
\begin{equation}
\frac{1}{\lambda^{(i)}_t}\,\left[\xi_2^{(i)} F(m_{H1},m_{H2}) + \xi_3^{(i)}
  F(m_{H1},m_{H3})\right]\,,
\end{equation}
where the
unitarity condition $\sum_{j=1}^{3}\xi_j^{(i)} =0$ has been used, $F$ denotes a function of mirror fermion masses $m_{Hj}$ ($j=1,2,3$), and
$\lambda^{(i)}_t$ are the well-known combinations of CKM elements, with $i=K,
d, s$ standing for $K$, $B_d$ and $B_s$ systems, respectively.

It is important to note that mirror fermion contributions are enhanced by a
factor $1/\lambda_t^{(i)}$ and are different for
$K$, $B_d$ and $B_s$ systems, thus breaking universality.
As  $\lambda_t^{(K)}\simeq 4\cdot 10^{-4}$, whereas
$\lambda_t^{(d)}\simeq 1\cdot 10^{-2}$ and $\lambda_t^{(s)}\simeq
4\cdot 10^{-2}$, the deviation from the SM prediction in the
$K$ system is found to be by more than an order of magnitude larger than in the
$B_d$ system, and even by two orders of magnitude larger than in the
$B_s$ system. Analogous statements are valid for the $Y$
and $Z$ functions.

 Other LHT peculiarities are the rather small number of new particles and
parameters (the SB scale $f$, the parameter $x_L$ describing $T_+$ mass and
interactions, the mirror fermion masses and $V_{Hd}$ and $V_{H\ell}$
parameters) and the
absence of new operators in addition to the SM ones.
On the other hand, one has to recall that Little Higgs models are low
energy non-linear sigma models, whose unknown UV-completion introduces a
theoretical uncertainty reflected by a logarithmically enhanced cut-off dependence~\cite{Buras:2006wk,Blanke:2006eb} in 
$\Delta F=1$ processes that receive contributions from Z-penguin and box
diagrams.
See~\cite{Buras:2006wk,Blanke:2006eb} for a detailed discussion of this issue.

\subsubsubsection{Phenomenological results}

We conclude this section with a summary of the main results found in recent
LHT phenomenological studies~\cite{Hubisz:2005bd,Blanke:2006eb,Blanke:2006sb,Choudhury:2006sq,Blanke:2007db}.

In the quark sector~\cite{Hubisz:2005bd,Blanke:2006eb,Blanke:2006sb},
the most evident departures from the SM predictions are found for
    CP-violating observables that are strongly suppressed in the SM. 
These are the branching ratio for $K_L \to \pi^0 \nu \bar \nu$ and
    the CP-asymmetry $S_{\psi \phi}$, that can be enhanced by an order of
    magnitude relative to the SM predictions. Large departures from SM expectations are also possible for $Br(K_L \to
  \pi^0 \ell^+ \ell^-)$  and $Br(K^+ \to \pi^+ \nu \bar \nu)$ and the
    semileptonic CP-asymmetry $A^s_\text{SL}$, that can be enhanced by an
    order of magnitude.
The branching ratios for $B_{s,d} \to \mu^+ \mu^-$ and $B \to X_{s,d}
    \nu \bar \nu$, instead,  are modified by at most $50\%$ and $35\%$,
    respectively, and the effects of new electroweak penguins in $B \to \pi K$
    are small, in agreement with the recent data.
The new physics effects in $B\to X_{s,d}\gamma$ and $B\to X_{s,d}\ell^+\ell^-$
turn out to be below $5\%$ and $15\%$, respectively, so that agreement 
with the data can easily be obtained.
Small, but still significant effects have been found in $B_{s,d}$ mass
    differences. In particular, a $7\%$ suppression of $\Delta M_s$ is
    possible, thus improving the compatibility with the recent experimental
    measurement~\cite{Abulencia:2006ze,Lucchesi:2006di}.

The possible discrepancy between the values of $\sin 2\beta $
following directly from $A_{\rm CP}(B_d \to \psi K_S)$ and
indirectly from the usual analysis of the unitarity triangle
involving $\Delta M_{d,s}$ and $|V_{ub}/V_{cb}|$ can
be cured within the LHT model thanks to a new phase $\varphi_{B_d}\simeq
-5^o$.

The universality of new physics effects, characteristic for MFV models,
    can be largely broken, in particular between $K$ and $B_{s,d}$ systems.
In particular, sizable departures from MFV relations between $\Delta M_{s,d}$
    and $Br(B_{s,d} \to \mu^+ \mu^-)$ and between $S_{\psi K_S}$ and the $K \to
    \pi \nu \bar \nu$ decay rates are possible.
Similar results have been recently obtained in a model with $Z'$-contributions~\cite{Promberger:2007py}.

More recently, the most interesting lepton flavour violating decays have also been studied~\cite{Choudhury:2006sq,Blanke:2007db}.
These are 
 $\ell_i
\rightarrow \ell_j \gamma$ analyzed
in~\cite{Choudhury:2006sq,Blanke:2007db} and $\tau \rightarrow \mu P$ (with $P=\pi,
\eta, \eta'$) , $\mu^- \rightarrow e^-
e^+ e^-$, the six three-body decays $\tau^- \rightarrow \ell_i^- \ell_j^+ \ell_k^-$, 
the rate for $\mu-e$ conversion in nuclei, and the $K$ or $B$ decays
$K_{L,S} \rightarrow \mu e$, $K_{L,S}
\rightarrow \pi^0 \mu e$, $B_{d,s} \rightarrow \mu e$, $B_{d,s} \rightarrow
\tau e$ and $B_{d,s} \rightarrow \tau \mu$ studied in~\cite{Blanke:2007db}.
It was found that essentially all the rates considered can reach or approach
present experimental upper bounds~\cite{Eidelman:2004wy}.
In particular, in order to suppress the $\mu \rightarrow e \gamma$ and $\mu^-
\rightarrow e^- e^+ e^-$ decay rates and the $\mu-e$ conversion rate below 
the experimental upper bounds, the
$V_{H\ell}$ mixing matrix has to be rather hierarchical, unless mirror
leptons are quasi-degenerate. 
One finds~\cite{Blanke:2007db} that the pattern of the branching ratios
   for LFV processes differs significantly from the one encountered in
   supersymmetry~\cite{Paradisi:2005fk,Paradisi:2005tk,Paradisi:2006jp}. This is welcome as the distinction between
   supersymmetry and LHT models will be non-trivial in high energy collider
   experiments.
Finally, the muon anomalous magnetic moment $(g-2)_\mu$ has also been 
considered~\cite{Choudhury:2006sq,Blanke:2007db}, finding the result $a_\mu^{LHT} < 1.2
\cdot 10^{-10}$,
even for the scale $f$ as low as $500 \gev$.
This value is roughly a factor $5$ below the current experimental
uncertainty, implying that the possible discrepancy between the SM prediction
and the data cannot be solved in the LHT model. 

\subsubsection{Universal Extra Dimensions}

Since the work of Kaluza and Klein \cite{Kaluza:1921tu,Klein:1926tv} models with more than three spatial dimensions often have been used to unify the forces of nature. More recently, inspired by string theory, extra dimensional models have been proposed to explain the origin of the TeV scale \cite{Antoniadis:1990ew,Lykken:1996fj,Witten:1996mz,Horava:1996ma,Horava:1995qa,Caceres:1996is,Arkani-Hamed:1998rs,Antoniadis:1998ig,Arkani-Hamed:1998nn,Randall:1999ee}.

A simple extension of the SM including additional space dimensions is the ACD model \cite{Appelquist:2000nn} with one universal extra dimension (UED). Here all the SM fields are democratically allowed to propagate in a flat extra dimension compactified on an orbifold $S^{1}/Z_{2}$ of size $10^{-18}$ m or smaller. In general UED models there can also be contributions from terms residing at the boundaries. Generically, these terms  would violate bounds from flavour and CP violation. To be consistent with experiment, we will assume the minimal scenario where these terms vanish at the cut-off scale. The only additional free parameter then compared to the SM is the compactification scale $1/R$. Thus, all the tree level masses of the KK particles and their interactions among themselves and with the SM particles can be described in terms of $1/R$ and the parameters of the SM.
In the effective four dimensional theory there are, in addition to the
ordinary SM particles, denoted as zero $(n=0)$ modes, corresponding infinite towers of KK modes $(n \geq 1)$ with masses
$m^2_{(n),{\rm KK}}=m^2_0 + m^2_n$, where $m_n=n/R$ and
$m_0$ is the mass of the zero mode.


A very important property of UEDs is the conservation of KK parity that implies the absence of tree level KK contributions to low energy processes taking place at scales $\mu \leq 1/R$. Therefore the flavour-changing neutral current (FCNC) processes like particle-antiparticle mixing, rare $K$ and $B$ decays and radiative decays are of particular interest. Since these processes first appear at one-loop in the SM and are strongly suppressed, the one-loop contributions from the KK modes to them could in principle be important.
Also, due to conservation of KK parity the GIM mechanism significantly improves the convergence of the sum over KK modes and thus removes the sensitivity of the calculated branching ratios to the scale $M_s \gg 1/R$ at which the higher dimensional theory becomes non-perturbative, and at which the towers of the KK particles must be cut off in an appropriate way.
Since the low energy effective Hamiltonians are governed by local operators
 already present in the SM and the flavour and CP violation in this model is
 entirely governed by the SM Yukawas the UED model belongs to the class of
 models with 
CMFV \cite{Buras:2000dm,D'Ambrosio:2002ex}. This has automatically the following important consequence for the FCNC processes considered in \cite{Buras:2002ej,Buras:2003mk,Colangelo:2006vm,Colangelo:2006gv}: the impact of the KK modes on the processes in question amounts only to the modification of the Inami-Lim one-loop functions \cite{Inami:1980fz}, i.e. each function, which in the SM depends only on $m_t$, now also becomes a function of $1/R$:
\begin{equation}
F\left(x_t, 1/R\right) = F_0\left(x_t\right) + \sum \limits_{n=1}^{\infty} F_{n}\left(x_t, x_n\right), \quad x_t = \frac{m_t^2}{m_W^2}, \quad x_n = \frac{m_n^2}{m_W^2}\,.
\end{equation}

\newpage 

\newpage
\subsection{Tools for flavour physics and beyond}
\label{sec:tools}



\subsubsection{Tools for flavour physics}
\label{sec:flavortools}

An increasing number of calculations of flavour (related) observables is
appearing, including more and more refined approaches and methods. 
It is desirable to have these calculations in the form of computer codes
at hand. This allows to easily use the existing knowledge for 
checks of the parameters/models for a phenomenological/experimental analysis, 
or to check an independent calculation. 

As a first step in this direction we present here a collection of
computer codes connected to the evaluation of flavour related
observables.  (A different class of codes, namely fit codes for the
CKM triangle, are presented later in Section~\ref{sec:fittools}.) Some
of these codes are specialized to the evaluation of a certain
restricted set of observables at either low or high energies (the
inclusion of codes for high-energy observables is motivated by the
idea of testing a parameter space from both sides, i.e.\ at flavour
factories and at the LHC). Others tools are devoted to the evaluation
of particle spectra including NMFV effects of the MSSM or the
2HDM. Some codes allow the (essentially) arbitrary calculation of
one-loop corrections including flavour effects. Finally tools are included
that faciliate the hand-over of flavour parameters and observables.  
Following the general idea of providing the existing
knowledge to the community, only codes that are either already
publicly available, or that will become available in the near future
are included.  In order to be useful for the high-energy physics
community, it is mandatory that the codes provide a minimum of user
friendlyness and support.

As a second step it would be desirable to connect different codes
(working in the same model) to each other. This could go along the
lines of the SUSY Les Houches Accord~\cite{Skands:2003cj,Allanach:2008qg}, 
i.e.\ to define a common language, a common set of input parameters. It 
would require the continuous effort of the various authors of the codes 
to comply with these definitions. Another, possibly simpler approach is
to implement the tools as sub-routines, called by a master code that
takes care of the correct defintion of the input parameters. This is
discussed in more detail in Section \ref{sec:mastertool}. It will
facilitate the use of the codes also for non-experts.

\begin{table}[tbh!]
\renewcommand{\arraystretch}{1.2}
\begin{center}
{\small
\begin{tabular}{|l|l|c|}
\hline
name & short description & av.\\ \hline
\num\ no name & $K \bar K$ mixing, $B_{(s)} \bar B_{(s)}$ mixing,
  $b \to s \gamma$, $b \to s\, l^+ l^-$ in NMFV MSSM & $o$\\
\num\ no name & $B$~physics observables in the MFV MSSM & $+$\\
\num\ no name & rare $B$ and $K$ decays in/beyond SM & $o$\\
\num\ {\tt SusyBSG} & $B\to X_s\gamma$ in MSSM with MFV & $+$\\
\num\ no name & FCNC observables in MSSM & $o$\\
\num\ no name & FC Higgs/top decays in 2HDM I/II & $o$\\
\num\ no name & squark/gluino production at LO for NMFV MSSM & $+$\\
\num\ {\tt FeynHiggs} & Higgs phenomenology in (NMFV) MSSM & $+$\\
\num\ {\tt FCHDECAY} & FCNC Higgs decays in NMFV MSSM & $+$\\
\num\ {\tt FeynArts/FormCalc} & (arbitrary) one-loop corrections 
                                                    in NMFV MSSM & $+$\\
\num\ {\tt SLHALib2} & read/write SLHA2 data, i.e.\ NMFV/RPV/CPV 
                                                     MSSM, NMSSM & $+$\\
\num\ {\tt SoftSUSY} & NMFV MSSM parameters from GUT scale input & $+$\\
\num\ {\tt SPheno}   & NMFV MSSM parameters from GUT scale input & $+$\\
\hline
\end{tabular}
\caption{
Overview about codes for the evaluation of flavour related observables; 
\newline
av.\ $\equiv$ availability: $+$ = available, $o$ = planned
}
\label{tab:tools}
}
\end{center}
\renewcommand{\arraystretch}{1}
\vspace{-1em}
\end{table}

An overview of the available codes is given in Table~\ref{tab:tools}. 
To give a better idea of the properties of each code we also provide a list
summarizing the authors, a short description, the models included, the input
and output options, as well as the available literature:

\newpage

\begin{enumerate}

\item\ul{no name}
\begin{itemize}
\item[Authors:] M.~Ciuchini \textit{et al.} 
     \cite{Ciuchini:2002uv,Silvestrini:2005zb,Ciuchini:2006dx}
\item[Description:] calculation of 
  $K \bar K$ mixing, $B_{(s)} \bar B_{(s)}$ mixing,
  $b \to s \gamma$, $b \to s\, l^+ l^-$
\item[Models:] NMFV MSSM
\item[Input:] electroweak-scale soft SUSY-breaking parameters
\item[Output:] see Description, no special format
\item[Availability:] available from the authors in the near future
\end{itemize}

\item\ul{no name}
\begin{itemize}
\item[Authors:] G.~Isidori, P.~Paradisi
     \cite{Isidori:2006pk}
\item[Description:] calculation of $B$~physics observables
\item[Models:] MFV MSSM
\item[Input:] electroweak-scale soft SUSY-breaking parameters
\item[Output:] see Description, no special format
\item[Availability:] available from the authors upon request
\end{itemize}

\item\ul{no name}
\begin{itemize}
\item[Authors:] C.~Bobeth, T.~Ewerth, U.~Haisch
     \cite{Gambino:2004mv,Bobeth:2004jz,Bobeth:2005ck}
\item[Description:] calculation of 
  BR's, F/B asymmetries for rare $B$ and $K$ decays (in/exclusive) 
\item[Models:] SM, SUSY, CMFV
\item[Input:] SM parameters, SUSY masses, scales
\item[Output:] see Description, no special format
\item[Availability:] available from the authors in the near future
\end{itemize}

\item\ul{\tt SusyBSG}
\begin{itemize}
\item[Authors:] G.~Degrassi, P.~Gambino, P.~Slavich
     \cite{Degrassi:2007kj}
\item[Description:] Fortran code for $B(B\to X_s \gamma)$ 
\item[Models:] SM, MSSM with MFV
\item[Input:] see manual (SLHA(2) compatible)
\item[Output:] see Description, no special format
\item[Availability:] {\tt cern.ch/slavich/susybsg/home.html}, manual available 
\end{itemize}

\item\ul{no name}
\begin{itemize}
\item[Authors:] P.~Chankowski, S.~J\"ager, J.~Rosiek
     \cite{Buras:2004qb}
\item[Description:] calculation of various FCNC observables in the MSSM
  (computes 2-, 3-, 4-point Greens functions that can be used as building
  blocks for various amplitudes)
\item[Models:] MSSM
\item[Input:] MSSM Lagrangian parameters in super CKM basis (as in SLHA2)
\item[Output:] see Description, no special format
\item[Availability:] available from the authors in the near future
\end{itemize}

\item\ul{no name}
\begin{itemize}
\item[Authors:] S.~Bejar, J.~Guasch
     \cite{Bejar:2000ub,Bejar:2001sj,Bejar:2003em}
\item[Description:] calculation of FC decays: $\phi \to tc$, $\phi \to bs$, 
  $t \to c \phi$  ($\phi = h,H,A$)
\item[Models:] 2HDM type I/II (with $\lambda_5, \lambda_6$)
\item[Input:] similar to SLHA2 format
\item[Output:] similar to SLHA2 format
\item[Availability:] available from the authors in the near future
\end{itemize}

\item\ul{no name}
\begin{itemize}
\item[Authors:] G.~Bozzi, B.~Fuks, M.~Klasen
\item[Description:] SUSY CKM matrix determination through squark- and
  gaugino production at LO 
\item[Models:] NMFV MSSM
\item[Input:] MSSM spectrum as from SUSPECT (SLHA2 compliant)
\item[Output:] cross section (and spin asymmetry, in case) as
  functions of CKM parameters 
\item[Availability:] from the authors upon request
\end{itemize}

\item\ul{\tt FeynHiggs}
\begin{itemize}
\item[Authors:] S.~Heinemeyer, T.~Hahn, W.~Hollik, H.~Rzehak, G.~Weiglein
     \cite{Heinemeyer:1998np,Degrassi:2002fi,Heinemeyer:2004by}
\item[Description:] Higgs phenomenology (masses, mixings, cross
  sections, decay widths) 
\item[Models:] (N)MFV MSSM, CPV MSSM
\item[Input:] electroweak-scale soft SUSY-breaking parameters (SLHA(2)
  compatible) 
\item[Output:] Higgs masses, mixings, cross sections, decay widths
  (SLHA(2) output possible)
\item[Availability:] {\tt www.feynhiggs.de} , manual available
\end{itemize}

\item\ul{\tt FCHDECAY}
\begin{itemize}
\item[Authors:] S.~Bejar, J.~Guasch
     \cite{Bejar:2004rz,Bejar:2005kv,Bejar:2006hd}
\item[Description:] 
                    $\br(\phi \to bs, tc)$ ($\phi = h,H,A$), 
                    $\br(b \to s\gamma)$, masses, mixing matrices
\item[Models:] NMFV MSSM
\item[Input:] via SLHA2 
\item[Output:] via SLHA2 
\item[Availability:] {\tt fchdecay.googlepages.com} , manual available
\end{itemize}

\item\ul{\tt FeynArts/FormCalc}
\begin{itemize}
\item[Authors:] T.~Hahn
     \cite{Hahn:1998yk,Hahn:2000kx,Hahn:2005qi}
\item[Description:] Compute (essentially) arbitrary one-loop corrections
\item[Models:] NMFV MSSM, CPV MSSM
\item[Input:] Process definition 
\item[Output:] Fortran code to compute e.g. cross-sections, can be
  linked with SLHALib2 to obtain data from other codes
\item[Availability:] {\tt www.feynarts.de}, {\tt www.feynarts.de/formcalc},
                     manual available
\end{itemize}

\item\ul{\tt SLHALib2}
\begin{itemize}
\item[Authors:] T.~Hahn 
     \cite{Skands:2003cj,Hahn:2006nq}
\item[Description:] read/write SLHA2 data
\item[Models:] NMFV MSSM, RPV MSSM, CPV MSSM, NMSSM
\item[Input:] SLHA2 input file
\item[Output:] SLHA2 output file in the SLHA2 record
\item[Availability:] {\tt www.feynarts.de/slha} , manual available
\end{itemize}

\item\ul{\tt SoftSUSY}
\begin{itemize}
\item[Authors:] B.~Allanach
     \cite{Allanach:2001kg}
\item[Description:] evaluates NMFV MSSM parameters from GUT scale input
\item[Models:] NMFV MSSM
\item[Input:] SLHA2 input file
\item[Output:] SLHA2 output file
\item[Availability:] {\tt hepforge.cedar.ac.uk/softsusy} , manual available
\end{itemize}

\item\ul{\tt Spheno}
\begin{itemize}
\item[Authors:] W.~Porod
     \cite{Porod:2003um}
\item[Description:] evaluates NMFV MSSM parameters from GUT scale input
                    and some flavour obs.\
\item[Models:] NMFV MSSM
\item[Input:] SLHA2 input file
\item[Output:] SLHA2 output file
\item[Availability:] {\tt ific.uv.es/$\sim$porod/SPheno.html} , 
                     manual available
\end{itemize}

\end{enumerate}


\subsubsection{Combination of flavour physics and high-energy tools}
\label{sec:mastertool}

It is desirable to connect different codes (e.g.\ working in
the (N)MFV MSSM, as given in the previous subsection) to each other. 
Especially interesting is the combination of codes that provide the evaluation
of (low-energy) flavour observables and others that deal with high-energy 
(high $p_T$) calculations for the same set of parameters.
This combination would allow to test the ((N)MFV MSSM) parameter space with
the results from flavour experiments as well as from high-energy experiments
such as ATLAS or CMS. 

A relatively simple approach for the combination of different codes is their
implementation as sub-routines, called by a ``master code''.
This master code takes care of the correct defintion
of the input parameters for the various subroutines. 
This would enable e.g.\ experimentalists to test
whether the parameter space under investigation is in agreement with various
existing experimental results from both, flavour and high-energy experiments.  

A first attempt to develop such a ``master code'' has recently been
started~\cite{mastertool}. So far the flavour physics code
(2)~\cite{Isidori:2006pk}
and the more high-energy observable oriented code 
{\tt FeynHiggs}~\cite{Heinemeyer:1998np,Degrassi:2002fi,Heinemeyer:2004by}
have been implemented as subroutines. The inclusion of further codes is
foreseen in the near future 
(see~\cite{Buchmueller:2007zk} for the latest developments).

The application and use of the master code would change once experimental data
showing a deviation from the SM predictions is available. This can come
either from the on-going flavour experiments, or latest (hopefully) from ATLAS
and CMS. If such a ``signal'' appears at the LHC, it has to be determined to
which model and to wich parameters within a model it can correspond. Instead
of checking parameter points (to be investigated experimentally) for their
agreement with experimental data, now a scan over a chosen model could be
performed. Using the master code with its subroutines each scan point can be
tested against the ``signal'', and preferred parameter regions can be obtained
using a $\chi^2$ evaluation. It is obvious that the number of evaluated
observables has to be as large as possible, i.e.\ the number of subroutines
(implemented codes) should be as big as possible.


\subsubsection{Fit tools}
\label{sec:fittools}

The analysis of the CKM matrix or the Unitarity Triangle (UT) requires
to combine several measurements in a consistent way in order to bound
the range of relevant parameters.


\subsubsubsection{The UTfit package}

The first approach derives bounds on the parameters $\bar{\rho}$ and
$\bar{\eta}$, determining the UT. 
The various observables, in particular $\epsilon_K$, which
parameterizes CP violation in the neutral kaon sector,
the sides of the UT $\left | V_{ub}/V_{cb} \right |$,
$\Delta m_d$, $\Delta m_d/\Delta m_s$, and the angles $\beta$, $\alpha$
and $\gamma$, can be expressed as functions of $\bar{\rho}$ and $\bar{\eta}$,
hence their measurements individually define probability regions in the
($\bar{\rho}$, $\bar{\eta}$) plane. Their combination can be achieved in a
theoretically sound way in the framework of the Bayesian approach
\cite{Ciuchini:2000de}.

Each of the functions relates a constraint $c_j$
(where $c_j$ stands for $\epsilon_K$, $\left | V_{ub}/V_{cb} \right |$, etc.)
to $\bar{\rho}$ and $\bar{\eta}$, via a set of parameters ${\mathbf x}$,
where ${\mathbf x} =\{x_1, x_2, \ldots, x_N\}$ stands for all
experimentally determined or theoretically calculated
quantities on which the various $c_j$ depend,
\begin{equation}
c_j=c_j(\bar{\rho},\bar{\eta}; {\mathbf x}).
\label{eq:c_j}
\end{equation}
The quantities $c_j$ and ${\mathbf x}$ are affected by several uncertainties,
which must be properly taken into account.
The final p.d.f. obtained
starting from a flat distribution of $\bar{\rho}$ and $\bar{\eta}$ is
\begin{equation}
f(\bar{\rho},\bar{\eta}) \propto
 \int
\prod_{j=1,{\rm M}}f_j(\hat{c}_j\,|\,\bar{\rho},\bar{\eta},{\mathbf x})
\prod_{i=1,{\rm N}}f_i(x_i)\,\mbox{d}{x_i}\, .
\label{eq:flat_inf}
\end{equation}
The integration can be done by Monte Carlo methods. There are several ways
to implement a Monte Carlo integration, using different
techniques to generate events.

The {\rm UT}{\it fit} Collaboration has developed a software package, written
in C++, that implements such a Bayesian Monte Carlo approach with the
aim of performing the UT analysis. A considerable effort has been spent
in order to achieve an optimal Monte Carlo generation efficiency.
All the recent analyses published by the Collaboration are based on this
package \cite{Bona:2005vz,Bona:2005eu,Bona:2006sa,Bona:2006ah,Bona:2007qt}.

The {\rm UT}{\it fit} code includes an interface to import job options from a
set of configuration files, an interface for storing the relevant p.d.f.s
inside ROOT histograms, tools for generating input quantities, the p.d.f.s
of which cannot be expressed in simple analytical form but must be numerically
defined - e.g. the current
measurements of $\alpha$ and $\gamma$ - and tools for plotting
one-dimensional p.d.f.s and two-dimensional probability
regions in the ($\bar{\rho}$, $\bar{\eta}$ plane).
The {\rm UT}{\it fit} code can be easily re-adapted to solve
any kind of statistical problem that can be formalized in a Bayesian
inferential framework.


\subsubsubsection{The CKMFitter package}

Another, somewhat different approach is followed by
CKMFitter, an international group of experimental and theoretical
particle physicists. Its goal is the phenomenology of the CKM matrix by
performing a global analysis: 
\begin{itemize}
\item within the SM, by quantifying the agreement between the data
  and the theory, as a whole; 
\item within the SM, by achieving the best estimate of the
  theoretical parameters and the not yet measured observables; 
\item within an extended theoretical framework, e.g.\ SUSY, by
  searching for specific signs of new physics by quantifying the agreement
  between the data and the extended theory, and by pinning down additional
  fundamental and free parameters of the extended theory. 
\end{itemize}
The CKMfitter package is entirely based on the frequentist approach. The
theoretical uncertainties are modeled as allowed ranges (Rfit approach) and no
other \textit{a priori} information is assumed where none is available. More
detailed information is provided in Ref.~\cite{Charles:2004jd} and on the
CKMfitter website~\cite{CKMFitterWWW}. 

The source code of the CKMfitter package consists of more than 40,000 lines of
Fortran code and 2000 lines of C++ code. It is publicly available on the
CKMfitter website. Over the years, the fit problems became more and more
complex and the CPU time consumption increased. The global fit took about 20
hours (on one CPU). A year ago, it was decided to move to Mathematica [gain:
analytical vs. numerical methods]: the global fit takes now 12 minutes. For
the plots, we moved also from PAW with kumac macros to ROOT.



\newpage \section{Weak decays of hadrons and QCD}
\label{sec:hu}


\subsection{Overview}

QCD interactions, both at short and long distances, necessarily
modify the amplitudes of quark flavour processes. These interactions
need to be computed sufficiently well in order to determine the
parameters and mechanisms of quark flavour physics from the weak
decays of hadrons observed in experiment.
The standard framework is provided by the effective weak Hamiltonians
\begin{equation}\label{eq:heffcq}
{\cal H}_{eff}\sim\sum_i C_i Q_i\, , 
\end{equation}
based on the operator product expansion and the
renormalization group method. The Wilson coefficients $C_i$ include all
relevant physics from the highest scales, such as the weak scale $M_W$,
or some new physics scale, down to the appropriate scale of a given
process, such as $m_b$ for $B$-meson decays. This part is theoretically
well under control. Theoretical uncertainties are dominated by the
hadronic matrix elements of local operators $Q_i$.
Considerable efforts are therefore devoted to calculate, estimate,
eliminate or at least constrain such hadronic quantities in flavour
physics applications.

This section reviews the current status of theoretical methods to treat 
the strong interaction dynamics in weak decays of flavoured mesons,
with a particular emphasis on $B$ physics. Specific aspects of $D$-meson
physics will be discussed in \ref{sec:charm}, kaons will be considered
in \ref{sec:kaon}.

The theory of charmless two-body $B$ decays and the concept of factorization 
are reviewed in \ref{subsec:exbdec}.
The status of higher-order perturbative QCD calculations in this field
is described. Universal properties of electromagnetic radiative
effects in two-body $B$ decays, which influence precision studies and
isospin relations, are also discussed here.
Factorization in the heavy-quark limit simplifies the matrix elements
of two-body hadronic $B$ decays considerably. In this framework certain
nonperturbative input quantities, for instance $B$-meson transition
form factors, are in general still required.
QCD sum rules on the light cone (LCSR) provide a means to compute
heavy-to-light form factors at large recoil ($B\to\pi$, $B\to K^*$, \etc). 
The results have applications for two-body hadronic as well 
as rare and radiative $B$-meson decays. This subject is treated
in section~\ref{subsec:lcsr}.
Complementary information can be obtained from lattice QCD,
a general approach, based on first principles,
to compute nonperturbative parameters of interest to quark flavour
physics. Decay constants and form factors (at small recoil) are
among the most important quantities. Uncertainties arise from the
limitations of the practical implementations of lattice QCD.
A critical discussion of this topic and a summary of results
can be found in section~\ref{subsec:latqcd}.

%
%
%
%
%
%
%
%
%
%
%
%
%
%
%
%

\newpage \subsection{Charmless two-body $B$ decays}\label{subsec:exbdec}

\subsubsection{Exclusive decays and factorization}
\label{subsubsec:fact}

The calculation of branching fractions and CP asymmetries for charmless 
two-body $B$ decays is rather involved, due to the interplay of various 
short- and long-distance QCD effects. 
Most importantly, the hadronic matrix elements of the relevant
effective Hamiltonian ${\cal H}_{eff}^{\Delta B=1}$ \cite{Buchalla:1995vs}
cannot readily be calculated from first principles. 
The idea of factorization is to disentangle short-distance
QCD dynamics from genuinely non-perturbative hadronic effects.
In order to quantify the hadronic uncertainties resulting from 
this procedure we have to
\begin{itemize}
 \item establish a factorization formula in quantum field theory,
 \item identify and estimate the relevant hadronic input parameters.
\end{itemize}

\subsubsubsection{Basic concepts of factorization}

We consider generic charmless $B$ decays into a pair of mesons,
$B\to M_1 M_2$, where we may think of $B\to\pi\pi$ as a typical example.
The operators $Q_i$ in the weak Hamiltonian
can be written as the local product of quark currents (and
electro- or chromomagnetic field strength tensors), generically denoted
as $J_i^{a,b}$. In naive factorization one assumes that also on 
the hadronic level the matrix element can be written as a product,
\begin{equation}
C_i(\mu) \, \langle M_1 M_2| Q_i |B\rangle
  \approx 
  {C_i(\mu)} \ \langle M_1| J_i^a |B\rangle
     \ \langle M_2 |J_i^b|0\rangle + (M_1 \leftrightarrow M_2)
\label{eq:naive}
\end{equation}
where $C_i(\mu)$ are Wilson coefficients, 
and the two matrix elements (if not zero) define the $B \to M$ form factor 
and the decay constant of $M$, respectively. The naive factorization formula
(\ref{eq:naive}) cannot be exact, because possible QCD interactions
between $M_2$ and the other hadrons are neglected.
On the technical level, this is reflected by an unmatched dependence on the
factorization scale $\mu$. 

\begin{figure}[ht]
\begin{center}
(a) \includegraphics[width=0.27\textwidth]{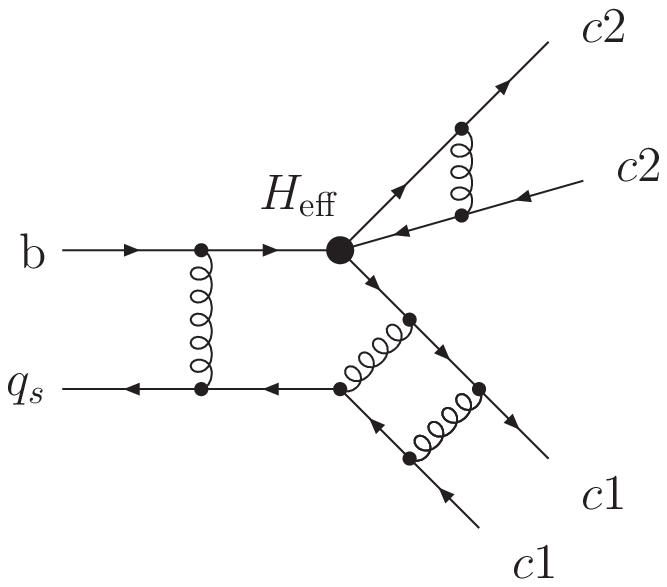}  \ \ \ \
(b) \includegraphics[width=0.27\textwidth]{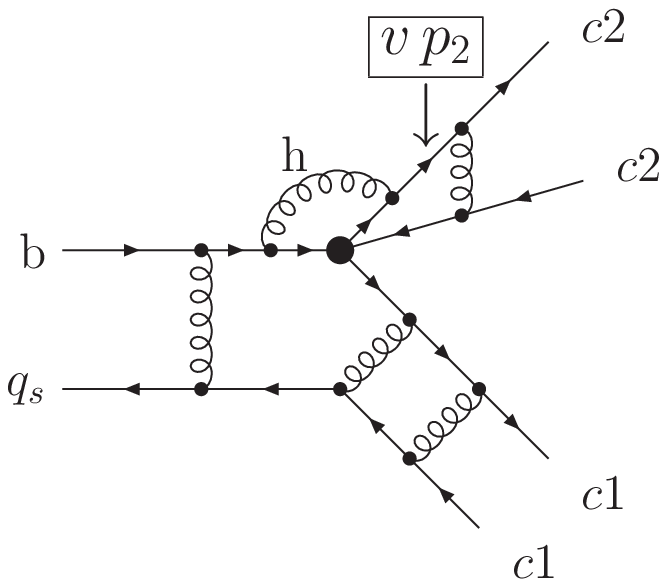}  \ \ \ \
(c) \includegraphics[width=0.27\textwidth]{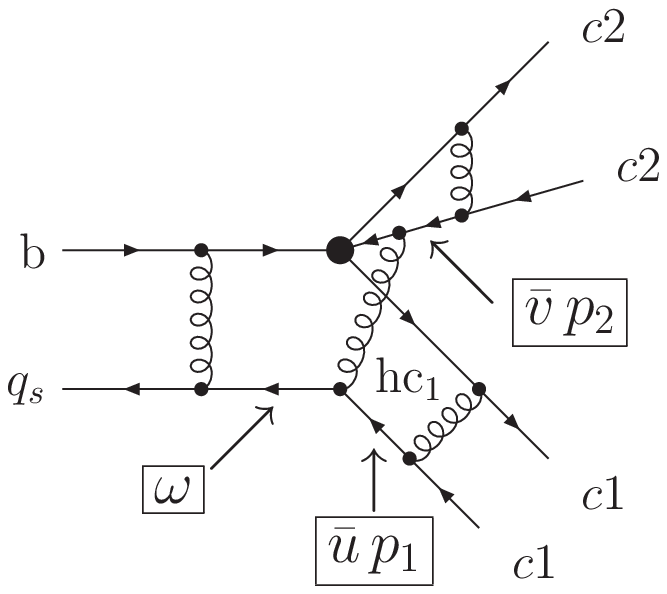} 
\caption{Sample diagrams for QCD dynamics in $B \to M_1 M_2$ transition:
(a) naive factorization, (b) vertex correction, sensitive to
the momentum fraction $v$ of collinear quarks inside the emitted pion, 
(c) spectator interactions, sensitive to the momenta of collinear
quarks in both pions and of the soft spectator in the $B$\/-meson.}
\label{fig:plots}
\end{center}
\end{figure}

In order to better understand the internal dynamics in the $B \to M_1 M_2$
transition, it is useful to classify the external degrees of freedom according 
to their typical momentum scaling in the $B$\/-meson rest frame:
\begin{center}
 \begin{tabular}{lcl}
    {heavy $b$ quark: $\ p_b \ \simeq \ m_b \, (1,0_\perp,0)$},
    && 
{constituents of $M_1$: $\ p_{c1} \simeq 
     u_i \, m_b/2 \, (1,0_\perp,+1)$} 
    \\
    {soft spectators:  $\ p_s \ \sim \ {\cal O}(\Lambda)$ },
    && 
{constituents of $M_2$: $\ p_{c2} \simeq v_i \, m_b/2 \, (1,0_\perp,-1)$} 
  \end{tabular}
\end{center}
where $\Lambda$ is a typical hadronic scale of the order of a
few 100~MeV. The index ${}_\perp$ denotes the directions in the
plane transverse to the two pion momenta and $u_i,v_i$ are momentum
fractions satisfying $0\leq u_i,v_i \leq 1$.
Interactions of particles with momenta 
$p_1$ and $p_2$ imply internal virtualites of order 
$(p_1 \pm p_2)^2$.
In Table~\ref{tab:modes} we summarize the situation
for the possible interactions between the $B$\/-meson and
pion constituents. We observe the emergence of two 
kinds of short-distance modes,
\begin{itemize}
  \item hard modes with invariant mass of order $m_b$,
  \item hard-collinear modes with energies of order $m_b/2$ and
        invariant mass of order $\sqrt{\Lambda m_b}$.
\end{itemize}
The systematic inclusion of these effects requires a simultaneous
expansion in $\Lambda/m_b$ and $\alpha_s$. The leading term in
the $\Lambda/m_b$ expansion can be written as 
\cite{Beneke:1999br,Beneke:2000ry}
\begin{eqnarray}      \label{eq:factform1}
     \langle M_1 M_2 | Q_i | B \rangle &=&
       F^{B M_1} f_{M_2} \int \! dv\, T^{\rm I}_i(v) \phi_{M_2}(v)
       \quad + (M_1 \leftrightarrow M_2)     \nonumber \\
&&       +\; \hat f_B f_{M_1} f_{M_2} \int \! d\omega\,du\,dv\, 
                           T^{\rm II}_i(u, v, \omega)
           \phi_{B_+}(\omega) \phi_{M_1}(u) \phi_{M_2}(v).
\end{eqnarray}
The functions $\phi_M$ and $\phi_{B_+}$
denote process-independent light-cone distribution amplitudes (LCDA) for
light and heavy mesons, respectively,
$f_M$, $\hat f_B$ are the corresponding decay constants, 
and $F^{BM}$ is a $B\to M$ QCD form factor at $q^2=0$. 
These quantities constitute the hadronic input.
The coefficient function $T_i^{\rm I}$ 
contains the effects of hard vertex corrections as in 
Fig.~\ref{fig:plots}(b).
$T_i^{\rm II}={\cal O}(\alpha_s)$ 
describes the hard and hard-collinear spectator interactions
as in Fig.~\ref{fig:plots}(c). 
The explicit scale dependence of the hard and hard-collinear 
short-distance functions $T_i^{\rm I}$, $T_i^{\rm II}$ matches the one 
from the Wilson coefficients and the distribution amplitudes. 
The formula~(\ref{eq:factform1}) holds for light
flavour-nonsinglet pseudoscalars or longitudinally polarized vectors
up to $1/m_b$ power corrections which do not, in general, factorize. 
Naive factorization, Fig.~\ref{fig:plots}(a),
is recovered in the limit $\alpha_s \to 0$ and $\Lambda/m_b \to 0$,
in which $T_i^{\rm I}$ reduces to $1$.

\begin{table}[t]
\caption{External momentum configurations and their interactions
  in $B \to M_1 M_2$.}
\label{tab:modes}
\begin{center}
\begin{tabular}{c|cccc}
\hline \hline
     & {heavy} &  {soft} & { coll$_1$} & 
    {coll$_2$} \\
\hline
 {heavy} 
     & --   & heavy & {hard} & {hard} \\
 {soft} & heavy & soft & {hard-coll$_1$} & {hard-coll$_2$} \\
 {coll$_1$}
     & {hard} & {hard-coll$_1$} & coll$_1$ & {hard} \\
 {coll$_2$}
     & {hard}  &{ hard-coll$_2$} & {hard} & coll$_2$ 
\\
\hline\hline
  \end{tabular}
\end{center}
\end{table}

\subsubsubsection{QCD factorization and soft-collinear effective theory (SCET)}

The factorization formula (\ref{eq:factform1}) can also be understood in
the context of an effective theory for soft-collinear interactions
(SCET), see for instance 
Refs.~\cite{Bauer:2000yr,Bauer:2001yt,Chay:2003zp,Bauer:2004tj}. 
Here the short-distance 
functions $T_i^{\rm I,II}$ arise as matching coefficients between
QCD and the effective theory. The effective theory for 
$B \to M_1 M_2$ decays is constructed in two steps. As a consequence, 
the short-distance function $T_i^{\rm II}$ can be further factorized into a 
hard coefficient $H_i^{\rm II}$ and a hard-collinear jet function $J$
\begin{equation}\label{eq:thj}
T^{\rm II}_i(u,v,\omega) = \int\!dz\,H^{\rm II}_i(v,z) J(z,u,\omega) .
\end{equation}
$H^{\rm II}_i$ and $J$ comprise (respectively) the contributions associated 
with the hard scale $\mu_b \sim m_b$
and the hard-collinear scale $\mu_{\rm hc} \sim \sqrt{m_b \Lambda}$
from Feynman diagrams that do involve the spectator and cannot be
absorbed into $F^{B M}$. 
The effective theory can be used to determine the hard-collinear
contributions and to resum, if desired, parametrically large logarithms
$\ln \mu_b/\mu_{\rm hc}$ by renormalization group methods.
We emphasize that the theoretical basis for the (diagrammatic)
factorization approach and SCET is {\em the same}. 
The factorization formula~(\ref{eq:factform1}) was originally derived by a 
power-counting analysis of momentum regions of QCD Feynman diagrams and the 
resulting convolutions~\cite{Beneke:1999br,Beneke:2000ry}.
However, in SCET the formulation of factorization proofs,
the classification of power corrections of order $\Lambda/m_b$, 
the emergence of approximate symmetries, etc.\ may be more 
transparent~\cite{Chay:2003ju,Bauer:2004tj}.

\subsubsubsection{QCD factorization vs.\ ``pQCD approach''}

The so-called ``pQCD approach'' 
\cite{Keum:2000wi} follows an alternative approach
to understand the strong dynamics in charmless $B$\/-decays. 
In contrast to QCD factorization, where the $B$ meson form factors
as well as a certain class of power corrections are identified 
as ``non-factorizable'' quantities of order $(\alpha_s)^0$,
the pQCD approach describes all contributions to the hadronic
matrix elements in terms of ${\cal O}(\alpha_s)$ hard-scattering 
kernels and non-perturbative wave functions. This is achieved by introducing
additional infrared prescriptions which include an exponentiation
of Sudakov logarithms and a phenomenological model for transverse
momentum effects. The discussion of parametric and systematic
theoretical uncertainties in the pQCD approach is more difficult, 
because a complete NLO (i.e.\ ${\cal O}(\alpha_s^2)$) analysis 
of non-factorizable effects has not yet been performed, 
and because independent information on the hadronic input functions 
is not available. We will therefore not attempt a detailed review
here, but instead refer to a recent phenomenological analysis 
\cite{Li:2005kt} for details.

\subsubsection{Theoretical uncertainties}

\subsubsubsection{Status of perturbative calculations}

The calculation of the coefficient functions 
$T_i^{\rm I,II}$ in SCET involves the determination of perturbative
matching coefficients as well as of anomalous dimensions for
effective-theory operators. 
The matching coefficients at order $\alpha_s$ have been calculated in
the original BBNS papers \cite{Beneke:1999br,Beneke:2001ev}. 
The 1-loop jet function entering $T_i^{\rm II}$ 
has been determined in 
\cite{Hill:2004if,Becher:2004kk,Kirilin:2005xz,Beneke:2005gs}. 
NLO results for the spectator scattering function at order $\alpha_s^2$
have been reported in \cite{Beneke:2005vv} and will be further discussed
in section \ref{subsubsec:hoqcd} below. One important
outcome of these investigations is that the perturbative expansion
at the hard-collinear scale seems to be reasonably well behaved, and
the uncertainty associated with the factorization-scale dependence
is under control.

\subsubsubsection{Hadronic input from non-perturbative methods}

Most of the theoretical information on $B$\/-meson form factors
(at large recoil) and light-cone distribution amplitudes comes from the 
QCD sum rule approach,
see Ref.~\cite{Colangelo:2000dp} and references
therein for a review. State-of-the art predictions for decays into
light pseudoscalars or vector mesons can be found
in Refs.~\cite{Ball:2004ye,Ball:2004rg,Ball:2005vx} and 
section \ref{subsec:lcsr}. Typically one finds
15-20\% uncertainties for form factors at $E=E_{\rm max}$ and
the $1/u$ moment of distribution amplitudes.
Recently, an alternative procedure has been proposed
\cite{DeFazio:2005dx} (see also Refs.~\cite{Chay:2005gh,Khodjamirian:2005ea}), 
where sum rules are derived {\em within}\/ SCET at 
the hard-collinear scale. In particular, this approach
allows us to separate the ``soft'' contribution to $B$\/-meson 
form factors, which is found to be dominating over the 
spectator-scattering term. 

Information on the light-cone distribution amplitude 
of the $B$\/-meson is encoded in the phenomenologically
relevant moments
\begin{equation}
\lambda^{-1}_B\equiv  \langle \omega^{-1}\rangle_B \equiv
  \int_0^\infty \frac{d\omega}{\omega} \, \phi_B(\omega,\mu)
\,, \qquad
  \sigma^{(n)}_B \,\langle \omega^{-1}\rangle_B
  \equiv 
 \int_0^\infty \frac{d\omega}{\omega} \, \ln^n\left[\frac{\mu}{\omega}\right] 
    \,\phi_B(\omega,\mu)
\end{equation}
A recent OPE analysis \cite{Lee:2005gz} finds
$ \lambda^{-1}_B = (2.09 \pm 0.24)~{\rm GeV}^{-1}$ and
$\sigma^{(1)}_B = 1.61 \pm 0.09$ at $\mu=1~$GeV. Similar results,
with somewhat larger uncertainties, have been obtained from sum
rules in Ref.~\cite{Braun:2003wx}.

\subsubsubsection{BBNS approach vs.\ BPRS approach}

So far, we have only considered the leading term in the $1/m_b$
expansion. Comparison with experimental data as well as
(model-dependent) estimates show that for certain decay
topologies power corrections may not be negligible. 
Different options for dealing with these (non-factorizable) 
contributions lead to some ambiguity in the phenomenological
analyses. The two main players are the ``BBNS approach''
\cite{Beneke:2001ev,Beneke:2002jn,Beneke:2003zv} and the
``BPRS approach'' \cite{Bauer:2004tj,Williamson:2006hb}.
A qualitative comparison of the different assumptions is
given in Table~\ref{tab:comp}. For more details see 
section \ref{subsubsec:hoqcd}, the original publications and the 
controversial discussion in \cite{Beneke:2004bn}.

\begin{table}[t]
\begin{center} 
\caption{Comparison of different phenomenological assumptions in
BBNS and BPRS approaches}
\label{tab:comp}
\begin{tabular}{p{0.25\textwidth}| p{0.33\textwidth} p{0.33\textwidth} }
\hline \hline
 & BBNS & BPRS \\
\hline
charm penguins
  & included in hard functions 
  & left as complex fit parameter $\Delta_P$
\\[0.2em]
spectator term & perturbative factorization    
& fit to data \newline 
    {\small (two real-valued quantities $\zeta$ and $\zeta_J$)}
\\[0.2em]
ext.\ hadronic input
   & form factor and LCDA \newline 
    {\small (different scenarios)} 
   & LCDA for light meson
\\[0.2em]
power corrections
  &  model-dependent estimate \newline
    {\small (complex functions $X_A$ and $X_H$)}
  &  part of systematic uncertainties
\\
\hline \hline
\end{tabular}

\end{center}

\end{table}
The main obstacle in this context is the quantitative
explanation of strong phases from final-state rescattering
effects. The factorization formula predicts these phases
to be either perturbative (and calculable) or power-suppressed.
This qualitative picture has also been confirmed by a recent
sum rule analysis \cite{Khodjamirian:2005wn}. 
However, a model-independent approach to calculate the genuinely
non-perturbative rescattering effects is still lacking.

\subsubsubsection{Flavour symmetries}

It is known for a long time
(see for instance \cite{Zeppenfeld:1980ex,Gronau:1990ka,Nir:1991cu})
that approximate flavour symmetries in QCD can
be used to relate branching fractions and CP asymmetries in different
hadronic decay channels. In this way the hadronic parameters can be
directly extracted from experiment. 
For instance, in case of $B \to \pi\pi, \pi\rho,\rho\rho$ decays, 
the isospin analysis provides a powerful tool to constrain the 
CKM angle $\alpha$ in the SM
(see Ref.~\cite{Charles:2006vd} for a recent discussion).
Isospin violation from the small quark mass difference $m_u-m_d$ and
QED corrections are usually negligible. Still one has to keep in mind
that long-distance radiative QED effects can be enhanced by large logarithms
$\ln M_B/E_\gamma$ and compete with short-distance isospin violation
from electroweak penguin operators in ${\cal H}_{eff}$. 
For instance, it has recently been shown 
\cite{Baracchini:2005wp} (see section \ref{subsubsec:qedcor} below) that the 
inclusion of soft photon radiation in charged $B \to \pi\pi, \pi K$ 
decays can give up to 5\% corrections, depending on the experimental cuts.
Including hadronic states with strange quarks, one can use 
flavour-SU(3) to get even more constraints. In general,
one expects corrections to the symmetry limit to be not larger 
than 30\% (with the possible exception of potentially large differences in
non-perturbative rescattering phases),
see for instance the sum-rule analysis in \cite{Khodjamirian:2003xk}.
In the long run, one should also aim to constrain first-order
$SU(3)$ corrections directly from experimental data.

\subsubsection{NNLO QCD corrections}\label{subsubsec:hoqcd}

NNLO QCD corrections to the heavy-quark expansion of hadronic matrix elements
for two-body charmless hadronic $B$-decays can be phenomenologically
relevant and are important to assess the validity and perturbative
stability of the factorization framework.
This section gives a concise account of available results and their
phenomenological impact.

\subsubsubsection{Hard and hard-collinear matching coefficients}
The hard coefficients $T^{\rm I}_i$ and $H^{\rm II}_i$ introduced
in \ref{subsubsec:fact} (eqs. (\ref{eq:factform1}) and (\ref{eq:thj}))
are found by matching the leading momentum
dependence of (respectively) QCD four- and five-point functions
with a $Q_i$ insertion
to operators in SCET$_{\rm I}$ given by products of a light (anti-)collinear
quark bilinear and a heavy-light current. Schematically,
\begin{eqnarray}
  Q_i &=& \int\! dt\, T^{\rm I}_i(t) [\bar \chi(t n_-) \chi(0)]
     \Big[C_{A0}\, [\bar \xi(0) h_v(0)] + \frac{1}{m_b}\int\!ds\, C_{B1}(s) [\bar \xi(0)
  \, D_{\perp \rm hc1}(s n_+) h_v(0)]\Big]
\nonumber \\
&&   + \frac{1}{m_b} \int\! dt\, ds\,
      H^{\rm II}_i(t,s)\, [\bar \chi(t n_-) \chi(0)] [\bar \xi(0)
  \, D_{\perp \rm hc1}(s n_+) h_v(0)] ,
\end{eqnarray}
where certain Wilson lines and Dirac structures have been suppressed.
The particular choice of heavy-light current in the first line is
designed to reproduce the full QCD (not SCET) form factors;
other choices of operator basis as, for instance,
in the ``SCET approach''~\cite{Bauer:2004tj},
simply result in a reshuffling of contributions between
the $T^{\rm I}_i$ and $H^{\rm II}_i$ terms. 
The product structure of either term together with the absence of
soft-collinear interactions from the SCET$_{\rm I}$ Lagrangian at leading power
suggests factorization of both terms' hadronic matrix elements
into a light-cone distribution amplitude $\langle M_2 | [\bar \chi
\chi] | 0 \rangle \propto \phi_{M_2}$ and (respectively) the QCD form
factor $F^{B M_1}$ and a SCET$_{\rm I}$ nonlocal
``form factor'' $\Xi^{B M_1}(s)$~\cite{Beneke:2003pa}. 
This expectation is indeed borne out by the finiteness of the
convolutions, found in all available computations.


The jet function $J$ (see eq. (\ref{eq:thj}))
arises in matching the $B1$-type
current from SCET$_{\rm I}$ onto SCET$_{\rm II}$ and is known to
NLO~\cite{Hill:2004if,Becher:2004kk,Kirilin:2005xz,Beneke:2005gs}.
This matching takes the form (in position space)
\begin{equation}
 \int \! d^{\,4} x\, T\Bigg({\cal L}^{(1)}_\mathrm{SCET_I}(x)
      [\bar \xi(0) D_{\perp \rm hc1}(s n_+) h_v(0)] \Bigg)=
\int \! dw\, dr J(s, r, w)  [\bar \xi(r n_+) \xi(0)] [\bar q_s(w n_-) h_v(0)],
\end{equation}
where we again have suppressed Dirac structures and Wilson lines.
Fourier transforming with respect to $s$, $r$, $w$ results in
$J(z,u,\omega)$ entering eq.~(\ref{eq:factform1}).

At leading power, all one-loop corrections to $H^{\rm II}_i$ and $J$ and 
part of the two-loop contributions to $T^{\rm I}_i$ are now available.
The current-current corrections to $H^{\rm II}_i$ for
the $V\!\!-\!A \times V\!\!-\!A$ operators ($i=1,2$)
have been found in Refs.~\cite{Beneke:2005vv,Kivel:2006xc,Pilipp2007}.
The imaginary parts of the corresponding two-loop contributions to 
$T^{\rm I}_i$ have been computed in Ref.~\cite{Bell:2007tv,Bell:2006tz}. 
These are sufficient to obtain the topological tree amplitudes $a_1$ and $a_2$,
involving the large Wilson coefficients $C_1 \sim 1.1$ and $C_2 \sim
-0.2$, at NNLO up to an ${\cal O}(\alpha_s^2)$ correction to the real
part of $T^{\rm I}_i$. In particular, the imaginary part of $a_{1,2}$
is now fully known at ${\cal O}(\alpha_s^2)$. 
As it is first generated at ${\cal O}(\alpha_s)$,
this represents a first step towards an NLO prediction of direct CP
asymmetries in QCD factorization.
Spectator-scattering corrections from the remaining 
$V\!\!-\!A \times V\!\!+\!A$ operators, as well as penguin contractions 
and magnetic penguin insertions, have been computed
in Ref.~\cite{Beneke:2006mk}. Together they
constitute the QCD penguin amplitudes $a_4^p$ ($p=u,c$) and the
colour-allowed and colour-suppressed electroweak penguin amplitudes 
$a^p_9 \pm a^p_7$ and $a^p_{10}$, where the sign in front of $a^p_7$ 
depends on the spins of the final-state mesons, and certain numerically 
enhanced power corrections ($a^p_{6,8}$, annihilation, etc.) are
omitted (see, however, section~\ref{subsubsubsec:pheno}).

\subsubsubsection{Phenomenological impact and final remarks}
\label{subsubsubsec:pheno}
Numerical estimates of the $a_i$ and their uncertainties require
estimating $1/m_b$ corrections, some of which are ``chirally
enhanced'' for pseudoscalars in the final state.
Of these, the scalar penguin $a_6^p$, and its electroweak analog $a_8^p$,
happen to factorize at ${\cal O}(\alpha_s)$.
NNLO corrections are not known and their factorization is an open
question. Here we use the known ${\cal O}(\alpha_s)$ results.
Annihilation and twist-3 spectator interactions do not factorize
already at LO (${\cal O}(\alpha_s)$). The former are not
included in any $a_i$ but enter the physical decay amplitudes. The
latter have flavour structure identical to the $a_i$ and are by
convention included as estimates.
For the colour-allowed and colour-suppressed tree amplitudes
$a_1$ and $a_2$, we find
\begin{eqnarray}
a_1(\pi\pi) &=& 1.015 + [0.025 + 0.012i]_V  + [? +  0.027 i]_{VV}
   \nonumber \\
   &&     -\,\left[\frac{r_{\rm sp}}{0.485} \right]
   \Big\{ [0.020]_{\rm LO} + [0.034 + 0.029i]_{HV} + [0.012]_{\rm tw3} \Big\}
   \nonumber \\
   &=& 0.975^{+0.034}_{-0.072} + (0.010^{+0.025}_{-0.051})i,  \label{eq:a1}
\\[0.2cm]
a_2(\pi\pi) &=& 0.184 - [0.153 + 0.077i]_V + [? - 0.049 i]_{VV}
   \nonumber \\
   && + \,\left[ \frac{r_{\rm sp}}{0.485} \right]
   \Big\{ [0.122]_{\rm LO} + [0.050 +0.053i]_{HV} + [0.071]_{\rm tw3} \Big\}
   \nonumber \\
   &=& 0.275^{+0.228}_{-0.135} + (-0.073^{+0.115}_{-0.082})i.  \label{eq:a2}
\end{eqnarray}
In each expression, the first line gives the form-factor (vertex)
contribution, the second line the  spec\-ta\-tor-scattering contribution,
and the third line their sum with an estimate of the theoretical
uncertainties due to hadronic input
parameters (form factors, LCDAs, quark masses), power corrections, and
neglected higher-order perturbative corrections as explained in
detail in Ref.~\cite{Beneke:2006mk}, where also the input parameter
ranges employed here are given.
The first two lines in Eqs.~(\ref{eq:a1}) and~(\ref{eq:a2})
are decomposed into the tree (naive factorization,
$\alpha_s^0$), one-loop ($V$), and two-loop ($VV$)
vertex correction (the question marks denote unknown real parts);
tree ($\alpha_s$, LO), one-loop ($\alpha_s^2$, $HV$), and
twist-3 power correction (tw3) to spectator scattering.
The prefactor $r_{\rm sp} = (9 f_{M_1} \hat f_B)/(m_b\, F^{B M_1}
\lambda_B)$ encapsulates
the bulk of the hadronic uncertainties of the spectator-scattering term.
Numerically, for $a_1$ the corrections are, both
individually and in their sum, at the few-percent level, such that
$a_1$ is very close to 1 and to the naive-factorization result.
On the other hand, individual corrections to $a_2$ are large, with a near
cancellation between naive factorization and the one-loop vertex
correction. $a_2$ is thus especially sensitive to spectator
scattering and to higher-order vertex corrections. That these are all
important is seen from the $VV$, LO, and $HV$ numbers
in eq.~(\ref{eq:a2}).

Analogous expressions can be given for the remaining amplitude
parameters $a_3^p \dots a_{10}^p$~\cite{Beneke:2006mk}, except that
no two-loop vertex corrections are known. Qualitatively,
NNLO spectator-scattering corrections are as important for the
leading-power, but small (electroweak) penguin amplitudes
$a_{3,5,7,10}^p$ as they are for $a_2$ but are found to be small for
the large electroweak penguin amplitude $a^p_{9}$.
Corrections to the QCD penguin amplitude $a_4^p$ are also small,
in spite of the involvement of the large Wilson
coefficient $C_1$. This is due to a numerical cancellation, which may
be accidental. The scalar QCD and electroweak penguin
amplitudes $a_6^p$ and $a_{8}^p$ are power suppressed but
``chirally enhanced''. NNLO corrections to them are currently
unknown but might involve sizable contributions proportional
to $C_1$, unless a similar numerical cancellation as in the case of
$a_4^p$ prevents this. This would be relevant for direct CP
asymmetries in the $\pi K$ system and elsewhere. For a more complete
discussion, see Ref.~\cite{Beneke:2006mk}.

A good fraction of NNLO corrections to the QCD
factorization formula are now available. While the perturbation expansion
is well-behaved in all cases, some of these corrections turn out
to be significant,
particularly those to the colour-suppressed tree and (electroweak) penguin
amplitudes. Further important corrections to the QCD and
colour-suppressed EW penguin amplitudes proportional to $C_1$ may
enter through the chirally-enhanced power corrections $a_6^p$ and
$a_{8}^p$, making their NNLO calculation an important goal.

\subsubsection{QED corrections to hadronic $B$ decays}\label{subsubsec:qedcor}

\subsubsubsection{Introduction}
The large amount of data collected so far at $B$ factories has allowed to 
reach a statistical accuracy on $B$ decays into pairs of (pseudo)scalars 
at a level where electromagnetic
effects cannot be neglected anymore\cite{Aubert:2006fh,Chao:2005ht}. 
On one hand, a correct simulation
of the unavoidable emission of photons from charged particles has to
be included in Monte Carlo programs in order to evaluate the correct
efficiency. On the other hand, a clear definiton of the effective cut
on (soft) photon spectra is essential for a consistent comparison both
between theory and experiments and between results from different experiments.

We discuss the theoretical and experimental treatment
of radiative corrections in hadronic $B$ decays. 
We present analytical expressions to describe the leading effects 
induced by both real and virtual (soft) photons in the generic 
process $H \rightarrow P_1 P_2(\gamma)$, where both $H$ and $P_{1,2}$ 
are scalar or pseudoscalar particles. We then discuss 
the procedures to be adopted in experimental 
analyses for a clear definition of the observables.

\subsubsubsection{The scalar QED calculation}
General properties of QED have been exploited in detail for most of the pure
electroweak processes or in general for processes that can be fully treated
in terms of perturbation theory. This is not the case of hadronic decays.
However, due to the universal character of infrared QED singularities, 
it is possible to estimate the leading ${\cal O}(\alpha)$ contributions to 
these processes within scalar QED, in the approximation of a point-like 
weak vertex. 

The most convenient infrared-safe observable related to the process 
$B \rightarrow P_1 P_2$ is the photon inclusive width
\begin{eqnarray}
\Gamma^{\rm incl}_{12}(E^{\rm max})= 
\Gamma(B\rightarrow P_1 P_2 +n\gamma)|_{\sum E_{\gamma} < 
E_{\gamma}^{\rm max} }= 
\Gamma_{12} + \Gamma_{12 + n \gamma}(E_{\gamma}^{\rm max})~,
\end{eqnarray}
namely, the width for the process $B \rightarrow P_1P_2$ accompanied by 
any number of (undetected) photons, with total missing energy less or equal to 
$E^{\rm max}$ in the $B$
meson rest frame. The infrared cut-off
$E_{\gamma}^{\rm max}$ can be the photon energy below which the state 
$| P_1 P_2 \rangle$ cannot be distinguished from the state 
$|P_1 P_2 +n \gamma \rangle$;
however, in principle it can also be chosen to be a high reference scale 
(up to the kinematical limit). 
At any order in perturbation theory we can decompose $\Gamma^{\rm incl}_{12}$ 
in terms of two theoretical quantities: the so-called non-radiative width, 
$\Gamma^0_{12}$, and the corresponding energy-dependent e.m.~correction 
factor $G_{12}(E_{\gamma}^{\rm max})$,
\begin{eqnarray}
\Gamma^{\rm incl}_{12}(E_{\gamma}^{\rm max}) = 
\Gamma^0_{12}(\mu) ~ G_{12}(E_{\gamma}^{\rm max} , \mu)~.
\label{eq:prod}
\end{eqnarray}
In the limit $E_{\gamma}^{\rm max} \ll M_B$ the electromagnetic 
correction factor can be reliably estimated 
within scalar QED.  
We define the non-radiative width $\Gamma^0_{12}(\mu)$ as
\begin{eqnarray}
\Gamma^0_{12}(\mu) &=& 
\frac{\beta}{16 \pi M_B}\left| {\cal A}_{B\rightarrow P_1P_2}(\mu) \right|^2~,
\label{eq:Gamma12} \\
\beta^2 &=& 
\left[ 1-\left(r_1+r_2\right)^2\right]\left[ 1- \left(r_1-r_2\right)^2\right]~,
\qquad r_{i} ~=~ \frac{m_i}{M_B}~,
\end{eqnarray}
namely the tree-level rate expressed in terms of the renormalized 
(scale-dependent) weak coupling.
Here the $m_i$ refer to the masses of the light mesons in the final state,
$M_B$ is the $B$-meson mass.   
The function $G_{12}(E_{\gamma}^{\rm max}, \mu)$  can be written as
\begin{eqnarray}
G_{12} (E, \mu) = 
1 + \frac{\alpha}{\pi} \left[ b_{12} \ln\left(\frac{ M_B^2 }{4E^2} \right)
+ F_{12} + \frac{1}{2} H_{12} + N_{12}(\mu) \right]~,
\label{eq:G12}
\end{eqnarray}
where $H_{12}$ represents the finite term arising from virtual corrections, 
and $F_{12}$ the energy-independent contribution generated by the 
real emission (here $E\equiv E^{\rm max}_\gamma$):
\begin{eqnarray}
\int_{E_\gamma < E }  ~\frac{d^3 \vec{k} }{(2\pi)^3 ~2 E_\gamma}~ 
\sum_{\rm spins} \left| 
\frac{{\cal A} (B \rightarrow P_1 P_2\gamma)}{
  {\cal A}(B \rightarrow P_1 P_2)} \right|^2
=~ \frac{\alpha}{\pi} \left[ 
b_{12} \ln\left(\frac{ m^2_\gamma }{4E^2 } \right)
+  F_{12} +{\cal O}\left( \frac{E}{M_B} \right)\,  \right]~.
\end{eqnarray}
As expected, after summing real and virtual corrections, the infrared 
logarithmic divergences cancel out in $G_{12} (E, \mu)$, giving rise
to the universal $\ln (M_{B}/E_{\gamma}^{\rm max})$ term.
The scale dependence contained in $N_{12}(\mu)$ cancels 
out in the product $\Gamma^0_{12} \times G_{12}$
due to the corresponding scale dependence of the weak coupling. 
For the explicit expressions of $F_{12}$,$H_{12}$ and $N_{12}$ and a more
detailed discussion of the $\mu$-dependence we refer to
\cite{Baracchini:2005wp}. The result thus obtained can be applied
to both $B$ and $D$ decays.

We finally give the results for $G_{+-}$ and $G_{+0}$ in the limit
$m_{1,2}$, $E\ll M_B$, which represents a convenient,
and very good approximation:
\begin{eqnarray}
G_{+-} &=& 1-\frac{\alpha}{\pi}
\left\{ \left[ 2\ln \epsilon
+1+\ln\left( 1- \delta^2 \right)  \right]
\ln\left( \frac{4{E}^{2}}{{M_B}^{2}} \right)
 -4\ln \epsilon +\frac{\pi^{2}}{3}+1 +{\cal O}(\delta)
\right\} \\
G_{+0} &=&  1 - \frac{\alpha}{\pi}
\left\{\left[ \ln \epsilon  +1
+\ln \left( 1+\delta \right) \right]
\ln  \left( \frac{4E^2}{M_B^2} \right)
-2\ln\epsilon +\frac{\pi^2}{6}-1 +{\cal O}(\delta)
\right\}
\end{eqnarray}
where
\begin{eqnarray}
\epsilon = \frac{m_1+m_2}{2 M_B}~, \qquad
\delta = \frac{m_1-m_2}{m_1 + m_2}~,
\end{eqnarray}
with $12=+-$, $+0$, respectively.
This approximation also serves to clarify the physical
relevance of the correction factors.
The logarithmic terms as well as the Coulomb correction
($\sim \pi^2$) are model-independent, well defined effects.
On the other hand, the remaining constant pieces ($\pm 1$) are not
meaningful in the absence of the proper UV matching, but they are
subdominant and numerically rather small.

\subsubsubsection{Inclusion of final state radiation effects in an 
experimental analysis}

We will discuss in particular the inclusion of final state radiation in
the analysis of rare $B$ decays at $B$ factories. In this kind of environment,
the efficiency is estimated through Monte Carlo simulation where QED effects
are taken into account using the PHOTOS simulation 
package\cite{Golonka:2005pn}.  
The first issue is then to check if the performances of the entire event
simulation chain are the ones expected from the theory. One can thus compare
the simulated $G_{12}(E^{\rm max}_{\gamma})$ function, as well as the energy
and angular distribution of the generated photons (whose analytical expression
can be found in \cite{Baracchini:2005wp}) and then, if needed, correct
the distributions on which efficiency and parametrization of the fit variables 
are evaluated. Then, particular care has to be taken in order to quote the
results in such a way that radiation effects can be disentagled. In principle, 
it would be necessary to select $B$ candidates with a specified maximum amount 
of ${\cal O} (100$ MeV) photon energy in the final states, a quantity which is
difficult to reconstruct in a $B$ factory context. Instead, one could define
the data sample selecting on an observed variable which can be clearly related
to the maximum allowed energy for photons $E_{\gamma}^{\rm max}$. The variable
$\Delta E = E^{\ast}_{B} - \sqrt{s}/2$, where $E^{\ast}_{B}$ is the 
reconstructed
$B$ candidate energy in the $e^+ e^-$ center of mass (CM) frame and 
$\sqrt{s}$ the total CM energy, is clearly suitable for this purpose.
The $\Delta E$ window chosen for
the analysis would then allow for the presence of radiated photons up to the 
chosen cut, providing the possibility of quoting results, \eg on branching 
fractions, with a defined cut on the soft photon spectrum. 
Once a result of this kind is
obtained, it is easy to extract the weak couplings -- which cannot be 
directly measured due to the intrinsic and unavoidable features of QED --
employing the theoretical calculation explained in the previous section.
This is very important, since the comparison between theoretical predictions
and experiments can be done more efficiently in terms of the weak couplings.
Moreover, a meaningful comparison between different experiments can only 
be done in terms of the weak couplings (non-radiative quantities) 
or in terms of the inclusive widths employing the same infrared cut-off.


\subsubsection{Outlook on future improvements}

The improvement of our quantitative understanding of hadronic
effects in charmless non-leptonic $B$\/-decays requires both
experimental and theoretical efforts:
\begin{itemize}
  \item Completion of the NNLO analyses for the 
        factorizable vertex and hard-scattering contributions
        to reduce the perturbative uncertainties.
  \item Further improvement in hadronic input parameters
        (form factors, LCDA) by non-perturbative methods,
        combined with experimental data on $B$\/- and
        $D$\/-meson decays.
  \item More systematic treatment of power-corrections.
  \item Better understanding of $SU(3)$\/-breaking effects
        in the analysis of $B_s$ and $B_{u,d}$ decays.
\end{itemize}
In the future, the main limitations will probably be due to
theoretical uncertainties in non-perturbative strong rescattering phases.

\newpage \subsection{Light-cone QCD sum rules}\label{subsec:lcsr}

\subsubsection{Distribution Amplitudes}
Light-cone wave functions or
distribution amplitudes (DA) are matrix elements defined near light-like 
separations connecting
hadrons to their partonic constituents. They are widely used in hard 
exclusive processes with high momentum transfer \cite{Chernyak:1983ej},
which are often dominated by light-like distances.
Formally they appear in the light-cone operator product expansion (LCOPE) and 
can be seen as the analogue of matrix elements of local operators in the 
operator product expansion (OPE). The terms in the OPE are ordered according 
to the dimension of the operators, the terms in the LCOPE according
to their twist, the dimension minus the spin. 
We shall discuss distribution amplitudes for light mesons, which are most 
relevant for the LHCb experiment
\footnote{There are of course other DA of interest. Baryon DA have recently 
been reviewed in \cite{Braun:2006hn}, the photon DA is treated in 
\cite{Ball:2002ps} and a recent lecture on the B-meson DA can be found in 
\cite{Grozin:2005iz}.}.
We shall take the $K(495)$ and the $K^*(892)$ as representatives for the 
light pseudoscalar and 
vector mesons\footnote{In the literature sometimes another phase convention
for the vector meson states is used, where 
$|V\rangle_{other} =i |V\rangle_{here}$.}.
 \begin{eqnarray}
    \langle 0 |\bar q(x) x_\mu\gamma^\mu\gamma_5 [x,0]s(0)
  |K(q)\rangle &=& i  f_K q \! \cdot \! x 
  \int_0^1 du\, e^{-i\bar uq  \cdot  x} \phi_K(u)\,  + 
                                        O(x^2,m_K^2) ,\nonumber\\
  \langle 0 |\bar q(x) x_\mu\gamma^\mu[x,0] s(0)
  |K^{*}(q,\lambda)\rangle &=& (\varepsilon^{(\lambda)}\cdot x)
  f_{K^*}  m_{K^*}\int_0^1 du\, e^{-i\bar uq \cdot x}
  \phi_K^\parallel(u) \,  + O(x^2,m_{K^*}^2) ,\label{eq:defDAs}  \\
  \langle 0 |\bar q(x)\sigma_{\mu\nu}[x,0]s(0)
  |K^{*}(q,\lambda)\rangle & = &
  i(\varepsilon^{(\lambda)}_\mu q_\nu -\varepsilon^{(\lambda)}_\nu q_\mu)
 f_{K^*}^\perp(\mu) \int_0^1 du\, e^{-i\bar uq \cdot x} \phi_K^\perp(u)\, + 
                                     O(x^2,m_{K^*}^2). \nonumber
 \end{eqnarray}
The vector $x_\mu$ is to be thought of as a vector close to the light-cone. 
The variable $u$ ($\bar u\equiv 1-u$) can be interpreted as the collinear 
momentum fraction carried by one of the constituent quarks in the meson.
Corrections to the leading twist come
from three sources: 
1. other Dirac-structures 
(e.g. $\langle 0 | \bar q(x) \gamma_5 [x,0] s(0) |K(q)\rangle$), 
2. higher Fock states 
(including an additional gluon) and 
3. mass and light-cone corrections as indicated in the equations above. 

The wave functions $\phi(u,\mu)$ are non-perturbative objects. 
Their asymptotic forms are known from perturbative QCD, 
$\phi(u,\mu) \stackrel{\mu \to \infty}{\to} 6 u\bar u$.
Use of one-loop conformal symmetry of massless QCD is made
by expanding in the eigenfunctions of the evolution kernel, the Gegenbauer
polynomials $C_n^{3/2}$,  
\begin{equation}
\phi(u,\mu) = 
6u\bar u\left(1 + \sum^\infty_{n=1} \alpha_n(\mu) C_n^{3/2}(2u-1)\right)\,,
\end{equation}
where the $\alpha_n$ are hadronic parameters, the Gegenbauer moments.
If $n$ is odd they vanish for particles with definite G-parity, 
e.g. $\alpha_{2n+1}(\pi) = 0$.
For the kaon $\alpha_{2n+1}(K) \neq 0$, which contributes to SU(3) breaking. 
In practice the expansion is truncated after a few terms. This is 
motivated by the fact that the hierarchy of anomalous dimensions 
$\gamma_{n+1} > \gamma_n > 0$ implies
$|\alpha_{n+1}| < |\alpha_n|$ at a sufficiently high scale.  
From concrete calculations and fits it indeed appears that
the hierarchy already sets in at typical hadronic scales
$ \sim 1\,{\rm GeV}$. Moreover, for smooth kernels the higher Gegenbauer 
moments give small contributions upon convolution much like in the familiar 
case of the partial wave expansion in quantum mechanics.

A different method is to model the wave-functions by using 
experimental and theoretical constraints. In \cite{Ball:2005ei} a recursive 
relation between the Gegenbauer moments was proposed, which involves only two 
additional parameters.
This constitutes an alternative tool especially in cases where the conformal 
expansion is converging slowly.

We shall not report on higher-twist contributions here but 
refer to the literature \cite{Ball:2006wn,Ball:2007rt}.
It should also be mentioned that higher-twist effects can be rather prominent 
such as in the time dependent CP asymmetry in $B \to K^* \gamma$ via soft 
gluon emission \cite{Ball:2006cv}.

\subsubsubsection{Decay constants}
The decay constants normalize the DA.
For the pseudoscalars $\pi, K$ they are well known form experiment. 
The decay constants of the $\eta$ and $\eta'$, and in general their wave
functions, are more complicated due to $\eta$-$\eta'$ mixing and the 
chiral anomaly and shall not be discussed here.
For the vector particles there are two decay constants as seen from 
\eqref{eq:defDAs}.
The longitudinal decay constants can be taken from experiment. For instance 
for $\rho^0$, $\omega$ and  $\phi$ they are taken from $V^0 \to e^+ e^-$ and 
for $\rho^-$ and $K^{*-}$ from $\tau^- \to V^- \nu_\tau$. 
It is worth noting  that the difference in 
$f_{\rho^0}$ and $f_{\rho-}$ seems consistent with the expected size of isospin breaking, whereas
some time ago there seemed to be a slight tension \cite{Ball:1996tb}.

For the transverse decay constants $f^\perp$ one has to rely on theory.
QCD sum rules provide both longitudinal and transverse decay constants 
\cite{Ball:2006nr,Ball:2005vx}
\begin{alignat}{4}
&f_\rho &=\,& (206 \pm 7)\, {\rm MeV} \qquad & & f_\rho^\perp(1\, {\rm GeV}) &=\,& 
 (165 \pm 9)\, {\rm MeV} \nonumber \\
&f_{K^*} &=\,& (222 \pm 8)\, {\rm MeV} \qquad & & f_{K^*}^\perp(1\, {\rm GeV}) &=\,&
 (185 \pm 10)\, {\rm MeV} \, .
\end{alignat} 
In lattice QCD there exist two quenched calculations of the ratio of decay
constants \cite{Braun:2003jg,Becirevic:2003pn}, which are consistent 
with the sum rule values above. Combining all these experimental, sum rule 
and lattice results we get \cite{Ball:2006eu}
\begin{alignat}{4}
&f_\rho &=\,& (216 \pm 2)\, {\rm MeV} \qquad & & f_\rho^\perp(1\, {\rm GeV}) &=\,& 
 (165 \pm 9)\, {\rm MeV} \nonumber \\
 &f_\omega &=\,& (187 \pm 5)\, {\rm MeV} \qquad & & f_\omega^\perp(1\, {\rm GeV}) &=\,& 
 (151 \pm 9)\, {\rm MeV} \nonumber \\
&f_{K^*} &=\,& (220 \pm 5)\, {\rm MeV} \qquad & & f_{K^*}^\perp(1\, {\rm GeV}) &=\,&
 (185 \pm 10)\, {\rm MeV}  \nonumber \\
 &f_{\phi} &=\,& (215 \pm 5)\, {\rm MeV} \qquad & & f_{\phi}^\perp(1\, {\rm GeV}) &=\,&
 (186 \pm 9)\, {\rm MeV}\, .
\end{alignat}
\subsubsubsection{The first and second Gegenbauer moment}
As mentioned before, the first Gegenbauer moment vanishes for particles 
with definite G-parity. Intuitively the first Gegenbauer moment of the kaon
is a measure of the average momentum fraction carried by the strange quark. 
Based on the constituent 
quark model it is expected that $\alpha_1(K) > 0$. A negative value of this 
quantity \cite{Ball:2003sc}  created some confusion and initiated 
reinvestigations. The sum rule used in that work is of the non-diagonal type 
and has a non-positive definite spectral function, which makes the extraction 
of any kind of residue very unreliable. 
Later on diagonal sum rules were used and 
stable values were obtained
\cite{Ball:2005vx,Khodjamirian:2004ga} ($\mu = 1\,{\rm GeV}$)
\begin{equation}
\label{eq:a1dSR}
\alpha_1(K,\mu) = 
 0.06\pm 0.03, \quad \alpha_1^\parallel(K^*,\mu) = 0.03\pm 0.02,\quad 
\alpha_1^\perp(K^*,\mu) = 0.04\pm 0.03,
\end{equation}
although with relatively large uncertainties. An interesting alternative 
method was suggested in \cite{Braun:2004vf}
where the first Gegenbauer moment was related to a quark-gluon matrix 
element via the equation of motion.
An alternative derivation and a completion for all cases was later given in 
\cite{Ball:2006fz}. 
The operator equation for the kaon is
\begin{equation*}
\frac{9}{5}\, \alpha_1(K)  =  -\frac{m_s-m_q}{m_s+m_q} +
4\,\frac{m_s^2-m_q^2}{m_{K}^2}  - 8 \kappa_{4}(K)\,, 
\end{equation*}
where the twist-4 matrix element $\kappa_4$ is defined as:
$\langle 0 | \bar q  (g G_{\alpha\mu}) 
i\gamma^\mu\gamma_5  s|K(q)\rangle = i q_\alpha f_{K} m_K^2 \kappa_{4}(K)$. 
Similar equations exist for the longitudinal and transverse case. 
It is worth stressing that those operator relations are
completely general and it remains to determine the twist-4 matrix elements. 
Attempts to determine them from 
QCD sum rules \cite{Braun:2004vf,Ball:2006fz} turn out to be consistent 
with the determinations from
diagonal sum rules \eqref{eq:a1dSR} but cannot compete in terms of the 
accuracy.
Later lattice QCD provided the first Gegenbauer moment for the kaon DA 
from domain-wall fermions \cite{Boyle:2006pw}
and Wilson fermions  \cite{Braun:2006dg} whose values agree very well with 
the central value
of $\alpha_1(K)$ in \eqref{eq:a1dSR}, but have significantly lower uncertainty.

The second Gegenbauer moment has also been determined from diagonal sum rules 
for the $\pi$ and $K$
\cite{Khodjamirian:2004ga,Ball:2006wn}
\begin{equation}
\alpha_2(\pi,1 \,{\rm GeV})  = 0.27 \pm 0.08 \qquad\quad  
\frac{\alpha_2(K) }{\alpha_2(\pi) } = 1.05 \pm 0.15
\end{equation}
It can be seen that the SU(3) breaking in the second moment is 
presumably moderate. Values of $\alpha_2$ for the vector mesons 
$\rho$, $K^*$ and $\phi$ have recently been updated in \cite{Ball:2007rt}.

\subsubsection{Heavy-to-light form factors from  LCSR}

Light-cone sum rules (LCSR) 
were developed to improve on some of the shortcomings of three-point
sum rules designed to describe meson-to-meson transition form factors. 
The problem is that for $B \to M$ transitions, where $M$ is a light
meson, higher order
matrix elements grow with $m_b$ rendering the OPE non-convergent. 
In the case $D \to M$ three-point sum rules and LCSR yield
comparable results. A review of the framework of LCSR can be found in 
\cite{Colangelo:2000dp}.

The form factors of $V$ and $A$ currents for $B$ to light pseudoscalar and
vector mesons are defined as ($q=p_B-p$)
\begin{eqnarray}
& &  \langle P(p)| \bar q\gamma_\mu  b|\bar B(p_B) \rangle \,=\, 
f_+(q^2)\,\left[(p_B+p)_{\mu}-
\frac{m_B^2-m_P^2}{q^2}\,q_{\mu}\right]
+ f_0(q^2)\,\frac{m_B^2-m_P^2}{q^2}\,q_{\mu}  \label{eq:FFPV}  \\
& & 
c_V\langle V(p,\varepsilon) | 
     \bar q\gamma_\mu(1-\gamma_5) b |\bar B(p_B)\rangle =  
\frac{2V(q^2)}{m_B+m_V}
\epsilon_{\mu\nu\rho\sigma}\varepsilon^{*\nu} p_B^\rho p^\sigma\,  
-2i m_V\, A_0(q^2)\,\frac{\varepsilon^*\cdot q}{q^2} q_\mu \label{eq:FFV-A} \\
& & -i(m_B+m_V)A_1(q^2)\,
   \left[\varepsilon^*_\mu-\frac{\varepsilon^*\cdot q}{q^2} q_\mu\right]
+iA_2(q^2)\,\frac{\varepsilon^*\cdot q}{m_B+m_V}\,
   \left[(p_B+p)_{\mu}-\frac{m_B^2-m_V^2}{q^2}\,q_{\mu}\right] \nonumber  
\end{eqnarray}
The factor $c_V$ accounts for the flavour content of 
particles: $c_V=\sqrt{2}$ for $\rho^0$,
$\omega$ and $c_V=1$ otherwise.
The tensor form factors, relevant for $B \to V \gamma$ or $B \to P(V) l^+l^-$, 
are defined as
\begin{eqnarray}
& &\langle  P(p)|\bar q \sigma_{\mu\nu} q^\nu  b|\bar B(p_B)\rangle   \,=\, 
\frac{i f_T(q^2)}{m_B+m_P}\left[ q^2(p+p_B)_{\mu}-
(m_B^2-m_P^2)q_{\mu}\right] \label{eq:FFPT} \\
& & c_V\langle V(p,\varepsilon) | 
      \bar q \sigma_{\mu\nu} q^\nu (1+\gamma_5) b |\bar B(p_B)\rangle = 
2 i\, T_1(q^2)\, \epsilon_{\mu\nu\rho\sigma}\, \varepsilon^{*\nu}
p_B^\rho p^\sigma \, \label{eq:FFpeng} \\
& & + T_2(q^2)\,\left[(m_B^2-m_{V}^2)\varepsilon^*_\mu -
        (\varepsilon^*\cdot q)(p_B+p)_\mu\right]
+ T_3(q^2)\,(\varepsilon^*\cdot q)\left[q_\mu-
     \frac{q^2}{m_B^2-m_{V}^2}(p_B+p)_\mu\right] \nonumber 
\end{eqnarray}
with  
$T_1(0)=T_2(0)$.
Note that the tensor form factors depend on the renormalization
scale $\mu$ of the matrix element. All form factors in 
(\ref{eq:FFPV}) - (\ref{eq:FFpeng}) are positive and $\epsilon^{0123}=-1$.

LCSR allow us to obtain the form factors from a suitable correlation function 
for virtualities of $0 < q^2 \la 14\,{\rm GeV}^2$. 
The residue in the sum rule is
of the type $(f_B f_+(q^2))_{\rm SR}$. 
Using a second sum rule for $(f_B)_{\rm SR}$ to the same accuracy, the
form factor is obtained as $f_+ = (f_B f_+(q^2))_{\rm SR}/(f_B)_{\rm SR}$,
where several uncertainties cancel. 
The final uncertainties of the sum rule results 
for the form factors are around $10\%$ and
slightly more for the $B\to K$ transitions due to the additional uncertainty 
in the first Gegenbauer moment.
The most recent and up-to-date calculation for $B\to M$ form factors, 
including twist-3 radiative corrections, 
can be found in \cite{Ball:2004ye,Ball:2004rg}. 
It is not obvious how the accuracy can be significantly improved by 
including further corrections.
One interesting option would be to calculate NNLO QCD corrections, which 
could first be attempted in the large-$\beta_0$ limit.

Another interesting question is whether it is possible to extend the 
form factor calculations to the
entire physical domain $0 < q^2 <  (m_B-m_{P(V)})^2$. 
It has been advocated by Becirevic and Kaidalov
\cite{Becirevic:1999kt} to write the form factor $f_+$ as a dispersion
relation in $q^2$ with a lowest-lying pole term plus a contribution from
multiparticle states, which in a minimal setup can be approximated
by an effective pole term at higher mass:
\begin{eqnarray}
f_+(q^2) &=& \frac{r_1}{1-q^2/m_1^2} + \int_{(m_B+m_P)^2}^\infty
ds\,\frac{\rho(s)}{s-q^2} \to
\frac{r_1}{1-q^2/m_1^2} + \frac{r_2}{1-q^2/m_{\rm fit}^2} \,.\label{eq:para}
\end{eqnarray}
In the past it has often been popular to adopt Vector Meson Dominance (VMD), 
i.e. to set $r_2 = 0$. 
BaBar measurements of semileptonic decay spectra with five bins in the 
$q^2$-distribution now strongly disfavour simple VMD \cite{Ball:2005tb}. 
Another important point is that the fits to the parametrisation (\ref{eq:para})
allow us to reproduce the results from LCSR extremely well 
\cite{Ball:2004ye,Ball:2004rg}. The parametrisation also passes
a number of consistency tests. The soft pion point 
$f_0(m_B^2) = f_B/f_\pi$ can be attained upon extrapolation, leading to 
a $B$-meson decay constant of $f_B \approx 205 \,{\rm MeV}$. 
This is well in the ballpark of expectations and consistent with the Belle 
measurement of $B \to \tau \nu$. Moreover the residue 
$(r_1)_{f_+}  =  (f_{B^*} g_{BB^*\pi})/(2m_{B^*})$, which is rather stable 
under the fits, agrees within ten percent with what is known from hadronic 
physics.
Representative results are given in Table~\ref{tab:lcsrformf}. 
\begin{table}[t]
\begin{center}
\caption{Form factors from light-cone sum rules.}
\label{tab:lcsrformf}
\begin{tabular}{cccccc}
\hline\hline
\vspace*{1mm}
$f^{B\to\pi}_+(0)$ &  
$T^{B\to\rho}_1(0)$ &
$V^{B\to\rho}(0)$ &  
$A^{B\to\rho}_0(0)$ &   
$A^{B\to\rho}_1(0)$ &
$A^{B\to\rho}_2(0)$\\  
\hline
$0.258\pm 0.031$ &
$0.267\pm 0.023$ &
$0.323\pm 0.030$ &
$0.303\pm 0.029$ &
$0.242\pm 0.023$ &
$0.221\pm 0.023$\\
\hline\hline
\end{tabular}
\end{center}
\end{table}
More form factors can be found in  
eq. (27) and Tab. 3 of  \cite{Ball:2004ye} for $B \to \pi,K,\eta$ and
in Tab. 8 of \cite{Ball:2004rg} for $B \to \rho, K^*, \phi, \omega$.
It has to be emphasized that the $B \to K,K^*$ transitions have been 
evaluated before the progress in the SU(3)-breaking was achieved. An update 
would be timely and will certainly
be undertaken for such important cases as $B \to K^* l^+ l^-$.
In particular for the $B \to K^* \gamma$ decay rate in the standard model 
(SM) it was emphasized by \cite{Bosch:2004nd,Beneke:2004dp} that within the 
framwork of QCD factorization $T_1(0)_{\rm SM-exp,QCDF} = 0.28 \pm 0.02$. 
An update of SU(3)-breaking effects yields $T_1(0) = 0.31 \pm 0.04$ 
\cite{Ball:2006fw}, which seems reasonably consistent.

In certain decay channels, such as $B \to K^* l^+l^-$, 
several form factors enter at the same time. Sometimes ratios of decay 
rates are needed, e.g. for the extraction of $|V_{td}/V_{ts}|$
from $B \to K^* \gamma$. Simply taking the uncertainties in the individual 
form factors and adding them linearly could be a drastic overestimate since 
parametric uncertainties, such as those from $m_b$, 
might cancel in the quantities of interest. In the former case no 
efforts have been undertaken. In the latter case a consistent evaluation 
\cite{Ball:2006nr} leads to the form factor ratio 
$\xi\equiv T_1^{B\to K^*}(0)/T_1^{B \to \rho}(0) = 1.17\pm 0.09 $.

\subsubsection{Comparison with heavy-to-light form factors from
relativistic quark models}

{\it W.~Lucha, D.~Melikhov, S.~Simula, B.~Stech}


Quark models have been frequently used in the past to estimate
hadronic quantities such as form factors. They may be applied
to complicated processes hardly accessible to lattice calculations
and they provide connections between different processes through
the wave functions of the participating hadrons.
\begin{table}[h]
\begin{center}
\caption{\label{Table:qm_B_decay}Examples of form factors for
$B\to \rho$ [$B_s\to K^*$] from the quark model
\cite{Melikhov:2000yu}.}
\begin{tabular}{cccccc}
\hline\hline
 $V(0)$ & $A_1(0)$ & $A_2(0)$ & $A_0(0)$ & $T_1(0)$ & $T_3(0)$ \\
\hline
 0.31 [0.44] & 0.26 [0.36] & 0.24 [0.32] & 0.29 [0.45] & 0.27 [0.39] & 0.19 [0.\
27] \\
\hline\hline
\end{tabular}
\end{center}
\end{table}
Relativistic quark models are based on a simplified picture of QCD: 
Below the
chiral symmetry breaking scale $\mu_{\chi}\approx1$ GeV, quarks
are treated as particles of fixed mass interacting via a
relativistic potential and~hadron wave functions and masses are
found as solutions of three-dimensional reductions of the
Bethe--Salpeter equation. The structure of the confining potential
is restricted by rigorous properties of QCD, such as heavy-quark
symmetry~for the heavy-quark sector
\cite{Faustov:1995xc,Melikhov:1997qk} and spontaneously broken
chiral symmetry for the light-quark sector \cite{Lucha:2006rq}.
The values of the constituent-quark masses and the parameters of
the potential are fixed by requiring~that the spectrum of observed
hadron states is well reproduced
\cite{Godfrey:1985xj,Ebert:1997nk}.


Various versions of the quark model were applied to the  
description of weak properties of heavy hadrons 
(see e.g.
\cite{Wirbel:1985ji,Isgur:1988gb,Scora:1995ty}).
For instance,
the weak transition form factors are given in the quark model
in \cite{Melikhov:1995xz} by relativistic double spectral
representations in terms of the wave functions of initial and
final hadrons and the double spectral density of the corresponding
Feynman diagrams with massive quarks. 
This approach led to very
successful predictions for $D$ decays
\cite{Melikhov:1999nv,Melikhov:2000yu}. Many results for various $B$ 
and $B_s$ decays have been obtained
\cite{Melikhov:2000yu,Beyer:1998ka,Beyer:2001zn,Kruger:2002gf,Melikhov:2004mk},
yielding an overall picture in agreement with other approaches,
such as QCD sum rules.
Table \ref{Table:qm_B_decay} gives examples of the results from
\cite{Melikhov:2000yu}. A comparison between various quark models
performed in \cite{Ivanov:2007cw} leads to a qualitative estimate
of the overall uncertainty of some 10--15$\%$.
The main limitation of the quark model approach
is the difficulty to provide rigorous
estimates of the systematic errors of the calculated hadron
parameters. In this respect, quark models cannot compete with lattice
gauge theory.


\newpage \subsection{Lattice QCD}\label{subsec:latqcd}

\subsubsection{Recent results}

In this section we give a summary of recent lattice results 
relevant to flavour physics.
The tables should be consulted with an eye on the systematics discussed 
in \ref{subsubsec:systematics}.
For a more complete coverage, see the review talks on heavy flavour
physics \cite{Okamoto:2005zg,Onogi:2006km,DellaMorte:2007ny} and kaon
physics \cite{Dawson:2005za,Lee:2006cm,Juttner:2007sn} at the last
few lattice conferences. 


%


\subsubsubsection{Decay constants}

The axial-vector decay constants relevant to the $\pi\to\ell\nu$ leptonic
decays
\beq
\langle 0|(\bar{d}\gamma_\mu\gamma_5u)(x)|\pi(p)\rangle=i f_\pi p_\mu e^{-i p\cdot x}
\eeq
(and analogously for $K,D,B$ mesons) may be evaluated on the lattice.
Some recent results are collected in Table~\ref{tab:decaycons}.
\begin{table}[ht]
\begin{center}
\caption{Decay constants from lattice QCD.}
\label{tab:decaycons}
\begin{tabular}{llccl}
\hline\hline
$f_K/f_\pi=1.24(2)$       &$\Nf=2+1$  & dom/dom & no&RBC/UKQCD \cite{Allton:2007hx}
\\
$f_K/f_\pi=1.218(2)(^{+11}_{-24})$  &$\Nf=2+1$&stag/dom & no&NPLQCD \cite{Beane:2006kx}
\\
$f_K/f_\pi=1.208(2)(^{+07}_{-14})$  &$\Nf=2+1$&stag/stag&yes&MILC \cite{Bernard:2006wx}
\\
$f_K/f_\pi=1.189(7)$                &$\Nf=2+1$&stag/stag&yes&HPQCD \cite{Follana:2007uv}

\\[2mm]

$f_{D_s}=242(09)(10)\MeV$           &$\Nf=0$  &--- /clov&yes&ALPHA \cite{Juttner:2003ns}
\\
$f_{D_s}=240 (5) (5)\MeV$           &$\Nf=0$  &--- /clov&yes&RomeII \cite{deDivitiis:2003wy}
\\
$f_{D_s}=249(03)(16)\MeV$           &$\Nf=2+1$&stag/stag&yes&FNAL/MILC/+ \cite{Aubin:2005ar}
\\
$f_{D_s}=238(11)(^{+07}_{-27})\MeV$ &$\Nf=2$  &clov/clov&yes&CP-PACS \cite{Okamoto:2005zg}
\\
$f_{D_s}=241(3)\MeV$                &$\Nf=2+1$&stag/stag&yes&HPQCD \cite{Follana:2007uv}
\\

$f_D=232(7)(^{+6}_{-0})(53)\MeV$    &$\Nf=0$  &--- /dom & no&RBC \cite{Lin:2006vc}
\\
$f_D=202(12)(^{+20}_{-25})\MeV$     &$\Nf=2$  &clov/clov&yes&CP-PACS \cite{Okamoto:2005zg}
\\

$f_{D_s}/f_D=1.05(2)(^{+0}_{-2})(6)$&$\Nf=0$  &--- /dom & no&RBC \cite{Lin:2006vc}
\\
$f_{D_s}/f_D=1.24(7)$               &$\Nf=2+1$&stag/stag&yes&FNAL/MILC/+ \cite{Aubin:2005ar}
\\
$f_{D_s}/f_D=1.162(9)$              &$\Nf=2+1$&stag/stag&yes&HPQCD \cite{Follana:2007uv}

\\[2mm]

$f_{B_s}=192(6)(4)\MeV$             &$\Nf=0$  &--- /clov&yes&RomeII \cite{deDivitiis:2003wy}
\\
$f_{B_s}=205(12)\MeV$               &$\Nf=0$  &--- /clov&yes&ALPHA \cite{DellaMorte:2003mn}
\\
$f_{B_s}=191(6)\MeV$                &$\Nf=0$  &--- /clov&yes&ALPHA \cite{Guazzini:2006bn}
\\
$f_{B_s}=242(9)(51)\MeV$            &$\Nf=2$  &clov/clov&yes&CP-PACS \cite{AliKhan:2001jg}
\\
$f_{B_s}=217(6)(^{+37}_{-28})\MeV$  &$\Nf=2$  &stag/wils&yes&MILC \cite{Bernard:2002pc}
\\
$f_{B_s}=260(7)(26)(8)\MeV$         &$\Nf=2+1$&clov/clov& no&HPQCD \cite{Wingate:2003gm}
\\

$f_{B_s}/f_B=1.179(18)(23)$         &$\Nf=2$  &clov/clov&yes&CP-PACS \cite{AliKhan:2001jg}
\\
$f_{B_s}/f_B=1.16(1)(3)(^{+4}_{-0})$&$\Nf=2$  &stag/wils&yes&MILC \cite{Bernard:2002pc}
\\
$f_{B_s}/f_B=1.13(3)(^{+17}_{-02})$ &$\Nf=2$  &clov/clov& no&JLQCD \cite{Aoki:2003xb}
\\
$f_{B_s}/f_B=1.20(3)(1)$            &$\Nf=2+1$&stag/stag&yes&HPQCD \cite{Gray:2005ad}
\\
$f_{B_s}/f_B=1.29(4)(6)$            &$\Nf=2$  & dom/dom & no&RBC \cite{Gadiyak:2005ea}
\\
\hline\hline
\end{tabular}
\end{center}
\end{table}
The first column gives the statistical and systematic errors.
The second column says whether the simulations are quenched ($\Nf=0$), or
dynamical with a common $m_{ud}$ mass only ($\Nf=2$), or with strange quark
loops included ($\Nf=2+1$).
The remaining columns indicate the light quark formulation in the sea and
valence sectors and whether a continuum extrapolation has been attempted.
To the quenched results, an extra 5\% scale setting error should be added
(see \ref{subsubsubsec:burn}).
Generally, the lattice results compare favourably to the recent experimental
determinations (using the appropriate CKM element from another process)
$f_D\!=\!223(17)(03)\MeV$ at CLEO \cite{Artuso:2005ym},
$f_{D_s}\!=\!282(16)(7)\MeV$ at CLEO \cite{Artuso:2006kz},
$f_{D_s}\!=\!283(17)(16)\MeV$ at BaBar \cite{Aubert:2006sd} and
$f_B\!=\!229(^{+36}_{-31})(^{+34}_{-37})\MeV$ at Belle \cite{Ikado:2006un}.
One may also form the ratio
$\sqrt{M_{D_s}}f_{D_s}/\sqrt{M_D}f_D$
and compare
to the result $1.30(12)$, implied by the CLEO and BaBar numbers.


\subsubsubsection{Form factors}
\label{subsubsubsec:formf}

The vector form factors of semi-leptonic decays like $B\to\pi\ell\nu$ 
or $D\to K\ell\nu$, defined in (\ref{eq:FFPV}), 
can be calculated in the range $q_\mr{min}^2<q^2<q_\mr{max}^2$, where
$q_\mr{max}^2=(M_B-M_\pi)^2,(M_D-M_K)^2$, respectively, while $q_\mr{min}^2$ is
a soft bound (set by the cut-off effects and noise one considers tolerable).
Often $f_+(0)=f_0(0)$ is used and a parametrisation is employed to
extrapolate.
Among the most popular are those of Be\'cirevi\'c-Kaidalov
\cite{Becirevic:1999kt} and Ball-Zwicky \cite{Becirevic:1999kt,Ball:2005tb}
\bea
f_+^\mr{BK}(q^2)&=&\frac{f}{(1-\til q^2)(1-\alpha\til q^2)}\;,\qquad
f_0^\mr{BK}(q^2)=\frac{f}{1-\til q^2/\beta}
\\
f_+^\mr{BZ}(q^2)&=&\frac{f}{1-\til q^2}+
\frac{r\til q^2}{(1-\til q^2)(1-\alpha\til q^2)}\label{eq:fpbz}
\eea
where $\til q^2=q^2/M_{B^*}^2$ (or $q^2/M_{D^*}^2$ for $D$-decays), with
the parameters $f=f_+(0),\al$ (BK,BZ) and $r$ (BZ).
The expression in (\ref{eq:fpbz}) is equivalent to the approximate
form in (\ref{eq:para}).
Some recent results, with the same meaning of the columns as before, 
are given in Table~\ref{tab:formf}.
\begin{table}[ht]
\begin{center}
\caption{Form factors from lattice QCD.}
\label{tab:formf}
\begin{tabular}{llccl}
\hline\hline
$f_+^{K\to\pi}(0)=0.960(5)(7)$ &$\Nf=0$  &--- /clov& no&RomeI-Orsay \cite{Becirevic:2004ya}
\\
$f_+^{K\to\pi}(0)=0.952(6)$    &$\Nf=2$  &clov/clov& no&JLQCD \cite{Tsutsui:2005cj}
\\
$f_+^{K\to\pi}(0)=0.968(9)(6)$ &$\Nf=2$  & dom/dom & no&RBC \cite{Dawson:2006qc}
\\
$f_+^{K\to\pi}(0)=0.9680(16)$  &$\Nf=2+1$& dom/dom & no&UKQCD/RBC \cite{Antonio:2006ev}
\\
$f_+^{K\to\pi}(0)=0.962(6)(9)$ &$\Nf=2+1$&stag/clov& no&FNAL/MILC/+ \cite{Okamoto:2004df}

\\[2mm]

$f_+^{D\to\pi}(0)=0.64(3)(6)$  &$\Nf=2+1$&stag/stag& no&FNAL/MILC/+ \cite{Aubin:2004ej}

\\[2mm]

$f_+^{D\to K}(0)=0.73(3)(7)$   &$\Nf=2+1$&stag/stag& no&FNAL/MILC/+ \cite{Aubin:2004ej}

\\[2mm]

$f_+^{B\to\pi}(0)=0.23(2)(3)$  &$\Nf=2+1$&stag/stag& no&FNAL/MILC/+ \cite{Okamoto:2004xg}
\\
$f_+^{B\to\pi}(0)=0.31(5)(4)$  &$\Nf=2+1$&stag/stag&yes&HPQCD \cite{Dalgic:2006dt}

\\[2mm]

$\mathcal{F}^{B\to D}(1)=1.074(18)(16)$&$\Nf=2+1$&stag/stag& no&FNAL/MILC/+ 
\cite{Okamoto:2004xg}
\\
$\mathcal{F}^{B\to D}(1)=1.026(17)$    &$\Nf=0$  &--- /clov&yes&RomeII
\cite{deDivitiis:2007ui}
\\
\hline\hline
\end{tabular}
\end{center}
\end{table}
The definition of $\cal{F}$ is given in \cite{Okamoto:2004xg}.
Earlier work on the $B\to\pi\ell\bar\nu$ form factors can be found in
\cite{Bowler:1999xn,Abada:2000ty,ElKhadra:2001rv,Aoki:2001rd}.
For $D\to K\ell\nu$ and $D\to\pi\ell\nu$ the $q^2$-dependence of the
form factors has been traced out by the FNAL/MILC/+ collaboration
\cite{Aubin:2004ej} and compared to experimental results by the BES
\cite{Ablikim:2004ej} and FOCUS \cite{Link:2004dh} collaborations.
For $B\to\pi\ell\nu$ the $q^2$-dependence, as determined by the HPQCD and
FNAL/MILC/+ collaborations, is in reasonable agreement
\cite{Okamoto:2005zg}.
For a generic comment why the form factor at $q^2=0$ is not always the best
thing to ask for from the lattice, see section \ref{subsubsec:prospects}. 


\subsubsubsection{Bag parameters}

On the lattice, the SM bag parameters $B_K(\mu)$ and $B_B(\mu)$ for neutral
kaon and $B$-meson mixing
\bea
\langle\bar{K}^0|(\bar{s}d)_{V\!-\!A}(\bar{s}d)_{V\!-\!A}|K^0\rangle&=&
\frac{8}{3}M^2_K f_K^2 B_K 
\\
\langle\bar{B}^0_q|(\bar{b}q)_{V\!-\!A}(\bar{b}q)_{V\!-\!A}|B^0_q\rangle&=&
\frac{8}{3}M^2_{B_q}f_{B_q}^2 B_{B_q}\qquad(q=d,s)
\eea
are extracted indirectly.
The measured quantities are $f_B^2B_B$ and $f_B$; then the ratio is taken
to obtain the quoted $B_B$ (similar for $B_K$).
Therefore, it makes little sense to combine $B_B$ from one group and $f_B$ from
another to come up with a lattice value for $f_B\sqrt{B_B}$.
On the other hand
\beq
\xi=\frac{f_{B_s}^{}\sqrt{B_{B_s}^{}}}{f_{B_d}\sqrt{B_{B_d}} }
\eeq
is benevolent, from a lattice viewpoint, since it follows from the
ratio of the same correlator with two different quark masses (in practice, an
extrapolation $m_d\to m_d^\mr{phys}$ is needed).
Many systematic uncertainties cancel in such ratios, but the chiral
extrapolation error is not reduced.
It would make sense to quote the renormalisation scheme and scale independent
quantity
\beq
\hat B_X=\lim_{\mu\to\infty}\;
\al_{s}(\mu)^{2/\beta_0}\Big[1+\frac{\al_{s}}{4\pi}J_\Nf+...\Big]\;
B_X(\mu)
\eeq
with known $J_\Nf$.
From a perturbative viewpoint $B_X$ and $\hat B_X$ are equivalent, 
but from a lattice perspective the latter is much better defined.
Recent results for $B_K=B_K(2\GeV)$ and $B_B=B_B(m_b)$ are quoted in
Table~\ref{tab:bagpar}.
\begin{table}[ht]
\begin{center}
\caption{Bag parameters from lattice QCD.}
\label{tab:bagpar}
\begin{tabular}{llccl}
\hline\hline
$B_K=0.5746(061)(191)$    & $\Nf=0$   & --- /dom  & yes & CP-PACS \cite{AliKhan:2001wr}
\\
$B_K=0.55(7)$             & $\Nf=0$   & --- /over & yes & MILC \cite{DeGrand:2003in}
\\
$\hat B_K=0.96(10)$  [hat]& $\Nf=0$   & --- /wils & yes & Becirevic et al.\ 
\cite{Becirevic:2004aj}
\\
$B_K=0.563(21)(49)$       & $\Nf=0$   & --- /dom  & yes & RBC \cite{Aoki:2005ga}
\\
$B_K=0.563(47)(07)$       & $\Nf=0$   & --- /over & yes & BMW \cite{Babich:2006bh}
\\
$\hat B_K=0.789(46)$ [hat]& $\Nf=0$   & --- /twis & yes & ALPHA \cite{Dimopoulos:2006dm}
\\

$B_K=0.49(13)$            & $\Nf=2$   & clov/clov &  no & UKQCD \cite{Flynn:2004au}
\\
$B_K=0.495(18)$           & $\Nf=2$   &  dom/dom  &  no & RBC \cite{Aoki:2004ht}
\\
$B_K=0.618(18)(135)$      & $\Nf=2+1$ & stag/stag &  no & HPQCD/UKQCD \cite{Gamiz:2006sq}
\\
$B_K=0.557(12)(29)$       & $\Nf=2+1$ &  dom/dom  &  no & RBC/UKQCD \cite{Antonio:2007pb}

\\[2mm]

$B_{B_s}=0.940(16)(22)$               & $\Nf=0$ & --- /over &  no & Orsay \cite{Becirevic:2005sx}
\\
$B_B=0.836(27)(^{+56}_{-62})$         & $\Nf=2$ & clov/clov &  no & JLQCD \cite{Aoki:2003xb}
\\
$B_{B_s}/B_B=1.017(16)(^{+56}_{-17})$ & $\Nf=2$ & clov/clov &  no & JLQCD \cite{Aoki:2003xb}
\\
$B_{B_s}/B_B=1.06(6)(4)$              & $\Nf=2$ &  dom/dom  &  no & RBC \cite{Gadiyak:2005ea}

\\[2mm]

$f_{B_s}\sqrt{\hat{B}_{B_s}}=281(21) \MeV$ & $\Nf=2+1$ & stag/stag &  no & HPQCD \cite{Dalgic:2006gp}

\\[2mm]

$\xi=1.14(3)(^{+13}_{-02})$           & $\Nf=2$ & clov/clov &  no & JLQCD \cite{Aoki:2003xb}
\\
$\xi=1.33(8)(8)$                      & $\Nf=2$ &  dom/dom  &  no & RBC \cite{Gadiyak:2005ea}
\\
\hline\hline
\end{tabular}
\end{center}
\end{table}
Note that these values refer to bag parameters with spinor structure $VV+AA$
in the 4-fermion operator, as they appear in the SM.


%


%


\subsubsubsection{BSM matrix elements}

There are several hadronic matrix elements for BSM operators available from the
lattice.
Kaon-mixing matrix elements with $VV-AA,SS+PP,SS-PP,TT$ spinor structure in the
4-fermion operator are found in
\cite{Donini:1999nn,Becirevic:2004qd,Aoki:2005ga,Babich:2006bh,Nakamura:2006eq}
and $\langle\pi^0|Q_\gamma^+|K^0\rangle$ is being addressed in \cite{Becirevic:2000zi}.
In the literature, they go by the name of ``SUSY matrix elements'', 
but the idea
is that only the (perturbatively calculated) Wilson coefficient refers to the
specific BSM theory, while the (lattice evaluated) matrix element is fully
generic.
Thanks to massless overlap fermions \cite{Neuberger:1997fp,Neuberger:1998wv}
obeying the Ginsparg-Wilson relation \cite{Ginsparg:1981bj} and hence enjoying
a lattice analogue of chiral symmetry \cite{Luscher:1998pq}, it is now possible
to avoid admixtures of operators with an unwanted chirality structure.


\subsubsubsection{CKM matrix elements}

In his Lattice 2005 write-up \cite{Okamoto:2005zg}, Okamoto quantifies the
magnitudes of all CKM matrix elements, except $|V_{td}|$, using
\emph{exclusively lattice results} (and experimental data, of course).
They are collected in Table~\ref{tab:ckmlat}.
\begin{table}[t]
\begin{center}
\caption{CKM matrix elements from lattice QCD.}
\label{tab:ckmlat}
\begin{tabular}{ccccc}
\hline\hline
$|V_{us}|_\mr{Lat05}$ & $|V_{ub}|_\mr{Lat05}$ & $|V_{cd}|_\mr{Lat05}$ &
$|V_{cs}|_\mr{Lat05}$ & $|V_{cb}|_\mr{Lat05}$\\
\hline
$0.2244(14)$ & $3.76(68)\,10^{-3}$ & $0.245(22)$ & $0.97(10)$ & 
$3.91(09)(34)\,10^{-2}$\\
\hline\hline
\end{tabular}
\end{center}
\end{table}
The magnitudes $|V_{ud}|,|V_{ts}|,|V_{tb}|$ may be subsequently determined,
if one assumes unitarity of $V_\mr{CKM}$. This gives
$|V_{ud}|_\mr{Lat05}^\mr{SM}=0.9743(3)$,
$|V_{ts}|_\mr{Lat05}^\mr{SM}=3.79(53)\,10^{-2}$ and
$|V_{tb}|_\mr{Lat05}^\mr{SM}=0.9992(1)$.


\subsubsection{Scale setting and systematic effects}
\label{subsubsec:systematics}


\subsubsubsection{Burning $\Nf\!+\!1$ observables in $\Nf$ flavour QCD}
\label{subsubsubsec:burn}

In a calculation with, say, a common $ud$ and separate $s,c$ quark masses, four
observables must be used to set the lattice spacing and to adjust
$m_{ud},m_s,m_c$ to their physical values (with $m_{ud}$ there is a practical
problem, but this is immaterial to the present discussion).
In general, $\Nf\!+\!1$ lattice observables cannot be used to make predictions,
since LQCD establishes a connection
\bdm
\underbrace{
\left(
\begin{array}{c}
M_p\\M_\pi\\M_K\\M_D\\ \hline M_B
\end{array}
\right)
}_{\mbox{experiment}}
\quad\Longleftrightarrow\quad\;
\underbrace{
\left(
\begin{array}{c}
\Lambda_\mr{QCD}\\m_{ud}\\m_s\\m_c\\ \hline m_b
\end{array}
\right)
}_{\mbox{parameters}}
+
\underbrace{
\left(
\begin{array}{c}
f_\pi\\f_K,B_K\\f_D,f_{D_s}\\...\\ \hline f_B,f_{B_s}\\B_B,B_{B_s}
\end{array}
\right)
}_{\mbox{predictions}}
\;.
\edm
With infinitely precise data it would not matter which observables are
sacrificed to specify the bare parameters in a given run (every observable
depends a bit on each of the $\Nf\!+\!1$ parameters).
In practice, the situation is different.
To adjust the bare parameters in a controlled way,
it is important to single out $\Nf\!+\!1$ observables 
that are easy to measure, do not show tremendous cut-off effects
and depend strongly on one physical parameter but as weakly as 
possible on all other.
By now it is clear that one should not use any broad resonance (e.g.\ the
$\rh$), since this introduces large ambiguities \cite{Gattringer:2003qx}.

Frequently, the Sommer radius $r_0$ \cite{Sommer:1993ce} is used as an
intermediate  scale-setting quantity; e.g.\ the continuum limit is taken for
$f_{B_s}r_0$.
But the issue remains what physical distance should be identified with $r_0$.
Typically, a quenched lattice study converts a value for $f_{B_s}r_0$ with
specified statistical and systematic errors into an MeV result for $f_{B_s}$,
assuming that $r_0$ is exactly $0.5\fm$ (the preferred value from charmonium
spectroscopy), or exactly $0.47\fm$ (from the proton mass), or exactly
$0.51\fm$ (from $f_K$).
If one is interested in quenched QCD, any of these values is fine.
However, if one intends to use the result for phenomenological purposes, it is
more advisable to attribute a certain error to $(r_0\MeV)$ itself.
For instance, one might use $r_0=0.49(2)\fm$.
This is where the suggestion to add an extra 5\% scale-setting ambiguity to
most quenched results comes from.
In principle, such ambiguities persist in $\Nf=2+1$ QCD, but they get smaller
as one moves towards realistic quark masses.


\subsubsubsection{Perturbative versus non-perturbative renormalisation}

On the lattice, there are two types of renormalisation.
Obviously, any operator which ``runs'' requires renormalisation.
For instance, when calculating a bag parameter, the lattice result is
$B_X^\mr{glue,ferm}(a^{-1})$, where the superscript indicates the specific
cut-off scheme defined by the gluon and fermion actions that have been used.
In order to obtain an observable with a well-defined continuum limit, this
object needs to be converted into a scheme where the pertinent scale $\mu$ is
not linked to the cut-off $a^{-1}$.
Consequently, the conversion factor in
$B_X^{\MSbar}(\mu)=C(\mu a)B_X^\mr{glue,ferm}(a^{-1})$
would diverge in the continuum limit, but this is immaterial, since $C(\mu a)$
is not an observable.

Besides, a finite renormalisation is used for many quantities of interest.
For instance, to measure $f_\pi$, one multiplies the point-like axial-vector
current $A_\mu=\bar{d}\gamma_\mu\gamma_5u$ with a renormalisation factor $Z_A$.
Asymptotically (for large $\beta$), this factor behaves like
$Z_A=1+\mr{const}/\beta+O(\beta^{-2})$.
Accordingly, $Z_A(\beta)$ may be calculated either in weak coupling perturbation
theory or non-perturbatively.
For some actions both avenues have been pursued, and sometimes it was found
that within perturbation theory  it is difficult to estimate the error (there
may be big shifts when going from 1-loop to 2-loop and/or all perturbative
calculations of $Z_A(\beta)$ may differ significantly from the outcome of a
non-perturbative determination).
The results with $\Nf=2+1$ staggered quarks rely on perturbation theory and
some experts fear that some of the renormalisation factors may be less
precisely known than what is currently believed.
On the other hand one might argue that these actions involve UV-filtering
(``link-fattening'')
and may be less prone to such uncertainties than unfiltered (``thin-link'')
actions.
These issues are under active investigation.


\subsubsubsection{Summary of extrapolations}

Lattice calculations are done in a euclidean box $L^3 \times T$ with a finite
lattice spacing $a$.
From a field-theoretic viewpoint only the $T\to\infty$ limit is needed to
define particle properties (to locate the pole of an Euclidean Green's function
and to extract the residue, the $t\to\infty$ behaviour of the correlation
function $C(t)$ needs to be studied).
All other limits are taken subsequently in the physical observables.
A summary of all extrapolations involved is:
\begin{itemize}
\vspace{-2pt}
\itemsep-2pt
\item[1)]
$T \to \infty$ or removal of excited states contamination
(in practice, choosing $T\!\gg\!L$ is sufficient)
\item[2)]
$a \to 0$ or removal of discretisation effects
(at fixed $V\!=\!L^3$ and fixed $M_\mr{had}L$)
\item[3)]
$V \to \infty$ or removal of (spatial) finite-size effects
(at fixed renormalised quark masses)
\item[4)]
$m_{ud} \to m_{ud}^\mr{phys}$ or chiral extrapolation
\item[5)]
$m_b \to m_b^\mr{phys}$ or heavy-quark extrapolation/interpolation
(not with Fermilab formulation)
\vspace{-2pt}
\end{itemize}
Extrapolations 1-3 are standard in the sense that one knows how to control
them.
The chiral extrapolation is far from innocent, since it is not really
justified to use chiral perturbation theory \cite{Gasser:1983yg,Gasser:1984gg}
if one cannot clearly identify chiral logs in the data, and it is hard to tell
such logs from lattice artifacts and finite-size effects.
The entries with the smallest error bars among the $\Nf=2+1$ data quoted above
stem from simulations with the staggered action.
In such studies the extrapolations 2 and 4 are performed by means of staggered
chiral perturbation theory \cite{Aubin:2003mg,Aubin:2003uc}, using a large
number of fitting parameters.
This makes it hard to judge whether
the quoted error is realistic, but at least the ``post
processing'' is done in a field-theoretic framework (no modelling).
The fifth point depends on the details of the heavy-quark formulation (NRQCD,
HQET, Fermilab) employed, but eventually, with $a^{-1}\simeq10\GeV$ and higher,
one could use a standard relativistic action.


\subsubsubsection{Conceptual issues}

Besides these practical aspects, there might be conceptual issues regarding the
theoretical validity of certain steps.
In the past, the so-called quenched approximation has been used, where the
functional determinant is neglected.
While fundamentally uncontrolled, it seems to have little impact on the final
result of a phenomenological study --- as long as no flavour singlet quantity
is measured, final-state interactions are not particularly important and the
long-distance physics involved does not exceed $\sim\!1\fm$ (i.e.\ for
$\Mpi>200\MeV$, which still is the case in present simulations).
State-of-the-art calculations use the partially quenched framework
\cite{Bernard:1993sv,Sharpe:1997by,Sharpe:2000bc}, which, despite its name, is
\emph{not} a half-way extrapolation from quenched to unquenched.
It amounts to having, besides $m_{ud}^\mr{sea}=m_{ud}^\mr{val}$, also data with
$m_{ud}^\mr{sea}>m_{ud}^\mr{val}$ which typically stabilise the extrapolation
to $m_{ud}^\mr{sea}=m_{ud}^\mr{val}=m_{ud}^\mr{phys}$.
But even with the determinant included, things remain somewhat controversial.
The rooting procedure with staggered quarks (to obtain $\Nf=2+1$, the
square-root of $\det(D^\mr{st}_{m_{ud}})$ and the fourth-root of
$\det(D^\mr{st}_{m_s})$ is taken) has been the subject of a lively debate.
Much theoretical progress on understanding its basis has been achieved ---
for a summary see the plenary talks on this point at the last three
lattice conferences \cite{Durr:2005ax,Sharpe:2006re,Creutz:2007rk,
Kronfeld:2007ek}. 


\subsubsection{Prospects of future error bars}
\label{subsubsec:prospects}



Future progress on the precision of lattice calculations of QCD
matrix elements will hopefully come from a variety of improvements, including
a growth in computer power, the development of better algorithms,
the construction of better interpolating fields, and the design of better 
relativistic and heavy quark actions.
Some of these factors are easier to forecast than others.
For instance, the amount of CPU power is a rather monotonic function of time
(for the lattice community as a whole, not for an individual collaboration).
By contrast, progress at the algorithmic frontier comes in evolutionary steps
-- we have just witnessed a dramatic improvement of full QCD hybrid Monte
Carlo algorithms \cite{Clark:2006wq}.
The last two points are somewhere in between; here, every collaboration has
its own preferences, which are largely driven by the kind of physics it wants
to address.
Below, some estimates for future error bars on quantities relevant to flavour
physics will be given, but it is important to keep in mind two caveats.

The first caveat is a reminder that the anticipated percentage errors
quoted below belong to a rather restricted class of observables.
In the foreseeable future lattice methods can only be competitive for processes
where the following conditions hold simultaneously:
\begin{itemize}
\vspace{-2pt}
\itemsep-2pt
\item
only one hadron in initial and/or final state,
\item
all hadrons stable (none near thresholds),
\item
all valence quarks in connected graphs,
\item
all momenta significantly below cut-off scale $2\pi/a$.
\end{itemize}
This is the case for the quantities discussed below, but it means that quick
progress on other interesting quantities, such as $f^{B\to\rh}(q^2)$, is not
likely.

The second caveat concerns the role of the theoretical uncertainties, as
discussed in the previous paragraph.
For instance, some of the estimates given below assume that certain (finite)
renormalisation (i.e.\ matching) factors will be known at the 2-loop level.
Such calculations are tedious and rely on massive computer algebra (the lattice
regularisation reduces the full Lorentz symmetry, resulting in a proliferation
of terms).
Accordingly, future progress of such calculations is difficult to predict.
In the same spirit one should mention that in the predictions discussed below
it is assumed that for $\Mpi=250...350\MeV$ one is in a regime where chiral
perturbation theory applies and can be used to further extrapolate the lattice
data to the physical pion mass.
In the unlikely event that for some specific process this is not the case, the
corresponding prediction would undergo substantial revision.

With these caveats in mind it is interesting to discuss the projected error
bars as they are released by some lattice groups.
For instance MILC has a detailed ``road-map'' of their expected percentage
errors (including statistical and theoretical uncertainties) for a number of
matrix elements.
They are collected in the following Table~\ref{tab:latterr},
which they kindly provide.
\begin{table}[t]
\begin{center}
\caption{Prospects for lattice uncertainties (MILC Collab.). The
$B\to\pi\ell\nu$ form factor is taken at $q^2 = 16\GeV^2$.}
\label{tab:latterr}
\begin{tabular}{c|cccc} 
\hline\hline
                         &  Lat'06 &   Lat'07  & 2-3\,yrs. & 5-10\,yrs.\\
\hline
$f_{D_s},f_{B_s}$        &    10   &     7     &     5     &    3-4    \\
$f_D,f_B$                &    11   &    7-8    &     5     &     4     \\
$f_B\sqrt{B_B}$          &    17   &    8-13   &    4-5    &    3-4    \\
$\xi$                    &    --   &     4     &     3     &    1-2    \\
$(B,D)\to(K,\pi)\ell\nu$ &    11   &     8     &     6     &     4     \\
$B\to (D,D^*)\ell\nu$    &     4   &     3     &     2     &     1     \\
\hline\hline
\end{tabular}
\end{center}
\end{table}
By far the most ambitious plans are those of HPQCD.
They have just released numbers
for $f_{D_s}$ and $f_{D_s}/f_D$ with a claimed accuracy of 1.3\% and
0.8\%, respectively \cite{Follana:2007uv}.
They plan on computing $f_{B_s}$ and $f_B$ as well as the $B\to\pi$ form factor
at $q^2\!\simeq\!16\GeV^2$ to 4\%.
Finally, they envisage releasing the ratio $f_{B_s}/f_B$ with 2\% accuracy and
$\xi$ with 3\% accuracy by the end of 2007.

In this context it is worth pointing out that progress in other fields, in
particular in experiment, has the potential to ease the task for the lattice
community.
For instance, quoting the vector form factor $f_+$ for semileptonic
$B\to\pi\ell\nu$ decay at $q^2\!=\!0$ is not the best thing to ask for from the
lattice, since a long extrapolation is needed (see \ref{subsubsubsec:formf}).
Still, in the past this was common practice, since there was very limited
experimental information available.
In the meantime the situation has changed.
Now, rather precise information on the shape of this form factor (via binned
differential decay rate data $d\Gamma/dq^2$) is available, and only the
absolute normalisation is difficult to determine in experiment (see e.g.\
\cite{Flynn:2006vr} for a detailed analysis).
As a result MILC and HPQCD give the future lattice precision attainable at 
$q^2\!=\!16\GeV^2$, i.e.\ at a momentum transfer which can be reached in the
simulation.






%
\newpage \section{New physics in benchmark channels}
\label{sec:npbc}

\subsection{Radiative penguin decays}
\label{sec:radpeng}


The flavour changing neutral current (FCNC) 
transitions $b\to s\gamma$ and $b\to d\gamma$ are among 
the most valuable probes of flavour physics. They place stringent 
constraints on a variety of New Physics models, in particular on those
where the flavour-violating transition to a right-handed $s$- or $d$-quark
is not suppressed, in contrast to the Standard Model (SM).
Assuming the SM to be valid, the combination of these two processes offers 
a competitive way to extract the ratio of  
CKM matrix elements $|V_{td}/V_{ts}|$. This determination is 
complementary to the one from $B$ mixing and to the one of 
the SM unitarity triangle
based on the tree-level observables $|V_{ub}/V_{cb}|$ and the angle $\gamma$. 
Other interesting observables are the CP and isospin asymmetries and
 photon polarization.
Radiative $B$ decays are also characterized by the large impact
of short-distance QCD corrections \cite{Bertolini:1986th}.
Considerable  effort has gone into the calculation
of these corrections, which are now approaching next-to-next-to-leading
order (NNLO) accuracy \cite{Misiak:2006zs,Bobeth:1999mk,Misiak:2004ew,Gorbahn:2004my,Gorbahn:2005sa,Czakon:2006ss,Melnikov:2005bx,Blokland:2005uk,Asatrian:2006ph,Asatrian:2006sm,Asatrian:2006rq,Bieri:2003ue,Misiak:2006ab}. On the experimental side,  
both exclusive and inclusive $b\to s\gamma$ branching ratios are known
with good accuracy, $\sim 5\%$ for $B\to K^*\gamma$ and $\sim 7\%$ for
$\bar B\to X_s\gamma$ \cite{Barberio:2007cr}, while the situation is less favourable
for $b\to d\gamma$ transitions: measurements are only available for
exclusive channels.
Here, we shall discuss first the inclusive modes, then the exclusive ones.
We shall begin with an overview of the current status of the SM 
calculations and later consider the situation for models of New 
Physics.

\subsubsection{$\bar B \to X_{(s,d)} \gamma$ inclusive (theory)}
The inclusive decay rate of the $\bar B$-meson
($\bar B = \bar B^0$ or $B^-$) is known to be well approximated by the
perturbatively calculable partonic decay rate of the $b$-quark:
\begin{equation} \label{bsgam.relation}
\Gamma\left(\bar B \to X_s \gamma\right)_{E_\gamma > E_0} =
\Gamma\left(b \to X_s^{\rm parton} \gamma\right)_{E_\gamma > E_0}
+ {\cal O}\left( \frac{\Lambda^2}{m_b^2}, \frac{\Lambda^2}{m_c^2},
  \frac{\Lambda\alpha_s}{m_b} \right)
\end{equation}
with $\Lambda \sim \Lambda_{\scriptscriptstyle\rm QCD}$
and $E_0$ the photon energy cut in the $\bar{B}$-meson rest frame.
The non-perturbative corrections on the r.h.s. of the above equation
were analyzed in Refs.~\cite{Falk:1993dh,Bigi:1992ne,Buchalla:1997ky,Voloshin:1996gw,Khodjamirian:1997tg,Ligeti:1997tc,Grant:1997ec,Lee:2006wn}. There are also
additional non-perturbative effects that become important when $E_0$ 
becomes too large
($E_0 \sim m_b/2-\Lambda$)~\cite{Becher:2006pu,Andersen:2006hr,Bigi:2002qq}
or too small ($E_0 \ll m_b/2$) \cite{Kapustin:1995fk,Ligeti:1999ea}.
\begin{figure}[h]
\begin{center}
\includegraphics[width=4cm,angle=0]{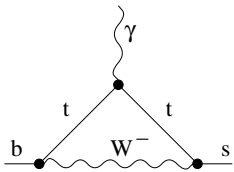}
\caption{\sf Sample LO diagram for the $b\to s\gamma$ transition.\label{fig:LOdiag}} 
\end{center} 
\end{figure}

It is convenient to consider the perturbative contribution first. At the
leading order (LO), it is given by one-loop diagrams like the one in
Fig.~\ref{fig:LOdiag}. Dressing this diagram with one or two virtual gluons
gives examples of the next-to-leading order (NLO) and the
NNLO diagrams,  respectively. The gluon and light-quark
bremsstrahlung must be included as well. The current experimental accuracy 
(see Eq. (\ref{bsgamexp})) can be matched on the theoretical side only after 
including the NNLO QCD corrections \cite{Misiak:2006zs}.

At each order of the perturbative series in $\alpha_s$, large logarithms
$L=\ln M_W^2/m_b^2$ are resummed by employing a low-energy
effective theory that arises after decoupling the top quark and the 
electroweak bosons. 
For example, the  LO includes  all  $\alpha_s^n L^n$ terms,
the NLO all  $\alpha_s^n L^{n-1}$ terms, etc.
Weak interaction vertices (operators) in this theory are
either of dipole type  ($\bar s \sigma^{\mu\nu} b F_{\mu\nu}$
 and  $\bar s
\sigma^{\mu\nu} T^a b G^a_{\mu\nu}$ ) or contain four quarks ($[\bar s \Gamma
b][\bar q \Gamma' q]$).  Coupling constants at these vertices (Wilson
coefficients) are first evaluated at the electroweak renormalization scale
$\mu_0 \sim m_t, M_W$ by solving the so-called~ {\em matching}~ conditions.
Next, they are evolved down to the low-energy scale $\mu_b \sim m_b$ according
to the effective theory renormalization group equations (RGE).  The RGE are
governed by the operator~ {\em mixing}~ under renormalization.  Finally, one
computes the~ {\em matrix elements}~ of the operators, which in the
perturbative case amounts to calculating on-shell diagrams with single
insertions of the effective theory vertices.

The NNLO matching and mixing are now completely known
\cite{Bobeth:1999mk,Misiak:2004ew,Gorbahn:2004my,Gorbahn:2005sa,Czakon:2006ss}.  The same refers to those matrix
elements that involve the photonic dipole operator alone
\cite{Melnikov:2005bx,Blokland:2005uk,Asatrian:2006ph,Asatrian:2006sm,Asatrian:2006rq}. Matrix elements involving other operators are known at
the NNLO either in the so-called large-$\beta_0$ approximation
\cite{Bieri:2003ue} or in the formal $m_c \gg m_b/2$ limit
\cite{Misiak:2006ab}. The recently published NNLO estimate
\cite{Misiak:2006zs}:
\begin{equation} \label{bsgam.estimate}
{\cal B}({\bar B}\to X_s\gamma)_{E_\gamma > 1.6\;\rm GeV} = (3.15 \pm 0.23) \times 10^{-4}
\end{equation}
is based on this knowledge. The four types of uncertainties:
non-perturbative (5\%), parametric (3\%), higher-order (3\%) and
$m_c$-interpolation ambiguity (3\%) have been added in quadrature in
(\ref{bsgam.estimate}) to obtain the total error.
The main uncertainty is due to unknown $O(\alpha_s\Lambda/m_b)$ 
non-perturbative effects related to the matrix elements of four-quark 
operators (see \cite{Buchalla:1997ky}) 
for which no  estimate exists. Similar effects related to dipole 
operators have been recently estimated
in the vacuum insertion approximation  \cite{Lee:2006wn}.

As far as inclusive $b\to d \gamma$ decays are concerned, 
their measurement  is quite challenging. Moreover, due to
non-perturbative effects that are suppressed only by 
$\Lambda_{\scriptscriptstyle\rm QCD}/m_b$,
their  theoretical accuracy is not much better than in the exclusive case. 
On the other hand, the experimental
prospects in the exclusive case are brighter.


\subsubsection{$\bar B \to X_{(s,d)} \gamma$ inclusive (experiment)}

\subsubsubsection{Present status}

The inclusive $b\to s\gamma$ branching fraction has been measured by BaBar,
BELLE and CLEO using both a sum of exclusive modes and a fully inclusive
method~\cite{Abe:2001hk,
Koppenburg:2004fz,Aubert:2005cua,Aubert:2006gg}. The inclusive measurement utilizes the continuum
subtraction technique using the off-resonance data sample. In order
to suppress the continuum contribution the BaBar measurement uses lepton tags.
The analyses of BELLE and CLEO are untagged and their systematic errors are 
dominated
by continuum subtraction. The accuracy of the BaBar measurement is limited
by the subtraction of backgrounds from other $B$ decays. 
The BELLE measurement extends
the minimum photon energy down to 1.8${\rm GeV}$, which covers 95\% of the
entire photon spectrum. All $b\to s\gamma$ branching fractions measured by
BaBar, BELLE and CLEO using both exclusive and inclusive methods agree 
well, giving a new world average of \cite{Barberio:2007cr}
\begin{equation}\label{bsgamexp}
{\cal B}({\bar B}\to X_s\gamma)_{E_\gamma > 1.6\;\rm GeV} =
(3.55\pm 0.30)\times 10^{-4} . 
\end{equation}
This is a bit high compared to the recent NNLO calculation in
(\ref{bsgam.estimate}).
 
The published measurements are based on only a fraction of the available
statistics, but improvements with the full data set will be limited by
systematic errors: from the fragmentation of the hadronic $X_s$ in the
sum of exclusive modes, and from the subtraction of backgrounds
in the fully inclusive method. A new method measures the spectrum of
photons recoiling against a sample of fully reconstructed
decays of the other $B$. This is currently statistics limited, but
should eventually have lower systematic errors. A final accuracy
of 5\% on the inclusive $b\to s\gamma$ branching fraction looks achievable.
As for the $b\to d\gamma$ inclusive branching fraction, the measurement
using a sum of exclusive modes is under study and looks to be 
feasible with the full datasets from the B factories.
Preliminary results have appeared in \cite{Aubert:2007se}.

We note that the $b\to s\gamma$ spectral shape also provides valuable
information on the shape functions in $B$ meson decays. This 
information has been used as an input in the extraction of $V_{ub}$
from inclusive $b\to u\ell\nu$ decays~\cite{Aubert:2006qi,Aubert:2005mg}. 

Measurements of the direct CP asymmetries,
published for inclusive $b\to s\gamma$ by BaBar~
\cite{Aubert:2004hq} and BELLE~\cite{Nishida:2003yw}, show no deviation
from zero. All these measurements will be statistics limited
at current $B$-factories, and will not reach the sensitivity to probe
the SM prediction.

\subsubsubsection{Future prospects}

One would expect a substantial improvement of the experimental precision
for inclusive measurements at future $B$-factories.
Studies have been performed for SuperKEKB/Belle with $50\;{\rm ab}^{-1}$
data, assuming the existing Belle detector~\cite{Hashimoto:2004sm}.  This is
probably a reasonable assumption in many cases since the expected
improvements in the detector, especially in the calorimeter, would be
just sufficient to compensate for the necessity
to cope with the increased background.

For the measurements that are fully statistics dominated now, it is
straightforward to extrapolate to a larger integrated luminosity.  The
error for the direct asymmetry measurement of $b\to s\gamma$ would be
$\pm0.009{\rm(stat)}\pm0.006{\rm(syst)}$ for $5\;{\rm ab}^{-1}$ or
$\pm0.003{\rm(stat)}\pm0.002{\rm(syst)}\pm0.003{\rm(model)}$ for
$50\;{\rm ab}^{-1}$. A small systematic error implies that kaon charge
asymmetries are well under control.
The size of the total error is still much larger than the SM estimate, 
but a few percent deviation from
zero due to New Physics could be identified.  

One would also expect a better measurement of the branching fraction of
$B\to X_s\gamma$.  Although the background level is more and more
severe, it would be possible to lower the $E_\gamma$ bound by $0.1\;{\rm
GeV}$ with roughly twice more data, and it would be possible to measure
the branching fraction for $E_\gamma>1.5\;{\rm GeV}$ with a few ${\rm
ab}^{-1}$.  Beyond that, one may need to make use of the $B$-tag events or
$\gamma\to e^+e^-$ conversion to suppress backgrounds from continuum and
neutral hadrons.  Another challenging measurement would be inclusive
$b\to d\gamma$ to improve our knowledge on $|V_{td}/V_{ts}|$ besides the
$\Delta m_s$ measurement, since the one from exclusive $B\to\rho\gamma$
will hit the theory limit soon.  The first signal may be measured with
$5\;{\rm ab}^{-1}$ using the sum-of-exclusive method, with a total error
of $\sim25\%$, of which the systematic error would already be
dominant.

\subsubsection{\boldmath Exclusive $b\to (s,d)\gamma$ transitions (theory)}
Whereas the inclusive modes can be essentially computed  perturbatively, 
the treatment of
exclusive channels is more complicated. QCD factorisation
\cite{Ali:2001ez,Ali:2004hn,Beneke:2004dp,Bosch:2001gv,Bosch:2004nd,Becher:2005fg} has provided a consistent framework allowing one to
write the relevant hadronic matrix elements as
\begin{equation}\label{eq1}
\langle V\gamma|Q_i| B\rangle =
\left[ T_1^{B\to V}(0)\, T^I_{i} +
\int^1_0 d\xi\, du\, T^{II}_i(\xi,u)\, \phi_B(\xi)\, \phi_{2;V}^\perp(v)\right]
\cdot\epsilon\,.
\end{equation}
Here $\epsilon$ is the photon polarisation four-vector, $Q_i$ is one of
the operators in the  effective Hamiltonian for $b\to (s,d)\gamma$ transitions,
$T_1^{B\to V}$ is a $B\to V$ transition form factor,
and $\phi_B$, $\phi_{2;V}^\perp$ 
are leading-twist light-cone distribution amplitudes
of the $B$ meson and the vector meson $V$, respectively.
These quantities are universal non-perturbative objects and
describe the long-distance dynamics of the matrix elements, which
is factorised from the perturbative short-distance interactions
included in the hard-scattering kernels $T^I_{i}$ and $T^{II}_i$
(see Sec. \ref{sec:hu} for a more general discussion).

Eq.~(\ref{eq1}) is sufficient to calculate observables that are of
$O(1)$  in the heavy quark expansion, like 
${\cal B}(B\to K^*\gamma)$. For ${\cal B}(B\to
(\rho,\omega)\gamma)$, on the other hand,
power-suppressed corrections play an important
r\^ole, for instance weak annihilation, which is mediated by a
tree-level diagram. In this case, the parametric suppression by one
power of $1/m_b$ is alleviated by an enhancement factor $2\pi^2$
relative to the loop-suppressed contributions at leading order in
$1/m_b$. Power-suppressed contributions also determine the
time-dependent CP asymmetry in $B\to V\gamma$, see
Refs.~\cite{Atwood:1997zr,Grinstein:2004uu,Grinstein:2005nu,Ball:2006cv}, as well as isospin asymmetries \cite{Kagan:2001zk} ---
all observables with a potentially large contribution from new
physics. A more detailed analysis of power corrections in 
$B\to V\gamma$, including also $B_s$ decays, was given in \cite{Ball:2006eu}.

The non-perturbative quantities entering Eq.~(\ref{eq1}), i.e.\
$T_1^{B\to V}$ and the light-cone distribution amplitudes, 
at present are not provided by lattice, although this may change in
the future. The most up-to-date predictions come from QCD sum rules on
the light-cone, which are discussed in section \ref{subsec:lcsr}.
In Ref.~\cite{Ball:2006nr}, the following
  result was obtained  for the branching fraction
  ratio:
\begin{equation}
R \equiv  \frac{\overline{\cal B}(B\to (\rho,\omega)\gamma)}{
\overline{\cal B}(B\to K^*\gamma)} = \frac{|V_{td}|^2}{|V_{ts}|^2} (0.75\pm
    0.11(\xi) \pm 0.02(\mbox{UT param., O($1/m_b$)}))\, ,\label{R}
\end{equation}
where
$\xi \equiv T_1^{B\to K^*}(0)/T_1^{B\to\rho}(0) = 1.17\pm 0.09$
(Sec. \ref{subsec:lcsr}).
The error of $\xi$ is dominated by that of the tensor decay constants
$f_{\rho,K^*}^\perp$, which currently are known to about 10\% accuracy
\cite{Ball:2006nr}; a new determination on the lattice is under way
\cite{private}, which will help to reduce the error on $\xi$ to $\pm
0.05$.  Concerning Eq.~(\ref{R}) two remarks are in order.
First, the smallness of the $1/m_b$ correction are due to an accidental CKM suppression.
Second, the $1/m_b$ corrections have a dependence on  $|V_{td}/V_{ts}|$ 
as well, originating from a
discrimination in the $u$ and $c$-loops.
Eq.~(\ref{R}) allows one to determine $|V_{td}/V_{ts}|$ from
experimental data; at the time of writing (February 07), HFAG quotes
$R_{\rm exp} = 0.028\pm 0.005$, from which one finds
$ \left| V_{td}/V_{ts} \right|^{\rm HFAG}_{B\to V\gamma} =
0.192\pm 0.014({\rm th}) \pm 0.016({\rm exp}) $
which agrees very well with the results from global fits
\cite{Bona:2006ah,Charles:2004jd}. 
The branching ratios
themselves carry a larger uncertainty because the individual $T_1^{B\to V}$ are
less accurately known than their ratio.  The explicit results can be found in 
\cite{Ball:2006eu}.
The isospin asymmetry in $B\to K^*\gamma$ was first studied in
Ref.~\cite{Kagan:2001zk} and found to be very sensitive to penguin
contributions; it was updated in \cite{Ball:2006eu} with the result
\begin{equation}
A_I(K^*) = \frac{\Gamma(\bar B^0\to\bar K^{*0}\gamma) - \Gamma(B^-\to
 K^{*-}\gamma)}{\Gamma(\bar B^0\to\bar K^{*0}\gamma) + \Gamma(B^-\to
 K^{*-}\gamma)} = (5.4\pm 1.4)\% \,;
\end{equation}
the present (February 07)
experimental result from HFAG \cite{Barberio:2007cr} is $(3\pm4)\%$. 
 The isospin asymmetry for $B \to \rho \gamma$ depends rather
crucially on the angle $\gamma$ \cite{Ball:2006eu}.
The last observable in exclusive $B\to V\gamma$ transitions to be
discussed here is the time-dependent CP asymmetry, which is sensitive
to the photon polarisation. Photons produced from the short-distance
process $b\to (s,d)\gamma$ are predominantly left-polarised, with the
ratio of right to left-polarised photons given by the helicity
suppression factor $m_{s,d}/m_b$. For $B\to K^*\gamma$, where direct
CP violation is doubly CKM suppressed, the CP asymmetry is given by
\begin{equation}\label{-1}
A_{CP}(t)  = \frac{\Gamma(\bar B^0(t)\to \bar K^{*0}\gamma) -
               \Gamma(     B^0(t)\to      K^{*0}\gamma)}{
               \Gamma(\bar B^0(t)\to \bar K^{*0}\gamma) +
               \Gamma(     B^0(t)\to      K^{*0}\gamma)}
= C \cos(\Delta m_B t ) +   S \sin(\Delta m_B t )\,,
\end{equation}
with $S_{K^*\gamma} = -(2+O(\alpha_s))\sin (2\beta) m_s/m_b+\dots\approx -3\%$ being the
contribution induced by the electromagnetic dipole operator
$O_7$. The dots denote additional contributions induced by $b\to
s\gamma g$, which are not helicity suppressed, but involve higher
(three-particle) Fock states of the $B$ and $K^*$ mesons. 
The dominant contributions to the latter, due to $c$-quark loops,
have been calculated in Ref.~\cite{Ball:2006cv} from QCD sum rules on the
light-cone in an expansion in inverse powers of the charm mass
and updated for all other channels in \cite{Ball:2006eu}.
A calculation of the charm-loop contribution without reference
to a $1/m_c$ expansion is in preparation  \cite{new_charm}
and shows that there is a large strong phase.
The $u$-quark loop contributions are essential for $b \to d$ transitions 
since they
are of the same CKM-order as the $c$-quark loops:
a new method for their estimation was devised in \cite{Ball:2006eu},
building on earlier ideas developed for $B \to \pi\pi$ 
\cite{Khodjamirian:2000mi}.
\begin{center}
\begin{tabular}{l | l | l | l | l | l}
$S_{V\gamma}$ & $B \to \rho$ & $B \to \omega$ & $B \to K^*$ & $B_s \to \bar K^*$ & $B_s \to \phi$
\\ \hline
in \% & $0.2 \pm 1.6$ & $0.1 \pm 1.7$ & $-(2.3\pm 1.6)$ &
$0.3 \pm 1.3$ & $-(0.1 \pm 0.1)$
\end{tabular}
\end{center}
 This class of observables is interesting
because any experimental signal much larger than 2\% will
constitute an unambiguous signal of New Physics. Scenarios beyond the SM
that do modify $S$ must include the possibility of a spin-flip on the
internal line which removes the helicity suppression of
$\gamma_R$. Examples include left-right symmetric models and non-MFV
SUSY. To date the experimental result is $S_{\rm HFAG} = -(28\pm 26)\%$.

\newpage

\subsubsection{\boldmath Exclusive $b\to (s,d)\gamma$ transitions (experiment)}

\subsubsubsection{Present status}

Many exclusive $b\to (s, d)\gamma$ modes have been studied by 
BaBar, Belle and CLEO.
Results for several important channels are collected in the following
table \cite{Barberio:2007cr}: 
\begin{center}
\begin{tabular}{l|c|c|c|c|c}
decay & $B^+\to K^{*+}\gamma$ & $B^0\to K^{*0}\gamma$ &
 $B^+\to\rho^+\gamma$ & $B^0\to\rho^0\gamma$ & $B^0\to\omega\gamma$
\\ \hline
 & & & & & \\
BR/$10^{-6}$ & $40.3\pm 2.6$  & $40.1\pm 2.0$ & $0.88^{+0.28}_{-0.26}$  &
 $0.93^{+0.19}_{-0.18}$ & $0.46^{+0.20}_{-0.17}$ 
\end{tabular}
\end{center}
The results on the $B\to \rho\gamma$, $B\to\omega\gamma$
branching fractions are still statistics limited, but by the end of the
$B$ factories it is likely that the theoretical uncertainties will be the
most significant factor.  

Direct CP asymmetries have been published for $B\to K^*\gamma$ and 
$B\to K^+\phi\gamma$ decays~\cite{Aubert:2004te,Nakao:2004th,Aubert:2006he}. 
The time-dependent CP asymmetry has been measured~\cite{Ushiroda:2006fi,
Ushiroda:2005sb,Aubert:2005bu}
using the technique
of projecting the $K_S$ vertex back to the beam axis for
a large sample of $B\to K^{*0}\gamma \to K_S^0\pi^0\gamma$ and
$B\to K_S^0\pi^0\gamma$ decays in the high $K\pi$-mass range. 
In the near future, similar measurements using
other exclusive radiative decay modes such as $B^0\to K_S^0\phi\gamma$,
for which $\phi\to K^+K^-$ provides the $B$-decay vertex measurement,
could provide similar constraints.

\subsubsubsection{Future prospects}

A systematic study of CP violation in radiative penguin $B$ decays
will be performed at LHCb using a dedicated high-$p_T$ photon 
trigger~\cite{Belyaev:2007zz}.
Due to small branching ratios of order 10$^{-5}$ - 10$^{-6}$
their reconstruction requires a drastic suppression
of backgrounds from various sources, in particular combinatorial
background from $b\bar b$ events, containing primary and secondary vertices
and characterized by high charged and neutral multiplicities. 

The background suppression exploits the generic properties of beauty
production in $pp$ collisions. The large mass of beauty
hadrons results in hard transverse momentum spectra of secondary particles.
The large lifetime, $\langle \beta \gamma c \tau \rangle \sim 5~
{\mathrm{mm}}$, results in a good isolation of the ${B}$ decay vertex
and in the inconsistency of tracks of ${B}$-decay products with the
reconstructed ${pp}$-collision vertex.   

\begin{figure}[h]
\begin{center}
\setlength{\unitlength}{1mm}
\begin{picture}(110,70)
\put(0,-3){\epsfig{file=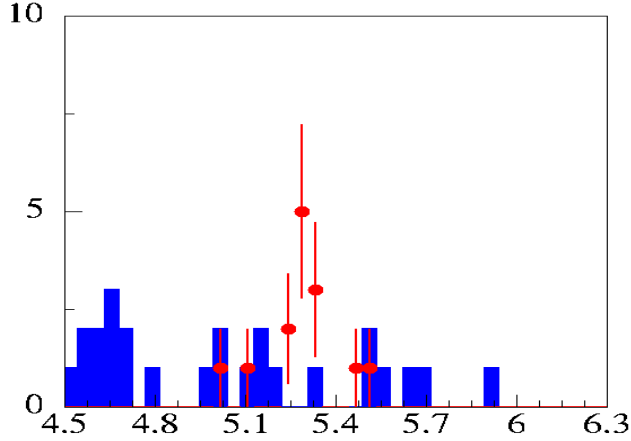,height=65mm,width=90mm}}
\end{picture}
\caption{\sf The invariant mass distribution for selected $B^0\to K^{*0}\gamma$
             candidates from a $b\bar B$ inclusive sample. The points indicate
             true $B^0\to K^{*0}\gamma$ events and the filled histogramm
             represents combinatorial background.\label{fig:kstgmass}} 
\end{center} 
\end{figure}

The selection procedure was optimized on the example of 
$B^0\to K^{*0}\gamma\to K^+\pi^-\gamma$
decay~\cite{lhcb-bkstgam}, 
which LHCb considers as a control channel for the study of systematic
errors common for radiative penguin decays. The selection cuts, based
on using the two-body kinematics and various geometrical cuts on the 
primary and secondary vertices, were applied to 34 million fully
simulated ${b\bar b}$ events. 
The invariant mass distribution for the selected events, 
shown in Fig.~\ref{fig:kstgmass}, corresponds to a data sample
collected in 13 $min$ of LHCb running at nominal luminosity of
$2\times 10^{32} cm^{-1} s^{-1}$.
LHCb expects the yield for $B^0\to K^{*0}\gamma$ decays to be 36k signal events
per 1${\rm fb^{-1}}$ of accumulated data with background to signal 
ratio $0.78\pm 0.11$.
For $B_s\to \phi\gamma$ decays, the corresponding yield is estimated to be
6k with the background to signal ratio less than 0.95 at 95\% CL.
The measurement of $B^0\to K^{*0}\gamma$ decay looks also feasible at 
ATLAS~\cite{atlas-bkstgam}.

Similar to $B^0\to K^0_s\pi^0\gamma$ decays
the time-dependent CP-asymmmetry sensitive to the photon polarisation can 
also be measured in $B_s\to \phi\gamma$ decays provided that the proper time 
resolution is sufficient to resolve $B_s$--$\bar B_s$ oscillations. The proper
time resolution depends on the kinematics and topology of particular
$B_s$ candidates, mainly on the opening angle between kaons from $\phi$ decays.
The sensitivity of this measurement is presently under study at LHCb.
 
For the future $B$-factory, scaling the error of the measured
time-dependent CP violation asymmetry for the $B^0\to K^0_s\pi^0\gamma$
channel, one would expect a statistical accuracy of 
about 0.1 at $5\;{\rm ab}^{-1}$, or 0.03 at
$50\;{\rm ab}^{-1}$.

LHCb also studied the possibility to measure the photon polarisation
in the radiative decays of polarized beauty baryons, like $\Lambda_b
\to \Lambda\gamma$, using the angular asymmetry between the $\Lambda_b$
spin and the photon momentum combined with the $\Lambda^0\to p\pi$
decay polarisation~\cite{lhcb-barpen,Hiller:2007ur}.

\subsubsection{New Physics calculations and tools}
New Physics  affects
the matching conditions for the Wilson coefficients of the operators
in the low-energy effective theory 
and may even induce sizable coefficients for operators that have
 negligible or vanishing coefficients in the SM.
The theoretical accuracy of the predictions for radiative $B$ decays
in extensions of the SM is  far from the accuracy achieved in the SM.  
Complete NLO matching conditions  are available  for 
the MSSM with Minimal Flavour Violation (MFV) and/or large $\tan\beta$,
as well as for a class of non-supersymmetric models  \cite{Bobeth:1999ww}
that includes 
Multi-Higgs-Doublet-Models and Left-Right symmetric (LR) models. 
The unknown NNLO contributions to the matching conditions
beyond the SM are unlikely to be numerically relevant at present.

\subsubsubsection{Summary of New Physics calculations}
Here is a brief  summary of recent calculations and analyses 
in the most popular New Physics scenarios. 

\begin{itemize}
\item [$\bullet$] {\bf 2HDMs} have been studied in full generality at NLO
 \cite{Ciafaloni:1997un,Ciuchini:1997xe,Borzumati:1998tg}. In the type-II 2HDM ${\cal B}(\bsg)$
places a strong bound on the mass of the charged Higgs boson,
$M_{H^+}>295$~GeV at 95\% CL, independently of the other 2HDM parameters 
\cite{Misiak:2006zs}.
This is much stronger than other available direct and indirect constraints
on $M_{H^+}$.

\item [$\bullet$] {\bf MSSM} 
The complete LO contributions 
in the MSSM have been known since the early nineties
\cite{Bertolini:1990if,Barbieri:1993av,Oshimo:1992zd,Diaz:1993km,Okada:1993sx,Garisto:1993jc,Borzumati:1993zg,Barger:1994ki} but the NLO analysis is still incomplete to date. New
sources of flavour violation generally arise in the MSSM, making a
complete analysis quite complicated even at the LO
\cite{Borzumati:1999qt}. While $\bsg$ does place important constraints 
on the MSSM parameter space, they depend sensitively on the exact SUSY 
scenario and are hard to summarize because of the large number of parameters.

\begin{itemize}
\item {\bf MFV}
In the MFV scenario the NLO
QCD calculation of $\bsg$ is now complete: the two-loop
diagrams involving gluons were computed in ref.~\cite{Ciuchini:1998xy,Bobeth:1999ww}, and the
two-loop diagrams involving gluinos were more recently computed in
ref.~\cite{Borzumati:2003rr,Degrassi:2006eh}.  Since weak interactions affect the squark and quark mass
matrices in a different way, their simultaneous diagonalization is not
RG-invariant  and MFV can be imposed
only at a certain renormalization scale. The results of \cite{Borzumati:2003rr,Degrassi:2006eh}
therefore depend explicitly on the MFV scale, which is determined by the 
mechanism of SUSY breaking.

\item {\bf Large {\boldmath $\tan\beta$}.} 
In the limit of heavy superpartners, the Higgs sector of the MSSM is
modified by non-decoupling effects and can differ substantially from
the type-II 2HDM. Large higher order contributions to $\bsg$ in that limit
originate from terms enhanced by $\tan\beta$ factors, and can be taken
into account to all orders in an effective lagrangian approach
\cite{Degrassi:2000qf,Carena:2000uj,D'Ambrosio:2002ex,Buras:2002vd,Gomez:2006uv}. In fact, large  $\tan\beta$ and 
logs of $M_{susy}/M_W$ have been identified in \cite{Degrassi:2000qf}
as dominant NLO QCD contributions in MFV with heavy squarks.
 Ref.~\cite{Lunghi:2006uf} recently studied the
$\tan\beta$-enhanced effects when  MFV is
valid at the GUT scale and additional flavour violation in the
squark sector is generated by the RGE of
the soft SUSY-breaking parameters down to the weak scale. 

\item {\bf Beyond MFV.} 
  In the more general case of arbitrary flavour structure in the
  squark sector, experimental constraints on $b\to s$ transitions have
  been recently studied at LO~\cite{Besmer:2001cj,Ciuchini:2002uv} 
  and including $\tan\beta$-enhanced NLO
  effects~\cite{Okumura:2003hy,Foster:2004vp,Foster:2005wb,Foster:2005kb}:
  radiative decays play a central role in these analyses, and the
  constraints are quite strong for some of the flavour-violating
  parameters.

\end{itemize}

\item [$\bullet$] {\bf Large extra dimensions}. In these models
the contribution to $\bsg$ from the Kaluza-Klein excitations of the SM
particles can induce bounds on the size of the additional
dimension(s).  This has been studied in ref.~\cite{Agashe:2001xt,Buras:2003mk} for the
case of flat extra dimensions and in ref.~\cite{Kim:2002kkb,Agashe:2004ay,Agashe:2004cp} for the case
of warped extra dimensions. 

\item [$\bullet$] {\bf Little Higgs.}
In these models  the Higgs boson
is regarded as the pseudo-Goldstone boson of a global symmetry that is
broken spontaneously at a scale much larger than the weak scale. The
most extensively studied version of the model, the Littlest Higgs,
predicts the existence of heavy vector bosons, scalars and quarks.
The contribution to $\bsg$ from these new particles has been studied
in ref.~\cite{Huo:2003vd,Buras:2006wk} for the original Littlest Higgs model, and in
ref.~\cite{Blanke:2006sb} for the model in which an additional T-parity and
additional particles are introduced to preserve the SU(2) custodial
symmetry.

\item [$\bullet$] {\bf LR models.} 
The contributions of Left-Right symmetric models
to $\bsg$ are known at the NLO \cite{Bobeth:1999ww}, but 
no recent phenomenological analysis is available.

\end{itemize}

An alternative to the analysis of $\bsg$ in different models consists in
constraining the Wilson coefficients of the effective theory.
This {\bf model independent approach} has been applied combining various 
$B$ decay modes and neglecting operators that do not 
contribute in the SM  \cite{Ali:2002jg,Hiller:2003js}. While ${\cal B}(\bsg)$ 
fixes only $|C_7(m_b)|$,
the sign can be learned from $B\to X_s \ell^+\ell^-$ \cite{Gambino:2004mv}.

\subsubsubsection{MSSM tools for $\bsg$}
Several public codes (see also Sec. \ref{sec:tools})
that determine the MSSM mass spectrum
and other SUSY observables contain MSSM calculations of ${\cal B}(\bsg)$ in 
various approximations.  In {\tt micrOMEGAs} \cite{Belanger:2004yn} the
SM part of the calculation is performed at NLO, while the
MSSM contributions are implemented following \cite{Degrassi:2000qf}.
The calculation in {\tt SuSpect} \cite{Djouadi:2002ze} includes also the 
NLO gluon corrections to the
chargino contributions from \cite{Ciuchini:1998xy} in the case of 
light squarks. In contrast, {\tt
SPheno} \cite{Porod:2003um} and {\tt FeynHiggs} \cite{Heinemeyer:1998yj,Hahn:2005qi} include the
SUSY contributions only at LO, but they allow for a general flavour
structure in the squark sector. 
A computer code for the NLO QCD calculation of  
${\cal B}(\bsg)$ in the MSSM with MFV \cite{Borzumati:2003rr,Degrassi:2006eh}
has recently been published \cite{Degrassi:2007kj}.

\newpage %








\subsection{Electroweak penguin decays}

\subsubsection{Introduction}

In the SM, the electroweak penguin decays $b \to s(d) \ell^+\ell^-$
are only induced at the one-loop level, leading to small branching
fractions and thus a rather high sensitivity to contributions from new
physics beyond the SM. On the partonic level, the main contribution to
the decay rates comes 
from the semi-leptonic operators ${\cal O}_9$, ${\cal O}_{10}$
and from the electromagnetic dipole operator ${\cal O}_7^\gamma$ 
in the effective Hamiltonian for $|\Delta
B|=|\Delta S(D)|=1$ transitions \cite{Buchalla:1995vs}.  Radiative
corrections induce additional sensitivity to the current-current
and strong penguin operators ${\cal O}_{1-6}$ and ${\cal O}_8^g$. Part
of these effects are process-independent and can be absorbed into
effective Wilson coefficients. In certain regions
of phase-space and for particular exclusive and inclusive observables,
hadronic uncertainties are under reasonable control and the
corresponding short-distance Wilson coefficients in and beyond the SM
can be tested with sufficient accuracy.

Because of their small branching fractions these decays are
experimentally challenging.  Their detection requires
excellent triggering and identification of leptons, with low
misidentification rates for hadrons.  Combinatorial backgrounds from
semileptonic $B$ and $D$ decays must be managed, and backgrounds from
long-distance contributions, such as $B \to J/\psi X_{s}$, must be
carefully vetoed.  Once identified, their interpretation (particularly
the angular distributions) requires disentangling the contributing
hadronic final states.  Most of these experimental problems can be
managed by confining studies to the simplest exclusive decay modes.
Leptonic states are restricted to $e^+e^-$ and $\mu^+\mu^-$, and
hadronic states are the simplest one- or two-particle varieties,
typically $K$, $K^*$, $\phi$, or $\Lambda$.  More inclusive studies
are significantly less sensitive but have the advantage of a simpler
theoretical interpretation.  Fortunately, measuring fully
reconstructed decays to final states with leptons (especially muons)
is a strength of all future proposed $B$ physics experiments, hence
all are capable of contributing to this topic in the LHC era.

\subsubsection{Theory of electroweak penguin decays}

\label{sec:exth}

\paragraph{Inclusive decays}

The heavy quark expansion and the operator product expansion in the
theory of inclusive $\bar{B} \to X_s \ell^+ \ell^-$ decays allow to
calculate radiative QCD and QED corrections to the partonic decay rate
and to pa\-ra\-me\-trize and estimate power corrections to the hadronic
matrix elements in a systematic way.  The calculation of NNLO QCD
corrections has (essentially) been completed recently
\cite{Bobeth:1999mk,Asatryan:2001zw,Asatrian:2001de,Asatrian:2002va,Ghinculov:2003qd,Gambino:2003zm,Gorbahn:2004my,Bobeth:2003at}.
These reduce the perturbative uncertainties below 10\%.  Also subleading
$\Lambda_{\rm QCD}^2/m_c^2$ and $\Lambda_{\rm QCD}^2/m_b^2$,
$\Lambda_{\rm QCD}^3/m_b^3$
corrections~\cite{Buchalla:1997ky,Buchalla:1998mt,Falk:1993dh,Ali:1996bm,Bauer:1999za,Bauer:1999kf}
as well as finite bremsstrahlung effects~\cite{Asatryan:2002iy,
  Asatrian:2003yk} are available in the literature.

At this level of accuracy, QED effects become important, too.  For
instance, the scale ambiguity from $\alpha_{\rm em}(\mu)$ between
$\mu=M_W$ and $\mu=m_b$ alone results in an uncertainty of about $\pm
4\%$.
QED corrections to the Wilson coefficients have been calculated in
Ref.~\cite{Bobeth:2003at}, and the results for the two-loop anomalous
dimension matrices have been confirmed in \cite{Huber:2005ig}.  QED
brems{}strahlung contributions where the photon is collinear with one
of the outgoing leptons are enhanced by $\ln(m_b^2/m_\ell^2)$.  They
disappear after integration over the whole available phase space but
survive and remain numerically important when $q^2$ is restricted to
either low or high values.

A numerical analysis \cite{Huber:2005ig}, done under the assumption of
perfect separation of electrons and energetic collinear photons,
results in the following branching ratios integrated in
the range $1$~GeV$^2 < m^2_{\ell\ell} < 6$~GeV$^2$:
\begin{eqnarray}
\hskip -0.5cm
{\cal B}(\bar{B} \to X_s \mu^+ \mu^-) & = & \label{eq:muonBR} 
\Big[ 
 1.59 
\pm 0.08_{\rm scale}
\pm 0.06_{m_t} 
\pm 0.024_{ C,m_c }
\pm 0.015_{m_b} 
\pm 0.02_{\alpha_s(M_Z)}\nonumber \\
&& \hspace*{28.5pt} \pm 0.015_{\rm CKM} 
\pm 0.026_{{\rm BR}_{sl}} 
\Big] \times 10^{-6} = (  1.59  \pm 0.11 ) \times 10^{-6} \;, \\
\hskip -0.5cm
{\cal B}(\bar{B} \to X_s e^+ e^-) & = & \label{eq:electronBR} 
\Big[ 
 1.64  
\pm 0.08_{\rm scale} 
\pm 0.06_{m_t} 
\pm 0.025_{ C,m_c }
\pm 0.015_{m_b} 
\pm 0.02_{\alpha_s(M_Z)} \nonumber \\
&& \hspace*{28.5pt} \pm 0.015_{\rm CKM}
\pm 0.026_{{\rm BR}_{sl}}
\Big] \times 10^{-6}  = (  1.64   \pm 0.11) \times 10^{-6} \;,
\end{eqnarray}
where the error includes the parametric and
perturbative uncertainties only.
For central values and error bars of the input parameters see Table~1
of Ref.~\cite{Huber:2005ig}.  The electron and muon channels receive
different contributions because of the $\ln(m_b^2/m_\ell^2)$ present
in the bremsstrahlung corrections. The difference gets reduced when
the BaBar and Belle angular cuts are included. One should also
keep in mind that the contributions of the intermediate $\psi$ and
$\psi^{\prime}$ are assumed to be subtracted on the
experimental side. A numerical formula
that gives the branching ratio for non-SM values of the relevant
Wilson coefficients is given in Eqs.~(12) and~(13) of
Ref.~\cite{Huber:2005ig}.

The differential branching ratio (BR) is sensitive to the interference
of the Wilson coefficients $C_{7}$ and $C_{9}$. The forward-backward
asymmetry (FBA) for the charged leptons is sensitive to the products
$C_{7} \, C_{10}$ and $C_{9} \, C_{10}$. For instance, reversing the
sign of $C_{7}$ makes the zero of the FBA disappear~\cite{Ali:2002jg}
and leads to an enhancement of the low-$q^2$ integrated
BR:
\begin{eqnarray}
&&
{\cal B}(\bar{B} \to X_s \mu^+ \mu^-) =   3.11 \cdot 10^{-6} \label{eq:muonBRsignreversed}  \,,\qquad
{\cal B}(\bar{B} \to X_s e^+ e^-) =  3.19 \cdot 10^{-6} \;. \label{eq:electronBRsignreversed}
\end{eqnarray}
(A similar value for that case has been found in \cite{Gambino:2004mv}.)

\paragraph{Exclusive decays}


\def\calAslash{\rlap{\hspace{0.08cm}/}{{\EuScript A}}}
\def\nbarslash{\rlap{\hspace{0.02cm}/}{\bar n}}
\def\nslash{\rlap{\hspace{0.02cm}/} {n}}
\def\polslash{\rlap{\hspace{0.02cm}/} {\varepsilon}}


We focus on the theoretical description of $B \to K^*
\ell^+\ell^-$ decay as one of the phenomenologically most important examples. 
The double-differential spectrum may be parametrized as \cite{Lee:2006gs}
\begin{equation}
 \frac{d^2 \Gamma}{d q^2 \, d\cos\theta_\ell} =
\frac38 \left[ (1+\cos^2\theta_\ell) \, H_T(q^2)
 + 2 \, \cos\theta_\ell \, H_A(q^2) + 2 \, (1-\cos^2\theta_\ell) \, H_L(q^2) \right] \,.
\label{klldecayamplitude}
\end{equation}
Here, for $\bar B^0$ or $B^-$ decays, $\theta_\ell$ is the angle between
the $\ell^+$ and the $B$\/-meson 3-momentum in the $\ell^+\ell^-$ c.m.s.\footnote{Different
sign conventions are used in the literature.} 
and $q^2=m_{\ell\ell}^2$ is the invariant mass of the lepton pair.
Alternatively, the functions $H_X(q^2)$ can be expressed in terms of 
transversity amplitudes \cite{Kruger:2005ep}
\begin{eqnarray}
 H_T(q^2) &=& |A_{\perp,L}|^2+ |A_{\perp,R}|^2 + |A_{\parallel,L}|^2+ |A_{\parallel,R}|^2 \,,
\\
 H_L(q^2) &=& |A_{0,L}|^2+ |A_{0,R}|^2 \,,
\\
 H_A(q^2) &=& 2 \, {\rm Re} \left[A_{\parallel,R}  A_{\perp,R}^* - A_{\parallel,L}  A_{\perp,L}^*\right] \,.
\end{eqnarray}

If the invariant mass of the lepton pair is sufficiently 
below the charm threshold at $q^2=4m_c^2$ and above the 
real-photon pole at $q^2=0$, the transversity amplitudes can be
estimated within the QCD factorization approach
 \cite{Beneke:2001at,Beneke:2004dp,Ali:2006ew}
\begin{eqnarray}
 A_{\perp, L/R}\simeq- A_{\parallel, L/R} &\simeq& \sqrt2 \, N \, m_B \, \left(1-\frac{q^2}{m_B^2} \right)
  \left[ {\cal C}_9^\perp(q^2) \mp C_{10} \right] \, \zeta_\perp(q^2) \,,\\
 A_{0,L/R} &\simeq& - \frac{N m_B^2}{\sqrt{q^2}} \left(1-\frac{q^2}{m_B^2}\right)
  \left[ {\cal C}_9^\parallel(q^2) \mp C_{10} \right] \, \zeta_\parallel(q^2)
\end{eqnarray}
where the normalization factor $N$ is defined in Eq.~(3.7) 
in \cite{Kruger:2005ep}.
The functions ${\cal  C}_{9,10}^\perp(q^2)$ can be calculated perturbatively
in the heavy-quark limit, requiring $q^2 \lsim \Lambda m_b \ll 4 m_c^2$
\cite{Beneke:2001at,Beneke:2004dp}.
Large logarithms can be resummed using renormalization-group techniques in
soft-collinear effective theory \cite{Ali:2006ew}.
The form factors $\zeta_{\perp,\parallel}(q^2)$ have to be estimated
from experimental data or theoretical models.\footnote{The conventions
  to define the form factors $\zeta_{\perp,\parallel}$ in
  \cite{Ali:2006ew} are different from those of
  Ref.~\cite{Beneke:2001at}. Therefore the explicit expressions for
  ${\cal C}_{9}^{\perp,\parallel}$ also differ.}  
$1/m_b$ power corrections may be sizeable and currently constitute 
a major source of theoretical uncertainty.

Similarly, in the region far above the charm resonances, the
helicity amplitudes can be treated within heavy-quark effective
theory, based on an expansion in $\Lambda/m_b$ and $4m_c^2/q^2$
\cite{Grinstein:2004vb}. 
To first approximation one finds
\begin{eqnarray}
 A_{\perp, L/R} &\simeq& - \sqrt2 \, N \, m_B \, \left(1-\frac{q^2}{m_B^2} \right)
  \left[ {\cal C}_9^{\rm eff}(q^2)+ \frac{2m_b m_B}{q^2} \, C_7^{\rm eff} \mp C_{10} \right] \, 
  m_B \, g(q^2)
\,,\\
A_{\parallel, L/R} &\simeq& 
- \sqrt2 \, N \, m_B \,
  \left[ {\cal C}_9^{\rm eff}(q^2)+ \frac{2m_b m_B}{q^2} \, C_7^{\rm eff} \mp C_{10} \right] \, 
  \frac{f(q^2)}{m_B}
\,,\\
 A_{0,L/R} &\simeq& - N \, m_B \, \frac{m_B^2-q^2}{2 m_{K^*}\, \sqrt{q^2}} 
  \left[ {\cal C}_9^{\rm eff}(q^2)+ \frac{2m_b}{m_B} \, C_7^{\rm eff} \mp C_{10} \right] 
 \frac{f(q^2)+(m_B^2-q^2) \, a_+(q^2)}{m_B} \,.
\end{eqnarray}
Here $f(q^2),g(q^2),a_+(q^2)$ are the leading HQET form factors \cite{Grinstein:2004vb}.
The effective ``Wilson coefficients'' $C_9^{\rm eff}$
are functions of the lepton invariant mass $q^2$, and combine
short-distance dynamics encoded in Wilson coefficients and
(non-trivial) long-distance dynamics at the scale $m_b$.
In the naive factorization approximation, they 
are related to ${\cal C}_9^{\perp,\parallel}(q^2)$ via
\begin{eqnarray}
  {\cal C}_9^{\perp}(q^2) &\approx& C_9(\mu) + 
Y(q^2,\mu) +  \frac{2 m_b m_B}{q^2} \, C_7^{\rm eff}(\mu) + \ldots =
C_9^{\rm eff}(q^2) + \frac{2 m_b m_B}{q^2} \, C_7^{\rm eff}
 + \ldots
\\
 {\cal C}_9^{\parallel}(q^2) &\approx& C_9(\mu) + 
Y(q^2,\mu) +\frac{2 m_b}{m_B} \, C_7^{\rm eff}(\mu) =
  C_9^{\rm eff}(q^2) + \frac{2 m_b}{m_B} \, C_7^{\rm eff} + \ldots
\end{eqnarray}
(In the following, we will also use the notation
 $C_{9,10}(\mu=m_b)=A_{9,10}$ and $C_7^{\rm eff}(\mu=m_b)=A_7$.)

It is to be stressed that the theoretical systematics in the kinematic regions
$q^2 \ll 4 m_c^2$ and $q^2 \gg 4 m_c^2$ is quite different, due to the different
short-distance effects to be accounted for in the calculation of 
${\cal C}_9^{\perp,\parallel}(q^2)$ or $C_{7,9}^{\rm eff}$, the independent 
hadronic form factors in SCET/HQET, and the different nature of 
(non-factorizable) power corrections.

Experimentally, the dilepton invariant mass spectrum and the
forward-backward (FB) asymmetry are the observables of principal
interest. Their theoretical expressions can be easily derived from
Eq.~(\ref{klldecayamplitude}). In particular, the forward-backward
asymmetry vanishes at $q_0^2$, if $ {\rm Re}\left[{\cal
    C}_9^{\perp}(q_0^2)\right] = 0 \,, $ which turns out to be very
sensitive to the size and relative sign of the electroweak Wilson
coefficients $C_7$ and $C_9$ \cite{Burdman:1998mk, Ali:1999mm}.  The
theoretical predictions depend on the strategy to fix the hadronic
input parameters, and on the scheme to organize the perturbative
expansion in QCD. The authors of \cite{Beneke:2001at,Beneke:2004dp}
fix the hadronic form factors from QCD sum rules \cite{Ball:1998kk}
and calculate the short-distance coefficients in fixed-order
perturbation theory. For the partially integrated branching fraction
they find
\begin{eqnarray}
  \int\limits_{1~{\rm GeV}^2}^{6~{\rm GeV}^2} dq^2 \,
  \frac{d{\rm Br}[B^+ \to K^{*+}\ell^+\ell^-]}{dq^2}
  &=& 
  \left(\frac{\zeta_\parallel(4~{\rm GeV}^2)}{0.66} \right)^2
  \ (3.33^{+0.40}_{-0.31}) \cdot 10^{-7}
\label{eq:BRBFS}
\end{eqnarray}
where the leading dependence on one of the $B \to K^*$ form factors has
been made explicit. For neutral $B$ mesons the result is about 10\%
smaller.  The forward-backward asymmetry zero in this scheme comes out
to be
\begin{equation}
   q_0^2[K^{*0}] = 4.36^{+0.33}_{-0.31}~{\rm GeV}^2 \,,\qquad
   q_0^2[K^{*+}] = 4.15^{+0.27}_{-0.27}~{\rm GeV}^2 \,,
\label{eq:AFBBFS}
\end{equation}
with an additional uncertainty from power corrections estimated to be
of the order of 10\%.

The authors of \cite{Ali:2006ew} fix the form factor $\zeta_\perp(0)$
by comparing the experimental results on $B \to K^*\gamma$ with the
theoretical predictions at NLO at leading power and assuming a simple
energy dependence of the form factor. Furthermore, the leading
perturbative logarithms in SCET are resummed.  They get a somewhat
smaller value for the partially integrated branching
fraction\footnote{Notice that the upper limit of integration in
  (\ref{eq:BRAli}) is slightly larger than those in (\ref{eq:BRBFS}).}
\begin{equation}
\int \limits_{1\mbox{\scriptsize ~GeV}^2}^{7\mbox{\scriptsize
~GeV}^2} d q^2 \frac{d Br(B^+ \to K^{\ast +} \ell^+
\ell^-)}{dq^2}=(2.92^{+0.57}_{-0.50}
\vert_{\zeta_\parallel}~^{+0.30}_{-0.28} \vert_{\mbox{\scriptsize
CKM}} ~^{+0.18}_{-0.20})\times 10^{-7}~,
\label{eq:BRAli}
\end{equation}
which is mainly due to a smaller default value for the 
$B \to K^*$ form factor $\zeta_\parallel$  taken from  \cite{Ball:2004rg}. 
The forward-backward asymmetry zero now reads
\begin{equation}
q^2_0=(4.07^{+0.16}_{-0.13})~ \mbox{GeV}^2~\,,
\end{equation}
where the smaller parametric uncertainties compared to
(\ref{eq:AFBBFS}) are traced back to the renormalization-group
improvement of the perturbative series and the different strategy to
fix $\zeta_\perp(q^2)$. Isospin-breaking effects between charged and
neutral $B$ decays, and potentially large hadronic uncertainties from
power corrections have not been specified in \cite{Ali:2006ew}.

As has been pointed out in \cite{Grinstein:2005ud}, the $K^*$ meson is
always observed through the resonant $B \to (K\pi) \ell^+\ell^-$
decay. Depending on the considered phase-space region in the Dalitz
plot, this may induce further corrections to the position of the
asymmetry zero. On the other hand, it allows for an analysis of
angular distributions.  Following Ref.~\cite{Kruger:2005ep}, one can
consider the polarization fractions
\begin{equation}
  F_L(q^2) = \frac{H_L(q^2)}{H_L(q^2)+H_T(q^2)} \,, \qquad
  \qquad
  F_T(q^2) = \frac{H_T(q^2)}{H_L(q^2)+H_T(q^2)}
\label{eq:FLT}
\end{equation}
and the $K^\ast$-polarization parameter $\alpha_{K^\ast}(q^2) =2
F_L/F_T-1 $.  Like the FBA, these observables have smaller hadronic
uncertainties (for small values of $q^2$), as the hadronic
form-factors cancel in the ratios to first approximation
\cite{Kruger:2005ep}. Introducing the angle $\theta_K$ of the $K$
meson relative to the $B$\/-momentum in the $K^*$ rest frame,
the triple differential decay rate reads
\begin{eqnarray}
  \frac{d^3\Gamma}{dq^2 \, d\cos\theta_l \, d\cos\theta_K}
&=& \left\{
  \frac{9}{8} \, F_L \, \cos^2 \theta_K \, \sin^2 \theta_{\ell} 
+ \frac{9}{32} \, (1 - F_L) \, \sin^2 \theta_K \, (1 + \cos^2 \theta_{\ell}) \right\}
\frac{d\Gamma}{dq^2} \cr
&& {} + \frac{3}{4}  \, \sin^2 \theta_K \cos \theta_{\ell} 
 \left(\frac{d\Gamma_{F}}{dq^2}- \frac{d\Gamma_{B}}{dq^2} \right) \,.
\label{eq:triplerate}
\end{eqnarray}
Finally, the remaining angle, $\phi$, between the decay planes of the lepton
pair and $K^*$ meson defines the distribution \cite{Kruger:2005ep} 
\begin{eqnarray}
\frac{d^2\Gamma}{dq^2 \, d\phi} & = & \frac{1}{2\pi} \left( 1 + \frac{1}{2} \, (1 - F_L) \, A^{(2)}_T \, \cos 2\phi 
    +  A_{\rm Im} \, \sin 2\phi \right) \frac{d\Gamma}{dq^2} \,, 
\label{eq:At2}
\end{eqnarray}
where the asymmetry $A^{(2)}_T(q^2)$ is sensitive to new physics from right-handed currents, 
and the amplitude $A_{\rm Im}$ is sensitive to complex phases in the hadronic matrix elements. 
In the SM, the asymmetry $A^{(2)}_T$ and the amplitude $A_{\rm Im}$ are negligble at low $q^2$, 
so the measurement of either is a precision null test.

The differential decay rate for $B \to K \ell^+\ell^-$ can be found in
\cite{Beneke:2001at}.  Within the SM the FB asymmetry in $B \to K
\ell^+\ell^-$ is highly suppressed.  At hadron colliders, also the
decay modes $B_s \to \phi \ell^+\ell^-$ and $B_s \to \eta^{\,(')}
\ell^+\ell^-$ can be studied.  Their theoretical description is
analogous to the $B \to K^*(K)$ case, but accurate numerical studies
require better knowledge of the hadronic parameters entering the
$B_s$, and $\phi(\eta,\eta')$-meson wave functions.

Baryonic decay channels, $\Lambda_b \to \Lambda^0 \ell^+\ell^-$, are
theoretically less well understood. So far, they have only been
discussed within the (naive) factorization approximation, based on
symmetry relations and model estimates for the $\Lambda_b \to
\Lambda^0$ form-factors (see e.g.\
\cite{Hiller:2001zj,Chen:2001zc,Aliev:2002tr}).  Besides the $q^2$
spectrum and the FBA, the baryonic $b \to s \ell^+\ell^-$ decays offer
the possibility to study various asymmetry parameters and $\Lambda^0$
polarization effects, which exhibit a particular dependence on NP
effects
\cite{Aliev:2002ww,Giri:2005yt,Turan:2005cw,Aliev:1999ap,Turan:2005pf,Giri:2005mt,Chen:2002rg}.
Also a possible initial $\Lambda_{b}$ polarisation can be accounted
for \cite{Aliev:2005np}.


\paragraph{Charmonium resonances in $b \to s \ell \ell$}

The calculation of inclusive and exclusive observables in $b \to s
\ell^+\ell^-$ decays is complicated by the presence of long-distance
contributions related to intermediate $c \bar c$ pairs from the
4-quark operators in the effective Hamiltonian. 
The effect depends on the invariant mass $q^2$ of the lepton pair.

For the inclusive rate, the charm quarks can be
integrated out perturbatively within an OPE based on an expansion in
$\alpha_s$ and $(1/m_c,1/m_b)$ (with the ratio $m_c/m_b$ kept fixed).
Below the charm threshold $q^2 \ll 4 m_c^2$, the expansion in
$1/m_c^2$ still converges, and the inclusive decay spectrum can be
described in terms of a local OPE
\cite{Voloshin:1997gw,Ligeti:1997tc,Grant:1997ec,Chen:1997dj,Buchalla:1997ky,Buchalla:1998mt}. Similarly, for exclusive
decays it is possible to integrate out the intermediate charm loops
perturbatively, leading to non-local operators whose matrix elements
can be further investigated using QCDF, SCET or (light-cone) sum
rules, see the discussion in Sec.~\ref{sec:hu} and
\cite{Khodjamirian:1997tg,Ball:2006cv} (for the case $q^2=0$).

Approaching the charm threshold at $q^2 \sim 4m_c^2$, the heavy-quark
expansion breaks down, both in inclusive and exclusive decays. A
pragmatic solution is to ignore the $c\bar c$ resonance region
completely by introducing ``appropriate'' experimental cuts on
$q^2$. Alternatively, one may attempt to model a few resonances
explicitly (in practice the $J/\psi$ and the $\psi(2S)$), see e.g.\
\cite{Ali:1999mm} and references therein.  However, this method bears
the danger of double-counting when combined with the OPE result, which
can be avoided by using dispersion relations for the electromagnetic
vacuum polarization \cite{Kruger:1996cv}. Still, non-factorizable soft
interactions between the resonating charmonium system and the $B \to
X_s$ transition cannot be accounted for in a systematic way at
present.

For values of $q^2$ above the charm threshold, the invariant mass of
the hadronic final state is small, and the decay rate is dominated by
a few exclusive states. To trust the OPE result for the inclusive
spectrum, one has to smear the experimental spectrum over a
``sufficiently'' large $q^2$ range and rely on the (semi-local)
duality approximation.  For the description of the exclusive channels
in that region, one has to rely on an expansion in terms of $4m_c^2/q^2$
within HQET \cite{Grinstein:2004vb}. In summary, to avoid contamination from
charmonium or light vector resonances, one should consider the range
$1~{\rm GeV}^2 \leq q^2 \leq 6~{\rm GeV}^2$.

Finally, one has to mention that light-quark loops need a similar
investigation in order to assess the role of light vector resonances
at small values of $q^2$.  We also should stress that while analyzing
the $\bar{c}c$ background in inclusive $B\to X_s l^+l^-$ transitions, 
special care should be taken of the chain of $B\to J/\psi X_s$,
$J/\psi\to l^+l^- X$ decays, mimicking $b\to s l^+l^-$ with
$q^2 < m^2_{J/\psi}$.


\subsubsection{Experimental studies of electroweak penguin decays}

\paragraph{Measurements (prospects) at (Super-)$B$ factories}

The $B$-factory experiments BaBar and Belle have succeeded in measuring
the $b \to s \ell^+ \ell^-$ process in $B$ decays, both
exclusively~\cite{Abe:2004ir,Ishikawa:2006fh,Aubert:2006vb} and
inclusively~\cite{Iwasaki:2005sy,Aubert:2004it}.  Measured observables include:
total branching fractions; direct CP asymmetries; partial branching
fractions vs. the dilepton $q^2$ and the hadronic $X_s$ mass; and, for $B \to
K^* \ell^+\ell^-$, the dilepton angular asymmetry $A_{FB}$ vs. the dilepton
$q^2$, the $K^*$ longitudinal polarization vs. the dilepton $q^2$, and fits of
the $d^2 \Gamma/d\cos\theta \, dq^2$ distribution to extract
experimentally $A_9/A_7$ and $A_{10}/A_7$.  Upon accumulation of more
data in current $B$ factories or the proposed super $B$ factories, it
should be possible to extract most of the observables described in
Section~\ref{sec:exth}, in increasingly finer binning and precision.
The expected experimental sensitivity of 50 ab$^{-1}$ of $B \to
K^*\ell^+\ell^-$ data at a super $B$ factory is comparable to 3.3
fb$^{-1}$ of $B^0 \to K^{*0}\mu^+\mu^-$ data at LHCb, as described
below.

The optimal measurement technique is to completely reconstruct the
signal $B$ decay: selection of events with an electron or muon pair,
selection of all hadrons of the appropriate $X_s$ system ($K$ or $K^*$
mesons for the exclusive case, and a $K$ plus 1, 2, 3 or 4 pions
for the inclusive case), and then application of the standard
kinematic requirements in mass and energy for the resulting $B$
candidate.  Partial or full reconstruction requirements for the recoil
$B$ are in general suboptimal.  Triggering signal events is fully
efficient and particle identification is both efficient (typically
80-90\% per particle) and pure (negligible fake rates for electrons,
percent level fake rates for muons and kaons) down to low particle lab
momenta ( 0.3 GeV/$c$ for electrons and 0.7 GeV/$c$ for muons).
Charmonium background can be efficiently vetoed by the lepton-pair
mass and does not significantly contaminate the $q^2$ regions dominated
by the short-distance physics of interest.  The remaining combinatorial
background, mostly from semileptonic $B$ and $D$ decays, is
significant, but it can be reliably separated from signal by
extrapolation from distributions in kinematic sidebands, typically via
an unbinned maximum likelihood fit.  Branching fraction results are
shown in Table~\ref{tab:bfactorykll}.  The effective signal to
background ratio for these results varies from 1:2 (inclusive) up to 2:1
(Belle $K^*\ell\ell$).  Comparable sensitivity is attained for both
electron and muon decay channels.

    \begin{table}[t!]
\centering
\begin{tabular}{lrrrr}%
     Result  & \multicolumn{1}{c}{$\int \cal{L}$ (fb$^{-1}$)} &  \multicolumn{1}{c}{yield} &
\multicolumn{1}{c}{efficiency (\%)} & \multicolumn{1}{c}{$\cal{B}$ ($10^{-6}$)} \\%
      \hline
BaBar $B\to K\ell\ell$~\cite{Aubert:2006vb}    & 208 & $46\pm10$ & $15\pm1$ & $0.34\pm0.07\pm0.02$\\
Belle $B\to K\ell\ell$~\cite{Abe:2004ir}    & 253 & $79\pm11$ & $13\pm1$ & $0.55\pm0.08\pm0.03$\\
HFAG $B\to K\ell\ell$~\cite{Barberio:2006bi}      & & & & $0.44\pm0.05$ \\ \hline
BaBar $B\to K^*\ell\ell$~\cite{Aubert:2006vb}  & 208 & $57\pm14$ & $7.9\pm0.4$ & $0.78\pm0.19\pm0.11$\\
Belle $B\to K^*\ell\ell$~\cite{Abe:2004ir}  & 253 & $82\pm11$ & $4.6\pm0.2$ & $1.65\pm0.23\pm0.11$\\
HFAG $B\to K^*\ell\ell$~\cite{Barberio:2006bi}   & & & & $1.17\pm0.16$\\ \hline
BaBar $B \to X_s \ell\ell$~\cite{Aubert:2004it}  & 82 & $40\pm10$ & $2.0\pm0.4$ & $5.6\pm1.5\pm1.3$\\
Belle $B\to X_s \ell\ell$~\cite{Iwasaki:2005sy}    & 140 & $68\pm14$ & $2.7\pm0.5$ & $4.1\pm0.8\pm0.9$\\
HFAG $B\to X_s \ell\ell$~\cite{Barberio:2006bi}     & & & & $4.5\pm1.0$\\
\end{tabular}
\caption{\label{tab:bfactorykll}%
Branching fraction measurements at $B$ factories for $b \to s \ell^+ \ell^-$ decays, including integrated luminosity, signal yield, detection efficiency, 
and the measured branching fraction over the full $q^2$ range. 
The HFAG averages are also included.}
    \end{table}%

    Assuming HFAG branching fractions, and the efficiencies
    and backgrounds observed in the Belle results, the expected signal
    yields (and their statistical precision) per 1 ab$^{-1}$ are
    $229\pm16$ (7\%), $215\pm16$ (7\%), and $486\pm24$ (5\%), for
    $K\ell\ell$, $K^*\ell\ell$, and $X_s \ell\ell$, respectively.  The
    experimental uncertainty for total branching fractions should
    therefore be less than or comparable to current Standard Model
    theoretical uncertainties, using $B$-factory data alone.  Direct CP
    violation will be bounded at the level of 5-7\% with 1 ab$^{-1}$,
    and thus a Super $B$ factory would obtain a high precision test
    ($\sim1\%$) of the null result expected in the Standard Model.
    Similar precision is expected for measuring differences in
    branching fractions between electron and muon channels, which is
    also an interesting null test of the Standard
    Model~\cite{Hiller:2003js,Yan:2000dc}.  A possible complicating
    factor for the inclusive $X_s \ell\ell$ (partial) branching
    fractions is the necessity of an aggressive requirement on the
    mass $M_{X_{s}}$ to be less than 1.8 GeV/$c^2$.  Such a tight cut
    may introduce significant shape function effects into the
    interpretation of the results, in the same manner as a photon
    energy cut does for $B\to X_s
    \gamma$~\cite{Lee:2005pk,Lee:2005pw}.  A looser $M_{X_{s}}$
    requirement will have poorer precision, and thus Super $B$ factory
    samples may be required to compare with the most precise
    predictions.

The $B$ factories have also succeeded in accumulating large enough $B\to
K^*\ell\ell$ samples to perform angular analyses as a function of
dilepton mass.  The angles analyzed thus far include the angle,
$\theta_{\ell}$, between the positive (negative) lepton and the
$B$ ($\overline{B}$) momentum in the dilepton rest frame, and the
angle, $\theta_{K}$, of the $K$ meson relative
to the $B$ momentum in the $K^*$ rest frame.  
The integrated longitudinal
$K^*$ polarization $F_L$ and the forward-backward asymmetry $A_{FB}$
are related to the decay products' angular distribution via 
Eq.~(\ref{eq:triplerate}),
which upon integration of one of the angular variables reduces to
\begin{eqnarray}
\frac{d^2\Gamma}{dq^2 \, d\cos\theta_{K}} 
& = &
\left\{ \frac{3}{2} \, F_L \, \cos^2 \theta_K  + \frac{3}{4} \, (1 - F_L) \, \sin^2 \theta_K
\right\} \frac{d\Gamma}{dq^2} \,,
 \\
\frac{d^2\Gamma}{dq^2 \, d\cos\theta_{\ell}} 
& = &
\left\{
\frac{3}{4} \,F_L \, \sin^2 \theta_{\ell} + \frac{3}{8} \, (1 - F_L) \, (1 + \cos^2 \theta_{\ell})
+ A_{FB} \, \cos\theta_\ell \right\} \, \frac{d\Gamma}{dq^2} \,.
\end{eqnarray}
From the singly- or doubly-differential angular distributions (in a given $q^2$\/-bin)
it is then possible to infer $A_{FB}(q^2)$ and $F_{L}(q^2)$ simultaneously. 
There is also the remaining angle, $\phi$, between the decay planes of the lepton
pair and $K^*$ meson, which has yet to be analyzed, see Eq.~(\ref{eq:At2}).

BaBar has measured $A_{FB}$ and $F_L$, in two bins of $q^2$ (above and
below 8.4 GeV/$c^2$), via unbinned maximum likelihood fits to the
singly-differential distributions of $\cos\theta_{\ell}$ and
$\cos\theta_K$, which take into account signal efficiency as a
function of angle as well as background angular distributions (which
are in general non-uniform and forward-backward
asymmetric)~\cite{Aubert:2006vb}.  Table~\ref{tab:bfactoryafb} shows the
expected precision for these observables extrapolated to Super B
luminosities, assuming HFAG branching fractions and Standard Model
predictions for $d\Gamma/dq^2$.  The ultimate 50 ab$^{-1}$ precision
of the $A_{FB}$ of $B\to K^* \ell\ell$, integrated over the
theoretically preferred range of 1-6 GeV$^2/c^4$, is 2.6\%.  If this
region is extended more aggressively to the original BaBar choice of
0.1-8.4 GeV$^2/c^4$, the signal statistics are doubled, and the
precision improves to 1.8\%.  Similar precision is expected for
$F_{L}$.  Measuring integrated angular observables of these types has
the advantages of model independence in their interpretation; the
underlying relation between these measurements, the Wilson
coefficients, and the form factors can change without necessitating
revision of the measurement.  The averaging of multiple experimental
results is also very straightforward.
 
\begin{table}[t!]
\centering
\begin{tabular}{llrrrr}%
\multicolumn{1}{c}{$\int \cal{L}$ (ab$^{-1}$)} 
&  &  \multicolumn{1}{c}{1} &\multicolumn{1}{c}{5} &
\multicolumn{1}{c}{10} & \multicolumn{1}{c}{50} \\%
      \hline
$K^*\ell\ell: \ A_{FB}$ &
  $q^2$ in 1-6 GeV$^2/c^4$ & 18 & 8.2 & 5.8 & 2.6\\
& $q^2 >10$ GeV$^2/c^4$ & 11 & 4.7 & 3.3 & 1.5\\
& All & 7.9 & 3.5 & 2.5 & 1.1\\ \hline
$K^*\ell\ell \ F_{L}$ &
  $q^2$ in 1-6 GeV$^2/c^4$ & 12 & 5.3 & 3.7 & 1.7\\
& $q^2 >10$ GeV$^2/c^4$ & 9.4 & 4.2 & 3.0& 1.3\\
& All & 7.2 & 3.2 & 2.3 & 1.0\\ \hline
$K^+ \ell\ell \ A_{FB}$ &
  All & 8.4 & 3.7 & 2.6 & 1.2\\
\end{tabular}
\caption{\label{tab:bfactoryafb}Expected statistical precision 
of a Super $B$ factory, in percent, for the angular observables 
$A_{FB}$ and $F_L$ versus the integrated luminosity, 
integrated over various ranges of $q^2$.}
\end{table}%

\begin{figure}[ht!]
\begin{center}
  \includegraphics[width=0.5\textwidth]{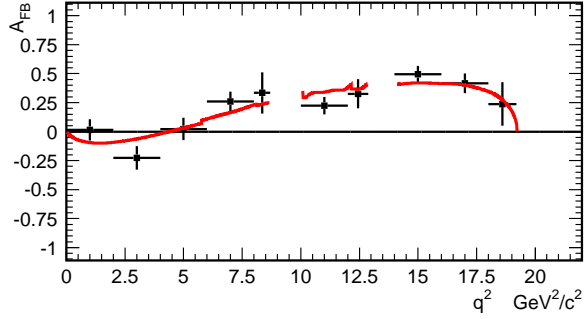}
  \caption{Expected measurement of $dA_{FB}/dq^2$ for $B \to K^* \ell^+ \ell^-$
(points) with 
5 ab$^{-1}$ of data from a Super $B$ factory; 
the best fit of that data for Wilson 
coefficients $A_9$ and $A_{10}$ is superimposed (solid line)~\cite{hl6afb}.}
\label{fig:SuperBAfb}
\end{center}
\end{figure}

Alternatively, Belle has analyzed the doubly-differential distribution
$d^{2}\Gamma/d\cos\theta_{\ell} dq^2$ and then performed a maximum
likelihood fit to extract the Wilson coefficient ratios $A_9/A_7$ and
$A_{10}/A_7$ from the data~\cite{Ishikawa:2006fh}.  Using the
theoretical approximation in Ref.~\cite{Ali:2002jg}, and assuming the
form factor model of Ref.~\cite{Ali:1999mm}, they find
\begin{eqnarray}
A_9/A_7 & \simeq & -15.3^{+3.4}_{-4.8}\pm1.1\nonumber \\
A_{10}/A_7 & \simeq & 10.3^{+5.2}_{-3.5}\pm1.8,
\end{eqnarray}
where the $A_i$ are the leading order Wilson coefficients.  This is in
agreement with the LO Standard Model predictions of -12.3 and 12.8,
respectively.  The dominant systematic uncertainty is from theoretical
model dependence, particularly the form factor model and parametric
uncertainty from $m_b$.  This method has been studied for Super 
$B$-factory luminosities, as discussed in Ref.~\cite{hl6afb}.
Figure~\ref{fig:SuperBAfb} shows a projection of $dA_{FB}/dq^2$ from a
likelihood fit to the Wilson coefficients, for a simulated sample of 5
ab$^{-1}$, compared to $A_{FB}$ integrated over various bins in $q^2$
measured from the same sample.  Employing the entire range of $q^2$,
the expected statistical precision is shown in
Table~\ref{tab:belleci}.  With 5-10 ab$^{-1}$, the expected
statistical uncertainty will be less than the current systematic
uncertainty.  The expected ultimate statistical sensitivity for 50
ab$^{-1}$ is about 4\% for each coefficient.  These fits extract
essentially the same information as that obtained from measuring the
zero $q^2_0$ of $dA_{FB}/dq^2$ (a theoretically clean estimator of
$A_9/A_7$), except that the distribution is analyzed globally and not just
in the vicinity of $q^2_0$; equivalent uncertainties for $q^2_0$ are
identical to those of $A_9$.  In order to control theoretical
uncertainties, it may be necessary to restrict the fit to 1-6
GeV$^2/c^4$. For that measurement the price in experimental statistics
is roughly a factor of 0.6, with an even larger sacrifice in
sensitivity for $A_{10}$, which is most relevant at high $q^2$.

\begin{table}[ht!]
\centering
\begin{tabular}{lrrrr}%
\multicolumn{1}{c}{$\int \cal{L}$ (ab$^{-1}$)} &  \multicolumn{1}{c}{1} &\multicolumn{1}{c}{5} &
\multicolumn{1}{c}{10} & \multicolumn{1}{c}{50} \\%
      \hline
$A_9$ &25 & 11 & 7.8 & 3.5\\
$A_{10}$ &29 & 13 & 9.2& 4.1\\
\end{tabular}
\caption{\label{tab:belleci} Expected statistical precision 
for a Super $B$ factory, in percent, for Wilson coefficients $A_9$ and $A_{10}$
versus the integrated luminosity, integrated over the entire range of $q^2$.}
\end{table}%

With more data, it could also be possible to bound other Wilson
coefficients which are negligible in the Standard Model, such as those
corresponding to scalar operator products or products with flipped
chirality.  Fitting triply- or quadruply-differential distributions
with the additional decay angles $\cos \theta_K$ and $\phi$, as is
currently done for large samples of $B \to VV$ decays, will also be
possible.

Measuring the angular distribution of inclusive $B \to X_s \ell\ell$
decays has not yet been attempted, however with thousands of
events expected at a Super $B$ factory there will be sufficient
statistics for a precise measurement of $A_{FB}$~\cite{Hewett:2004tv}.
This is an attractive measurement, as observables such as $q^2_0$ are
predicted more precisely than for the exclusive case ($\sim 5\%$).
Scaling from the expected yield per ab$^{-1}$ of $486\pm24$, and
assuming the same sensitivity to $A_9/A_7$ per event as for the $B \to
K^*\ell\ell$ Wilson coefficient fits, a 5\% statistical precision for
$A_9/A_7$ (and hence $q^2_0$) could be achieved with roughly 10
ab$^{-1}$, although again a critical issue for the precision is how
wide a range of $q^2$ is appropriate for such fits.  Understanding
systematic uncertainties from a sum-of-exclusive-modes analysis will
be challenging, in particular the effect of imprecise $X_s$
fragmentation modeling on the multiply-differential efficiency.

\paragraph{$B_d \rightarrow K^{*0} \mu^+ \mu^-$ at LHCb}

The exclusive $B_d \rightarrow K^{*0} \mu^+ \mu^-$ decay can be triggered and
reconstructed in LHCb with high efficiency due to the clear di-muon signature
and K/$\pi$ separation provided by the RICH detector~\cite{LHCb-2007-038}.

The selection criteria including the trigger have an efficiency of $1.1\%$ for
signal. The trigger accepts 89\% of the Monte Carlo signal events, which are
reconstructed offline. In $2\rm\:fb^{-1}$ of integrated luminosity this
selection gives an estimated signal of 7200 events with a total background of
3500 events in a $\pm50\rm \,Me\kern -0.08em V\!/c^2$ mass window around the
$B$ mass and $\pm100\rm \,Me\kern -0.08em V\!/c^2$ window around the $K^{*0}$
mass. The branching ratio for $B_d \rightarrow K^{*0} \mu^+ \mu^-$ was assumed
to be $1.22\times10^{-6}$. The irreducible non-resonant $B_d \rightarrow K^+
\pi^- \mu^+ \mu^-$ background was estimated at 1730 events; the branching
ratio used for this was set using a 90\% upper limit estimate found from the
sidebands of the $K^{*0}$ mass in~\cite{Aubert:2006vb}. Other large components
of the background are 1690 from events with two semileptonic $B$ decays, 640
of which are from semileptonic decays of both the $b$ and the $c$ quarks
within the same decay chain. Exclusive backgrounds from other $b \rightarrow s
\mu^+ \mu^-$ decays were considered and contribute at a very low level of 20
events.

The selection efficiency as a function of $q^2$ is flat in the region
$4m_\mu^2$ to $9\rm \,Ge\kern -0.08em V^2\!/c^4$ due to the high boost of the
$B_d$. For high $q^2$ values the selection efficiency as a function of
$\theta_l$ is flat while for low $q^2$ the efficiency is highest around
$\theta_l= \pi/2$~\cite{LHCb-2007-039}.

In addition to the well-known forward-backward asymmetry, $A_{FB}$, LHCb will
be able to extract information about the differential decay rate
$d\Gamma/ds$ and the transversity amplitudes $A_0$, $A_\parallel$, and
$A_\perp$ through the asymmetry $A_T^{(2)}$ and the $K^{*0}$ longitudinal
polarisation $F_L$, see Eqs~(\ref{eq:triplerate}) and (\ref{eq:At2}).

\begin{figure}[htbp]
  \centering
  \ifpdf
  \includegraphics[width=0.4\linewidth]{sll/LHCbAFBToyMC.pdf} \
  \includegraphics[width=0.4\linewidth]{sll/LHCbAFBZeroPoint.pdf}
  \else
  \includegraphics[width=0.4\linewidth]{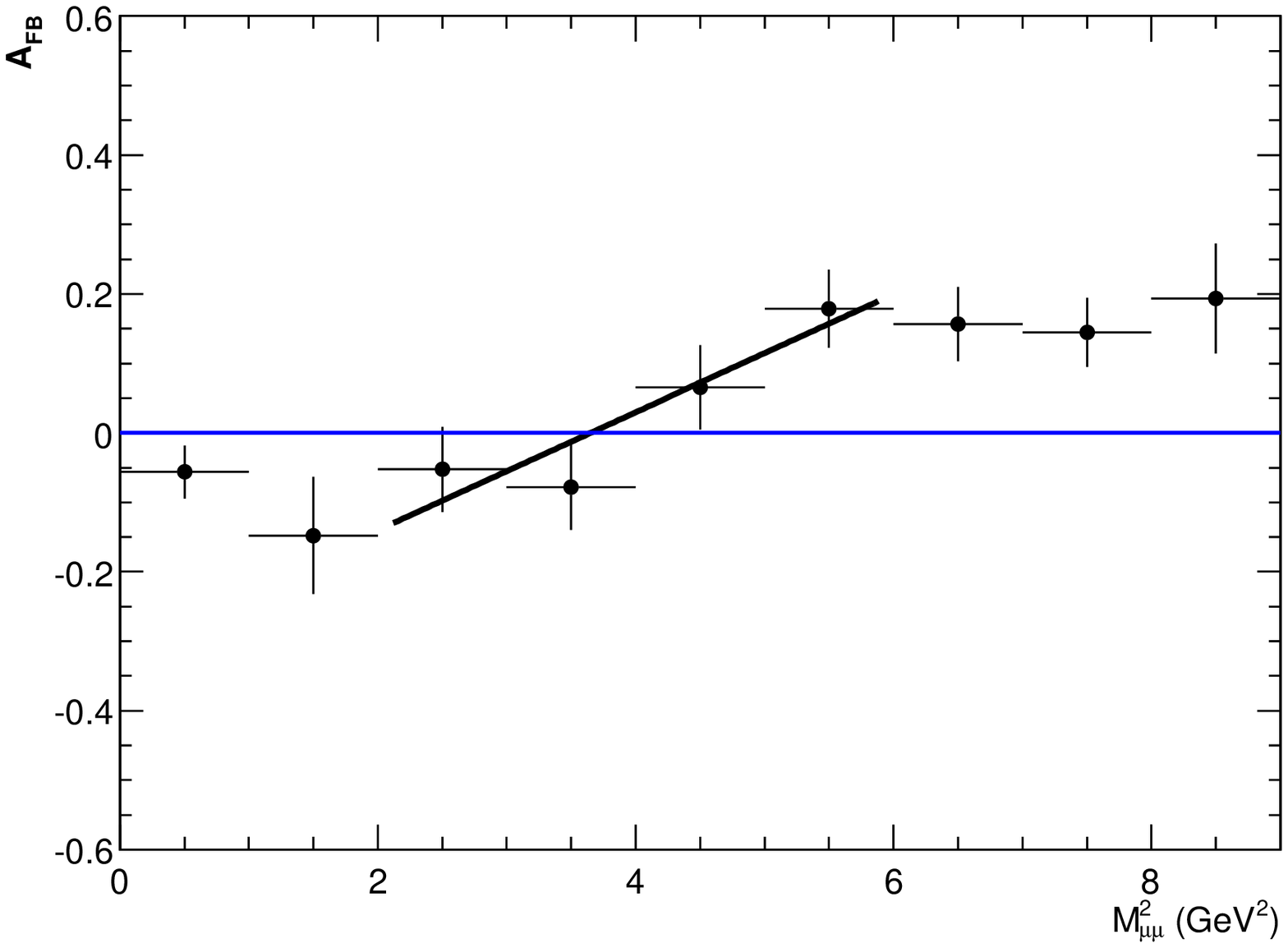} \
  \includegraphics[width=0.4\linewidth]{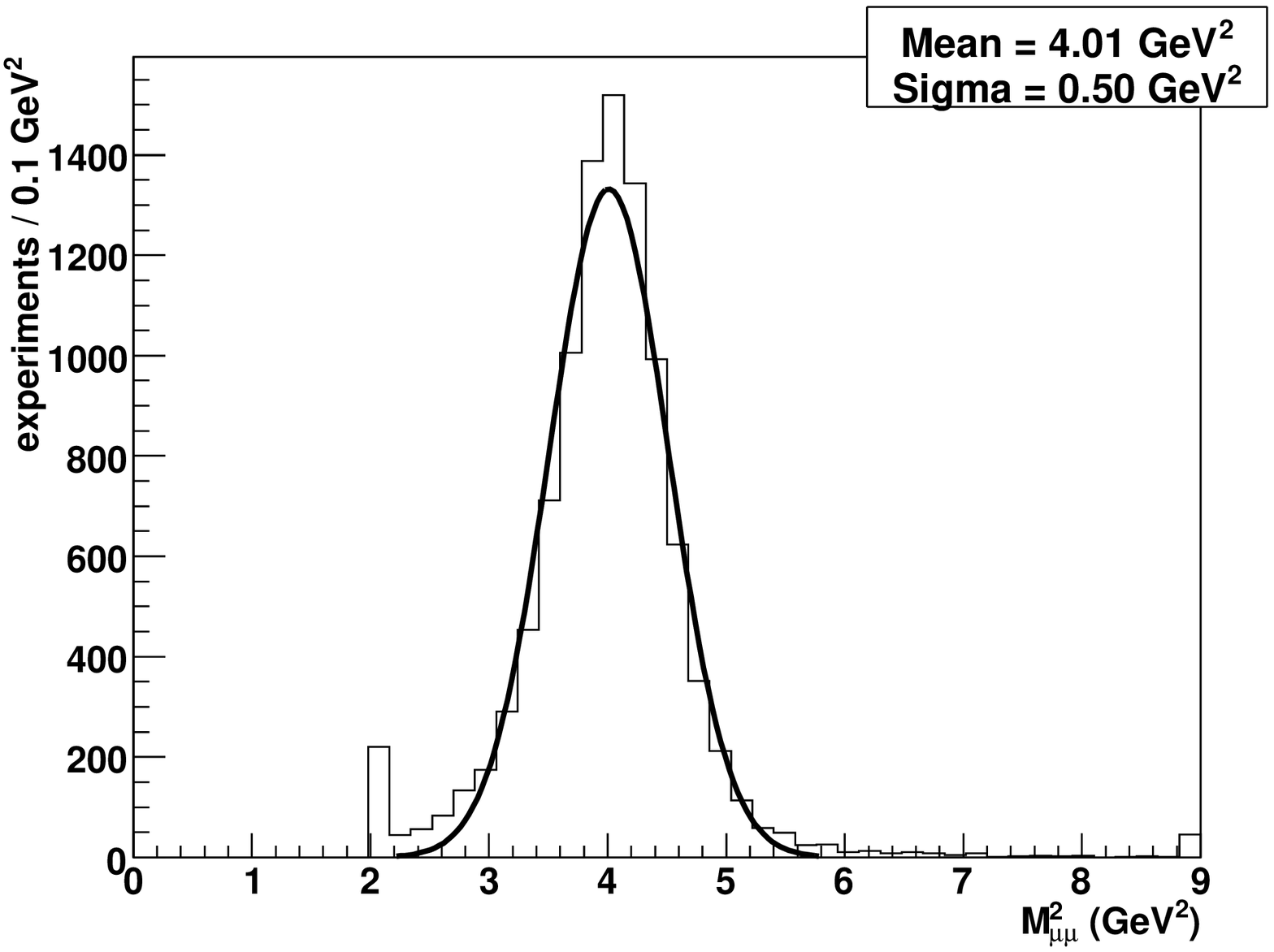}
  \fi
  \caption{The forward-backward asymmetry in $B_d \rightarrow K^{*0} \mu^+
    \mu^-$ with $2 \rm\:fb^{-1}$ of integrated luminosity at LHCb. To the left
    the forward-backward asymmetry as a function of $q^2$ in a single toy
    Monte Carlo experiment and to the right the fitted zero point location for
    an ensemble of Monte Carlo experiments. The peaks at 2 and 9 correspond to
    fits where the zero point was outside this region.}
  \label{fig:LHCbAFB}
\end{figure}

For measuring the zero point in $A_{FB}$, a linear fit is performed to the
measured $A_{FB}$ in the region $2-6\rm \,Ge\kern -0.08em V^2\!/c^4$ as
illustrated in Fig.~\ref{fig:LHCbAFB}. For the resolution in the zero point of
$A_{FB}$~\cite{LHCb-2007-039} we estimate $0.50 (0.27)\rm \,Ge\kern -0.08em
V^2\!/c^4$ with $2 (10) \rm\:fb^{-1}$ of integrated luminosity. If the
background is ignored the resolution is $0.43 (0.25)\rm \,Ge\kern -0.08em
V^2\!/c^4$.

The statistical errors for $A_{FB}$, $A_T^{(2)}$ and $F_L$ have been
estimated by performing simultaneous fits to the $\theta_l$, $\theta_K$ and
$\phi$ projections of the full angular distribution in 3 bins of $q^2$ below
the $\psi$ resonances~\cite{LHCb-2007-057}. In the theoretically favoured
region of $1 < q^2 < 6\rm \,Ge\kern -0.08em V^2\!/c^4$ the resolution in
$A^{(2)}_T$ is $0.42 (0.16)$ with $2 (10) \rm\:fb^{-1}$ of integrated 
luminosity.
See Table~\ref{tab:LHCbKstarmumu} for estimated statistical errors on all the
parameters. In particular the resolution on $A_T^{(2)}$ would improve if the
theoretically comfortable region could be expanded upwards from $6\rm
\,Ge\kern -0.08em V^2\!/c^4$.
\begin{table}[htbp]
  \centering
  \begin{tabular}{c | c c c c c c}%
    $q^2$ region & 
      \multicolumn{2}{c}{$A_{FB}$} & 
      \multicolumn{2}{c}{$A_T^{(2)}$} &
      \multicolumn{2}{c}{$F_L$} \\%
    ($\rm \,Ge\kern -0.08em V^2\!/c^4$) & 
      $2\rm\:fb^{-1}$ & $10\rm\:fb^{-1}$ &
      $2\rm\:fb^{-1}$ & $10\rm\:fb^{-1}$ &
      $2\rm\:fb^{-1}$ & $10\rm\:fb^{-1}$ \\%
    \hline%
    $0.05-1.00$& $0.034$ &$0.017$ &$0.14$ &$0.07$ & $0.027$ &$0.011$ \\%
    $1.00-6.00$& $0.020$ &$0.008$ &$0.42$ &$0.16$ & $0.016$ &$0.007$ \\%
    $6.00-8.95$& $0.022$ &$0.010$ &$0.28$ &$0.13$ & $0.017$ &$0.008$ \\%
  \end{tabular}%
  \caption{  \label{tab:LHCbKstarmumu} The expected resolution for measurements of the parameters
    $A_{FB}$, $A_T^{(2)}$ and $F_L$, for the
    $B_d \rightarrow K^{*0} \mu^+ \mu^-$ decay at LHCb in regions of the
    squared di-muon mass $q^2$ with $2$ and $10\rm\:fb^{-1}$ of integrated
    luminosity.}
\end{table}


\paragraph{$R_K$ at LHCb}


Reconstructing $B^+\rightarrow K^+e^+e^-$ as well as 
$B^+\rightarrow K^+\mu^+\mu^-$ 
allows us to extract the ratio $R_K$ of the two branching fractions, 
integrated over a given di-lepton mass range. 
The same reconstruction requirements are applied to 
$B^+\rightarrow K^+\mu^+\mu^-$ and $B^+\rightarrow K^+ e^+e^-$ decay. 
A proper bremsstrahlung correction is essential in the latter channel. 
The correction for the lower reconstruction
and trigger efficiency in the electron mode is extracted from
$B^+\rightarrow J/\psi K^+$ decays. The di-lepton mass range is chosen to be
$4m_\mu^2 < q^2 < 6\:\rm GeV^2/c^4$ in order to avoid 
$c\bar c$ resonances (especially in the $e^+e^-$ mode) and threshold effects 
due to the higher $\mu$ mass. The event yields are extracted from a 
fit to the $K\ell^+\ell^-$ mass distributions. 
Peaking backgrounds from $B^+\rightarrow J/\psi K^+$ and 
$B_d\rightarrow K^\ast\ell^+\ell^-$ are measured using control samples
and included in the fit.

\begin{figure}[t]
\begin{center}
\includegraphics[width=0.38\textwidth]{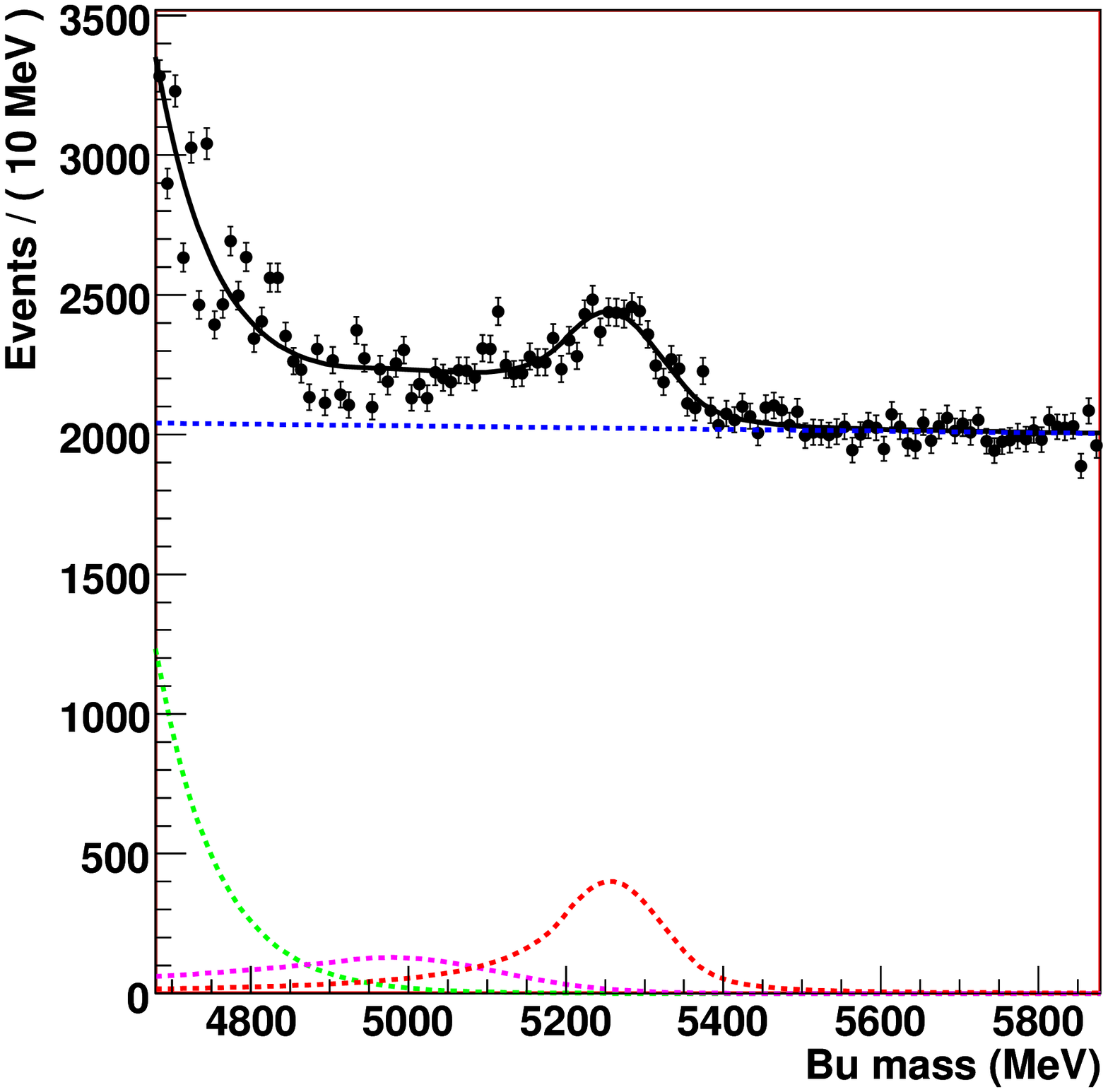} \
\includegraphics[width=0.38\textwidth]{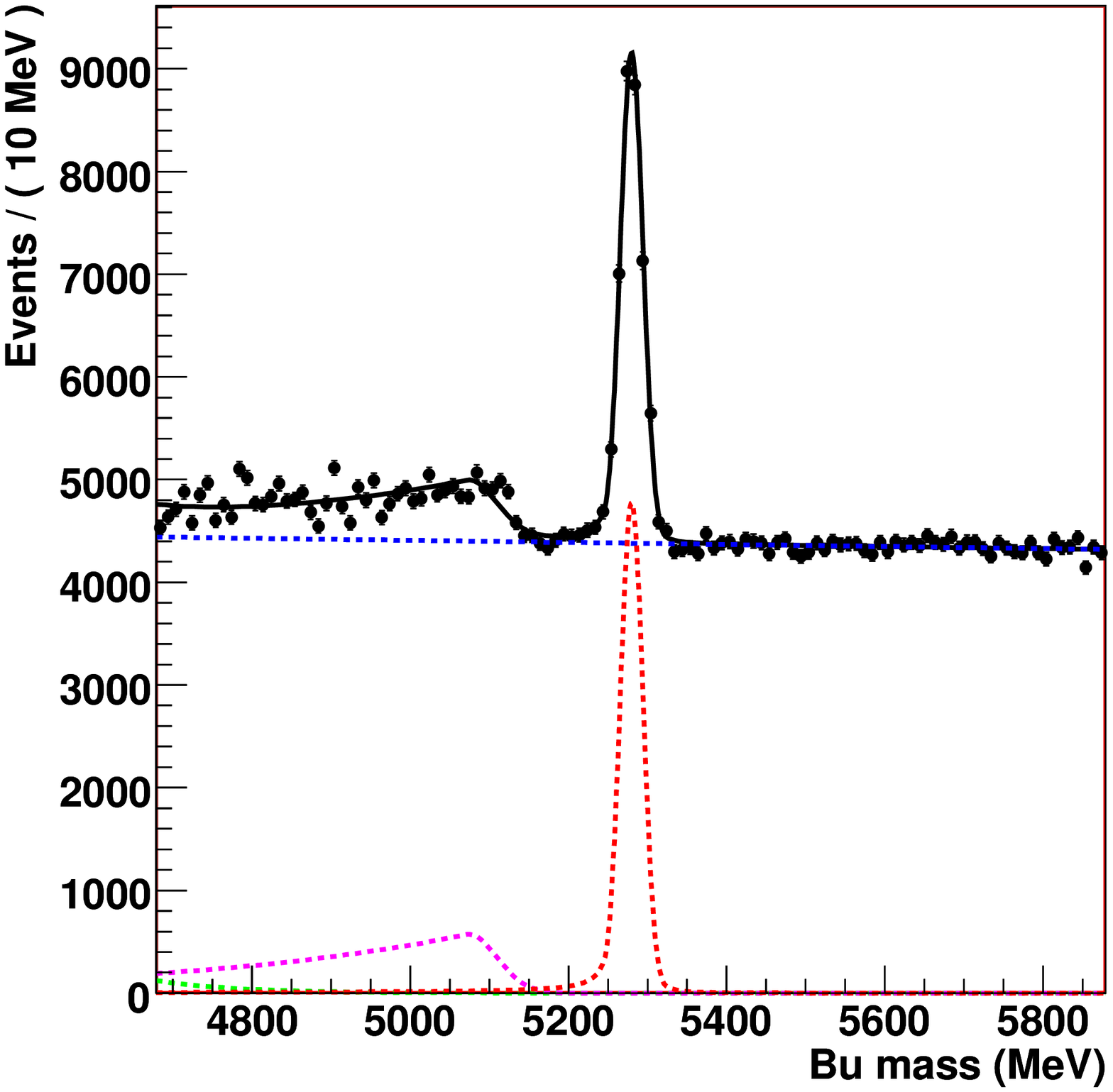}
\caption{Expected $B^+$ candidate mass distributions in the 
$B^+\rightarrow K^+e^+e^-$ (left)
and $B^+\rightarrow K^+\mu^+\mu^-$ (right) modes
for $10\rm\:fb^{-1}$ at LHCb. The dotted lines show the
contributions from signal and 
specific backgrounds as extracted from the fit
(see text).}
\label{figRK}
\end{center}
\end{figure}

The expected $B$ candidate mass distributions are shown in 
Fig.~\ref{figRK} for five years ($10\rm\:fb^{-1}$) of data taking. 
The yields returned by the fit are given in the table below. 
They are compatible with the number of true MC events.
The $B/S$ ratios are given for the full signal box within
$\pm600\:\rm MeV$ around the $B_u$ mass (shown in Fig.~\ref{figRK}).

\begin{center}
    \begin{tabular}{l r r r}%
       & \multicolumn{1}{c}{Yield} & $B/S$ &
       \multicolumn{1}{c}{$\sigma(m_{B_u})$} \\%
      \hline%
      $B^+\rightarrow K^+\mu^+\mu^-$ & $18\:774\pm 230$ & $ \sim 29$ & $14\:\rm MeV/c^2$ \\%
      $B^+\rightarrow K^+e^+e^-$      & $9\:240\pm 380$ & $ \sim 30$ & $68\:\rm MeV/c^2$  \\%
    \end{tabular}%
\end{center}

The errors on the yields are the statistical error returned by the fit. Using these
errors one gets an error on $R_K$ of $4.3\%$ for $10\rm\:fb^{-1}$.

\paragraph{Semileptonic rare $B$ decays at ATLAS}



With the ATLAS experiment, new physics effects in $b\to sl^+l^-$ transitions will be
searched for in the branching ratio and forward-backward asymmetry 
$A_{FB}(q^2)$ 
between $b$-hadron and $l^+$ momenta.
With baryonic decays ($\Lambda_b\to\Lambda^0\mu^+\mu^-$) new physics effects can also be extracted from 
$\Lambda^0$ polarisation and asymmetry parameters (Figs.~2,3,4 from \cite{Giri:2005yt}), 
but influence of possible initial $\Lambda_b$ polarisation has to be accounted for \cite{Aliev:2005np}.
Note that the measurement of the di-lepton mass spectrum is more sensitive to the ATLAS detector efficiency than 
to new physics.

The main part of $B$-physics studies will be performed in the initial LHC low-luminosity stage 
(3 years at $L=10^{33}\;\mathrm{cm}^{-2}\mathrm{s}^{-1}$).
It is expected that the luminosity will vary by a factor of $\sim 2$ during beam-coast and there will be $2-3$ 
interactions per collision.
The production rate of $b\bar{b}$ pairs at ATLAS is $\sim 500\;$kHz, which implies 
having $5\cdot10^{12}$ $b\bar{b}$ pairs per year ($10^7\;$seconds).

Experimental feasibility studies for rare decays of $B^0_d$, $B^0_s$, $B^{+}$ and $\Lambda_b$ 
at ATLAS have been performed using the full detector simulation chain \cite{Rimoldi:2003kf}.
The decay kinematics was defined via
matrix elements included into the $b$-physics Pythia interface \cite{Smizanska:2003PB}
($B^0_d$, $B^0_s$) or using the EvtGen decay tool \cite{Lange:2001uf,Smizanska:2004EG} 
($B^{+}$, $\Lambda_b$) with matrix elements taken from theoretical publications in
\cite{Melikhov:1997wp,Melikhov:2000yu,Ali:2002jg,Aliev:2002ww,Chen:2001zc}.
The $pp$ interactions were generated using Pythia6 \cite{Sjostrand:2003wg} tuned for correct $b$-quark
production \cite{Smizanska:2003PB}.
Events were filtered at generator level to emulate the di-muonic LVL1 trigger cuts 
(see below) and charged tracks from the $B$-decays were required to fit in ATLAS tracking system 
capabilities ($p_\mathrm{T}\gsim0.5\;\mathrm{GeV}$, $|\eta|<2.5$ \cite{unknown:1997fs}).
These cuts influence the $q^2$ spectrum and $A_{FB}$ shape.
Study of the sample of $\Lambda_b\to\Lambda^0\mu^+\mu^-$ events have shown that higher di-muon mass 
values are preferred (fraction of events with $q^2$ below $J/\psi$ mass decreased from $67\;$\% 
to $58\;$\%) and $A_{FB}$ is affected in the $q^2/M_b^2<0.1$ region 
(suppression by $40\;$\% of $|A_{FB}|$ was found).

The trigger system at ATLAS consists of three levels: 
Level 1 trigger (LVL1), Level 2 trigger (LVL2) and Event Filter (EF) \cite{unknown:2003hw}.
LVL1 stage is based on the detection of two high-$p_\mathrm{T}$ muons by the fast 
muon trigger chambers 
($p_{\mathrm{T}\mu_1}>6\;$GeV, $p_{\mathrm{T}\mu_2}>4\;$GeV and $|\eta_{\mu_{1,2}}|<2.5$ driven by detector acceptance).
A preliminary study of the di-muonic LVL1 performance was shown in \cite{Lagouri:2006TR}.
The LVL1 rate is dominated by real di-muons giving a rate of $\sim150\;$Hz, 
but also by events with a single muon, doubly counted due to overlap of trigger chambers.
In order to suppress the fake di-muon triggers, a system of overlap flags was introduced.
The study indicated that signal rejection due to this overlap-removal algorithm is less then $0.5\;\%$.
Efficiency suppression due to small di-muonic opening angles was also studied, 
finding the effect below $1\;$\%. Overall $(75-80)\;$\% single muon 
and $\sim60\;$\% di-muon trigger efficiency was found for the sample of 
$\Lambda_b\to\Lambda^0\mu^+\mu^-$ events.
At the second level, the muon $p_\mathrm{T}$ measurement will be confirmed in the Muon Precision Chambers, 
Tile Calorimeter and extrapolated to the Inner Detector in order to reject muons from $K/\pi$ decays.
The di-muon specific detailed LVL2 and EF strategies have not yet been set up.
The purpose of LVL2 is to select preliminary candidates for the $B$-hadrons rare decay, 
based on track parameters and fast calculations.
A secondary fast vertex fit can optionally be used at LVL2 level to achieve a satisfactory background rejection.
At the EF level, offline-like selection cuts will be applied.

The key signature of rare decays is the presence of the opposite-charge muon pair.
The di-muon pair is likely to form a secondary vertex which is detached from the primary vertex.
The identification of this vertex, if particularly close to the interaction point, requires well 
reconstructed leptons.
The event selection is done in the following order: 
muon and di-muon identification; secondary hadron selection; $B$-hadron selection.
The analysis has to rely on topological variables as vertex quality, vertex separation 
($c\tau_B\ge 0.5\;$ps) and pointing to primary vertex constraint on the $B$-hadron momentum.
The vertexing algorithm used is the one adopted from the CDF 
collaboration \cite{Marriner:1996VX}.
Simple vertex fits are used to select secondary hadrons and di-muon 
candidates, 
while for the $B$-hadron the whole cascade decay topology is fitted at once.

Due to low signal BRs, great background suppression has to be achieved.
The main background source comes from beauty decays producing a muon pair in the 
final state.
The present study based on a sample of $b\bar{b}\to X\mu_{p_\mathrm{T}>6(4)\mathrm{\,GeV}}\mu_{p_\mathrm{T}>4\mathrm{\,GeV}}$ 
events, provides upper limits for fake events as sketched in Table~\ref{tab:ATLAS_rare_summary}.

\begin{table}[ht]
\begin{minipage}[b]{0.45\textwidth}
\begin{tabular}{lcc}
\hline
\hline
Decay                                & Signal & Background \\
\hline
$B^0_d     \to K^{0*}    \mu^+\mu^-$ & $2500$ &  $12000$   \\
$B^0_s     \to \phi      \mu^+\mu^-$ &  $900$ &  $10000$   \\
$B^+       \to K^{*+}    \mu^+\mu^-$ & $2300$ &  $12000$   \\
$B^+       \to K^+       \mu^+\mu^-$ & $4000$ &  $12000$   \\
$\Lambda_b \to \Lambda^0 \mu^+\mu^-$ &  $800$ &   $4000$   \\
\hline
\hline
\end{tabular}
\caption{Expected number of events for signal and background upper limit after $30\;$fb$^{-1}$ measurement.}
\label{tab:ATLAS_rare_summary}
\vspace{3mm}
\end{minipage}
\hspace{4mm}
\begin{minipage}[b]{0.50\textwidth}
\begin{tabular}{lccc}
\hline
\hline
Interval of $q^2/M^2_B$ & $-^{0.00}_{0.14}$ & $-^{0.14}_{0.33}$ & $-^{0.55}_{0.71}$ \\
\hline
Number of events        & $570$             & $540$             & $990$             \\
$A_{FB}$       & $11.8\%$          & $-6.1\%$          & $-13.7\%$         \\
Statistical error       &  $4.2\%$          &  $4.3\%$          &   $3.2\%$         \\
SM prediction           &   $10\%$          &  $-14\%$          &   $-29\%$         \\
\hline
\hline
\end{tabular}
\caption{Averaged $A_{FB}$ of $B^0_d\to K^{0*}\mu^+\mu^-$ from ATLAS simulations (not corrected for detector effects and background) at $L_{\mathrm{int}}=30\;$fb$^{-1}$, its statistical precision and comparison to SM prediction.}
\label{tab:ATLAS_Bd_AFB}
\vspace{-2.22mm}
\end{minipage}
\end{table}

In Table~\ref{tab:ATLAS_Bd_AFB} the reconstructed $A_{FB}$ is presented for $B^0_d\to K^{0*}\mu^+\mu^-$ decay.
We divide the $q^2/M^2_B$--region into three intervals:
the first interval from $(2m_{\mu}/M_B)^2$ to the so-called ``zero-point'' \cite{Burdman:1998mk},
the second interval from the ``zero-point'' to the lower boundaries 
of the $J/\psi$ and $\psi'$ resonances, and
the last interval from the resonance area to $(M_B-M_{K^*})^2/M_B^2$.
Data collected in 3 years of LHC operations, corresponding to 
$30\;$fb$^{-1}$ of integrated luminosity, will be enough to confirm the 
Standard Model or to set strong limits on SM extensions.

An attempt to estimate the statistical errors of the branching ratio 
measurements has been made 
for $B^+\to K^+\mu^+\mu^-$ and $B^+\to K^{*+}\mu^+\mu^-$ decays 
\cite{Policicchio:2007BR}.
They were $\sim 3.5\;$\% and $\sim 6.5\;$\%, 
respectively for $B^+\to K^+\mu^+\mu^-$ and $B^+\to K^{*+}\mu^+\mu^-$ decays.
These errors on the branching ratio measurements are much smaller than the 
current experimental and theoretical ones.



\subsubsection{Phenomenological implications and new physics constraints}




\paragraph{New Physics in exclusive $b \to s \ell^+ \ell^-$ induced decays }

The potential of Standard Model (SM) tests and New Physics (NP)
searches with $b \to s \ell^+ \ell^-$ transitions has been stressed
and explored in several works, e.g.,
\cite{Hewett:2004tv,Hiller:2003di}, and references therein.  Of
particular interest for the LHC are the exclusive decays (i) $B_s \to
\ell^+ \ell^-$, (ii) $B\to K^{(\ast)} \ell^+ \ell^-$, $B_s\to \phi
\ell^+ \ell^-$, $ B_s\to \eta^{(\prime)} \ell^+ \ell^-$ and (iii)
$\Lambda_b\to \Lambda \ell^+ \ell^- $, where $\ell=e, \mu, (\tau)$.
Decays involving additional photons, such as $B_s \to \ell^+ \ell^-
\gamma$ \cite{Dincer:2001hu} are more sensitive to the hadronic QCD
dynamics than the modes {(i--iii)}.  They are briefly considered
in Sec. \ref{sec:rare}.
Lepton flavor violating (LFV) decays such as
$b \to s e^\pm \mu^\mp$ are discussed e.g. in
\cite{Dedes:2002rh,Fujihara:2005uq} and will not be considered further here.
We stress that FCNCs with final state $\tau$-leptons are poorly
constrained experimentally to date, and
it would be highly desirable to fill this gap since they test third
generation couplings.  The latter feature is also shared by the
di-neutrino final states discussed, e.g., in \cite{Buchalla:2000sk}
and in Sec. \ref{sec:neutrinos}.

The presence of NP can lead to modified values for the short-distance
coefficients $C_i$, including new CP-violating phases, and the
generation of new operators in the weak effective Hamiltonian.  These
could include chirality flipped versions of the SM operators
${\cal O}_i^\prime$ (down by $m_s/m_b$ within the SM) from right-handed
currents or scalar operators from Higgs exchanges $ {\cal O}_{S,P}$ (down
by $m_\ell m_{b}/m_W^2$ within the SM), or tensor currents.
Scenarios with {\it light} NP particles require additional operators,
build out of the latter, see \cite{Becher:2002ue} for the MSSM with
light sbottom and gluino.
Model-independent information on $C_{7,8,9,10}^{\rm (eff)}$ has been
previously extracted from combined analysis of $b \to s \ell^+ \ell^-$
and radiative $b \to s \gamma,s g$ data
\cite{Ali:1999mm,Ali:2002jg,Ishikawa:2006fh}, also including
(pseudo)-scalar contributions $C_{S,P}$
\cite{Bobeth:2001sq,Hiller:2003js}.  In this program the study of
correlations between decays and observables is an important
ingredient, which enables identification of a possible SM breakdown
and its sources.

The leptonic decay $\Bqll$ is a smoking gun for neutral Higgs effects
in SUSY models with large $\tan \beta$ and is discussed in detail
in Section~\ref{sec:rare}.
A clean test of minimal flavour violation (MFV, see
section~\ref{sect:MFV}) is the 
$B_d$-$B_s$-ratio $R_{\ell\ell} \equiv
\BR(\bar{B}^0_{d} \to \ell^+ \ell^-) / \BR(\bar{B}^0_{s} \to \ell^+
\ell^-)$.  In the SM and within MFV models 
$0.02 \lsim  R_{\ell\ell} {}_{|_{\rm SM}} \lsim 0.05 $, 
whereas in non-MFV scenarios  $R_{\ell\ell}$ can be ${\cal{O}}(1)$~\cite{Bobeth:2002ch}.
Phases in $C_{S,P}$ are probed with {time-dependent and integrated
  CP-asymmetries} requiring lepton-polarization measurements
\cite{Huang:2000tz,Dedes:2002er,Chankowski:2004tb}.

Besides the measurement of branching ratios, the $\BKll$ and
$\BKastll$ decays offer a number of orthogonal observables.
For instance, the latest experimental results from Belle and BaBar for
these modes \cite{Abe:2005km, Ishikawa:2006fh, Aubert:2006vb} already
include first investigations of angular distributions.
The {dilepton mass ($q^2$) spectra} of $\bar B \to K^{(*)} \ell^+
\ell^-$ are sensitive to the sign of $Re({C_7^{\rm eff}}^\ast C_9^{\rm
  eff})$ and to NP contributions in $C_{9,10}$, and flipped
$C'_{9,10}$ \cite{Aliev:1999gp} -- however, with rather large hadronic
uncertainties from form factors and non-factorizable long-distance
effects (see Sec.~\ref{sec:exth}).  Using constraints on $|C_{S,P}|$
from $B_s \to \mu^+ \mu^-$ \cite{Bobeth:2001sq} shows that $\bar B \to
K^{(*)} \ell^+ \ell^-$ spectra are rather insensitive to NP effects in
$C_S$ and $C_P$.

%
%
%
%
The {forward-backward asymmetry} for decays into light pseudoscalars,
$A_{FB}(\BKll)$, vanishes in the SM.  Beyond the SM it is proportional
to the lepton mass and the matrix elements of the new scalar and
pseudoscalar penguin operators.  The BaBar measurement of the angular
distribution \cite{Aubert:2006vb} is consistent with a zero FB
asymmetry.  Using model-independent constraints on $|C_{S,P}|$ from $B_s
\to \mu^+ \mu^-$ \cite{Bobeth:2001sq} one expects $A_{FB}(B \to K \mu^+
\mu^-) < 4\%$. Moreover, in the MSSM with large $\tan \beta$ one has
$C_S \simeq -C_P$, and the FB asymmetry comes out even smaller,
$A_{FB}(B \to K \ell^+ \ell^-) \lsim 1 \, (30) \% $ for $\ell = \mu
(\tau)$ \cite{Yan:2000dc, Demir:2002cj,Choudhury:2003xg}.
In contrast, for decays into light vector mesons, $A_{FB}(\BKastll)$ is
non-zero in the SM and exhibits a characteristic zero $q_0^2$, whose
position is relatively free of hadronic uncertainties, see
Sec.~\ref{sec:exth}.  In a general model-independent NP analysis
\cite{Aliev:1999gp, Cornell:2005kb} the position of the zero, the
magnitude and shape of $A_{FB}(\BKastll)$ are found to depend on the
modulus and phases of all Wilson coefficients.
Note that also $\Lambda_b \to \Lambda \ell^+ \ell^-$ decays share the
universal SM $A_{FB}$-zero in lowest order of the $1/m_b$ and
$\alpha_s$ expansion \cite{Hiller:2001zj}.  In off-resonance $B \to K
\pi \ell^+ \ell^-$ decays the analogous $A_{FB}$ zero is also
sensitive to NP effects \cite{Grinstein:2005ud}.
The {CP-asymmetry} for the FB asymmetry in $\BKastll$ is a
quasi-null test of the SM \cite{Buchalla:2000sk}, with
$A_{FB}^{CP}|_{\rm SM}< 10^{-3}$.
Sizable values can arise beyond the SM, for instance from 
non-standard CP-violating $Z$-penguins, contributing to  $\arg[C_{10}]$.
%

%
%
The (CP-averaged) {isospin asymmetry} in $\BKastll$ is defined from
the difference between charged and neutral $B$ decays
\cite{Feldmann:2002iw}.  It vanishes in naive factorization (assuming
isospin-symmetric form factors).  A non-zero value arises from
non-factorizable interactions where the photon couples to the
spectator quark. For small values of $q^2$, the isospin asymmetry can
be analyzed in QCDF \cite{Feldmann:2002iw}. The largest contributions
are induced by the strong penguin operators ${\cal O}_{3-6}$, and the
sign of the asymmetry depends on the sign of $C_7^{\rm eff}$. Within
the SM and minimal-flavour violating MSSM scenarios, the isospin
asymmetry is found to be small. Sizable deviations of $A_I(\BKastll)$
from zero would thus signal NP beyond MFV.

%

Following Ref.~\cite{Kruger:2005ep}, one can construct further
observables from an angular analysis of the decay $\BKastKpill$, see
(\ref{eq:FLT},\ref{eq:At2}).
The SM predictions are consistent with the existing experimental data
for the (integrated) value of the longitudinal $K^*$ polarization
$F_L$ \cite{Aubert:2006vb}.
%
A model-independent analysis with flipped ${\cal O}'_7$ shows some
sensitivity of the angular observables to right-handed currents
\cite{Kruger:2005ep}, see also \cite{Aliev:1999gp}.
The shapes of the transverse asymmetries $A_T(q^2)$ depend strongly
on $C_7$ and $C'_7$ whereas NP effects in $C_{9,10}$ are rather small
taking into account constraints from other $B$-physics data.
Moreover, the zeros of $A_T^{(1,2)}(q^2)$ are sensitive to $C'_7$.  NP
can give large contributions to the polarization parameter
$\alpha_{K^\ast}(q^2)$ and $F_{L,T}(q^2)$ in extreme scenarios,
however the influence of $C_9$ and $C_{10}$ is stronger and
theoretical errors are larger than in ${A}_T^{(1,2)}$.

%
%
%
The {muon-to-electron ratios}
\begin{equation}
  R_H \equiv \int_{q_1}^{q_2} dq^2 \, \frac{d\Gamma(B\to H \MM)}{dq^2} \Bigg/
             \int_{q_1}^{q_2} dq^2 \, \frac{d\Gamma(B\to H \EE)}{dq^2}, 
\qquad H = \{K, K^\ast\}
\end{equation}
are probing for non-universal lepton couplings, for instance from
Higgs exchange or R-parity violating interactions in SUSY models.
Kinematic lepton-mass effects are tiny, ${\cal O}(m_\mu^2/m_b^2)$.
Taking the same integration boundaries for muon and electrons, the SM
predictions are rather free of hadronic uncertainties
\cite{Hiller:2003js}
\begin{equation}
  R_H^{\rm SM} = 1 + {\cal O}(m_\mu^2/m_b^2), 
  \quad {\rm with} \quad
  R_K^{\rm SM} = 1 \pm 0.0001, \quad
  R_{K^\ast}^{\rm SM} = 0.991 \pm 0.002,
\end{equation}
and agree with the measurements 
$R_K = 1.06 \pm 0.48 \pm 0.08$ and  $R_{K^\ast} = 0.91 \pm 0.45 \pm 0.06$
\cite{Aubert:2006vb}.

Studying correlations between different observables, one may be able
to discriminate between different NP models.  For instance,
non-trivial correlation effects appear between $R_K$ and
${\cal{B}}(B_s \to \mu^+ \mu^-)$, since $\BKll$ depends on $ C_{S,P} +
C'_{S,P}$ whereas $\BR(\Bqll)$ on $C_{S,P} -C'_{S,P}$
\cite{Hiller:2003js}.
Also, ${\cal{B}}(B_s \to \mu^+ \mu^-)$ and $\Delta m_s$ are strongly
correlated in the minimal-flavour violating MSSM at large $\tan \beta$
\cite{Buras:2002wq}, whereas no such correlation occurs in models with
an additional gauge singlet, like the NMSSM studied in
\cite{Hiller:2004ii}.
A summary of all observables with central results is given in Table~\ref{tab:sum:obs}.

\begin{table}[h]
\begin{center}
  \caption{Summary of observables in $\BKll$, $\BKastll$ and 
$\bar B^0_q \to \ell^+ \ell^-$ decays. }
\label{tab:sum:obs}
\begin{tabular}{p{4cm}l}
\hline\hline
\textbf{Observable} & \textbf{comments} \\
\hline
$d\Gamma(\bar B \to K^{(*)} \ell^+ \ell^-)/dq^2$ &
   Hadronic uncertainties (form factors, non-factorizable effects, $c\bar c$) \\
 & SM: depends on $|C_{7,9,10}^{\rm eff}|$ and 
   ${\rm Re}({C_7^{\rm eff}}^\ast C_9^{\rm eff})$\\
 & NP: sensitive to $Z$-penguins, $C'_{9, 10}$, ${\rm sgn}(C_7^{\rm eff})$, 
       but not to $C_{S,P}^{(')}$ 
\\
\hline
$A_{FB}(\BKll)$ &
   SM: $ \simeq 0$ (quasi null test)   \\
 & NP: sensitive to  $C_S + C'_S$ \\
 & using $B_s \to \mu^+ \mu^-$ constraint: $<$(few $\%$ for $\mu^+\mu^-$)
\\
\hline
$dA_{FB}(\BKastll)/dq^2$ &
   Hadronic uncertainties \\
(shape and magnitude) & NP: sensitive to ${\rm sgn}(C_7^{\rm eff})$, 
   ${\rm sgn}(C_{10}^{\rm eff})$, $Z$-penguins \\[0.2em]
FB asymmetry zero & 
   Smaller uncertainties (test of the SM) \\[0.2em]
$A_{FB}^{CP}$ &
 SM: $ < 10^{-3}$ (quasi null test)  \\
 & NP: CP-phase in  $C_{10}$  (+ dynamic strong phase)
\\
\hline
$dA_I(\BKastll)/dq^2$ &
   Hadronic uncertainties \\
 & SM: ${\cal O}(+10 \%)$ for $q^2 \leq 2 \GeV^2$; depends on $C_{5,6}$ 
   (c.f.\ $A_I(\BKastgamma)$) \\ 
 & \phantom{SM:} ${\cal O}(-1 \%)$ for $2 \leq q^2 \leq 7 \GeV^2 $; depends on $C_{3,4}$  \\
 & NP: sensitive to strong penguin operators; ${\rm sgn}(C_7^{\rm eff})$ \\
\hline
${A}_T^{(1,2)}$, ${\alpha}_{K^\ast}$,
${F}_{L,T}$ & 
   Smaller uncertainties (test of SM) \\
 & NP: right-handed currents, e.g., $C_7^\prime$
\\
\hline
$R_{K^{(*)}}$ & 
   Tiny uncertainties: $< \pm 1 \%$ \\
 & SM: $1+{\cal{O}}(m_\mu^2/m_b^2)$ (common cuts) \\ 
 & NP: non-universal lepton couplings; $C_{S,P}^{(')}$, neutral Higgs exchange \\
\hline
${\cal{B}}(\bar B^0_q \to \ell^+ \ell^-)$ & 
   Uncertainties: $f_{B_q}$ \\
 & SM: depends on $|C_{10} \, V_{tq}|$ \\
 & NP: lepton-mass effects; $C_{S,P}^{(')}$, neutral Higgs exchange  \\
$ R_{\ell\ell}$ & 
   Uncertainties: $f_{B_d}/f_{B_s}$ \\
 & SM: $\sim|V_{td}|^2/|V_{ts}|^2 f_{B_d}^2/f_{B_s}^2$ \\ 
 & NP: test of MFV \\
\hline\hline
\end{tabular}
\end{center}
\end{table}



\paragraph{$B \to K^*\ell\ell$ and universal extra dimensions}

FCNC $B$ decays are  sensitive to new physics scenarios involving
extra dimensions.  As an example,  we
discuss here  the possibility to constrain the 
model proposed in  \cite{Appelquist:2000nn} (ACD model),  
which is an extension of the SM by a fifth (universal) extra dimension. 
The extra dimension is compactified to the orbifold $S^1/Z_2$, and  
all the SM fields are allowed to propagate in all dimensions.  
This model only requires a single additional  parameter with respect to
the SM, namely the  radius $R$  of the compactified extra dimension.
The Standard Model is recovered in the limit $1/R \to \infty$ where the predicted 
extra Kaluza-Klein particles decouple from the low energy theory.

The effective Hamiltonian inducing $b \to s \ell^+ \ell^-$,  
$b \to s \nu \bar \nu$ and $b \to s \gamma$
transitions in ACD has been computed in \cite{Buras:2002ej,Buras:2003mk}. 
In the case of the exclusive modes $B \to K^{(*)} \ell^+ \ell^-$, 
$B \to K^{(*)} \nu \bar \nu $ and $B \to K^*\gamma$ there are several
observables sensitive to $1/R$ that can be used to  probe this scenario
 \cite{Colangelo:2006vm,Colangelo:2006gv}.  
At present, the most stringent experimental bound
on $1/R$ comes from $B \to K^* \gamma$, leading to
$\displaystyle{1 / R} \ge 300-400$ GeV, depending on the assumed hadronic uncertainties.

\begin{figure}[ht]
\begin{center}
\includegraphics[width=0.38\textwidth] {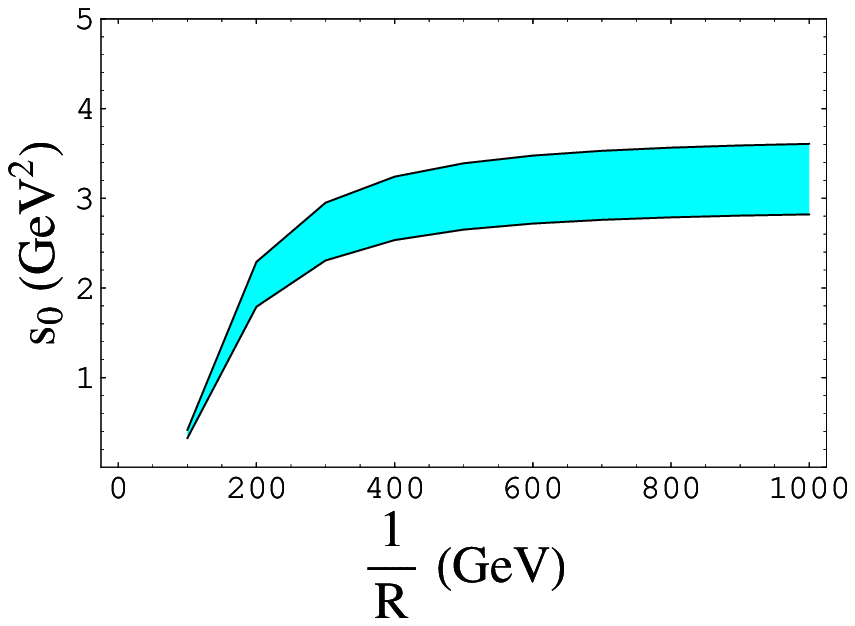} \hspace{0.8cm}
 \includegraphics[width=0.37\textwidth] {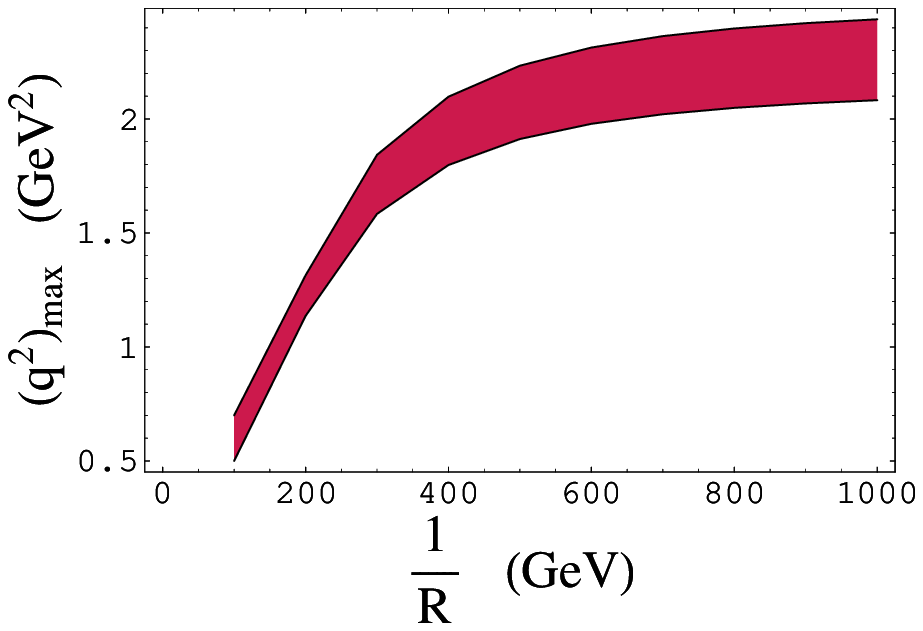}
\end{center}
\caption{Position of the zero, $s_0\equiv q^2_0$, 
of $A_{FB}$
(left) and of the maximum of the longitudinal $K^*$ helicity
fraction (right) in $B \to K^* \ell^+ \ell^-$
as a function of $1/R$ in the ACD extra dimension scenario.  
$R$ is the radius of the compactified extra dimension.
The uncertainties only include the $B \to K^*$ form-factor dependence;
non-factorizable corrections have not been taken into account.
} \vspace*{1.0cm}
\label{fig:afb}
\end{figure}

For values of $1/R$ of the order of a few hundred GeV, one expects
an enhancement  of $B(B \to K^{(*)} \ell^+ \ell^-)$ and $B(B \to K^{(*)} \nu \bar \nu)$ 
with respect to the SM (of the order of 20\%  for $1/R=300$ GeV) and a suppression  
of  $B(B \to K^* \gamma)$ (at the same level for  $1/R=300$ GeV). In general,
the sensitivity to $1/R$ is masked  by the uncertainty of the hadronic 
$B \to K^{(*)}$ matrix elements.  A useful observable with smaller hadronic
uncertainties is the  position of the forward-backward asymmetry zero
in $B \to K^*\ell^+\ell^-$,
which in ACD is shifted to  smaller values as $1/R$ decreases, 
as shown in Fig.~\ref{fig:afb} (left).
Another interesting quantity, which however has a more pronounced dependence on 
hadronic uncertainties is the position $(q^2)_{max}$ of the maximum of the longitudinal
helicity fraction of  $K^*$ in the same process; its sensitivity 
 to $1/R$ is also shown in Fig.~\ref{fig:afb} (right).

In the case of $B \to K^{(*)} \tau^+ \tau^-$ decays, 
 $\tau$-polarization asymmetries can  be considered, in which the hadronic
form factor dependence drops out for large $K^*$ recoil energies.
The transverse asymmetry decreases as $1/R$ is decreased, whereas
the branching fraction increases.
The combined observation of this pattern of deviations from SM
results would represent a signature of the ACD scenario.

%

\newpage \subsection{Neutrino modes}
\label{sec:neutrinos}



\newcommand{\bsnn}{\mbox{$B \to X_s\,\nu\,\bar\nu$}}
\newcommand{\bsntnt}{\mbox{$B \to X_s\,\nu_\tau\,\bar\nu_\tau$}}
\newcommand{\bsll}{\mbox{$B \to X_s\,\ell^+\,\ell^-$}}
\newcommand{\ct}{\mbox{$\widetilde C$}}
\newcommand{\pk}{p_{_K}}
\newcommand{\Ks}{{K^*}}
\newcommand{\Btaun}{{B^- \to \tau^- \bar\nu}}

Here we discuss the so called neutrino modes.  In particular, we talk
about the rare SM modes $B \to X_s\,\nu\,\bar\nu$ and $B \to
\tau\,\nu$. Experimentally, these modes are similar since
both are associated with large missing energy. In  $B \to
X_s\,\nu\,\bar\nu$ there are the two neutrinos, in $B \to
\tau\,\nu$ the $\tau$ decays very fast, yielding a final state with
two neutrinos as well. Theoretically these two modes are different. 
$B\to X_s\,\nu\,\bar\nu$ is a FCNC process and thus occurs at one loop
in the SM. $B \to \tau\,\nu$, on the other hand, occurs at tree
level, but it is strongly suppressed for several reasons: helicity,
a small CKM factor and the decay mechanism by weak annihilation $\sim 1/m_B$.

\subsubsection{Neutrino modes: theory}

\subsubsubsection{Inclusive $b\to s\nu\bar{\nu}$ decays}

Here we follow \cite{Grossman:1995gt} with necessary updates.  The
FCNC decay \bsnn\ is very sensitive to extensions of the SM and
provides a unique source of constraints on some NP scenarios which
predict a large enhancement of this decay mode.  In particular, the
\bsntnt\ mode is very sensitive to the relatively unexplored couplings
of third generation fermions.

{}From the theoretical point of view, the decay \bsnn\ is a very clean
process. Both the perturbative $\alpha_s$ and the non-perturbative
$1/m_b^2$ corrections are known to be small.  Furthermore, in contrast
to the decay \bsll, which suffers from (theoretical and experimental)
background such as $B\to X_s\,J/\psi\to X_s\,\ell^+\,\ell^-$, there
are no important long-distance QCD contributions. 
Therefore, the decay \bsnn\ is well suited to search for and constrain NP
effects.

Another advantage of the \bsnn\ mode is that the missing energy
spectrum can be calculated essentially in a model independent way.
Thus, one can directly compare experimental data with the theoretical
expressions as derived in specific models.  Under the only assumption
of two-component left-handed neutrinos the most general form of the
four-fermion interaction responsible for $B \to X_q\,\nu_i\,\bar\nu_j$
reads
\beq \label{Lgeneral}
  {\cal L} = C_L\, O_L + C_R\, O_R \,,
\eeq
where
\beq \label{OL-OR}
O_L = [\bar q_L \, \gamma_\mu\, b_L]\,
  [\bar \nu^i_L\, \gamma^\mu \nu^j_L] \,, \qquad
O_R = [\bar q_R \, \gamma_\mu\, b_R]\,
  [\bar \nu^i_L\, \gamma^\mu \nu^j_L]\,.
\eeq
Here $L$ and $R$ denote left- and right-handed components, $q=d,s$,
and $i,j=e,\mu,\tau$.  As the flavours of the decay products are not
detected, in certain models more than one final state can contribute
to the observed decay rate.  Then, in principle, both $C_L$ and $C_R$
carry three indices $q,\,i,\,j$, which label the quark and neutrino
flavours in the final state.

In the SM, \bsnn\ proceeds via $W$-box and $Z$-penguin diagrams and 
only $O_L$ is present. The corresponding coefficient reads
\beq \label{HeffSM}
C_L^{\rm SM} \simeq \frac{\sqrt2\,G_F\,\alpha}{\pi \sin^2\theta_W}\,
V_{tb}^*\,V_{ts}\, X_0(x_t) \,,\qquad
X_0(x) = \frac x8\, \left[\frac{2+x}{x-1}+\frac{3x-6}{(x-1)^2}\,\ln x \right].
\eeq
where $x_t=m_t^2/m_W^2$.
The leading $1/m_b^2$ and $\alpha_s$ corrections
to the SM result 
are known. Thus, the theoretical uncertainties in the SM rate are rather
small, less than $O(5\%)$. They come mainly from the
uncertainties in $m_t$, $|V_{ts}|$ and unknown higher order
corrections.
At lowest order, the missing energy spectrum in the $B$ rest-frame is
given by
\cite{YG-book}
\beq \label{shape}
\frac{{\rm d}\Gamma(B \to X_q\,\nu_i\,\bar\nu_j)}{{\rm d}x} =
\frac{m_b^5}{96\pi^3}\,
  \left(|C_L|^2+|C_R|^2\right)\, {\cal S}(r,x) \,.
\eeq
Here we have not yet summed over the neutrino flavours.
The function
${\cal S}(r,x)$ describes the shape of the missing energy spectrum
\beq \label{shapefn}
{\cal S}(r,x) = \sqrt{(1-x)^2-r}\,
  \Big[ (1-x)\,(4x-1) + r\,(1-3x) - 6\eta\sqrt{r}\,(1-2x-r) \Big] \,.
\eeq
The dimensionless variable $x={E_{\rm miss}/m_b}$ can range between
$(1-r)/2\leq x\leq1-\sqrt{r}$, and $r={m_s^2/m_b^2}$.  The parameter
$\eta=-{\rm Re}(C_L\,C_R^*)/(|C_L|^2+|C_R|^2)$ ranges between
$-\frac12\leq\eta\leq\frac12$. Since $r$ is very small, in practice the
spectrum is independent of the relative size of $C_L$ and $C_R$
and therefore immune to the presence of new physics.

It is convenient to define two ``effective" coefficients $\ct_L$ and
$\ct_R$, which can be computed in terms of the parameters of any model
and are directly related to the experimental measurement.  To remove
the large uncertainty in the total decay rate associated with the
$m_b^5$ factor, it is convenient to normalize $B(\bsnn)$ to the
semileptonic rate $B(B\to X_c\,e\,\bar\nu)$.  The contribution from
$B\to X_u\,e\,\bar\nu$, as well as possible NP effects on the
semileptonic decay rate are negligible.  In constraining NP, we can
also set $m_s=0$ and neglect both order $\alpha_s$ and $1/m_b^2$
corrections.  This is justified, since when averaged over the spectrum
these effects are very small, and would affect the numerical bounds on
the NP parameters only in a negligible way.  For the total $B \to
X_q\,\nu_i\,\bar\nu_j$ decay rate into all possible $q=d,s$ and
$i,j=e\,,\mu\,,\tau$ final state flavours, we then obtain
\beq \label{BRNP}
{B(B \to X \,\nu\,\bar\nu)
\over B(B\to X_c\,e\,\bar\nu)} =
  {\ct^2_L + \ct^2_R \over|V_{cb}|^2\, f(m_c^2/m_b^2)}\,,
\eeq
where $f(x)=1-8x+8x^3-x^4-12x^2\ln x$ is the usual phase-space factor,
and we defined
\beq \label{ctLR}
\ct^2_L = {1\over 8 G_F^2}\, \sum_{q,i,j} {\left|C_L^{qij}\right|}^2 , \qquad
\ct^2_R = {1\over 8 G_F^2}\, \sum_{q,i,j} {\left|C_R^{qij}\right|}^2 .
\eeq
Note that channels with a different lepton flavour in the final state
do not interfere. Thus, the sum among different channels is in the
rate and not in the amplitude.
The SM prediction, including NLO QCD corrections 
\cite{Buchalla:1995vs,Misiak:1999yg,Buchalla:1998ba}, is
$B^{\rm SM}(\bsnn) = 4 \times 10^{-5}$.


New physics can generate new contributions to $C_L$ and/or to $C_R$.
Many new physics models were studied in \cite{Grossman:1995gt}. In
general, there are bounds from other processes, in particular, $b \to
s \ell^+ \ell^-$. In all models where these two processes are related,
the NP contribution to the neutrino modes is bounded to be below the
SM expectation. In that case one needs to measure the neutrino mode at
high precision in order to be able to probe these models of new
physics.

The other case may be more interesting. In some models there is an
enhancement of the couplings to the third generation. Then \bsnn\ is
related only to $b \to s \tau^+ \tau^-$. This mode is very hard to
measure and thus there is no tight bound on these models. In that
cases NP could enhance the rate much above the SM rate. That is, if we
find that the rate of \bsnn\ is much above the SM rate, it will be an
indication for models where the third generation is different.

\subsubsubsection{Exclusive $b\to s\nu\bar{\nu}$ decays}

%

In principle, the theoretically cleanest
observables are provided by inclusive decays, on the other hand, the
exclusive variants will be more readily accessible in
experiment. Despite the sizable theoretical uncertainties in the
exclusive hadronic form factors, these processes could therefore give
interesting first clues on deviations from what is expected in the
Standard Model \cite{Buchalla:2000sk}.  This is particularly true if
those happen to be large or if they show striking patterns.
In the following, we discuss integrated observables
and distributions in the invariant mass of the dilepton system, $q^2$,
for the three-body decays $B\to M  \nu \bar \nu$, with $M=K$, $K^*$.
The kinematical range of $q^2$ is given  by $0\leq q^2 \leq (m_B-m_M)^2$.
In the $B\to M \nu \bar \nu$ decays, $q^2$ is not directly measurable but 
it is related to the kaon energy in the $B$-meson rest frame, $E_M$,
by the relation $q^2 = m^2_B+m^2_M-2 m_B E_M$, where
$m_M\leq E_M\leq (m^2_B+m^2_M)/(2 m_B)$.

\subsubsubsection*{$B\to K \nu \bar{\nu}$}

The dilepton spectrum of this mode is particularly simple and it is
sensitive only to the combination $|C_L^\nu +C_R^\nu|^2$
\cite{Colangelo:1996ay,Melikhov:1998ug}.  This is in contrast to the
inclusive case where only the combination $|C_L^\nu|^2 +|C_R^\nu|^2$ 
entered the decay rate. In the
inclusive case all the interference terms average to zero when we sum
over all the possible hadronic final states. In this way exclusive
processes are natural grounds where to perform tests of right-handed
NP currents, given their interference with the purely left-handed SM
current. Finally, the dilepton spectrum is
\cite{Colangelo:1996ay,Melikhov:1998ug}
\beq
  \frac{d \Gamma (B \to K \nu \bar{\nu})}{d s}  =  
  \frac{G_F^2  \alpha^2  m_B^5}{256 \pi^5} 
      \left| V_{ts}^\ast  V_{tb} \right|^2  \lambda_K^{3/2}(s) 
\, f^2_+(s)\, |C_L^\nu +C_R^\nu|^2\,,
\label{eq:dBKnunu}
\eeq
where we have defined the dimensionless variables
$s =q^2/m_B^2$ and $r_M =m_M^2/m^2_B$, and the function
\beq
\lambda_M(s) =  1 + r_M^2 + s^2 - 2s - 2r_M - 2r_Ms~.
\eeq
In the case of $M=K$ the hadronic matrix elements needed for our analysis 
are given by (\ref{eq:FFPV}) with $P=K$.  
Up to small isospin breaking effects, which we shall neglect,
the same set of form factors describes both charged ($B^- \to K^-$)
and neutral ($\bar B^0 \to \bar K^0$) transitions. Thus in the
isospin limit we get
\beq
  \Gamma(B \to K \nu \bar{\nu}) \equiv \Gamma(B^+ \to K^+ \nu
  \bar{\nu})= 2 \Gamma(B^0 \to K_{L,S} \nu \bar{\nu})~.
\eeq
The absence of absorptive final-state interactions in this process also leads 
to $\Gamma(B \to K \nu \bar{\nu})=\Gamma(\bar B \to \bar K \nu \bar{\nu})$,
preventing the observation of any direct CP violating effect.
Integrating Eq.~(\ref{eq:dBKnunu}) over the full range of $s$ leads to
\beq
\label{eq:BRKnunu}
{\cal{B}}(B\to K \nu \bar{\nu})=(3.8^{+1.2}_{-0.6}) \times 10^{-6}
~\left|\frac{C_L^\nu +C_R^\nu }{C_L\vert_{SM}^\nu} \right|^2 \, , 
\eeq
where the error is due to the uncertainty in the form factors.


If the experimental sensitivity on $B(B\to K \nu \bar{\nu})$ 
reached the $10^{-6}$ level, then the uncertainty due the form factors 
would prevent a precise extraction of $|C_L^\nu +C_R^\nu|$ from 
(\ref{eq:BRKnunu}). 
This problem can be substantially reduced by relating the differential distribution 
of $B\to K \nu \bar{\nu}$ to the one of $B \to \pi e \nu_e$ \cite{Falk:1993fr,Aliev:1997se}:
\beq
\frac{d \Gamma(B \to K \nu \bar{\nu})/ds }
{d \Gamma(B^0 \to \pi^- e^+ \nu_e) /ds  }=
\frac{3 \alpha^2}{4 \pi^2} \left|\frac{ V_{ts}^\ast  V_{tb}}{V_{ub}}\right|^2
\left(\frac{\lambda_K(s)}{\lambda_\pi(s)}\right)^{3/2}
\left\vert \frac{ f_+^K(s)}{f_+^\pi(s)} \right\vert^2 |C_L^\nu +C_R^\nu|^2~.
\label{eq:Bknn_r}
\eeq
Indeed $f_+^K(s)$ and $f_+^\pi(s)$ coincide up to 
$SU(3)$ breaking effects, which are expected to be 
small, especially far from the endpoint region.
An additional uncertainty in (\ref{eq:Bknn_r}) is induced by 
the CKM ratio $| V_{ts}^\ast  V_{tb}|^2/|V_{ub}|^2 $ which,
however, can independently  be determined from other processes.

\subsubsubsection*{$B\to \Ks \nu \bar{\nu}$}

A great deal of information can be obtained from the channel 
$B\to \Ks \nu \bar{\nu}$ investigating, together with the lepton invariant 
mass distribution, also the forward-backward (FB) asymmetry in the dilepton 
angular distribution. This may reveal effects beyond the Standard Model that 
could not be observed in the analysis of the decay rate.
The dilepton invariant mass spectrum of $B\to \Ks \nu \bar{\nu}$ decays is 
sensitive to both combinations $|C_L^\nu - C_R^\nu|$ 
and  $|C_L^\nu + C_R^\nu|$ \cite{Colangelo:1996ay,Melikhov:1998ug,Kim:1999wa}:
\beqa
  \frac{d \Gamma (B\to \Ks \nu \bar{\nu})}{d s} & = & 
    \frac{G_F^2  \alpha^2  m_B^5}{1024 \pi^5} 
    \left| V_{t s}^\ast  V_{tb} \right|^2  \lambda_\Ks^{1/2}(s)
    \Bigg\{ \frac{ 8 s \lambda_\Ks(s)V^2(s) }{(1+\sqrt{r_\Ks})^2} 
    \left|C_L^\nu +C_R^\nu\right|^2 \Bigg.  \no\\
&& +\frac{1}{r_\Ks} \Bigg[ (1+\sqrt{r_\Ks})^2 \left( \lambda_\Ks(s)+
   12 r_\Ks s \right) A_1^2(s) 
  +\frac{\lambda_\Ks^2(s)A_2^2(s)}{(1+\sqrt{r_\Ks})^2} \Bigg. \no\\
&& \Bigg.\Bigg.  \qquad\qquad - 2 \lambda_\Ks(s) (1-r_\Ks-s) 
  A_1(s) A_2(s) \Bigg] \left| C_L^\nu -C_R^\nu \right|^2  \Bigg\}~,\ 
   \label{eq:dBKstnunu}
\eeqa
where the form factors $A_1(s)$, $A_2(s)$ and $V(s)$ are defined in 
(\ref{eq:FFV-A}).
Integrating  Eq.~(\ref{eq:dBKstnunu}) over the full range of $s$ leads to
\begin{eqnarray}
B(B\to \Ks \nu \bar{\nu})
&=& (2.4^{+1.0}_{-0.5}) \times 10^{-6} 
~\left|\frac{C_L^\nu +C_R^\nu }{C_L\vert_{SM}^\nu} \right|^2
+ (1.1^{+0.3}_{-0.2}) \times 10^{-5}  
~\left|\frac{C_L^\nu - C_R^\nu }{C_L\vert_{SM}^\nu} \right|^2~, 
\label{eq:BRKstnunu} \\
B(B\to \Ks \nu \bar{\nu})\Big|_{\rm SM} &=& (1.3^{+0.4}_{-0.3}) \times 10^{-5}~.
\label{eq:BRKstnunuSM}
\end{eqnarray}

A reduction of the error induced by the poor knowledge of the 
form factors can be obtained by normalizing the dilepton distributions of
$B\to K^* \nu \bar{\nu}$ to the one of $B \to \rho e \nu_e$ 
\cite{Ligeti:1995yz,Aliev:1997se}. 
This is particularly effective in the limit $s\to 0$, where the contribution 
proportional to $|C_L^\nu + C_R^\nu|$ (vector current) drops out.
%

\subsubsubsection{$B \to \ell\,\nu$}

Recently, the Belle \cite{Ikado:2006un} and BaBar \cite{Aubert:2006fk} 
collaborations have observed the purely leptonic decays 
$B^- \to \tau^-\,\bar \nu$, (\ref{btaunubelle}) and (\ref{btaunubabar}). 
Even if both measurements are still affected by large uncertainties,
the observation of the $\Btaun$ transition represents a fundamental
step forward towards a deeper understanding of both flavour and
electroweak dynamics. The precise measurement of its decay rate could
provide clear evidence of New Physics, such as a non-standard Higgs sector
with large $\tan\beta$ \cite{Hou:1992sy}.

Due to the $V-A$ structure of the weak interactions,
the SM contributions to $B \to \ell\,\nu$ are helicity suppressed.
Hence, these processes are very sensitive to non-SM effects
(such as multi-Higgs effects) which might induce an effective pseudoscalar
hadronic weak current \cite{Hou:1992sy}.
In particular, charged Higgs bosons ($H^\pm$) appearing in any model with
two Higgs doublets (including the SUSY case) can contribute at tree level to
the above processes. The relevant four-Fermi interaction for the decay of
charged mesons induced by $W^\pm$ and $H^\pm$ has the following form:
\beq
\frac{4G_F}{\sqrt{2}}V_{ub}
\left[(\,\overline{u}\gamma_{\mu}P_Lb\,)
(\,\overline{\ell}\gamma^{\mu}P_L\nu\,)
-\tan^{\!2}\!\beta\left(\frac{m_{b} m_{\ell}}{m^{2}_{H^\pm}}\right)
(\,\overline{u}P_Rb\,)(\,\overline{\ell}P_L\nu\,)\right]
\eeq
where $P_{R,L}=(1\pm \gamma_5)/2$. Here we keep only the $\tan\beta$ 
enhanced part of the $H^\pm ub$ coupling, namely the $m_b\tan\beta$ term.
The decays $B\rightarrow \ell\nu$ 
proceed 
via the axial-vector part of the $W^\pm$ coupling and via the pseudoscalar 
part of the $H^\pm$ coupling. 
The amplitude then reads
\beq
\mathcal{A}_{B\rightarrow \ell\nu}=\frac{G_F}{\sqrt{2}}V_{ub}f_B
\left[m_{\ell}-m_{\ell}\,\,\tan^{2}\!\beta
\frac{m^{2}_B}{m^{2}_{H^\pm}}\,\right]\overline{l}(1-\gamma_{5})\nu.
\eeq
We observe that the SM term is proportional to $m_{\ell}$ because of the 
helicity suppression while the charged Higgs term is proportional to 
$m_{\ell}$ because of the Yukawa coupling.

The SM expectation for the $\Btaun$ branching fraction is 
\begin{equation}
 \label{eq:BR_B_taunu}
{\cal B}(\Btaun)^{\rm SM} = 
\frac{G_{F}^{2}m_{B}m_{\tau}^{2}}{8\pi}\left(1-\frac{m_{\tau}^{2}}
{m_{B}^{2}}\right)^{2}f_{B}^{2}|V_{ub}|^{2}\tau_{B}
=(1.59\pm 0.40) \times 10^{-4  ~,}
\end{equation}
where we used
$|V_{ub}| = (4.39 \pm 0.33) \times 10^{-3}$ from inclusive $b\to
u$ semileptonic decays \cite{Barberio:2007cr}, $\tau_{B} = 1.643\pm 0.010$~ps,
and the recent unquenched lattice result $f_B = 0.216\pm 0.022$ GeV
\cite{Gray:2005ad}.

The inclusion of scalar charged currents leads to the following
expression \cite{Hou:1992sy}:
\beq \label{rH}
R_{B\tau\nu}=\frac{{\cal B}(\Btaun)}{{\cal B}(\Btaun)^{\rm SM}}= r_H=
\left[1-\tan^2\beta\,\frac{m^{2}_B}{m^{2}_{H^\pm}}\right]^2\,,
\eeq
Interestingly, in models
where the two Higgs doublets are coupled separately to up- and
down-type quarks, the interference between $W^{\pm}$ and $H^{\pm}$
amplitudes is necessarily {\em destructive}. For a natural choice of
the parameters ($30\la \tan\beta\la 50$, $0.5 \la
M_{H^\pm}/\rm{TeV} \la 1$) Eq.~(\ref{rH}) implies a (5-30)\%
suppression with respect to the SM.  The corresponding expressions for
the $K \to \ell \nu$ channels are obtained with the replacement $m_B
\to m_K$, while for the $D \to \ell \nu$ case $m^{2}_{B} \to (m_s/m_c)
m^{2}_{D}$. It is then easy to check that a $30\%$ suppression of
$B(B \to \tau \nu)$ should be accompanied by a $0.3\%$ suppression
(relative to the SM) in $B(D \to \ell \nu)$ and $B(K \to \ell
\nu)$.  At present, the theoretical uncertainty on the corresponding
decay constants does not allow to observe such effects.

Apart from the experimental error, one of the difficulties 
in obtaining a clear evidence of a possible deviation of $R_{B\tau\nu}$ 
from unity is the large parametric uncertainty induced by $|f_{B}|$ and $|V_{ub}|$.
An interesting way to partially circumvent this problem is obtained by 
normalizing 
$B(\Btaun)$ to the $B^0_d$--$\bar B^0_d$ mass difference ($\Delta M_{B_d}$)
\cite{Isidori:2006pk}.
Neglecting the tiny isospin-breaking differences in masses, life-times and decay 
constants, between $B_d$ and $B^-$ mesons, we can write \cite{Isidori:2006pk}
\beqa
\left. \frac{B(\Btaun)}{\tau_B \Delta M_{B_d} } \right|^{\rm SM} &=& 
\frac{3\pi}{4 \eta_B S_0(m_t^2/M_W^2) {\hat B_{B_d}} } \frac{m^2_\tau}{M_W^2}
\left(1-\frac{m^{2}_\tau}{m^{2}_{B}}\right)^2 \left|\frac{V_{ub}}{V_{td}}\right|^2~, \\
&=& 1.77 \times 10^{-4} \left( \frac{|V_{ub}/V_{td}|}{0.464}\right)^2 
\left(\frac{0.836}{\hat B_{B_d}}\right)~. 
\label{eq:Btn_DMB}
\eeqa
Following standard notation, we have denoted by $S_0(m_t^2/M_W^2)$,
$\eta_B$ and $B_{B_d}$ the Wilson coefficient, the QCD correction
factor and the bag parameter of the $\Delta B=2$ operator within the
SM (see e.g.~Ref.\cite{Buras:2002vd}),  using the unquenched lattice
result $\hat B_{B_d} = 0.836 \pm 0.068$ \cite{Aoki:2003xb} and
$|V_{ub}/V_{td}|=0.464 \pm 0.024 $ from the UTfit collaboration
\cite{Bona:2005eu}.

The ratio $R^\prime_{B\tau\nu}=B(\Btaun)/\tau_B \Delta M_{B_d}$ could become 
a more stringent test of the SM in the near future, with higher statistics on 
the $\Btaun$ channel. In generic extensions of the SM the New Physics 
impact on $R_{B\tau\nu}$ and 
$R^\prime_{B\tau\nu}$ is not necessarily the same. However, it should
coincide if the non-SM contribution to $\Delta M_{B_d}$ is negligible, which is
an excellent approximation in the class of models 
considered in \cite{Isidori:2006pk}.

For consistency, the $|V_{ub}/V_{td}|$ combination entering in
$R^\prime_{B\tau\nu}=B(\Btaun)/\tau_B \Delta M_{B_d}$ should be
determined without using the information on $\Delta M_{B_d}$ and
$\Btaun$ (a condition that is already almost fulfilled).
In the near future one could determine this ratio with negligible
hadronic uncertainties using the relation
$|V_{ub}/V_{td}|=|\sin\beta_{_{\rm CKM}}/\sin\gamma_{_{\rm CKM}}|$.

From Eq.~(\ref{rH}), it is evident that such tree level NP 
contributions, namely the $r_H$ factor, do not introduce any 
lepton flavour dependent correction and thus departures from the 
SM lepton universality are not introduced. 
However, as pointed out in Ref.~\cite{Masiero:2005wr}, this is no longer true 
in realistic supersymmetric frameworks if the model contains sizable
sources of flavour violation in the lepton sector (a possibility 
that is well motivated by the large mixing angles in the neutrino sector).
In the last case, we can expect observable deviations from the SM
in the ratios 
\beq
R_P^{\ell_1/\ell_2} = \frac{B(P\to \ell_1 \nu) }{B(P \to \ell_2 \nu)}~.
\eeq
with $P=\pi,K,B$ and $\ell_{1,2}=e,\mu,\tau$.  The lepton-flavour violating
(LFV) effects can be quite large in $e$ or $\mu$ modes, while in first
approximation they are negligible in the $\tau$ channels.  In the most
favourable scenarios, taking into account the constraints from LFV
$\tau$ decays \cite{Paradisi:2005tk,Paradisi:2005fk}, spectacular
order-of-magnitude enhancements for $R_B^{e/\tau}$ and $\cO(100\%)$
deviations from the SM in $R_B^{\mu/\tau}$ are allowed
\cite{Isidori:2006pk}.
The key ingredients that allow visible non-SM contributions in $R_P^{\mu/e}$ 
within the MSSM are large values of $\tan\beta$ and sizable mixing angles in 
the right-slepton sector, such that the $P\to \ell_i \nu_{j}$ rate 
(with $i\not=j$) becomes non negligible.

\subsubsection{Neutrino modes: experiment}

Experimental prospects for neutrino modes, such as $b \to s\,\nu\,\bar\nu$, 
$B \to \tau\,\nu$ and $b \to c\,\tau\,\nu$, are discussed.
Because of the missing multiple neutrinos in the final state, these
decays lack kinematic constraints, which could be used to surpress 
background processes.
The $e^+e^-$ $B$-factories, where background is relatively low and can
be reduced by reconstructing the accompanying $B$ meson, would be the 
ideal place to measure these decays.
We also discuss the prospect for $B \to \mu\,\nu$, which can be
used to test the lepton universality in comparison to $B \to \tau \nu$.

Belle and BaBar have used hadronic decays to reconstruct the
accompanying $B$ (hadronic tags), for which the tagging efficiency
is about 0.3(0.1)\% for the charged (neutral) $B$ meson.
BaBar has used also semileptonic decays $B \to D^{(*)} \ell\,\nu$
(semileptonic tags) to increase the efficiency at the expense of the 
signal-to-noise ratio.
 
The present $e^+e^-$ $B$-factory experiments are starting to measure
some of these decays, as demonstrated by the first evidence of 
$B \to \tau\,\bar \nu$, which was recently reported by Belle.
However, precision measurements and detection of very difficult modes,
such as $b \to s\, \nu\,\bar\nu$, require at least a couple of tens
ab$^{-1}$ data, which can be reached only at the proposed super 
$B$-factories.
  
\subsubsubsection{$b \to s \nu\bar\nu$}

\begin{figure}[!b]
\begin{center}
\includegraphics[width=.4\textwidth]{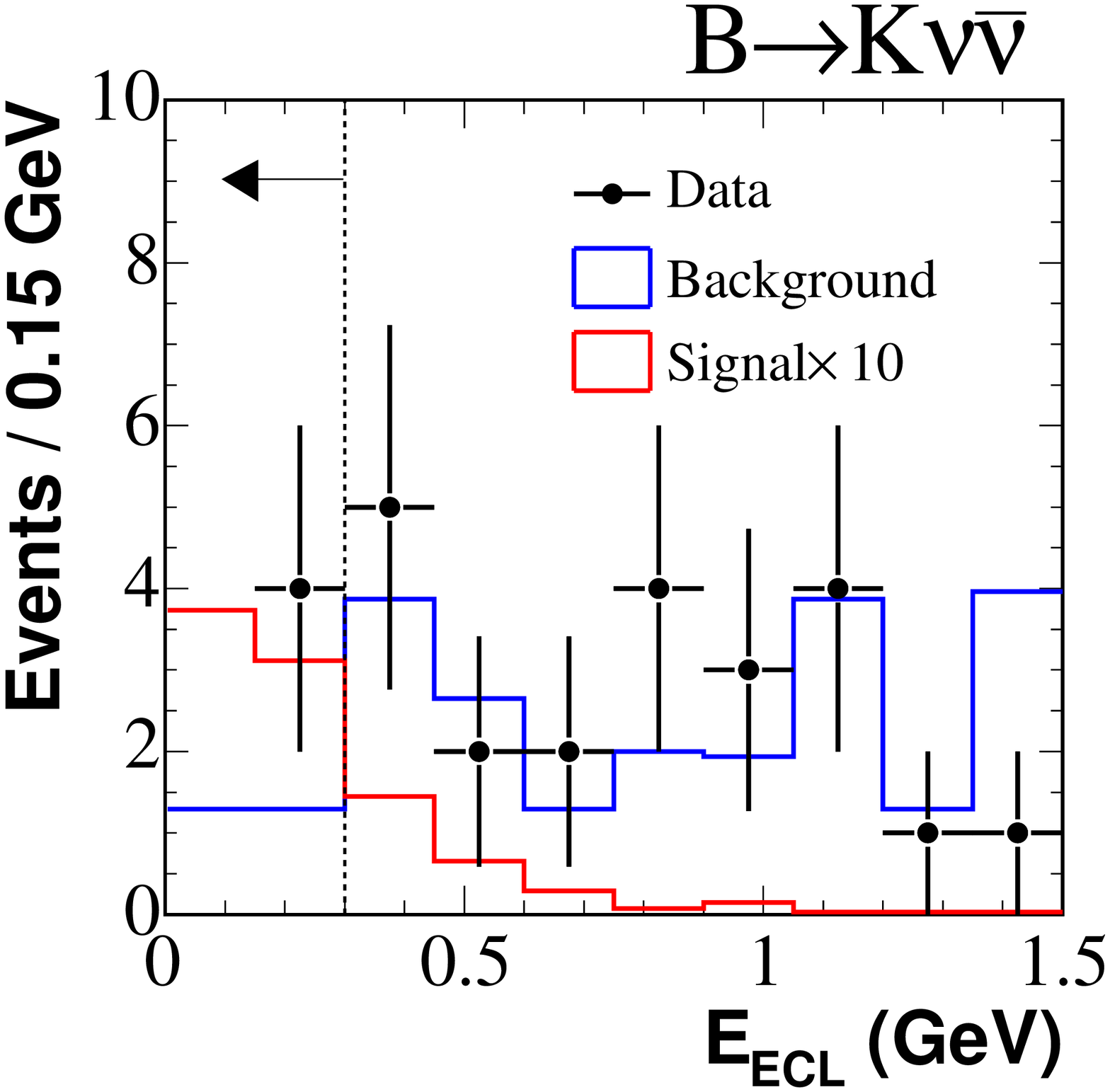}
\includegraphics[width=.4\textwidth]{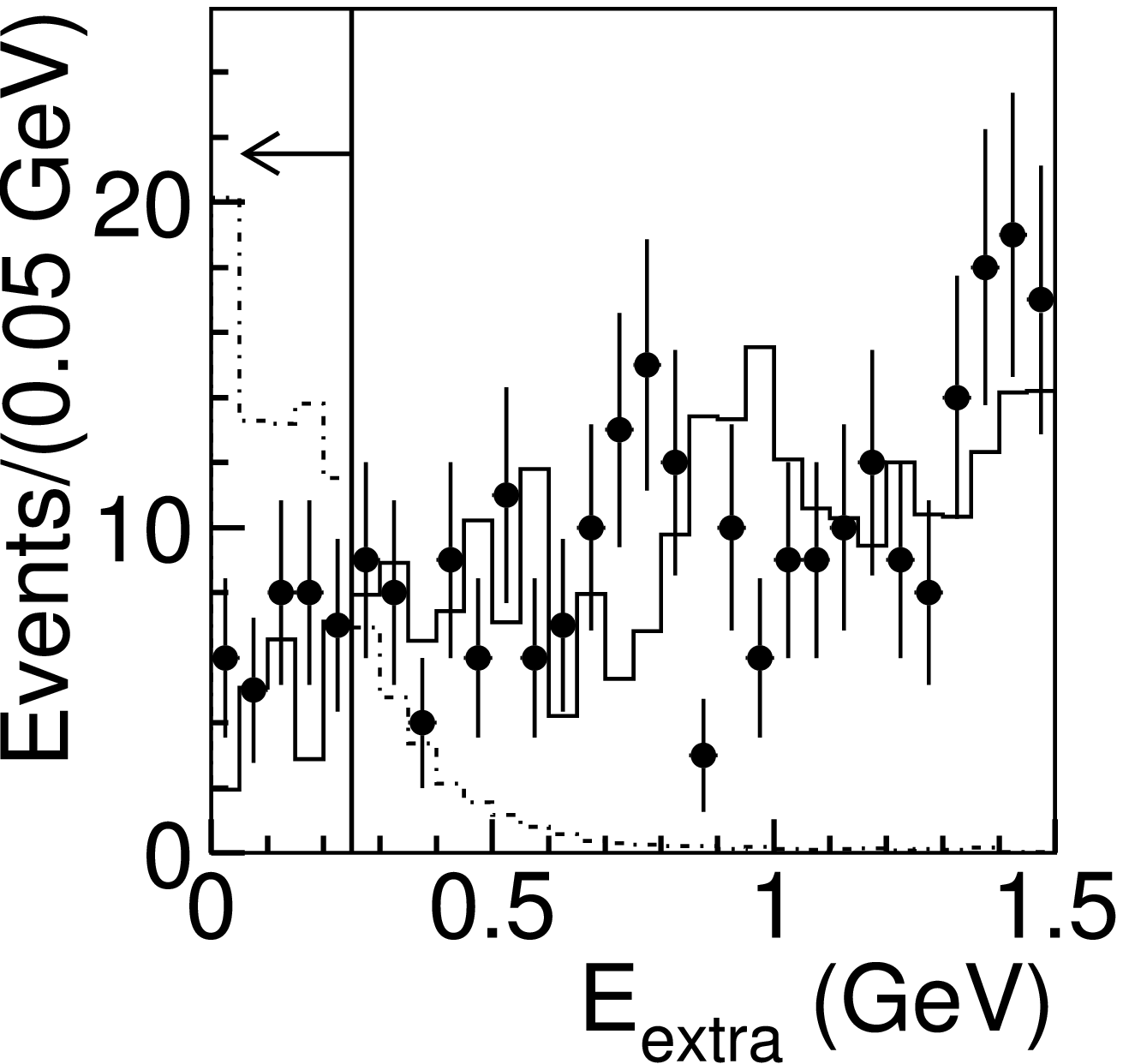}
\caption{Distribution of remaining energy for $B^- \to K^- \nu\,\bar\nu$ 
candidates; a) from Belle's analysis using the hadronic tag on a 
253\,fb$^{-1}$ data sample, and b) from BaBar's analysis using the
semileptonic tag on a 82\,fb$^{-1}$ data sample.}
\label{fig:Knunu}
\end{center}   
\end{figure}

Presently, experimental limits on exclusive $b \to s \nu \bar\nu$ 
modes are available from Belle and BaBar.
Belle has reported the result of a search for $B^- \to K^- \nu \bar\nu$
using a 253 fb$^{-1}$ data sample~\cite{Abe:2005bq}.
The analysis utilizes the hadronic tags, and requires that the event 
has no remaining charged tracks nor neutral clusters other than the 
$K^-$ candidate.
Fig.~\ref{fig:Knunu}~a) shows the distribution of remaining neutral
cluster energy recorded in the electromagnetic calorimeter ($E_{ECL}$)
after all the selection cuts are applied.
The signal detection efficiency is estimated to be $43\%$ for the
tagged events.
In the signal region, defined as $E_{ECL} < 0.3$ GeV, the expected number
of signals is 0.70, assuming the Standard Model branching fraction of
${\cal B}(B \to K^- \nu\,\bar\nu) = 4 \times 10^{-6}$, while 
the number of background estimated from the sideband data is $2.6 \pm 1.6$.
The deduced upper limit (90\% C.L.) on the branching fraction is 
${\cal B}(B^- \to K^- \nu \bar\nu) < 3.6 \times 10^{-5}$.
More recently, Belle has reported an upper limit of 
${\cal B}(B^0 \to K^{*0} \nu\,\bar\nu) < 3.4 \times 10^{-4}$,
from a similar analysis on a 492 fb$^{-1}$ data sample~\cite{Abe:2006vg}.

BaBar has reported ${\cal B}(B^- \to K^- \nu\,\bar\nu) < 5.2 \times 10^{-5}$,
by combining the hadronic and semileptonic tag events from a 82 fb$^{-1}$ 
data sample~\cite{Aubert:2004ws}.
Fig.~\ref{fig:Knunu}~b) shows the distribution of the remianing energy
($E_{extra}$ in BaBar's notation) for the semileptonic tag sample.
Because of the large $B^- \to D^{(*)} \ell\,\bar\nu$ branching fractions,
the semileptonic tag method has a factor 2 to 3 higher efficiency than the 
hadronic tag method. 

Based on a simple-minded extrapolation from the Belle analysis with the
hadronic tags, the required integrated luminosity for observing the 
$B^- \to K^- \nu\,\bar\nu$ decay with 3(5) $\sigma$ statistical 
significance is 12(33) ab$^{-1}$.
The statistical precision for the branching fraction measurement will 
reach 18\% at 50 ab$^{-1}$.
Addition of the semileptonic tag sample may improve the sensitivity
(this is under investigation).

It is extremely difficult to perform an inclusive search for 
$b \to s \nu \bar\nu$. No serious studies have been made yet.

\subsubsubsection{$B \to \tau\nu$}

Detection of $B^- \to \tau^-\,\bar\nu$ is very similar to that of 
$B \to K^{(*)} \nu\,\bar\nu$, and it requires that the event has
no extra charged tracks nor neutral clusters other than those from 
the $\tau$ decay and the accompanying $B$ decay. 

\begin{figure}[hb]
\begin{center}
\includegraphics[width=.44\textwidth]{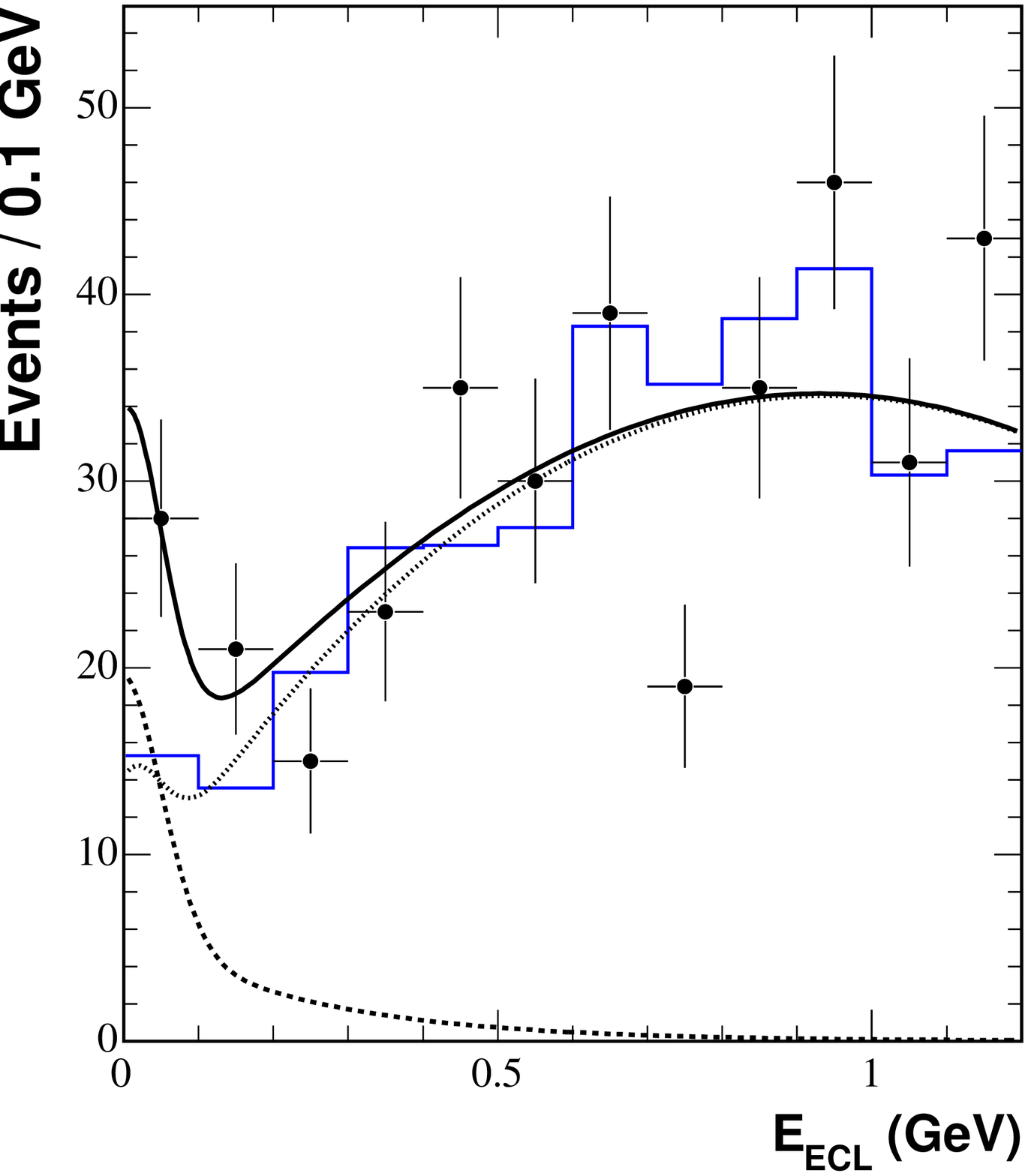} 
\includegraphics[width=.37\textwidth]{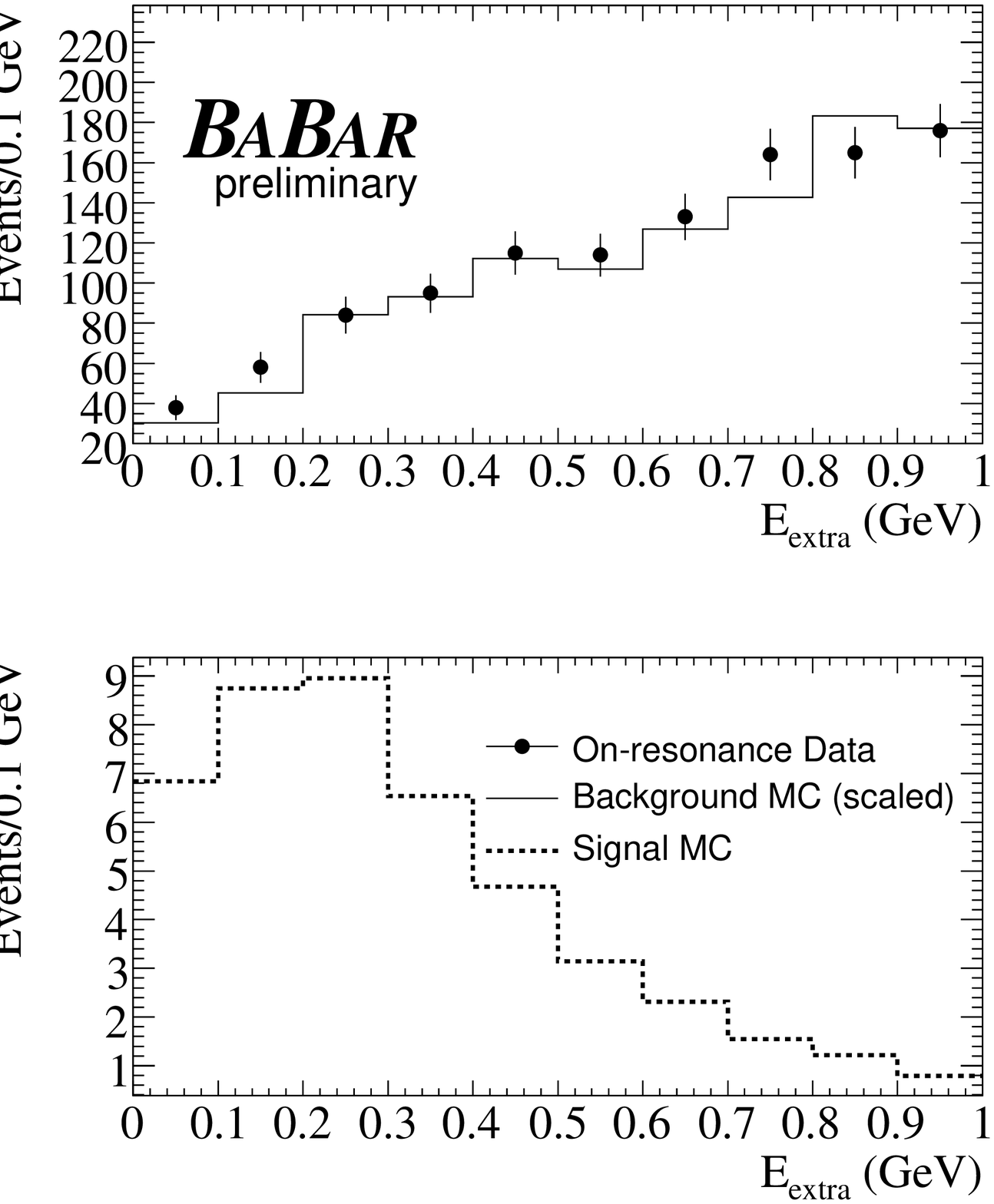} 
\caption{Distribution of the remaining energy for $B^- \to \tau^-\,
\bar\nu$ candidates;
a) from Belle's analysis using the hadronic tag on a 414\,fb$^{-1}$ 
data sample, and
b) from BaBar's analysis using the semileptonic tag on a 288\,fb
$^{-1}$ data sample.}
\label{fig:ecl_taunu}
\end{center}   
\end{figure}

Recently Belle has reported the first evidence for 
$B^- \to \tau^-\,\bar\nu$ by applying the hadronic tag on a 
414\,fb$^{-1}$ data sample~\cite{Ikado:2006un}.
The reconstructed $\tau$ decay modes are 
$\tau^- \to e^-\,\bar{\nu}_e\,\nu_{\tau}$, 
$\mu^-\,\bar{\nu}_{\mu}\,\nu_{\tau}$, 
$\pi^-\,\nu_{\tau}$, 
$\pi^-\,\pi^0\,\nu_{\tau}$, 
$\pi^-\,\pi^+\,\pi^-\,\nu_{\tau}$.
Fig.~\ref{fig:ecl_taunu}~a) presents the $E_{ECL}$ distribution,
combined for all the $\tau$ decay modes, which shows an excess of 
events near $E_{ECL} = 0$.
The number of signal ($N_s$) and background events ($N_b$) in the signal
region are determined to be $N_s = 17.2^{+5.3}_{-4.7}$ and 
$N_b = 32.0 \pm 0.7$ by an unbinned maximum likelihood fit.
The significance of the excess is $3.5 \sigma$ including both statistical 
and systematic uncertainties.
The obtained branching fraction is~\cite{Ikado:2006un} 
\begin{equation}\label{btaunubelle}
{\cal B}(B^- \to \tau^-\,\bar\nu)=
(1.79 ^{+0.56}_{-0.49}\mbox{(sta)} ^{+0.46}_{-0.51}\mbox{(sys)})
\times 10^{-4}.
\end{equation}

BaBar has reported results of a $B^- \to \tau^-\,\bar\nu$ search using
the semileptonic tag on a 288 fb$^{-1}$ data sample~\cite{Aubert:2006fk}.
The tag reconstruction efficiency is about 0.7\%, depending slightly 
on run periods.
When all the analyzed $\tau$ decay modes are combined, 213 events are
observed, while the background is estimated to be $191.7 \pm 11.7$.
Since the excess is not significant, they provide an upper limit of 
${\cal B}(B^- \to \tau^-\,\bar\nu) < 1.8 \times 10^{-4}$ (90\% C.L.), 
and also quote the value~\cite{Aubert:2006fk} 
\begin{equation}\label{btaunubabar}
{\cal B}(B^- \to \tau^-\,\bar\nu) = (0.88 ^{+0.68}_{-0.67}\mbox{(sta)} 
\pm 0.11 \mbox{(sys)}) \times 10^{-4}.
\end{equation}
The semileptonic tag gives roughly two times higher efficiency 
than the hadronic tag, but introduces more backgrounds. 

Within the context of the Standard Model, the product of the $B$ 
meson decay constant and the magnitude of the CKM matrix element 
$|V_{ub}|$ is determined to be
$f_B |V_{ub}| = (10.1 ^{+1.6}_{-1.4}\mbox{(sta)}
^{+1.3}_{-1.4}\mbox{(sys)}) \times 10^{-4}$\,GeV
from the Belle result.
Using the value of $|V_{ub}|=(4.39 \pm 0.33) \times 10^{-3}$ from
inclusive charmless semileptonic $B$ decay data~\cite{Barberio:2007cr},
we obtain $f_B = 0.229 ^{+0.036}_{-0.031}\mbox{(sta)}
^{+0.034}_{-0.037}\mbox{(sys)}$ GeV.

\begin{figure}[thb]
\begin{center}
\includegraphics[width=.4\textwidth]{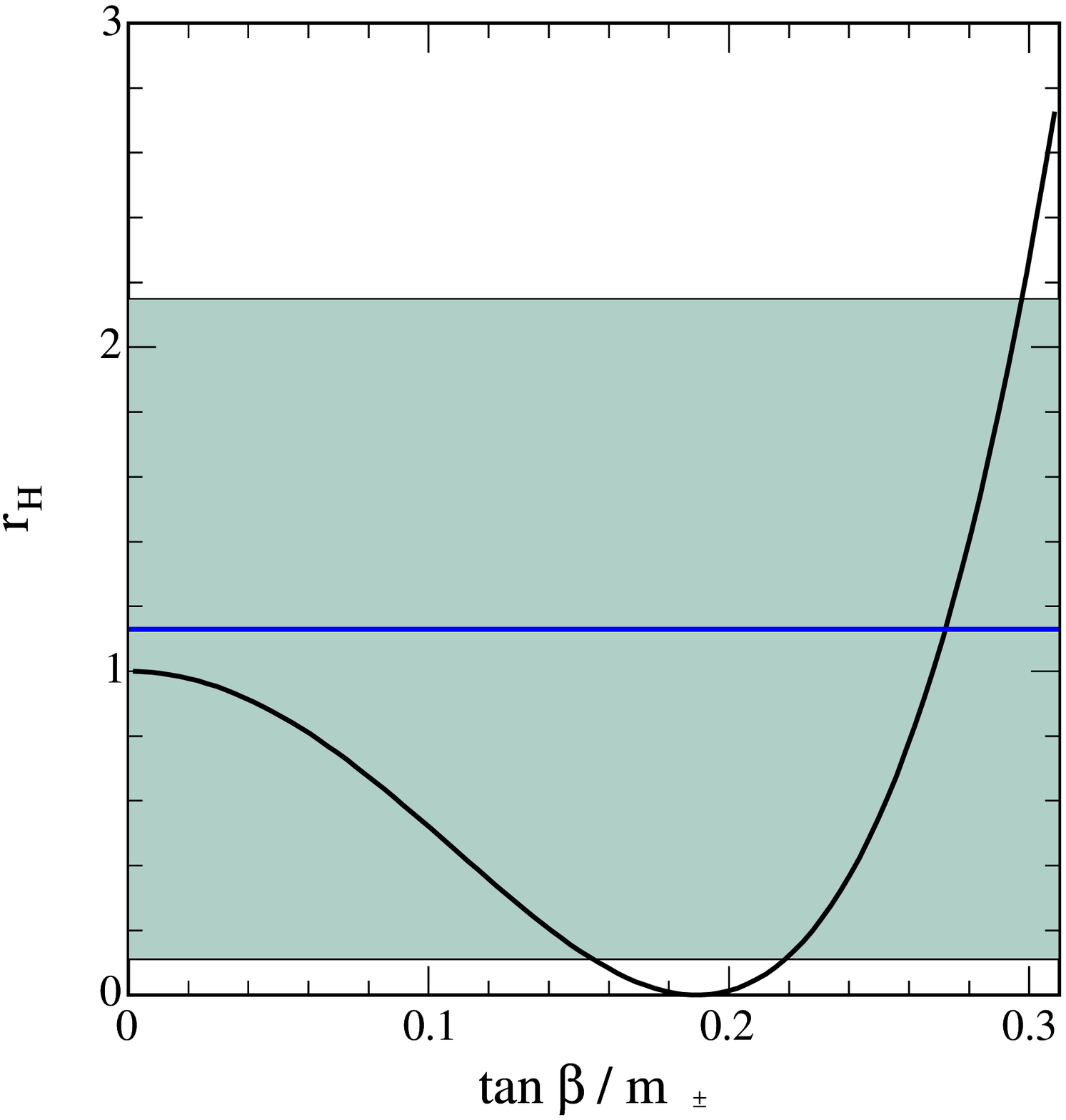} 
\includegraphics[width=.4\textwidth]{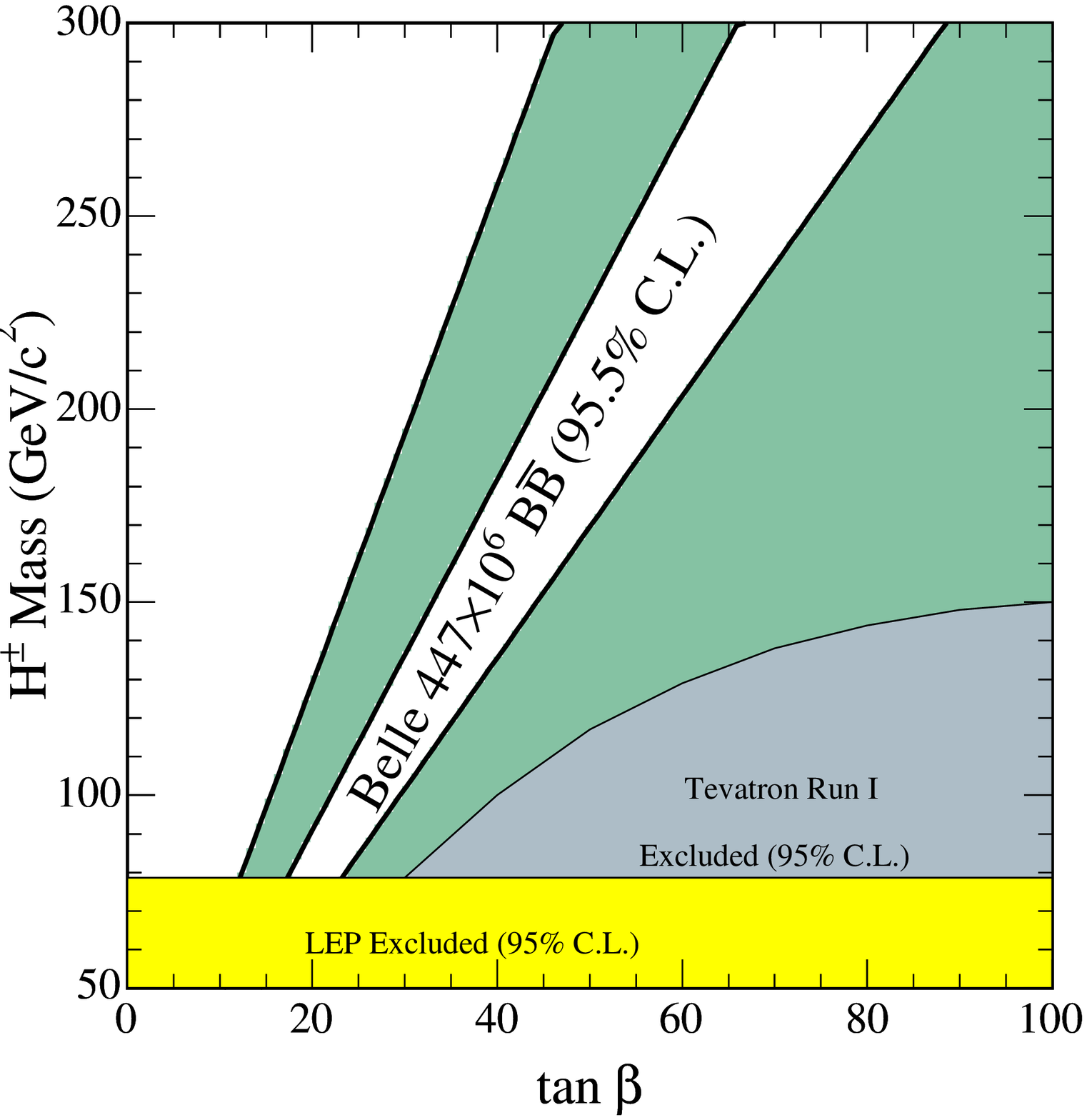} 
\includegraphics[width=.4\textwidth]{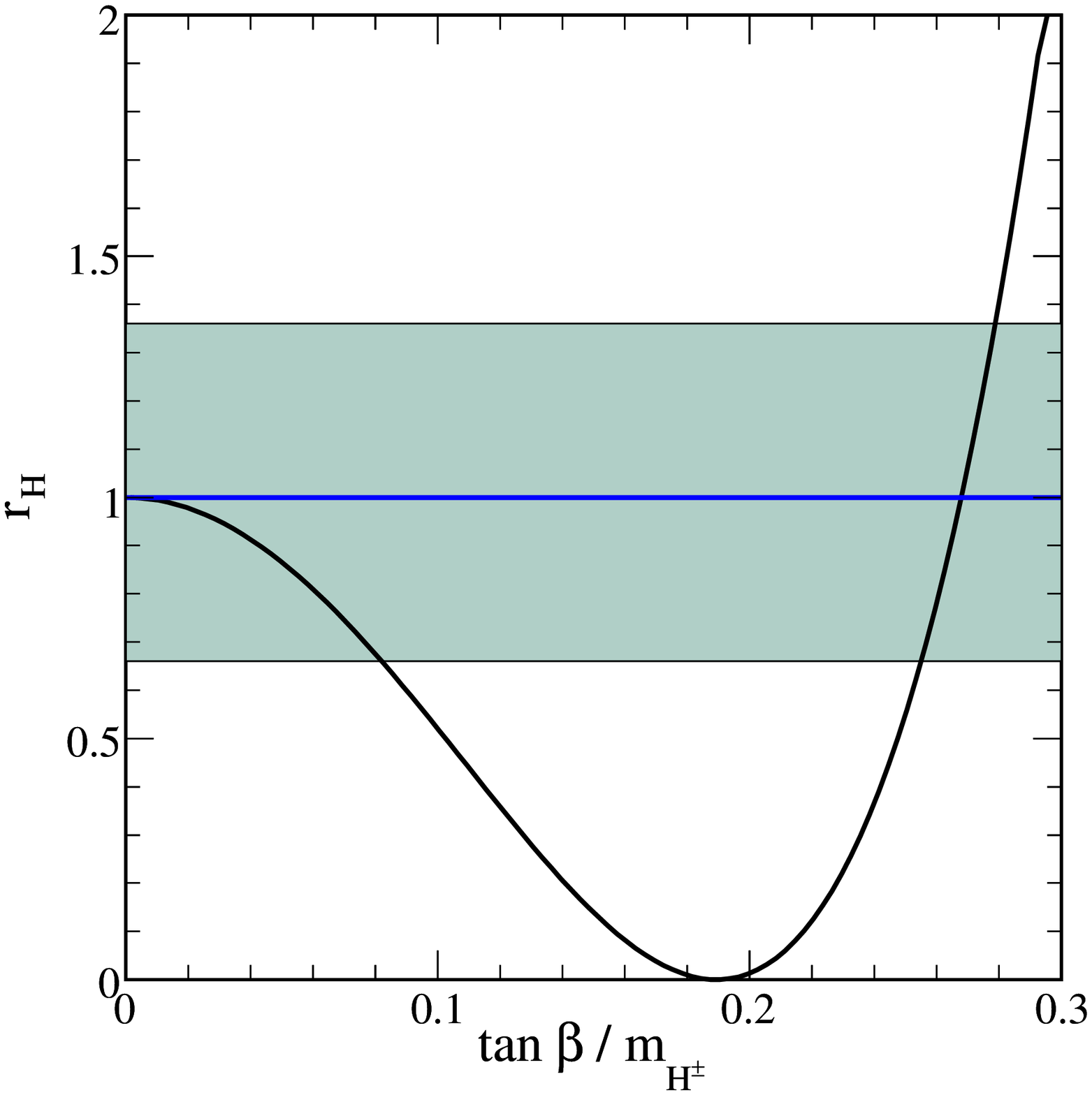} 
\includegraphics[width=.4\textwidth]{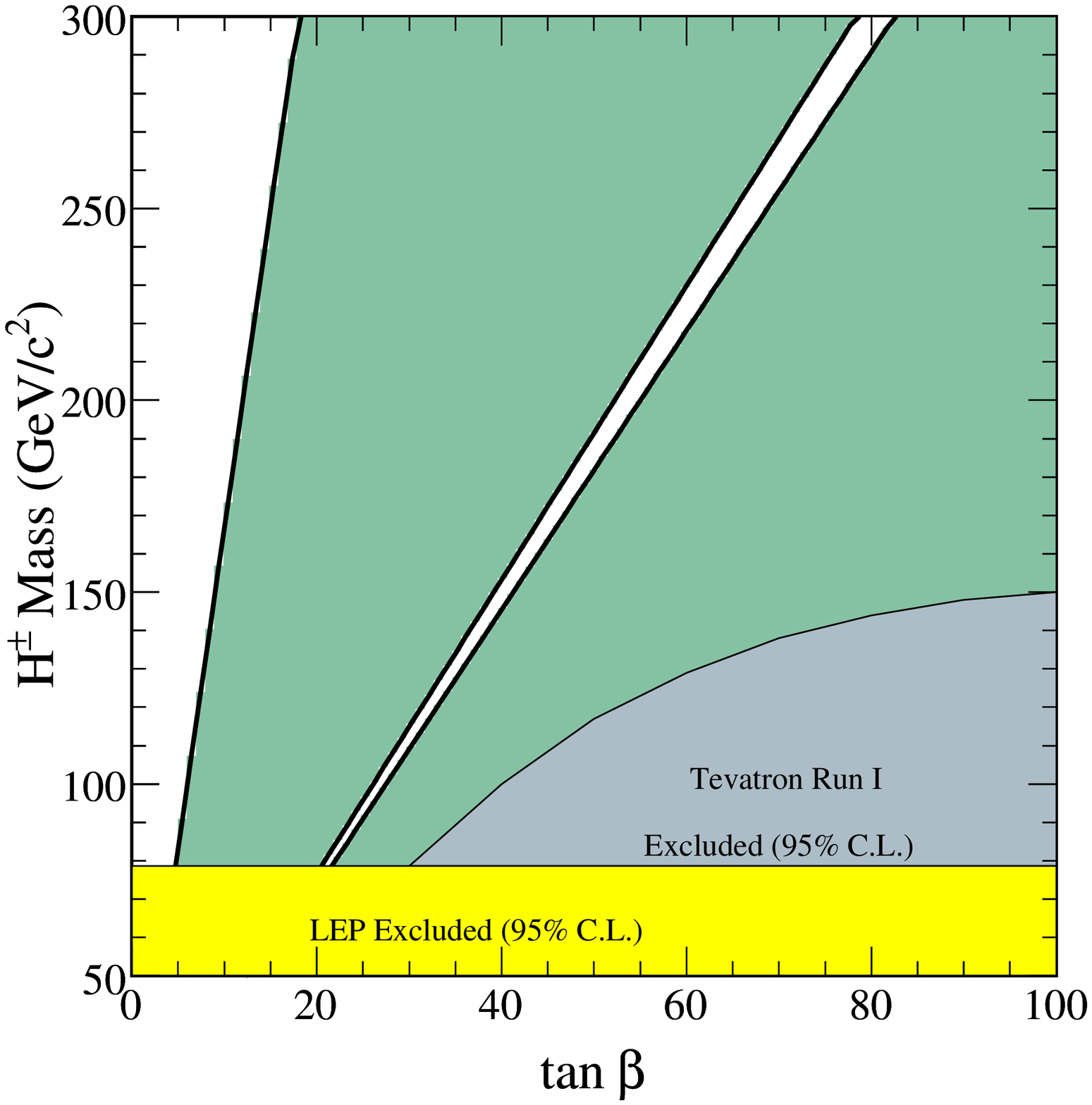} 
\caption{
The constraint on the charged Higgs;
$\pm 1\sigma$ boundary in the ratio $r_H$ (left) and the $95.5\%$ C.L. 
exclusion boundaries in the $(M_{H^{+}},\tan\beta)$ plane (right).
The top figures show the constraint from the present Belle result.
The bottom figures show the expected constraints at 5ab $^{-1}$.}
\label{fig:higgs_taunu}
\end{center}   
\end{figure}

The charged Higgs can be constrained by comparing the measured 
branching fraction (${\cal B}^{\rm exp}$) to the Standard 
Model value of ${\cal B}^{SM} = (1.59 \pm 0.40) \times 10^{^4}$, 
which is deduced from the above $|V_{ub}|$ value and 
$f_B = (0.216 \pm 0.022)$\,GeV obtained from lattice QCD calculations
~\cite{Gray:2005ad}.
Using the Belle result, the ratio (\ref{rH}) is $r_H = 1.13 \pm 0.53$, 
which then constrains the charged Higgs in the $(M_{H^+}, \tan\,\beta)$ 
plane, as shown in Fig.~\ref{fig:higgs_taunu}~(top).
The hatched area indicates the region excluded at a confidence level
of 95.5\%.

\begin{figure}[hbt]
\begin{center}
\includegraphics[width=.6\textwidth]{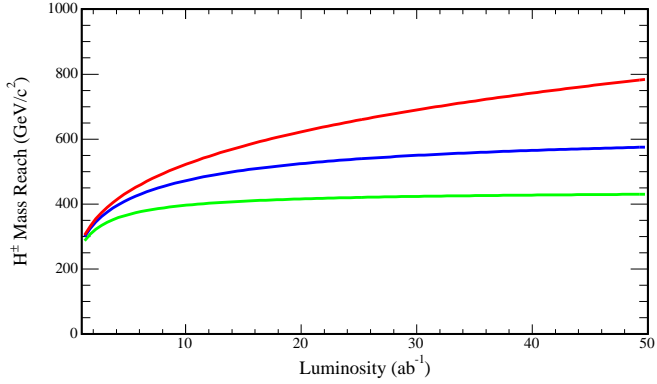} 
\caption{
Expected $M_{H^+}$ reach at $\tan \beta =30$ as a function of the 
integrated luminosity. The three curves correspond to 
$(\Delta |V_{ub}|/|V_{ub}|,\Delta f_B / f_B) =$
red:(0\%,0\%), blue:(2.5\%,2.5\%) and green:(5\%,5\%).}
\label{fig:higgs_taunu_pro}
\end{center}   
\end{figure}
%

Further accumulation of data helps to improve on both the statistical
and systematic uncertainty of the branching fraction.
Some of the major systematic errors, such as ambiguities in the 
reconstruction efficiency and the signal and background shapes, 
come from the limited statistics of a control sample.
On the other hand, the error in the ratio $r_H$ depends on the 
errors in the determination of $|V_{ub}|$ and $f_B$.
Fig.~\ref{fig:higgs_taunu}~(bottom) shows the expected constraint 
at 5 ab$^{-1}$, assuming the scaling of the experimental error by
$1/\sqrt{L}$ (L is the luminosity) and 5\% relative error for 
both $|V_{ub}|$ and $f_B$.
Fig.~\ref{fig:higgs_taunu_pro} presents the $M_{H^+}$ reach at 
$\tan\,\beta = 30$ as a function of the integrated luminosity.
Here the $M_{H^+}$ reach is defined as the upper limit of the 95.5\% 
excluded region at a given $\tan\,\beta$.
The figure shows the expectation for three cases, 
$(\Delta |V_{ub}|/|V_{ub}|,\Delta f_B / f_B) = $
(0\%,0\%), (2.5\%,2.5\%) and (5\%,5\%).
Precise determination of $|V_{ub}|$ and $f_B$ is desired to maximize
the physics reach.

\subsubsubsection{$B \to D^{(*)} \tau\nu$}

\begin{figure}[hbt]
\begin{center}
\includegraphics[width=.5\textwidth]{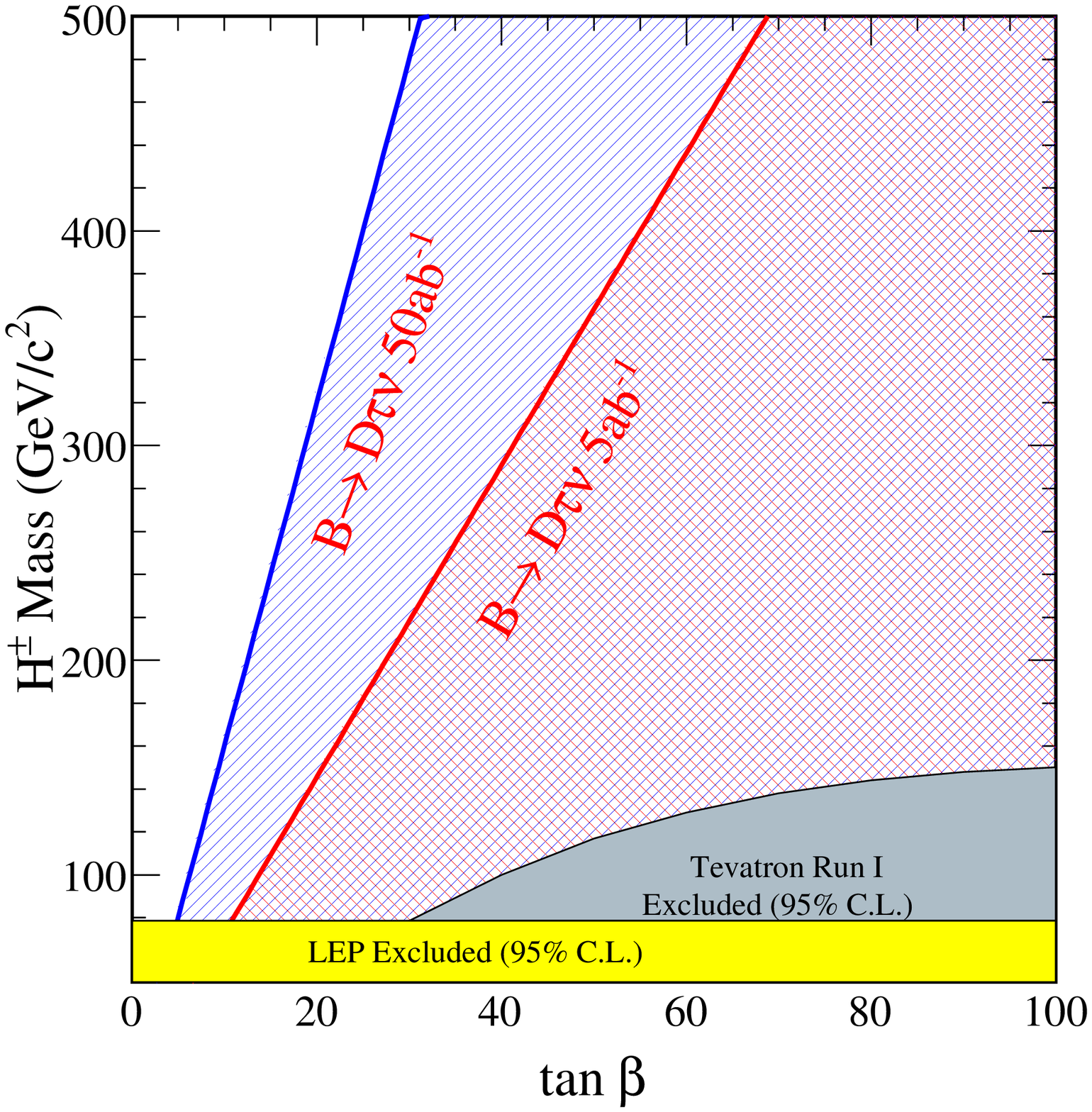} 
\caption{
Expected constraint on the charged Higgs from measurements of the 
$B \to D \tau\,\bar\nu$ branching fraction at 5 and 50 ab$^{-1}$.
}
\label{fig:higgs_Dtaunu}
\end{center}   
\end{figure}

The semileptonic $B$ decay into $\tau$ final state, 
$B \to D^{(*)} \tau\,\bar\nu$, 
is also a sensitive probe for the charged Higgs.
In the SM, the branching fractions are expected to be
about $8 \times 10^{-3}$ for 
$B \to D \tau\,\bar\nu$  
and  $1.6 \times 10^{-2}$ for 
$B \to D^* \tau\,\bar\nu$, respectively.
Because of the presence of at
least two neutrinos in the final state, the reconstruction of these modes requires the
reconstruction of the other B meson in the event, 
and hence requires a larger data sample with respect to 
that used to measure $B \to D^{(*)} \ell\,\bar\nu$ where $\ell = \mu, \; e$.
Fig~\ref{fig:higgs_Dtaunu} presents the expected future constraint in 
the $(M_{H^{+}},\tan\,\beta)$ plane  for a Super $B$ factory with
a 5 and 50~ab$^{-1}$ data sample.

\subsubsubsection{$B \to \mu \nu$}

Contrary to the $B^- \to \tau\,\bar\nu$ case, the $B^- \to \mu^-\,\bar\nu$ 
decay has more kinematic constraint because it has only one neutrino 
in the final state and the charged lepton at a fixed energy in the
$B$ rest frame.
Therefore, present analyses by Belle and BaBar take a conventional
approach, where one looks for a single high momentum lepton, and 
then inclusively reconstructs the accompanying $B$ via a 4-vector sum 
of everything else in the event.
The lepton momentum is smeared in the center-of-mass frame due to $B$
momentum to give a couple of hundred MeV/c width.

\begin{figure}[hbt]
\begin{center}
\includegraphics[width=.40\textwidth]{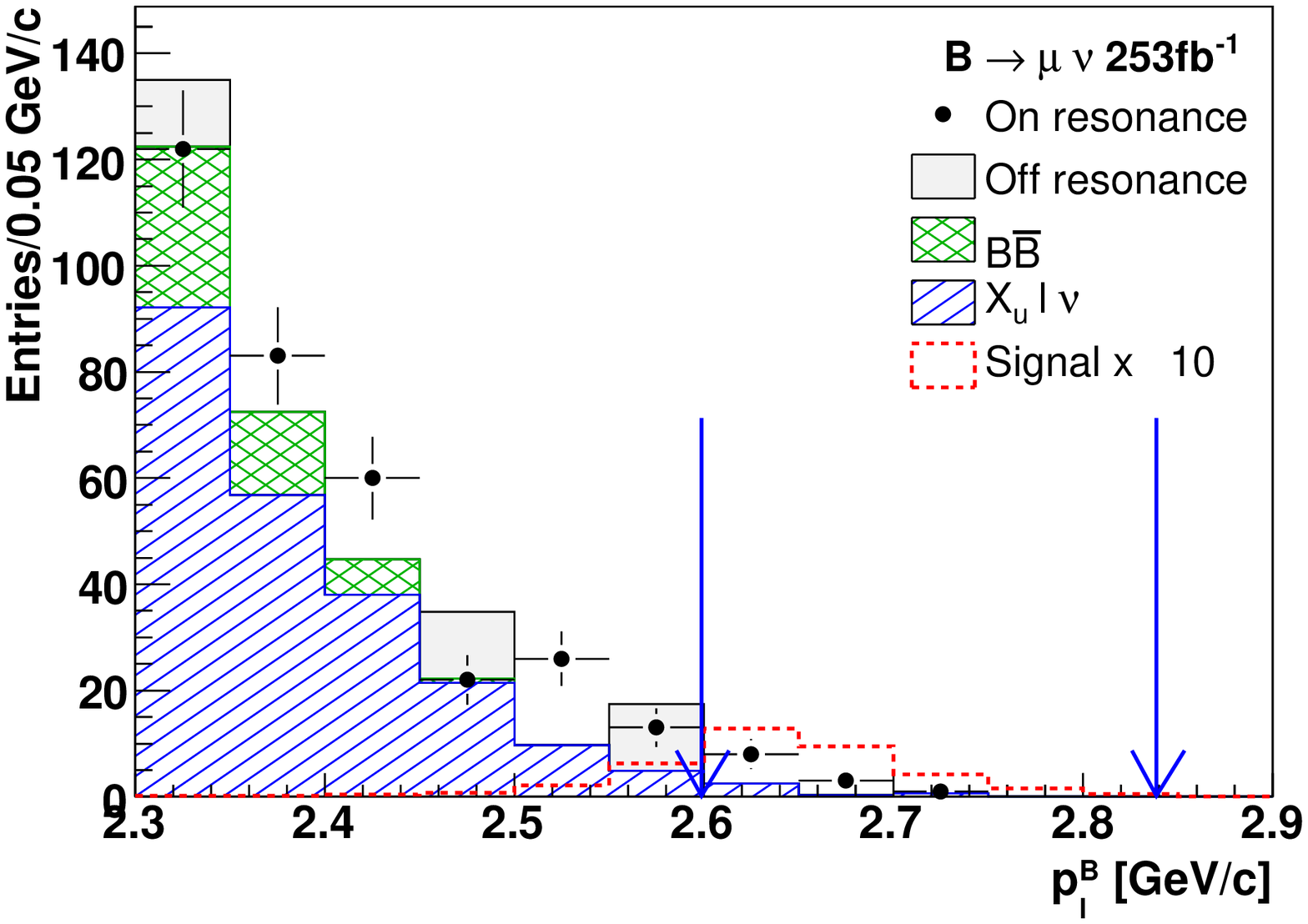} 
\includegraphics[width=.45\textwidth]{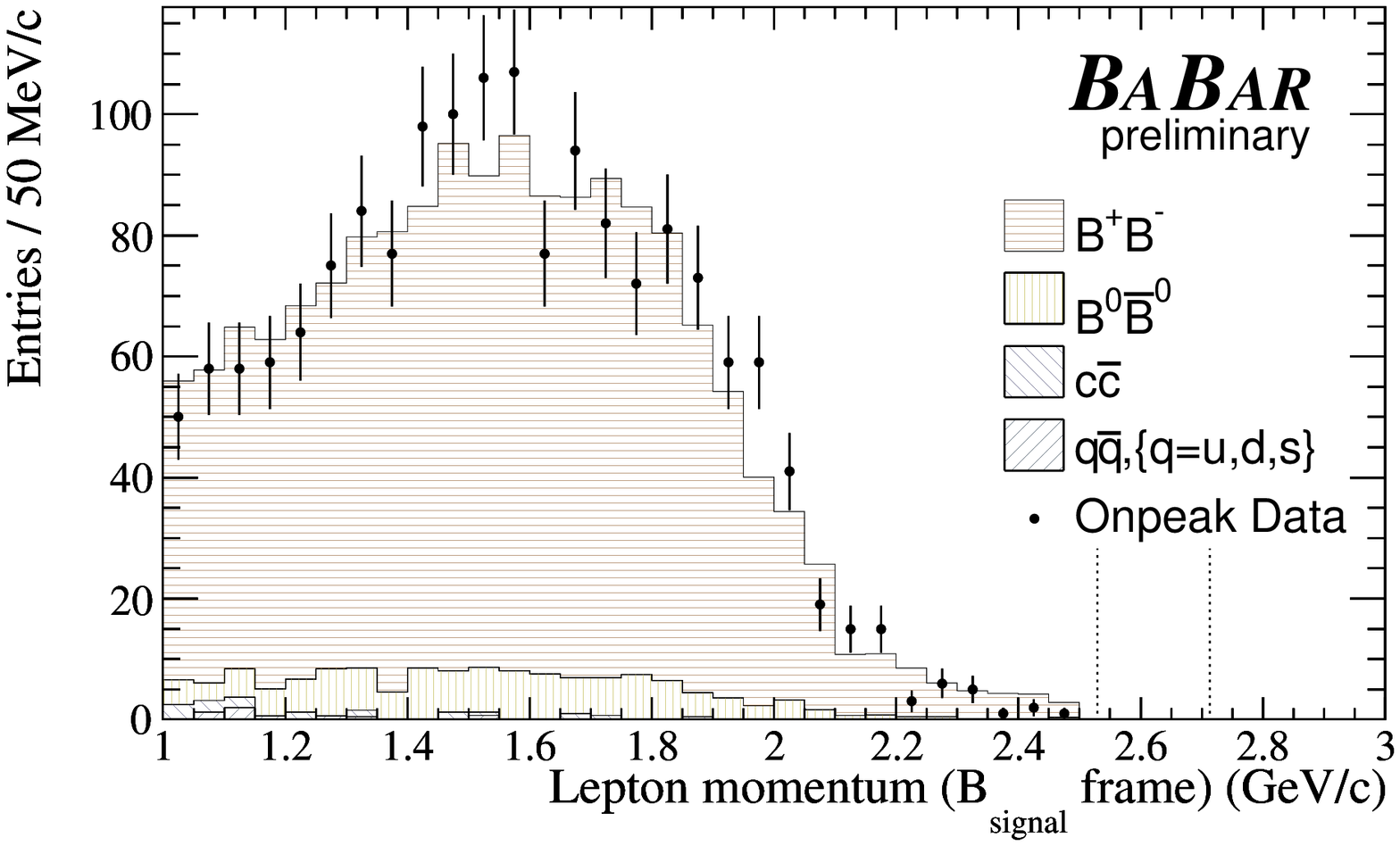} 
\caption{
a) Muon momentum distribution from the Belle analysis 
using an inclusive reconstruction of the accompanying $B$
for a 253\,fb $^{-1}$ data sample. 
b) The same distribution from the BaBar analysis using the hadronic
tags on a 208.7 fb$^{-1}$ data sample.
}
\label{fig:el_munu}
\end{center}   
\end{figure}

Fig.~\ref{fig:el_munu}~a) shows the muon momentum distribution from
the Belle analysis to search for the $B^- \to \mu^-\,\bar\nu$ decay
using the conventional approach on a 253\,fb$^{-1}$ data sample.  The
signal detection efficiency is 2.2\%.  The expected number of signals
based on the Standard Model branching fraction ($7.1 \times 10^{-7}$)
is 4.2, while the estimated background is 7.4. The reported upper
limit is ${\cal B}(B^- \to \mu^-\,\bar\nu) \leq 1.7 \times
10^{-6}$(90\% C.L.)~\cite{Satoyama:2006xn}.

Recently BaBar has reported a result of the $B \to \mu\,\nu$
search using the hadronic tags on a 208.7\,fb$^{-1}$ data sample.  In
this case, as the $B$ momentum is determined by the full
reconstruction, there is no smearing in the lepton momentum.
Fig.~\ref{fig:el_munu}~b) is the muon momentum distribution after
all the selection cuts are applied.  The signal detection efficiency is
about 0.15\%, an order of magnitude lower than for the conventional
analysis.  The reported upper limit is ${\cal B}(B^- \to
\mu^-\,\bar\nu) \leq 7.9 \times 10^{-6}$(90\%
C.L.)~\cite{Aubert:2006at}.

\begin{figure}[hbt]
\begin{center}
\includegraphics[width=.6\textwidth]{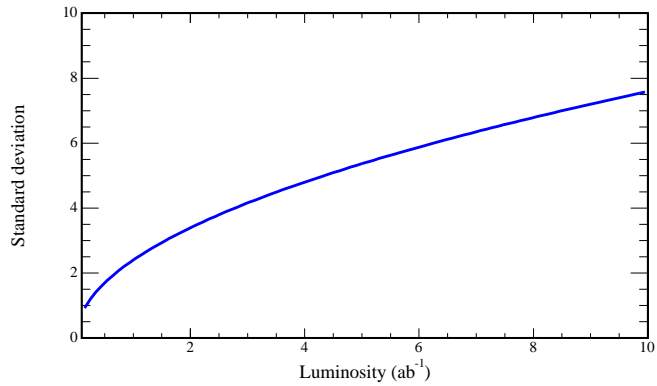} 
\caption{Expected sensitivity for $B^- \to \mu^-\,\bar\nu$ as a 
function of the integrated luminosity.}
\label{fig:munu_pro}
\end{center}   
\end{figure}

Fig.~\ref{fig:munu_pro} shows the expected statistical significance 
as a function of the integrated luminosity, based on a simple 
extrapolation from the present Belle result.
Accumulation of 1.6\,(4.3) ab$^{-1}$ data will allow us to detect the
$B^- \to \mu^-\,\bar\nu$ signal with 3\,(5) statistical significance.
The 50\,ab$^{-1}$ data at super $B$-factories will allow us to detect
about 800 signal events and measure the branching fraction with about 6\% 
statistical precision.

There are some points which need to be further studied.
\begin{itemize}

\item Optimization of the tagging; 
there may be some improvement by using the semileptonic tag in 
addition to the hadronic tag, especially for $B^- \to K^- \nu\,\bar\nu$, 
for which the impact of additional neutrinos seems to be relatively 
small.

\item Effects of backgrounds in a high luminosity environment;
future prospects are discussed so far by extrapolation from the present 
results, which may be too simple.
In particular, the impact of higher backgrounds to the tagging efficiency
and the missing energy resolution have to be more carefully examined.
   
\end{itemize}

%

\newpage 
\subsection{Very rare decays}
\label{sec:rare}
%



\subsubsection{Theory of $B_q \to \ell^+ \ell^-$ and related decays}
A particularly important class of very rare decays are the leptonic 
FCNC decays of a $B_d$ or a $B_s$ meson. In addition to the electroweak-loop
suppression the corresponding decay rates are helicity suppressed in the SM
by a factor of $m_{\ell}^2/m_B^2$, where $m_\ell$ and $M_B$ are the masses 
of lepton and $B$ meson, respectively. 
The effective
$|\Delta B|=|\Delta S| =1$ Hamiltonian, which describes $b\to s$ decays,
already contains 17 different operators in the Standard Model, in a
generic model-independent analysis of new physics this number will
exceed 100. One virtue of purely leptonic $B_s$ decays is their
dependence on a small number of operators, so that they are accessible
to model-independent studies of new physics. These statements, of
course, equally apply to $b\to d$ transitions and leptonic $B_d$ decays.
While in the Standard Model all six $B_q\to \ell^+ \ell^- $ decays (with
$q=d$ or $s$ and $\ell=e,\mu$ or $\tau$) are related to each other in a
simple way, this is not necessarily so in models of new physics.
Therefore all six decay modes should be studied. 

Other very rare decays, such as 
$B_q\to\ell^+\ell^-\ell^{\prime +}\ell^{\prime -}$,
$\ell^+\ell^-\gamma$, $e^+\mu^-$, are briefly considered in
Sec. \ref{otherrare} below.

 
\boldmath
\subsubsubsection{$B_q \to \ell^+ \ell^-$ in the Standard Model} 
\unboldmath

\begin{figure}[htb]
\begin{center}
  \includegraphics[width=.3\textwidth]{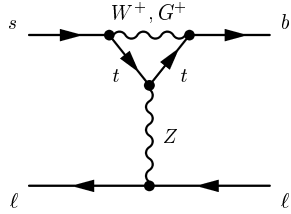}
\caption{Left: Z-penguin contribution to $B_s\to \ell^+\ell^-$.}\label{fig:smp}
\end{center}
\end{figure}

Photonic penguins do not contribute to $B_q \to \ell^+ \ell^-$, because 
a lepton-anti-lepton pair with zero angular momentum has charge conjugation 
quantum number $C=1$, while the photon has $C=-1$. 
The dominant contribution stems from the Z-penguin diagram and is shown in 
Figure~\ref{fig:smp}.

There is also a box diagram with two W bosons, which is suppressed by a
factor of $M_W^2/m_t^2$ with respect to the Z-penguin diagram. These
diagrams determine the Wilson coefficient $C_A$ of the operator 
\begin{eqnarray}
Q_A & = & \ov b_L \gamma^{\mu} q_L \, 
          \ov \ell \gamma_{\mu} \gamma_5 \ell . \label{defqa} 
\end{eqnarray}
We will further need operators with scalar and pseudoscalar couplings 
to the leptons:  
\begin{eqnarray}
Q_S & = & m_b \ov b_R q_L \, 
          \ov \ell \ell , \qquad\qquad
Q_P \; = \; m_b \ov b_R q_L \, 
          \ov \ell \gamma_5 \ell . \label{defsc} 
\end{eqnarray}
Their coefficients $C_S$ and $C_P$ are determined from 
penguin diagrams involving the Higgs or 
the neutral Goldstone boson, respectively. While $C_S$ and $C_P$ are
tiny and can be safely neglected in the Standard Model, the situation
changes dramatically in popular models of new physics discussed below. 
The effective Hamiltonian reads 
\begin{eqnarray}
       H & = & \frac{G_F}{\sqrt{2}} \, 
        \frac{\alpha}{\pi \sin^2\theta_W} \, V_{tb}^* V_{tq} \, 
        \left[ \, C_S Q_S + C_P Q_P + C_A Q_A \, \right] +\, h.c.
        \label{hami}
\end{eqnarray} 
The operators $Q'_S$, $Q'_P$ and $Q'_A$, where the chiralities of the 
quarks in the $\bar b q$ currents are flipped with respect to those in 
(\ref{defqa}), (\ref{defsc}), may also become relevant in
general extensions of the SM.
 
$C_A$ has been determined in the  next-to-leading order (NLO) of QCD
\cite{Buchalla:1993bv,Misiak:1999yg,Buchalla:1998ba}. The NLO corrections are in the percent range and
higher-order corrections play no role. $C_A$ is commonly expressed in
terms of the $\ov{\rm MS}$ mass of the top quark, $\ov m_t$. A pole mass
of $m_t^{\rm pole}=171.4 \pm 2.1 \, \gev $ corresponds to
$\ov{m}_t=163.8\pm 2.0\, \gev $.  An excellent approximation to the NLO
result for $C_A$, which holds with an accuracy of $5 \cdot 10^{-4}$ for
$149 \, \gev < \ov m_t < 179 \,\gev$, is  
\begin{eqnarray}
 C_A (\ov m_t)  & = &  0.9636 
   \lt[ \frac{80.4 \, \gev}{M_W} 
        \frac{\ov m_t}{164 \, \gev}\rt]^{1.52} \label{defca}
\end{eqnarray}
In the literature $C_A(\ov m_t)$ is often called $Y(\ov m_t^2/M_W^2)$. 
The exact expression can be found e.g.\ in Eqs.~(16-18) of \cite{Buchalla:1998ba}.  
The branching fraction can be compactly expressed in terms of the Wilson
coefficients $C_A$, $C_S$ and $C_P$: 
\begin{eqnarray}%
B\lt( B_q \to \ell^+\ell^- \rt) &=&
\frac{G_F^2\,\alpha^2}{64\, \pi^3 \sin^4\theta_W} \, 
      \lt| V_{tb}^* V_{tq} \rt|^2 \, \tau_{B_q}  
     \, M_{B_q}^3 f_{B_q}^2 \,
      \sqrt{1-\frac{4 m_{\ell}^2}{M_{B_q}^2}} \nonumber \\
&& \times \lt[ \lt( 1- \frac{4 m_{\ell}^2}{M_{B_q}^2} \rt) 
    \, M_{B_q}^2 \, C_S^2 \; +\; 
    \lt( M_{B_q} C_P - \frac{2 m_{\ell}}{M_{B_q}} C_A \rt)^2 \rt] 
  \label{defbr}. 
\end{eqnarray}
Here $f_{B_q}$ and $\tau_{B_q}$ are the decay constant and the lifetime
of the $B_q$ meson, respectively, and $\theta_W$ is the Weinberg angle. 
Since $B_q \to \ell^+\ell^-$
is a short-distance process, the appropriate value of 
the fine-structure constant 
is $\alpha=\alpha(M_Z)=1/128$. With \eq{defca} and $C_S=C_P=0$
\eq{defbr} gives the following Standard Model predictions:
\begin{eqnarray}%
B\lt( B_s \to \tau^+\tau^- \rt) &=& 
  \lt( 8.20\pm 0.31 \rt) \cdot 10^{-7\phantom{4}} \, \times
  \frac{\tau_{B_s}}{1.527\, \mbox{ps}}  \, 
  \lt[ \frac{\lt| V_{ts}\rt|}{0.0408} \rt]^2 \,  
  \lt[ \frac{f_{B_s}}{240\,\mev } \rt]^2 \label{stnum} \\[1mm]
B\lt( B_s \to \mu^+\mu^- \rt) &=& 
  \lt( 3.86\pm 0.15 \rt) \cdot 10^{-9\phantom{4}} \, \times
  \frac{\tau_{B_s}}{1.527\, \mbox{ps}}  \, 
  \lt[ \frac{\lt| V_{ts}\rt|}{0.0408} \rt]^2 \, 
  \lt[ \frac{f_{B_s}}{240\,\mev } \rt]^2 \label{smunum} \\[1mm]
B\lt( B_s \to e^+ e^- \rt) &=& 
  \lt(9.05 \pm 0.34 \rt) \cdot 10^{-14} \, \times 
    \frac{\tau_{B_s}}{1.527\, \mbox{ps}}  \, 
  \lt[ \frac{\lt| V_{ts}\rt|}{0.0408} \rt]^2 \,  
  \lt[ \frac{f_{B_s}}{240\,\mev } \rt]^2 \label{senum}  \\[1mm]
B\lt( B_d \to \tau^+\tau^- \rt) &=& 
  \lt( 2.23 \pm 0.08 \rt) \cdot 10^{-8\phantom{0}} \, \times
  \frac{\tau_{B_d}}{1.527\, \mbox{ps}}  \, 
  \lt[ \frac{\lt| V_{td}\rt|}{0.0082} \rt]^2 \,  
  \lt[ \frac{f_{B_d}}{200\,\mev } \rt]^2 \label{dtnum} \\[1mm]
B\lt( B_d \to \mu^+\mu^- \rt) &=& 
  \lt( 1.06\pm 0.04 \rt) \cdot 10^{-10} \, \times
  \frac{\tau_{B_d}}{1.527\, \mbox{ps}}  \, 
  \lt[ \frac{\lt| V_{td}\rt|}{0.0082} \rt]^2 \, 
  \lt[ \frac{f_{B_d}}{200\,\mev } \rt]^2 \label{dmunum} \\[1mm]
B\lt( B_d \to e^+ e^- \rt) &=& 
  \lt( 2.49\pm 0.09 \rt) \cdot 10^{-15} \, \times 
    \frac{\tau_{B_d}}{1.527\, \mbox{ps}}  \, 
  \lt[ \frac{\lt| V_{td}\rt|}{0.0082} \rt]^2 \,  
  \lt[ \frac{f_{B_d}}{200\,\mev } \rt]^2 \label{denum} 
\end{eqnarray}
The dependences on the decay constants, which have sizable theoretical
uncertainties, and on the relevant CKM factors have been factored out.
While $|V_{ts}|$ is well-determined through the precisely measured 
$|V_{cb}|$, the determination of $|V_{td}|$ involves the global fit to
the unitarity triangle and suffers from larger uncertainties. The
residual uncertainty in \eqsto{stnum}{denum} stems from the   
2$\,\gev$ error in $\ov{m}_t$. 

Alternatively, within the standard model, 
the CKM dependence as well as the bulk of the hadronic uncertainty 
may be eliminated by normalizing to the well-measured meson mass differences 
$\Delta M_{B_q}$, thus trading $f^2_{B_q}$ for a (less uncertain) bag 
parameter $\hat B_q$ \cite{Buras:2003td}:
\begin{equation}\label{eq:Bllnormalized}
 B(B_q \to \ell^+ \ell^-) = C \frac{\tau_{B_q}}{\hat B_q} 
\frac{Y^2(\ov m^2_t/M^2_W)}{S(\ov m^2_t/M^2_W)}\Delta M_q ,
\end{equation}
where $S$ is a perturbative short-distance function, 
$C=4.36\cdot 10^{-10}$ includes a normalization
and NLO QCD corrections, and $\ell = e, \mu$.
This reduces the {\em total} uncertainty within the SM below
the 15 percent level. (A similar formula may be written for $\ell=\tau$.)

\boldmath
\subsubsubsection{$B_q \to \ell^+ \ell^-$ and new physics} 
\unboldmath%


{\bf Additional Higgs bosons}

\noindent The helicity suppression factor of $m_\ell/M_{B_q}$ in front of $C_A$ in
\eq{defbr} makes $B(B_q \to \ell^+ \ell^-)$ sensitive to physics
with new scalar or pseudoscalar interactions, which contribute to $C_S$
and $C_P$. This feature renders $B_q \to \ell^+ \ell^-$ highly
interesting to probe models with an extended Higgs sector. Practically
all weakly coupled extensions of the Standard Model contain extra Higgs
multiplets, which puts $B(B_q \to \ell^+ \ell^-)$ on the center
stage of indirect new physics searches. Higgs bosons couple to fermions
with Yukawa couplings $y_f$. In the Standard Model $y_b\propto m_b/M_W$
and $y_\ell\propto m_\ell/M_W$ are so small that Higgs penguin diagrams,
in which the Z-boson of Figure~\ref{fig:smp} is replaced by a Higgs boson, play
no role. In extended Higgs sectors the situation can be dramatically
different. Models with two or more Higgs multiplets can not only
accommodate Yukawa couplings of order one, they also generically contain
tree-level FCNC couplings of neutral Higgs bosons. In simple
two--Higgs--doublet models these unwanted FCNC couplings are usually
switched off in an ad-hoc way by imposing a discrete symmetry on the
Higgs and fermion fields, which leads to the celebrated
two-Higgs-doublet models of type I and type II. Here we only discuss the
latter model, in which one Higgs doublet $H_u$ only couples to up-type
fermions while the other one, $H_d$, solely couples to down-type
fermions \cite{Gunion:1992hs}. The parameter controlling the size of the
down--type Yukawa coupling is $\tan\beta=v_u/v_d$, the ratio of the
vacuum expectation values acquired by $H_u$ and $H_d$. The Yukawa
coupling $y_f$ of $H_d$ to the fermion $f$ 
satisfies $y_f\sin\beta = m_f \tan\beta/v$
with $v=\sqrt{v_u^2+v_d^2}=174\, \gev$.  Hence $y_b\approx 1$ for $\tan
\beta \approx 50$. The dominant contributions to $C_S$ and $C_P$ for
large $\tan\beta$ involve charged and neutral Higgs bosons, but the
final result can be solely expressed in terms of $\tan\beta$ and the
charged Higgs boson mass $M_{H^+}$ \cite{Logan:2000iv}
\begin{eqnarray}%
        C_S \;=\; C_P &=& \frac{m_{\ell}}{4 M_W^2} \tan^2 \beta \,\,
        \frac{\ln r}{r-1}\qquad\qquad\quad \mbox{with}
\quad r\;=\; \frac{M_{H^+}^2}{\ov m_t^2} .
        \label{wc}
\end{eqnarray}
while $C_A$ remains the same as in the SM.
Although for very large values of $\tan\beta/M_{H^+}$ the branching
fraction can be enhanced, the contributions in \eq{wc} typically reduce
$B(B_q \to \ell^+ \ell^-)$ with respect to the Standard Model
value. The decoupling for $M_{H^+}\to \infty$ is slow, e.g.\ for 
$\tan\beta=60$ and $ M_{H^+}=500\, \gev$ the new Higgs contributions
reduce $B(B_q \to \ell^+ \ell^-)$ by 50\%!  


\noindent {\bf Supersymmetry}  

\noindent The generic Minimal Supersymmetric Standard Model (MSSM) 
contains many
new sources of flavour violation in addition to the Yukawa couplings.
These new flavour violating parameters stem from the
supersymmetry--breaking terms and their effects could easily exceed
those of the CKM mechanism. In view of the success of the CKM
description of flavour--changing transitions one may supplement the
MSSM with the hypothesis of \emph{Minimal Flavour Violation (MFV)},
which can be formulated systematically using symmetry arguments 
\cite{D'Ambrosio:2002ex}.  In the
MFV--MSSM the only sources of flavour violation are the Yukawa
couplings, just as in the Standard Model. In this section the MSSM is
always understood to be supplemented with the assumption of MFV. While in
MFV scenarios the contributions from virtual supersymmetric particles to
FCNC processes are normally smaller than the Standard Model
contribution, the situation is very different for $B_q \to
\ell^+\ell^-$.

The MSSM has two Higgs doublets.  At tree-level the couplings are as in
the two-Higgs-doublet model of type II, because the holomorphy of the
superpotential forbids the coupling of $H_u$ to down-type fermions and
that of $H_d$ to up-type fermions. At the one-loop level, however, the
situation is different, and both doublets couple to all fermions.  The
loop-induced couplings are proportional to the product of a
supersymmetry-breaking term and the $\mu$ parameter.  If $\tan\beta$ is
large, the loop-induced coupling of $H_u^*$ and the tree-level coupling
of $H_d$ give similar contributions to the masses of the down-type
fermions, because the loop suppression is compensated by a factor of
$\tan \beta$ \cite{Hall:1993gn}. In this scenario the Higgs sector is that of a
\emph{general}\ two-Higgs-doublet model, which involves FCNC Yukawa
couplings of the heavy neutral Higgs bosons $A^0$ and $H^0$ \cite{Hamzaoui:1998nu}.
The Wilson coefficients $C_S$ and $C_P$ differ from those in \eq{wc} in
two important aspects: they involve three rather than two powers of
$\tan \beta$ and they depend on the mass $M_{A^0}\sim M_{H^0}$ instead
of the charged Higgs boson mass. The branching ratios scale as
\begin{eqnarray}%
B(B_q\to \ell^+\ell^-)_{\rm SUSY} &\propto &
        \frac{m_b^2 m_\ell^2\, \tan^6\beta}{M_{A^0}^4}\nonumber 
\end{eqnarray}
and could, in principle, exceed the Standard Model results in
\eqsto{stnum}{denum} by a factor of $10^3$ \cite{Babu:1999hn}. Thus the
experimental upper limit on $B(B_s\to \mu^+\mu^-)$ from the Tevatron,
which is larger than $B(B_s\to \mu^+\mu^-)_{\rm SM}$ in \eq{smunum} by
a factor of 25, already severely cuts into the parameter space of the
MSSM.  $B(B_s\to \mu^+\mu^-)$ in MSSM scenarios with large $\tan\beta$
has been studied extensively \cite{Babu:1999hn,Chankowski:2000ng,Isidori:2001fv,Isidori:2002qe,Buras:2002vd,Bobeth:2001sq,Buras:2002wq,Dedes:2001fv}.

Very popular special cases of the MSSM are the minimal supergravity
model (mSUGRA) \cite{Nilles:1982ik,Chamseddine:1982jx,Barbieri:1982eh,Hall:1983iz,Soni:1983rm} and the Constrained Minimal
Supersymmetric Standard Model (CMSSM). While the MSSM contains more than
100 parameters, mSUGRA involves only 5 additional parameters and is 
therefore much
more predictive. In particular correlations between $B(B_s\to\mu^+\mu^-)$ 
and other observables emerge, for example with the
anomalous magnetic moment of the muon and the mass of the lightest
neutral Higgs boson \cite{Dedes:2001fv}.  Other well-motivated variants of the
MSSM incorporate the parameter constraints from grand unified theories
(GUTs).  $B(B_s\to \mu^+\mu^-)$ is especially interesting in GUTs
based on the symmetry group SO(10) \cite{Dedes:2001fv,Dermisek:2003vn,Dermisek:2005sw}. In the minimal
SO(10) GUT the top and bottom Yukawa couplings $y_b$ and $y_t$ unify at
a high scale implying that $\tan\beta$ is of order 50. While realistic
SO(10) models contain a non--minimal Higgs sector, any experimental
information on the deviation of $y_b/y_t$ from 1 is very desirable, as
it probes the Higgs sectors of GUT theories. In conjunction with other
observables like the mass difference in the 
$B_s$ - $\bar B_s$ system \cite{Buras:2002wq} or
$B(B^+\to \tau^+ \nu_\tau)$ \cite{Hou:1992sy,Akeroyd:2003zr,Isidori:2006pk}, 
which depend in different ways
on $\tan\beta$ and the masses of the non-Standard Higgs bosons and the
supersymmetric particles, the measurement of $B(B_s\to \mu^+\mu^-)$ at
the LHC will, within the MSSM, answer the question whether the top and
bottom Yukawa couplings unify at high energies.

\boldmath
\subsubsubsection{Other very rare decays}\label{otherrare}
\unboldmath%

The decays $B_q \to \ell^+ \ell^- \gamma$ and $B_q \to \ell^+ \ell^-
\ell^{\prime +} \ell^{\prime -}$ are of little interest from a theoretical
point of view. First, they are difficult to calculate, since they involve
photon couplings to quarks and are thereby sensitive to soft hadron dynamics.
Second, they are not helicity--suppressed, because the (real or virtual)
photon can recoil against a lepton pair in a $J=1$ state. This implies that
they probe operators of the effective Hamiltonian which can more easily be
studied from $B_q\to X \gamma$ and $B\to X \ell^- \ell^-$ decays. However, the
absence of a helicity suppression makes $B_q \to \ell^+ \ell^- \gamma$ a
possible threat to $B_q \to \ell^+ \ell^-$ as will be discussed in the
experimental sections.  A naive estimate gives $B(B_s\to \mu^+\mu^-
\gamma)\sim (m_B^2/m_\mu^2)\, \alpha/(4 \pi)\, B(B_s\to\mu^+\mu^-)
\sim B(B_s\to\mu^+\mu^-) $, while a more detailed analysis even finds $B(B_s\to
\mu^+\mu^- \gamma) > B(B_s\to \mu^+\mu^-)$ \cite{Melikhov:2004mk}.

Lepton-flavour violating (LFV) decays like $B_q \to \ell^\pm \mu^\mp$,
$\ell=e$, $\tau$, are negligibly small in the Standard Model. 
They are suppressed by
two powers of $m_\nu/M_W$, where $m_\nu$ denotes the largest neutrino
mass. However, this suppression factor is absent in certain models of
new physics.
In supersymmetric theories
with R parity (such as the MSSM) their branching ratios are smaller than
those of the corresponding lepton-flavour conserving decay, e.g.\ $B_q
\to \mu^+ \mu^-$. Large effects, however, are possible in models
that contain LFV tree-level couplings or leptoquarks. Here
supersymmetric theories without R parity and the Pati-Salam model should
be mentioned. Supersymmetry without R parity involves a plethora of new
couplings, which are different for all combinations of quark and lepton 
flavour involved, so that no other experimental constraints prevent
large effects in $B_q \to \ell^\pm \mu^\mp$. 
Flavour physics in the Pati-Salam model has been studied in 
\cite{Valencia:1994cj}.

\subsubsection{Present experimental status of $B_q\rightarrow \ell^+ \ell^-$ 
decays}

The experimental searches for $B_q\rightarrow \ell^+ \ell^-$ have
focused on $B_s \rightarrow \mu^+ \mu^-$ and $B_d \rightarrow \mu^+ \mu^-$. 
For the $e^+ e^-$ final states, the branching fractions are suppressed with respect to 
$B(B \rightarrow \mu^+ \mu^-)$ by $m_e^2/m^2_\mu = 2.3 \times 10^{-5}$. 
The best limit that has been set is 
$B(B \rightarrow e^+ e^-) < 61 \times 10^{-9} ~@ ~90\%$ 
confidence level (CL) \cite{Aubert:2004gm}. Though the branching fraction of 
the $\tau^+ \tau^-$ mode is enhanced by a factor of 212 with respect to that 
 of the $\mu^+ \mu^-$ mode, 
the only experimental upper limit from BaBar is 
$B(B_d \to \tau^+ \tau^-)  < 4.1 \times 10^{-3} ~@~90\%$~CL~\cite{Aubert:9:2qw}.
This is less sensitive than the decay $B \to \mu^+ \mu^-$.
 Due to at least two missing neutrinos in the decays of the two $\tau$s the 
 reconstruction of this mode is rather difficult, since no kinematic 
constraint can be employed to eliminate backgrounds. At an $e^+ e^-$ super $B$ 
 factory the $B_d \rightarrow \tau^+ \tau^-$ mode may be observable by 
 fully reconstructing one $B$ meson in a hadronic mode and then searching 
 for $B_d \rightarrow \tau^+ \tau^-$ in the recoil system.

\begin{table}[t]
\begin{center}
\caption{Branching fraction upper limits  $@ 90\%$ confidence level for $B_s \rightarrow \mu^+ \mu^-$ from different experiments.}
\begin{tabular}{|l|c|c|c|c|}
\hline \textbf{Experiment} & \textbf{Year} & \textbf{Limit $[10^{-9}]$} & Process &
\textbf{Reference}
\\
\hline 
D0 & 2007 & 75 & $ p \bar p$ at $\rm 1.96 \,TeV$ & \cite{d03}  \\
CDF & 2006 & 80  & $ p \bar p$ at $\rm 1.96 \,TeV$ & \cite{cdf1,Bernhard:2006fa}  \\
\hline
CDF & 2005 & 150  & $ p \bar p$ at $\rm 1.96\, TeV$ & \cite{Abulencia:2005pw}  \\
D0 & 2005 & 410 & $ p \bar p$ at $\rm 1.96\, TeV$ & \cite{Abazov:2004dj}  \\
CDF & 2004 & 580& $ p \bar p$ at $\rm 1.96\, TeV$ & \cite{Acosta:2004xj}  \\
CDF & 1998 & 2,000 & $p \bar p$ at $\rm 1.8\, TeV$ & \cite{Abe:1998ah}  \\
L3 & 1997 & 38,000 & $ e^+ e^- \rightarrow Z$ & \cite{Acciarri:1996us} \\
 \hline
\end{tabular}
\label{tab:bstomumu}
\end{center}
\end{table}

\begin{table}[t]
\begin{center}
\caption{Branching fraction upper limits at $ 90\%$ confidence level for
$B_d \rightarrow \mu^+ \mu^-$ from different experiments.}
\begin{tabular}{|l|c|c|c|c|}
\hline
\textbf{Experiment} & \textbf{Year} & \textbf{Limit $[10^{-9}]$} & Process &
\textbf{Ref}
\\
\hline
CDF & 2006 & 23  & $ p \bar p$ at $\rm 1.96\, TeV$ & \cite{cdf1,Bernhard:2006fa}  \\
\hline
CDF & 2005 & 39  & $ p \bar p$ at $\rm 1.96\, TeV$ & \cite{Abulencia:2005pw}  \\
BaBar   & 2005 & 83 & $e^+ e^- \rightarrow \Upsilon(4S)$ & \cite{Aubert:2004gm} \\
CDF & 2004 & 150 & $ p \bar p$ at $\rm 1.96\, TeV$ & \cite{Acosta:2004xj}  \\
Belle & 2003 & 160 & $e^+ e^- \rightarrow \Upsilon(4S)$ & \cite{Chang:2003yy} \\
CLEO & 2000 & 610 & $e^+ e^- \rightarrow \Upsilon(4S)$ & \cite{Bergfeld:2000ui} \\
D0 & 1998 & 40,000 & $ p \bar p$ at $\rm 1.8\, TeV$ & \cite{Abbott:1998hc}  \\
CDF & 1998 & 680 & $p \bar p$ at $\rm 1.8\, TeV$ & \cite{Abe:1998ah}  \\
L3 & 1997 & 10,000 & $ e^+ e^- \rightarrow Z$ & \cite{Acciarri:1996us} \\
UA1 & 1991 & 8,300 & $ p \bar p $ at $\rm 630\, GeV$ & \cite{Albajar:1991ct} \\
ARGUS & 1987 & 45,000 & $e^+ e^- \rightarrow \Upsilon(4S)$ & \cite{Albrecht:1987rj} \\
CLEO & 1987 & 77,000 & $e^+ e^- \rightarrow \Upsilon(4S)$ & \cite{Avery:1987cv} \\
\hline
\end{tabular}
\label{tab:btomumu}
\end{center}
\end{table}

\begin{figure*}[thb]
\centering
\mbox{
  \epsfig{file=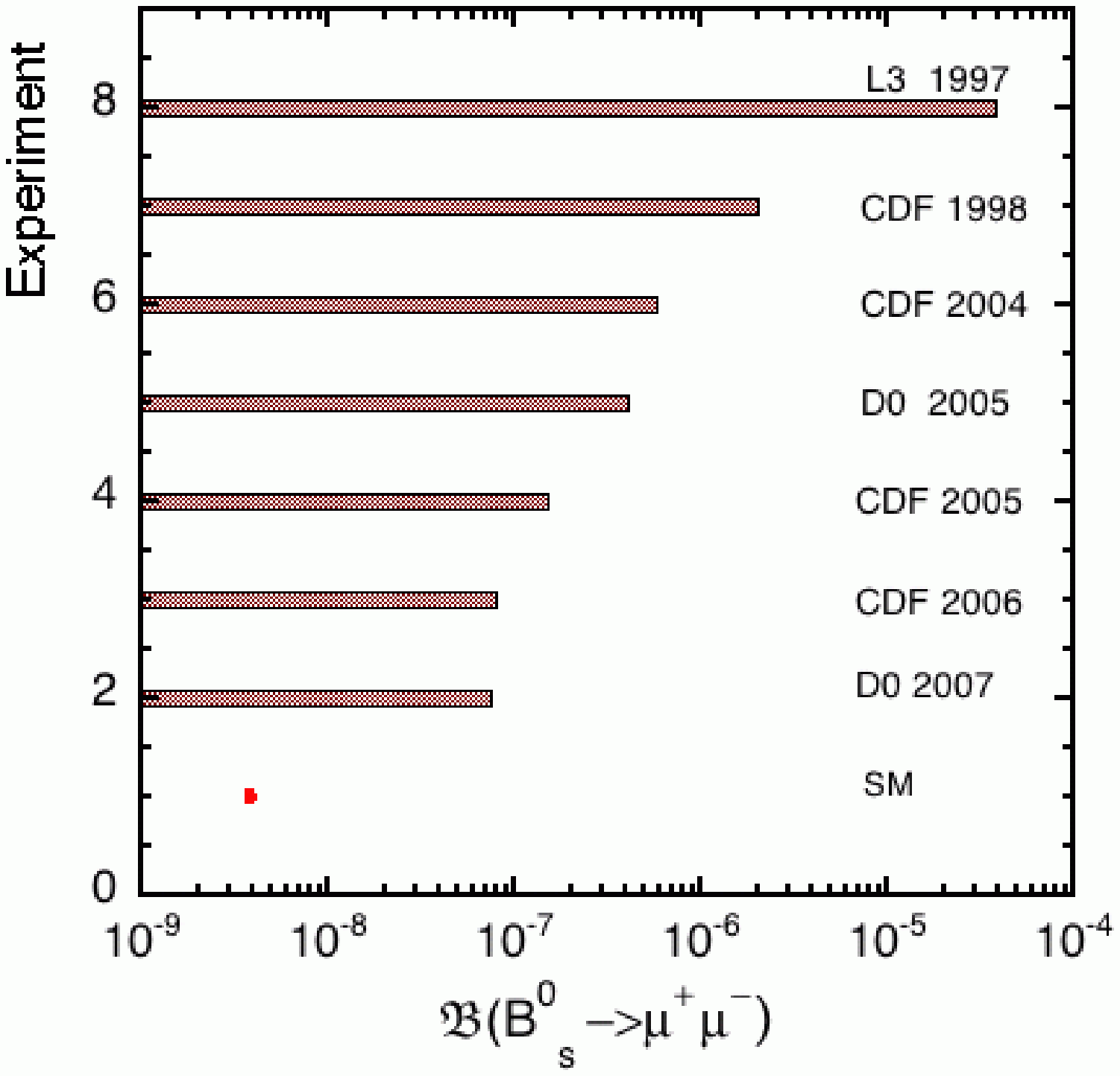,width=7.5cm, height=7.5cm}
  \epsfig{file=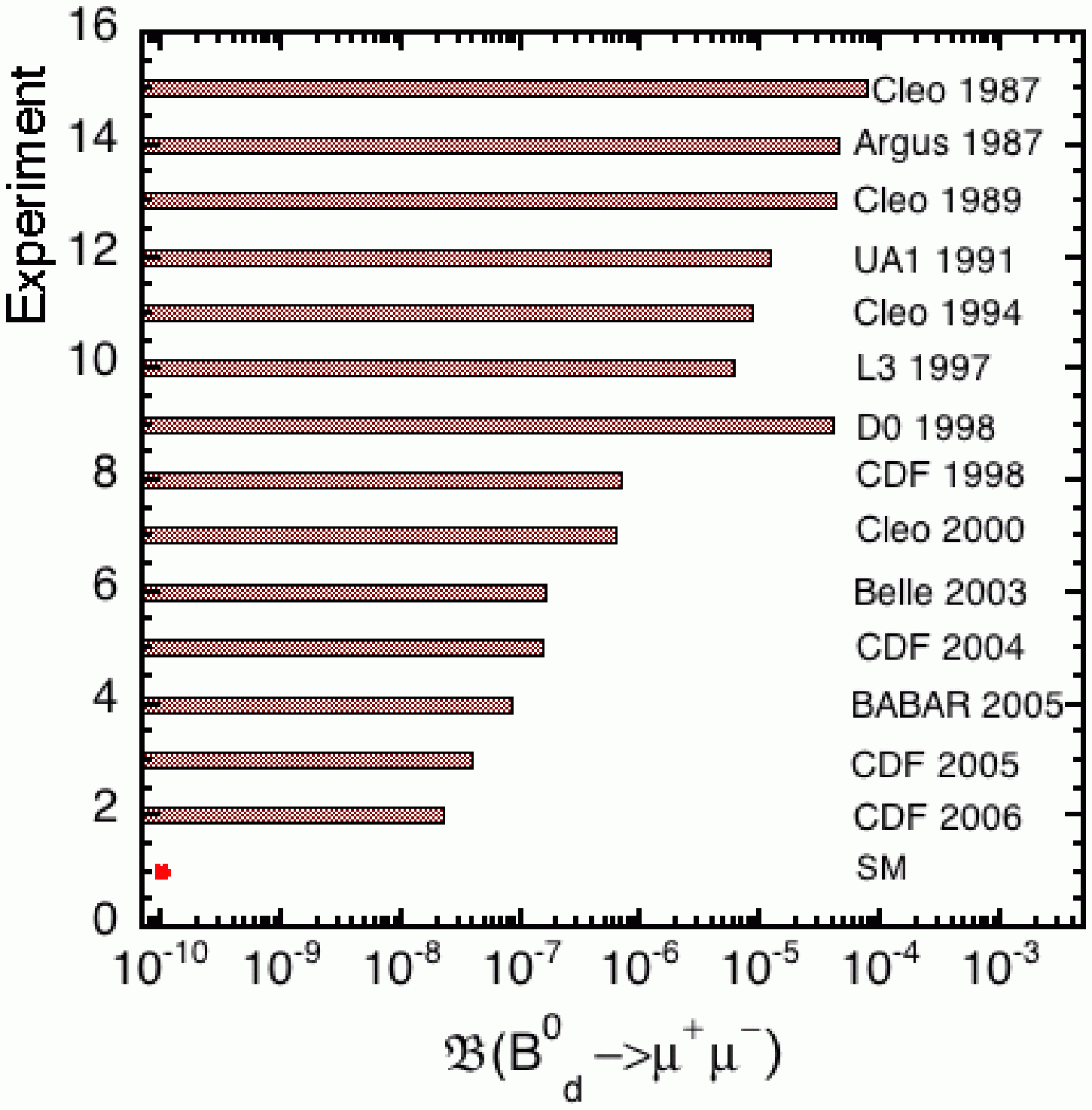,width=7.5cm, height=7.5cm}} 
\caption{Compilation of $90\%$ confidence level upper limits 
for $B(B_s \rightarrow \mu^+ \mu^-)$ (left) 
and $B(B_d \rightarrow \mu^+ \mu^-)$ (right)
from different experiments in comparison to the SM prediction.}
 \label{fig:bdmumu}
\end{figure*}

Thus, $B_{d,s} \rightarrow \mu^+ \mu^-$ are the most 
promising modes to test the Standard Model.
Table~\ref{tab:bstomumu} summarizes the searches 
for $B_s \rightarrow \mu^+ \mu^-$ by different experiments 
in the past two decades. The $90\%$ CL upper limits are shown in 
Figure~\ref{fig:bdmumu} in comparison to the SM prediction. 
The lowest limit of $B(B_s\rightarrow \mu^+ \mu^-) < 93 \times 10^{-9} \  @ \  95\%$ CL is obtained by 
the D0 experiment using $\rm about~2\,fb^{-1}$ of $p \bar p$ data \cite{d03}. 
Using $\rm 780\,pb^{-1}$ of $p \bar p$ data CDF achieved a branching 
fraction upper limit of $B(B_s\rightarrow \mu^+ \mu^-) < 
100 \times 10^{-9} \  @ \  95\%$ CL \cite{cdf1,Bernhard:2006fa}. 
The corresponding searches for $B_d \rightarrow \mu^+ \mu^-$ are summarized 
in Table~\ref{tab:btomumu}. Here, the lowest limit of 
$B(B_d \rightarrow \mu^+ \mu^-) < 30 \times 10^{-9} \  @ \ 95\%$ CL 
is obtained by the CDF experiment using $\rm 780\,pb^{-1}$ of $p \bar p$ 
data \cite{cdf1,Bernhard:2006fa}. The $90\%$ CL upper limits 
are also shown in Figure~\ref{fig:bdmumu} in comparison to the SM prediction. 

In the present CDF $B_s \rightarrow \mu^+ \mu^-$ analysis, 
the background level is at about one event, while the branching fraction 
upper limit $@~90\%$ CL 
lies about a factor of 20 above the SM value. 
Thus, any analysis attempting to reach a sensitivity at the level of the 
SM prediction needs a significant improvement in background rejection. 
Scaling the present CDF result to a luminosity of  $\rm 10\,fb^{-1}$ yields 
branching fraction upper 
limits at $90\%$ confidence level of $6.2 \times 10^{-9}$ 
for $B_s\to\mu^+\mu^-$ and $1.8 \times 10^{-9}$ for $B_d\to\mu^+\mu^-$. 
A simple scaling of the BaBar result to $\rm 1\,ab^{-1}$ 
yields  $B( B_d \rightarrow \mu^+ \mu^-) < 9 \times 10^{-9} \  @ \ 90\%$ CL.

\subsubsection{LHC preparations  for measurements of the very rare $B$ decays}

Three LHC experiments, LHCb, ATLAS and CMS, are  aiming for 
the measurement of very rare $B$ decays. Differences in the detector 
layouts lead to different strategies in data-taking, 
triggers  
and the offline selections to maximize the gain  of signal events. 

\subsubsubsection{Luminosity conditions and triggers}

 Whilst the nominal LHCb  luminosity
will be  $\rm (2-5) \times 10^{32}\,cm^{-2}s^{-1}$, the forward muon stations
 can identify muons with low values of transverse momenta, allowing the 
 first level trigger (L0) to collect events with one or two muons with 
 $p_{T}$ values as low as $\rm 1.1 \,GeV/c$ \cite{LHCb:2003pe}. 
 Because the beauty cross section grows rapidly at small transverse momenta, 
 the lower LHCb luminosity is compensated by higher $b$-production.  
 ATLAS and CMS  will start to collect the exclusive di-muon $B$ decays
 at a luminosity of few times $\rm 10^{33}\, cm^{-2} s^{-1}$ 
and will later continue  at the  nominal LHC luminosity 
of $\rm 10^{34}\,cm^{-2} s^{-1}$.
 Thus rare $B$-decays will be recorded at all LHC luminosities.
However the central detector geometries will allow muons to be recorded
only above $\rm \psubt \sim \,(3-6)\,GeV/c$ at 
the first trigger level (L1) ~\cite{Panikashvili:2006ka,CMS:2002}. 

First level triggers for the exclusive di-muon $B$ decays in 
LHCb, ATLAS and CMS  are summarized in Table~\ref{tab:L1}.
In LHCb the strategy relies on both the single muon trigger 
with $\rm \psubt  \ge 1.1\, GeV/c$ 
and di-muon trigger streams  with $\rm \Sigma \psubt (\mu\mu) \ge 1.3\, GeV/c$.
 ATLAS and CMS will collect the majority of their 
 signal events at $\rm 2 \times 10^{33}\, cm^{-2} s^{-1}$ through the 
 di-muon trigger with the muon transverse momentum thresholds 
 $\rm 6 \,GeV/c$ and  $\rm 3\, GeV/c$, respectively. 
Such triggers will result in output  rates of 
about $\rm 700\, Hz$ and $\rm 3500\, Hz$ for ATLAS and CMS, respectively, 
and about $\rm 200\, kHz$ for LHCb.

\begin{table}[h]
\begin{center}
\caption{L1(0) trigger  $p_T$ thresholds. The output trigger rates 
are given for a luminosity of
$\rm 2 \times 10^{32}\,cm^{-2}s^{-1}$ (LHCb) and  
$\rm 2 \times 10^{33}\,cm^{-2}s^{-1}$ (ATLAS/CMS).}
\label{tab:L1}
\begin{tabular}{p{3cm}cc}
\hline\hline
\textbf{Experiment}  & \textbf{L1(0) momentum cut} & \textbf{L1(0) Rate} \\
\hline
ATLAS  $2\mu$        & $\rm \psubt (\mu) \ge 6.0\, GeV/c$           &  $\rm0.7\, kHz$  \\
CMS   $2\mu$         & $\rm \psubt (\mu) \ge3.0\, GeV/c$            &  $\rm3.8\, kHz$ \\
LHCb $1\mu$          & $\rm \psubt (\mu) \ge1.1\, GeV/c$            &  $\rm110\, kHz$ \\
LHCb $2\mu$          & $\rm \Sigma \psubt (\mu\mu) \ge 1.3\, GeV/c$ &  $\rm145\, kHz$ \\

\hline\hline
\end{tabular}
\end{center}
\end{table}

The high level trigger (HLT) strategy is similar for all three experiments.
First, one confirms the presence of trigger muon(s) 
by reconstructing tracks within the so called region of interest (RoI) 
around a muon candidate and by matching reconstructed tracks in the
inner detector with tracks from the muon system.  
Further, cuts are applied to the muons requiring the $\psubt $ values to be 
above $\rm 3\, GeV/c$ for LHCb and above
 $\rm 4 \, GeV/c$ and $\rm6\, GeV/c$ for CMS and ATLAS, respectively. 
Then, primary and secondary vertices are reconstructed. 
Cuts on vertex quality $\chi^2 \le 20$\
 and on the flight path of $B_s$ candidates $L_{xy} \ge 200\, \mu m$ 
 (ATLAS) and $L_{3D} \ge 150\, \mu m$ (CMS) are applied. 
LHCb (single muon stream) uses an impact parameter cut $IP(\mu ) \ge\
 3 \sigma_{IP}$ and for the di-muon stream the secondary vertex quality cut
  $\chi^2 \le 20$.
Finally, a cut on the invariant mass of the two muons is applied,
${\rm 4 \,GeV/c^2} \le M_{\mu \mu} \le {\rm 6 \, GeV/c^2}$ (ATLAS), 
$ M_{\mu \mu} \ge 2.5\, \rm GeV/c^2$ (LHCb di-muon stream), or a mass window 
around the nominal $B_s$ mass of $\pm 150\, \rm MeV/c^2$ (CMS).
The HLT rate is less than $\rm 1.7\, Hz$ for CMS and about $\rm 660\, Hz$ 
for LHCb. 
A detailed description of trigger algorithms can be found 
in~\cite{LHCb:2003pe,Panikashvili:2006ka,CMS:2002}.

\subsubsubsection{Offline performance and signal selection}

After the trigger the offline analysis faces the challenge of 
selecting a signal from backgrounds of similar topology. 
The most important offline performance parameters 
for the di-muon events in the kinematic ranges accepted by triggers 
are given in Table \ref{Tab2}. The  differences lead consequently to 
 different selection strategies.
 
 \begin{table}[h]
\begin{center}
\caption{LHC detector performance parameters for $B\to\mu^{+}\mu^{-}$ events in 
the kinematic ranges of trigger acceptances. $\sigma_{Im}$ is the muon track 
impact parameter resolution, $\sigma_{M_{\mu\mu}}$ is 
the $B_{s}\to\mu^{+}\mu^{-}$ mass resolution.}
\label{Tab2}
\begin{tabular}{p{3cm}ccc}
\hline\hline
\textbf{Experiment}  & \textbf{LHCb} & 
\textbf{ATLAS} & \textbf{CMS} \\
\hline

 $\psubt^{\mu}$, GeV/c  & $> 3$ & $> 6$ & $>4$ \\
\hline
$\sigma_{Im}$, $\rm \mu m$  & $14-26$ & $25-70 $ & $30-50$ \\
\hline
$\sigma_{M_{\mu\mu}}$, $\rm MeV/c^2$  & $18  $ & $84  $  & $36 $ \\
\hline\hline
\end{tabular}
\end{center}
\end{table}

In ATLAS the reconstructed di-muon invariant mass $ M_{\mu\mu}$ is 
required to be 
within an interval of ($-70\rm\, MeV/c^2$, $+140\rm \, MeV/c^2$) 
around the $B_s$ mass.
The isolation cut in the ATLAS experiment requires no charged 
tracks with $\psubt \ge \rm 0.8\, GeV/c$ 
in an angular cone $\theta \le 15^{\circ}$ around the $B_s$ candidate. 
For the reconstructed vertices 
the significance of the reconstructed flight path in the 
transverse plane defined as $L_{xy}/\sigma_L$ is required to be larger than
11 and the vertex  reconstruction quality parameter $\chi^2 \le 15$. 
The space separation between two muon candidates 
is $\Delta R = \sqrt {\Delta \phi^2 + \Delta \eta^2 } \le 0.9$.
Details of the study can be found in~\cite{Nikitin:2006jz}.

In CMS isolation is defined as 
\begin{equation}\label{idef}
I = \frac{p_T(B_s^0)}{p_T(B_s^0)+\Sigma_{trk} |p_T|} \ge 0.85 \, .
\end{equation}
A value of $\Sigma_{trk} |p_T|$ is calculated for all 
 charged tracks in a cone with $\Delta R = 1$ around the $B_s$ candidate. 
 For the muon separation the value of $\Delta R$ should be in 
the range (0.3, 1.2). The vertex cuts are the following:  
$L_{xy} / \sigma_L \ge 18$ and $\chi^2 \le 1$. The momentum of the $B_s$ 
candidate should point to the primary vertex: $\cos\alpha \ge 0.995$, 
where $\alpha$ is the angle between the momentum of the $B_s$ candidate 
and the vector connecting the primary and secondary vertices 
$\vec{V}_{sec}-\vec{V}_{prim}$.
A tight mass cut is applied: $|M_{\mu \mu}- M_{B_s}| \le 100 \rm \, MeV/c^2$. 
Details of the study are given in~\cite{CMS:2006}.

In LHCb the selection is divided into several steps~\cite{Martinez:2007mi}. 
First the following soft selection cuts are applied: 
$| M_{\mu \mu} - M_{B_s}| \le  600 \rm \, MeV/c^2$, 
vertex quality cut $\chi^2 \le 14$, 
$IP / \sigma_{IP} \le 6$ for the  $B_s$ candidate,  
secondary and primary vertex separation 
$|Z_{sec} - Z_{prim}|/\sigma_V \ge 0$, pointing angle 
$\alpha < 0.1\, \rm rad$, soft muon identification for both 
candidates ($\epsilon_{\mu} =$95\% and $\epsilon_{\pi} = $1\%).
Further on three categories of discriminant variables are introduced: 
Geometry (G; lifetime, $B_s$ and $\mu$ impact parameter,
distance of closest approach 
(DOCA) and isolation), PID (particle identification) and IM (invariant mass). 
These variables are used to compute the S/B ratio event by event, while 
no further cuts are applied. Each event is weighted with its S/B ratio in 
the signal sensitivity calculation.
Using this method it is expected to 
reconstruct about 70 signal events per 
$2\rm \,fb^{-1}$~\cite{Martinez:2007mi}. 
If the previous method is combined with the requirement G $>0.7$, 
with no background events left, this leads to an estimate  of 
20 signal events to be reconstructed in the same period as above.

In Table~\ref{tab:num} the number of signal events is shown for 
each experiment for different integrated luminosities. 
For ATLAS/CMS the number for $2 \rm \,fb^{-1}$ is simply scaled from 
the one for $10 \rm \,fb^{-1}$. In the same way the LHCb number for  
 $10 \rm \,fb^{-1}$ is obtained by scaling the number for $\rm 2 \,fb^{-1}$. 
  The CMS and ATLAS studies for $100 \rm \, fb^{-1}$ were published
  in~\cite{Nikitenko:1999ak} and~\cite{Atlas:1999fr}, respectively.
  In the CMS study harder selection criteria have been applied for high
luminosity, hence the reconstruction efficiency for signal events is lower with
respect to lower luminosity.

\begin{table}[h]
\begin{center}
\caption{Number of signal events as a function of integrated luminosity.  
The time after which the corresponding luminosity will be delivered
is indicated in parentheses.}
\label{tab:num}
\begin{tabular}{p{3cm}ccccc}
\hline\hline
\textbf{Experiment}  & \textbf{\boldmath{$\rm 2\, fb^{-1}$}} & 
\textbf{\boldmath{$10 \rm \,fb^{-1}$}} & 
\textbf{\boldmath{$30 \rm \,fb^{-1}$}} & 
\textbf{\boldmath{$100 \rm \,fb^{-1}$}} &
\textbf{\boldmath{$130 \rm \,fb^{-1}$}}
\\
\hline
ATLAS                & 1.4  & 7.0 & 21.0 & 92 & 113 (4 years)\\
CMS                  & 1.2  & 6.1 & 18.3 & 26 & 44 (4 years)\\
LHCb             & 20  & 100 (5 years)  & -    & - &\\
\hline\hline
\end{tabular}
\end{center}
\end{table}


\subsubsubsection{Background studies}

The search for $B_s\rightarrow \mu^+ \mu^-$ has to deal 
with the problem of an enormous level of background. 
  
The largest contribution is expected to come from combinatorial background.
These events consist  predominantly of beauty decays,
where the di-muon candidates originate either from semileptonic
decays of $b$ and $\bar{b}$ quarks or from cascade decays of
one of the $b\bar{b}$ quarks. To determine the contribution of this background 
LHCb simulated a sample of inclusive $ b \bar{b}$ events, requiring that 
 both $b$-quarks have $|\theta| < \rm 400\, mrad$,
to match, on the safe side, the LHCb acceptance of $\rm 300 \, mrad$.
 Nevertheless, the sample of 34 million events corresponds
to only 0.16$\rm \, pb^{-1}$. The study of this sample, however, showed
 that in the sensitive region of phase space, the relevant
background contains two real muons from $b$-decays. 
Hence, a specific sample of 8 million events was generated,
 corresponding to an effective luminosity of 30$\rm \,pb^{-1}$, where for both  $b$-hadron decays
 a muon is required among the decay products. 
 LHCb uses this sample to evaluate the background and extrapolates the result 
to a given integrated luminosity, for instance, 2 fb$^{-1}$. 
In the sensitive region ($G > 0.7$)~\cite{Martinez:2007mi},
no background event was selected, 
hence an upper limit of 125 events is estimated at 90$\%$ CL.
ATLAS simulated $b \bar{b}$ events with two muons, requiring to have 
transverse momenta $\rm \psubt > 6\, (4)\, GeV/c $ for the first (second) 
muon. In CMS the cut for both muons was $\rm \psubt >3\, GeV/c $.
The pseudorapidity of each of the muons was required to be 
in the range $|\eta| < 2.4$ in agreement with the trigger acceptances. 
Additionally the di-muon mass was required to be in the interval  
$ M_{\mu\mu} < 8 {\rm \, GeV/c^2}$ and 
$5\, {\rm GeV/c^2} <  M_{\mu\mu} < 6\, {\rm GeV/c^2}$ 
in ATLAS and CMS, respectively. 
The number of background events
generated with these cuts corresponds to $\rm 10\, (8)\, pb^{-1}$ 
for ATLAS (CMS).
Both experiments evaluated the background using these samples, 
and extrapolated the results to
a given integrated luminosity. At $\rm 10\, fb^{-1}$ ATLAS 
expects 20$\pm 12$ events~\cite{Atlas:bkg} and 
CMS 14$\pm^{22}_{14}$ events~\cite{CMS:2006}.

Due to the high  sensitivity of the LHC experiments, the background composition
may be changed relative to the situation at the Tevatron. In addition to 
combinatorial background, contributions from
topologically similar rare exclusive decays 
as well as misidentification effects may  become important.
We give a classification of the different types 
of these potential backgrounds
and several estimates of their contribution.

 First, let us consider the very rare 
 decays $B^{0\pm}\to(\pi^{0\pm},\gamma)\mu^{+}\mu^{-}$ 
 with branching ratios expected to be 
$\sim\,2 \times 10^{-8}$~\cite{Melikhov:2004mk}.
A background contribution may arise when the $\pi$/$\gamma$ is soft 
and escapes detection.
The di-muon invariant mass distribution has been  
modeled in ATLAS and CMS for cases when a
 $\pi^{\pm}$ 
 is not reconstructed in the inner tracker,
or  a $\pi^{0}(\gamma)$ with $E_{T}\leq\,\mathrm{(2-4)}$ GeV escapes
detection in the electromagnetic calorimeter.
Based on a full detector simulation  CMS concluded that  
 neither of the processes $B^0\to\gamma\mu^{+}\mu^{-}$,  
$B^{\pm}\to\pi^{\pm}\mu^{+}\mu^{-}$ or  $B^{0}\to\pi^{0}\mu^{+}\mu^{-}$ 
will contribute significantly in the signal region. 
ATLAS reached similar conclusions for the first two processes,
while they plan to do a detailed study for the third decay.
These very rare decay channels are worth studying in their own right,
since some  properties
(for example the di-muon invariant mass spectrum) are also sensitive to
NP contributions~\cite{Melikhov:2004mk}.

Decays into four leptons, such as 
$B^{+}_{(c)}\to\mu^{+}\mu^{-}\ell^{+}\nu_{\ell}$, 
are another possible background source to $B_{s}\to\mu^{+}\mu^{-}$.
If the $p_{T}$ of one of the leptons
is below the detector reconstruction capabilities, 
then there are only two tracks observed from the $B$-meson vertex and the 
invariant mass of the di-lepton pair can be close to the $B_{d,s}$ mass.
The expected branching fractions of these decays are 
$5 \times 10^{-6}$ and $8 \times 10^{-5}$ \
for $B^{+}$ and $B^{+}_{c}$, respectively~\cite{Nikitin:2007zh}.
Using the fast simulation tool (ATLFAST),
ATLAS showed that the number of background
events from $B^{+}\to\mu^{+}\mu^{-}\mu^{+}\nu$ can be as high as 50~\%  
of the accepted signal events from $B_{s}\to\mu^{+}\mu^{-}$ with a SM rate. 
In CMS the analysis showed that the contribution from this source is 
negligible. 
The difference is due to different mass resolutions of ATLAS and CMS. 
LHCb simulated  a resonant mode of the four-lepton channel 
$B^{+}_{(c)}\to(J/\psi\to\mu^{+}\mu^{-})\mu \nu$ in which two muons are coming 
from $J/\psi$. The study led to the conclusion  
that the background from this channel in the mass region 
$\rm \pm 60\, MeV/c^2$ around the $B_s$ mass  is less than 10\% of a 
$B_{s}\to\mu^{+}\mu^{-}$ signal within the SM.

The last category considered are backgrounds 
from $B$ decay channels where secondary hadrons are misidentified as muons. 
The simplest backgrounds come from the two-body hadronic decays
 $B_{d,s}\to K^{\pm}\pi^{\mp}$, $B_{d,s}\to K^{\pm}K^{\mp}$
and $B_{d,s}\to\pi^{\pm}\pi^{\mp}$.
The background contribution can be estimated  by
 assigning to each of the final-state hadrons 
 a probability that it would be registered as a muon. This probability was
  obtained from  full detector simulations of large samples of beauty events. 
Such a study has been performed at LHCb, resulting in  $\sim 2$ events per
$2 \rm \,fb^{-1}$ (in a $\pm 2\sigma$ mass window). CMS concluded that these 
backgrounds are negligible. ATLAS studies are in progress.
Fake signal events can also be generated by semileptonic $B$ decays
such as $B^{0}\to\pi^{-}\mu^{+}\nu_{\mu}$ which have a 
branching ratio $\sim\,10^{-4}$.
As in the previous case, background can arise from $\pi-\mu$ 
misidentification and a soft neutrino escaping an indirect identification.
Similar channels to be accounted for are $B_{s}\to K^{-}\mu^{+}\nu_{\mu}$ 
and $B^{+}\to K^{+}\mu^{+}\mu^{-}$. 

\subsubsubsection{LHC reach for $B_{s}\to\mu^{+}\mu^{-}$}

\begin{figure}
\centering
 \epsfig{file=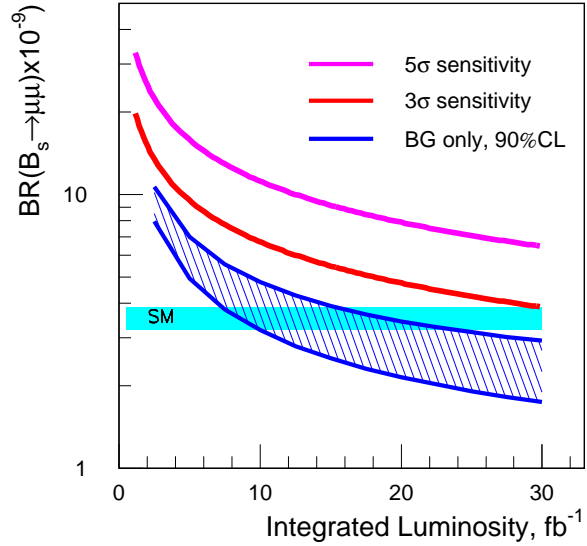,width=8.cm, height=8.cm} 

\caption{Branching ratio of $B_{s}\to\mu^{+}\mu^{-}$ 
observed (3$\sigma$) or discovered (5$\sigma$)
as a function of integrated luminosity for  ATLAS/CMS. }
\label{reach1}
\end{figure}

\begin{figure}
\centering
 \epsfig{file=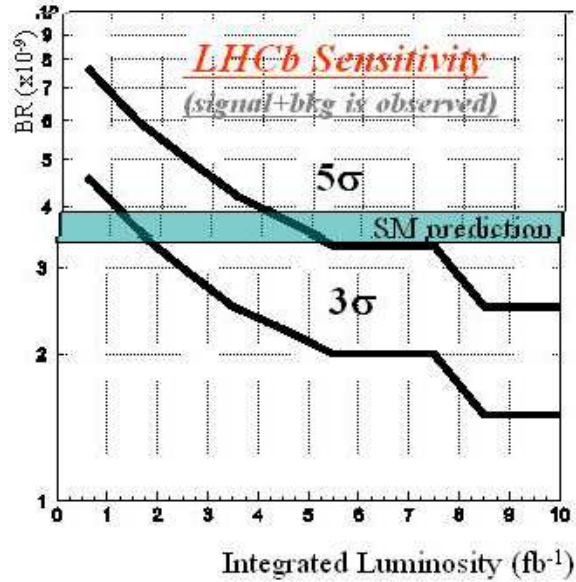,width=8cm, height=8cm} 
\caption{Branching ratio of $B_{s}\to\mu^{+}\mu^{-}$ 
observed (3$\sigma$) or discovered (5$\sigma$)
as a function of integrated luminosity for LHCb. }
\label{reach2}
\end{figure}

The results of the signal and background studies described in the previous 
sections were finally used to estimate  upper 
limits on the branching ratio of $B_{s}\to\mu^{+}\mu^{-}$, 
which are shown in Figures~\ref{reach1} and~\ref{reach2}. 
ATLAS and CMS used the algorithms of \cite{Eidelman:2004wy}, 
while LHCb developed the new approach published in~\cite{Martinez:2007mi}. 
In all cases the results were given at 90\% confidence level 
as a function of integrated luminosity. The theory prediction for
 $B(B_{s} \to \mu^+ \mu^-)$ shown in Figures~\ref{reach1} and~\ref{reach2}  
uses the value of $f_{B_s} =(230 \pm 9)\, \rm MeV$
 extracted from the CDF measurement of $\Delta M_{B_s} = 17.8
 \pm \rm 0.1 \,ps^{-1}$. The prediction therefore assumes that new physics
 neither affects $B_{s} \to \mu^+ \mu^-$ nor $\Delta M_{B_s}$.
Note that the above value for $f_{B_s}$ is also consistent
with direct QCD lattice calculations (see section \ref{subsec:latqcd}).
 
After one year of LHC the expected results from  LHCb will allow to exclude 
or discover NP in  $B_{s}\to\mu^{+}\mu^{-}$. 
ATLAS and CMS will reach this sensitivity after 
three years. After LHC  achieves its nominal 
luminosity, the ATLAS and CMS statistics will increase substantially. 
After five years  all three experiments will be in a position to 
provide a measurement of the branching ratio of 
$B_{s}\to\mu^{+}\mu^{-}$.  

\subsubsection{Conclusions}

The very rare decays $B_q\to\mu^+\mu^-$ are special in many respects. 
Their branching ratios are small in the
Standard Model, but can be enhanced significantly in the widely
studied Minimal Supersymmetric Standard Model (MSSM). Leptonic meson
decays belong to the physics topics that can be experimentally
studied by three of the four major LHC experiments, namely LHCb, ATLAS
and CMS. The LHC experiments will probe the branching fraction of 
$B_s\to \mu^+\mu^-$ down to the Standard Model value and possibly reveal a
smoking gun signal of new physics well ahead of the direct searches using
high-\psubt physics.  Irrespectively of whether $B(B_s \to
\mu^+\mu^-)$ is found in agreement with the Standard Model prediction or
not, the measurement will severely constrain the Higgs sector of the
MSSM and will provide valuable input for LHC Higgs physics: any sizable
enhancement of $B(B_s \to \mu^+\mu^-)$ implies a large value of
$\tan\beta$, so that the non-standard Higgs bosons couple strongly to
$b$-quarks and $\tau$-leptons. Then these Higgs bosons will be dominantly 
produced in association with $b$-jets and will decay dominantly into 
$b$-hadrons and $\tau$-leptons. 



%

\newpage \subsection{UT angles from tree decays}




\subsubsection{Introduction}

\label{sec:intro} 
It is very fortunate that the $B$ system allows an almost pristine
determination of all the three angles from ``tree''
decays. $\beta(\phi_1)$ from $J/\psi K_S$-like modes and $\gamma
(\phi_3)$ from $DK$-type modes are genuine tree decays and are
theoretically very clean. The irreducible theory error (ITE) for
$\beta$ is expected to be less than 1\% and may be even considerably
less than that~\cite{Boos:2004xp}.\footnote{For a more conservative
(but data driven) estimate see, {\it e.g.},
ref.~\cite{Ciuchini:2005mg}.} On $\gamma$ the ITE is estimated at
O(0.1\%). For $\alpha (\phi_2)$ the situation with regard to theory
error is a bit more complicated. Isospin analysis allows, in
principle, extraction of $\alpha (\phi_2)$ from $\pi \pi$, $\rho \pi$,
or $\rho \rho$, but electroweak penguin contributions (EWP) do not
respect isospin. So, in each of the three channels the EWP
contributions and other isospin violations are difficult to ascertain
rigourously. But given that there are three channels it seems
reasonable that the theory error even for $\alpha$ will be small,
O(few\%) (see, {\it e.g.}, ~\cite{Gronau:2005pq}).  Given that we now
have theoretical methods that will allow us to quite precisely
determine all the three angles, which are fundamental parameters of
the SM, it is clearly important to determine them with accuracy
roughly commensurate with what the theoretical methods promise. In
this brief report we will summarize the current status as to our
attempts to extract these three angles directly from data collected
primarily through the spectacular successes of the two asymmetric $B$
factories, followed by our guess estimates for the potential of a
Super $B$ factory (SBF) with regard to this goal. Of course, LHCb will
soon begin operation, and our expectations for the precisions on
tree-level angle determinations from LHCb are also presented.

\subsubsection{Angles from $B$ factories of today \& of tomorrow}
\label{sec:angles_bfac}

\subsubsubsection{$\beta (\phi_1)$}
\label{sucsec:beta}

Measurements of $C\!P$ asymmetries in the proper-time distribution
of neutral $B$ decays to $C\!P$ eigenstates mediated by 
$b \to c\overline{c} s$ transition provide a direct measurement of 
$\sin 2\beta$ (= $\sin 2\phi_1$). 
%
%
The time-dependent decay-rate asymmetry for decays to $C\!P$ eigenstates 
containing a charmonium and a $K^0_S$ meson is given by 
\begin{equation}
A_{C\!P}(t)= S_{b \to c\overline{c} s} \sin (\Delta m_d t) - C_{b \to c\overline{c} s} 
\cos (\Delta m_d t).
\label{eq:timedep}
\end{equation}
where $\Delta m_d$ is the mass difference between the two $B^0$ mass eigenstates.
Since these decays are dominated by a single (tree level)
amplitude~\footnote{The same processes can be described by
a penguin diagram which brings corrections at order $\sim \lambda^4$.},
one expects to a very good approximation 
$S_{b \to c\overline{c} s} = - \eta_{C\!P} \sin 2\beta$ and $C_{b \to c\overline{c} s} = 0$
where $\eta_{C\!P}$ is the $C\!P$ eigenvalue of the final state.

In 2001,
both BaBar and Belle collaborations established $C\!P$ violation
in the $B$ system through the $\sin 2\beta$ measurements in
$b \to c\overline{c} s$ decays~\cite{Aubert:2001nu,Abe:2001xe}.

\begin{figure}[htb]
  \begin{center}
    \resizebox{0.55\textwidth}{!}{
      \includegraphics{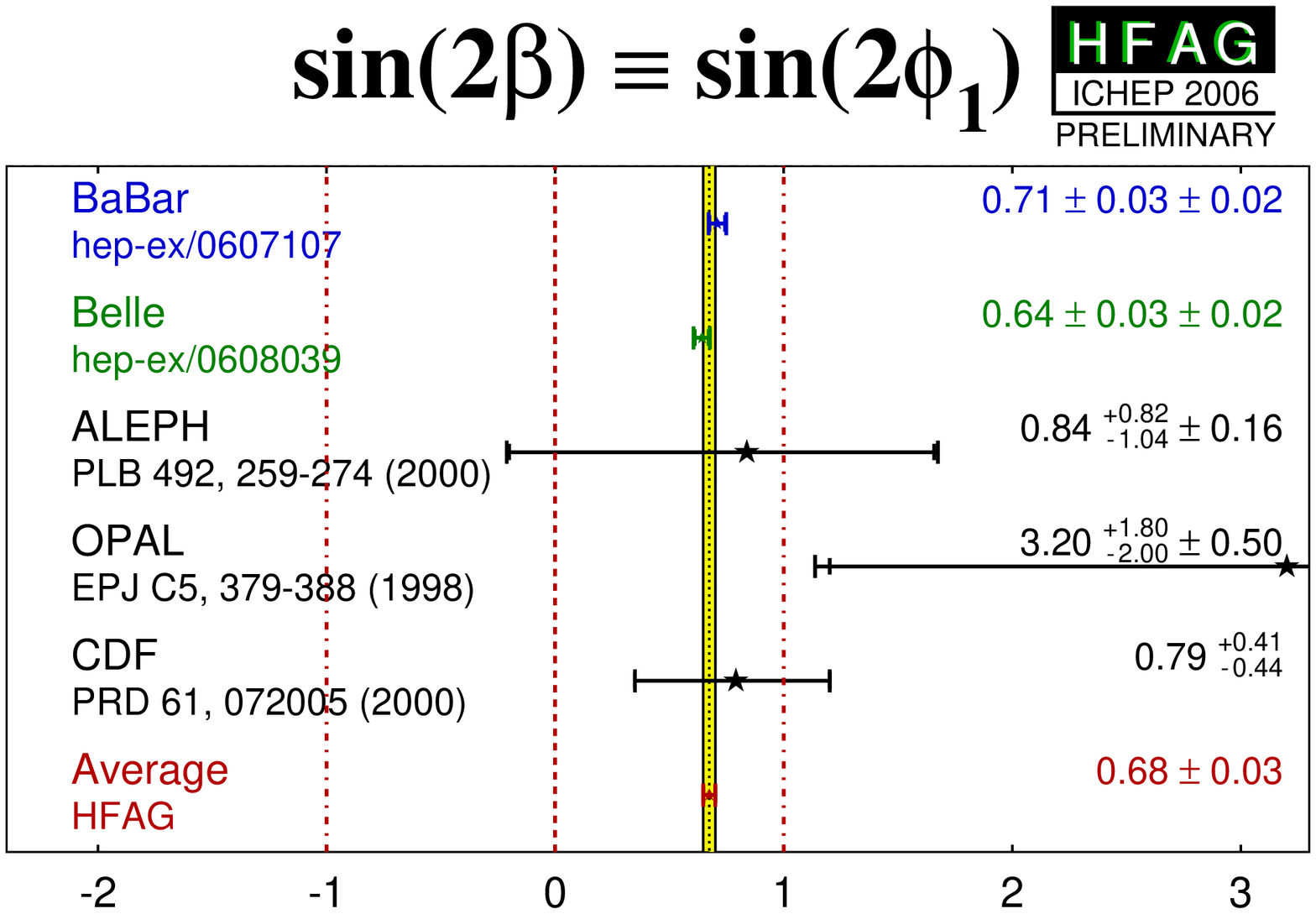}
    }
    \hfill
    \resizebox{0.44\textwidth}{!}{
      \includegraphics{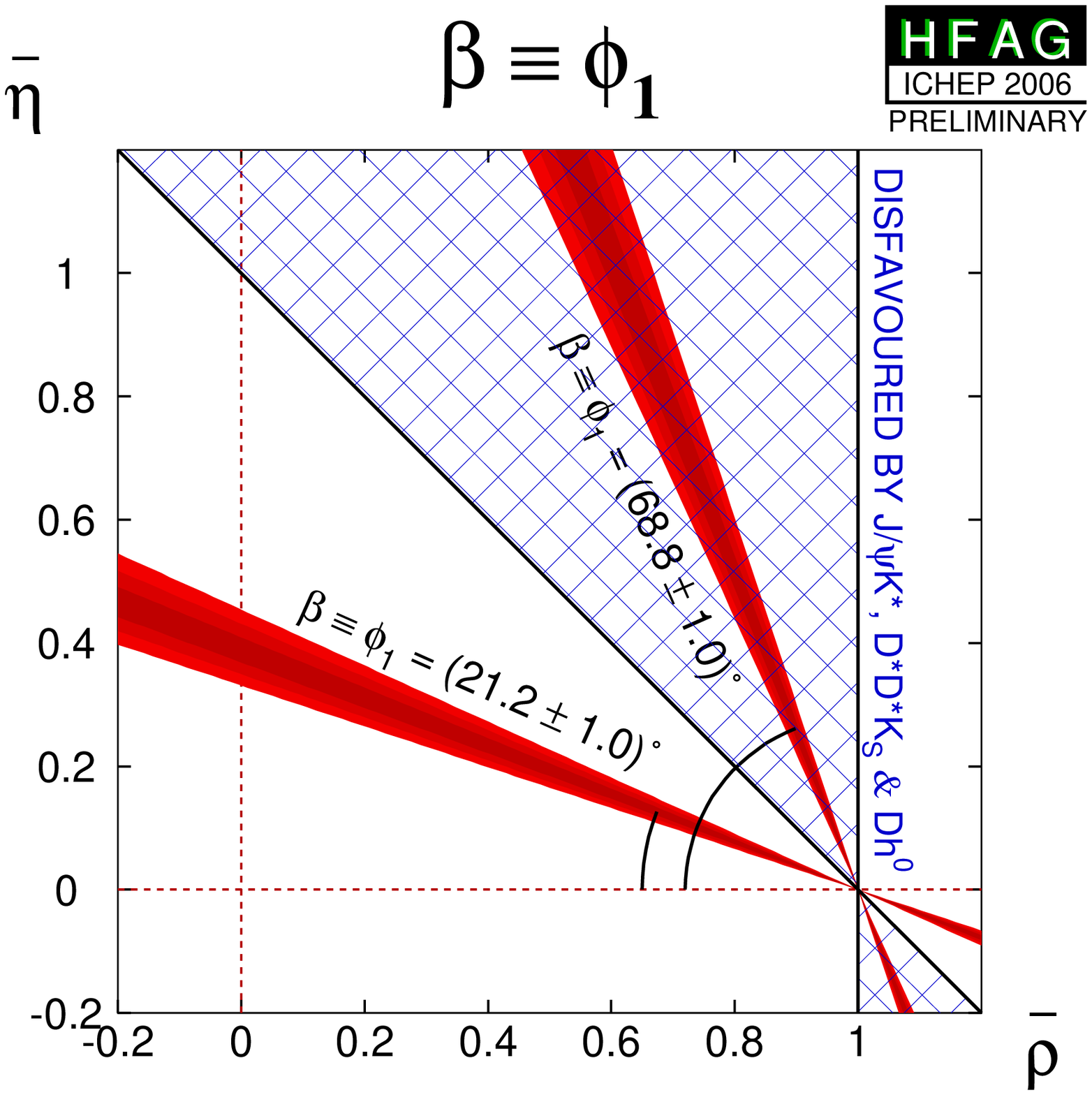}
    }
  \end{center}
  \vspace{-0.5cm}
  \caption{
  (Left) World average of measurements of $S_{b \to c\overline{c} s}$
  as calculated by HFAG~\cite{Barberio:2006bi}.
  (Right) Constraints on the $(\overline{\rho},\overline{\eta})$ plane,
  obtained from the average of $-\eta_{C\!P} S_{b \to c\overline{c} s}$
  and Eq.~\ref{eq:betaresult}.
  }
  \label{fig:cp_uta:ccs}
\end{figure}

In the latest results, the BaBar collaboration~\cite{Aubert:2006aq}, using a 
$348$ million $B\overline{B}$ events, includes the $C\!P$-odd ($\eta_{C\!P} = -1$)
final states $J/\psi K^0_S$, $\psi(2S)K^0_S$, $\chi_{c1}K^0_S$ and 
$\eta_c K^0_S$
as well as the $C\!P$-even ($\eta_{C\!P} = +1$) $J/\psi K^0_L$ final state.
In addition, the vector-vector final state $J/\psi K^*$ with 
$K^* \to K^0_S \pi^0$, which is found from 
an angular analysis to have $\eta_{C\!P}$ close to $+1$~\cite{Aubert:2001pe},
is used.
The Belle collaboration~\cite{Chen:2006nk} uses a 
sample of $535$ million $B\overline{B}$ events where only $J/\psi K^0_S$ and
$J/\psi K^0_L$ ($golden$ modes) are analysed.
The results for $-\eta_{C\!P} S_{b \to c\overline{c} s}$
and $C_{b \to c\overline{c} s}$ are given in Table~\ref{tab:cp_uta:ccs} and
in Fig.~\ref{fig:cp_uta:ccs} and are at the $5\%$ level for each
collaboration.

\begin{table}[hhh]
  \begin{center}
    \caption{Results for the
      $C\!P$-violating parameters in the $b \to c\overline{c} s$ decays:
      $S_{b \to c\overline c s}$ and $C_{b \to c\overline{c} s}$.
      The $B$ factory averages are given after ICHEP 2006 as calculated by
      HFAG~\cite{Barberio:2006bi}. 
      The final world averages include also the results from
      ALEPH, OPAL and CDF (which use only the $J/\psi K^0_S$ final state).}
    \vspace{0.2cm}
    \setlength{\tabcolsep}{0.0pc}
    \begin{tabular*}{\textwidth}{@{\extracolsep{\fill}}lrcc} \hline
      \multicolumn{2}{l}{Experiment} &
      $- \eta_{C\!P} S_{b \to c\overline{c} s}$ & $C_{b \to c\overline{c} s}$ \\
      \hline
      BaBar & \cite{Aubert:2006aq} &
      $0.710 \pm 0.034 \pm 0.019$ & $0.070 \pm 0.028 \pm 0.018$ \\
      Belle & \cite{Chen:2006nk} &
      $0.642 \pm 0.031 \pm 0.017$ & $-0.018 \pm 0.021 \pm 0.014$ \\
      \hline
      \multicolumn{2}{l}{\bf \boldmath $B$ factory average} &
      $0.674 \pm 0.026$ & $0.012 \pm 0.022$ \\
      \multicolumn{2}{l}{\small Confidence level} &
      \small $0.18$ & \small $0.02$ \\
      \hline
      \multicolumn{2}{l}{\bf Average} &
      $0.675 \pm 0.026$ & $0.012 \pm 0.022$ \\
      \hline
    \end{tabular*}
    \label{tab:cp_uta:ccs}
  \end{center}
\end{table}

The world average computed by the Heavy Flavor Averaging Group 
(HFAG)~\cite{Barberio:2006bi} includes also
the results obtained by the ALEPH, OPAL and CDF experiments and is 
\begin{equation}
  \sin 2\beta = 0.675 \pm 0.026
  \label{eq:betaresult}
\end{equation}
where most of the systematic uncertainties have been treated as uncorrelated.
This result suggests that on the time scale of 2008, when an integrated 
luminosity of order of 2 fb$^{-1}$ is expected from the $B$ factories, the 
total uncertainty on $\sin 2\beta$ will be around 0.02.

The actual $\sin 2\beta$ result gives a precise constraint 
on the $(\overline{\rho},\overline{\eta})$ plane, as shown in Fig.~\ref{fig:cp_uta:ccs}
and can be compared with the expected value 
obtained with other constraints from $C\!P$ conserving quantities,
and with $C\!P$ violation in the kaon system, in the form of the parameter 
$\epsilon_K$. Such comparisons have been performed by phenomenological
groups: for example, the result from the global UT fit without
the measurement of $\sin 2\beta$ is obtained by CKMfitter~\cite{Charles:2004jd} 
to be $0.823 ^{+0.018}_{-0.085}$
or by UTfit~\cite{Bona:2005vz} to be $0.759 \pm 0.037$. It is clear that the
increased precision in the $\sin 2\beta$ measurement is now revealing
some tension with the rest of the fit. This is mainly due to the actual
$V_{ub}$ value, and in particular to the inclusive one, strikingly in
countertendency with respect to the relatively {\it low} value of
$\sin 2\beta$~\cite{Bona:2006ah}.

With $\sin 2\beta$ being now a precision measurement, other
analyses are being performed in order to remove the two-fold ambiguity
unavoidable with a sine determination.

Considering the $B$ meson decays to the vector-vector final state $J\psi K^{*0}$,
in the case of a final state not flavour-specific ($K^{*0} \to K^0_S \pi^0$),
a time-dependent transversity analysis can be performed allowing sensitivity
to both $\sin 2\beta$ and $\cos 2\beta$~\cite{Dunietz:1990cj}.
Such analyses have been performed by both $B$ factory experiments:
from Table~\ref{tab:cosbeta} we can remark that at present the results are
dominated by large and non-Gaussian statistical errors, but nevertheless it 
can be said that $\cos 2\beta >0$ is preferred by the experimental data 
in $J\psi K^*$.

Finally, decays of $B$ mesons to final states such as $D\pi^0$ are governed by
$b \to c\overline{u}d$ transitions. If the final state is a $C\!P$ eigenstate,
{\it i.e.} $D_{C\!P}\pi^0$, the usual time-dependence formulae are recovered,
with the sine coefficient sensitive to $\sin 2\beta$.
Since there is no penguin contribution to these decays, there is even less
associated theoretical uncertainty than for $b \to c\overline{c}s$ decays like
$B \to J\psi K^0_S$.
When multi-body $D$ decays, such as $D \to K^0_S\pi^+\pi^-$, are used,
a time-dependent analysis of the Dalitz plot of the neutral $D$ decay
allows a direct determination of the weak phase: 
$2\beta$~\cite{Bondar:2005gk}.
Such analyses have been performed by both $B$-factory experiments.
The decays $B \to D\pi^0$, $B \to D\eta$, $B \to D\omega$,
$B \to D^*\pi^0$ and $B \to D^*\eta$ are used.
The daughter decays are $D^* \to D\pi^0$ and $D \to K^0_S\pi^+\pi^-$.
The results are shown in Table~\ref{tab:cosbeta}.
Again, it is clear that the data prefer $\cos 2\beta >0$.
Taken in conjunction with the $J\psi K^*$ results,
$\cos 2\beta <0$ can be considered to be ruled out at approximately 
$2.3\sigma$~\cite{Bona:2005vz}. Time-dependent analysis of the decay
$B \to D^{*+}D^{*-}K^0_S$ also prefers $\cos 2\beta > 0$.

\begin{table}
  \begin{center}
    \caption{
      Results from the $B$ factories together with the HFAG 
      averages~\cite{Barberio:2006bi} from the $B^0 \to J\psi K^{*0}$  and 
      the $B^0 \to D^{(*)}h^0$ analyses. 
    }
    \vspace{0.2cm}
    \setlength{\tabcolsep}{0.0pc}
    \begin{tabular*}{\textwidth}{@{\extracolsep{\fill}}lrcc} \hline
      \multicolumn{2}{l}{$B^0 \to J\psi K^{*0}$} & $\sin 2\beta$ & $\cos 2\beta$ \\
      \hline
      BaBar & \cite{Aubert:2004cp} &
      $-0.10 \pm 0.57 \pm 0.14$ & $ 3.32 \, ^{+0.76}_{-0.96} \pm 0.27$ \\
      Belle & \cite{Itoh:2005ks} &
      $-0.24 \pm 0.31 \pm 0.05$ & $ 0.56 \pm 0.79 \pm 0.11 $ \\
      \hline
      \multicolumn{2}{l}{\bf Average} &
       $0.16 \pm 0.28$ & $1.64 \pm 0.62 $ \\
      \hline \hline
      \multicolumn{2}{l}{$B^0 \to D^{(*)}h^0$} & $\sin 2\beta$ & $\cos 2\beta$ \\
      \hline
      BaBar & \cite{Aubert:2006an} &
      $0.45 \pm 0.36 \pm 0.05 \pm 0.07$ & $0.54 \pm 0.54 \pm 0.08 \pm 0.18$ \\
      Belle & \cite{Krokovny:2006zm} &
      $0.78 \pm 0.44 \pm 0.22$ & $1.87 ^{+0.40+0.22}_{-0.53-0.32}$ \\
      \hline
      \multicolumn{2}{l}{\bf Average} &
       $0.57 \pm 0.30$ & $1.16 \pm 0.42$ \\
      \hline
    \end{tabular*}
    \label{tab:cosbeta}
  \end{center}
\end{table}

\subsubsubsection{$\alpha(\phi_2)$}
\label{sucsec:alpha}

\begin{figure}[!h]
\vspace*{-7mm}
\begin{center}
\begin{picture}(350,100)(0,0)
\Oval(30,25)(15,5)(0)
\Oval(130,25)(15,5)(0)
\Oval(130,60)(10,3)(0)
\ArrowLine(130,10)(30,10)
\ArrowLine(30,40)(80,40)
\Vertex(80,40){2}
\ArrowLine(80,40)(130,40)
\DashLine(80,40)(105,60){4}
\Vertex(105,60){2}
\ArrowLine(105,60)(130,70)
\ArrowLine(130,50)(105,60)
\Text(75,50)[]{$\lambda^{3}$}
\Text(100,70)[]{$\lambda^{0}$}
\Text(15,25)[]{$\overline{B}{}^{0}$}
\Text(153,25)[]{$\pi^{+},\rho^{+}$}
\Text(153,60)[]{$\pi^{-},\rho^{-}$}
\Text(35,45)[]{$b$}
\Text(80,3)[]{$\overline{d}$}
\Text(105,45)[]{$u$}
\Text(115,50)[]{$\overline{u}$}
\Text(115,70)[]{$d$}
\Oval(200,40)(30,5)(0)
\Oval(302,22)(12,3)(0)
\Oval(302,58)(12,3)(0)
\ArrowLine(200,10)(302,10)
\ArrowLine(230,70)(200,70)
\Vertex(230,70){2}
\Vertex(272,70){2}
\ArrowLine(302,70)(272,70)
\ArrowArc(251,70)(21,180,0)
\DashLine(272,70)(230,70){4}
\Gluon(252,49)(275,40){-3}{3}
\Vertex(275,40){2}
\ArrowLine(275,40)(302,46)
\ArrowLine(302,34)(275,40)
\Text(230,79)[]{$\lambda^{0}$}
\Text(272,79)[]{$\lambda^{3}$}
\Text(251,57)[]{$t,c,u$}
\Text(250,38)[]{$g$}
\Text(205,75)[]{$b$}
\Text(250,3)[]{$\overline{d}$}
\Text(285,75)[]{$d$}
\Text(288,49)[]{$\overline{u}$}
\Text(288,31)[]{$u$}
\Text(185,40)[]{$\overline{B}{}^{0}$}
\Text(323,24)[]{$\pi^{+},\rho^{+}$}
\Text(323,60)[]{$\pi^{-},\rho^{-}$}
\end{picture}
\end{center}
\vspace*{-5mm}
\caption{The tree (left) and penguin (right) diagrams contributing to 
``charmless" $B$ decays such as $B\to \pi\pi$,
$B\to \rho\rho$ and $B\to \rho\pi.$} 
\label{fig:pipi_diagrams}
\end{figure}
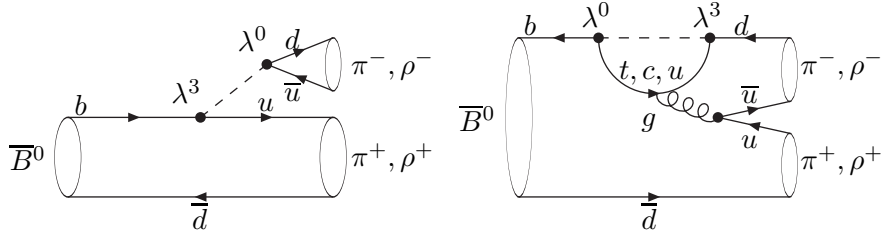

\begin{figure}[htb]
  \centering
  \begin{minipage}[c]{0.55\textwidth}
    \centering

\hspace*{-1.0cm}
\begin{tabular}{cccc}
\hline Decay mode & BR($\times 10^{6}$)& $\;\; S_f \;\;$ &  $C_f$ (or 
$A_{C\!P}$ for $B^+$)  \\
             &    &    &   \\
\hline
$B^0\to \pi^+\pi^-$ & $5.2 \pm 0.2$ & $-0.59\pm 0.09$  & $-0.39\pm 0.07$      \\
$B^+\to \pi^+\pi^0$ & $5.7 \pm 0.4$  & -         & $0.04 \pm 0.05$  \\
$B^0\to \pi^0\pi^0$ & $1.3 \pm 0.2$& -         & $0.36^{+0.33}_{-0.31}$ \\
$B^0\to \rho^+\rho^-$ & $23.1^{+3.2}_{-3.3}$         &          &                 \\
                    &   [$f_L = 0.968 \pm 0.023$]       &  $-0.13\pm 0.19$         & $-0.06\pm 0.14$    \\
$B^+\to \rho^+\rho^0$ & $18.2 \pm 3.0$                             &          &                 \\
                    &   [$f_L = 0.912^{+0.044}_{-0.045}$]       &   -       & $-0.08 \pm 0.10$  \\
$B^0\to \rho^0\rho^0$ & $1.16 \pm 0.46$                    &          &                 \\
                    &   [$f_L = 0.86^{+0.12}_{-0.14}$]  &    -      &   -             \\
\hline
\end{tabular}

  \end{minipage}
  \begin{minipage}[c]{0.35\textwidth}
    \hspace*{2.cm}
    \includegraphics[width=0.86\textwidth]{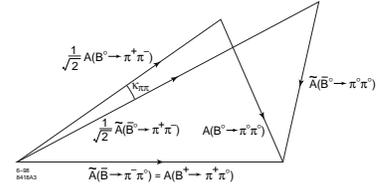}
  \end{minipage}
 \caption{Table: Summary of measured decay properties of the
 $B\to \pi\pi$ and $B\to \rho\rho$ decays that are relevant
 to the determination of the CKM unitarity angle $\alpha$.
 We quote here the averages updated after ICHEP 2006 as given by
 HFAG~\cite{Barberio:2006bi}
 with a total of $882$ million $B\overline{B}$ pairs from BaBar ($347$ million
 events~\cite{Aubert:2006ap})
 and Belle ($535$ million events~\cite{Abe:2006cc}) experiments. Figure:
 Isospin triangles for the $B\to \pi\pi$ system.}
 \label{fig:pipirhorhoresults}
\end{figure}

The CKM unitarity angle $\alpha (=\phi_2)$, defined as
$\alpha = \mbox{arg} \left[ -\frac{V^{ }_{td}V^*_{tb}}{V^{ }_{ud}V^*_{ub}}\right]$, 
is a measure of the relative phase of the CKM elements $V_{ub}$ and $V_{td}$
in the usual parameterization of the CKM unitarity matrix.  
Most of the experimental information on $\alpha$ is extracted from measurements 
of the charmless decays $B\to \pi\pi$, $B\to \rho \pi$ and $B\to \rho \rho$, 
which can arise from the tree-level transition $b\to u(\overline{u} d)$, 
carrying the CKM element $V_{ub}$ (left diagram in Fig.~\ref{fig:pipi_diagrams}). 
In a simple world, where a decay mode such as $B\to \pi^+\pi^-$ is dominated 
by a single tree diagram, one needs only to measure the time-dependent $C\!P$ asymmetry 
$S_{\pi\pi}=\sin2\alpha$.  
However, a complication to this picture arises from the presence of 
loop (penguin) processes (right diagram in Fig.~\ref{fig:pipi_diagrams}), 
involving different CKM matrix elements, but leading to the same final states.
The interference of the two diagrams then obscures the connection 
between the $C\!P$ observables and the angle $\alpha$, requiring
a ``tree and penguin disentanglement" strategy in the experimental program.  
This involves a larger set of experimental observables for the 
determination of the angle $\alpha$ that includes the time-dependent 
$C\!P$ asymmetries $S_f$ and $C_f$ in $B^0$ decays, and the branching 
fractions and direct $C\!P$ asymmetries in both neutral and charged $B$ decays. 
The net effect of the penguin amplitude is to introduce the possibility 
of direct $C\!P$ violation ($C_f\neq 0$) and a nonzero value of 
$\Delta\alpha^f=\alpha^f_{\it eff}-\alpha$, where $\alpha^f_{\it eff}$ is 
determined from the relation $S_f=\sqrt{1-C_f^2}\sin2\alpha^f_{\it eff}$. 
For the $B\to \pi\pi$ decays, the penguin correction $\Delta\alpha^{\pi\pi}$ 
can be determined from an isospin analysis~\cite{Gronau:1990ka} of the 
decay amplitudes of the $B\to\pi\pi$ and $\overline{B} \to \pi\pi$ decays.
(See Fig.~\ref{fig:pipirhorhoresults}.) 
A key element of this analysis is the branching fraction for the decay 
$B\to \pi^0\pi^0$, which is an indicator of the size of the penguin 
effects and consequently of the penguin correction $\Delta \alpha^{\pi\pi}$, 
which is bounded~\cite{Grossman:1997jr} by  
$\sin^2\Delta \alpha^{\pi\pi} < \frac{\overline B(B^0\to \pi^0\pi^0)}{B(B^\pm\to \pi^\pm\pi^0)}$.
Ref.~\cite{Bona:2007qt} proposes to add information on the
hadronic amplitudes to the isospin analysis, for example by using
the branching ratio of $B_s \to K^+K^-$ to constraint the penguin
contribution (even allowing SU(3) breaking effects as large as 100\%).
This would help constraining the
value of $\alpha$, in particular eliminating the solutions at
$\alpha \sim 0$.

A system analogous to that of the $B\to \pi\pi$ decays is the 
family of the $B\to \rho\rho$ decays 
($B^0\to\rho^+\rho^-$, $B^+\to\rho^+\rho^0$, $B^0\to\rho^0\rho^0$ ). 
While in general the $B^0\to \rho\rho$ decays can be a mixture of $C\!P$-even 
and $C\!P$-odd components, the angular analysis of the decay $B^0\to \rho^+\rho^-$ 
(and also $B^+\to \rho^+\rho^0$) has shown that the $C\!P$-even component 
(longitudinal polarization) is dominant, hence significantly simplifying the 
time-dependent $C\!P$ analysis of the process~\cite{Aubert:2005nj,Abe:2005ft}. 
As in the case of $B\to \pi\pi$, time-dependent $C\!P$ asymmetries $S^L_{\rho\rho}$ 
and $C^L_{\rho\rho}$ are used to determine $\alpha^{\rho\rho}_{\it eff}$.  
The branching ratio for $B^0\to \rho^0\rho^0$  relative to 
$B\to \rho^+\rho^-$ and $B\to \rho^+\rho^0$ sets the scale 
of the penguin correction $\Delta\alpha^{\rho\rho} =\alpha^{\rho\rho}_{\it eff}-\alpha$, 
which can be determined from an isospin analysis of the decay amplitudes.  

In Table~\ref{fig:pipirhorhoresults} we present the current 
status of measurements used in the determination of $\alpha$ in 
the $B\to \pi\pi$ and $B\to \rho\rho$ systems~\cite{Barberio:2006bi}. 
Nearly all components of the isospin analysis in the $B\to \pi\pi$ 
system are now measured, albeit with varying degrees of precision. 
Also the current measurements allow for the isospin triangles to
close in both systems~\footnote{This was not the case for the 
$B\to \rho\rho$ system with the pre-2006 measurements.}.

The fact that the branching fraction for the decay $B\to \pi^0\pi^0$ 
is of the same order as the branching fractions for $B^+\to \pi^+\pi^0$ 
and $B^0\to \pi^+\pi^-$ is indicative of significant contributions 
from penguin amplitudes in this channel. 
Currently the $B\to \rho^0\rho^0$ search is giving the first evidence
of a signal (BaBar reporting a $3\sigma$ effect~\cite{Aubert:2006ae}) and
thus a very preliminary measurement of the rate. Still, the major
advantage of the $B\to \rho\rho$ system over the $\pi\pi$ one is clearly
evident from the suppression of $B\to \rho^0\rho^0$ relative to $B\to
\rho^+\rho^-$ and $B\to \rho^+\rho^0$ decays, implying a much smaller
$\Delta \alpha$ correction and smaller related uncertainties from this
source.
The current $\Delta \alpha$ correction upper limits are
$\Delta \alpha_{\pi\pi}< 41^\circ$ at $90\%$ C.L. from BaBar and 
$\Delta \alpha_{\rho\rho}< 21^\circ$ at $90\%$ C.L. from BaBar.

One other advantage of the $\rho\rho$ system is that,
in contrast to $\pi^0\pi^0$, a time dependent $C\!P$-asymmetry
analysis of the $\rho^0\rho^0$ final state will be possible
as soon as enough statistics are available.  This
feature will allow both $S^{00}$ and $C^{00}$ to be accessed.
From a feasibility study we can foresee for the $\rm 2 \, ab^{-1}$
scenario an error of 0.3 on $S^{00}$ and 0.25 on $C^{00}$.
This information will greatly help in reducing the
ambiguities in the $\alpha$ extraction from this system.

The $B\to \rho \pi$ system presents a special case with the possibility of 
additional handles: the final states $\rho^+\pi^-$ and $\rho^-\pi^+$, which can be
reached by both $B^0$ and $\overline{B}{}^0$, have substantial overlap in the Dalitz plot;
thus their amplitudes interfere and generate additional
dependence on $\alpha$ and the strong phases of the final states. 
Quinn and Snyder~\cite{Snyder:1993mx} have shown that the interference 
effect can be exploited to extract the angle $\alpha$ even in the presence of penguins. 
This involves the amplitude analysis of the $3\pi$ Dalitz distribution.

The $\rho^{\pm}\pi^{\mp}$ final states are not $C\!P$ eigenstates, and four
flavour-charge configurations $(B^0(\overline{B}{}^0) \to \rho^{\pm}\pi^{\mp})$ must be considered.
Both experiments assume that the amplitudes corresponding to these final states
are dominated by the three resonances $\rho^+$, $\rho^-$ and $\rho^0$. The $\rho$ resonances
are assumed to be the sum of the ground state $\rho(770)$ and the radial excitations
$\rho(1450)$ and $\rho(1700)$. Possible contributions to the $B^0\to\pi^+\pi^-\pi^0$
decay other than the $\rho$'s are studied as part of the systematic uncertainties.
The time-dependent analyses use a general parameterization~\footnote{See for details
Refs.~\cite{Aubert:2006fg},~\cite{Abe:2006yg} and~\cite{Barberio:2006bi}.}
that allows to describe the differential decay width as a linear combination
of independent functions, whose coefficients are the 26 free parameters of the fit.

\begin{table}
    \caption{Summary of measured $C\!P$-asymmetry parameters of the 
      $\rho\pi$ system following the convention used in~\cite{Aubert:2004iu}.
      We quote here the averages updated after ICHEP 2006 as given by
      the HFAG~\cite{Barberio:2006bi}
      with a total of $796$ million $B\overline{B}$ pairs from BaBar
      ($347$ million events~\cite{Aubert:2006fg})
      and Belle ($449$ million events~\cite{Abe:2006yg}) experiments.}
  \begin{center}
    \begin{tabular}{ccccc}
      \hline
        \multicolumn{5}{c}{$\rho^{\pm}\pi^{\mp}$ Q2B/Dalitz plot analysis} \\
        \hline
        $S_{\rho\pi}$ & $C_{\rho\pi}$ & $\Delta S_{\rho\pi}$ & 
	$\Delta C_{\rho\pi}$ & ${\cal A}_{C\!P}^{\rho\pi}$ \\
        $ 0.03 \pm 0.09\;$ & $\; 0.03 \pm 0.07\;$ &
        $\;-0.02 \pm 0.10\;$ & $\; 0.36 \pm 0.07\;$ & $\; -0.13 \pm 0.03\;$ \\
        \hline
        & ${\cal A}^{+-}_{\rho\pi}$ & & ${\cal A}^{-+}_{\rho\pi}$ & \\
        & $ 0.11 \pm 0.06$ & & $-0.19 \pm 0.13$ & \\
	\hline
      \end{tabular}
    \label{tab:rhopiresults}
  \end{center}
\end{table}

From the bilinear coefficients, both experiments extract the quasi-two-body (Q2B)
parameters. Considering only the charged bands in the Dalitz plot, the Q2B analysis
involves 5 different parameters
$S_{\rho\pi}$, $C_{\rho\pi}$, $\Delta S_{\rho\pi}$, $\Delta C_{\rho\pi}$ and
${\cal A}_{C\!P}^{\rho\pi}$.
The first two parameterize mixing-induced $C\!P$ violation related to the angle $\alpha$
and flavour-dependent direct $C\!P$ violation, respectively.
The second two are insensitive to $C\!P$ violation: $\Delta S_{\rho\pi}$ is related to
the strong-phase difference between the amplitudes contributing to $B^0 \to \rho\pi$
decays, and $\Delta C_{\rho\pi}$ describes the asymmetry between the rates
$\Gamma(B^0\to\rho^+\pi^-)+\Gamma(\overline{B}{}^0\to\rho^-\pi^+)$ and 
$\Gamma(B^0\to\rho^-\pi^+)+\Gamma(\overline{B}{}^0\to\rho^+\pi^-)$.
Finally, ${\cal A}_{C\!P}^{\rho\pi}$ is the time-independent charge asymmetry.
$C\!P$ symmetry is violated if either one of the following conditions is true:
${\cal A}_{C\!P}^{\rho\pi} \neq 0$, $C_{\rho\pi} \neq 0$ or $S_{\rho\pi} \neq 0$.
The first two correspond to $C\!P$ violation in the decay, while the last condition
is $C\!P$ violation in the interference of decay amplitudes with and without $B^0$
mixing.
In Table~\ref{tab:rhopiresults}, we report the HFAG averages of the Q2B
parameters provided by the experiments, which should be equivalent to determining
average values directly from the averaged bilinear coefficients.
One can transform the experimentally motivated $C\!P$ parameters ${\cal A}_{C\!P}^{\rho\pi}$
and $C_{\rho\pi}$ into the direct $C\!P$ violation parameters
$\mathcal{A}_{\rho \pi}^{+-}$ and $\mathcal{A}_{\rho \pi}^{-+}$
defined in~\cite{Aubert:2004iu}.
${\cal A}_{\rho \pi}^{-+}$ (${\cal A}_{\rho \pi}^{+-}$) describes $C\!P$ violation
in $B^0$ decays where the $\rho$ is emitted (not emitted) by the spectator interaction.
Both experiments obtain values for ${\cal A}_{\rho \pi}^{-+}$ and ${\cal A}_{\rho \pi}^{+-}$ which are averaged in the
Table~\ref{tab:rhopiresults}.
In addition to the $B^0 \to \rho^{\pm} \pi^{\mp}$ Q2B contributions
to the $\pi^+\pi^-\pi^0$ final state, there can also be a $B^0 \to \rho^0 \pi^0$
component. 
Belle and BaBar have extracted the Q2B parameters associated
with this intermediate state which average to:  $S_{\rho^0\pi^0} = 0.30 \pm 0.38 $ and
$C_{\rho^0\pi^0} = 0.12 \pm 38 $ (HFAG Summer 2007).

\begin{figure}
\center
\hspace*{-0.5cm}
\includegraphics[width=5.0truecm]{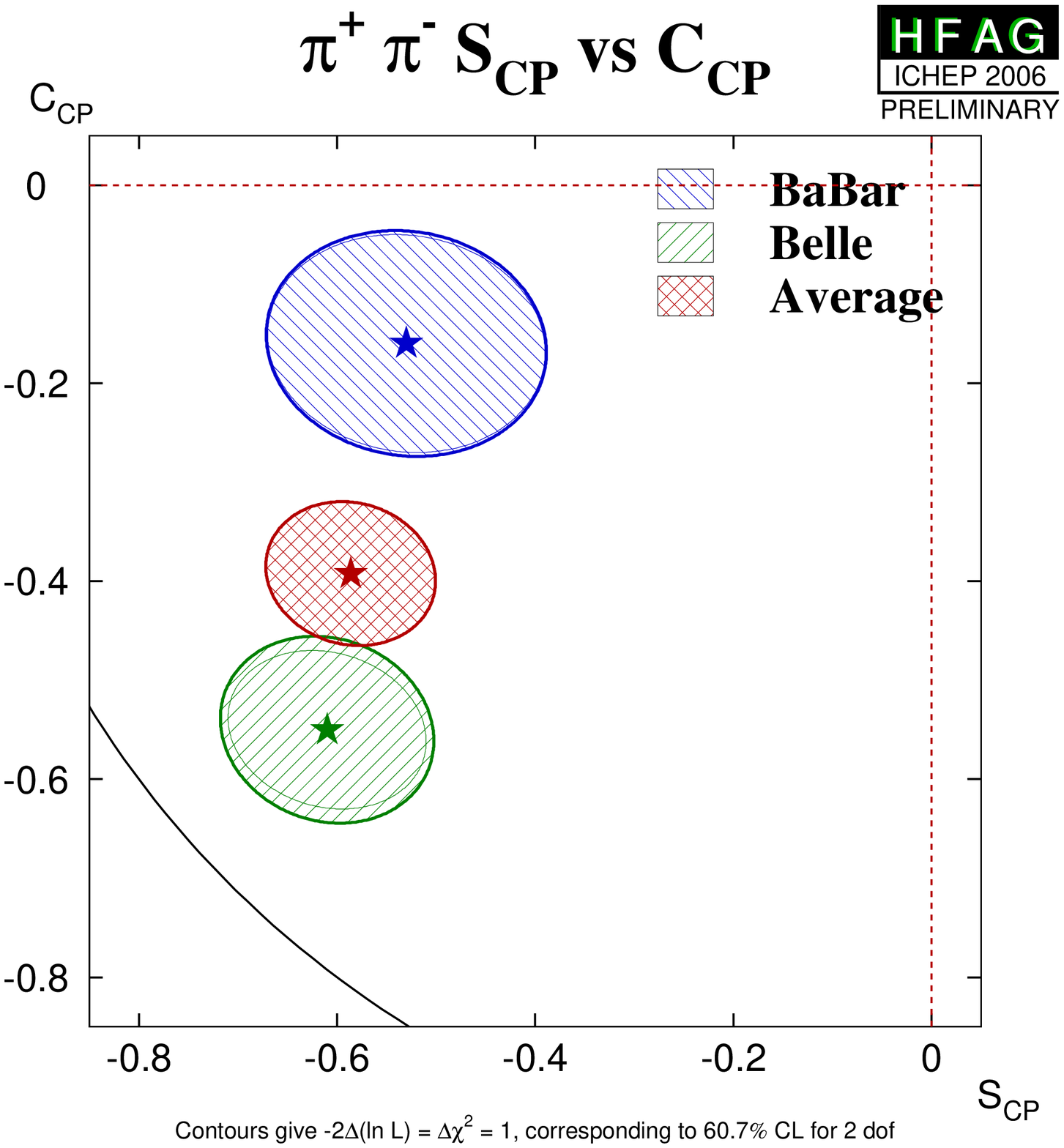}
\includegraphics[width=5.0truecm]{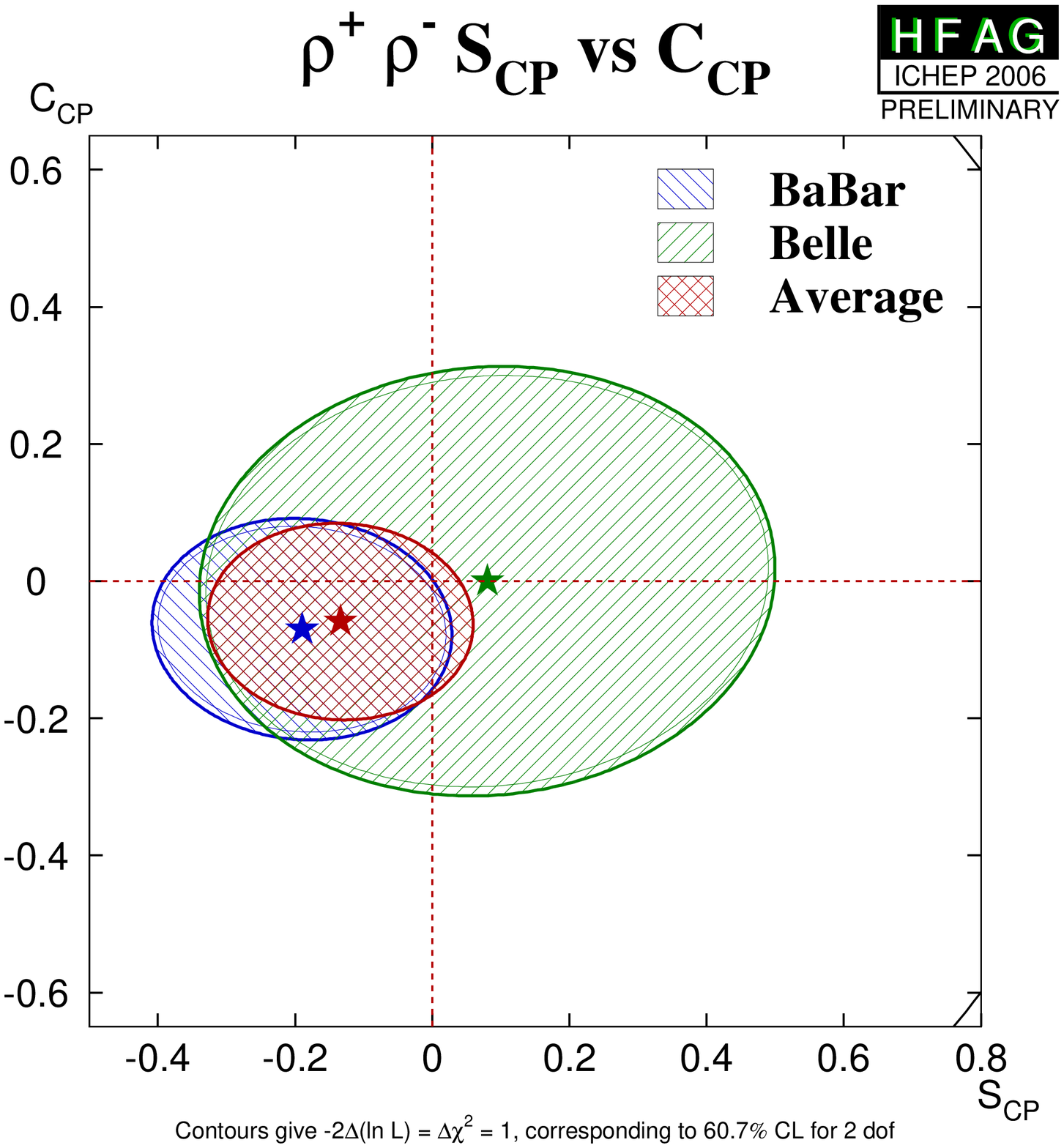}
\includegraphics[width=5.0truecm]{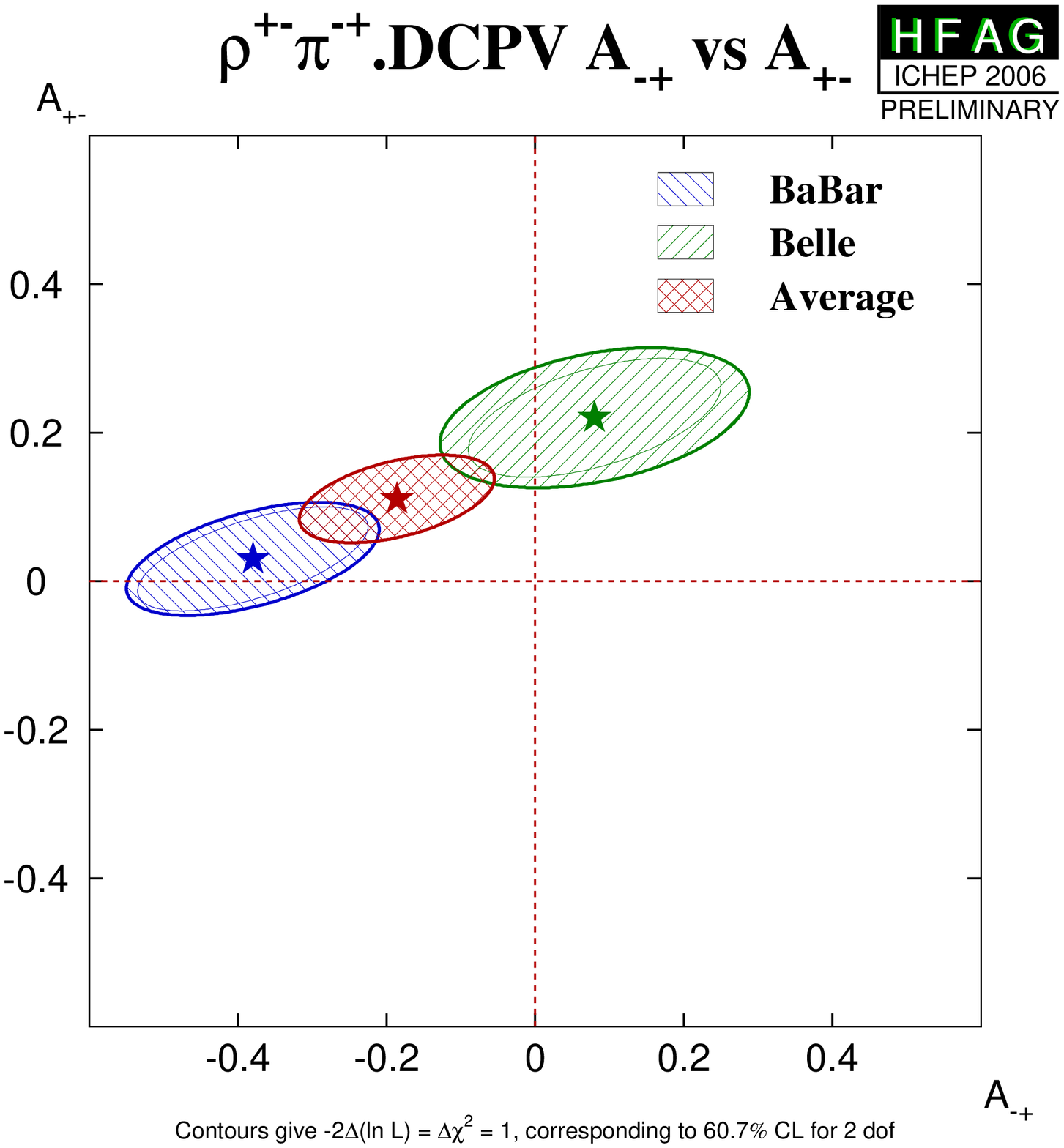}
\caption{The experimental results on the $C\!P$ asymmetry parameters in the $\pi\pi$
(left), $\rho\rho$ (center) and $\rho\pi$ (right) systems, as summarized by
HFAG~\cite{Barberio:2006bi}.} 
\label{fig:SvsC}
\end{figure}

In Fig.~\ref{fig:SvsC}, the plots of the averages and the separate results 
on the various $C\!P$-violating parameters are shown:
it can be seen that the two collaborations, BaBar and Belle, are still
discrepant at the level of $2 \sigma$ ($1.5\sigma$) in the $B\to \pi^+\pi^-$ 
($\B\to \rho^{\pm}\pi^{\pm}$) system.
In the $\rho\rho$ system, though, some updates
to the entire currently available statistics are still missing.

We can get an estimate of the current experimental value of $\alpha$ putting together
all the analyses in all the modes. The results on the Standard Model (SM) solution
from the two fitting groups are:
$(92 \pm 7) ^\circ$ for the bayesian approach~\cite{Bona:2005vz} and 
$(93^{+11}_{-9})^\circ$ for the frequentist approach~\cite{Charles:2004jd}.
From the same analyses we can also extract the SM $\alpha$ values using
the UT fit constraints and without using the $\alpha$ information:
$(93 \pm 6) ^\circ$ for the bayesian approach and 
$(98^{+5}_{-19})^\circ$ for the frequentist one.
We can remark how the current values are in very good agreement
with the expected SM values.


\subsubsubsection{$\gamma (\phi_3)$}
\label{sucsec:gamma}


{\bf Measurement of $\gamma$ from $B$ decays to open charm}

\begin{figure}[htb]
\vspace*{-7mm}
\begin{center}
\begin{picture}(350,100)(0,0)
\Oval(50,25)(15,5)(0)
\Oval(150,25)(15,5)(0)
\Oval(150,60)(10,3)(0)
\ArrowLine(50,10)(150,10)
\ArrowLine(100,40)(50,40)
\Vertex(100,40){2}
\ArrowLine(150,40)(100,40)
\DashLine(100,40)(125,60){4}
\Vertex(125,60){2}
\ArrowLine(125,60)(150,70)
\ArrowLine(150,50)(125,60)
\Text(95,50)[]{$V_{cb}^{*}$}
\Text(120,70)[]{$V_{us}$}
\Text(35,25)[]{$B^{+}$}
\Text(165,25)[]{$\overline{D}{}^{0}$}
\Text(165,60)[]{$K^{+}$}
\Text(55,45)[]{$\overline{b}$}
\Text(100,5)[]{$u$}
\Text(125,45)[]{$\overline{c}$}
\Text(135,50)[]{$\overline{s}$}
\Text(135,70)[]{$u$}
\Oval(200,25)(15,5)(0)
\Oval(300,20)(10,3)(0)
\Oval(300,57.5)(12.5,3)(0)
\ArrowLine(200,10)(300,10)
\ArrowLine(250,40)(200,40)
\Vertex(250,40){2}
\ArrowLine(300,70)(250,40)
\DashLine(250,40)(270,30){4}
\Vertex(270,30){2}
\ArrowLine(270,30)(300,45)
\ArrowLine(300,30)(270,30)
\Text(245,50)[]{$V_{ub}^{*}$}
\Text(260,25)[]{$V_{cs}$}
\Text(205,45)[]{$\overline{b}$}
\Text(250,5)[]{$u$}
\Text(275,65)[]{$\overline{u}$}
\Text(290,35)[]{$c$}
\Text(280,25)[]{$\overline{s}$}
\Text(185,25)[]{$B^{+}$}
\Text(315,20)[]{$K^{+}$}
\Text(315,60)[]{$D^{0}$}
\end{picture}
\end{center}
\vspace*{-5mm}
\caption{Feynman diagram of the $B^{+}\rightarrow
\overline{D}{}^{0}K^{+}$ and $B^{+}\rightarrow D^{0}K^{+}$ decays.
\label{fig:feyDK}}
\end{figure}
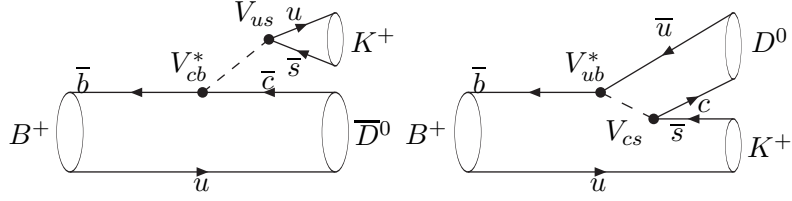

The possibility of observing direct $C\!P$ violation in $B \to DK$
decays was first discussed by I.Bigi, A.Carter and
A.Sanda~\cite{Bigi:1988ym,Carter:1980hr}. Since then, various methods
to measure the weak angle $\gamma$ (= $\phi_3$) using $B \to DK$
decays have been proposed. All these methods are based on two key
observations: neutral $D^0$ and $\overline{D}{}^0$ mesons can decay to
a common final state, and the decay $B^+ \to DK^+$ can produce neutral
$D$ mesons of both flavours via $\overline{b} \to \overline{c} u
\overline{s}$ and $\overline{b} \to \overline{u} c \overline{s}$
transitions (Fig.~\ref{fig:feyDK}), with a relative phase $\theta_{+}$
between interfering amplitudes that is the sum, $\delta_B + \gamma$,
of strong and weak interaction phases. For the decay, $B^- \to DK^-$,
the relative phase is $\theta_{-} = \delta_B - \gamma$, so both
$\delta_B$ and $\gamma$ can be extracted from measurements of such
charge conjugate $B$ decay modes. The feasibility of the $\gamma$
measurement crucially depends on the size of $r_B$, the ratio of the
$B$ decay amplitudes involved ($r_B = |A(B^+ \to D K^+)|/|A(B^+ \to
\overline{D} K^+)|$).  The value of $r_B$ is given by the ratio of the
CKM matrix elements $|V_{ub}^{*} V_{cs}^{}|/|V_{cb}^{*} V_{us}^{}|$
and the colour suppression factor, and is estimated to be in the range
0.1-0.2~\cite{Gronau:2002mu}.  These methods are theoretically clean because
the main contributions come from tree-level diagrams
(Fig.~\ref{fig:feyDK})~\footnote{$D$-$\overline{D}$ mixing is
  neglected in the current analyses. This effect can be included
  though~\cite{Atwood:2003jb} and is shown to be very small within the
  SM~\cite{Grossman:2005rp}.}.
Various methods have been proposed to exploit this strategy using
different combinations of final states. These approaches include using
the branching ratios of decays to $C\!P$ eigenstates (GLW
method~\cite{Gronau:1990ra,Gronau:1991dp,Gronau:1998vg}) or using
doubly Cabibbo suppressed $D$ modes (ADS method~\cite{Atwood:1996ci}).  A
Dalitz plot analysis of a three-body final state of the $D$ meson
allows one to obtain all the information required for the
determination of $\gamma$ in a single decay
mode~\cite{Atwood:2000ck,Giri:2003ty,bondar}.  Three-body final states such as $K^0_S
\pi^+ \pi^-$~\cite{Giri:2003ty,bondar} have been suggested as promising modes
and give today the best estimate of the angle $\gamma$.

%
\begin{figure*}[htb]
\begin{center}
\begin{tabular}{cc}
\includegraphics[width=15.5pc]{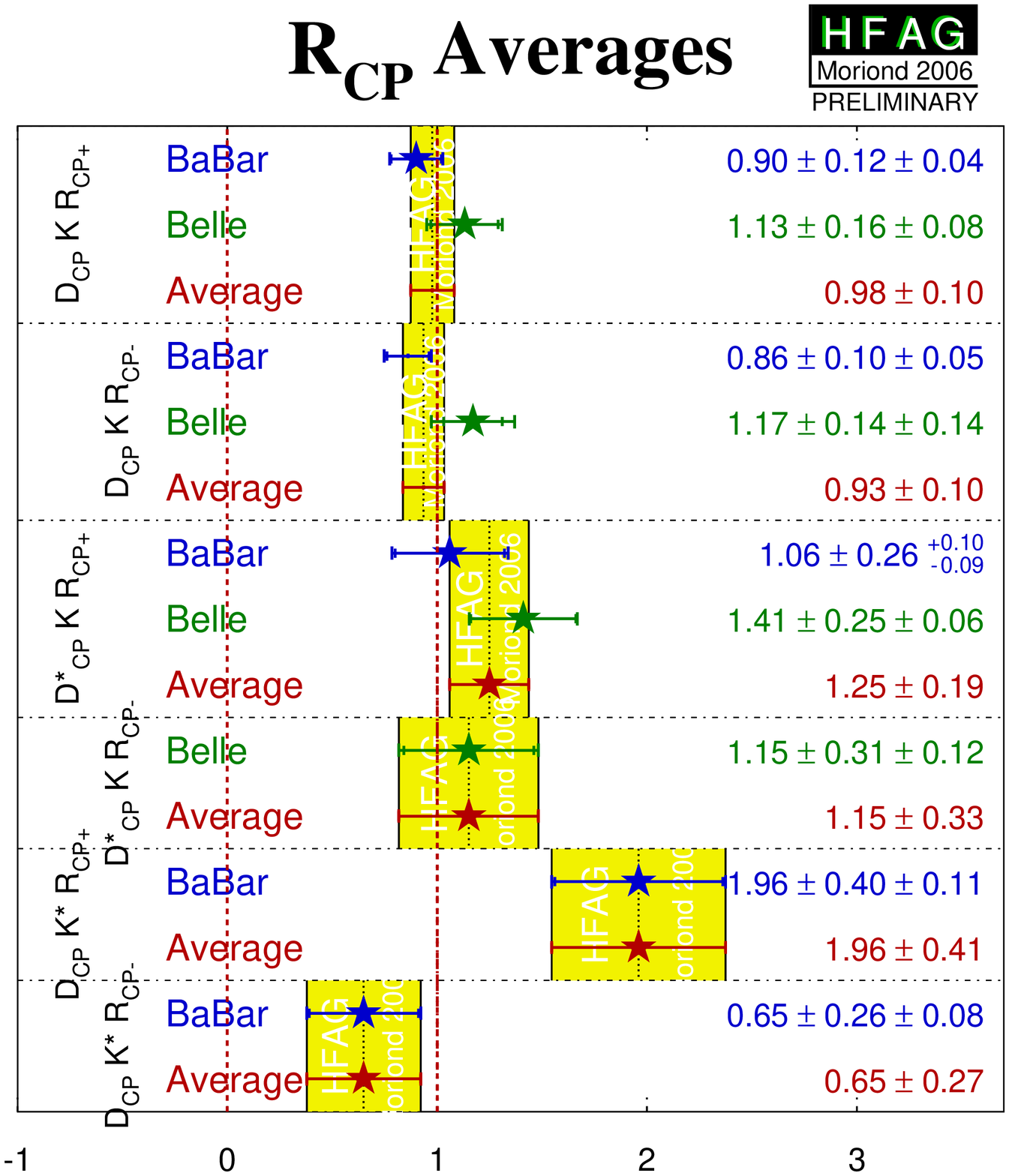}& 
\includegraphics[width=15.5pc]{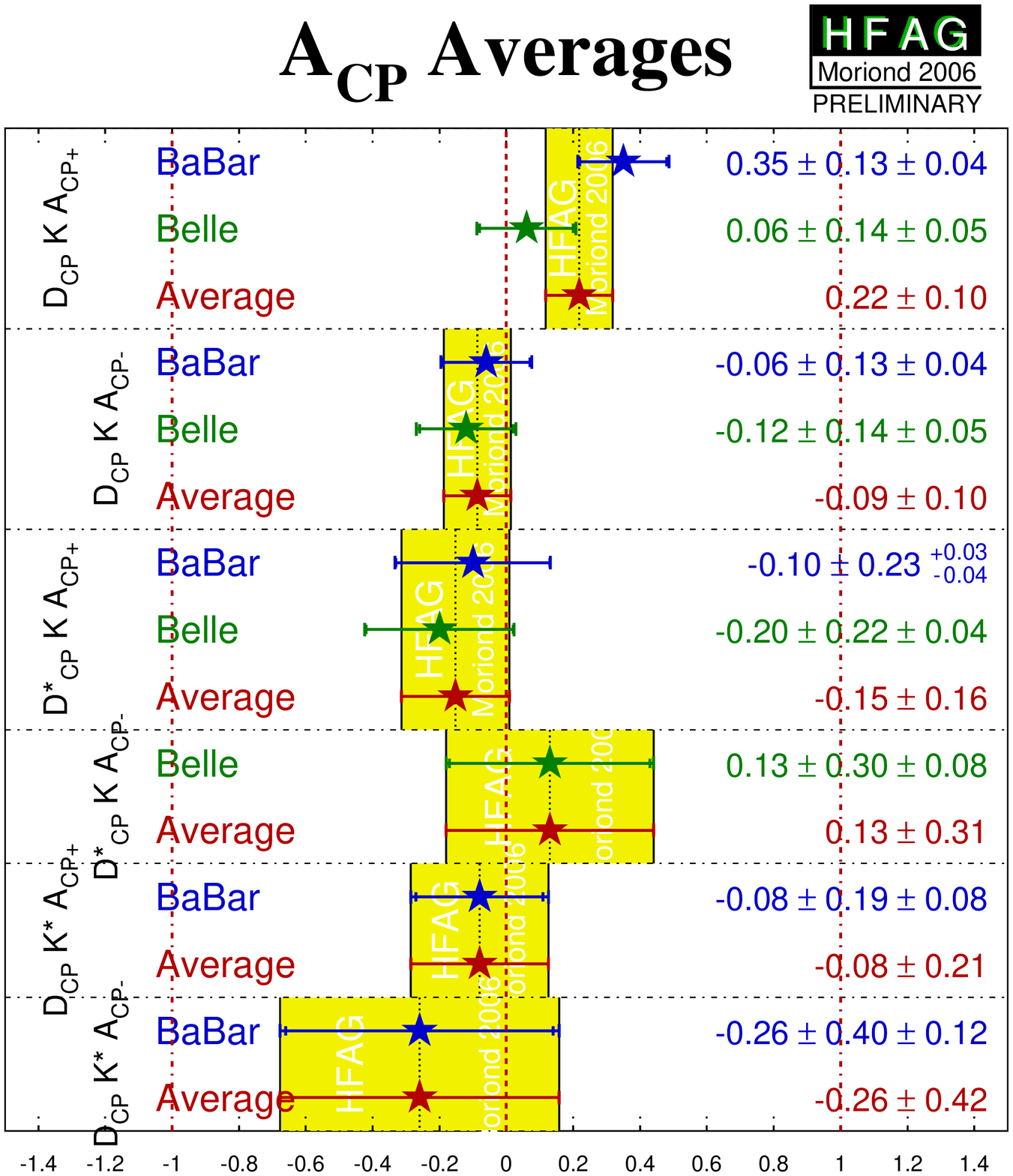}
\end{tabular}
\end{center}
\caption{$R_{C\!P_{\pm}}$ and $A_{C\!P_{\pm}}$ averages obtained by the 
$B$ factories~\cite{Barberio:2007cr}.}
\label{fig:glw_results_bfactory}
\end{figure*}
\begin{figure*}[htb]
\begin{center}
\begin{tabular}{c}
\includegraphics[width=15.5pc]{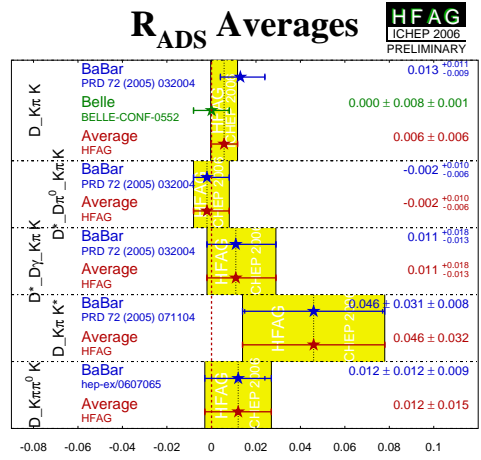} 
\end{tabular}
\end{center}
\caption{$R_{ADS}$ averages obtained by the 
$B$ factories~\cite{Barberio:2007cr}.}
\label{fig:ads_results_bfactory}
\end{figure*}

In the GLW method, the $D$ is reconstructed through its decay to
$C\!P$ eigenstates. The experimental observables are 
the ratio of charge averaged partial rates, $R_{C\!P_{\pm}}$, and 
the charge asymmetry, $A_{C\!P_{\pm}}$ which
are related to the model parameters
through the relations $R_{C\!P_{\pm}} = 1 + r_B^2 \pm 2r_B \cos \delta_B
\cos \gamma$ and $A_{C\!P_{\pm}} = \pm 2 r_B \sin \delta_B \sin \gamma / 
R_{C\!P_{\pm}}$.
$C\!P_+$ refers to the $C\!P$-even final states, $\pi^+ \pi^-$ and 
$K^+ K^-$, and $C\!P_-$ refers to the $C\!P$-odd final states, 
$K_S^0 \pi^0$, $K_S^0 \phi$, $K_S^0 \omega$...
Results are available from both BaBar and Belle
in the decay modes $B^\pm \to DK^\pm, B^\pm \to D^*K^\pm$ 
and $B^\pm \to DK^{*\pm}$ (Fig.~\ref{fig:glw_results_bfactory}). 
The errors for $R_{C\!P_{\pm}}$ and $A_{C\!P_{\pm}}$
are typically 10\% for the most promising mode, $B^\pm \to DK^\pm$.
A 3$\sigma$ significance for the charge asymmetry of the $B \to DK$ mode
seems to be within reach in the near future, when 
1~$\rm ab^{-1}$ of data will be collected by each experiment.
For the ADS method, using a suppressed $D \to f$ decay
($D^0 \to K^+ \pi^-$, $K^+ \rho^-$, $K^* \pi^-$...),
the measured quantities are the partial rate asymmetry, $A_{ADS}$, 
and the charge averaged rate, 
$R_{ADS} = \Gamma (B^- \to [f]_D K^-)/\Gamma (B^- \to [\overline{f}]_D K^-)$.
$R_{ADS}$ is related to the physical parameters by the expression 
$r_B^2 + r_D^2 + 2 r_B r_D \cos (\delta_B + \delta_D) \cos \gamma$.
The overall effective
branching ratio is expected to be small ($\sim 10^{-7}$), but the
two interfering diagrams are of the same order of magnitude and
large asymmetries are therefore expected. 
The method has four unknowns: $\gamma$, $r_B$, $\delta_B+\delta_D$ and 
the amplitude ratio $r_D$. However, the value of $r_D$ can be measured
using decays of $D$ mesons of known flavour. 
If one wants to use the ADS method alone, two modes need to be used. 
Of course, one can also combine one ADS mode (as an example) 
with one GLW $C\!P$ eigenstate.
No significant signal has been yet observed for the ADS modes 
at the $B$ factories so 
only $R_{ADS}$ has been measured so far for the 
$D^{(*)}K^{(*)}$ modes (Fig.~\ref{fig:ads_results_bfactory}). 
These measurements will bring soon valuable constraints on $r_B$.

In the Dalitz method, $D^0$ and $\overline{D}{}^0$ mesons decay into 
the same final state $K^0_S \pi^+ \pi^-$~\cite{Giri:2003ty,bondar} 
(or $K^+ \pi^- \pi^0$~\cite{Atwood:2000ck}). 
Assuming no $C\!P$ asymmetry in neutral $D$ decays, 
the amplitude of decay as a function of Dalitz plot variables 
$m^2_+$ = $m^2_{K^0_S \pi^+}$ and $m^2_-$ = $m^2_{K^0_S \pi^-}$ is 
$M_{\pm} = f(m^2_{\pm},m^2_{\mp}) + 
r_B e^{\pm i \gamma + i \delta_B} f(m^2_{\mp},m^2_{\pm})$,
where $f(m^2_{+},m^2_{-})$ is the amplitude of the 
$\overline{D}{}^0 \to K^0_S \pi^+ \pi^-$ decay.
The method has a second ambiguous solution: ($\gamma + 180^{\circ}, 
\delta_B + 180^{\circ}$), since this transformation does not change
the sum or difference of phases that are actually measured. \\
Results from the two $B$ factories Belle and BaBar are available.
The Belle collaboration uses a data sample 
of $386 \times 10^6 B\overline{B}$ pairs~\cite{Poluektov:2006ia} where the
reconstructed states are $B^+ \to D K^+$,
$B^+ \to D^* K^+$ with $D^* \to D\pi^0$ and $B^+ \to D K^{*+}$ with 
$K^{*+} \to K^0_S \pi^+$. Analysis by the
BaBar collaboration~\cite{Aubert:2006am} is based on $347 \times 10^6 B\overline{B}$ 
pairs using $B^+ \to D K^+$ and 
$B^+ \to D^* K^+$ with two $D^*$ channels: $D^* \to D\pi^0$ and 
$D^* \to D\gamma$ (the previous BaBar~\cite{Aubert:2005iz} publication 
includes also the $B^+ \to D K^{*+}$ channel but this mode is not included 
in the recent update). 
The number of reconstructed signal events in the Belle's data are 
$331 \pm 23$, $81 \pm 11$ and $54 \pm 8$ for the $B^+ \to D K^+$,
$B^+ \to D^* K^+$ and $B^+ \to D K^{*+}$ channels, respectively. BaBar
finds $398 \pm 23$, $97 \pm 13$ and $93 \pm 12$ signal events in the 
$B^+ \to D K^+$, $B^+ \to D^*[D\pi^0] K^+$ and $B^+ \to D^*[D\gamma] K^+$ 
channels respectively.
The amplitude $f$ is parametrized as a coherent sum of two-body
decay amplitudes (16 for BaBar, 18 for Belle) plus a non-resonant decay amplitude
and is determined directly in data from a large and clean sample of flavour-tagged
decays produced in continuum $e^+ e^-$ annihilation. 
For example, Belle includes five Cabibbo-allowed amplitudes: 
$K^{*}(892)^{+}\pi^-$, $K^{*}(1410)^{+}\pi^-$,
$K^{*}_0(1430)^{+}\pi^-$, $K^{*}_2(1430)^{+}\pi^-$ and $K^*(1680)^+\pi^-$, 
their doubly Cabibbo-suppressed partners, and eight channels with a $K^0_S$ 
and a $\pi\pi$ resonance: $\rho$, $\omega$, $f_0(980)$, $f_2(1270)$, 
$f_0(1370)$, $\rho(1450)$, $\sigma_1$ and $\sigma_2$ . 
The parameters of the $\sigma$ resonances obtained in the fit are 
$M_{\sigma_1} = 519 \pm 6$ MeV/$c^2$, $\Gamma_{\sigma_1} = 
454 \pm 12$ MeV/$c^2$, $M_{\sigma_2} = 1050 \pm 8$ MeV/$c^2$ and  
$\Gamma_{\sigma_2} = 101 \pm 7$ MeV/$c^2$ (the errors are statistical only), 
while the parameters of the other resonances are taken to be the same as in 
the CLEO analysis~\cite{Muramatsu:2002jp}. 
The agreement between the data and the fit result is satisfactory 
for the purpose of measuring $\gamma$ and the discrepancy is taken into 
account in the model uncertainty.

\begin{figure}[htb]
\begin{center}
\includegraphics[width=4.75truecm]{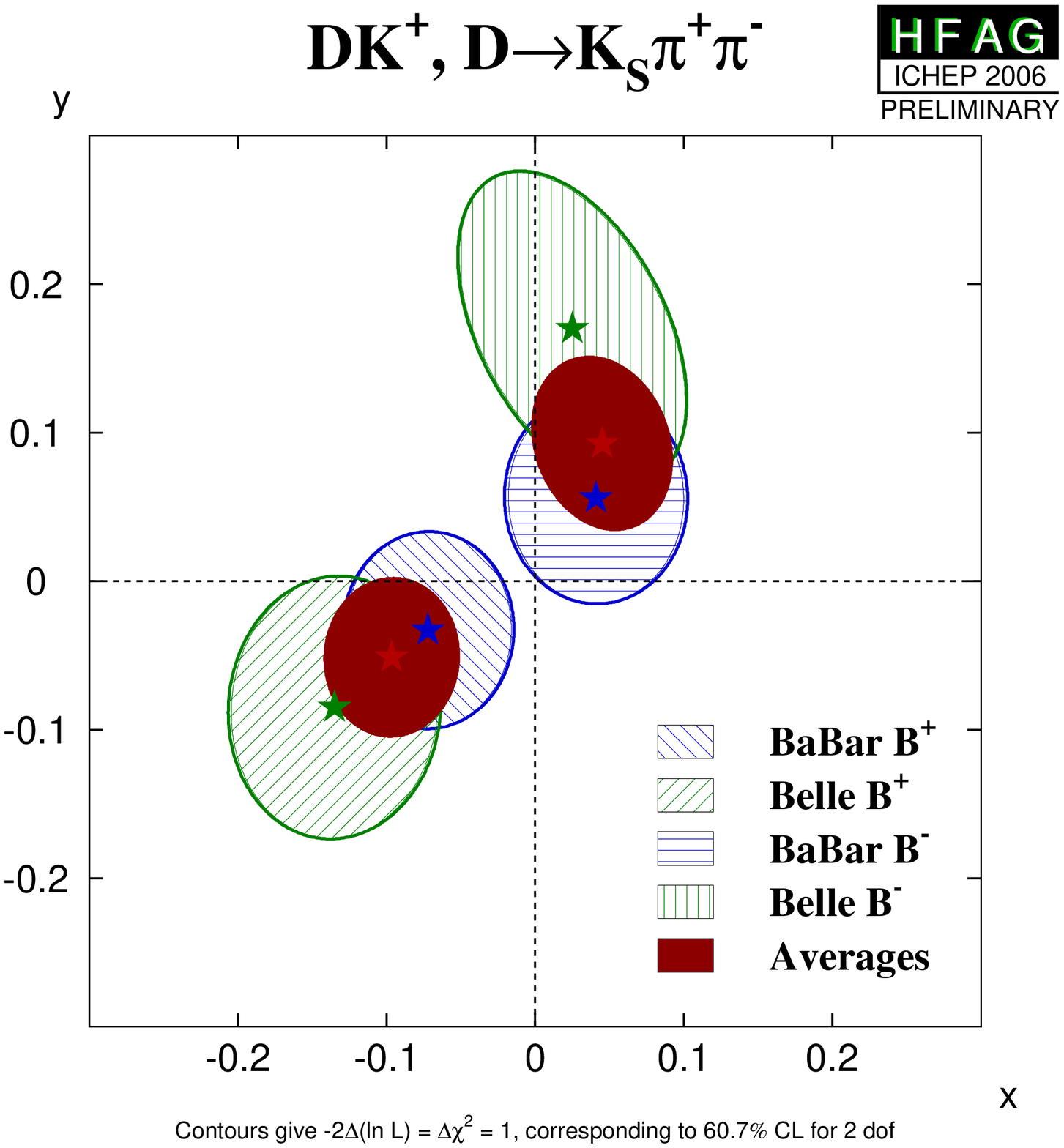}
\includegraphics[width=4.75truecm]{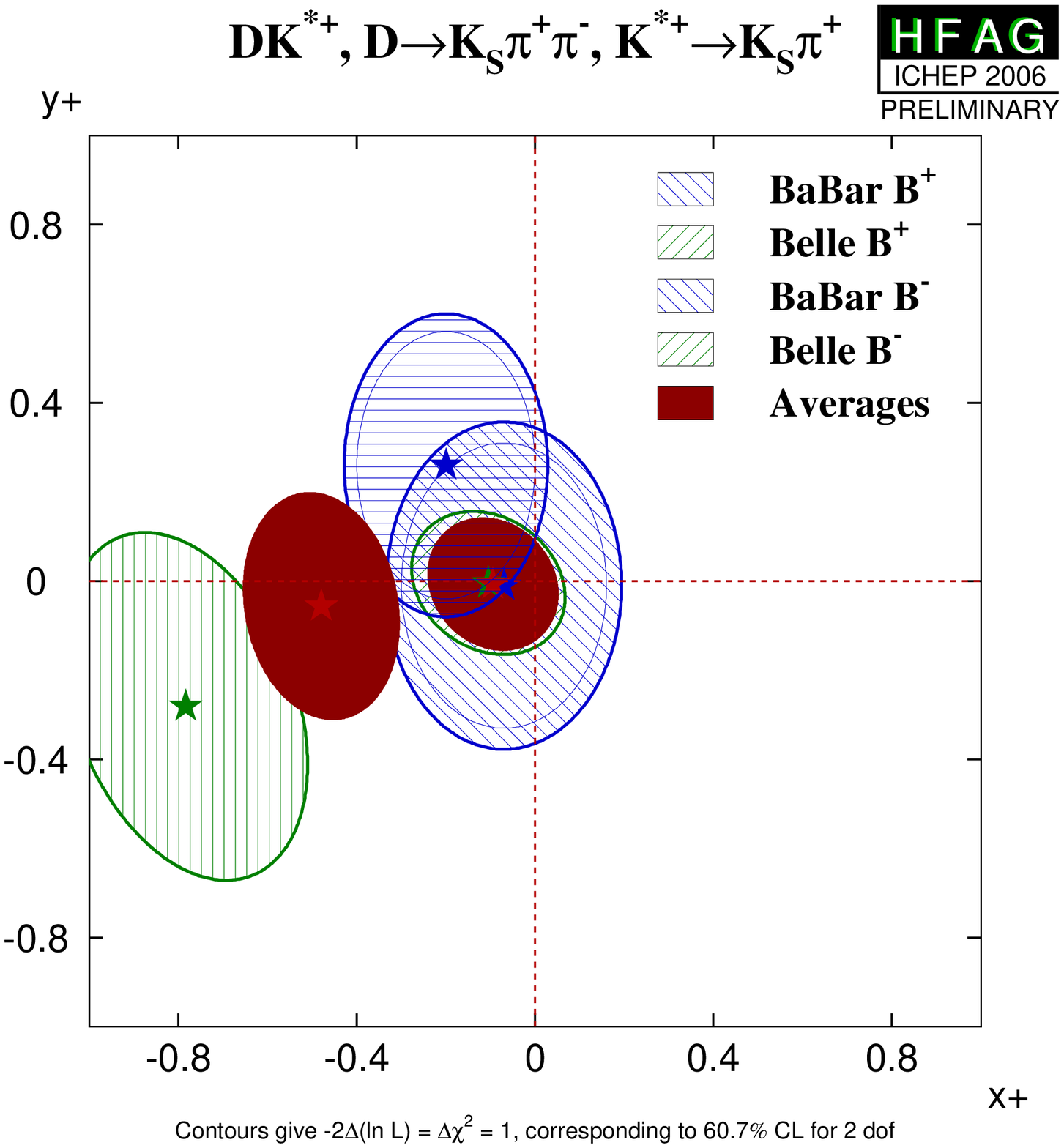}
\includegraphics[width=4.75truecm]{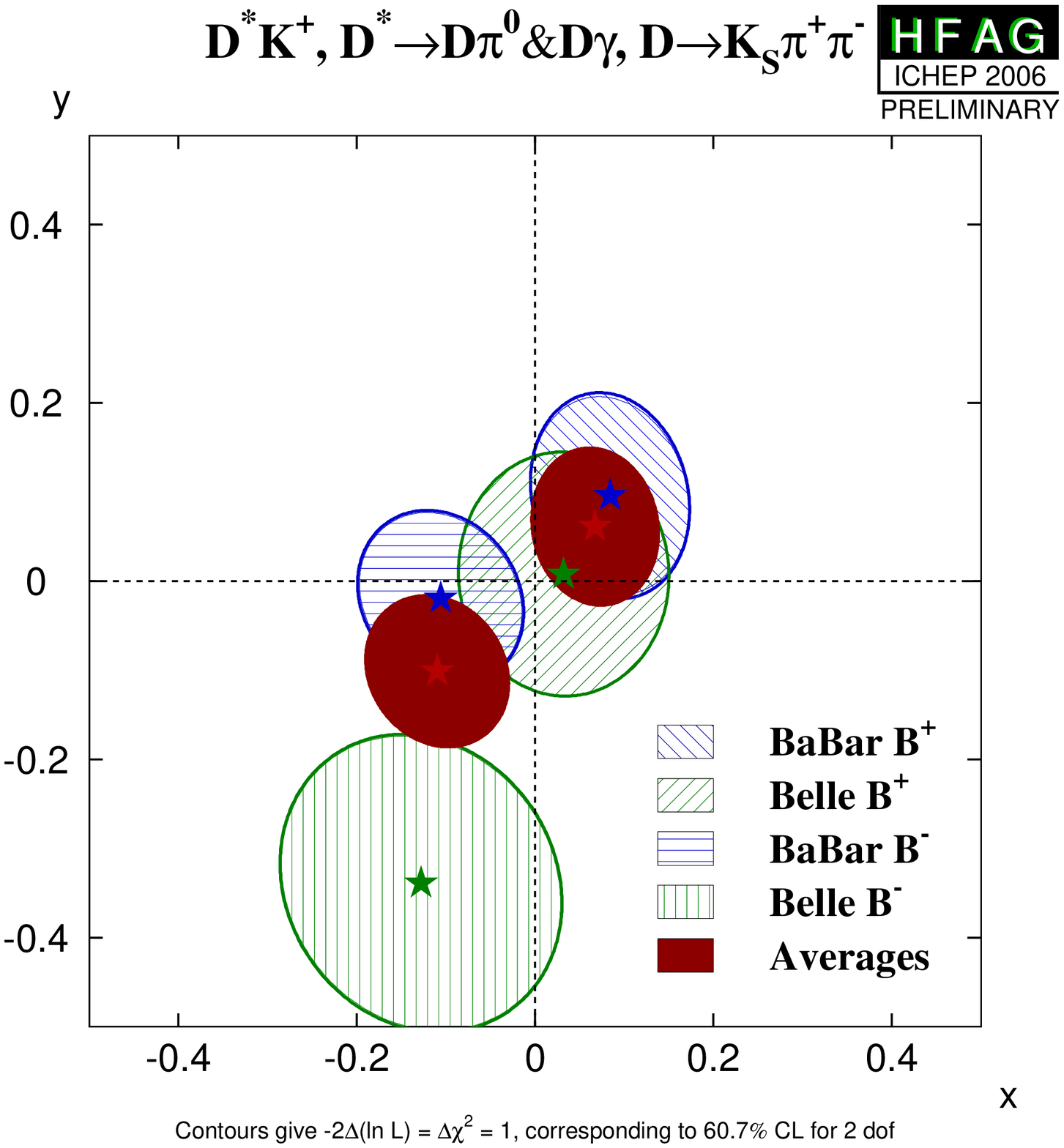}
\end{center}
\caption{Results of signal fits with free parameters $x_{\pm} = 
r \cos \theta_{\pm}$ and $y_{\pm} = r \sin \theta_{\pm}$ for 
$B^\pm \to D K^\pm$, $D^* K^\pm$ and $D K^{*\pm}$ modes from 
the BaBar and Belle latest publications~\cite{Poluektov:2006ia,Aubert:2006am}. 
The contours indicate one standard deviation.}
\label{fig:xy_dalitz}
\end{figure}

Once $f$ is determined, a fit to $B^{\pm}$ data is performed to obtain
the Cartesian parameters, $x_{\pm} = r_{\pm} \cos (\pm \gamma + \delta_B)$ 
and $y_{\pm} = r_{\pm} \sin (\pm \gamma + \delta_B)$, which have the
advantage to be Gaussian-distributed, uncorrelated and unbiased ($r_B$ 
is positive definite and hence exhibits a fit bias toward larger values
when its central value is in the vicinity of zero) and simplify the 
averaging of the various measurements.
Figure~\ref{fig:xy_dalitz} shows the results of the separate $B^+$ and
$B^-$ data fits for $B \to D K$, $D^* K$ and $D K^{*}$ modes in 
the $x-y$ plane for the BaBar and Belle collaborations. Confidence intervals 
were then calculated by each experiment using a frequentist technique
(the so-called Neyman ordering in the BaBar case, the Feldman and Cousins 
ordering~\cite{Feldman:1997qc} in the Belle case). The central values
for the parameters $\gamma$, $r_B$ and $\delta_B$ from the combined fit 
(using the $(x_{\pm}, y_{\pm})$ obtained for all modes) with their one 
standard deviation intervals are presented in Table~\ref{tab:res_dp_gamma}.
Note that there are large correlations
between the fit parameters $\gamma$ and $r_B$. 
With the available data 
the statistical error on $\gamma$ increases with decreasing $r_B$ and thus
it depends strongly on the central value
of $r_B$ as determined by the fit.
The uncertainties in the model used to parametrize the 
$\overline{D}{}^0 \to K^0_S \pi^+ \pi^-$ decay amplitude lead to an
associated systematic error in the fit result.
These uncertainties arise from the fact that there is no unique
choice for the set of quasi-2-body channels in the decay, as 
well as the various possible parameterizations 
of certain components, such as the non-resonant amplitude. 
To evaluate this uncertainty several alternative models
have been used to fit the data. 
\begin{table*}[htb]
\caption{Results of the combination of $B^+ \to D K^+$, $B^+ \to D^* K^+$,
and $B^+ \to D K^{*+}$ modes for BaBar and Belle analyzes. 
The first error is statistical, the
second is systematic and the third one is the model error.
In the case of BaBar, one standard deviation constraint is given for 
the $r_B$ values.}
\label{tab:res_dp_gamma}
\newcommand{\m}{\hphantom{$-$}}
\newcommand{\cc}[1]{\multicolumn{1}{c}{#1}}
\renewcommand{\tabcolsep}{2pc} 
\renewcommand{\arraystretch}{1.2} 
\begin{center}
\begin{tabular}{@{}lcc}
\hline
Parameter & BaBar & Belle \\
\hline
$\gamma$ & $(92 \pm 41 \pm 11 \pm 12)^\circ$ & 
           $(53^{+15}_{-18} \pm 3 \pm 9)^\circ$ \\
$r_B(DK)$ & $< 0.140$ & 
$0.159^{+0.054}_{-0.050} \pm 0.012 \pm 0.049$ \\
$\delta_B(DK)$ & 
$(118 \pm 63 \pm 19 \pm 36)^\circ$ & 
$(146^{+19}_{-20} \pm 3 \pm 23)^\circ$ \\
$r_B(D^*K)$ & $0.017 - 0.203$ & 
$0.175^{+0.108}_{-0.099} \pm 0.013 \pm 0.049$ \\
$\delta_B(D^*K)$  & 
$(-62 \pm 59 \pm 18 \pm 10)^\circ$ & 
$(302^{+34}_{-35} \pm 6 \pm 23)^\circ$ \\
$r_B(DK^*)$ & & 
$0.564^{+0.216}_{-0.155} \pm 0.041 \pm 0.084$ \\
$\delta_B(DK^*)$ & & 
$(243^{+20}_{-23} \pm 3 \pm 49)^\circ$ \\
\hline
\end{tabular}
\end{center}
\end{table*}

Despite similar statistical errors being obtained for $(x_{\pm}, y_{\pm})$ in 
both experiments, the resulting $\gamma$ error is much smaller in 
Belle's analysis.
Since the uncertainty on $\gamma$ scales roughly as $1/r_B$, the difference
is explained by noticing that the BaBar $(x_{\pm}, y_{\pm})$ measurements
favour values of $r_B$ smaller than the Belle results. 

All methods (GLW, ADS and Dalitz) are sensitive to the same parameters
of the $B$ decays,
and can therefore be treated in a combined fit to extract $\gamma$. 
Such comparisons have been performed by various phenomenological groups,
such as CKMfitter~\cite{Charles:2004jd} and UTfit~\cite{Bona:2005vz}.
The CKMfitter group using a frequentist statistical framework 
obtains $(77 \pm 31)^{\circ}$ whereas the UTfit group with 
a bayesian approach obtains $(82 \pm 19)^{\circ}$.
This is in agreement with the prediction from the global CKM fit (where
the direct $\gamma$ measurement has been excluded from the fit).
As mentioned earlier, the size of the $r_B$ parameters play a crucial role 
in the $\gamma$ determination and they are found to be $r_B(DK) < 0.13 $,
$r_B(D^*K) < 0.13$ and  $r_B(DK^*) < 0.27$ at 90\% C.L. by 
Ref.~\cite{Charles:2004jd}
and $r_B(DK) < 0.10 $,
$r_B(D^*K) < 0.12$ and  $r_B(DK^*) < 0.26$ at 90\% C.L.
by Ref.~\cite{Bona:2005vz}.
All values are in agreement with the naive expectation from CKM 
and colour suppression.

Clearly, the precision on $\gamma$ will improve with more data. 
However, the dependence of the sensitivity on the value of $r_B$
means that we should be careful when extrapolating the present results
to a higher statistics scenario. Assuming a value of $r_B$ in the 
range of 0.1-0.15, the statistical error obtained by the end of the $B$
factories (2 ab$^{-1}$) will be 10-15 degrees.
The way to improve the $\gamma$ sensitivity in the near future 
is to include more $D^0$ (and use of $D^{*0}$) modes,
with combined strategies~\cite{Atwood:2003jb}, use of differential spectra~\cite{Atwood:2002vw},
many body modes, charm factory inputs~\cite{Atwood:2003mj}, along with the use
of $B^0$ modes~\cite{Kayser:1999bu,Atwood:2002vw}. 
Although at present (and until the end of $B$ factories era) 
the $\gamma$ accuracy in the $K^0_S \pi^+ \pi^-$ analysis 
is dominated by the statistical uncertainty, the model error will eventually 
dominate in the context of a Super $B$ factory. 
Model independent ways to extract $\gamma$ have been 
proposed~\cite{Atwood:2000ck,Giri:2003ty,Bondar:2005ki}. 
One way to implement this is to notice that
in addition to flavour tagged $\overline{D}{}^0 \to K^0_S \pi^+ \pi^-$ decays,
one can use $C\!P$ tagged decays to $K^0_S \pi^+ \pi^-$ from the 
$\psi(3770) \to D\overline{D}$ process. Combining the two data sets, the 
amplitude and phase could be measured for each point on the Dalitz
plot in a model independent way. Study with MC simulations (assuming $r=0.2$)
indicates that with 50 ab$^{-1}$ of data $\gamma$ can be measured 
with a total accuracy of few degrees~\cite{Bondar:2005ki}. 
Combining all the methods with the statistics anticipated at
a Super $B$ factory (50 ab$^{-1}$), it is expected that
an error of about two degrees is obtainable (chapter 4).
%

{\bf Measurement of $\sin 2\beta + \gamma $ from $B$ decays 
to open charm}

\begin{figure*}[htb]
\begin{center}
\begin{tabular}{cc}
\includegraphics[width=15.5pc]{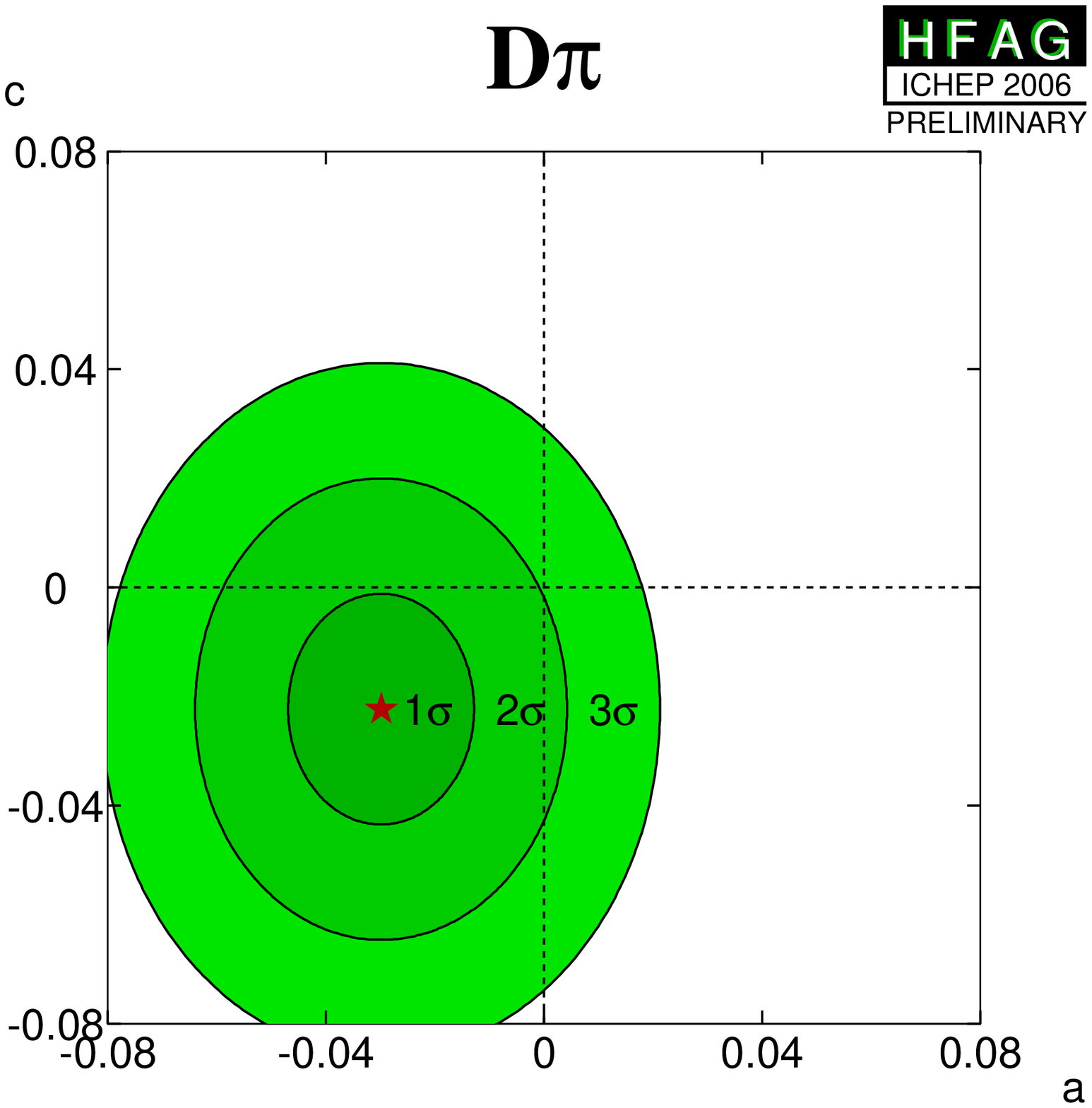}& 
\includegraphics[width=15.5pc]{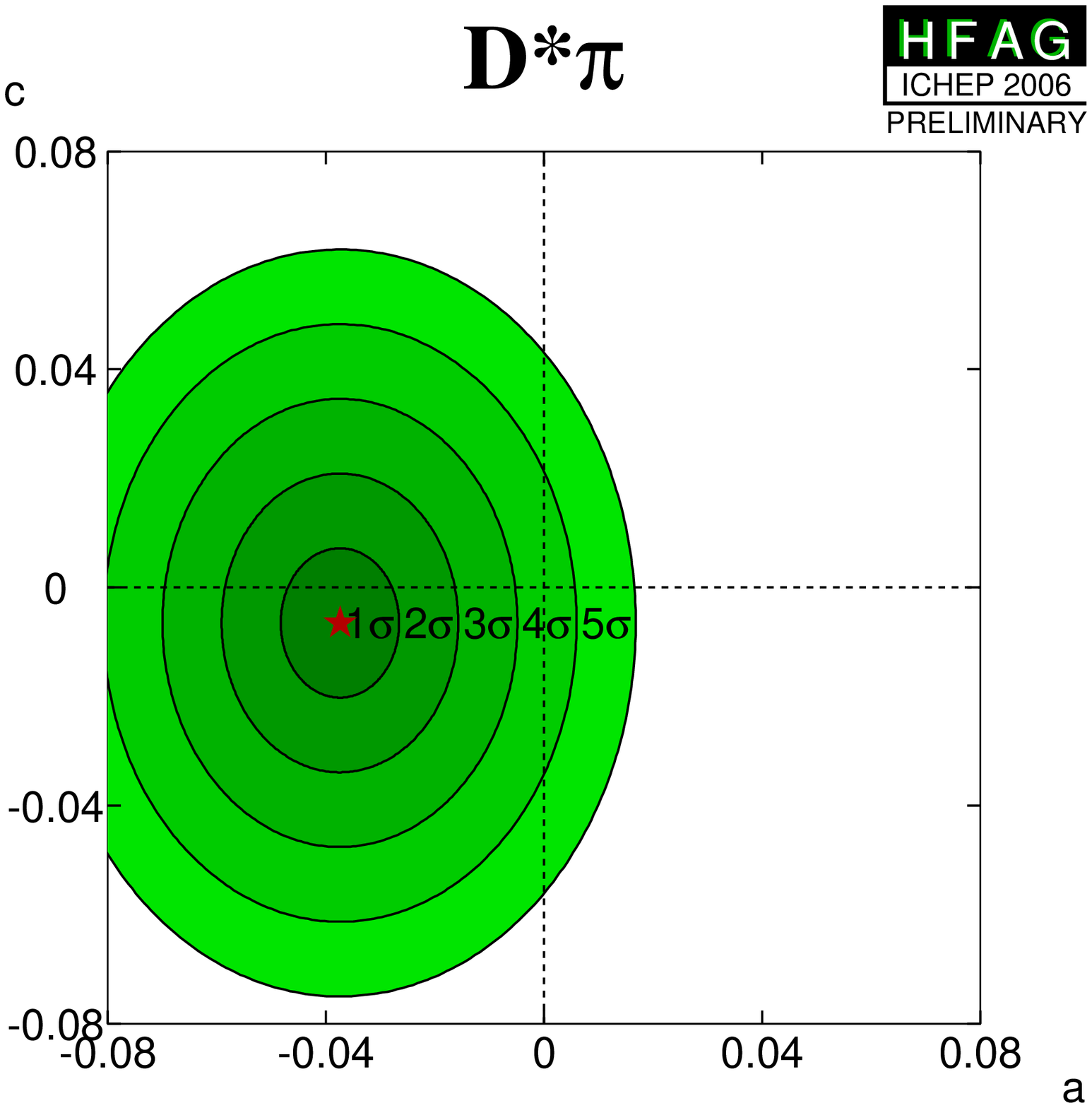}
\end{tabular}
\end{center}
\caption{Results of the $a$ and $c$ measurements for the $D\pi$ (left)
and $D^{*}\pi$ (right) modes.}
\label{fig:sin2phi1pphi3}
\end{figure*}

Interference between decays with and without mixing can occur in the
non-$C\!P$ eigenstates $B^0 \to D^{(*)\pm} \pi^{\mp}(\rho^{\mp})$. The
Cabibbo-favoured $\overline{b} \to \overline{c}$ decay amplitude
interferes with the Cabibbo-suppressed $b \to u$ decay amplitude with
a relative weak phase shift $\gamma$. These modes have the advantage
of a relatively large branching fraction but a small ratio $r$ of
suppressed to favored amplitudes. Time-dependent asymmetries in these
modes can be used to constraint $\sin (2\beta +
\gamma)$~\cite{Dunietz:1997in}: the coefficient of the $\sin (\Delta m
\Delta t)$ term can be written, to a very good approximation, as
$S^{\pm} = 2 r \sin (2\beta + \gamma \pm \delta)$, where $\delta$ is
the strong phase shift due to final state interaction between
the decaying mesons.\\
Potential competing $C\!P$ violating effects can arise from $b \to u$
transitions on the tag side if a kaon is used to tag the flavour on
the other $B$ in the event, resulting in an additional $\sin$ term
$S^{'\pm} = 2 r' \sin (2\beta + \gamma \pm
\delta')$~\cite{Long:2003wq}.  Here, $r'$ ($\delta'$) is the effective
amplitude (phase) used to parameterize the tag side interference. To
account for this term, one can rewrite $S^{\pm}$ as $S^{\pm} = (a\pm
c)+b$, where $a = 2 r \sin (2\beta + \gamma) \cos \delta$, $c = \cos
(2\beta + \gamma) [ 2r \sin \delta + 2r' \sin \delta' ]$ and $b = 2r'
\sin(2\beta + \gamma) \cos \delta'$.  The results from $B$
factories~\cite{Aubert:2005yf,Aubert:2006tw,Ronga:2006hv} are shown
for $D\pi$ and $D^{*}\pi$ modes in terms of $a$ and $c$ in
Fig.~\ref{fig:sin2phi1pphi3}. $C\!P$ violation would appear as $a \neq
0$.
External information is however needed to determine $r$ or $\delta$.
Naively, one can estimate $r \sim |V_{cd}^* V_{ub}/V_{ud} V_{cb}^*| \simeq 
0.02$. One popular choice is the use of SU(3) symmetry to obtain $r$
by relating decay mode to $B$ decays involving $D_s$ mesons~\cite{Dunietz:1997in}.
%


\subsubsection{Expectations from LHCb}
\label{angles_lhcb}

\subsubsubsection{Introduction}

This section summarises the outlook for measurements of CKM
angles through tree-level processes at LHCb.  All estimates
are given for $\rm 2 \, fb^{-1}$ of integrated luminosity,
which is a canonical year of LHCb operation.  
(In the summary section, extrapolations are also made to
$\rm 10  \, fb^{-1}$, which represents five years of operation.)
Background
estimates have been made using 34 million simulated 
generic $ b\overline{b}$ events and, where appropriate, with 
specific samples of known dangerous topologies.  Full details
may be found in the cited LHCb notes and other references.

\subsubsubsection{Measuring $\beta$ with $B^0 \to J/\psi K^0_S$ }

The channel $B^0\rightarrow J/\psi K^0_S$, with the $ J/\psi$
decaying to $ \mu^+\mu^-$, is relatively easy to trigger on
and reconstruct at LHCb.  In order to minimise systematic effects
selection cuts have been developed 
which impose the least possible bias on the lifetime distribution
of the decaying $B^0$.

It is estimated that 333k untagged triggered events will be collected per
2 fb$^{-1}$ of integrated luminosity. Background studies have been
performed using a large sample of generic $ b\overline{b}$ events
and a dedicated sample of prompt $ J/\psi$ events.  The results
indicate that the expected B/S ratio from the two sources is 
1.1 and 7.3 respectively.   The high background from prompt $ J/\psi$'s
has little consequence for the $\sin 2 \beta$ sensitivity, as the
events are restricted to low proper times.    
The performance of the flavour tag is determined from the similar
topology  $B^0\rightarrow J/\psi K^{*0}$ control channel.
The statistical precision on $\sin{2\beta}$ with 2 fb$^{-1}$
is estimated to be 0.015.  More information may be found
in~\cite{SANDRA}.

\subsubsubsection{Measuring $\alpha$ with 
$B^0\to\rho \pi$ and $B^0\to\rho\rho$ at LHCb}

The potential of LHCb in the decay $B^0 \to\rho \pi \to \pi^+\pi^-\pi^0$ 
has been studied extensively~\cite{LHCBALPHA}.  
The hard spectrum of the $ \pi^0$, together 
with the vertex constrains on the $ \pi^+\pi^-$ pair means that the decay
can be well isolated from background, even in the high multiplicity 
environment of the LHC.   A multivariate variable is built up to exploit
all available discriminating variables.    It is estimated that 
$1.4 \times 10^4$ events will be accumulated per 2~$\rm fb^{-1}$ of 
integrated luminosity.   The acceptance for these events is fairly uniform
over Dalitz space,  apart from in the region of low ($m_{\pi^+\pi^0}^2$,
$m_{\pi^-\pi^0}^2$),  which is depopulated due to the minimum
energy requirement on the $\pi^0$.  

The background has been studied with large simulated
samples of generic $ b \overline{b}$ events and with specific charmless
decay channels.   It is concluded that the $\rm B/S$ ratio should not
exceed one,  a value which has been assumed for the subsequent sensitivity
studies.

\begin{figure}
\begin{center}
\epsfig{file=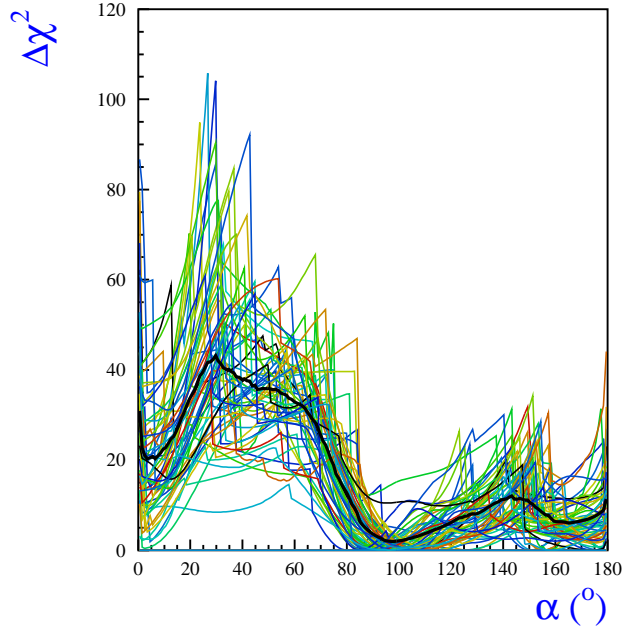,width=0.6\textwidth}
\end{center}
\caption{Change in $\chi^2$ with $\alpha$ for a fit to 
simulated experiments assuming the LHCb performance with
1000 signal events and a B/S ratio of 1.  Each curve corresponds
to a different experiment.   Superimposed in black is the average of all
experiments.  The input value of $\alpha$ is $97^\circ$.}
\label{fig:lhcb_dalitz_fit}
\end{figure}

The expected precision on the angle $\alpha$ has been estimated using
a toy Monte Carlo,   taking the resolutions and acceptances from
the full simulation,  and modelling the background as a combination
of non-resonant and resonant contributions.  Repeated toy experiments 
are performed, each of which has 10000 signal events.  Various scenarios
have been considered for the relative values of the penguin and tree
amplitudes contributing to the final state.
The results shown here assume the `strong penguin' case~\cite{BARBARPHYS}.
An unbinned log likelihood
fit is used to extract the physics parameters of interest,  in particular
$\alpha$.    The achievable precision on $\alpha$ varies between
amplitude scenarios, and fluctuates experiment to experiment. The statistical
error is below $10^\circ$ for about 90\% of experiments.
The mean value is around $8^\circ$.   On about 15\% of occasions the fit
converges to a pseudo-mirror solution,  but these effects diminish
with larger data sets.  Figure~\ref{fig:lhcb_dalitz_fit} shows the variation
in $\chi^2$ for fits to many toy experiments as a function of $\alpha$,
and the average of these curves,  with a clear minimum seen at the 
input value of $\alpha=97^\circ$.
Studies of potential systematic uncertainties
indicate that it will be important to have good understanding of the
$ \rho$ lineshape.

The performance of LHCb has also been investigated in the modes
$B^0 \to\rho^\pm \rho^\mp$ and $  B^\pm \to\rho^\pm \rho^0$.
It is concluded that although significant numbers of events can be 
accumulated,  the total event samples are similar in size to those
that will come from the $B$ factories.    More promising is the 
decay $B^0 \to\rho^0 \rho^0$ which can be used in an isospin
analysis to constrain the bias on $\alpha$ arising from penguin
contamination in the channel $B^0 \to\rho^\pm \rho^\mp$.
1200 events will be obtained per 2~$\rm fb^{-1}$, assuming a branching 
ratio of $1.2 \times 10^{-6}$.   More details on this analysis,
and estimates of its impact on the $\alpha$ extraction within
possible scenarios can be found in~\cite{LHCBALPHA}.



\subsubsubsection{Measuring $\gamma$ with $B\to D K$ strategies at LHCb}

In principle all $B\to D K$ channels, where the $ D$ decays 
hadronically, carry information on the angle $\gamma$.  LHCb has
investigated several modes, with the emphasis on those where
the decays involve charged tracks only.   The presence of one or more
kaons in the final state makes these decays particularly suited
to LHCb, on account of its RICH system.  The estimated event yields
for the modes so far considered are summarised in 
Table~\ref{tab:lhcb_dkyield}.  Background studies
have been carried out using large simulation samples of generic 
$ b\overline{b}$ events, as well as specific channels which are
potential sources of contamination, for example $B\to D \pi$.
In all cases it is concluded that the background levels can be reduced
to an acceptable level.   More information can be found in the referenced
notes.
Many of the strategies that have been investigated are common to those pioneered at 
the $B$ factories and discussed in Section~\ref{sucsec:gamma}.

\begin{table}
\caption{Expected event yields and estimated background for $2\, \rm fb^{-1}$
in $B\to D K$ decay modes so far considered at LHCb. In the rows where two signal yields
are listed, the background corresponds to that expected in either channel.
All numbers come from typical scenarios presented in the references quoted in the text.
The background in the $ D(K^0_S K^+ K^-) K^\pm$ final
state has not yet been studied,  but it is expected to be
significantly smaller than that in the $ D(K^0_S \pi^+ \pi^-) K^\pm$   
mode.}
\label{tab:lhcb_dkyield}
\begin{center} 
\begin{tabular}{lcc} \hline
Decay Mode                                        &  Signal  &  Background \\ \hline
$B^\pm\to D (K^+K^-) K^\pm$                   &  $2600$, $3200$    & $3700 \pm 1000$ \\
$B^\pm\to D (\pi^+\pi^-) K^\pm$               &  $900$,  $1100$    & $3600 \pm 1500$ \\
$B^\pm\to D (K^\pm \pi^\mp ) K^\pm$           &  $28000$, $28300$  & $17500 \pm 1000$   \\
$B^\pm\to D (K^\mp \pi^\pm ) K^\pm$           &  $10$, $400$       & $800 \pm 500$   \\
$B^\pm\to D (K^\pm \pi^\mp \pi^+\pi^-) K^\pm$ &  $30400$, $30700$ &  $20200 \pm 2500$      \\
$B^\pm\to D (K^\mp \pi^\pm \pi^+\pi^-) K^\pm$ &  $20$, $410$ &  $1200 \pm 360$      \\
$B^\pm\to D (K^0_S \pi^+ \pi^-) K^\pm$              &  $5000$ &  $1000-5000$ (90\% C.L.)  \\
$B^\pm\to D (K^0_S K^+ K^-) K^\pm$                  &  $1000$ &   /    \\
$B^\pm\to D (K^+K^- \pi^+ \pi^-) K^\pm$             &  $1700$ &  $1500 \pm 600$     \\
$B^\pm\to (D \pi^0) (K^\pm \pi^\mp ) K^\pm$   &  $16800$, $16600$ & $34300 \pm 11500$ \\
$B^\pm\to (D \pi^0) (K^\mp \pi^\pm ) K^\pm$   &  $350$, $100$     & $4800 \pm 3800$   \\
$B^\pm\to (D \gamma) (K^\pm \pi^\mp ) K^\pm$  &  $9400$, $9300$   & $34300 \pm 11500$ \\
$B^\pm\to (D \gamma) (K^\mp \pi^\pm ) K^\pm$  &  $10$, $140$      & $4800 \pm 3800$   \\
$B^0, \overline{B}{}^0 \to D (K^+K^-) K^{* 0}, \overline{K}{}^{* 0}$           
                                                        &   $240$, $450$ &  $<1000$ (90\% C.L.)   \\
														
$B^0, \overline{B}{}^0 \to D (\pi^+\pi^-) K^{* 0}$       &    $70$, $140$ &  $<1000$ (90\% C.L.)   \\
$B^0, \overline{B}{}^0 \to D (K^\pm \pi^\mp) K^{* 0}, \overline{K}{}^{* 0}$                    
                                                        &  $1750$, $1670$ &  $<1700$ (90\% C.L.)   \\
$B^0, \overline{B}{}^0 \to D (K^\mp \pi^\pm) K^{* 0}, \overline{K}{}^{* 0}$    
                                                        &   $350$, $260$  &  $<1700$ (90\% C.L.)   \\
\hline
\end{tabular}
\end{center}
\end{table}

The simplest topologies are $B\to D K$ decays where the
$ D^0 \, (\overline{D}{}^0)$ decays to a $C\!P$-eigenstate such
as $ K^+K^-$ or $ \pi^+\pi^-$,  or to $ K^\pm \pi^\mp$.   Of particular
interest is the subset of highly suppressed `ADS' decays
$B^\pm\to D (K^\mp \pi^\pm ) K^\pm$ where the interference
effects are highest.   The exact number of expected events in
this mode depends on the assumption for $r_B$, the ratio
of the interfering $B$ decay amplitudes.   Assuming a value
of $r_B=0.08$ leads to the expectation of around 400 events,
integrated over $B^+$ and $B^-$ channels,
with a variation dependent on the value of the strong phase difference
between the diagrams involved in both the B and D decays~\cite{MITESH1}.

The 3-body Dalitz analysis of $ K^0_S \pi^+\pi^-$ in 
$B\to D K$ decays has been successfully pioneered at the 
$B$ factories.  Here too LHCb expects to make a significant contribution
with 5000 triggered and reconstructed decays 
per 2~$\rm fb^{-1}$~\cite{JIMCRIS}.  
A technical challenge in selecting
these events is presented by those $ K^0_S$'s which decay downstream of
the VELO region; these decays account for around two thirds of the 
total sample.  
Although such events can be successfully reconstructed
offline,  this procedure is challenging to perform in the high
level trigger, where the existing track-search algorithm for $ K^0_S$ 
daughters
does not fit within the allocated CPU budget. It is hoped that
this difficulty will be overcome.  The problem is not so critical
for the sister 3-body mode $ D \to  K^0_S K^+K^-$, where the
two kaons offer the possibility of devising an inclusive high
level trigger selection not dependent on the finding of the $ K^0_S$.

The 4-body modes $ D \to K^\pm \pi^\mp \pi^+ \pi^-$ and 
$ D \to K^+ K^- \pi^+ \pi^-$ are particularly attractive to
LHCb  as all the decay products are prompt charged tracks.
Dependent on the charge of the decaying $B$, and the charges of 
the particles in the $ D$ decay, the $ K\pi\pi\pi$ channel
accesses four possible final states, of which the rarest two, 
$B^\pm\to D (K^\mp \pi^\pm \pi^+\pi^-) K^\pm$, possess
large interference effects through the ADS mechanism.
The expected sample size integrated over these two channels is about 400
events~\cite{ANDREW}.
Provided that the sub-resonant decay structure can be fitted in a 
four-body amplitude
analysis these suppressed channels will provide high sensitivy 
to $\gamma$, either
in isolation, or in conjunction with the other ADS modes.
An analysis of the 4-body Dalitz space of $ K^+K^-\pi^+\pi^-$ 
accesses $\gamma$ in a similar way to the 3-body self-conjugate mode
$ K^0_S \pi^+\pi^-$.  Here 1700 events are expected~\cite{ANDREW}.

Extensions of the standard $B\to DK$ strategies have also been considered
at LHCb.  Detailed studies have been performed of $B^0 \to D K^{\ast 0}$,
where the charge of the kaon in the $ K^{\ast 0} \to K^\pm \pi^\mp$
decay chain tags the flavour of the decaying $B^0$~\cite{KAZU}.  
Here both the 
interfering $B^0$ decay diagrams are colour suppressed, and hence 
the interference effects are higher than in the $B^\pm$ case, although 
the branching ratios are lower.    Another method under study is
$B^\pm \to D^\ast K^\pm$,  where the $ D^\ast$ decays either
through $ D^0 \pi^0$ or $ D^0 \gamma$.   As there is a $C\!P$-conserving
phase difference of $\pi$ between these two paths, separation of the
respective modes gives powerful additional constraints in the analysis. 
At LHCb the energy of the neutral particles is too low to permit efficient
selection.  However,  sufficient constraints exist in the decay topology
to allow a full reconstruction using the charged tracks alone.
Preliminary results indicate a promising performance, although there
are at present insufficient Monte Carlo statistics to make a meaningful
background estimate~\cite{MITESH2}.

Assuming the 2~$\rm fb^{-1}$  event yields listed in 
Table~\ref{tab:lhcb_dkyield}, and 
the background estimates coming out of the Monte Carlo studies, full 
sensitivity studies have been performed for several of the analyses.
The precision on $\gamma$ depends on the parameters assumed.
Taking $r_B=0.08$, the statistical undertainty
is found to be $6-10^\circ$ for a combined $B^\pm \to DK^\pm$ analysis
involving the two-body D decay modes, and $ D \to K \pi\pi\pi$, where
the resonant substructure of the latter decay is so-far 
neglected~\cite{MITESH1}.   A similar
sensitivity is found for the $B^0\to D K^{\ast 0}$ study involving
two body modes only, where the ratio of the interfering diagrams 
is taken to be $0.4$~\cite{KAZU}.
Estimates have also been made of the $\gamma$ sensitivity in    
$ K^0_S \pi^+ \pi^-$~\cite{JIMCRIS}.  Including acceptance
effects and background gives a typical sensitivity of $15^\circ$,
again taking $r_B=0.08$.
At present the only available studies of $K^+K^-\pi^+\pi^-$~\cite{Rademacker:2006zx}
are for signal events only.  A background free analysis with
the LHCb annual signal yield would have a statistical
uncertainty of $14^\circ$, also with $r_B=0.08$.
Systematic effects have not yet been considered,  but it is already
known from the $B$ factories that work is needed to improve the confidence 
in the $ D \to K^0_S \pi^+ \pi^-$ decay model,  an issue which
is likely to be important for all the 3 and 4 body D decays.

Other decay modes remain to be investigated,  for example $B^\pm \to
D K^{\ast \pm}$, $ K^{\ast \pm} \to K^0_S \pi^\pm$.   The full power
of the $B\to DK$ sensitivity will only come with a combined analysis
of all accessible decay modes.   The preliminary indications suggest
that $B \to DK$ decays will provide LHCb's most precise value 
of $\gamma$,  with
a few degrees uncertainty being achievable with 2~$\rm fb^{-1}$ of data.
There is no reason to expect that the experimental systematics will
significantly limit this sensitivity,  although more detailed studies are 
required.   It is clear, however, that residual uncertainties associated
with the understanding of the $D$ decay in the 3 and 4 body modes
could be important.  A possible scenario is presented in the Summary
section based on arbitrary assumptions concerning this source of uncertainty.

\subsubsubsection{Measuring $\gamma$ with $B_s, \,\overline{B}_s \to D_s^\pm K^
\mp$ and $B^0, \, \overline{B}{}^0 \to D^\pm \pi^\mp$}

The isolation of $B_s \to D_s^\pm K^\mp$ decays is  
experimentally very challenging,  because of the
low branching ratio and the order-of-magnitude more prolific 
 $B_s \to D_s \pi$ decay
mode.   The LHCb trigger system gives good performance for fully hadronic
modes and selects $B_s \to D_s^\pm K^\mp$ events with 
an efficiency of $29\%$.   The $ \pi-K$
discrimination of the RICH system reduces the 
$B_s \to D_s \pi$ contamination to
$\sim 10\%$.   It is estimated that the experiment will accumulate 6.2k
events per 2 $\rm{fb^{-1}}$ of integrated luminosity, with a combinatoric 
background to signal level of $<$~0.6~\cite{DSKRECSENS}.   
The excellent $\sim 30 \, \rm{fs}$
proper time precision provided by the silicon Vertex Locator will ensure 
that the  $B_s$ oscillations will be well resolved, and hence allow the 
$C\!P$ asymmetries to be measured.   It is estimated that the statistical 
precision on $\gamma$ from this channel alone 
will be $10^\circ$ for 2~$\rm {fb^{-1}}$,  
assuming $\Delta m_s = 17.5 \, {ps^{-1}}$, 
$|\Delta \Gamma_s |/ \Gamma_s = 0.10$~\cite{DSKRECSENS}.  
Note that this extraction requires 
knowledge of the weak mixing phase in the $B_s$ system,  which
is imported from parallel LHCb studies performed with 
$B^0 \to J/\psi \phi$ decays.

A potential difficulty with the $B_s \to D_s^\pm K^\mp$ $\gamma$
extraction arises from ambiguities.   In the limit that $\Delta \Gamma_s$
is very small the analysis returns an 8-fold ambiguity.
A non-zero value of $\Delta \Gamma_s$ in principle 
ameliorates the problem, reducing the number of true ambiguities to
four only,  but even in this case the eliminated solutions may
in practice remain as false minima, on account of the limited 
experimental resolution.  An attractive way to circumvent this 
difficulty is to make a combined analysis of the observables in
the $B_s$ decay and those in the U-spin symmetric 
$B^0 \rightarrow D^\pm \pi^\mp$ channel~\cite{Fleischer:2003yb}.
This approach has the added bonus of exploiting 
$B^0 \rightarrow D^\pm \pi^\mp$ decays in a manner which 
does not require knowledge of the ratio between the interfering
tree diagrams,  which in the $B^0$ system is known to be
very small, and hence hard to determine experimentally.
LHCb will accumulate 1730k events per 2 $\rm fb^{-1}$ in this channel~\cite{DPIREC}.
The combined analysis has the potential to reach a statistical
precision of $5^\circ$, depending on the values of the parameters 
involved.  Any bias associated with the U-spin symmetry assumption
also has a varying impact on the measurement, depending on the position
in parameter space.   In many scenarios the effect is expected to be
below the statistical uncertainty~\cite{DSKDPIUSPIN}.

\subsubsection{Summary}

Table~\ref{tab:summary} presents a summary of the current status
and the outlook for future direct measurements of the angles of the
unitarity triangle from tree dominated $B$ decays. 
The last column of this table is an estimate of the ITE, 
which is the intrinsic error coming purely
from theoretical limitations of the methods being used. 
It seems that for $\sin 2\beta$,  at the end of the $B$ factory
era with an estimated $\approx \rm 2 \, ab^{-1}$ of data, the
experimental
determination will be close to the expected theory error.
In fact the theory error ($ \lesssim 1\%$) is somewhat smaller
but apparently our current understanding is
that experimental systematics are difficult to reduce 
below about 2-3\%. Measurement of $\sin 2 \beta$ at LHCb
also looks very promising so far as the statistical error
goes.

For $\alpha$ although each of the three methods,
$\pi \pi$, $\rho \pi$, and $\rho \rho$ will have a
residual theory error due to isospin violation
by EWP and/or from other sources, it is quite likely
that once the experimental information with high statistics
on all the three modes becomes available the remaining intrisic theory error 
will be small, O(few\%). The current $B$ factories
and LHCb are expected to be able to determine $\alpha$ to an accuracy
around $5^{\circ}-8^{\circ}$, {\it i.e.} considerably
worse than the ITE. A Super $B$ factory should be able to
attain the level of accuracy O(2\%) $\approx$ ITE.

Unfortunately a precise determination of the angle $\gamma$
is likely to remain a challenge for a long time to come.
Admittedly we have been somewhat cautious in our projections
for the $B$ factories and there is some chance that we will
gain more from combined strategies, compared to projections
in this table, as additional
data becomes available in the next year or two. Indeed
 LHCb should however be able to do at least five times better than this
  ({\it i.e.} an accuracy
of about 2.6 degrees), 
      with a final uncertainty dependent on the errors associated with
      the
      knowledge of
	   the D decay structure in the modes exploited in the $B
	   \to DK$
	   channels.
It is interesting to note that with a SBF, 
and the very high statistics associated
with an LHCb upgrade, the experimental error on $\gamma$ could approach 1 degree,
but would still be larger than that of the associated ITE. 
	  
\begin{table}[h]
\caption{Unitarity Triangle from trees decays: 
Current status and future prospects. ITE means irreducible theory error;
see text especially regarding the LHCb projections.}
\label{tab:summary}
\begin{center} 
\begin{tabular}{lcccccc}
\hline
 & $\;$ BF (Now) $\;$ & $\;$ BF(End '08)$\;$ & 
   $\;\;\;$ LHCb
$\;\;\;$ &  $\;$ LHCb
$\;$  & SBF & ITE \\
$\int {\cal L} dt$ & $\sim$ 1 ab$^{-1}$ & 2 ab$^{-1}$ & 2 fb$^{-1}$ & 
10 fb$^{-1}$ &  50 ab$^{-1}$ & \\
\hline
$\sigma(\alpha)$ & $10^{\circ}$ ($11\%$) & $7^{\circ}$ ($8\%$) & 
$8.1^{\circ}$ ($9\%$) & $4.6^{\circ}$ ($5\%$) & $1.5^{\circ}$ ($1.6\%$) &
O(few \%) \\
$\sigma(\sin 2 \beta)$ & $0.026$ ($4\%$)  & $0.023$ ($3.3\%$) $\;$ & 
$\; 0.015$ ($2.1\%$) $\;$ & 
$\; 0.007$ ($1\%$) $\;$ & $0.013$ ($2\%$) &
$ \lesssim 1\%$ \\
$\sigma(\gamma)$ & $30^{\circ}$ ($46\%$) & $15^{\circ}$ ($23\%$) & 
$4.5^{\circ}$ ($7\%$) & $2.4^{\circ}$ ($4\%$) & 
$2^{\circ}$ ($3\%$) & O(0.1\%) \\
\hline
\end{tabular}
\end{center}
\end{table}

Lastly, we must caution the reader that 
the LHCb numbers in table~\ref{tab:summary} are merely illustrative values,
extrapolated from present simulation studies,  together with certain
(in some cases) arbitrary assumptions about systematic errors.
The estimated precisions for $\sin 2 \beta$ contain statistical
uncertainties only,  as the experimental systematics are impossible
to estimate properly in advance of first data.   The values for $\alpha$
are dominated by the input from the $B^0 \to \rho \pi$ analysis,
with the conservative assumption of a limiting systematic of $6^\circ$,
associated with issues in the Dalitz analysis and the understanding
of the $\rho$ lineshape.   The $\gamma$ estimates includes inputs
from the $B_s \to D_s K^\pm$, $B^\pm \to D^{(\ast)}(hh,\, hhhh) K^\pm$,
$B^\pm \to D (K^0_S \pi \pi) K^\pm$  and $B^0 \to D (hh) K^\ast(K^+\pi^-)$
analyses.   Here it is assumed that progress with the understanding
of the $D$ decay structure will result in systematics of $3^\circ$
for the $D \to  K^0_S \pi \pi$ mode, and twice this for the 4-body
decays.  An arbitrary $5^\circ$ error is assigned to the $B^0$ channel
to account for the possibility of other amplitudes contributing the
$D(hh) K^+\pi^-$ final state. 
The $B^\pm \to D K^\pm$ inputs
assume an $r_B$ value of 0.08. The assumed quantities for other
parameters are given elsewhere in the text and references.

%

\newpage 
\subsection{$B$-meson mixing}



\subsubsection{Introduction}

During this workshop there has been a breakthrough in the experimental 
study of
\bsmix mixing with the measurement of the following quantities:
the oscillation frequency \dms by the CDF
collaboration~\cite{Abulencia:2006ze}, the time-integrated
untagged charge asymmetry in semileptonic \Bs
decays $A_{SL}^{s,unt}$ and the dimuon asymmetry \asl by
D\O~\cite{Abazov:2007nw,Abazov:2006qw}, the \Bs lifetime from
flavour-specific final 
states~\cite{Barberio:2006bi,Buskulic:1996ei,Abe:1998cj,Abreu:2000sh,
Ackerstaff:1997qi,Abazov:2006cb},
\dgg from the time-integrated angular analysis of $B_s \to J/\psi \phi$ decays
by CDF~\cite{Acosta:2004gt}, supplemented by the three-dimensional constraint 
on $\Gamma_\text{s}$, \dg, and the \bsmix mixing phase from the time-dependent
angular analysis of \bspsiphi decays by D\O~\cite{Abazov:2007tx}.
These measurements can be compared with the Standard Model (SM) predictions and
used to constrain New Physics (NP) contributions to the \bsmix mixing 
amplitude.

In this section we first discuss the
theoretical predictions within the SM and their uncertainties. We then present
the results of a model-independent analysis of NP in \bsmix mixing. We
discuss the implications of the experimental data for SUSY models by either
allowing new sources of flavour and CP violation in the \Bs sector or by
considering a constrained Minimal Flavour Violation SUSY scenario.
The remainder of the section is devoted to the experimental aspects of the 
measurements listed above and gives an outlook for the LHC.

\subsubsection{Standard model predictions}
\label{sec:SMpredictions}
The neutral $B_d$ and $B_s$ mesons mix with their antiparticles leading to
oscillations between the mass eigenstates. The time evolution of the neutral
meson doublet is described by a Schr\"odinger equation with an effective $2
\times 2$ Hamiltonian
\be
i\frac{d}{dt} \begin{pmatrix} B_q \\ {\bar B}_q \end{pmatrix} =
\left[ 
\begin{pmatrix} M_{11}^q & {M_{12}^q} \\ {M_{12}^q}^* & M_{11}^q\end{pmatrix}
-\frac{i}{2} \begin{pmatrix} \Gamma_{11}^q & {\Gamma_{12}^q} \\
{\Gamma_{12}^q}^* & \Gamma_{11}^q\end{pmatrix} \right]
\begin{pmatrix}B_q\\ {\bar B}_q\end{pmatrix}\,,
\ee
with $q=d,s$. The mass difference $\Delta m_q$ and the width difference $\Delta
\Gamma_q$ are defined as
\be
\Delta m_q=m^q_H-m^q_L\,,\quad
\Delta\Gamma_q=\Gamma^q_L-\Gamma^q_H\,,
\ee
where $H$ and $L$ denote the Hamiltonian eigenstates with the heavier and
lighter mass eigenvalue, respectively. These states can be written as
\be
\vert B_q^{H,L}\rangle =\frac{1}{\sqrt{1+\vert (q/p)_q \vert^2}}\,\left(
\vert B_q\rangle \pm  \left(q/p\right)_q\vert {\bar B}_q\rangle
\right)\,.
\ee

Theoretically, the experimental observables $\Delta m_q$, $\Delta \Gamma_q$ and
$\vert(q/p)_q\vert$ are related to $M^q_{12}$ and $\Gamma^q_{12}$. In the
$B_d-\bar B_d$ and $B_s-\bar B_s$ systems, the ratio $\Gamma^q_{12}/M^q_{12}$ 
is of ${\cal O}(m_b^2/m_t^2)\simeq 10^{-3}$ and, neglecting terms of ${\cal O}
(m_b^4/m_t^4)$, one has
\be
\Delta m_q= 2\,\vert M^q_{12}\vert\,,\quad
\frac{\Delta\Gamma_q}{\Delta m_q}
=-{\mathrm{Re}}\left(\dfrac{\Gamma^q_{12}}{M^{q}_{12}} \right)\,,\quad
1-\left\vert\left(\dfrac{q}{p}\right)_q\right\vert =
\dfrac{1}{2}\,{\mathrm{Im}}
\left( \dfrac{\Gamma^q_{12}}{M^q_{12}}\right)\,.
\ee

The matrix elements $M^q_{12}$ and $\Gamma^q_{12}$ are related to the 
dispersive and the absorptive parts of the $\DB=2$ transitions, respectively. 
Short distance
QCD corrections to these matrix elements have been computed at the NLO for both
$M^q_{12}$~\cite{Buras:1990fn} and
$\Gamma^q_{12}$~\cite{Beneke:1998sy,Beneke:2003az,Ciuchini:2003ww}. The long
distance effects are contained in the matrix elements of four-fermion operators
which have been computed with lattice QCD using various approaches to treat the
$b$ quark 
(HQET, NRQCD, QCD)~\cite{Dalgic:2006gp,Gimenez:2000jj,Hashimoto:2000eh,
Aoki:2002bh,Becirevic:2000sj,Lellouch:2000tw,Yamada:2001xp}. The
corresponding bag parameters $B$ are found to be essentially insensitive to the
effect of the quenched approximation (see sec. \ref{subsec:latqcd}).

The quantity ${\rm Im}(\Gamma^q_{12}/M^q_{12})$ can be measured through the
CP asymmetry in $B_q$ decays to flavour-specific final states.
An important example is the semileptonic asymmetry 
\begin{equation}
A^{s}_{SL}={\rm Im}\left(\frac{\Gamma^q_{12}}{M^q_{12}}\right)=
\frac{N(\bar{B}_{s}\to l^{+} X) - N(B_{s} \to l^{-}
X)}{N(\bar{B}_{s}\to l^{+} X) + N(B_{s} \to l^{-} X)}.
\label{eq:A_SL_s}
\end{equation}

Two updated theoretical predictions for $\Delta \Gamma_{s}/\Gamma_{s}$ and for
the semileptonic asymmetry $A_{SL}^s$, obtained by
including NLO QCD and ${\cal O}(1/m_b)$~\cite{Beneke:1996gn} corrections, are
\bea
& \Delta \Gamma_{s}/\Gamma_{s} = (7 \pm 3)\cdot 10^{-2} \quad ,
& A_{SL}^s = (2.56 \pm 0.54)\cdot 10^{-5} \quad
\mbox{\cite{Ciuchini:2003ww}} \,, \nn \\
& \Delta \Gamma_{s}/\Gamma_{s} = (13 \pm 2)\cdot 10^{-2} \quad ,
& A_{SL}^s = (2.06 \pm 0.57)\cdot 10^{-5} \quad
\mbox{\cite{Lenz:2006hd}} \,.
\eea
The difference in the central values of $\Delta \Gamma_{s}/\Gamma_{s}$ 
is mainly
due to a different choice of the operator basis~\cite{Lenz:2006hd} and it is
related to unknown ${\cal O}(\alpha_s^2)$ and ${\cal O}(\alpha_s/m_b)$
corrections. Although the basis chosen in ref.~\cite{Lenz:2006hd} leads to
smaller theoretical uncertainties, the shift observed in the central values may
signal that the effect of higher-order corrections on $\Delta \Gamma_{s}
/\Gamma_{s}$ is larger than what could have been previously estimated. We take
into account this uncertainty by quoting, as final theoretical predictions in
the SM, the more conservative estimate~\cite{Tarantino:2007nf}
\be
\Delta \Gamma_{s}/\Gamma_{s} = (11 \pm 4)\cdot 10^{-2} \quad , \quad
A_{SL}^s = (2.3 \pm 0.5)\cdot 10^{-5} ~.
\ee

Concerning $\Delta m_s$, the SM predictions obtained by the UTfit and CKMfitter
Collaborations are
\be
\Delta m_s = (18.4 \pm 2.4)~\mathrm{ps}^{-1} 
\mbox{\cite{Bona:2006ah}} \quad , \quad
\Delta m_s = (18.9^{+5.7}_{-2.8})~\mathrm{ps}^{-1} 
\mbox{\cite{Charles:2004jd}}.
\label{eq:dms-sm}
\ee

\subsubsection{$B_s - \bar B_s$ mixing beyond the SM}
\label{sec:NP}

We now discuss the analysis of $B_s - \bar B_s$ mixing in the presence of
new physics (NP) contributions to the $\DB=2$ effective Hamiltonian.
These can
be incorporated in the analysis in a model independent way, parametrising the
shift induced in the mixing frequency and phase with two parameters, $C_{B_s}$
and $\phi_s\equiv 2\phi_{B_s}$, having in the SM expectation values of 1 and 0,
respectively
\cite{Soares:1992xi,Deshpande:1996yt,Silva:1996ih,Cohen:1996sq,
Grossman:1997dd}:
\be \label{phibsdef}
C_{B_s}  e^{i \phi_s} \equiv C_{B_s}  e^{2 i \phi_{B_s}} =
\frac{\left(M^s_{12}\right)^\mathrm{SM+NP}}
{\left(M^s_{12}\right)^\mathrm{SM}} \, .
\ee

As for the absorptive part of the $B_s -\bar B_s$ mixing amplitude, which is
derived from the double insertion of the $\DB=1$ effective Hamiltonian, it 
could be affected by NP effects in $\DB=1$ transitions through
penguin contributions. Such NP contributions were considered in 
\cite{Bona:2005eu,Bona:2006sa}. 
We shall neglect them in the present discussion.
In this approximation, which is followed by most authors, NP enters
\bsmix mixing only through the two parameters defined in (\ref{phibsdef}).

Since the SM phase of $\Gamma^s_{12}/M^s_{12}$ is small
in comparison with the current experimental sensitivity, we shall
assume in the following that CP violation in $B_s$ mixing
is dominated by the NP mixing phase $\phi_s$.
We then have 
\begin{equation}\label{alsphis}
A^{s}_{SL}= \frac{\Delta\Gamma_{s}}{\Delta M_{s}} \tan \phi_{s}
\end{equation}
and the same NP phase $\phi_s$ will also govern mixing-induced
CP violation in the exclusive channel \bspsiphi. Note that the phases in 
$A^s_{SL}={\rm Im}(\Gamma^s_{12}/M^s_{12})$ and in the
\bspsiphi asymmetry are different from each other in the SM,
where ${\rm arg}(-\Gamma^s_{12}/M^s_{12})\approx -0.004$
while the phase measured in \bspsiphi decay is 
$-2\beta_s\approx -2\lambda^2\eta\approx -0.04$ 
(see e.g. \cite{Lenz:2006hd}).


Making use of the experimental information described in sect.\ref{sec:EXP}, 
it is possible to constrain $C_{B_s}$ and
$\phi_{B_s}$~\cite{Grossman:2006ce,Bona:2005eu,Bona:2006sa,Lenz:2006hd,
Ligeti:2006pm,Ball:2006xx,Bona:2007vi}. We report here the results obtained in
ref.~\cite{Bona:2007vi}.

The use of $\Delta \Gamma_s /\Gamma_s$ from the time-integrated angular 
analysis of $B_s \to J/\psi \phi$ decays is described for instance in
ref.~\cite{Bona:2006sa}. Here we use only the CDF
measurement~\cite{Acosta:2004gt} as input, since the D\O\ analysis is now
superseded by the new time-dependent study~\cite{Abazov:2007tx}.  The latter provides
the first direct constraint on the $B_s$--$\bar B_s$ mixing phase, but also a
simultaneous bound on $\Delta \Gamma_s$ and $\Gamma_s$. The time-dependent
analysis determines the $B_s$--$\bar B_s$ mixing phase with a four-fold
ambiguity. First of all, being untagged, it is not directly sensitive to $\sin
\phi_s$, resulting in the ambiguity $(\phi_s,\cos \delta_{1,2}) \leftrightarrow
(-\phi_s,-\cos \delta_{1,2})$, where $\delta_{1,2}$ represent the strong phase
differences between the transverse polarisation and the other ones. Second, at
fixed sign of $\cos \delta_{1,2}$, there is the ambiguity $(\phi_s,\Delta
\Gamma_s) \leftrightarrow (\phi_s+\pi,-\Delta \Gamma_s)$. One could be tempted
to use factorisation \cite{Lenz:2006hd} or $B_d \to J/\psi K^*$ with SU(3)
\cite{rescigno} to fix the sign of $\cos \delta_{1,2}$. Unfortunately, neither
factorisation nor SU(3) are accurate enough to draw firm conclusions on these
strong phases. This is confirmed by the fact that the two approaches lead to
opposite results. Waiting for future, more sophisticated experimental analyses,
which could resolve this ambiguity with a technique similar to the one used by
BaBar in $B_d \to J/\psi K^*$ \cite{Aubert:2004cp}, we prefer to be 
conservative and keep the four-fold ambiguity.

\begin{figure}[ht]
\begin{center}
\includegraphics[width=0.45\textwidth]{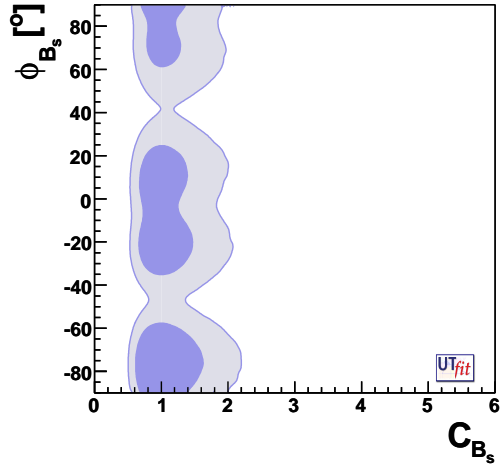}
\caption{Constraints on  $\phi_{B_s}$ vs. $C_{B_s}$ from the NP generalised
analysis of ref.~\cite{Bona:2007vi}.}
\label{fig:cbsphibs}
\end{center}
\end{figure}

Compared to previous analyses, the additional experimental input discussed 
below improves considerably the determination of the phase of the 
$B_s - \bar B_s$
mixing amplitude. The fourfold ambiguity inherent in the untagged analysis of
ref.~\cite{Abazov:2007tx} is somewhat reduced by the measurements of $A_{SL}^s$
and $A_{SL}$ (see (\ref{eq:A_SL})), 
which slightly prefer negative values of $\phi_{B_s}$. The results
for $C_{B_s}$ and $\phi_{B_s}$, obtained from the general analysis allowing for
NP in all sectors, are
\be
C_{B_s} = 1.03 \pm 0.29 \quad , \quad
\phi_{B_s}= (-75 \pm 14)^\circ \cup (-19 \pm 11)^\circ \cup (9 \pm 10)^\circ
\cup (102 \pm 16)^\circ ~.
\ee
Thus, the deviation from zero in $\phi_{B_s}$ is below the $1 \sigma$ level,
although clearly there is still ample room for values of $\phi_{B_s}$ very far
from zero. The corresponding p.d.f. in the $C_{B_s}$-$\phi_{B_s}$ plane is
shown in fig.~\ref{fig:cbsphibs}.

\subsubsection{$B_s - \bar B_s$ in SUSY with non-minimal flavour violation}
\label{sec:SUSY}

The results on $C_{B_s}$ and $\phi_{B_s}$ obtained above can be used to
constrain any NP model. As an interesting example we discuss here the case of
SUSY with new sources of flavour and CP violation, following ref.~\cite{vale}.

To fulfill our task in a model-independent way, we use the mass-insertion
approximation to evaluate the gluino mediated contribution to $b\to s$
transitions. Treating off-diagonal sfermion mass terms as interactions, we
perform a perturbative expansion of FCNC amplitudes in terms of mass insertions.
The lowest non-vanishing order of this expansion gives an excellent 
approximation to the full result, given the tight experimental constraints 
on flavour-changing
mass insertions. It is most convenient to work in the super-CKM basis, in which
all gauge interactions carry the same flavour dependence as in the SM. In this
basis, we define the mass insertions $(\delta^d_{ij})_{AB}$ as the off-diagonal
mass terms connecting down-type squarks of flavour $i$ and $j$ and helicity $A$
and $B$, divided by the average squark mass (see sec. \ref{sec:susy}).

The constraints on $(\delta^d_{23})_{AB}$ have been studied in detail in
ref.~\cite{Ciuchini:2006dx} using as experimental input the branching ratios
and CP asymmetries of $b \to s \gamma$ and $b \to s \ell^+ \ell^-$ decays and
the first measurement of $B_s - \bar B_s$ mixing. We perform the same analysis
using the full information encoded in $C_{B_s}$ and $\phi_{B_s}$, and the
recently computed NLO corrections to the $\Delta B=2$ SUSY effective
Hamiltonian~\cite{Ciuchini:2006dw}. We refer the reader to ref.~\cite{vale} for
all the details of this analysis.
\begin{figure*}[p]
\begin{center}
\includegraphics[width=0.45\textwidth]{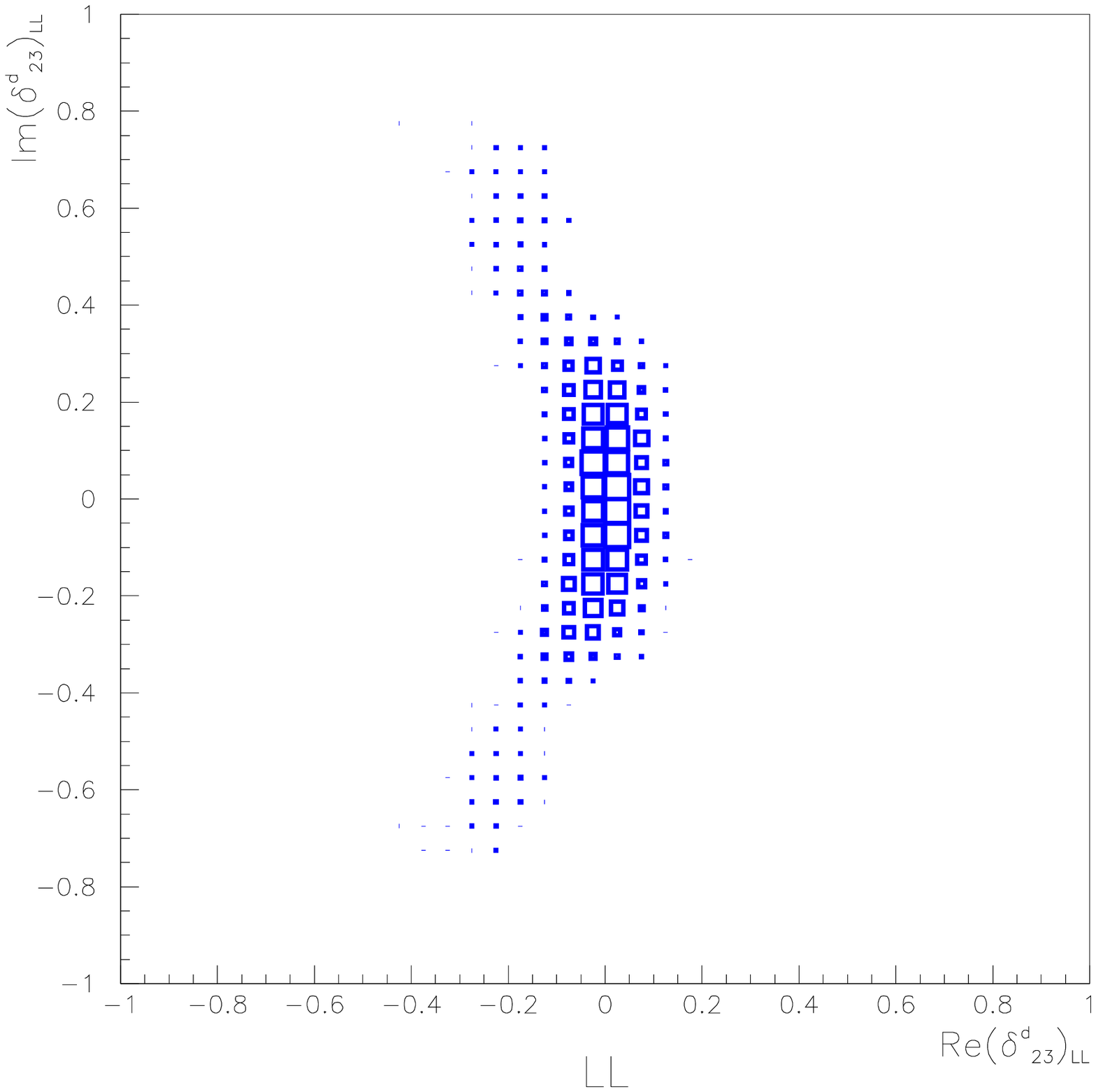} 
\includegraphics[width=0.45\textwidth]{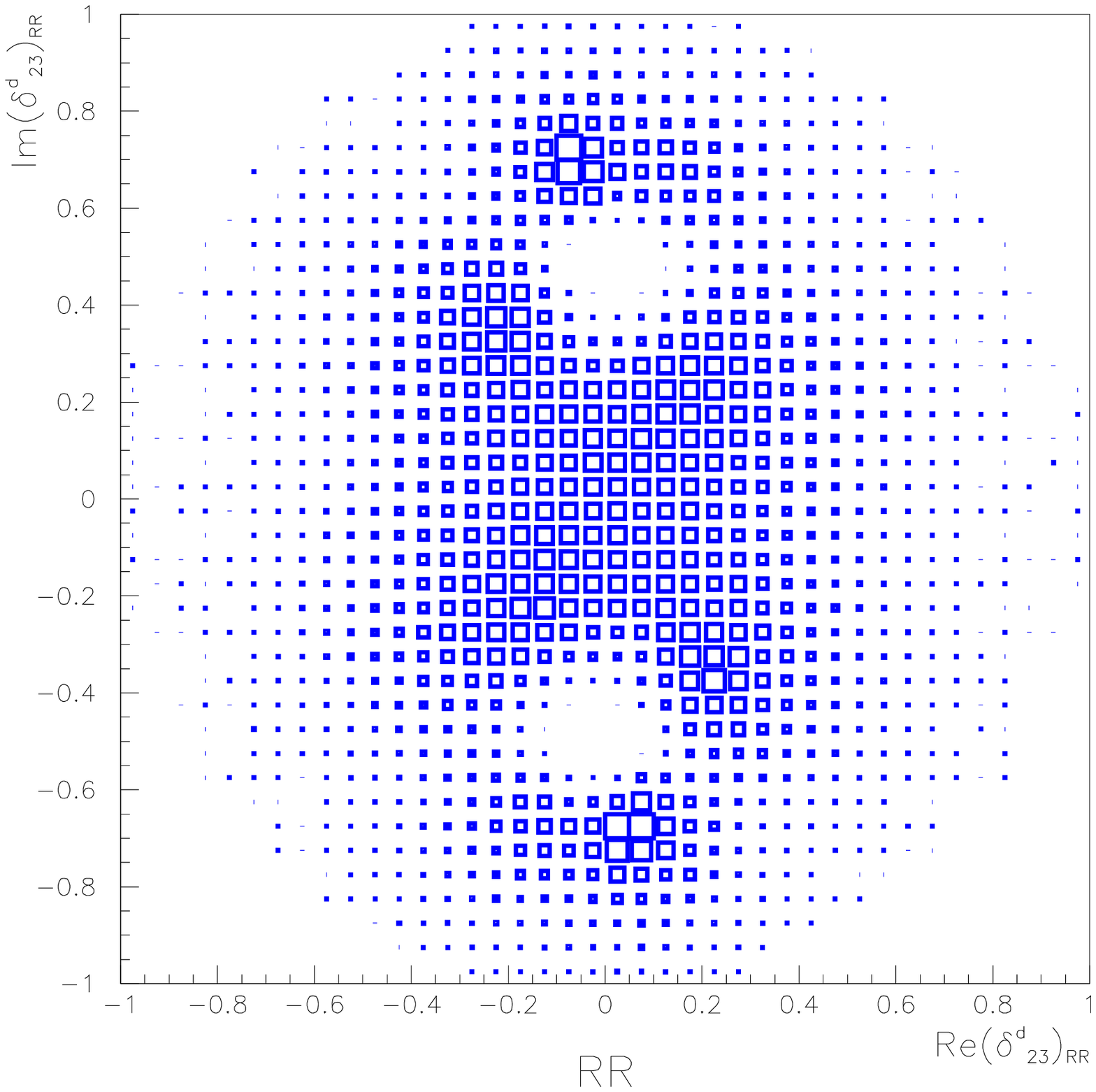} \\
\includegraphics[width=0.45\textwidth]{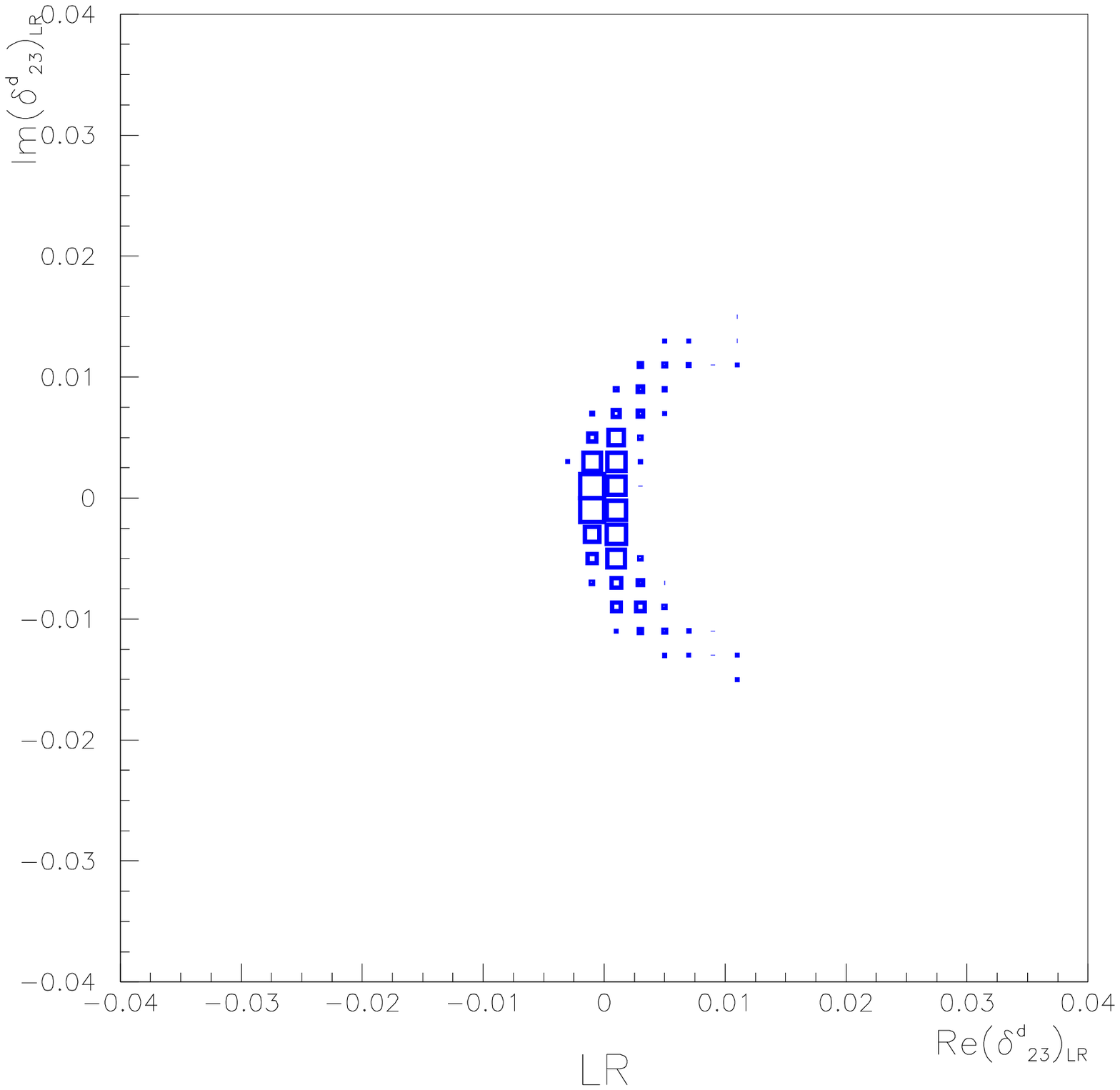} 
\includegraphics[width=0.45\textwidth]{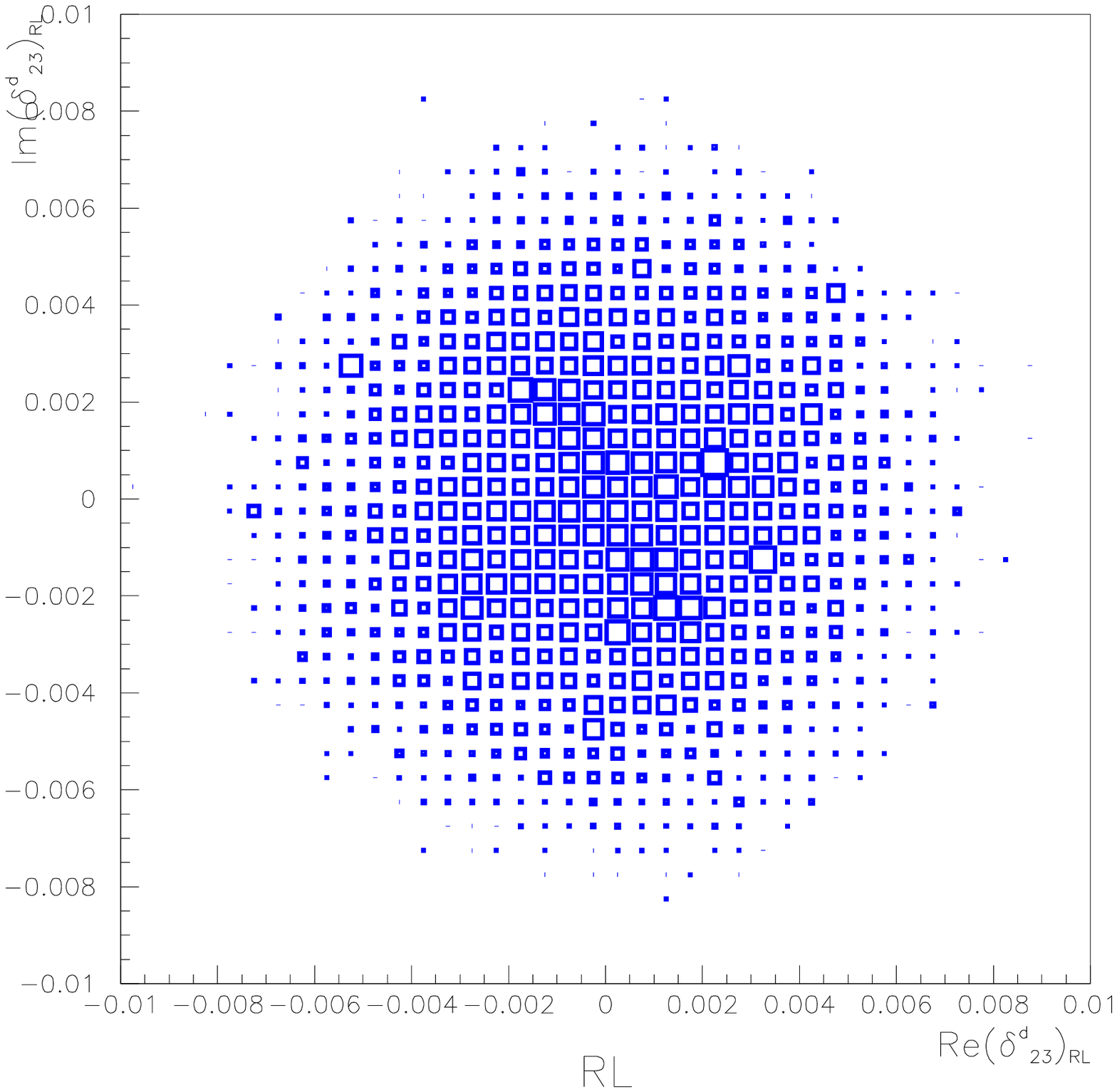} \\
\includegraphics[width=0.45\textwidth]{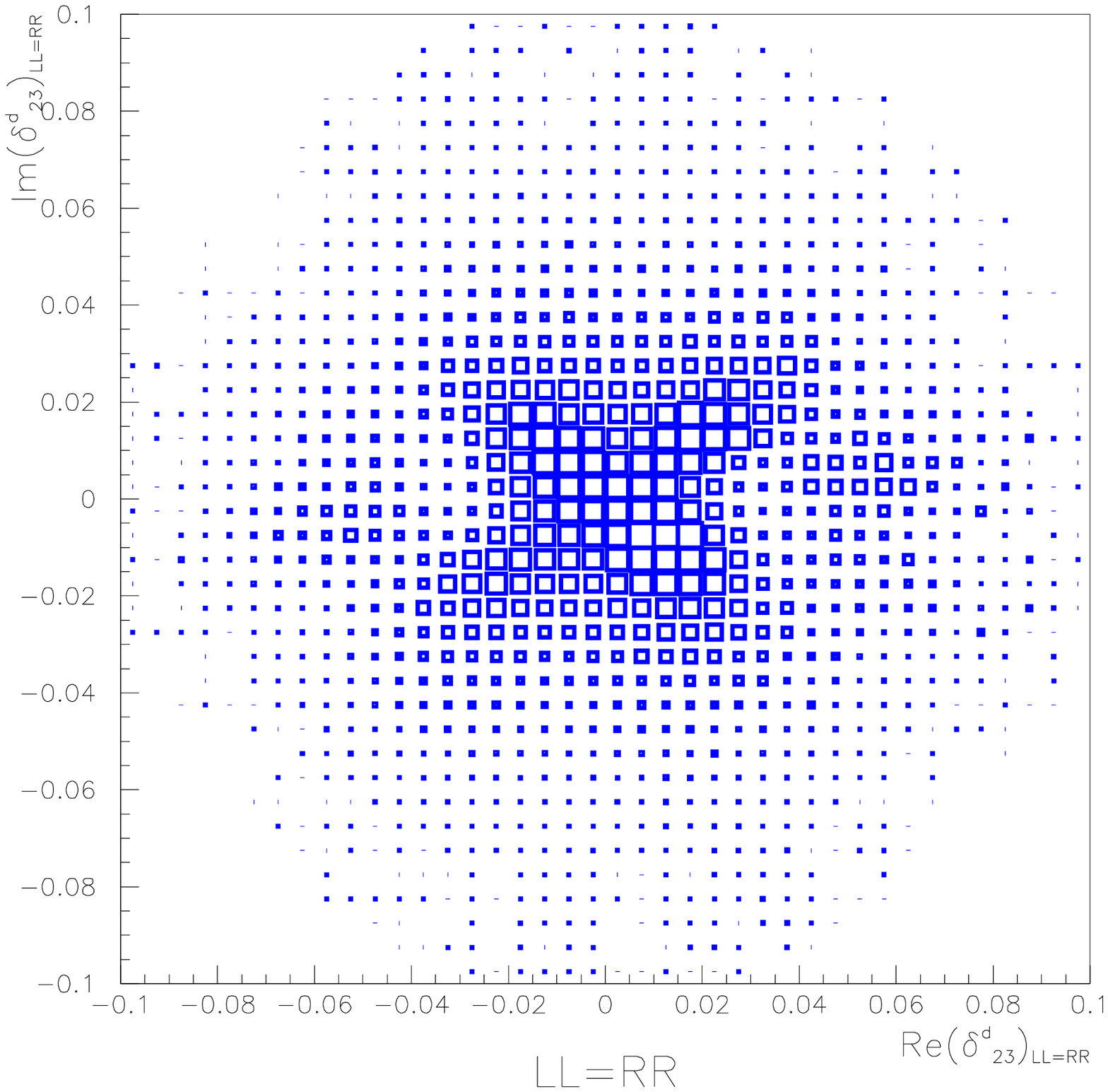}
\caption{Allowed range in the
Re$\left(\delta^d_{23}\right)_{AB}$-Im$\left(\delta^d_{23}\right)_{AB}$ plane,
with $AB=LL$ (top left), $AB=RR$ (top right), $AB=LR$ (middle left), $AB=RL$
(middle right) and $AB=LL$ with $\left(\delta^d_{23}\right)_{LL} =
\left(\delta^d_{23}\right)_{RR}$ (bottom).}
\label{fig:susy}
\end{center}
\end{figure*}

For definiteness, we present here the results obtained by choosing an average
squark mass of $350$ GeV, a gluino mass of $350$ GeV, $\mu=-350$ GeV and $\tan
\beta=3$. The dependence on $\mu$ and on $\tan \beta$ is induced by the
presence of a chirality flipping, flavour conserving mass insertion 
proportional
to $\mu\tan \beta$. In Fig.~\ref{fig:susy}, we show the allowed ranges in the
Re$\left(\delta^d_{23}\right)_{AB}$-Im$\left(\delta^d_{23}\right)_{AB}$ planes.
The corresponding upper bounds at 95\% probability are presented in
Table~\ref{tab:susy}.
\begin{table}[h]
\begin{center}
\caption{Upper bounds at 95\%  probability on the mass insertion parameters
$\vert\left(\delta^d_{23}\right)_{AB}\vert$, see the text for details.}
\label{tab:susy}
\begin{tabular}{|c|c|c|c|}
\hline
$ \left\vert\left(\delta^d_{23}\right)_{LL}\right\vert$ &
$ \left\vert\left(\delta^d_{23}\right)_{RR}\right\vert$ &
$ \left\vert\left(\delta^d_{23}\right)_{LL=RR}\right\vert$ &
$ \left\vert\left(\delta^d_{23}\right)_{LR,RL}\right\vert$ \\ 
\hline {\phantom{\huge{l}}}
$2\cdot 10^{-1}$ & $7 \cdot 10^{-1}$ & $5\cdot 10^{-2}$ & $5 \cdot 10^{-3}$\\
\hline
\end{tabular}
\end{center}
\end{table}

One finds that the constraints on $\left(\delta^d_{23}\right)_{LL}$ and
$\left(\delta^d_{23}\right)_{LL}= \left(\delta^d_{23}\right)_{RR}$ come from the
interplay of $B_s - \bar B_s$ mixing with $b\to s$ decays.
$\left(\delta^d_{23}\right)_{RR}$ is dominated by the information on $B_s - \bar
B_s$ mixing, while $\left(\delta^d_{23}\right)_{LR}$ and
$\left(\delta^d_{23}\right)_{RL}$ are dominated by $\Delta B=1$ processes.

\subsubsection{$B_s - \bar B_s$ in SUSY with minimal flavour violation}

As a second model-specific case for meson mixing we mention that of SUSY with
Minimal Flavour Violation (MFV). The MFV scenario is defined, in general, 
within
the effective field theory approach of ref.~\cite{D'Ambrosio:2002ex}. In the
specific case of SUSY, the soft squark mass terms, parametrised in the previous
section in terms of mass insertions, are expanded in terms of the SM Yukawa
couplings \cite{D'Ambrosio:2002ex,Hall:1990ac} and the relevant parameters
become the expansion coefficients. A detailed meson mixing study within this
approach has been performed in ref.~ \cite{Altmannshofer:2007cs} and for low
$\tan \beta$ shows that: (i) NP contributions are {\em naturally} small, for
$\Delta M_s$ of the order of $1/$ps; (ii) such contributions are always
positive; (iii) if $\mu$ is not small, gluino contributions enhance (even for
low $\tan \beta$) scalar operators, which then spoil the phenomenological
picture of (V-A)$\times$(V-A) dominated MFV \cite{Buras:2000dm}. In particular
item (i) emphasises the importance of precision determinations for lattice
parameters like $\xi$, if NP is of minimal flavour violating nature.

\subsubsection{Present experimental situation}
\label{sec:EXP}

New information concerning the \Bs mixing parameters became available 
during the workshop
{\it Flavour in the Era of the LHC}. The highlight was the measurement
of \dms by D\O\ and CDF. The D\O\ experiment used the semileptonic \bsmunux
decays with \dsphipi, and determined a 90\% confidence range for \dms : $ 17 <
\Delta m_s < 21\ {\rm ps}^{-1} $. The initial CDF result yielded a 3$\sigma$
observation of \bsmix mixing by making use of semileptonic and hadronic decay
modes \cite{Abulencia:2006mq}. Shortly after CDF published an improved analysis
\cite{Abulencia:2006ze}. In this analysis the signal yield was increased by
improving the particle identification and by using a neural network for the
event selection, which allows the use of additional decay modes. Moreover the
flavour tagging was improved by adding an opposite-side flavour tag based 
on the
charge of the kaons, and by the use of a neural network for the combination of
the kaon, lepton and jet-charge tags. The result for \dms equals 
\begin{equation}\label{bsmixexp}
\Delta m_s = 17.77 \pm 0.010 \pm 0.07\ {\rm ps}^{-1}. 
\end{equation}
The probability that a statistical fluctuation would
produce this signal is $8 \times 10^{-8}$ ($> 5\sigma$ evidence). 
This value for \dms is consistent with the SM expectation, 
see eq.~(\ref{eq:dms-sm}). The ratio
$|V_{td}/V_{ts}|$ was determined by CDF as well \cite{Abulencia:2006ze}, and
equals $0.2060 \pm 0.0007 (\dms) \apm{0.0081}{0.0060} (\dmd + {\rm theory})$.

Also information on the \Bs mixing phase became available \cite{Abazov:2007tx}. 
The D\O\ experiment performed two independent measurements of 
$A^{s}_{SL}$, defined in (\ref{eq:A_SL_s}), using
the same sign di\-muon pairs~\cite{Abazov:2006qw} and time-integrated
semileptonic decays $B_{s} \to \mu\nu D_{s}$ with $D_{s} \to
\phi\pi$~\cite{Abazov:2007nw}.

The same sign dimuon asymmetry in $B$ decays at Tevatron can be expressed
as~\cite{Grossman:2006ce}:
\begin{eqnarray}
A_{SL} & = & \frac{N(b\bar{b}\to \mu^{+} \mu^{+} X) - 
N(b\bar{b} \to \mu^{-} \mu^{-} X)}{N(b\bar{b}
\to \mu^{+} \mu^{+} X) + N(b\bar{b} \to \mu^{-} \mu^{-} X} 
= \frac{f_{d}Z_{d}A^{d}_{SL}+f_{s}Z_{s}A^{s}_{SL}}{f_{d}Z_{d}+f_{s}Z_{s}},
\label{eq:A_SL} \\
Z_{q} & = & \frac{1}{1-y^{2}_{q}} - \frac{1}{1+x^{2}_{q}}\ , \qquad
x_{q} = \Delta M_{q} / \Gamma_{q}\ , \qquad  y_{q} = \Delta\Gamma_{q}/
(2\Gamma_{q}). \nonumber
\end{eqnarray}
Here $f_{d} = 0.398\pm 0.012$ and $f_{s} = 0.103 \pm 0.014$ are the  \Bd  and
\Bs fragmentation fractions. The measured asymmetry $A_{SL}$ was presented by
D\O\  in Ref.~\cite{Abazov:2006qw}:
\begin{equation}
A_{SL}(\mathrm{D\O} )= A^{d}_{SL}+\frac{f_{s}Z_{s}}{f_{d}Z_{d}}A_{SL}^{s}
=-0.0092\pm 0.0044(stat.)\pm 0.0032(syst.).
\label{eq:A_SL_D0_1}
\end{equation}
Measurements of $A_{SL}^{d}$ were performed by the $b$ factories. The average
value of $A_{SL}^{d}$ is~\cite{Grossman:2006ce}:\\ 
\begin{equation}
 A_{SL}^{d} = +0.0011 \pm 0.0055\; .   
\end{equation}
This leads to the value of $A_{SL}^{s}$ from the same sign dimuon
asymmetry: 
\begin{equation}
 A_{SL}^{s} = -0.0064 \pm 0.0101 \;. 
\end{equation}
Recently D\O\ has also presented a time-integrated direct measurement of $A{SL}^{s}$  
using semileptonic $B_s \to D^{\pm}\mu^{\mp}\nu_{\mu}$ decays~\cite{Abazov:2007nw}.
They measure:
\begin{equation}
A_{SL}^{s} = +0.0245\pm 0.0193(stat.) \pm 0.0035(syst.).
\label{eq:A_SL_S_2}
\end{equation}
These two measurements of $A_{SL}^{s}$ are independent and their combination
gives the charge asymmetry in semileptonic \Bs decays: 
$ A_{SL}^{s} = 0.0001 \pm 0.0090$~\cite{Abazov:2007zj}.
The analysis of the time-dependent angular distributions in  
$B_{s}\to J/\psi \phi$ decays yields both the decay width difference
$\Delta\Gamma_{s}$ and CP violating phase $\phi_{s}$~\cite{Abazov:2007tx}:
\begin{eqnarray}
\Delta\Gamma_{s} & = 0.17 \pm 0.09 \pm 0.03~\mathrm{ps}^{-1} \; , \nonumber\\
\phi_{s}  & =- 0.79\pm 0.56 \pm 0.01\;.
\end{eqnarray}
 Combining the results for $ A_{SL}^{s}$,
$\Delta\Gamma_{s}$, $\phi_{s}$ and using the CDF result on the mass difference
\dms~\cite{Abulencia:2006ze} gives an improved estimate for $\phi_{s}$ and
$\Delta\Gamma_{s}$~\cite{Abazov:2007zj}:
: 
\begin{eqnarray}
\Delta\Gamma_{s} & = 0.13 \pm 0.09 ~\mathrm{ps}^{-1} \; , \nonumber\\
\phi_{s}  & =- 0.70^{+0.47}_{-0.39}\;. 
\end{eqnarray}
Also new results have been released recently concerning the \Bs lifetime and
\dg. At D\O\ the \Bs lifetime for \bsmunux was measured to be $1.398 \pm
0.044(stat) \apm{0.028}{0.025} (sys)$ ps$^{-1}$~\cite{Abazov:2006cb}. The
average \Bs lifetime equals $1.466 \pm 0.059$ ps$^{-1}$~\cite{Yao:2006px}. 
CDF published the measurement of 
$\dg=0.47\apm{0.19}{0.24} (stat) \pm 0.01 (sys)\, {\rm ps}^{-1}$ 
\cite{Acosta:2004gt}.


In the near future the LHC experiments LHCb, ATLAS and CMS
will start to provide information on \bsmix mixing. 
In the following sections the sensitivity of LHCb  
to the \Bs mixing parameters \dms, \dg, \phis and \asl 
and the prospects for CMS will be discussed.

\subsubsection{LHCb}

The LHCb experiment is designed as a single-arm forward spectrometer to study
$b$ decays and CP violation. Its main characteristics are precise vertexing,
efficient tracking and good particle identification. The high-precision
measurements at LHCb will enable further tests of the CKM picture, and probe
physics beyond the SM. This is in particular true for the measurement of \bsmix
mixing parameters such as \dms, \dg, $\Pphi_\Ps$ and \asl.

LHCb will run at a nominal luminosity of  ${\mathcal L} =
2\times10^{32}\:\text{cm}^{-2}\text{s}^{-1}$. Assuming a $b\bar b$ production
cross-section of $\sigma_{\Pb\Pab} = 500\:\mu\text{b}$, this will correspond to
an integrated luminosity of $2\:\text{fb}^{-1}$ per nominal year of
$10^7\:\text{s}$ of data taking. All event yields quoted below are for
$2\:\text{fb}^{-1}$. They have been obtained from a full Monte Carlo (MC)
simulation of the experiment, which included the following: pileup generation,
particle tracking through the detector material, detailed detector response
(including timing effects such as spillover), full trigger simulation, offline
reconstruction with full pattern recognition, and selection cuts.
High-statistics samples of signal events have been produced for a detailed 
study of resolutions and efficiencies. Combinatorial background has been 
studied using
a sample of $\sim 27$M inclusive $b\bar b$ events corresponding to about 10
minutes of data taking, while identified physics background sources have been
studied with large specific background samples.

\subsubsubsection{Sensitivity to \dms from \bsdspi}

\begin{figure}[ht]
\begin{center}
\includegraphics[width=0.48\textwidth]{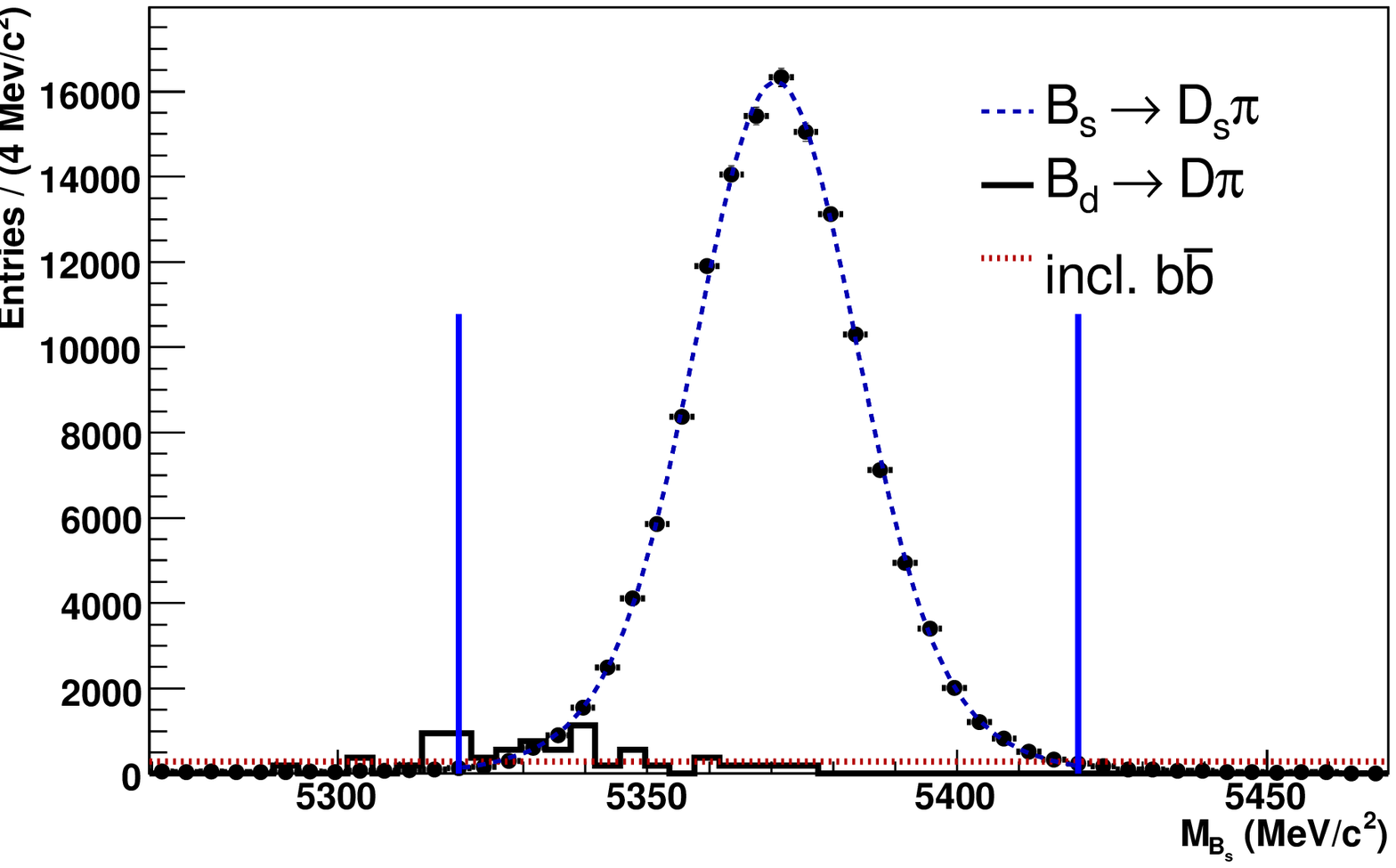} \hfill
\includegraphics[width=0.48\textwidth]{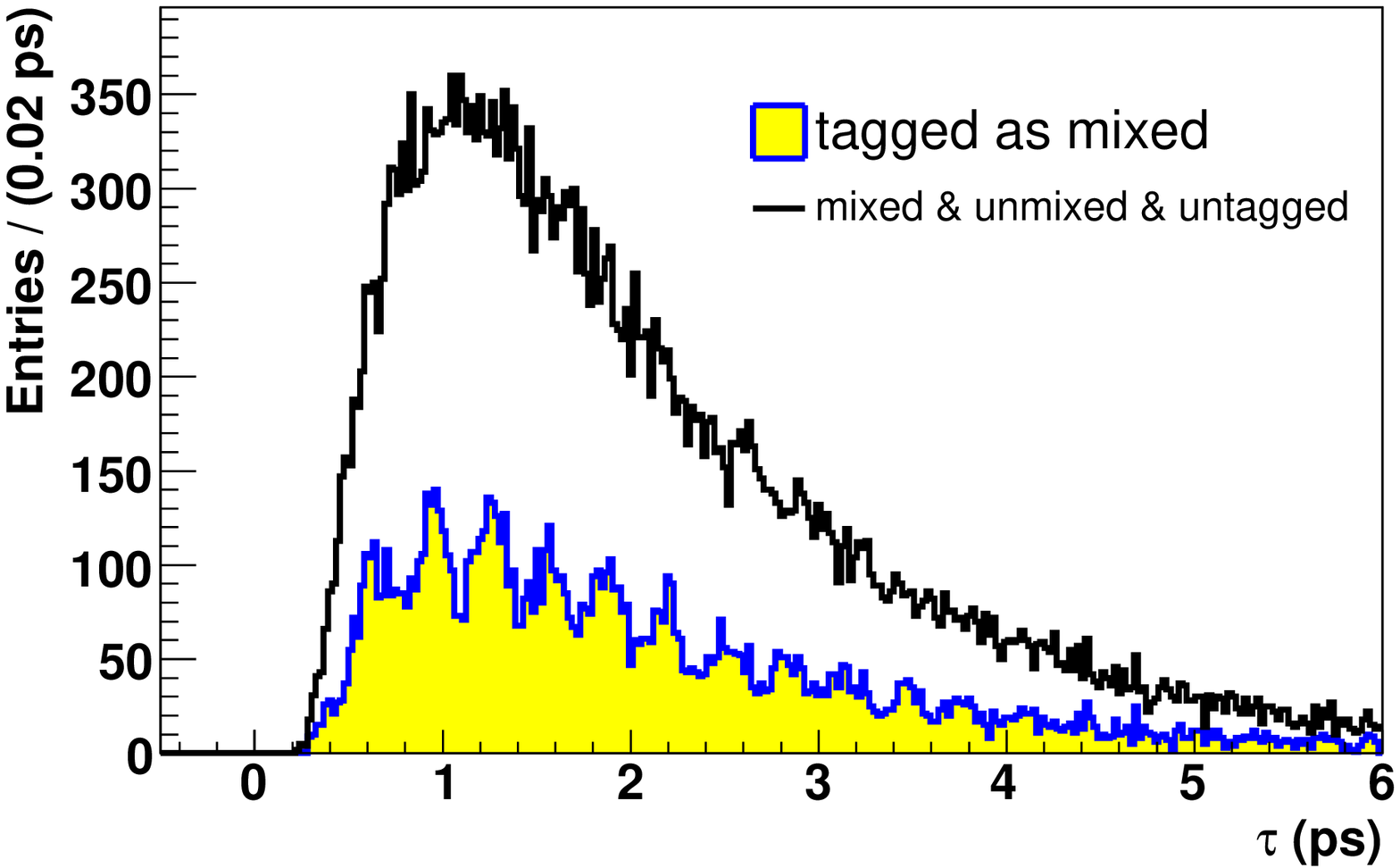}
\caption{\underline{Left}: Reconstructed \bsdspi mass distribution from full MC
simulation, after trigger and all selection cuts~\cite{bib:LHCb-2007-xxx_Borel}.
The points with error bars represent the signal (on an arbitrary vertical scale).
The histogram represents the $B\to D^-\pi^+$ background and the dotted flat
line represents the upper limit of the combinatorial background from $b\bar b$
events, normalised to the signal.
\underline{Right}: Reconstructed \bsdspi proper time distribution from full MC
simulation of the signal, corresponding to an integrated luminosity of
$0.5\:\text{fb}^{-1}$~\cite{bib:LHCb-2007-xxx_Borel}. The lower histogram
represents the events tagged as mixed. The background is not shown.}
\label{fig:LHCb/BsDspi_mass}
\end{center}
\end{figure}

The mass difference \dms between the mass eigenstates of the \bsmix system is
best measured as the frequency of the oscillatory behaviour of the proper time
distribution of flavour-tagged \PBs mesons decaying to a flavour-specific final
state. The best channel for this at LHCb is \BsDspi, with the subsequent 
$D^+_s$ decay to $K^+K^-\pi^+$, because of its easy topology with four 
charged tracks and its relatively large branching fraction of 
$B(\BsDspi)\times B(D^+_s\to K^+K^-\pi^+) = (1.77 \pm 0.48)\times
10^{-4}$~\cite{bib:LHCb-2007-xxx_Borel}. 
Such decays can be detected, triggered,
reconstructed and selected with a final mass resolution of $\sim 14$~MeV$/c^2$
(see Fig.~\ref{fig:LHCb/BsDspi_mass} left) and a total efficiency of about
0.4\%, leading to a yield of $\rm (140k \pm 40k)$ events in $2\:\text{fb}^{-1}$.
After the trigger and selection, the combinatorial background is expected to be
dominated by $b\bar b$ events, and has been estimated to be less than 5\% of the
signal at 90\% CL, in a $\pm 50$~MeV/$c^2$ mass window around the signal. Using
the same sample of simulated $b\bar b$ events, the background from partially
reconstructed $b$-hadron decays in the same mass window has been estimated 
to be less than 40\% at 90\% CL. 
This includes partially reconstructed $\Lambda_b$
and $B_d$ decays. A dedicated study showed that the background from
$B\to D^-\pi^+$ decays (where one of the charged pions from the $D$ decay
could be misidentified as a kaon) is approximately 5\% of the signal.

The proper time resolution, obtained on an event-by-event basis from the
estimated tracking errors, typically varies between 15~fs and 80~fs with an
average value of $\sim 40$~fs (dedicated studies are being done at LHCb to 
model the proper time resolution~\cite{LHCb_Peter} and to verify the 
estimated tracking errors~\cite{LHCb_Stefania} with data).
A flavour tagging power of $\epsilon D^2$ of at least 9\% 
is achieved on the MC signal, 
combining several tags in a neural network: a muon or electron from the
$b\to\ell$ decay of the other $b$-hadron, a charged kaon from the
$b\to c\to s$ decay of the other $b$-hadron, the vertex charge of the other
$b$-hadron, and a charged kaon accompanying the signal $B_s$ in the 
fragmentation chain~\cite{bib:lhcb-2007-058}.

The statistical uncertainty on the measurement of \dms using an integrated
luminosity of $2\:\text{fb}^{-1}$  is expected to
be $\pm 0.007$~ps$^{-1}$~\cite{DSKRECSENS}. 
It will be dominated by systematic uncertainties
related to the determination of the proper time scale.
Figure~\ref{fig:LHCb/BsDspi_mass} (right) shows the proper time distribution
from which such a measurement could be extracted. 

The \BsDspi sample will play a crucial role as a control sample in all
time-dependent \PsB analyses; indeed it can be used to measure directly the
dilution (due to flavour tagging and proper time resolution) on the $\sin(\dms
t)$ and $\cos(\dms t)$ terms in time-dependent CP asymmetries. It will also be
used as a normalisation channel for many measurements of $B_s$ branching 
fractions.
More details on the selection of  \BsDspi events can be found in 
Ref.~\cite{bib:LHCb-2007-xxx_Borel}.

\subsubsubsection{Sensitivity to \phis and \dg from exclusive \bccs decays}

The \bsmix mixing phase $\phi_s$ can be measured from the
flavour-tagged \Bs decays to CP eigenstates involving the \bccs quark-level
transition. The best mode for this at LHCb is \Bsjpphi. However, in this case,
the vector nature of the two particles in the final state causes their relative
angular momentum to take more than one value, resulting in a mixture of CP-even
and CP-odd contributions. An angular analysis is therefore required to separate
them on a statistical basis. This can be achieved with a simultaneous fit to the
measured proper time and so-called transversity angle of the reconstructed
decays. Such a fit is sensitive also to \dg because of the presence of the two
CP components.

The sensitivity to $\phi_s$ has been studied so far with the following modes:
\begin{itemize}
\item
$B_s\to J/\psi(\mu^+\mu^-)\phi(K^+K^-)$~\cite{bib:Fernandez_thesis,
bib:LHCb-2006-047_LF}
\item $B_s\to\eta_c(\pi^+\pi^-\pi^+\pi^-, \pi^+\pi^- K^+K^-,
K^+K^-K^+K^-)\phi(K^+K^-)$~\cite{bib:Fernandez_thesis,
bib:LHCb-2006-047_LF}
\item $B_s\to J/\psi(\mu^+\mu^-)\eta(\gamma\gamma,
\pi^+\pi^-\pi^0)$~\cite{bib:Fernandez_thesis,bib:LHCb-2006-047_LF}
\item $B_s\to J/\psi(\mu^+\mu^-)\eta'(\eta(\gamma\gamma)\pi^+\pi^-,
 \rho(\pi^+\pi^-)\gamma)$~\cite{bib:LHCb-2007-xxx_DV,bib:LHCb-2007-xxx_SJO}
\item
$B_s\to D^+_s(K^+K^-\pi^+)D^-_s(K^+K^-\pi^-)$~\cite{bib:Fernandez_thesis,
bib:LHCb-2006-047_LF}
\end{itemize}

\begin{table}[tbp]
\begin{center}
\caption{Characteristics of different exclusive \bccs modes for the measurement
of $\phi_s$. The first 6 columns of numbers are obtained from the full MC
simulation. They represent the expected number of triggered, reconstructed and
selected signal events with an integrated luminosity of 2\:$\text{fb}^{-1}$
(before tagging), the background-over-signal ratio determined mainly from
inclusive $b\bar b$ events, the \Bs mass resolution, the average value of the
estimated event-by-event \Bs proper time error scaled by the width of its pull
distribution, the flavour tagging efficiency, and the mistag probability. These
parameters have been used as input to a fast MC simulation to obtain the
sensitivity on $\phi_s$ given in the last column. The last line describes the
control channel (see text).}
\begin{tabular}{|l|r|l|c|c|c|c|c|}
\hline
Channel &  2\:$\text{fb}^{-1}$ & $B/S$ & 
$\sigma_{\rm mass}$ & $\sigma_{\rm time}$ & 
$\epsilon_{\text{tag}}$ & $\omega_{\text{tag}}$ & $\sigma(\phi_s)$ \\
& yield && [~\MeVctwo~] & [~fs~] &   [~\%~] & [~\%~] & [~rad~]  \\
\hline
\Bsjpphi          & 131k & 0.12 & 14 & 36 & 57 & 33 & 0.023 \\
\Bsetacphi        &   3k & 0.6  & 12 & 30 & 66 & 31 & 0.108 \\
\Bsjpetagg        & 8.5k & 2.0  & 34 & 37 & 63 & 35 & 0.109 \\
\Bsjpetappp       & 3k   & 3.0  & 20 & 34 & 62 & 30 & 0.142 \\
\Bsjpetaprimetapp & 2.2k & 1.0  & 19 & 34 & 64 & 31 & 0.154 \\
\Bsjpetaprimrhog  & 4.2k & 0.4  & 14 & 29 & 64 & 31 & 0.080 \\
\BsDsDs           & 4k   & 0.3  &  6 & 56 & 57 & 34 & 0.133 \\
\BsDspi           & 140k & 0.4  & 14 & 40 & 63 & 31 & ---   \\
\hline
\end{tabular}
\label{tab:LHCb_phis}
\end{center}
\end{table}

\begin{figure}[ht]
\begin{center}
\begin{tabular}{c}
\includegraphics[width=0.80\textwidth]{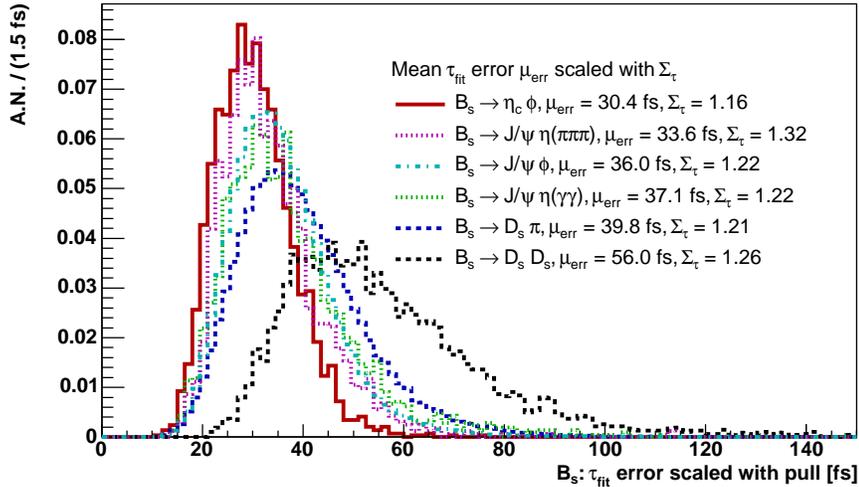} \hfill
\end{tabular}
\caption{Distribution of the event-by-event proper time resolution [fs]
for different $B_s$ channels, as obtained
from the full MC simulation. The normalisation is arbitrary.}
\label{fig:sensitivity/TauBsErr_All}
\end{center}
\end{figure}

The results are summarised in Table~\ref{tab:LHCb_phis}. For each signal event
in the full simulation the proper time and its error are estimated using a
least-squares fit. The distributions of the proper time errors (scaled with the
sigma of their pull distribution) are shown in
Fig.~\ref{fig:sensitivity/TauBsErr_All}. Most channels have a proper time
resolution below 40\:fs. A good proper time resolution is important 
for resolving the fast \bsmix oscillations.

The sensitivities to the \bsmix mixing parameters are determined by means
of fast parameterised simulations, 
with the results of Table~\ref{tab:LHCb_phis}
as inputs. A large number of experiments are generated assuming the following
set of parameters: $\Delta m_s = 17.5\ \text{ps}^{-1}$, $\phi_s =
-0.04\ \text{rad}$, $\Delta \Gamma_\Ps/\Gamma_{\Ps} = 0.15$, $1/\Gamma_{\Ps} =
1.45\ \text{ps}$, and a fraction of CP-odd component of $R_T = 0.2$ (for
\Bsjpphi). The different parameters are extracted by performing a 
likelihood fit
to the mass, proper time, and transversity angle (for \Bsjpphi) distributions,
including a background contribution. The \bccs likelihood is simultaneously
maximised with a similar likelihood for the \BsDspi control sample such as to
constrain \dms and the mistag fraction from the data. The background properties
are determined from the $B_s$ mass sidebands. The physics parameters, extracted
in the signal region with all other parameters fixed, are \phis, \dms,
$\Delta \Gamma_\Ps/\Gamma_{\Ps}$, $1/\Gamma_\Ps$, $\omega_{\text{tag}}$, and
$R_{T}$ (for \Bsjpphi).

The sensitivities to \phis for the different channels, obtained as the rms
of the distribution of the fit results, are given in the last column of
Table~\ref{tab:LHCb_phis}. They gently decrease with increasing $|\phi_s|$,
and do not depend much on $\Delta \Gamma_\Ps/\Gamma_{\Ps}$. For instance, the
statistical uncertainty on \phis for $\phi_s = -0.2\:\text{rad}$ is
$\pm 0.026\:\text{rad}$ from \Bsjpphi alone, with
$2\:\text{fb}^{-1}$~\cite{bib:Fernandez_thesis}. The best performance is
achieved with the \Bsjpphi sample, which also yields a statistical precision of
$\pm 0.0092$ on $\Delta \Gamma_\Ps/\Gamma_{\Ps}$ (2\:$\text{fb}^{-1}$). The
$\phi_s$ sensitivities obtained from the other modes (which are pure
CP-eigenstates) are not as good, but still interesting. Combining all modes, a
statistical uncertainty $\sigma(\phi_s) = \pm 0.0092\:\text{rad}$ is 
expected after $10\:\text{fb}^{-1}$.

LHCb has the potential to perform the first significant measurement of 
$\phi_s$, test the consistency with the SM expectations, and possibly
uncover New Physics that may be hiding in \bsmix mixing.

\subsubsubsection{Sensitivity to $A^{s}_{SL}$ from \bsmunux and \bsdspi }

The CP-violating charge asymmetry $A^{s}_{SL}$ is an important parameter 
to constrain new physics contributions in $B_s$ mixing, 
see Section~\ref{sec:NP}. 
$A^s_{SL}$ 
is accessible by measuring the charge 
asymmetry of the time-integrated rates of untagged $B_s$ 
decays to flavour-specific final 
states such as $D_s^- \mu^+ \nu X$ or $D_s^-\pi^+$~\cite{Nierste:2004uz}. 
In LHCb the asymmetry $A^s_{SL}$ is measured by  
fitting the time-dependent decay rates.
This method allows a determination of $A^s_{SL}$ also 
for a non-zero production asymmetry of 
$B_s$ and $\bar {B}_s$ mesons
which, at the LHC, is expected to be of ${\cal O}(1\%)$. 
Based on a large sample of fully 
simulated inclusive $b\bar{b}$ events and a dedicated signal sample, 
LHCb estimates 
a signal yield of 1M \bsmunux events in $2\:\text{fb}^{-1}$ of data, 
with a $B/S$ ratio of about 0.36~\cite{Xie1}. 
This leads to a statistical precision of 
$\pm 0.002$ on $A^s_{SL}$~\cite{Xie2}.
A similar analysis based on 140k \bsdspi events is expected 
to reach a precision of $\pm 0.005$ with the 
same integrated luminosity of $2\:\text{fb}^{-1}$~\cite{Xie2}. 
Systematic uncertainties are 
expected to be dominated by the detector charge asymmetry, 
which needs to be 
determined separately. 
A method is proposed to control the  detector charge asymmetry
by measuring the 
difference $A^s_{SL}-A^d_{SL}$ using \Bs and \Bd decays to the same final state, e.g. 
 $\Bs \rightarrow D_s^-\mu^+\nu X$ and $\Bd \rightarrow D^-\mu^+\nu X$, where 
$D_s^- \rightarrow K^+K^-\pi^-$ and $D^- \rightarrow K^+K^-\pi^-$.

\subsubsubsection{Correcting for trigger biases in lifetime fitting at LHCb}

Lifetime measurements at LHCb will help for the detector calibration and 
provide tests of theoretical predictions based on the heavy-quark expansion. 
In order to exploit the full range of decays available at LHCb, it is 
important to have a
method for fitting lifetimes in hadronic channels, which are biased by the
impact parameter cuts in the trigger. We have investigated a Monte-Carlo
independent method to take into account the trigger effects. 
The method is based
on calculating event-by-event acceptance functions from the decay geometry and
does not require any external input. Current results with the method are given
in ~\cite{vava-lhcb1}. The method is described, for the case of two-body 
decays, in~\cite{Rademacker:2005ay}.

The decay $B_{d} \to D^{-} \pi^{+}$ has an expected yield of 1.34M 
events per $2\:\text{fb}^{-1}$. 
The $S/B$ ratio is expected to be around 5~\cite{DPIREC}.
Fitting the $B_d$ lifetime with 60k toy Monte Carlo signal events achieves a
statistical precision of 0.007~ps, while fitting to 60k signal and 15k 
background
events achieves a precision of 0.009~ps (the current world average is 1.530 $\pm$
0.009 ps \cite{Yao:2006px}). A similar result is seen in data generated with the
full LHCb detector simulation~\cite{vava-lhcb1}. Therefore, although the
systematic errors associated with this method are unknown at the moment, we can
expect a very good measurement of the \Bd lifetime using the decay
$B_{d}\to D^{-} \pi^{+}$.


\subsubsection{CMS}

\subsubsubsection{Sensitivity to \dg}

Also at CMS the decay
$B_s \to J/\psi \, \phi \to \mu^+\mu^- K^+K^-$ 
is being studied~\cite{CMS_NOTE_2006-090}. Several important
background processes have been identified. The prompt \jpsi production is the
main source of background at trigger level, since it represents a dominant
contribution to the Level-1 dimuon trigger rate. For the off\-line selection, 
the main background is the inclusive decay $b\to\jpsi\, X$. 
The decay $B_d \to J/\psi \, K^{*0}\to \mu^+\mu^- K^+ \pi^-$ 
is of particular concern,
since the pion can be mistaken to be a kaon, and hence the decay be
misidentified as \bspsiphi\ . Furthermore, the final state of
this \Bd decay also displays a time-dependent angular distribution similar to
that of the \Bs decay under study, 
with different physical parameters. The \Bs decay
chain is selected at Level-1 by the dimuon trigger. 
The latter demands two muons
with a transverse momentum above $3\GeVc$, and the additional requirement that
these muons have opposite charge can be used.

In the HLT~\cite{DAQTDR}, $b$ candidates are identified by doing a partial
reconstruction of the decay products in the tracker in restricted tracking
regions and imposing invariant mass and vertex requirements~\cite{PTDR1}.

The HLT selection of the decay \bspsiphi\ has been separated in two steps. In
the first, called Level~2, \jpsi candidates with a displaced vertex are
identified. Tracks are then reconstructed in the tracking regions defined 
by the
Level~1 muon candidates, and all track pairs of opposite charge for which the
invariant mass is within $150\MeVcc$ of the world-average \jpsi mass are
retained. To remove the prompt \jpsi background, the two muon candidates are
then fitted to a common decay vertex and the significance of the transverse
decay length is required to be above $3$. With this selection, 
the accepted rate is reduced to approximately 15~Hz, with 80\% of the 
\jpsi originating in the decay of $b$ hadrons.

Next, at Level~3, a further reduction is achieved by doing a full
reconstruction of the \Bs decay. To reconstruct the kaons, the tracking region
is chosen around the direction of each \jpsi candidate. Assigning the kaon mass
to the reconstructed tracks, all oppositely charged track pairs for which the
invariant mass is within $20\MeVcc$ of the world-average mass of the $\phi$
meson are retained, for a resolution in the invariant mass of the $\phi$ meson
of $4.5\MeVcc$. With the two muon candidates, the four-track invariant mass is
required to be within $200\MeVcc$ of the world-average mass of the \Bs meson.
The resolution in the invariant mass of the \Bs meson is found to 
be $65\MeVcc$.
Here as well, a vertex fit of the four tracks is performed, imposing a similar
requirement as above. The total rate for this selection is well below 0.1~Hz,
and a yield of approximately $456000$ signal events can be expected within
$30\fbinv$ of data.

In the offline selection, candidates are reconstructed by combining two muons of
opposite charge with two further tracks of opposite charge. As CMS does not
possess a particle identification system suitable for this measurement, all
measured tracks have to be considered as possible kaon candidates, which adds a
substantial combinatorial background. A kinematic fit is made, where the four
tracks are constrained to come from a common vertex and the invariant mass of
the two muons is constrained to be equal to the mass of the \jpsi. With this
fit, a resolution on the invariant mass of the \Bs meson of $14\MeVcc$ is found.
The invariant mass of the two kaons is required to be within $8\MeVcc$ of the
world-average mass of the $\phi$ meson.

With this selection, a yield of approximately 327\,000 signal events can be
expected within $30\fbinv$ of data, with a background of 39\,000 events. These
do not include a requirement on the four-track invariant mass of the candidates,
since the sidebands could be used later in the analysis. However, only a small
fraction of these events are directly under the \Bs peak, and even a simple cut
will reduce the number of background events by a significant factor.

\providecommand{\apar}{\ensuremath{A_{\parallel}}}
\providecommand{\aperp}{\ensuremath{A_{\perp}}}
\providecommand{\azero}{\ensuremath{A_{0}}}
\providecommand{\gh}{\ensuremath{\Gamma_H}}
\providecommand{\gl}{\ensuremath{\Gamma_L}}

\providecommand{\psinv} {\mbox{\ensuremath{\mathrm{\; ps}^{-1}}}}
\providecommand{\pdl} {proper decay length}
\providecommand{\pdf}{p.d.f.}

The measurement of the width difference {\dg} can now be done on this sample of
untagged \Bs candidates. As mentioned earlier, the $J/ \psi\, \phi$ final state 
is an admixture of CP-even and CP-odd states, and an angular analysis is
required~\cite{Dunietz:2000cr}. As the CP-even and CP-odd components have
different angular dependences and different time evolutions, the different
parameters can be measured by performing an unbinned maximum likelihood fit on
the observed time evolution of the angular distribution. In the absence of
background and without distortion, the \pdf\ describing the data would be the
original differential decay rate. The distortion of this distribution by the
detector acceptance, trigger efficiency and the different selection criteria
must be taken into account by an efficiency function modelling the effect of 
the decay length requirements and the distortion of the angular distribution.

A sample corresponding to an integrated luminosity of $1.3\fbinv$ 
was considered, which allows us to have a realistic ratio of misidentified 
$B_d\to J/\psi\, K^*$ and signal events. 
With the low number of background events that remain after all
selection requirements, an accurate modelling of the background is not 
possible, neither of its angular distribution
nor of its time-dependent efficiency. Therefore the background events
are simply added to the data set and their expected distribution is not included
in the \pdf\ used in the fit. The \pdf\ then simply describes the \Bs
distribution. With such a fit, in which the invariant mass of the candidates is
not taken into account, a restriction on the invariant mass of the candidates
should obviously be made. Choosing a window of $\pm 36\MeVcc$ around the
world-average \Bs mass reduces the number of \Bd background events by another
59\%, while reducing the number of signal candidates by only 2.9\%. The
result of the fit is given in Table~\ref{tab:final}, where both the statistical
and expected systematic uncertainties are quoted. A first measurement of the
width difference of the weak eigenstates could thus be made with an
uncertainty of 20\%. On a larger sample, corresponding to an integrated
luminosity of $10\fbinv$, it is foreseen that the statistical uncertainty 
would be reduced to 0.011.

\begin{table*}[!tbh]
\caption{Results of the maximum likelihood fit for an integrated
luminosity of $1.3\fbinv$ (signal and background).}
\label{tab:final}
\begin{center}
\begin{tabular}{|c|c|c|c|c|c|c|c|}
\hline
Parameter	&Input value  & Result  & Stat. error & Sys. error & Total error
&  Rel. error \\
\hline
$|A_0(0)|^2$	& 0.57        &  0.5823   & 0.0061  &0.0152& 0.0163& 2.8\%\\
$|A_{||}(0)|^2$	& 0.217       &  0.2130   & 0.0077  &0.0063&0.0099& 4.6\%\\
$|A_\bot(0)|^2$	& 0.213       &  0.2047   &  0.0065&0.0099& 0.0118& 5.8\%\\
\gm		& 0.712\psinv & {0.7060}\psinv   & {0.0080}\psinv & 0.0227\psinv
& 0.0240\psinv &3.4\% \\
\dg		& 0.142\psinv & {0.1437 }\psinv  &{0.0255} \psinv & 0.0113\psinv
& 0.0279\psinv &19\% \\
\dgg		& 0.2         & 0.2036  & 0.0374 & 0.0173&0.0412&20\%\\
\hline
\end{tabular}
\end{center}
\end{table*}

\subsubsubsection{Missing particles in the reconstruction}

The best way to study the \bsmix oscillations is to have a fully
reconstructed final state of the \Bs decay. The disadvantage of such decay
channels is the limited statistics. Much more signal events can be 
collected in semileptonic decays as $\Bs \to \Dsm \ellp \nu$. 
Due to the missing
neutrino in this decay the \Bs momentum, and hence the proper-time resolution
for the \Bs, is less precise than in the fully reconstructed case, even if a
correction ($k$-factor) is applied.  
However, recently a new method ($\nu$-reco)
has been proposed \cite{Dambach:2006ha}, which allows us to calculate 
the neutrino momentum with the help of vertex information.

In order to verify the $\nu$-reco method a MC simulation has been 
developed to study \bsmix mixing in the semileptonic decay mode. 
Kinematical cuts, track parameters
and vertex positions (primary and secondary) have been simulated according to
typical hadron collider detector
conditions~\cite{Abe:1998cj,Abazov:2006cb,Starodumov:1997,atlas:1999fq}. 
The proper time resolution obtained is $\sigma = 132\fs$
with the $k$-factor method and $\sigma = 91\fs$ with the $\nu$-reco method.

%

\newpage \boldmath
\subsection{Hadronic $b\to s$ and $b\to d$ transitions}\label{sec:b2sandb2d}
\unboldmath%


Flavour-changing neutral current processes can occur only at the loop level in the Standard Model and therefore are potentially sensitive to new
virtual particles. In particular, hadronic FCNC $B$ decays are sensitive to new physics contributions to penguin operators. Among these decays, the penguin-dominated $b\to s\bar q q$ transitions are the most promising~\cite{Grossman:1996ke,Ciuchini:1997zp,London:1997zk}.
However, an accurate evaluation of the Standard Model amplitudes is required in order to disentangle new physics contributions.
Unfortunately hadronic uncertainties hinder a pristine calculation of 
the decay amplitudes. In this chapter, various theoretical approaches
to the calculation of the hadronic uncertainties are discussed. In addition, the present experimental status is presented together with prospects at $B$-factories and LHCb.

\subsubsection{Theoretical estimates of $\Delta S$ with factorization}
%
%
%


\noindent
In the following we quantify $\Delta S_f \equiv -\eta_f
S_f-\sin(2\beta)$, where $S_f$ is the sin-term of the 
time-dependent CP asymmetry, based on QCD factorization \cite{Beneke:1999br,Beneke:2000ry}
calculations of the $B\to f$ decay amplitudes. We may write the 
decay amplitude as
\begin{equation}
\label{ampl}
A(\bar B\to f) = V_{cb} V_{cs}^* \,a_f^c+ V_{ub} V_{us}^* \,a_f^u 
\propto 1 + e^{-i\gamma} \,d_f,
\end{equation}
where $d_f = \epsilon_{\rm KM} \,a_f^u/a_f^c \equiv 
\epsilon_{\rm KM} \hat d_f$ and 
$\epsilon_{\rm KM} = \left|V_{ub} V_{us}^*/(V_{cb} V_{cs}^*)\right| 
\sim 0.025$. The expectation that $\Delta S_f$ is small derives 
from the CKM suppression $\epsilon_{\rm KM}$ and the expectation 
that the ratio of hadronic amplitudes, $\hat d_f$, is not much 
larger than 1. Then 
\begin{equation}
\label{dSS}
\Delta S_f = 2 \,\epsilon_{\rm KM}\,
\mbox{Re}(\hat d_f) \cos(2\beta)\sin\gamma + O(d_f^2).
\end{equation}
QCD factorization calculations of $\Delta S_f$ for various final 
states have been performed at leading order \cite{Buchalla:2005us} 
and next-to-leading 
order \cite{Beneke:2003zv,Beneke:2005pu,Cheng:2005bg}. 
Other factorization-inspired 
calculations can be found in \cite{Williamson:2006hb, Li:2006jv}. 
The results are generally in good agreement with each other.
The following is primarily an update of \cite{Beneke:2005pu}. 
Ref.~\cite{Cheng:2005bg} also discusses an estimate of long-distance 
rescattering effects. Since the significance of the model underlying 
this estimate is unclear, these (small) effects will not be included here. 

\vskip0.2cm
The hadronic amplitudes $a_f^p$ are sums of 
``topological'' amplitudes, referring to colour-allowed tree ($T$), 
colour-suppressed tree ($C$), QCD penguin 
($P^p$), singlet penguin ($S^p$), electroweak penguin ($P_{\rm EW}^p, 
P_{{\rm EW},C}^p$) and annihilation contributions. 
The numerical analysis below takes into account all flavour
amplitudes following \cite{Beneke:2003zv}, but it suffices to focus on
a few dominant terms to understand the qualitative features of 
the result. Then, for the various final states, the relevant hadronic 
amplitude ratio is given by
\begin{equation}
\begin{array}{llcll}
\pi^0 K_S \phantom{spa} & 
\hat d_f\sim {\displaystyle \frac{[-P^u]+[C]}{[-P^c]}} & 
\phantom{sp}\phantom{spa} &
\rho^0 K_S  \phantom{spa}& 
\hat d_f\sim {\displaystyle \frac{[P^u]-[C]}{[P^c]}}
\\[0.6cm]
\eta^\prime K_S & 
\hat d_f\sim {\displaystyle \frac{[-P^u]-[C]}{[-P^c]}} & 
\phantom{sp} &
\phi K_S & 
\hat d_f\sim {\displaystyle \frac{[-P^u]}{[-P^c]}} 
\\[0.6cm]
\eta K_S & 
\hat d_f\sim {\displaystyle \frac{[P^u]+[C]}{[P^c]}} & 
\phantom{sp} &
\omega K_S & 
\hat d_f\sim {\displaystyle \frac{[P^u]+[C]}{[P^c]}} 
\end{array}
\end{equation}
The convention here is that quantities in square brackets have 
positive real part. (Recall from (\ref{dSS}) that $\Delta S_f$ 
mainly requires the real part of $\hat d_f$.) In factorization 
$\mbox{Re}\,[P^u/P^c]$ is near unity, 
roughly independent of the particular final 
state, hence $\Delta S_f$ receives a nearly universal, small and 
{\em positive} contribution of about $2 \epsilon_{\rm
  KM}\cos(2\beta)\sin\gamma
\approx 0.03$. On the contrary the magnitudes and 
signs of the penguin amplitudes' real parts can be very different. 
Hence the influence of the colour-suppressed tree amplitude 
$C$ determines the difference in $\Delta S_f$ 
between the different modes. For 
$(\pi^0, \eta, \omega)K_S$ the effect of $C$ is constructive, 
but for $(\rho, \eta^\prime)K_S$ it is destructive. However, 
the magnitude of $\mbox{Re}\,[P_c]$ is much larger for $ \eta^\prime
K_S$ than for $\rho K_S$, hence $\mbox{Re}\,(\hat d_f)$ remains small 
and positive for $ \eta^\prime K_S$, but becomes negative for 
$\rho K_S$. 

\begin{table}[t]
\vspace{0.5cm}
\begin{center}
\begin{tabular}{l|cc|l|cc}
Mode & $\quad \Delta S_f$ (Theory) & 
$\quad\Delta S_f$ [Range]$\quad$ &
Mode & $\quad \Delta S_f$ (Theory) & 
$\quad\Delta S_f$ [Range]$\quad$\\
\hline
&&&&& \\[-0.2cm]
$\pi^0 K_S$
 & $\phantom{-}0.07^{+0.05}_{-0.04}$ & $[+0.03,0.13]$ &
$\rho^0 K_S$
 & $-0.08^{+0.08}_{-0.12}$ & $[-0.29,0.01]$\\[0.2cm]
$\eta^\prime K_S$
 & $\phantom{-}0.01^{+0.01}_{-0.01}$ & $[+0.00,0.03]$ &
$\phi K_S$
 & $\phantom{-}0.02^{+0.01}_{-0.01}$ & $[+0.01,0.05]$\\[0.2cm]
$\eta K_S$
 & $\phantom{-}0.10^{+0.11}_{-0.07}$ & $[-0.76,0.27]$ &
$\omega K_S$
 & $\phantom{-}0.13^{+0.08}_{-0.08}$ & $[+0.02,0.21]$\\[0.2cm]
\end{tabular}
\end{center}
\caption{\label{tab1} Comparison of theoretical and experimental
  results for $\Delta S_f$.}
\end{table}

\vskip0.2cm
The result of the calculation of $\Delta S_f$ is shown in 
Table~\ref{tab1}. The columns labeled ``$\Delta S_f$ (Theory)'' 
use the input parameters (CKM parameters, strong coupling, quark masses, 
form factors, decay constants, moments of light-cone distribution 
amplitudes) summarized in Table~1 of~\cite{Beneke:2003zv}. 
The uncertainty estimate is computed 
by adding in quadrature the individual parameter uncertainties.  
The result displays the anticipated pattern. The variation of 
the central value from the nearly universal contribution of
approximately $\epsilon_{\rm KM}$ is due to $\mbox{Re}\,[C/P^c]$, and the 
error comes primarily from this quantity. It is therefore dominated 
by the uncertainty in the hard-spectator scattering contribution 
to $C$, and the penguin annihilation contribution to $P^c$. In 
general one expects the prediction of the asymmetry $S_f$ in 
factorization to be more accurate than the prediction of the 
direct CP asymmetry $C_f$, since $S_f$ is determined by 
$\mbox{Re}\,(a_f^u/a_f^c)$ which is large and calculated at
next-to-leading order. 
The resultant error on $\Delta S_f$ is roughly of the size of 
$\Delta S_f$ itself. 
Quadratic addition of theoretical errors may not always lead to a
conservative error estimate. Therefore we also 
perform a random scan of the allowed theory parameter 
space, taking the minimal and maximal value of an observable 
attained in this scan to define its predicted range. 
In doing so we discard all theoretical parameter
sets which give CP-averaged branching fractions not compatible 
within 3 sigma with the experimental data, that is we require 
$8.1 < 10^6\,\mbox{Br}\,(\pi^0 K^0)<11.8$,
$2.5 < 10^6\,\mbox{Br}\,(\rho^0 K^0)<8.2$,
$5.3 < 10^6\,\mbox{Br}\,(\phi K^0)<11.9$,
$2.9 < 10^6\,\mbox{Br}\,(\omega K^0)<7.5$,
$0.2 < 10^6\,\mbox{Br}\,(\eta K^0)< 2.4$. Note that 
we do not require the theoretical parameters to reproduce 
the $\eta^\prime K^0$ branching fraction for reasons 
explained in \cite{Beneke:2005pu}.
The resulting ranges for $\Delta S_f$ from a scan of 200000
theoretical parameter sets are shown in the columns labeled 
``$\Delta S_f$ [Range]'' in Table~\ref{tab1}. It is seen 
that the ranges are not much different from those 
obtained by adding parameter uncertainties in quadrature -- 
except for the $\eta K_S$ final state. For $\eta K_S$ large 
negative values of $\Delta S_f$ originate from 
small regions of the parameter space, where by cancellations 
the leading penguin amplitude $P_c$ becomes very small. This 
leads to large amplifications of $C/P^c$, and 
hence $\Delta S_f$. Except for the case of $\eta K_S$, 
these parameter space regions are excluded by the lower limits 
on the branching fractions. 

Factorization-based calculations of two-body final states with 
scalar mesons and three-body final states are on a less solid 
footing than the final states discussed above. The following 
estimates have been obtained for the three-kaon 
modes \cite{Cheng:2005ug}
\begin{equation}
\Delta S_{K^+ K^- K_S} = 0.06^{+0.08}_{-0.02}, 
\qquad 
\Delta S_{K_S K_S K_S} = 0.06^{+0.00}_{-0.00}. 
\end{equation}
The quoted error should be regarded with due caution.

\vskip0.2cm
In conclusion, QCD calculations of the time-dependent CP asymmetry 
in hadronic $b\to s$ transitions yield only small corrections 
to the expectation $-\eta_f S_f \approx \sin(2\beta)$. With 
the exception of the $\rho^0 K_S$ final state the correction 
$\Delta S_f$ is positive. The effect and theoretical 
uncertainty is particularly small for the two final states 
$\phi K_S$ and $\eta^\prime K_S$ \cite{Beneke:2003zv}. 
The final-state dependence of $\Delta S_f$ is 
ascribed to the colour-suppressed tree amplitude.
It appears difficult to constrain $\Delta S_f$ 
theory-independently by other observables. In particular, the 
direct CP asymmetries or the charged decays corresponding to 
$f=M K_S$ probe hadronic quantities other than those relevant 
to $\Delta S_f$, 
if these observables take values in the expected range. 
Here $M$ stands for a charged light meson.
Large
deviations from expectations such as large direct CP 
asymmetries would clearly indicate a defect in our understanding of
hadronic physics, but even then the quantitative implications for 
$S_f$ would be unclear. 
A hadronic interpretation of large 
$\Delta S_f$ would probably involve an unknown long-distance 
effect that discriminates strongly between the up- and charm-penguin 
amplitude resulting in an enhancement of the up-penguin 
amplitude. No model is known that could plausibly produce 
such an effect.

\subsubsection{Theoretical estimates of $\Delta S$ from three-body decays}
%
%
%
%
%
While a possibility of constraining the CKM weak phase from three-body $\Delta S=1$ $B$ decays has been raised
a long time ago \cite{Lipkin:1991st}, a discussion of three-body final states as probes of CKM phase has gained
more momentum only recently with the experimental advances. 
The present experimental situation 
that includes measurements of time-dependent CP asymmetries 
in $B^0\to K_SK_SK_S$, $B^0\to\pi^0\pi^0K_S$ and $B^0\to K^+K^-K_{S,L}$ 
is summarized in Table \ref{Table:Zupan}. 
The quoted CP asymmetries are phase space ($dps$) integrated quantities with
\beq
\label{s2beff}
S_{f}^\mathrm{3-body}\equiv (1-2f_+)\sin 2\beta^{\rm eff}=\frac{2\Im \int d\,ps\;( e^{-2 i\beta}A_f\bar A_f^*)}{\int d\,ps\; |A_f|^2+\int d\,ps\; |\bar A_f|^2}.
\eeq
Here  $f_+$ is the CP-even component fraction, while $A_f$ and $\bar A_f$ denote the $A(B^0\to f)$ and $A(\bar B^0\to f)$ amplitudes respectively.  While  $B^0\to K_S K_S K_S$ and $B^0\to \pi^0 \pi^0 K_S$ are decays into completely CP even final states \cite{Gershon:2004tk}, the decay $B^0\to K^+K^-K_{S}$ has both components, but is still mostly CP-even with  $f_+\sim 0.9$. This is obtained either from isospin analysis  from $B^+\to K_S K_S K_S$ decay assuming penguin dominance \cite{Grossman:2003qp,Abe:2002ms,Gronau:2003ep,Abe:2006gy,Gronau:2005ax}, or directly from angular analysis \cite{Aubert:2005ha}, in agreement with each other.


A $\Delta S=1$ $B$ decay amplitude can be in general decomposed in terms of "tree" ($\sim V_{ub}^* V_{us}$) and "penguin" ($\sim V_{cb}^* V_{cs}$) contributions as shown in Eq.~\ref{ampl} for the case of two-body $\bar B$ decays. An expression analogous to Eq.~\ref{dSS} holds for $\Delta S_f$,
here given by
\beq
\Delta S_f=\sin 2\beta^{\rm eff}-\sin 2\beta =2 \cos 2\beta \sin \gamma \Re(\xi_f),
\eeq
where $\sin 2\beta^{\rm eff}$ is defined in eq.~\ref{s2beff} and
the ratio
\beq
\xi_{f}\equiv \frac{V_{ub}^*V_{us}}{V_{cb}^*V_{cs}}\;\frac{\int d\, ps\; T_f^* P_f}{\int d\, ps\; P_f^* P_f},
\eeq
suitably averaged over the final phase space, replaces the ratio $d_f$ defined in the previous section for two-body decays. In addition, the
direct CP asymmetries are given by
\beq
C_f=-2 \sin \gamma \Im(\xi_f).
\eeq

\begin{table}
\caption{Measured CP asymmetries in $B^0\to 3P$ decays \cite{Barberio:2006bi}. }\label{Table:Zupan}
\begin{center}
\begin{tabular}{c c c} \hline \hline
Mode  & $\sin (2\beta^{\rm eff})$  & $C_f$ \\\hline
$K_SK_SK_S$ \cite{Aubert:2006ar,Chen:2006nk}& $0.51\pm 0.21$& $-0.23\pm 0.15$ \\
$\pi^0\pi^0K_S$ \cite{Aubert:2005id} & $-0.84 \pm 0.71 \pm 0.08$ &$0.27 \pm 0.52 \pm 0.13$ \\ 
$K^+K^-K_{S,L}$ \cite{Abe:2006gy,Aubert:2005ha}&$0.58 \pm 0.13^{+0.12}_{-0.09} $&$0.15 \pm 0.09  $\\
\hline \hline
\end{tabular}
\end{center}
\end{table}

The difference $\Delta S_f$ was analysed using SU(3) flavor symmetries \cite{Engelhard:2005ky,Grossman:2003qp,Engelhard:2005hu} 
and was calculated in a model-dependent way in Ref \cite{Cheng:2005ug}. 
The approach is based on flavor SU(3) and  exploits the fact 
that the related $\Delta S=0$ final states, $f'$, are more sensitive to 
the "tree" amplitudes which are CKM enhanced 
when compared to the $\Delta S=1$ amplitudes,
(because $V_{us} < V_{ud}$). However, "penguin" amplitudes are CKM suppressed  
(because $V_{cs}\to V_{cd}$). This then leads to a bound on $\xi_f$ of the form
\beq
\xi_f < \lambda \sum_{f'} a_{f'}\sqrt{\frac{Br(f')}{Br(f)}},
\eeq
where $\lambda=0.22$, $a_{f'}$ are the coefficients 
arising from $SU(3)$ Clebsch-Gordan coefficients,   
and the sum is over $\Delta S=0$ final states $f'$.  
The bounds are better if less modes enter the sum, which can be achieved through a 
dynamical assumption of small annihilation-like amplitudes. This then gives
\beq
\xi_{K^+K^-K^0}< 1.02 \text{~\cite{Engelhard:2005ky}},~~~\xi_{K_SK_SK_S}< 0.31 \text{~\cite{Engelhard:2005hu}},
\eeq
with bounds for a number of other modes listed in \cite{Engelhard:2005ky}. 
These are only very conservative upper bounds not at all indicative 
of the expected size $\xi_f\sim \lambda^2 T_f/P_f$. One also expects 
$\xi_{K^+K^-K^0}<\xi_{K_SK_SK_S}$, since in the latter case 
all the tree operator contributions are OZI suppressed 
as the final state does not contain valence $u$-quarks. 
This expectation was confirmed by a model-dependent calculation that 
combined QCD factorization with heavy-meson chiral perturbation theory \cite{Cheng:2005ug}. 
This approach is valid only in a region of phase space 
where one of the light mesons is slow and the other two are very energetic, 
while for the remaining phase space a model for the form factors was used. 
Ref. \cite{Cheng:2005ug} then obtains 
\beq
\Delta S_{K_S K_SK_S}=0.02,~~~~\Delta S_{K^+K^-K_S}\lesssim O(0.1).
\eeq
An argument exists that the latter could be smaller \cite{Chua:2006hr}, 
but one should also keep in mind the comment at the end of the previous section.

A different use of three-body final states is provided by the 
time-dependent Dalitz plot analysis with a fit to quasi-two body resonant modes. 
Interferences between resonances then fix relative strong phases 
giving additional experimental information. 
In this way BaBar was able to resolve the $\beta \to \pi/2 -\beta$ discrete ambiguity 
using a $B^0\to K^+K^- K_{S,L}$ Dalitz plot analysis~\cite{Aubert:2006av}.
The interference of CP-even and CP-odd contributions leads 
to a $\cos 2\beta^{\rm eff}$ term 
(with $\beta^{\rm eff}\to \beta$ in the limit of no tree pollution). 
Another example is measuring phases of $\Delta I=1$ amplitudes 
of $B\to(K^*\pi)_{I=1/2,3/2}$, $B_s\to (K^*\bar K)_{I=1}$ and $B_s \to (\bar
K^* K)_{I=1}$ from resonance interferences in $B\to K\pi\pi$ and $B_s\to
K\bar K\pi$. This then gives information on CKM parameters 
complementary to other methods~\cite{Ciuchini:2006kv,Ciuchini:2006st,Gronau:2006qn}.  
Using SU(3) hadronic uncertainties due to electroweak penguin operators $O_9$ and $O_{10}$ 
were shown to be very small in $B\to K\pi\pi$ and $B_s\to K\pi\pi$ and somewhat
larger in $B_s\to K\bar K\pi$ \cite{Gronau:2006qn}. 
The first processes imply a precise linear relation
between $\bar\rho$ and $\bar\eta$, with a measurable slope and
an intercept at $\bar\eta=0$ involving a theoretical error of 0.03. The decays $B_s\to K\pi\pi$ permit a measurement
of $\gamma$ involving a theoretical error below a degree. Furthermore, while time-dependence is required
when studying $B^0$ decays at the $\Upsilon(4S)$, it may not be needed when studying
$B_s$ decays at hadronic colliders.

\subsubsection{Flavour symmetries and estimates of $b\to s$ transitions}
Decomposing the $B\to MM$ amplitudes in terms of flavor SU(3) or isospin reduced matrix elements leads to relations between different amplitudes since the effective weak hamiltonian usually transform only under a subset of all possible representations \cite{Grinstein:1996us}. The group theoretical approach based on reduced matrix elements \cite{Zeppenfeld:1980ex,Savage:1989ub,Chau:1990ay} is equivalent to a diagrammatic approach of topological amplitudes \cite{Gronau:1994rj,Gronau:1995hm,Gronau:1995hn,Dighe:1995gq,Dighe:1997wj}. In the latter it is easier to introduce dynamical assumptions such as neglecting annihilation-like amplitudes. These were shown to be $1/m_b$ suppressed for decays into nonisosinglets \cite{Bauer:2004ck}, while not all of them are $1/m_b$ suppressed, if $\eta,\eta'$ occur in the final state 
(see Appendix C of \cite{Williamson:2006hb}). 

The SU(3) approach has been used in global fits to the experimentally measured $B\to PP$ and $B\to P V$ decays \cite{Chiang:2001ir,Chiang:2003rb,Chiang:2003pm,Chiang:2004nm,Chiang:2006ih,Zhou:2000hg,He:2000ys,Wu:2002nz,Wu:2004xx,Wu:2005hi} 
in which both the values of hadronic parameters as well as the value of weak phase $\gamma$ are determined. 
However, in order to obtain a stable fit a number of dynamical assumptions are needed. 
In the most recent fit to $B\to PP$ \cite{Chiang:2006ih} $t$-quark dominance in penguin amplitudes and negligible annihilation-like topologies (also for isosinglets) were assumed. Both $\beta$ and $\gamma$ were determined, with central values slightly above the CKMfitter and UTfit determinations. Allowing for a new weak phase in $P_{\rm EW}$ for $\Delta S=1$ modes leads to statistically significant reduction of $\chi^2$, while choosing this phase to be zero does give the size of $|P_{\rm EW}|$ in excellent agreement with 
the Neubert-Rosner relation \cite{Neubert:1998pt,Neubert:1998jq,Neubert:1998re,Gronau:1998fn}. A large strong phase difference $\arg(C/T)\sim -60^\circ$ was found, while expected to be $1/m_b$ suppressed from QCD factorization and SCET \cite{Beneke:2003zv,Bauer:2005kd,Bauer:2004tj}. As stressed in Ref. \cite{Gronau:2006ha} the direct CP asymmetries $A_{\rm CP}(B^0\to K^+\pi^-)$ and $A_{\rm CP}(B^+\to K^+\pi^0)$ would be of the same sign for $\arg(C/T)$ small, which is excluded at $4.7 \sigma$ at present.  

Assumption of negligible annihilation topologies used in SU(3) fits can be tested by comparing $B^0\to K^0\bar K^0$, $B^+\to K^+\bar K^0$, where annihilation is CKM enhanced,  with $B^+\to K^0\pi^+$ \cite{Hao:2006su,Gronau:1998gr}. SU(3) breaking has been addressed in \cite{Buras:2005cv,Chiang:2006ih} showing a small effect on the values of extracted parameters. Further tests of SU(3) breaking or searches of NP will be possible using $B_s$ decays \cite{Fleischer:2002zv,London:2004ej,Gronau:2000zy}, with the first CDF measurement of $Br(B_s\to K^+K^-)$ leading the way \cite{Tonelli:2005cc}.  Errors due to the dynamical assumptions can be reduced, if fits are made to only a subset of modes, e.g. to $\pi\pi, \pi K$ \cite{Buras:2005cv,Buras:2004th,Buras:2004ub,Buras:2003dj,Wu:2004xx,Chiang:2006ih}. Furthermore, 
 dynamical assumptions can be avoided entirely, if only a set of modes related through U-spin is used \cite{Soni:2006vi,Soni:2005ah}. This leads to stable fits, while giving $\gamma$ with a theoretical error of a few degrees \cite{Soni:2006vi}.  Further studies of SU(3) breaking effects are called for, though.

Because of the different CKM hierarchy of tree and penguin amplitudes in $\Delta S=1$ and $\Delta S=0$ decays, tree pollution in $\Delta S=1$ decays can be bounded using SU(3) related $\Delta S=0$ modes \cite{Grossman:2003qp}. Correlated bounds on $\Delta S_f$ and $C_f$ for $\eta'K_S$ and $\pi^0 K_S$ final states have been presented in \cite{Gronau:2006qh,Gronau:2004hp,Gronau:2003kx,Gronau:2005gz}. Such a model independent bound on $\Delta S_{\phi K_S}$ is not available at present, since many more $\Delta S=0$ modes enter, some of which have not been measured yet \cite{Raz:2005hu}. 

Very precise relations between $\Delta S=1$ $B\to \pi K$ CP asymmetries or decay rates can be obtained using isospin decompositions. The sum rule between decay widths $\Gamma(K^0\pi^+)+\Gamma(K^+\pi^-)=2\Gamma(K^+\pi^0)+2\Gamma(K^0\pi^0)$ \cite{Gronau:1998ep,Lipkin:1998ie} (equivalent to $R_n=R_c$ \cite{Gronau:2001cj}) is violated by CKM doubly suppressed terms calculable in $1/m_b$ expansion \cite{Bauer:2005kd,Bauer:2004tj,Beneke:2003zv,Williamson:2006hb}, while harder to calculate isospin-breaking corrections cancel to first order \cite{Gronau:2006eb}. The sum rule $\Delta(K^+\pi^-)+\Delta(K^0\pi^+)-2\Delta(K^+\pi^0)-2\Delta(K^0\pi^0)=0$ 
for the rate differences $\Delta(f)=\Gamma(\bar B\to \bar f)-\Gamma(B\to f)$ is valid in the isospin limit, and is thus violated by EWP. However, these corrections vanish in the SU(3), $m_b\to\infty$ limit making the sum rule very precise \cite{Gronau:2005kz}.

\subsubsection{Applications of $U$-spin symmetry to $B_d$ and $B_s$ decays}
%
%
%
%
%


The current data in $B$ physics suggests that $B_d$ decays agree well with
SM predictions, while $B_s$ decays remain poorly known and
might be affected by New Physics. Within the Standard Model, the CKM
mechanism correlates the electroweak part of these transitions, but 
quantitative predictions are difficult due to hadronic effects. The
latter can be estimated relying on the approximate $SU(3)$-flavour 
symmetry of QCD :
information on hadronic effects, extracted from data in one channel, 
can be exploited in other channels related by flavour symmetry, 
leading to more accurate predictions within the Standard Model.

In addition to isospin symmetry, an interesting theoretical tool
is provided by $U$-spin symmetry, which relates $d$- and $s$-quarks. 
Indeed, this symmetry holds for long- and short-distances
and does not suffer from electroweak corrections, 
making it a valuable instrument to analyse processes with significant 
penguins and thus a potential sensitivity to New Physics. 
However, due to the significant difference $m_s-m_d$, 
$U$-spin breaking corrections of order 30\% may occur,
depending on the processes.

As a first application of $U$-spin, relations were obtained between
$B_d\to \pi^+\pi^-$ and 
$B_s\to K^+ K^-$. This led to correlations among the observables
in the two decays such as branching ratios and CP 
asymmetries~\cite{Fleischer:1999pa,Fleischer:2002zv} and to
a prediction for 
$BR(B_s\to K^+ K^-)=(35^{+73}_{-20})\cdot 10^{-6}$~\cite{Buras:2004ub}.
These results helped to investigate the potential
of such decays to discover New Physics~\cite{London:2004ej,Baek:2005wx}.
Unfortunately, the accuracy of the method is limited not only by the
persistent discrepancy between Babar and Belle on
$B_d\to\pi^+\pi^-$ CP asymmetries, but also 
by poorly known $U$-spin corrections. In these analyses,
the ratio of tree contributions
$R_c=|T^s_{K\pm}/T^d_{\pi\pm}|$ was taken from 
QCD sum rules as $1.76\pm 0.17$~\cite{Khodjamirian:2003xk} (updated
to $1.52^{+0.18}_{-0.14}$~\cite{Khodjamirian:2004ga}). In addition,
the ratio of penguin-to-tree ratios
$\xi=|(P^s_{K\pm}/T^s_{K\pm})/(P^d_{\pi\pm}/T^d_{\pi\pm})|$
was assumed equal to $1$~\cite{Buras:2004ub} 
or $1\pm 0.2$~\cite{London:2004ej,Baek:2005wx}
in agreement with rough estimates within QCD 
factorisation (QCDF)~\cite{Safir:2004ua}.

Indeed QCDF may complement flavour symmetries by
a more accurate study of short-distance effects. However, QCDF cannot predict some significant $1/m_B$-suppressed long-distance effects, which
have to be estimated through models.
Recently, it was proposed to combine QCDF and $U$-spin in 
the decays mediated by penguin operators $B_d\to K^0 \bar{K}^0$ and 
$B_s\to K^0 \bar{K}^0$~\cite{Descotes-Genon:2006wc}.

First, tree ($T^{d0}$) and penguin ($P^{d0}$) contributions to 
$B_d\to K^0 \bar{K}^0$ can be determined by 
combining the currently available data with
$|T^{d0}-P^{d0}|$, which can be accurately computed in QCDF because
long-distance effects, seen as infrared divergences, cancel in this difference.
$U$-spin suggests accurate relations between these hadronic parameters in
$B_d\to K^0 \bar{K}^0$ and those in $B_s\to K^0 \bar{K}^0$. Actually,
we expect similar long-distance effects since the $K^0 \bar{K}^0$ final state
is invariant under the $d$-$s$ exchange. Short distances
are also related since the two processes are mediated by penguin operators
through diagrams with the same topologies. 
$U$-spin breaking arises only in a few places : factorisable corrections 
encoded in
$f=[M_{B_s}^2 F^{B_s\to K}(0)]/[M_{B_d}^2 F^{B_d\to K}(0)]$, 
and non-factorisable corrections from weak annihilation and 
spectator scattering. Because of these expected tight relations, 
QCDF can be relied upon
to assess $U$-spin breaking between the two decays. Indeed, up to
the factorisable factor $f$, penguin (as well as tree) contributions to
both decays are numerically very close. Penguins in 
$B_d\to K^0 \bar{K}^0$ and $B_s\to K^+ K^-$ should have very close values as 
well, whereas no such relation exists for the (CKM-suppressed) tree 
contribution to the latter, to be estimated in QCDF.

 These relations among hadronic parameters,
inspired by $U$-spin considerations and 
quantified within QCD factorisation, can be exploited to
determine the tree and penguin contributions to $B_s\to KK$ decays and
the corresponding observables. In particular, one gets 
$BR(B_s\to K^0 \bar{K}^0) = (18\pm 7\pm 4\pm 2)\cdot 10^{-6}$ and
$BR(B_s\to K^+ \bar{K}^-) = (20\pm 8\pm 4\pm 2)\cdot 10^{-6}$, in
very good agreement with the latest CDF measurement. 
The same method provides significantly improved
determinations of the $U$-spin breaking ratios $\xi=0.83\pm 0.36$ and
$R_c=2.2\pm 0.7$. These results have been exploited to determine
the impact of supersymmetric models on these decays~\cite{Baek:2006pb}.

New results on $B\to K$ form factors and
on the $B_d\to K^0 \bar{K}^0$ branching ratio and direct CP-asymmetry 
should lead to a significant 
improvement of the predictions in the $B_s$ sector.
The potential of other pairs of nonleptonic $B_d$ and $B_s$ decays remains
to be investigated.

\subsubsection{Applications of the RGI parametrization to $b\to s$ transitions}
Few general parametrizations of the $\Delta B=1$ hadronic amplitudes
exist in the literature. Here we use the parametrization proposed in Ref.~\cite{Buras:1998ra} which decomposes decay amplitudes in terms of Renormalization-Group-Invariant (RGI) parameters. For our purpose, we just
need to recall a few basic facts about the classification of RGI's.
First of all, we have six non-penguin parameters, containing only
non-penguin contractions of the current-current operators $Q_{1,2}$:
emission parameters $E_{1,2}$, annihilation parameters $A_{1,2}$ and
Zweig-suppressed emission-annihilation parameters
$\mathit{EA}_{1,2}$. Then, we have four parameters containing only
penguin contractions of the current-current operators $Q_{1,2}$ in the
GIM-suppressed combination $Q_{1,2}^{c}-Q_{1,2}^{u}$: $P_1^\mathrm{GIM}$ and
Zweig suppressed $P_{2-4}^\mathrm{GIM}$. Finally, we have four parameters
containing penguin contractions of current-current operators
$Q_{1,2}^{c}$ (the so-called charming penguins \cite{Ciuchini:1997hb})
and all possible contractions of penguin operators $Q_{3-12}$:
$P_{1,2}$ and the Zweig-suppressed $P_{3,4}$. In the following
Zweig-suppressed parameters are neglected. We refer the reader to the original reference for details.
We can then write schematically the $b\to s$ decay amplitude
as:
\begin{equation}
  \label{eq:penguindominated}
  \mathcal{A}(B \to F) = - V^*_{ub} V_{us} \sum
  \left(
    T_i + P_i^\mathrm{GIM}
  \right) -
  V^*_{tb} V_{ts} \sum P_i \,, 
\end{equation}
where $T_i = \{ E_i, A_i, \mathit{EA}_i\}$ are not present in pure-penguin
decays.

The idea developed in Refs.~\cite{Ciuchini:2001gv} is to write down the RGI parameters as the sum of their expression in the infinite mass
limit, for example using QCD factorization, plus an arbitrary
contribution corresponding to subleading terms in the power expansion.
These additional contributions are then determined by a fit to the
experimental data. In $b \to s$ penguins, the dominant
power-suppressed correction is given by charming penguins, and the
corresponding parameter can be determined with high precision from
data and is found to be compatible with a $\Lambda/m_b$ correction to
factorization~\cite{Ciuchini:2001gv}. However, non-dominant corrections, for
example GIM penguin parameters in $b \to s$ decays, can be extracted
from data only in a few cases (for example in $B \to K \pi$ decays).  Yet predictions for $\Delta S_f$ depend
crucially on these corrections, so that one needs external input to
constrain them. One interesting avenue is to extract the support of
GIM penguins from $SU(3)$-related channels ($b \to d$ penguins) in
which they are not Cabibbo-suppressed, and to use this support,
including a possible large $SU(3)$ breaking of $100\%$, in the fit of $b \to
s$ penguin decays. Alternatively, one can omit the calculation in
factorization and fit directly the RGI parameters from the
experimental data, instead of fitting the power-suppressed
corrections~\cite{Ciuchini:2005mg,Ciuchini:2007hx}.

Compared to factorization approaches, general parameterizations
have less predictive power but are more general.
In particular, they tend to overestimate the theoretical uncertainty and are thus best suited to search for NP in a conservative way.
In addition, these methods have the advantage that for several channels
the predicted $\Delta S$ decreases with the experimental uncertainty in
$BR$'s and CP asymmetries of $b \to s$ and $SU(3)$-related $b \to d$ penguins.

\begin{figure}[htb]
\begin{center}
\includegraphics[width=0.32\textwidth]{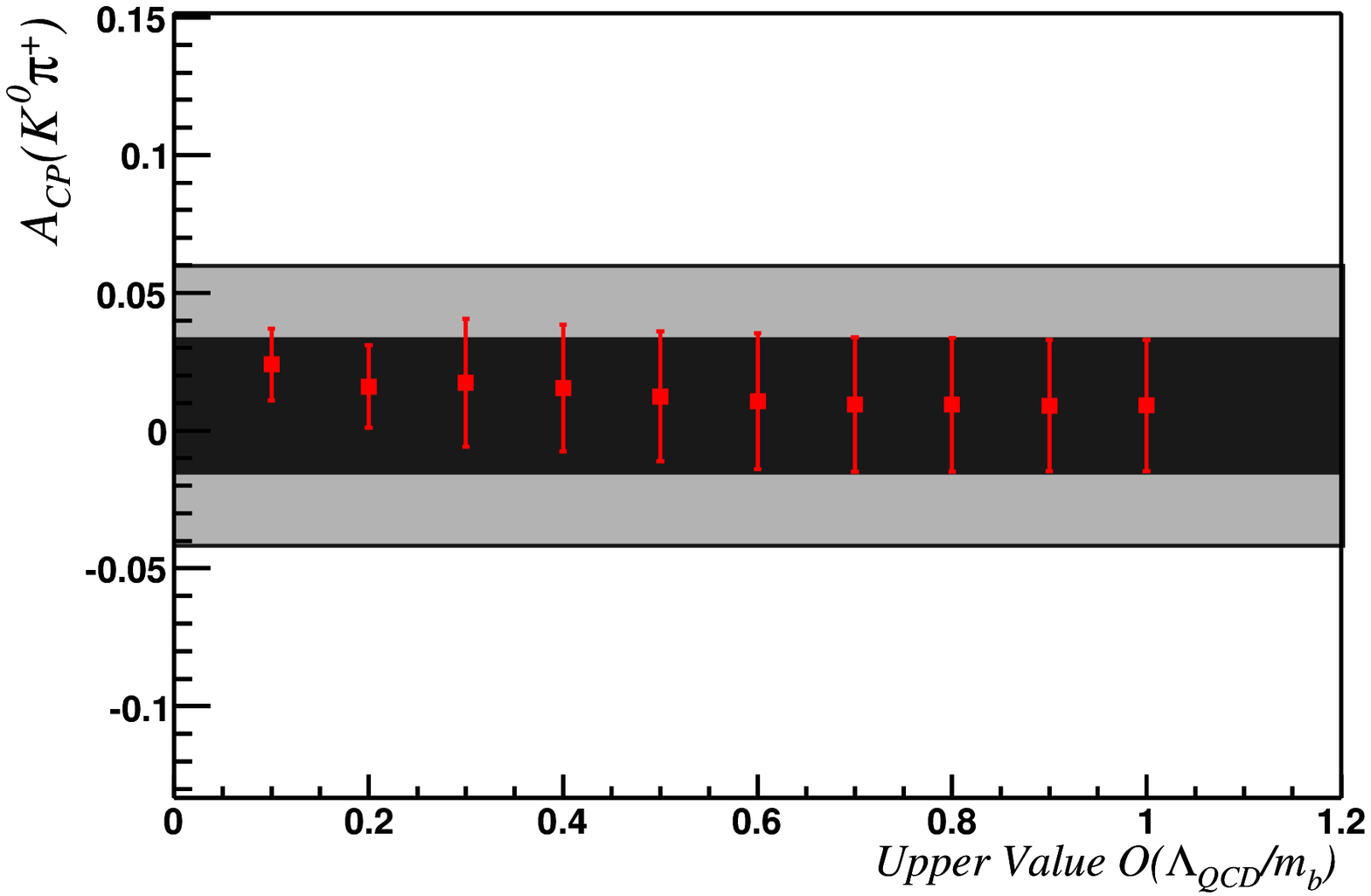} 
\includegraphics[width=0.32\textwidth]{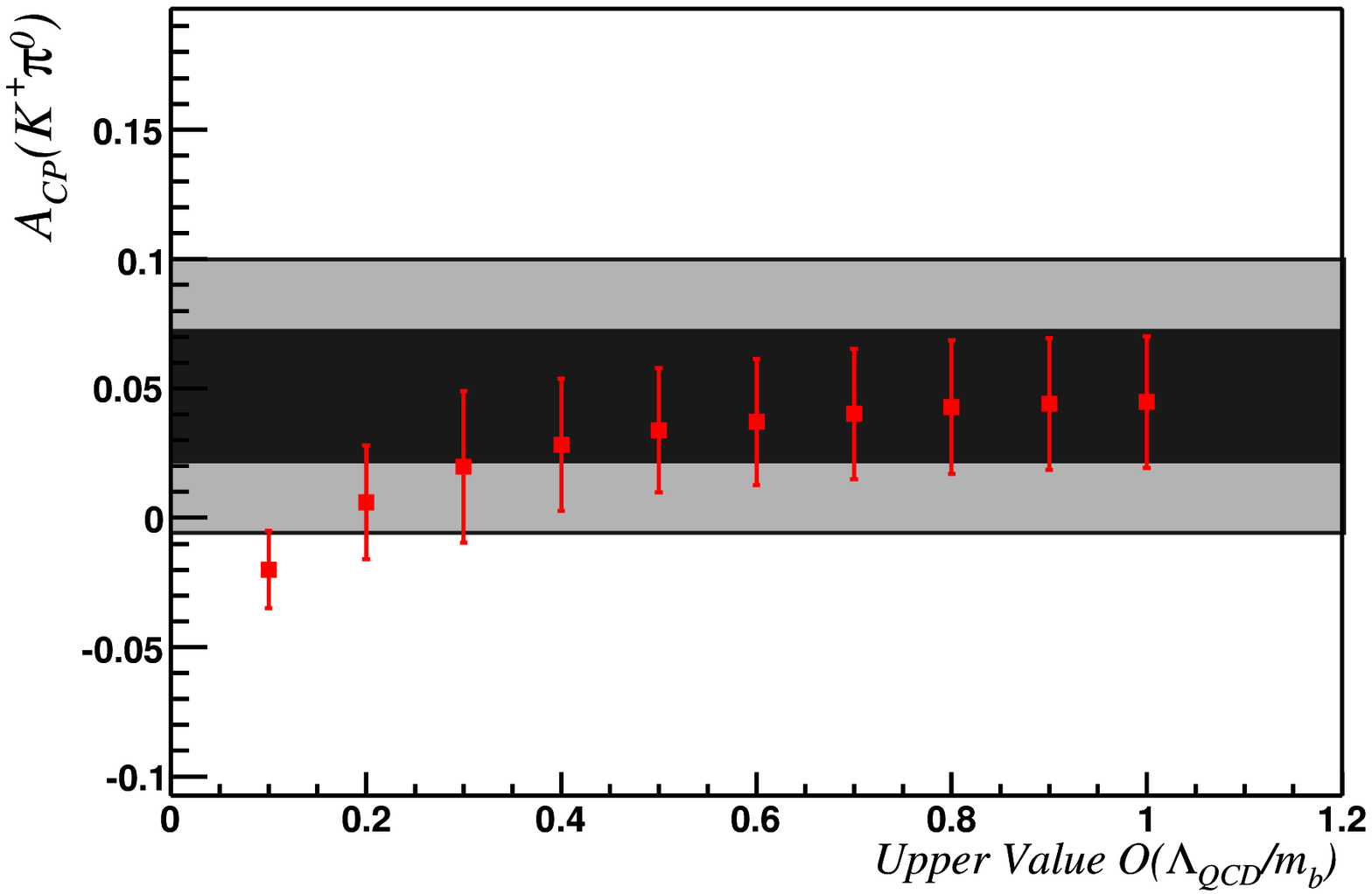} 
\includegraphics[width=0.32\textwidth]{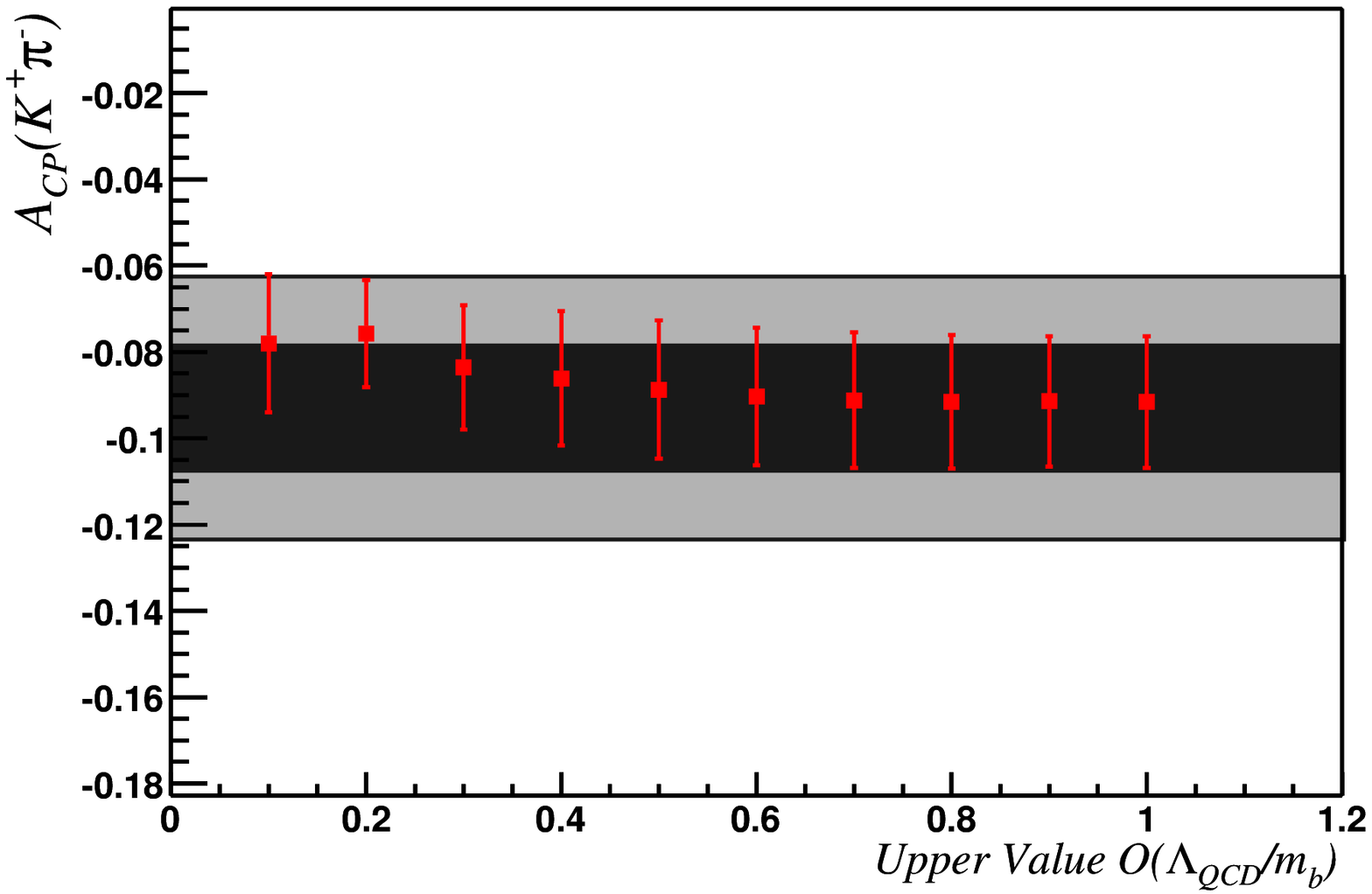}\\
\includegraphics[width=0.45\textwidth]{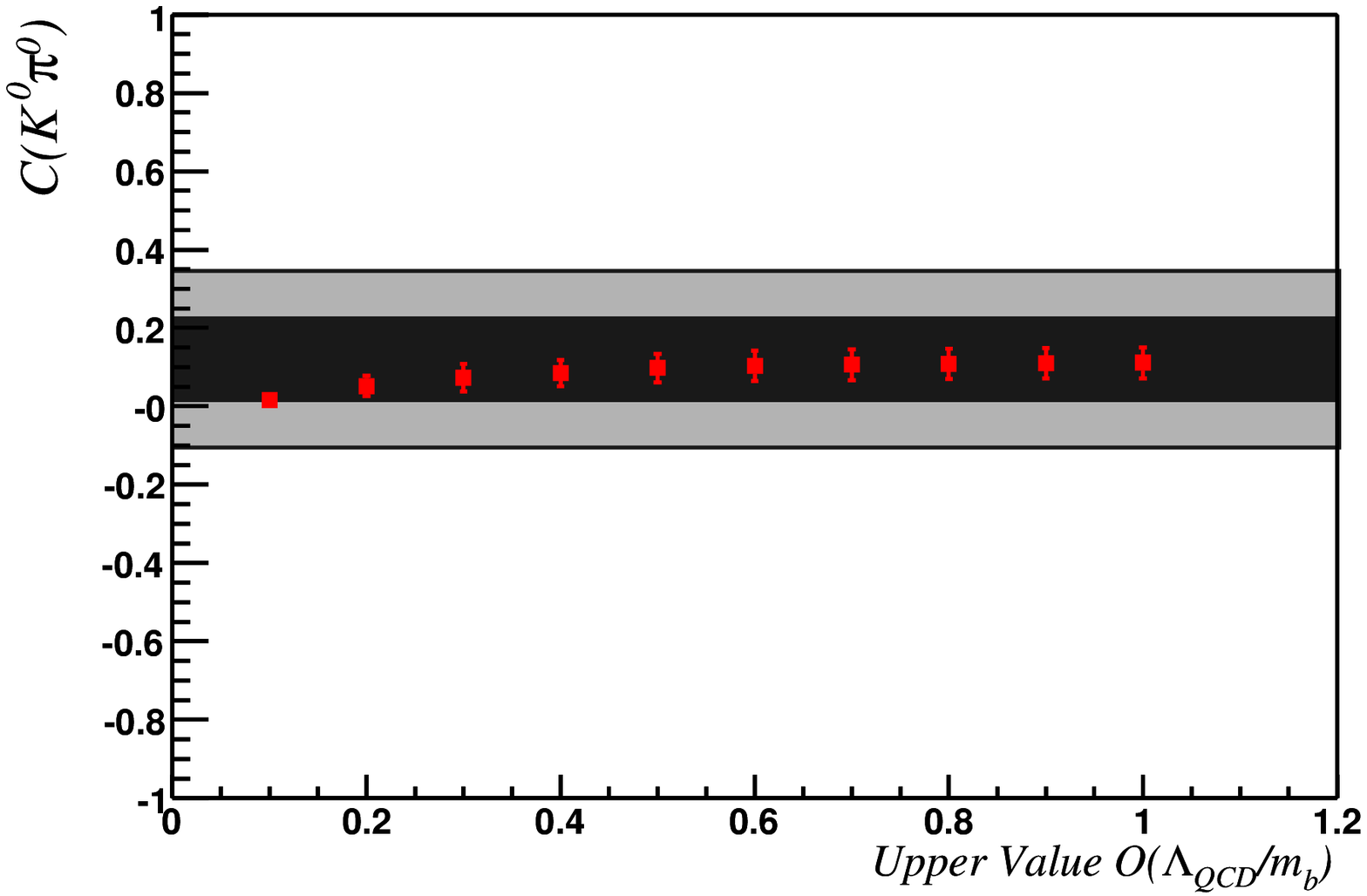} 
\includegraphics[width=0.45\textwidth]{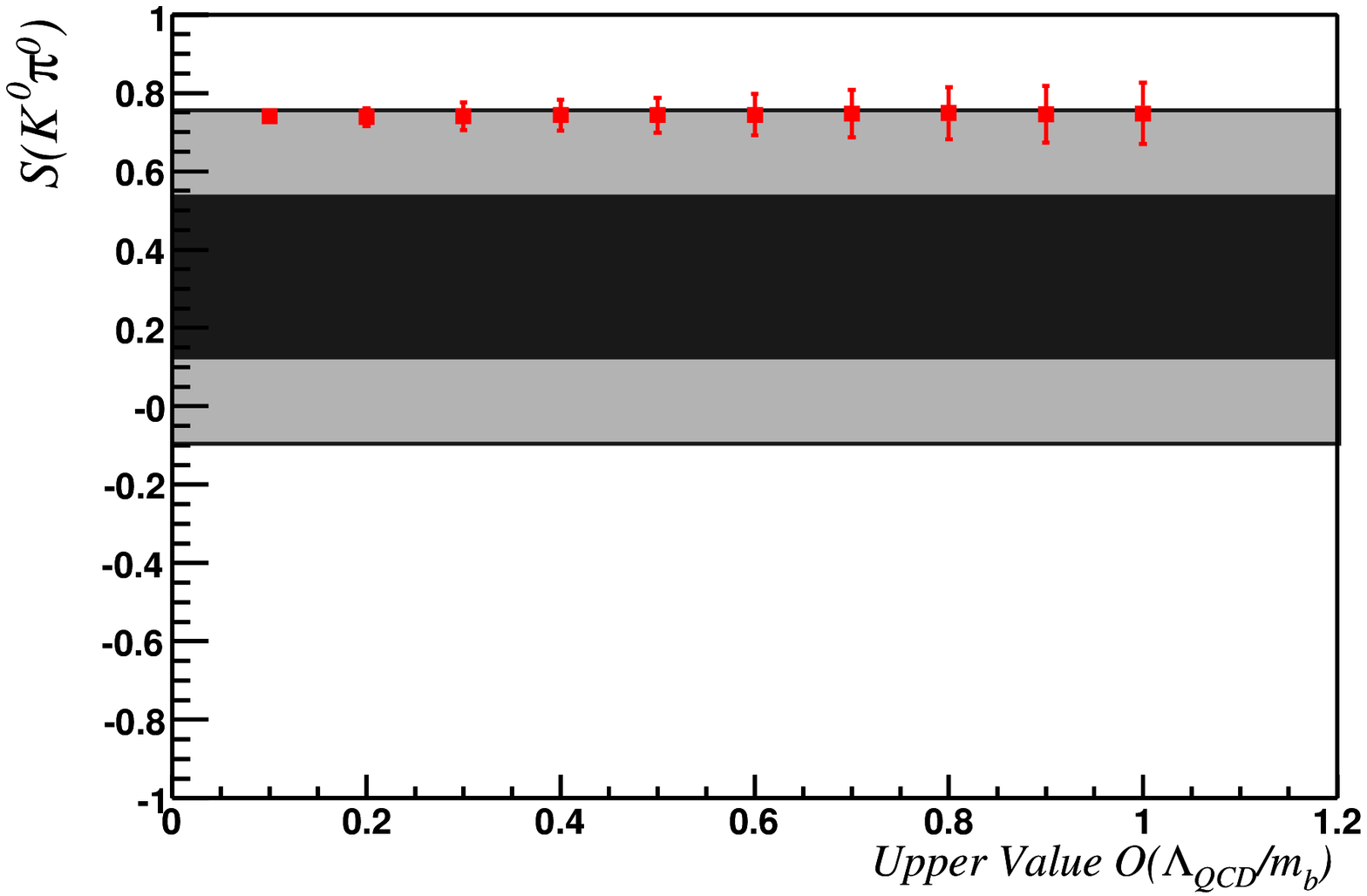} 
\caption{$CP$ asymmetries for $B \to K \pi$ decays, obtained varying subdominant contributions in the range [0, UV], 
with the upper value UV scanned between zero and one
(in units of $E_1$). For comparison, the experimental $68\%$ ($95\%$)
probability range is given by the dark (light) band.}
\label{fig:kpigp}
\end{center}
\end{figure}

In the analysis reported here~\cite{Silvestrini:2007yf,inprep}, we vary the absolute values of the subdominant amplitudes in the range $[0,UL]$ (while the phases are unconstrained) and study the dependence of the predictions on the upper limit $UL$.
For example we show in Fig.~\ref{fig:kpigp} the effect of changing
the upper limit of the range in which subdominant terms are varied
on the prediction of some observables in $B\to K\pi$ decays. It can be seen
that reasonable subdominant terms make any $K\pi$ puzzle
disappear. Furthermore, the prediction of $S_{\pi^0 K_S}$ has small theoretical error and is quite stable against the effect of subdominant terms.

In Table~\ref{tab:DeltaS} we collect predictions for $\Delta S_f$ obtained using the method sketched above for $UL=0.5$ (in units of the leading amplitude), as suggested by the $SU(3)$-related modes $B\to K K$. Notice that the theoretical uncertainty
is smaller for $B \to \pi^0 K_s$ because the number of observables
in the $B \to K \pi$ system is sufficient to constrain efficiently
the hadronic parameters. This means that the theoretical error can
be kept under control by improving the experimental data in these
channels. On the other hand, the information on $B \to \phi K_s$ is
not sufficient to bound the subleading terms and this results in a
relatively large theoretical uncertainty that cannot be decreased
without additional input on hadronic parameters. Furthermore, using
$SU(3)$ to constrain $\Delta S_{\phi K_s}$ is difficult because the
number of amplitudes involved is very large~\cite{Zeppenfeld:1980ex,Engelhard:2005hu,Engelhard:2005ky,Raz:2005hu}.

\begin{table}[htb]
  \caption{Predictions for $\Delta S_f$ using the RGI parametrization.}\label{tab:DeltaS} 
\begin{center}
\begin{tabular}{lr lr}
$\Delta S_{\pi^0 K_S}$ & $(2.4 \pm 5.9)\times 10^{-2}$ 
&$\Delta S_{\eta^\prime K_S}$ & $(-0.7 \pm 5.4)\times 10^{-2}$
\\
$\Delta S_{\phi K_S}$ & $(0.4 \pm 9.2)\times 10^{-2}$
&$\Delta S_{\rho^0 K_S}$ & $(-6.2 \pm 8.4)\times 10^{-2}$
\\
$\Delta S_{\omega K_S}$ & $(5.6 \pm 10.7)\times 10^{-2}$
& ~ & ~
\end{tabular}
\end{center}
\end{table}

The ideal situation would be represented by a pure penguin decay for
which the information on $P_i^\mathrm{GIM}$ is available with minimal
theoretical input. Such situation is realized by the pure penguin
decays $B_s \to K^{0(*)} \bar K^{0(*)}$. An upper bound for the
$P_i^\mathrm{GIM}$ entering this amplitude can be obtained from the
$SU(3)$-related channels $B_d \to K^{0(*)} \bar K^{0(*)}$. Then, even
adding a generous $100 \%$ $SU(3)$ breaking and an arbitrary strong
phase, it is possible to have full control over the theoretical error
in $\Delta S$ \cite{Ciuchini:2007hx}.

\subsubsection{$b\to s$ transitions in the MSSM}
In this section we discuss phenomenological effects of
the new sources of flavor and CP violation in $b \to s$ processes that
arise in the squark
sector~\cite{Gabbiani:1996hi,Barbieri:1997kq,Kagan:1998bh,Abel:1998iu,Kagan:1997sg,Fleischer:2001pc,Besmer:2001cj,Lunghi:2001af,Causse:2002mu,Hiller:2002ci,Khalil:2002fm,Kane:2002sp,Baek:2003kb,Agashe:2003rj,Cheng:2003im,Chakraverty:2003uv,Khalil:2003bi,Khalil:2003ng,Cheng:2004jf,Khalil:2004yb,Gabrielli:2004yi,Khalil:2004wp,Khalil:2005qg} of the Minimal
Supersymmetric Standard Model (MSSM).
In general, in the MSSM squark masses are neither flavor-universal,
nor are they aligned to quark masses, so that they are not flavor
diagonal in the super-CKM basis, in which quark masses are diagonal
and all neutral current vertices are flavor diagonal. The ratios of
off-diagonal squark mass terms to the average squark mass define four
new sources of flavor violation in the $b \to s$ sector: the mass
insertions $(\delta^d_{23})_{AB}$, with $A,B=L,R$ referring to the
helicity of the corresponding quarks. These $\delta$'s are in general
complex, so that they also violate CP. One can think of them as
additional CKM-type mixings arising from the SUSY sector. Assuming
that the dominant SUSY contribution comes from the strong interaction
sector, \textit{i.e.} from gluino exchange, all FCNC processes can be
computed in terms of the SM parameters plus the four $\delta$'s plus
the relevant SUSY parameters: the gluino mass $m_{\tilde g}$, the
average squark mass $m_{\tilde q}$, $\tan \beta$ and the $\mu$
parameter. The impact of additional SUSY contributions such as
chargino exchange has been discussed in detail in Ref.
\cite{Chakraverty:2003uv}. We consider only the case of small or moderate
$\tan \beta$, since for large $\tan \beta$ the constraints from $B_s
\to \mu^+ \mu^-$ and $\Delta m_s$ preclude the possibility of having
large effects in $b \to s$ hadronic penguin decays
\cite{Isidori:2001fv,Buras:2002vd,Baek:2003kb,Foster:2005kb,Foster:2006ze,Isidori:2006pk,Isidori:2007jw}.

Barring accidental cancellations, one can consider one single $\delta$
parameter, fix the SUSY masses and study the phenomenology. The
constraints on $\delta$'s come at present from $B \to X_s \gamma$, $B
\to X_s l^+ l^-$ and from the $B_s - \bar B_s$ mixing amplitude.
We refer the reader to refs.~\cite{Ciuchini:2002uv,Ciuchini:2006dx,Silvestrini:2007yf,inprep2} for all the details and results of this analysis.

Fixing as an example $m_{\tilde g}=m_{\tilde q}= \vert \mu \vert =$
350 GeV and $\tan \beta = 3$, one obtains the following constraints on
$\delta$'s:
\begin{equation}
\vert (\delta^d_{23})_{\mathrm{LL}}\vert < 2\times 10^{-1},\quad
\vert (\delta^d_{23})_{\mathrm{RR}}\vert < 7\times 10^{-1},\quad
\vert (\delta^d_{23})_{\mathrm{RL,LR}}\vert < 5\times 10^{-3}.
\end{equation}
Notice that all constraints
scale approximately linearly with the squark and gluino masses.

\begin{figure}[!ht]
\begin{center}
\includegraphics[width=0.24\textwidth]{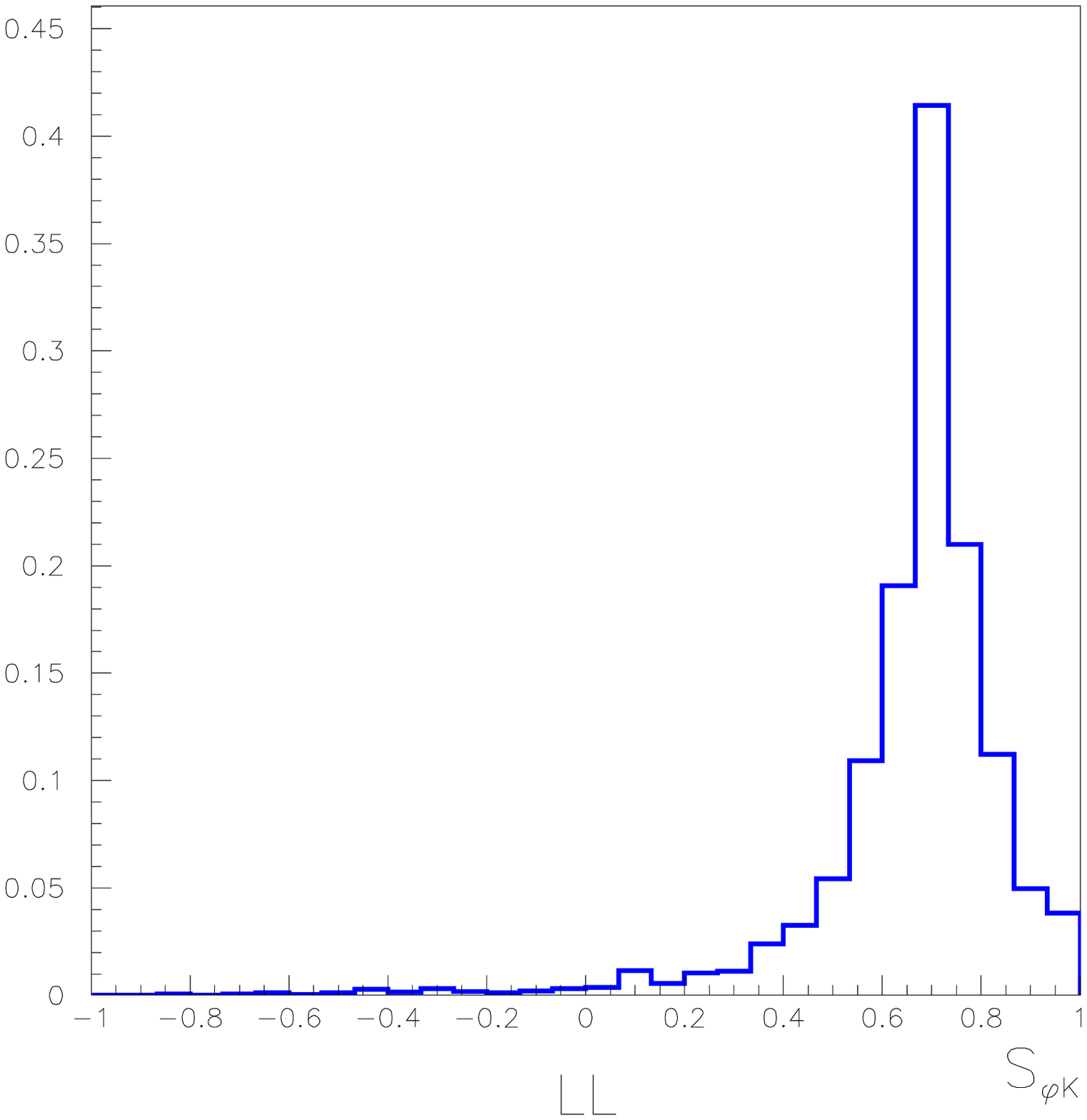} 
\includegraphics[width=0.24\textwidth]{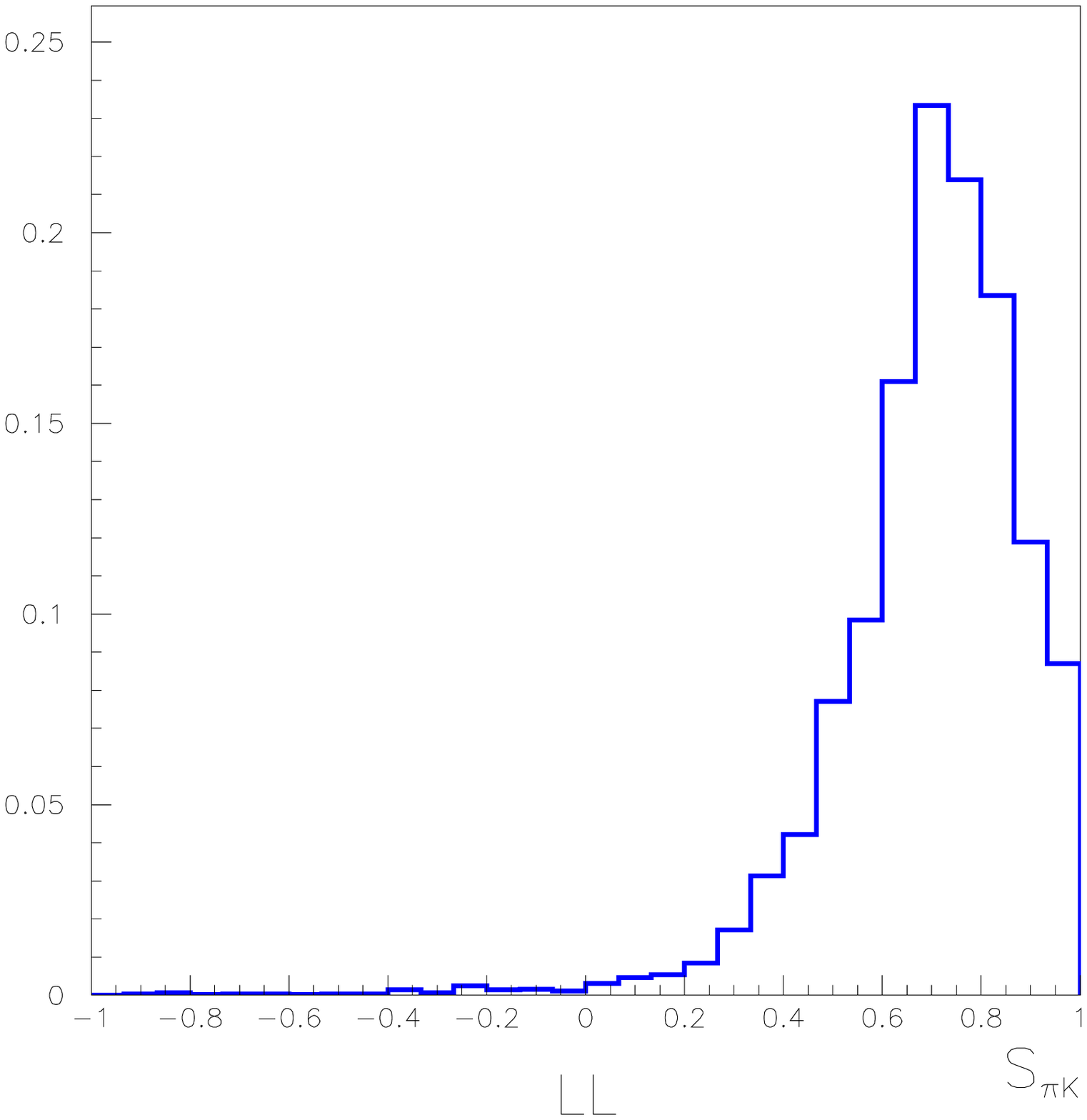} 
\includegraphics[width=0.24\textwidth]{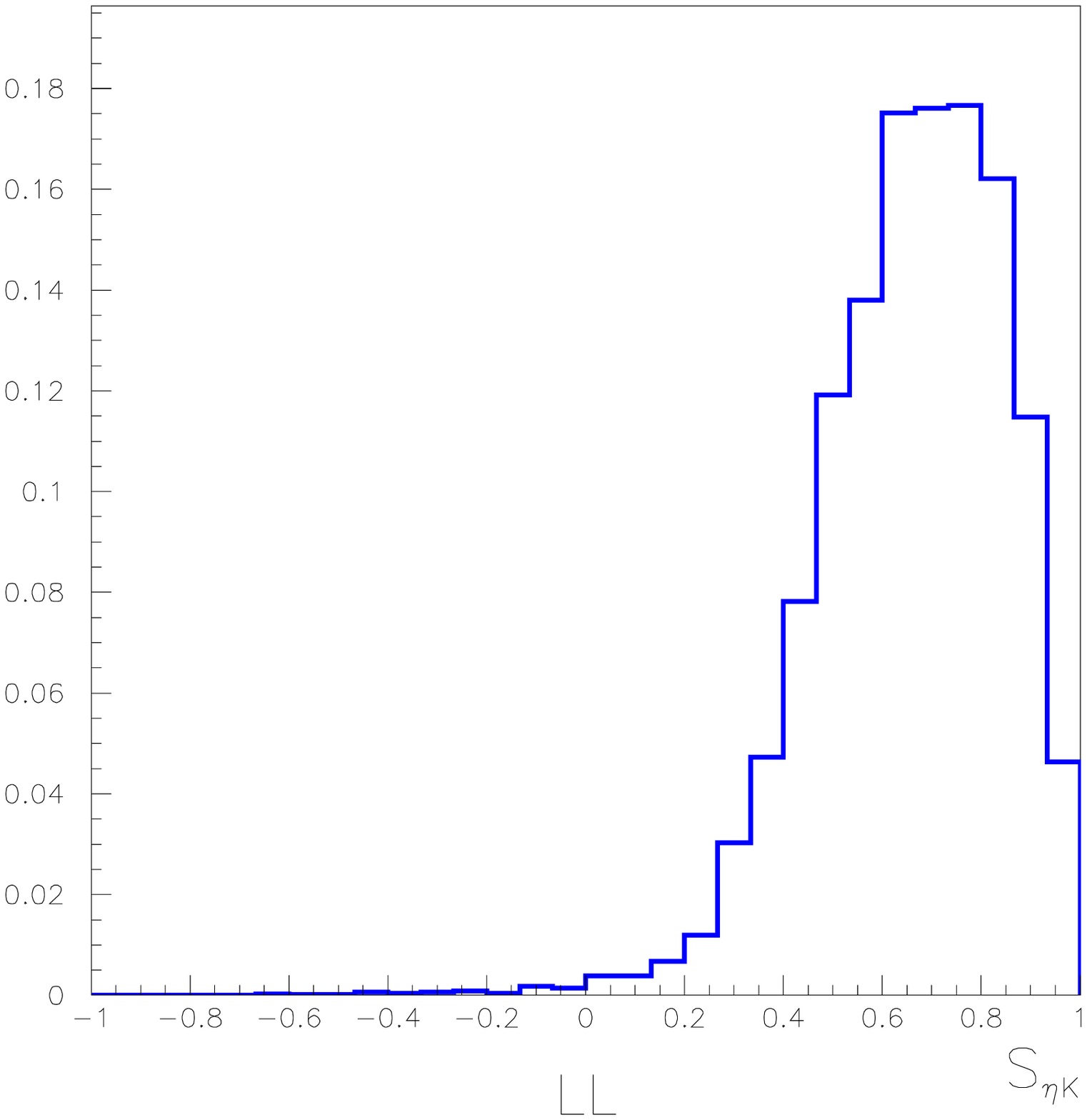} 
\includegraphics[width=0.24\textwidth]{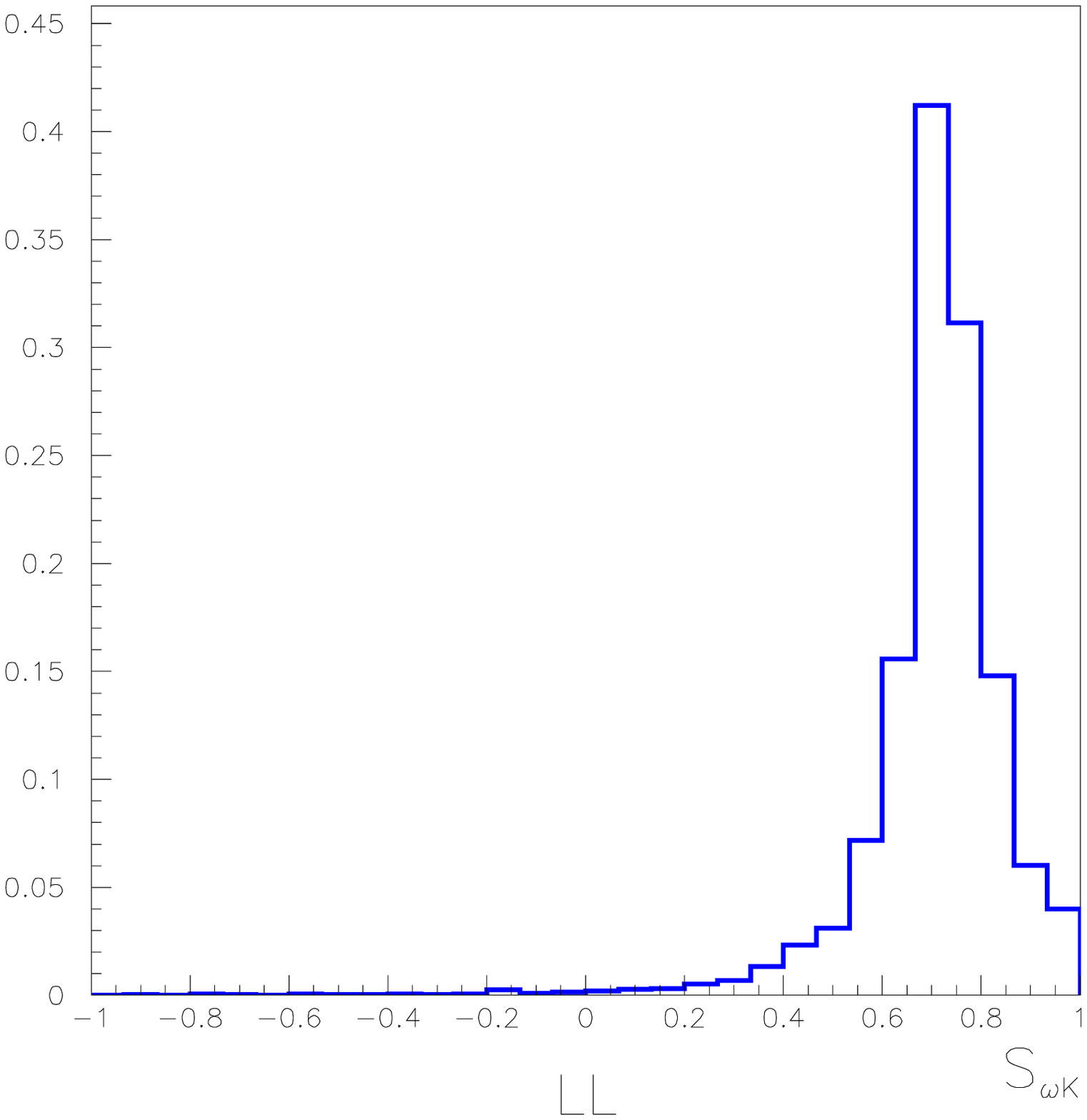} 
\includegraphics[width=0.24\textwidth]{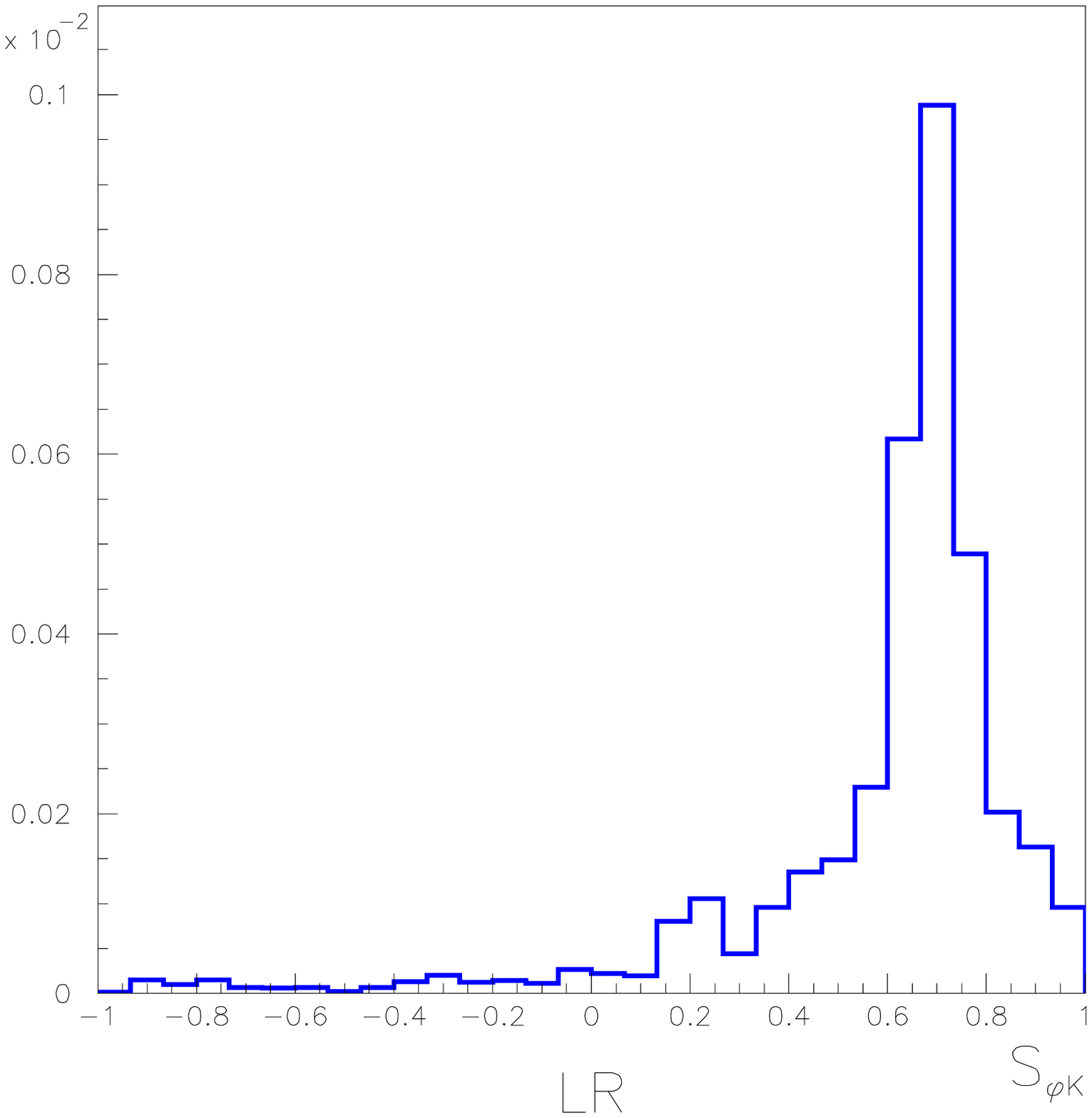} 
\includegraphics[width=0.24\textwidth]{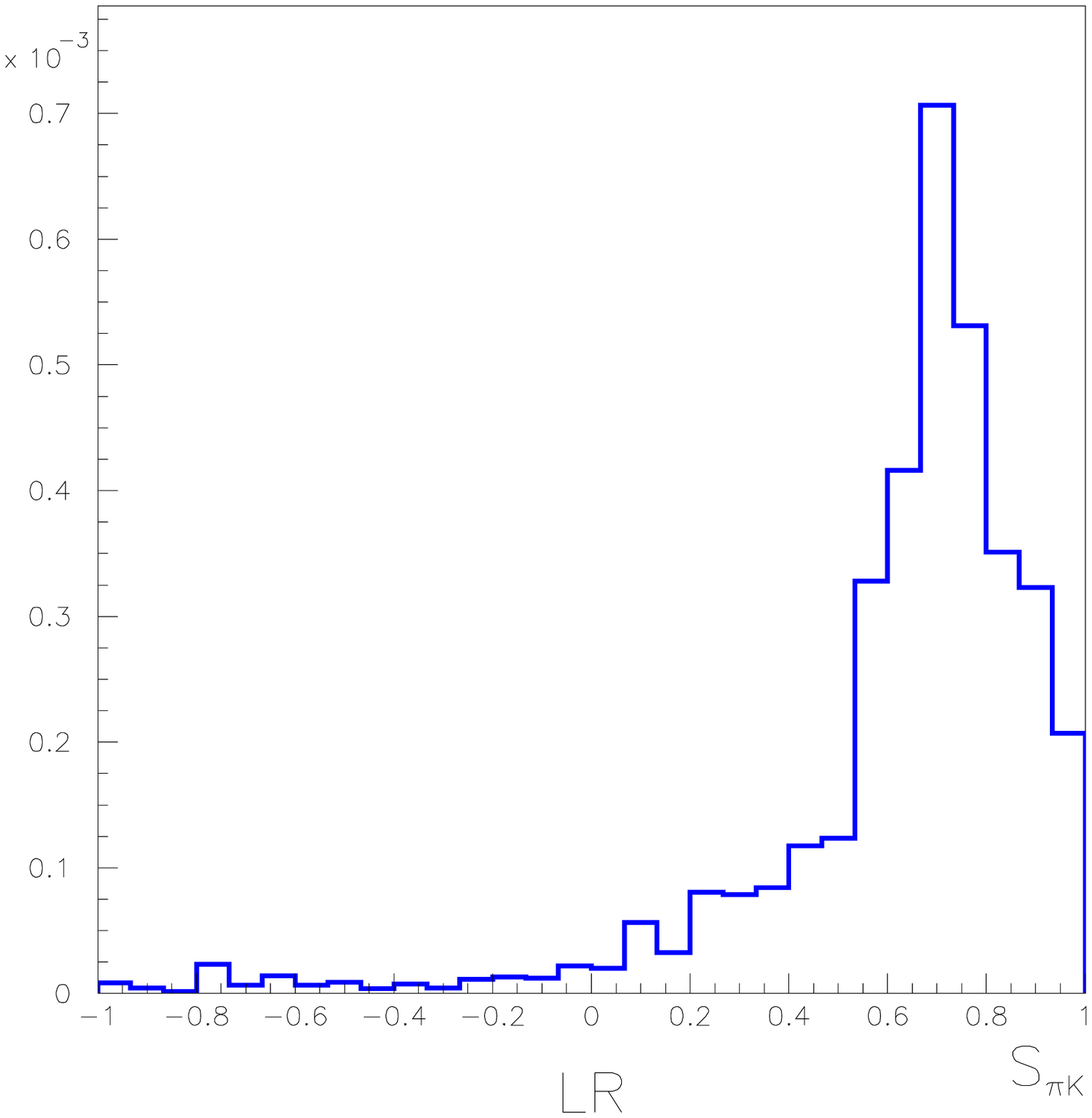} 
\includegraphics[width=0.24\textwidth]{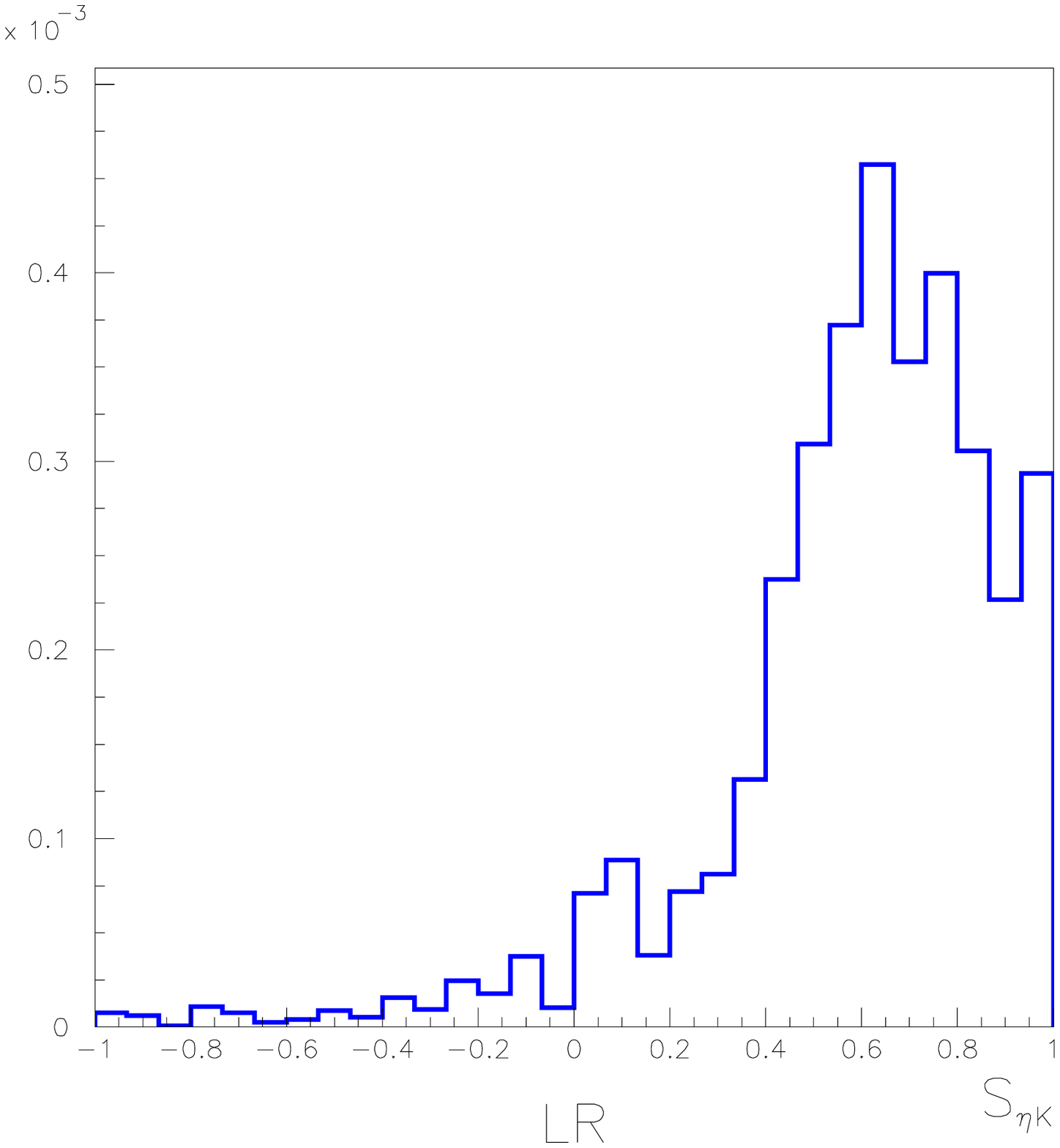} 
\includegraphics[width=0.24\textwidth]{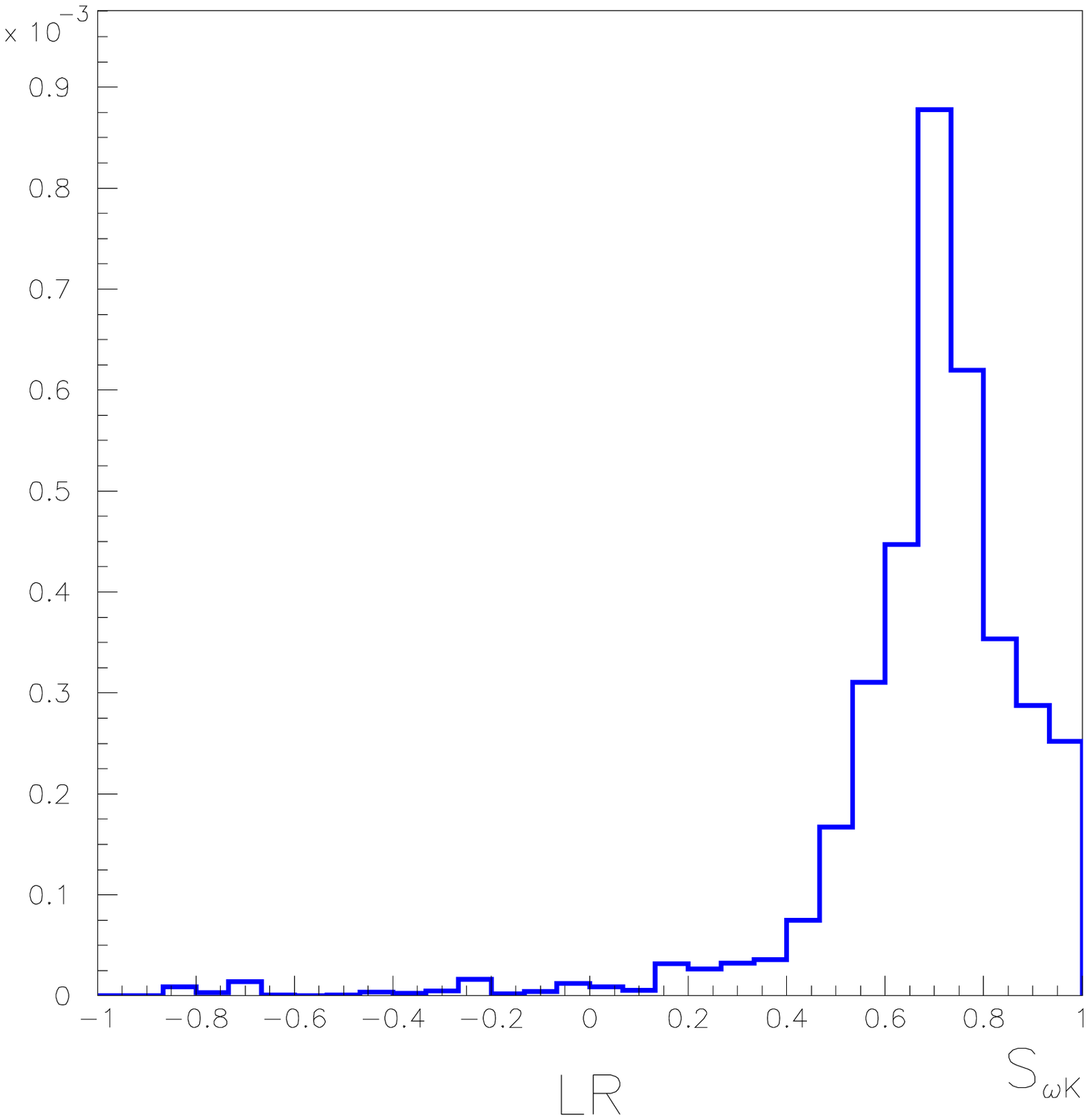} 
\includegraphics[width=0.24\textwidth]{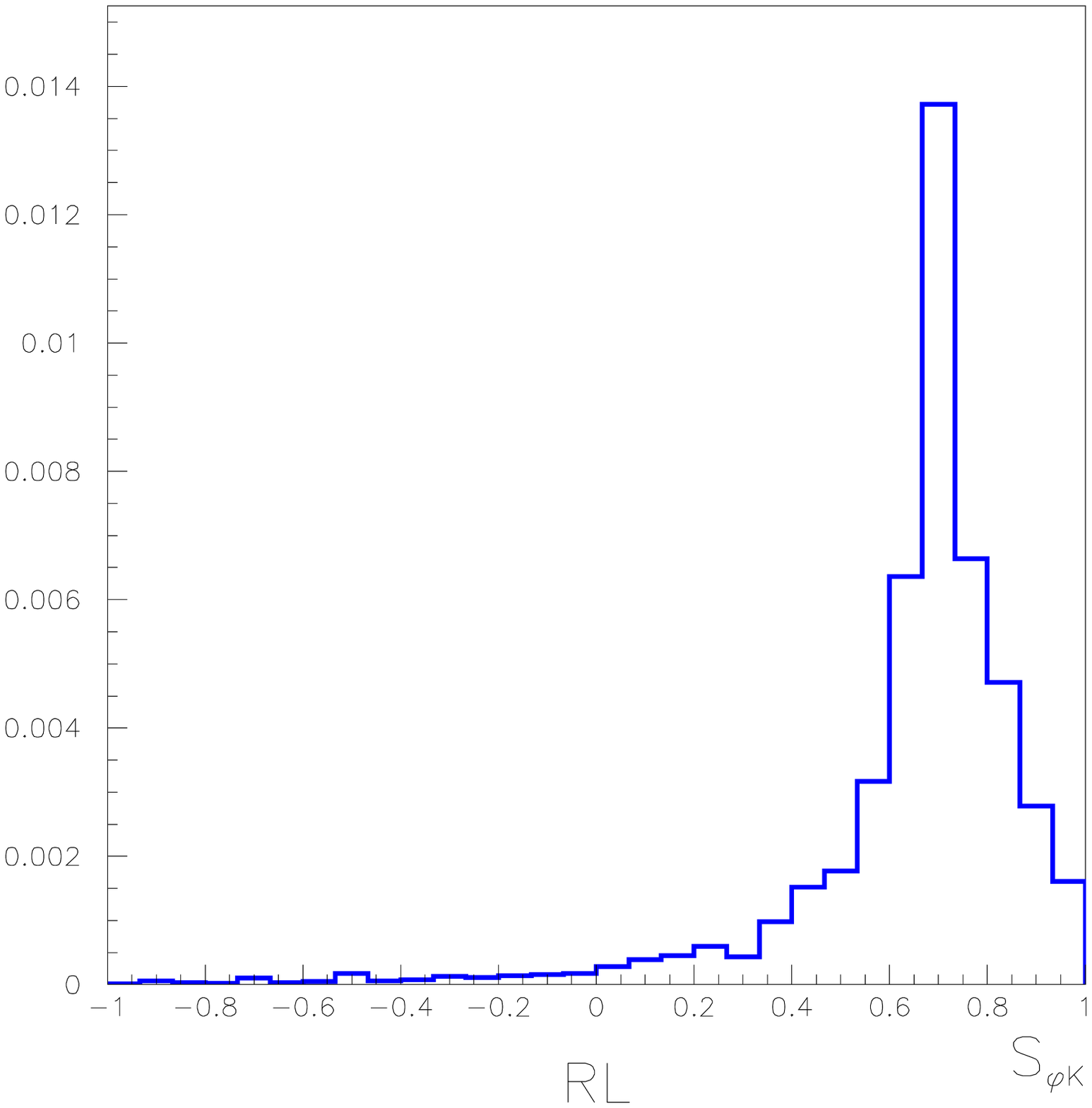} 
\includegraphics[width=0.24\textwidth]{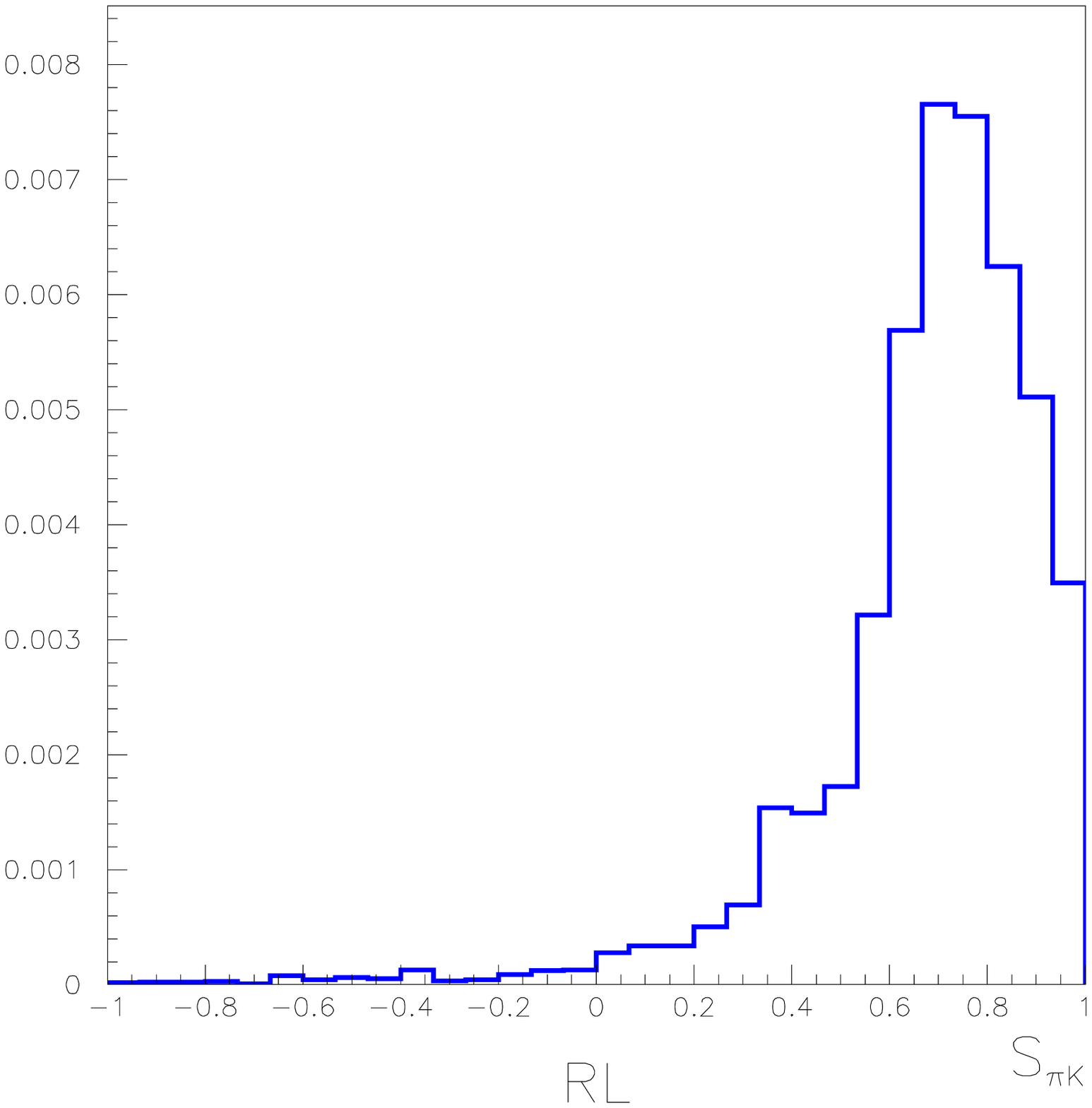} 
\includegraphics[width=0.24\textwidth]{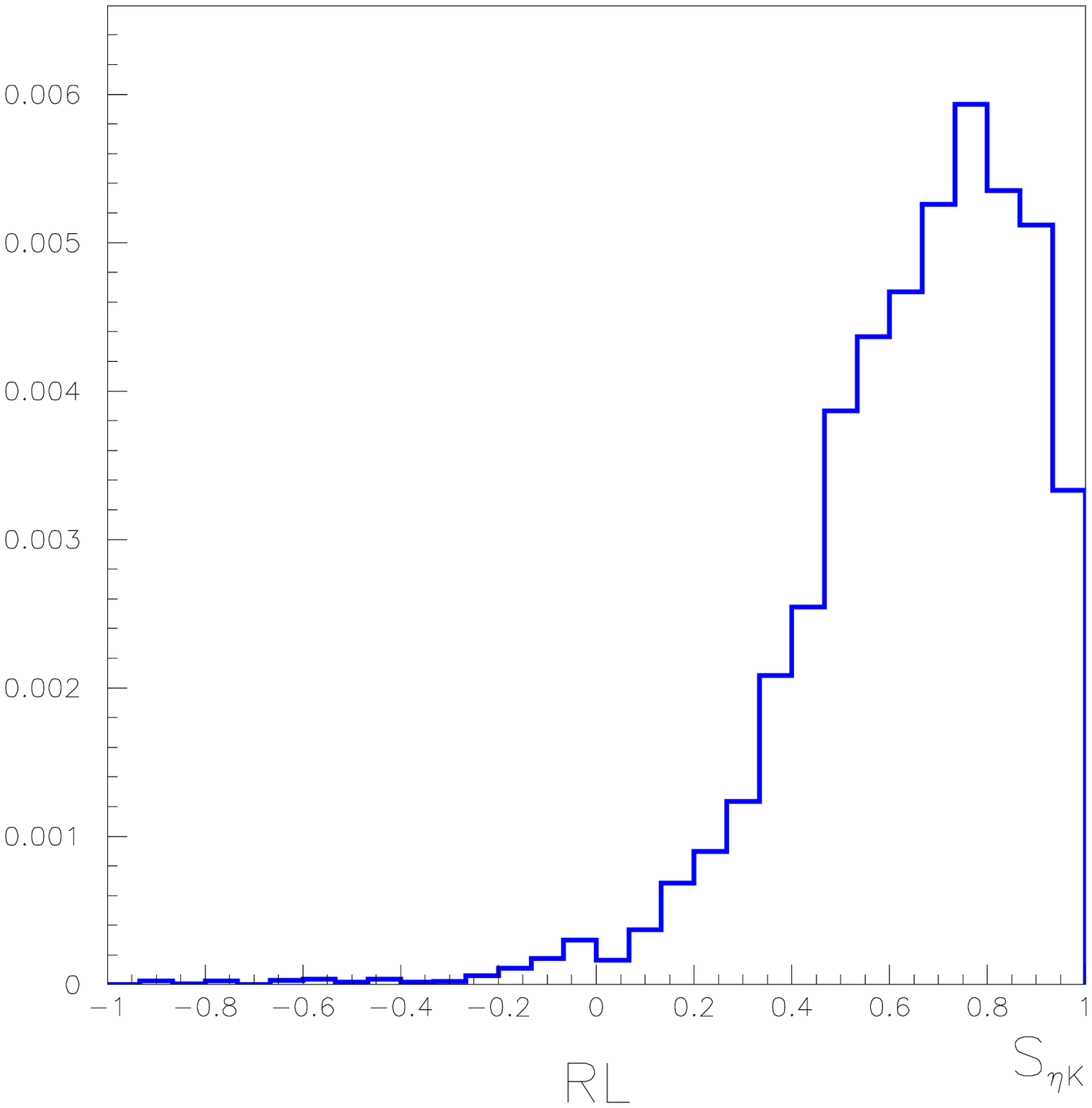} 
\includegraphics[width=0.24\textwidth]{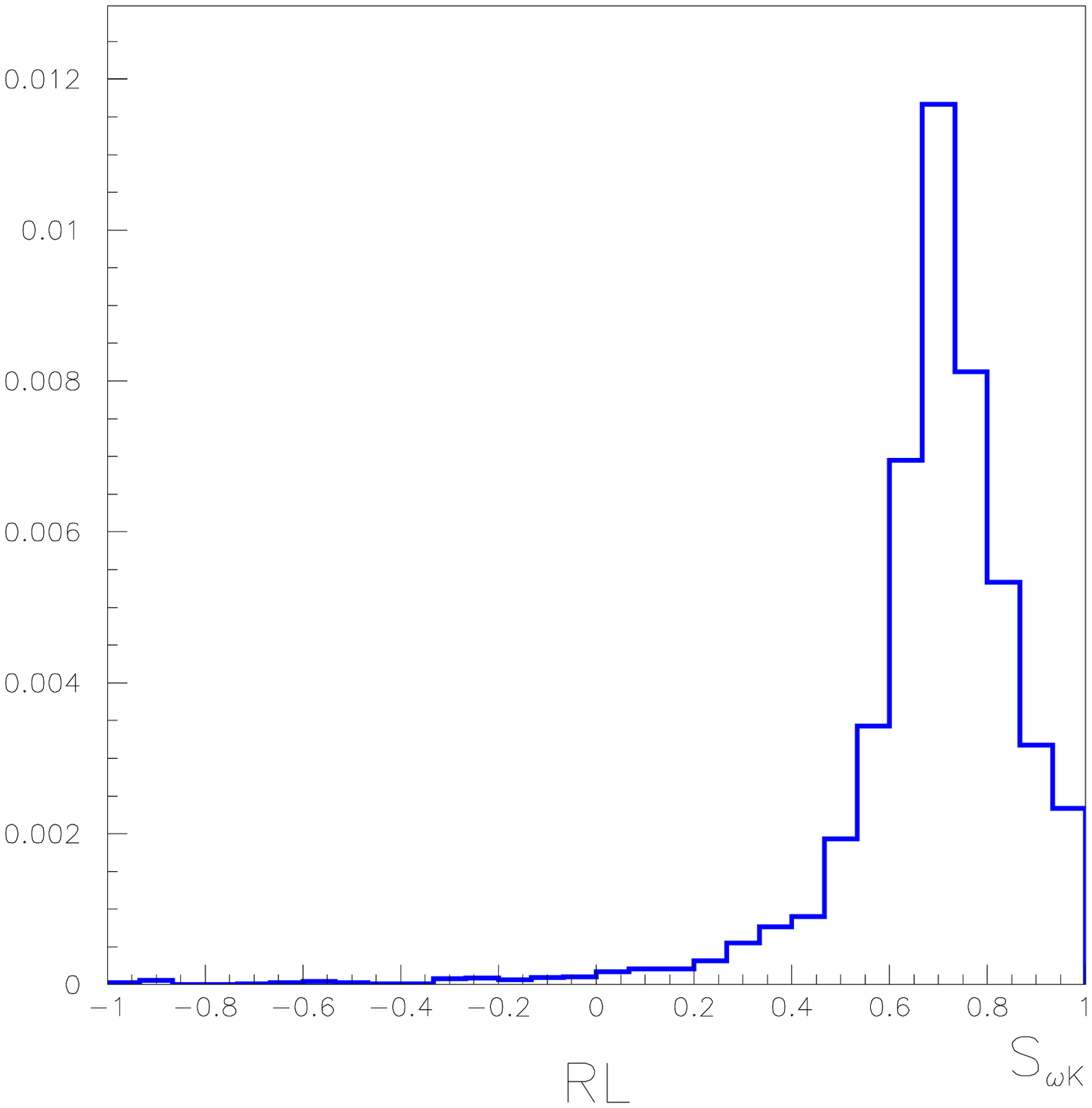} 
\includegraphics[width=0.24\textwidth]{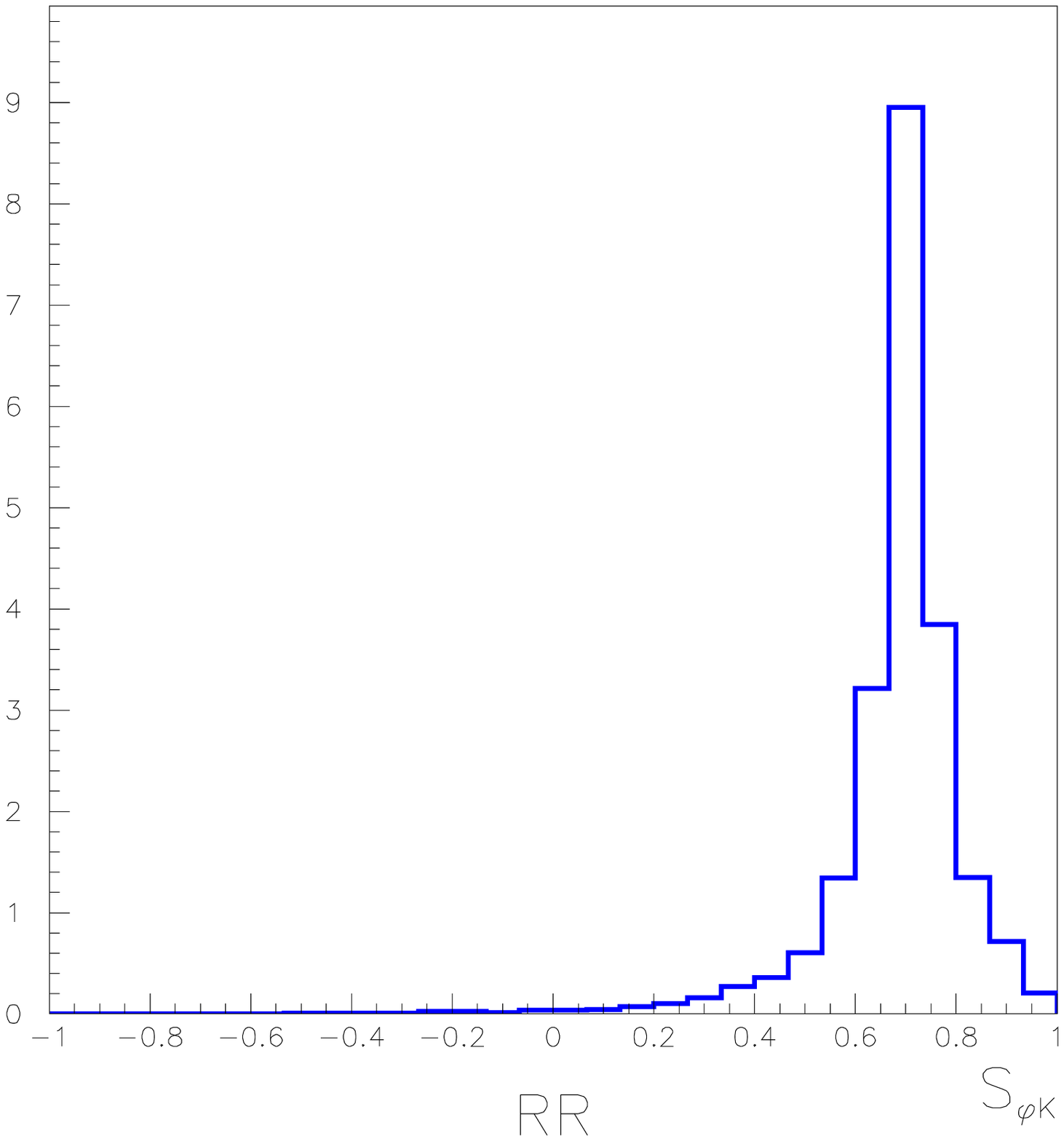} 
\includegraphics[width=0.24\textwidth]{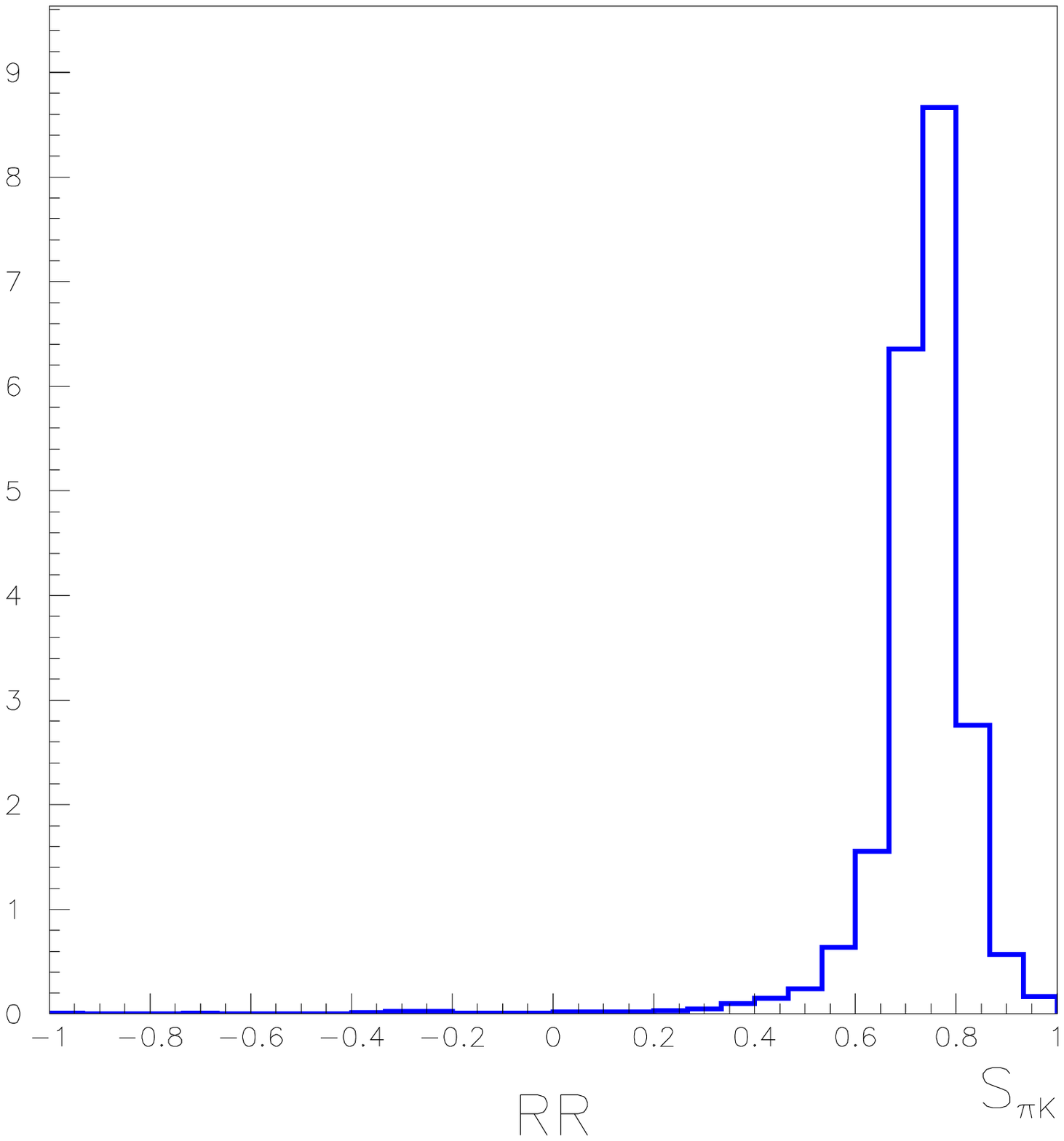} 
\includegraphics[width=0.24\textwidth]{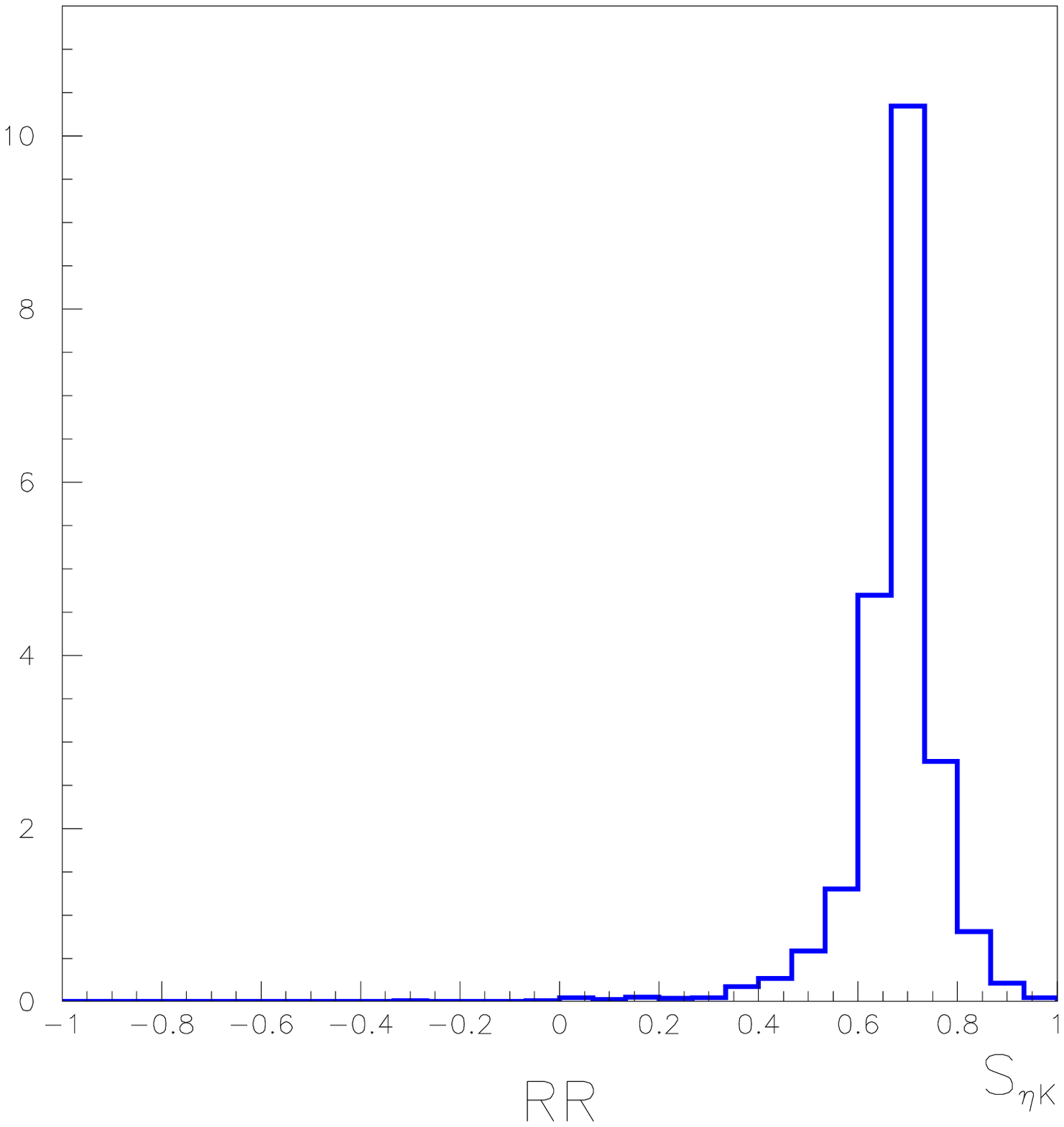} 
\includegraphics[width=0.24\textwidth]{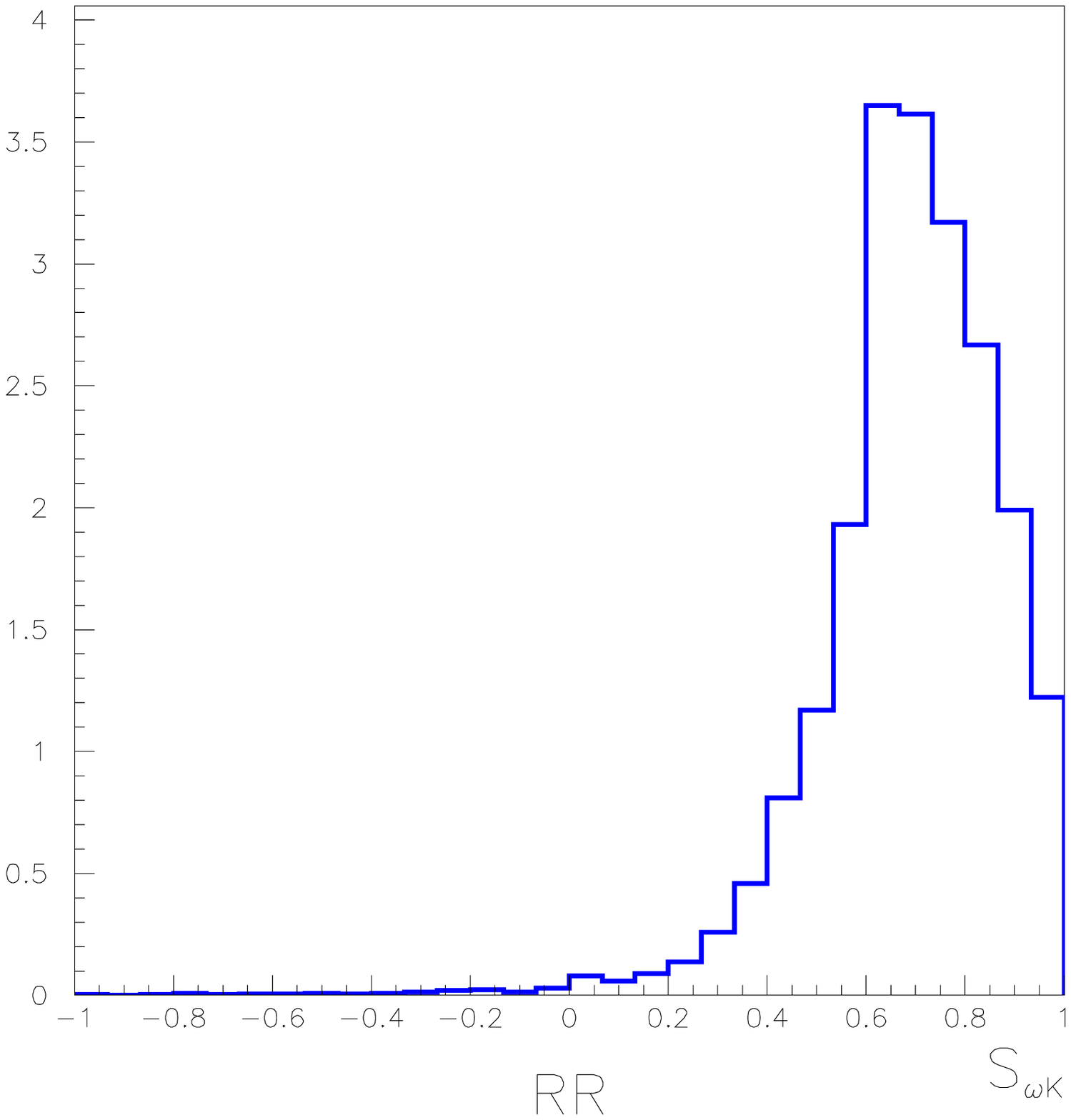} 
\caption{Probability density functions for $S_{\phi K_s}$, $S_{\pi^0
    K_s}$, $S_{\eta^\prime K_s}$ and $S_{\omega K_s}$ induced by
  $(\delta^d_{23})_{\mathrm{AB}}$ with $A,B=\{L,R\}$.}
\label{fig:S}
\end{center}
\end{figure}

Having the present experimental bounds on the $\delta$'s,
we can turn to the evaluation of the time-dependent CP asymmetries. The uncertainty in the calculation of SUSY effects is larger than the SM one. Following ref.~\cite{Ciuchini:2002uv}, we use QCDF enlarging the range for power-suppressed contributions to annihilation chosen in
Ref.~\cite{Beneke:2003zv} as suggested in Ref.~\cite{Ciuchini:2001zf}.
We warn the reader about the large theoretical uncertainties that
affect this evaluation.

\begin{figure}[!ht]
\begin{center}
\includegraphics[width=0.24\textwidth]{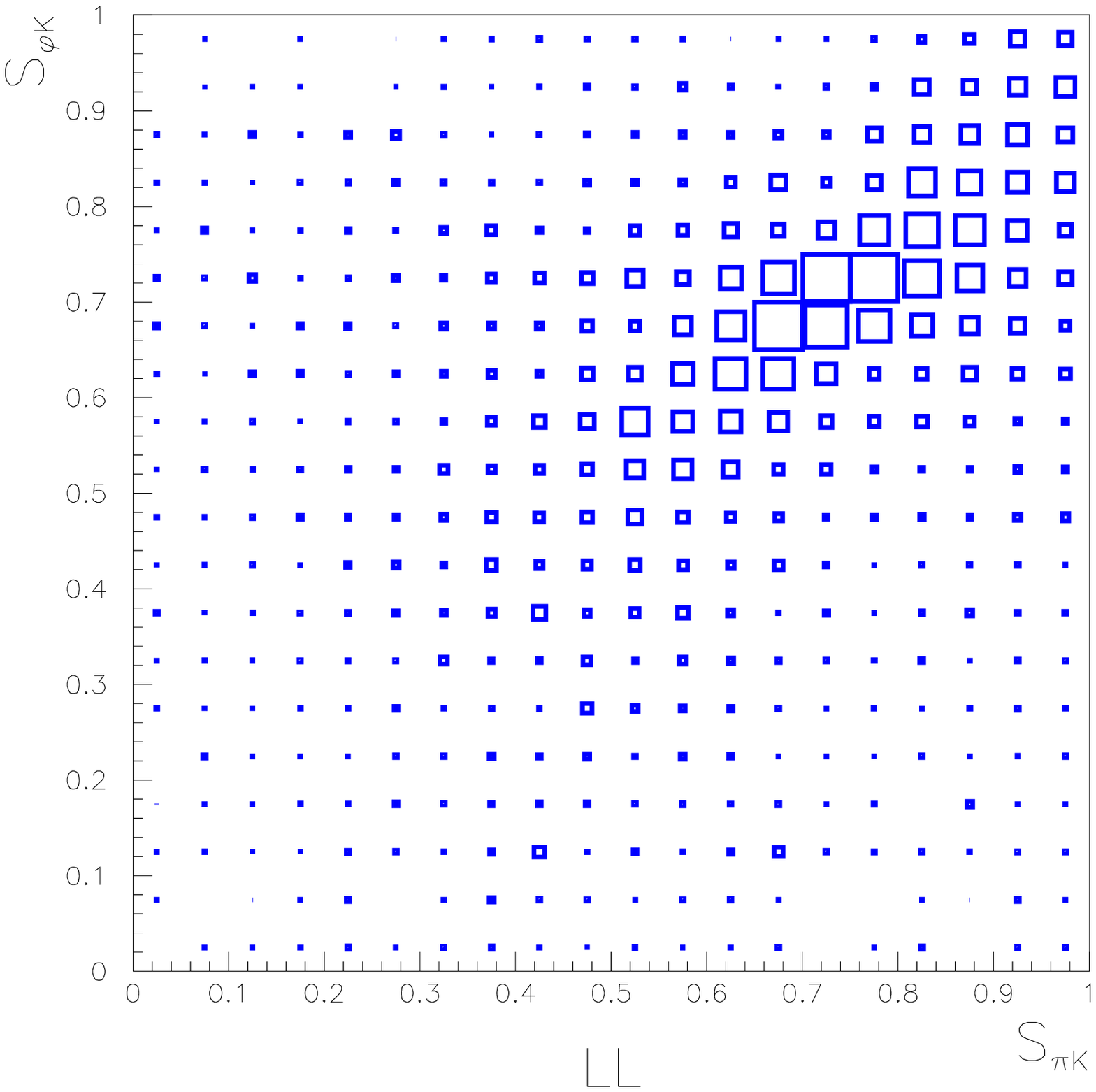} 
\includegraphics[width=0.24\textwidth]{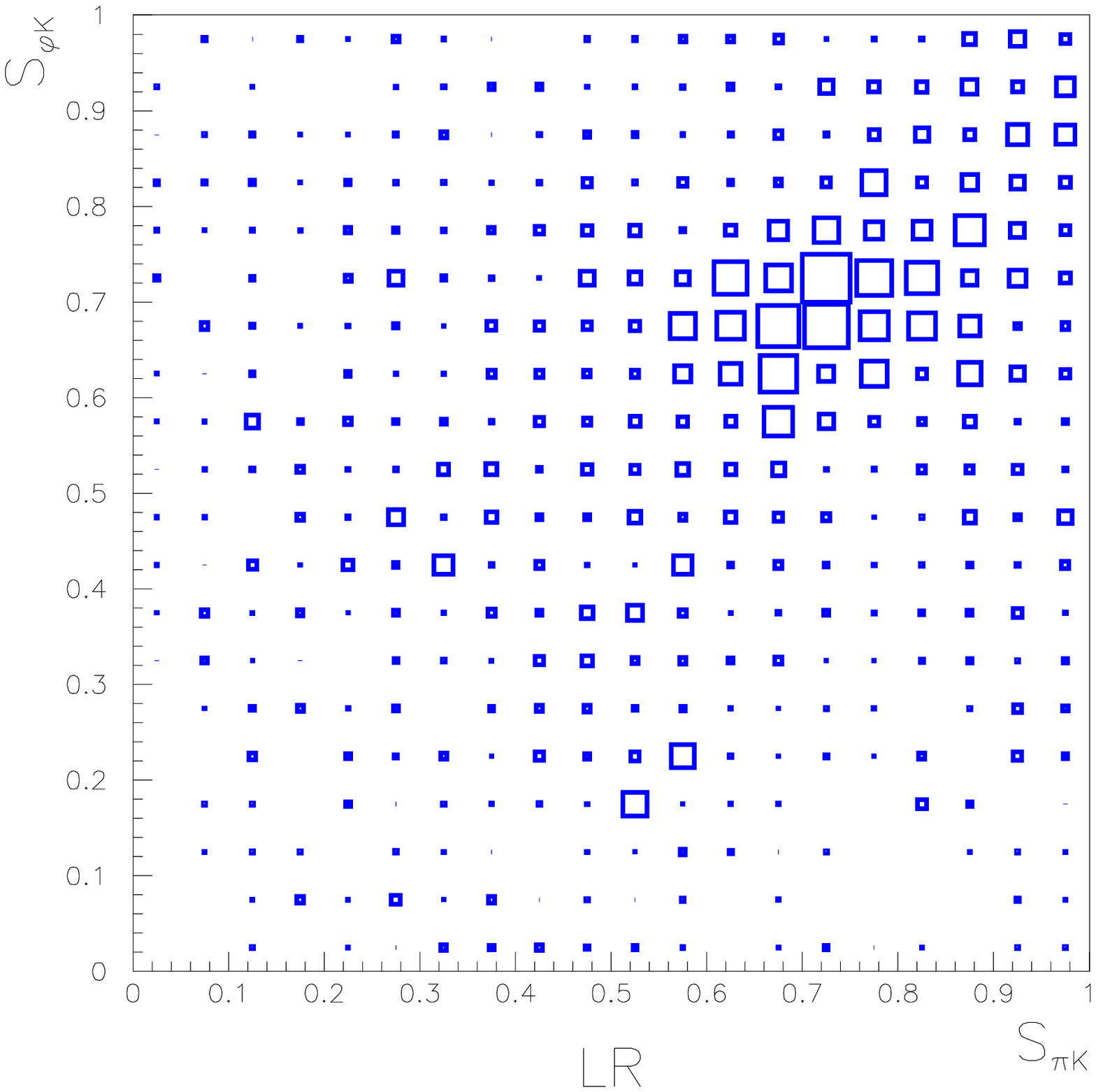} 
\includegraphics[width=0.24\textwidth]{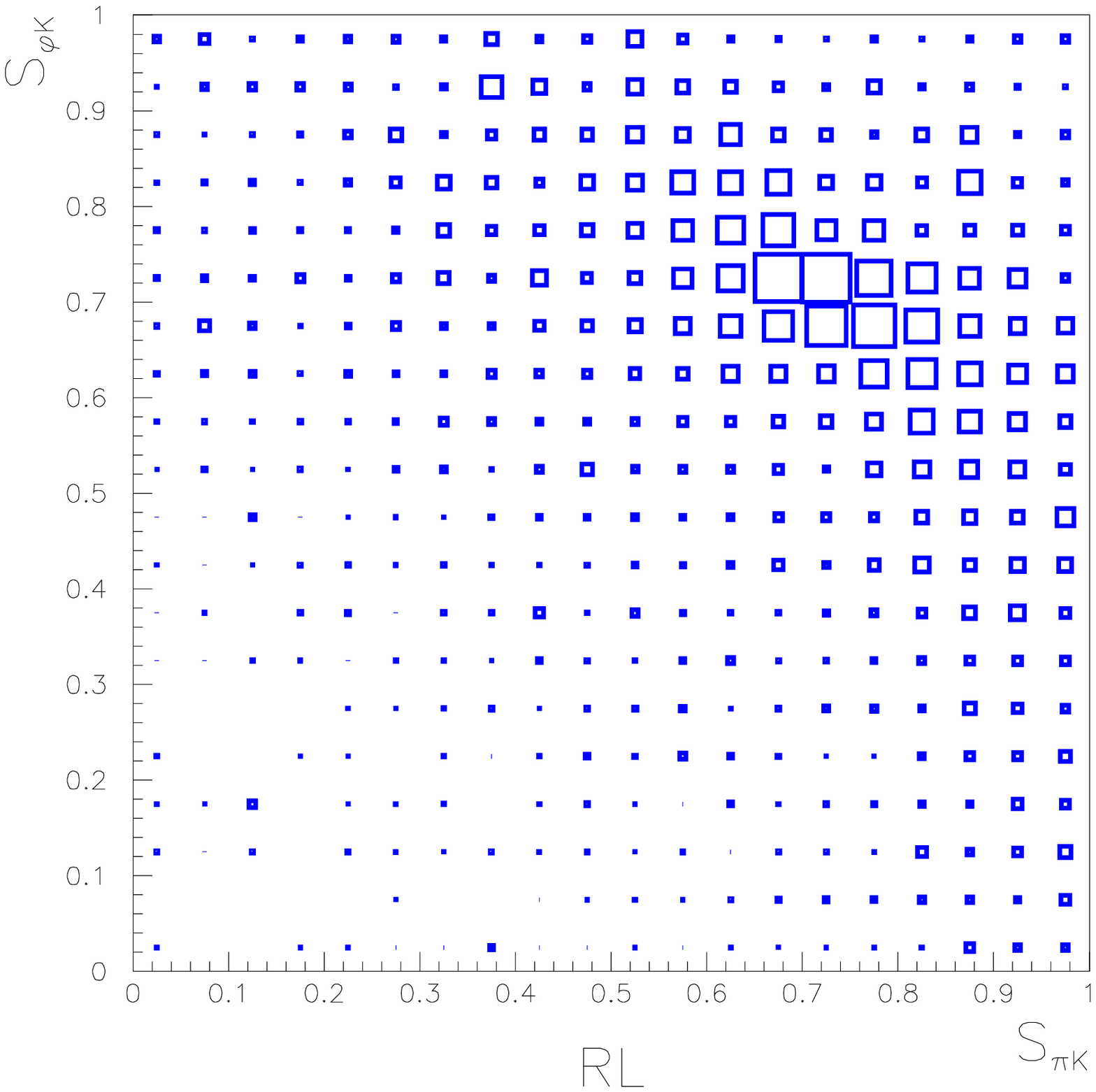} 
\includegraphics[width=0.24\textwidth]{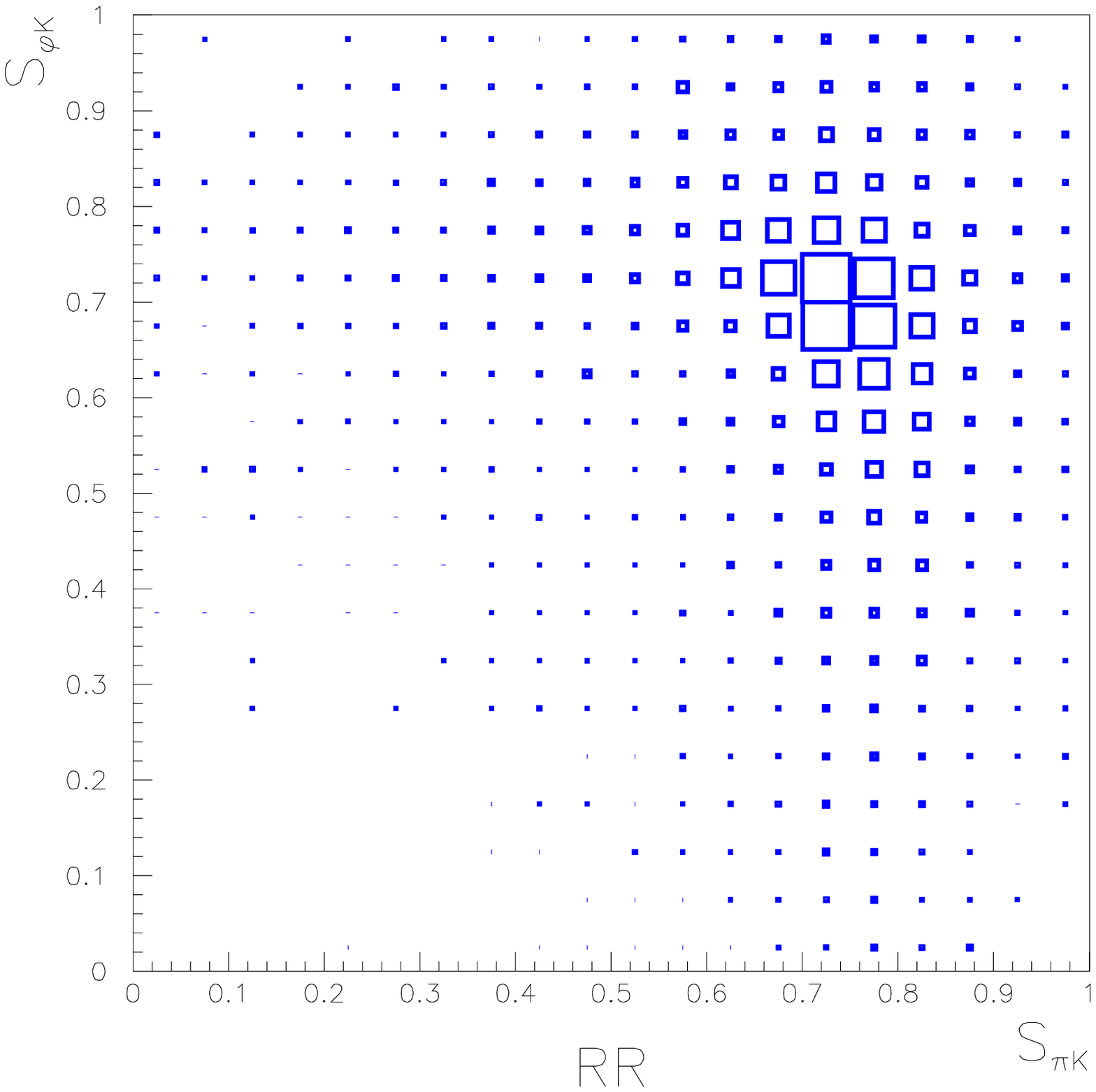} 
\caption{Correlation between $S_{\phi K_s}$ and $S_{\pi^0
    K_s}$ for $LL$, $LR$, $RL$ and $RR$ mass insertions.}
\label{fig:kphivskp0}
\end{center}
\end{figure}

In Fig.~\ref{fig:S} we present the results for $S_{\phi
  K_s}$, $S_{\pi^0 K_s}$, $S_{\eta^\prime K_s}$ and $S_{\omega K_s}$.
They do not show a sizable dependence on the sign of $\mu$ or on $\tan
\beta$ for the chosen range of SUSY parameters. We see that:
\begin{itemize}
\item deviations from the SM expectations are possible in all
  channels, and the present experimental central values can be reproduced;
\item deviations are more easily generated by $LR$ and $RL$ insertions, due
  to the enhancement mechanism discussed above;
\item as noticed in refs.~\cite{Kagan:2004ia,Endo:2004dc}, the
  correlation between $S_{PP}$ and $S_{PV}$ depends on
  the chirality of the NP contributions. For example, we show in
  Fig.~\ref{fig:kphivskp0} the correlation between $S_{K_S\phi}$
  and $S_{K_s\pi^0}$ for the four possible choices for mass
  insertions. We see that the $S_{K_S\phi}$ and $S_{K_s\pi^0}$ are correlated for $LL$ and $LR$ mass insertions, and
  anticorrelated for $RL$ and $RR$ mass insertions.
\end{itemize}

An interesting issue is the scaling of SUSY effects in $S_f$ with
squark and gluino masses. Similarly to the constraints
from other processes, the dominant SUSY contribution to $S_f$ scales linearly with SUSY masses as long as
$m_{\tilde g} \sim m_{\tilde q} \sim \mu$. This means that there is no
decoupling of SUSY contributions to $S_f$ as long as the
constraints from other processes can be satisfied with $\delta <1$. The bounds on $LL$ and $RR$ mass insertions quickly reach the physical boundary at $\delta=1$. On the other hand, $LR$ and $RL$ are well below that bound. Chirality flipping $LR$ and $RL$ mass insertions cannot become too large in order to avoid charge and color breaking minima and unbounded
from below directions in the scalar potential~\cite{Casas:1996de}.
Nevertheless, it is easy to check that the flavor bounds used above are
stronger for SUSY masses above the TeV scale. We conclude
that $LR$ and $RL$ mass insertions can give observable effects to $S_f$ for
SUSY masses within the reach of LHC and even above.

\subsubsection{Experimental status and future prospects for
time-dependent $CP$ violation in hadronic $b\to s(d)$ transitions}

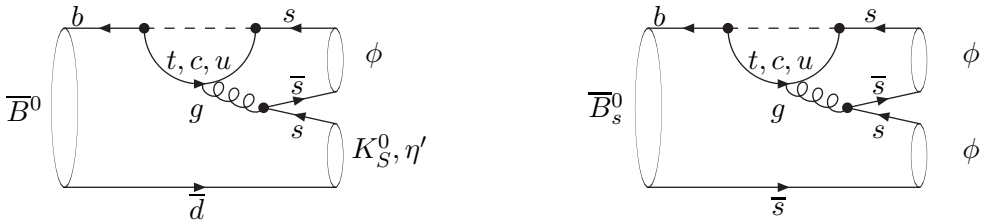
\begin{figure}[htb]
\begin{center}
\begin{picture}(400,100)(0,0)
\Oval(30,40)(30,5)(0)
\Oval(132,22)(12,3)(0)
\Oval(132,58)(12,3)(0)
\ArrowLine(30,10)(132,10)
\ArrowLine(60,70)(30,70)
\Vertex(60,70){2}
\Vertex(102,70){2}
\ArrowLine(132,70)(102,70)
\ArrowArc(81,70)(21,180,0)
\DashLine(102,70)(60,70){4}
\Gluon(82,49)(105,40){-3}{3}
\Vertex(105,40){2}
\ArrowLine(105,40)(132,46)
\ArrowLine(132,34)(105,40)
\Text(81,57)[]{$t,c,u$}
\Text(80,38)[]{$g$}
\Text(35,75)[]{$b$}
\Text(80,3)[]{$\overline{d}$}
\Text(115,75)[]{$s$}
\Text(118,49)[]{$\overline{s}$}
\Text(118,31)[]{$s$}
\Text(15,40)[]{$\overline{B}{}^{0}$}
\Text(153,24)[]{$\ks,\etap $}
\Text(147,60)[]{$\phi$}
\Oval(250,40)(30,5)(0)
\Oval(352,22)(12,3)(0)
\Oval(352,58)(12,3)(0)
\ArrowLine(250,10)(352,10)
\ArrowLine(280,70)(250,70)
\Vertex(280,70){2}
\Vertex(322,70){2}
\ArrowLine(352,70)(322,70)
\ArrowArc(301,70)(21,180,0)
\DashLine(322,70)(280,70){4}
\Gluon(302,49)(325,40){-3}{3}
\Vertex(325,40){2}
\ArrowLine(325,40)(352,46)
\ArrowLine(352,34)(325,40)
\Text(301,57)[]{$t,c,u$}
\Text(300,38)[]{$g$}
\Text(255,75)[]{$b$}
\Text(300,3)[]{$\overline{s}$}
\Text(335,75)[]{$s$}
\Text(338,49)[]{$\overline{s}$}
\Text(338,31)[]{$s$}
\Text(235,40)[]{$\overline{B}{}^{0}_{s}$}
\Text(373,24)[]{$\phi$}
\Text(373,60)[]{$\phi$}
\end{picture}
\end{center}
\vspace*{-5mm}
\caption{The penguin diagrams for the hadronic $\bz$ and $\B^0_s$ 
decays such as $\bz \to \phi \ks$, 
$\bz \to \etap \ks$ (left)
and $B^0_s \to \phi\phi$ (right).}
\label{fig:btos_diagrams}
\end{figure}

$CP$ asymmetries in $\bz$ and $B_s$ decays that are governed by
the $b\to s$ transition are very sensitive to new $CP$-violating
phases beyond the Standard Model (SM). There are a few golden modes
that are practically free from hadronic uncertainties;
examples include $\bz\to \phi\ks$, $\etap\ks$, $\ks\ks\ks$ and
$B^0_s\to\phi\phi$, see Figure~\ref{fig:btos_diagrams}.
Precise measurements for these decays have been among
the most important topics of quark flavor physics in
the last few years, and will also remain crucially important
in the future.

At the $B$ factories,
the decay chain $\Upsilon(4S)\to \bz\bzb \to f_{CP}f_{\rm tag}$
is used to measure time-dependent $CP$ asymmetries,
where one of the $B$ mesons decays at time $t_{CP}$ to a 
final state $f_{CP}$ 
and the other decays at time $t_{\rm tag}$ to a final state  
$f_{\rm tag}$ that distinguishes between $B^0$ and $\bzb$.
The rate of this decay chain
has a time dependence~\cite{Bigi:1988ym,Carter:1980hr}
given by
\begin{equation}
\label{eq:psig}
{\cal P}(\Delta{t}) = 
{e^{-|\Delta{t}|/{\taubz}}}{4{\taubz}}
\biggl\{1 + \fq\cdot 
\Bigl[ \cals\sin(\dmd\Delta{t})
   + \cala\cos(\dmd\Delta{t})
\Bigr]
\biggr\}.
\end{equation}
Here $\cals$ and $\cala$ are $CP$-violation parameters,
$\taubz$ is the $B^0$ lifetime, $\dmd$ is the mass difference 
between the two $B^0$ mass
eigenstates, $\Delta{t}$ = $t_{CP}$ $-$ $t_{\rm tag}$, and
the $b$-flavor charge $\fq$ = +1 ($-1$) when the tagging $B$ meson
is a $B^0$ ($\bzb$).
To a good approximation,
the SM predicts $\cals = -\xi_f\sinbb$ and $\cala =0$
for both tree transitions (e.g. $\btoccs$)
and penguin transitions (e.g. $\btosss$)
unless $V_{ub}$ or $V_{td}$ is involved in the decay amplitude.
Here $\xi_f = +1 (-1)$ corresponds to  $CP$-even (-odd) final states.

BaBar and Belle have accumulated
more than $10^9$ $B\bar{B}$ pairs with both experiments combined,
and have measured
time-dependent $CP$ asymmetries in various $\bz$
decays that are dominated by the $b\to s$ 
transition.
Details of the measurements are described 
elsewhere~\cite{Aubert:2005id,Aubert:2006ai,Aubert:2006ar,Aubert:2006ad,Aubert:2006av,Chen:2006nk,Abe:2006gy,Aubert:2006wv};
we here explain the essence of the measurements briefly.
Branching fractions for these charmless decay modes are
typically around $10^{-5}$ ignoring daughter
branching fractions. Efficient continuum suppression
using sophisticated techniques such as Fisher discriminants, likelihood ratios
and neural network has been performed
to keep a reasonable signal-to-noise ratio.
The flavor of the accompanying $B$ meson is identified
from inclusive properties of remaining particles;
information from primary and secondary leptons, charged kaons,
$\Lambda$ baryons, slow and fast pions is combined
by using a neural network (BaBar) or a lookup-table (Belle).
A typical effective efficiency for flavor tagging is 30\%
in both cases.
Good understanding of the vertex resolution function is
obtained by using large-statistics control samples
such as $B\to D^{(*)}\pi$, $D^*\ell\nu$ etc.
Lifetime and mixing measurements with a precision of
${\cal O}(1)$\% are obtained as byproducts.

\begin{figure}[htb]
\begin{center}
\includegraphics[width=7.9cm]{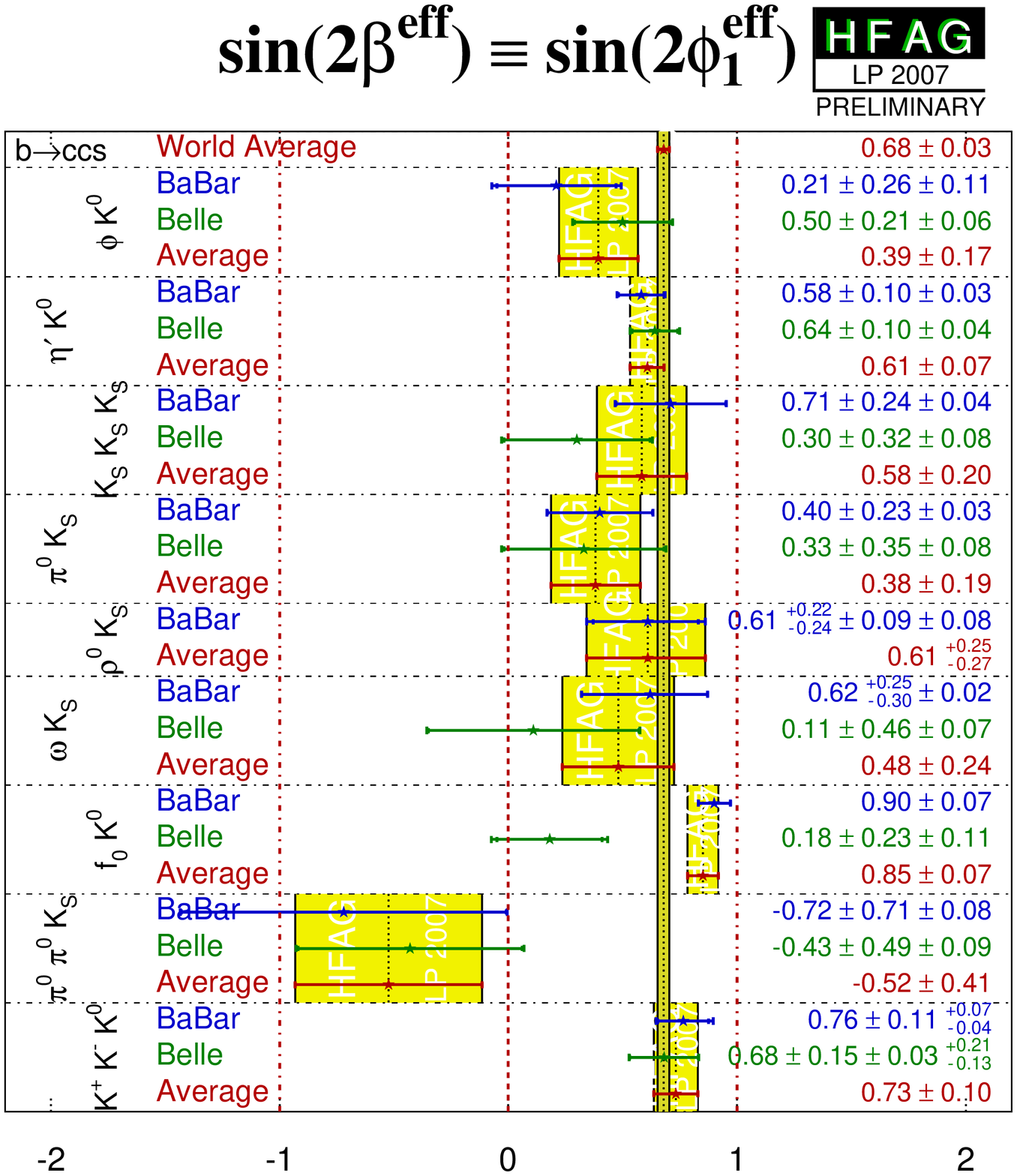}
\includegraphics[width=7.9cm]{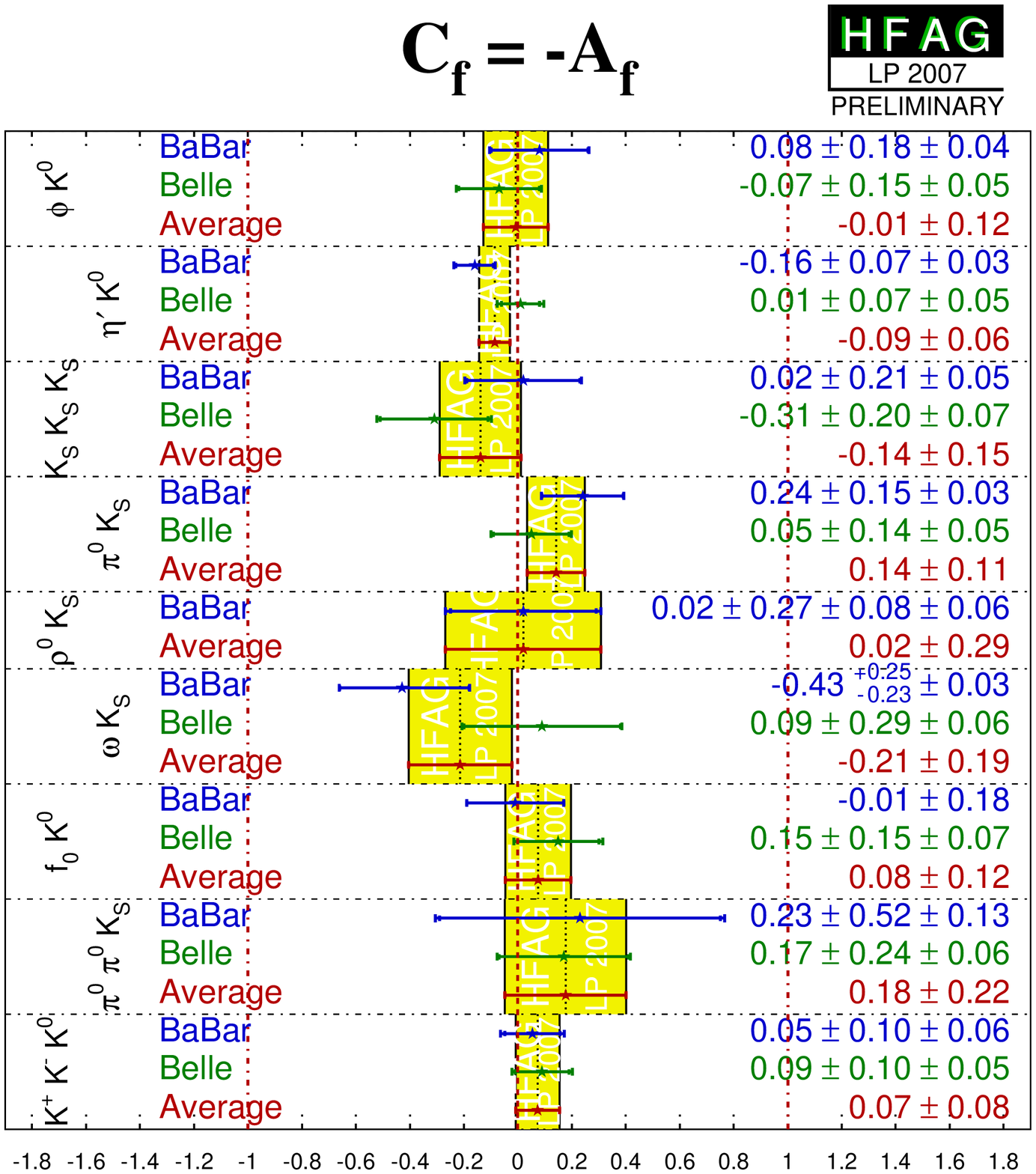}
\caption{Summary of experimental results on
time-dependent $CP$ asymmetries from BaBar and Belle
as of August 2007.}
\label{fig:b2s_now}
\end{center}
\end{figure}

The present status of the measurements is summarized in
Fig.~\ref{fig:b2s_now}. 
Although the result for each individual mode
does not significantly differ from
the SM expectation (i.e. $\cals_{\jpsi\kz}$),
most of the $\cals$ values are smaller
than the SM expectation.
When all the $b\to s$ modes are combined,
the result differs from the SM expectation
by 1.1$\sigma$.\footnote{%
Due to the highly non-Gaussian errors of the result from $B^0 \to f_0K^0_S$ 
with $f_0 \to \pi^+\pi^-$, and the fact that this result has a significant effect 
on the $\chi^2$ of the naïve $b\to s$ penguin average, 
this outlying point is excluded.
} 
Combining the results of all the $b\to s$ modes 
is naive as the theoretical uncerainties 
vary considerably amongst the modes.
Much more data are needed to firmly establish
a new $CP$-violating phase beyond the SM
for each golden mode.

Measurements of the $\cala$ terms yield
values consistent with zero, i.e. 
consistent with the SM at the moment.
Non-zero $\cala$ requires a strong phase difference
between the SM amplitude and the NP amplitude.
Therefore it is possible to observe
significant deviations from the SM for $\cals$
while $\cala$ is consistent with zero.
Also, since $\cala$ is not calculable precisely,
in general it is hard to obtain
quantitative information from the measurements
of $\cala$ terms.
An exception is the $\bz\to\kz\piz$ decay.
Thanks to a precise sum rule based on the isospin
symmetry~\cite{Gronau:2005kz}, the value for
$\cala_{\kz\piz}$ can be predicted within
the SM from measurements of branching fractions
and $CP$ asymmetries of the other $B\to K\pi$
decays; $\cala_{\kz\piz} = -0.16\pm 0.04$ is predicted
while measurements yield $\cala_{\kz\piz} = -0.12\pm 0.11$.

Due to further CKM-suppression, $CP$ asymmetry measurements
for modes dominated by the $b\to d$ transition require
even higher statistics than those required for
the studies of the $b\to s$ transition.
The only measurement available at the moment is
$\cals_{\bz\to\ks\ks} = -1.28^{+0.80}_{-0.73}{}^{+0.11}_{-0.16}$~\cite{Aubert:2006gm},
where the first error is statistic and the second error
is systematic.

In the near future the LHCb experiment will probe new CP violating phases beyond
the Standard Model in $b\to s$ transitions. 
With the copious production of 
$B^0_s$ mesons LHCb will be able to study $b\to s$ transitions 
using the the decay  $B^0_s\to\phi\phi$, see Figure~\ref{fig:btos_diagrams}.
In the Standard Model 
the CP violating phase $\cals_{\phi\phi}$ for $B^0_s\to\phi\phi$ 
is expected to be very close to zero 
as there is a cancellation of the $B^0_s$ mixing and 
decay phases~\cite{Raidal:2002ph}.

In the LHCb experiment the reconstruction efficiency for 
$B^0_s\to\phi\phi$ is expected to be larger 
than for $\bz \to \phi \ks$ 
which compensates for the four times smaller fraction of 
$b$-quarks to hadronise into a $B^0_s$ meson. In addition, flavour tagging
is also favourable for $B^0_s$ decays where the same-side kaon tagging
contributes significantly to the effective flavour tagging efficiency.
From a full simulation LHCb expects a yield of 3100 reconstruced   $B^0_s\to\phi\phi$ events
in a $2\fbinv $ data sample 
with a background to signal ratio $B/S < 0.8$ at 90\% C.L~\cite{ref:LHCbphiphi}.
The $\cals_{\phi\phi}$ sensitivity 
has been studied using a toy Monte Carlo,
taking the resolutions and acceptances from the full simulation.
A unbinned likelihood fit is performed on 500 toy data sets. 
This is used to extract $\cals_{\phi\phi}$ and all other physical parameters
which cannot be determined from elsewhere. 
In a  $2 \fbinv$ data set $\cals_{\phi\phi}$ can be measured 
with a precision of $\sigma(\cals_{\phi\phi}) = 0.11$ (statistical error only).
After about 5 years of data taking, LHCb is expected to accumulate a data 
sample of 10~\fbinv\ which will give a statistical uncertainty of 
$\sigma(\cals_{\phi\phi})=0.05$~\cite{ref:LHCbphiphi}.

\begin{table}[!bth]
\begin{center}
\caption{$CP$ reach at LHCb~\cite{ref:schneider} and at a Super-$B$ factory
for the $b\to s$ decay modes
that are theoretically cleanest. 
The estimated accuracy from the $B$ factories (2~ab$^{-1}$) is given for comparison.
We assume an integrated luminosity of 10 fb$^{-1}$ for LHCb
and 50 ab$^{-1}$ for a super $B$ factory, which are the goals of
the experiments. Errors for LHCb are statistical only.
Projections for the super $B$ factory
are from Ref.~\cite{SuperKEKB}  and include both statistical
and systematic uncertainties and
$\Delta\sinbb \equiv \sinbb^{\rm eff} - \sinbb$.}
\label{tab:b2s_future}
\begin{tabular}{ccccc}
\hline\hline
Mode              & Observable     & $B$ Factories & LHCb      & Super $B$ Factory\\
                  &                & 2 ab$^{-1}$   & 10~\fbinv & 50 ab$^{-1}$ \\
\hline
$\bz\to\phi\kz$   & $\Delta\sinbb$ &  0.13  & 0.10 & 0.029 \\
$\bz\to\etap\kz$  & $\Delta\sinbb$ &  0.05  & -    & 0.020 \\
$\bz\to\ks\ks\ks$ & $\Delta\sinbb$ &  0.15  & -    &  0.037 \\
$B^0_s\to\phi\phi$  & $\cals_{\phi\phi}$  & - & 0.05 & - \\
\hline\hline
\end{tabular}
\end{center}
\end{table}

In a similar study LHCb investigated the decay $\bz \to \phi\ks $.
A yield of 920 events is expected in $2~\fbinv$ of integrated luminosity 
with a background to signal ratio $0.3 < B/S <1.1$ at 90\% C.L.
The sensitivity for the CP violating asymmetry $\sin 2 \beta ^{\rm eff}$ 
is 0.23 (0.10) in a 2 (10)~\fbinv\ data sample~\cite{ref:LHCbphiks}.

Table~\ref{tab:b2s_future} lists the expected $CP$ reach
at LHCb and a Super-$B$ factory for the theoretically
cleanest $b\to s$ decay modes.
We expect that the precision will be better by
an order of magnitude than now. 
Such measurements will thus allow us to detect
effects from physics beyond the SM even if
the mass scale of the new physics is ${\cal O}(1)$ TeV.

\subsubsection{Two body hadronic $B$ decay results from the $B$-factories}

This class of $B$ decays manifests a wide range of interesting phenomenon, from direct $CP$ violation, broken SU(3) symmetry
constraints on the standard model uncertainties in measurements of the unitarity triangle angles, to the amplitude hierarchy found
in decays to final states containing two spin one particles (vector or axial-vector mesons, $V$ and $A$, respectively).

%
%
%
The only direct $CP$ violation signal observed by the $B$-factories is in the $B_d^0\to K^\pm\pi^\mp$ channel.  In contrast to
the small effect observed in kaon decay, the direct $CP$ asymmetry in $B_d^0\to K^\pm\pi^\mp$ is large:
$-0.093 \pm 0.015$~\cite{Aubert:2006ap,belle_ichep_summary}.  The quest for additional signals of direct $CP$ violation
in $B$ meson decays is ongoing in a plethora of different channels~\cite{Barberio:2006bi}.  The next goals of the
$B$-factories are to observe direct $CP$ violation in the decay of $B_u^\pm$ mesons and other $B_d^0$ channels.

%

The \B-factories have recently observed CPV in $B_d^0\to \etapr K^0$ decays~\cite{Aubert:2006wv,Chen:2006nk}.  These $b \to
s$ penguin processes are probes of NP, and have the most precisely measured time-dependent \CP\ asymmetry parameters of
all of the penguin modes.  Any deviation $\Delta S$ of the measured asymmetry parameter $S_{\eta^\prime K^0}$ from
$\sin 2\beta$ is an indication of NP (For example, see~\cite{Hewett:2004tv,Hashimoto:2004sm}). In addition to relying on
theoretical calculations of the SM pollution to these
decays~\cite{Beneke:2005pu,Cheng:2005ug,Williamson:2006hb}, it is possible to experimentally constrain the
SM pollution using SU(3) symmetry~\cite{Gronau:2006qh}.  This requires precision knowledge of the branching fractions
of the $B_d^0$ meson decays to the following pseudo-scalar pseudo-scalar (PP) final states $\piz\piz, \piz\eta, \piz\etapr,
\eta\eta, \etapr\eta, \etapr\etapr$ final states~\cite{Aubert:2006fy,Aubert:2006qd}.  The related decays
$B_{u,d}\to \eta^\prime \rho$ and $B_{u,d}\to \eta^\prime K^*$~\cite{Aubert:2006as,Wang:2006cb} can also be used to
understand the standard model contributions to $B_d^0\to \etapr K^0$ decays and the hierarchy of $\eta K^0$ to $\eta^\prime K^0$ decays.

%
%

The angular analysis of $B\to VV$ decays provides eleven observables (six amplitudes and five relative phases) that can
be used to test theoretical calculations~\cite{Dunietz:1990cj}. The hierarchy of $A_0$, $A_+$, and $A_-$ amplitudes obtained
from a helicity (or $A_0$, $A_\parallel$, and $A_\perp$ in the transversity basis) analysis of such decays allows one
to search for possible right handed currents in any NP contribution to the total amplitude. For low statistics studies
a simplified angular analysis is performed where one measures the fraction of longitudinally polarised events defined
as $f_L = |A_0|^2/\sum{|A_i|^2}$. Tree dominated decays such as $B^0_d\to \rho^+\rho^-$ have
$f_L \sim 1.0$~\cite{Aubert:2006af,Somov:2006sg}. Current data for penguin dominated processes ($\phi
K^*(892)$~\cite{Aubert:2006uk,Chen:2005zv}, $K^*(892)\rho$~\cite{Aubert:2006fs,Zhang:2005iz}) that are observed to have
non trivial values of $f_L$ can be accommodated in the SM.  In addition to this, one can search for T-odd \CP\
violating asymmetries in triple products constructed from the angular distributions~\cite{Datta:2003mj}. It has also
been suggested that non-standard model effects could be manifest in a number of other observables~\cite{London:2004ws}. The
measured rates of electroweak penguin dominated \B\ decays to final states involving a $\phi$ meson are also probes of
NP~\cite{Lu:2006nz}. The study of $B\to AV$ decays also provides this rich set of observables to study, however current results only
yield an upper limit on $B_d^0\to a_1^\pm\rho^\mp$ decays~\cite{Aubert:2006sw}.  \babar\ have recently studied the angular
distribution for the vector-tensor decay $B_d^0 \to \phi K^*(1430)$~\cite{Aubert:2006uk}.

\subsubsection{$B \rightarrow h^+h^{\prime -}$ decays at LHCb}

The charmless decays of $B$ mesons to two-body modes have been extensively 
studied at the $B$-factories.
Even if the current knowledge in the $B_d$ and $B_u$ sectors starts to be 
quite constrained, the $B_s$ sector still remains an open field. At 
present, by using a displaced vertex trigger, CDF has already collected 
an interesting sample of $B \rightarrow h^+h^{\prime -}$
decays \cite{Abulencia:2006psa}, providing a first 
observation of the two-body mode $B_s \rightarrow K^+K^-$. However it will 
most likely not be able to perform precision measurements of the time 
dependent CP asymmetry of the $B_s \rightarrow K^+K^-$ decay.

\begin{figure}
\begin{tabular}{cc}
\includegraphics[width=0.5\textwidth]{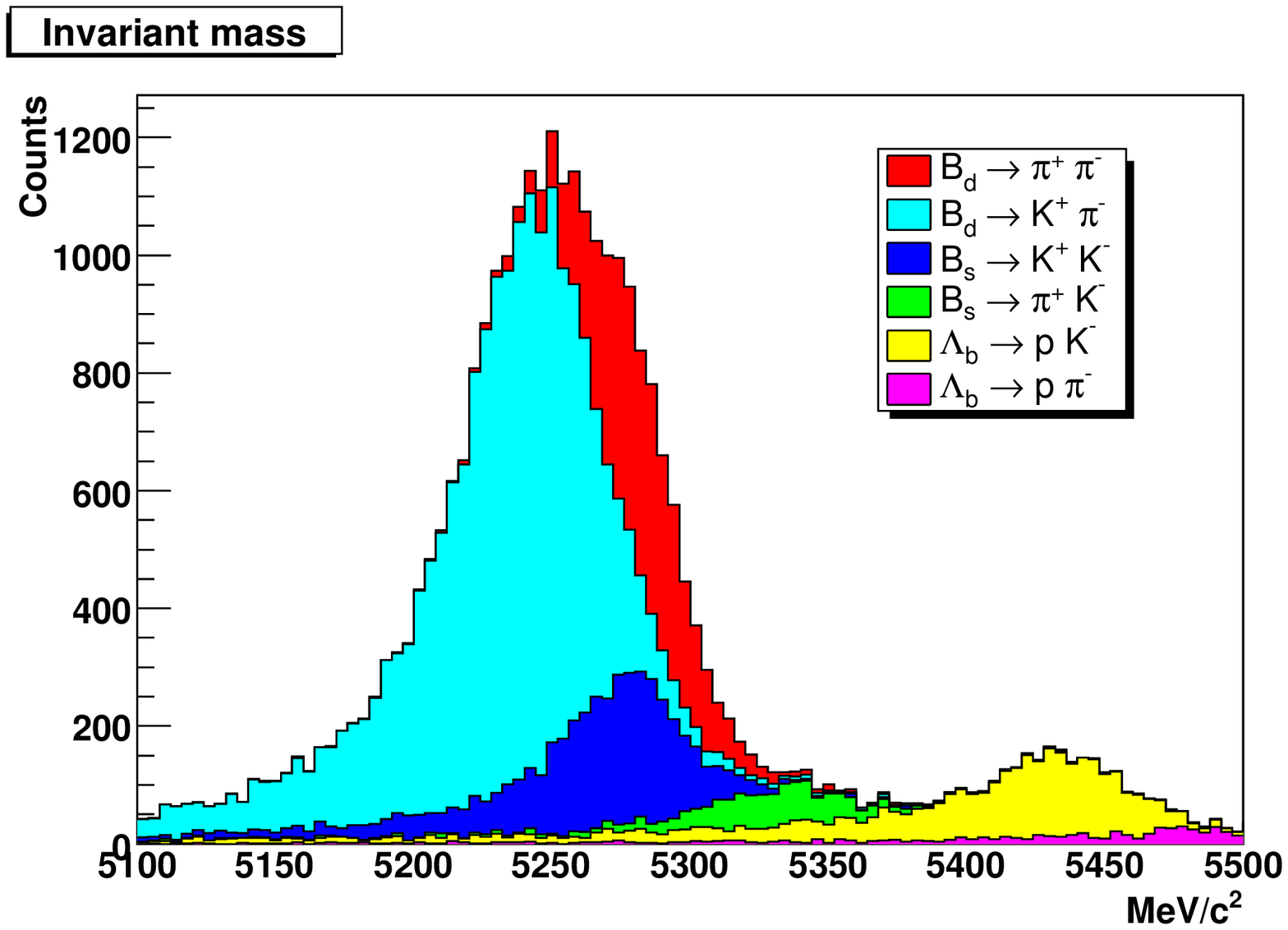}
\includegraphics[width=0.5\textwidth]{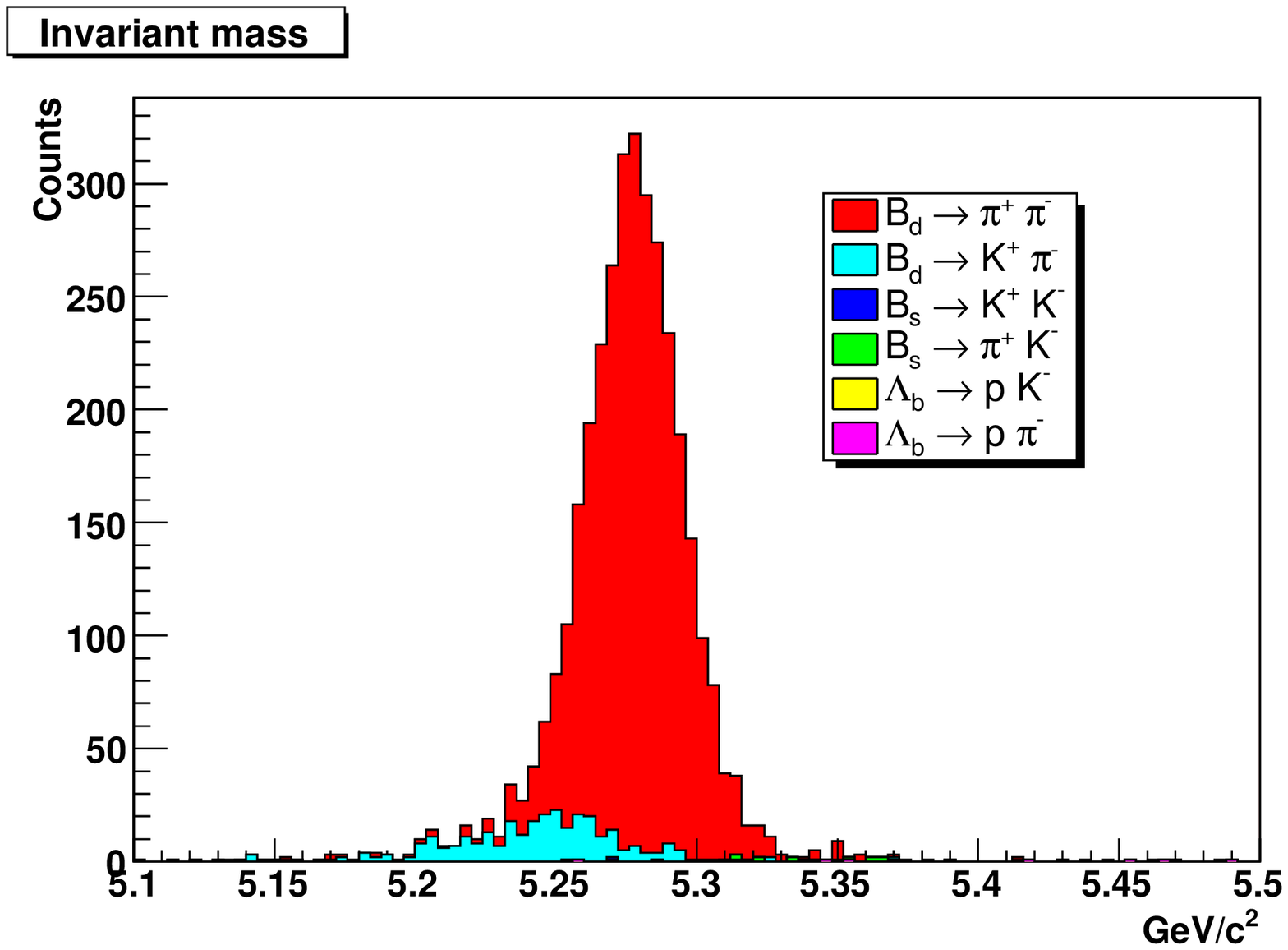}
\end{tabular}
\caption{\label{fig:hhmass}Left: $\pi^+\pi^-$ invariant mass distribution
for $B \rightarrow h^{+}h^{'-}$ decays expected at LHCb, obtained
without using PID information. Right: same plot after PID cuts are applied.}
\end{figure}

The LHCb experiment, thanks to the large beauty production cross section 
at the LHC and to its excellent vertexing and triggering capabilities, 
will be able to collect huge samples of $B \rightarrow h^+h^{\prime -}$ 
decays~\cite{ref:lhcb-2007-057}.
Furthermore, its particle identification system, composed in particular 
by two RICH detectors, will allow to disentangle the various
$B \rightarrow h^+h^{\prime -}$ 
modes with a purity exceeding 90\% as well as high efficiency. The PID 
capabilities of LHCb are clearly visible in Fig. \ref{fig:hhmass}, which 
shows the distribution of the $\pi^+\pi^-$ invariant mass from Monte Carlo
samples of $B \rightarrow h^+h^{\prime -}$ modes, before and after the employment of 
the PID information.

In order to calibrate the PID response, LHCb will make use of a dedicated 
trigger line - not making use of PID information in order not to 
introduce biases - intended to collect very large samples of $D^*$ 
decay chains to charged kaons and pions.
In order to reject combinatorial background, the event selection is based 
on a series of cuts, optimized by means of a multivariate 
technique, which include the transverse momenta and the impact 
parameter significances of the charged legs with respect to the primary 
vertex, the $\chi^2$ of the common vertex fit, the transverse momentum, 
the impact parameter significance and the distance of flight significance 
of the the candidate b-hadron and the invariant mass (the resolution for 
the $B \rightarrow h^+h^{\prime -}$ modes is expected to be about 18 
MeV/c$^2$). The event yields and background-to-signal
ratios estimated using a full GEANT4 based simulation are reported in 
Table~\ref{tab:hhyields}. 
\begin{table}
\begin{center}
\begin{tabular}{|c|c|c|c|c|}
\hline
Channel&
Assumed BR&
Annual yield&
B/S (combinatorial)&
B/S (two-body)\tabularnewline
\hline
$B_{d}^{0}\rightarrow\pi^{+}\pi^{-}$&
$4.8$&
36000&
$0.46$&
$0.08$\tabularnewline
\hline
$B_{d}^{0}\rightarrow K^{+}\pi^{-}$&
$18.5$&
138000&
$0.14$&
$0.02$\tabularnewline
\hline
$B_{s}^{0}\rightarrow\pi^{+}K^{-}$&
$4.8$&
10000&
$1.92$&
$0.54$\tabularnewline
\hline
$B_{s}^{0}\rightarrow K^{+}K^{-}$&
$18.5$&
36000&
$<0.06$&
$0.08$\tabularnewline
\hline
$\Lambda_{b}\rightarrow p\pi^{-}$&
$4.8$&
9000&
$1.66$&
$0.11$\tabularnewline
\hline
$\Lambda_{b}\rightarrow pK^{-}$&
$18.5$&
32000&
$<0.08$&
$0.02$\tabularnewline
\hline
\end{tabular}
\end{center}

\caption{\label{tab:hhyields} Annual yields and background-to-signal ratios
for $B \rightarrow h^+h^{\prime -}$ decays at LHCb~\cite{ref:lhcb-2007-057}.}
\end{table}

In order to measure CP violation from the time dependent CP asymmetries, 
other key ingredients are the tagging capability and the propertime 
resolution, the latter being particularly relevant to resolve
the fast $B_s$ oscillations. 
The effective tagging power for a $B_d$ decay at LHCb, according to full 
simulations, is expected to be about 5\%, while for a $B_s$ decay it 
is significantly larger, due to the larger efficiency of the same side 
kaon tagging, and is about 9\%. The calibration of the tagging power 
for $B \rightarrow h^+h^{\prime -}$ modes will be performed by 
using the flavour specific modes $B_d \rightarrow K^+\pi^-$ and 
$B_s \rightarrow \pi^+K^-$. As far as the propertime resolution is 
concerned, it is predicted by the full simulation to be about 40 fs, and 
it will be calibrated on data by using large samples of
$J/\psi \rightarrow \mu^+\mu^-$ decays, collected through a dedicated 
di-muon trigger line thought not to introduce biases in the $J/\psi$
propertime.

The direct CP asymmetries of the flavour specific
$B \rightarrow h^+h^{\prime -}$ modes can be 
measured without a time dependent fit, and without the need of
tagging the B meson. The 
statistical sensitivity on the charge asymmetry, corresponding to
a running time of $10^7~s$ at the nominal LHCb luminosity $2\cdot10^{32}\; \rm cm^{-2}\,s^{-1}$ ("one 
nominal LHCb year" in the following) is 0.003 for the
$B_d \rightarrow K^+\pi^-$ decay and 0.02 for the $B_s \rightarrow \pi^+K^-$ 
decay. In order to extract the direct (C) and mixing-induced (S) CP 
violation terms from the time dependent decay rates of the $B_d 
\rightarrow \pi^+\pi^-$ and $B_s \rightarrow K^+K^-$ and estimate the 
statistical sensitivity, we performed unbinned maximum likelihood fits on 
fast Monte Carlo data sets which parametrize the decay rates according to 
the outcomes of the full simulation. The expected sensitivity for 
C and S, corresponding to one nominal LHCb year, both for the $B_d 
\rightarrow \pi^+\pi^-$ and $B_s \rightarrow K^+K^-$ channels, is about 
0.04.

According to the method proposed in \cite{Fleischer:1999pa}, the employment 
of the U-spin symmetry allows to combine the measurements of C and S for 
the $B_d \rightarrow \pi^+\pi^-$ and $B_s \rightarrow K^+K^-$  modes in 
order to extract the $\gamma$ angle. 
Assuming a perfect U-spin symmetry, we predict a sensitivity on $\gamma$ 
for a nominal LHCb year around $5^\circ$. If a 20\% U-spin breaking is taken 
into account, the sensitivity deteriorates up to about $10^\circ$, still not 
spoiling the method of its predictive capabilities on $\gamma$. Being 
these modes characterized by the presence of loops inside the penguins, 
they could reveal New Physics effects, pointing to a value of $\gamma$ in 
contrast with the one determined from pure tree-level decays, such as 
$B \rightarrow DK$ modes.

In Table~\ref{tab:hhyields} LHCb also reports expected yields for 
$\Lambda_b$ baryon decays. An additional application of the $\Lambda_b$ baryon
that has been considered 
is testing CP and T symmetries using the decay modes $\Lambda_b \to \Lambda V$ 
where $V = J/\psi, \rho^0, \omega$.
This is discussed in Ref.~\cite{Leitner:2006sc}.

%
\newpage 

\clearpage

\subsection{Kaon decays}\label{sec:kaon}

%


\subsubsection{Introduction}

The rare decays $\kpn$ and $\klpn$ play an important role in the
search for the underlying mechanism of flavour mixing and CP
violation \cite{Buchalla:1996fp,Buras:2004uu,Isidori:2006yx,Bryman:2005xp}. 
As such they are excellent probes of physics beyond the
Standard Model (SM). Among the many rare $K$- and $B$-decays, the
$\kpn$ and $\klpn$ modes are unique since their SM branching ratios
can be computed to an exceptionally high degree of precision, not
matched by any other flavour-changing neutral current (FCNC) process
involving quarks.

The main reason for the exceptional theoretical cleanness of
{the} $\kpn$ and $\klpn$ decays is the fact that, within
the SM, these processes are mediated by electroweak amplitudes of
$\ord$$(G_F^2)$, described by $Z^0$-penguins and box diagrams
which exhibit a power-like GIM mechanism. This property implies a
severe suppression of non-perturbative effects, which is generally not
the case for meson decays receiving contributions of $\ord$$(G_F
\alpha_s)$ (gluon penguins) and/or $\ord$$(G_F \alpha_{em})$ (photon
penguins), {which therefore} 
have only a logarithmic GIM mechanism. A
related important virtue, following from this peculiar electroweak
structure, is the fact that $K\to \pi \nu \bar \nu$ amplitudes can be
described in terms of a single effective operator, namely
\be 
Q_{sd}^{\nu\bar \nu}= \left ( \bar s_L \gamma^\mu d_L \right ) \left (
  \bar \nu_L \gamma_\mu \nu_L \right ) \, .
\ee 
The hadronic matrix elements of $Q_{sd}^{\nu\bar \nu}$ relevant for
$K\to \pi \nu \bar \nu$ amplitudes can be extracted directly from the
well-measured $K^+\to \pi^0 e^+ \nu$ decay, including the leading
isospin breaking (IB) corrections \cite{Marciano:1996wy}.
The estimation of the matrix elements is improved and 
extended \cite{Mescia:2007kn}
beyond the leading order analysis. 

In the case of $\klpn$, which is CP-violating and dominated by the
dimension-six top quark contribution, the SM Short-Distance (SD)
dynamics is then encoded in a perturbatively calculable real function
$X$ that multiplies the CKM factor $\lambda_t=V_{ts}^* V_{td}$. In the
case of $\kpn$ also a charm quark contribution proportional to
$\lambda_c=V_{cs}^* V_{cd}$ has to be taken into account, but the
recent NNLO QCD calculation of the dimension-six charm quark
corrections \cite{Buras:2005gr, Buras:2006gb} and the progress in the
evaluation of dimension-eight charm and long-distance (LD) up quark
effects \cite{Isidori:2005xm} elevated the theoretical cleanness of
$\kpn$ almost to the level of $\klpn$. More details will be given in
\Sec{subsubsec:SM}.

The important virtue of {$K\to \pi \nu \bar \nu$} decays
is that their clean theoretical character remains valid in essentially
all extensions of the SM and {that $Q_{sd}^{\nu\bar
    \nu}$}, due to the special properties of the neutrinos, remains
the only relevant operator. Consequently, in most SM extensions the
New Physics (NP) contributions to $\kpn$ and $\klpn$ can be
parametrized in a model-independent manner by just two parameters, the
magnitude and the phase of the function \cite{Buras:1997ij}
\be 
X=\abs{X} e^{i\theta_X} \, ,
\ee 
that multiplies $\lambda_t$ in the relevant effective Hamiltonian. In
the SM, $\abs{X} = X_{\rm SM}$ and $\theta_X=0$.

\begin{figure}[t]
\vspace*{0.3truecm}
%
%


\begin{center}
  \includegraphics[width=15.8cm]{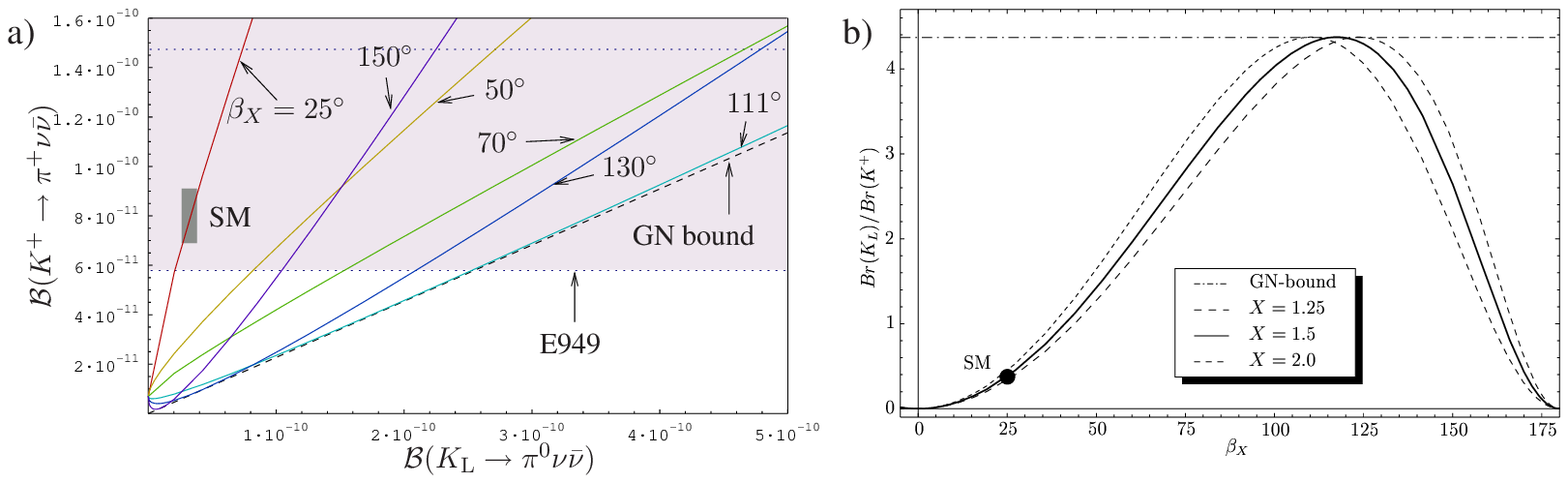}
\end{center}

\caption{a) $\BR(\kpn)$ vs. $\BR(\klpn)$ for various values of
  $\beta_X=\beta-\theta_X$ {(including E949 data)}
  \cite{Buras:2004ub}. The dotted horizontal lines indicate the
  lower part of the experimental 
range \cite{Anisimovsky:2004hr,Adler:2001xv,Adler:1997am} 
and the grey area the
  SM prediction. We also show the Grossman-Nir (GN) bound
  \cite{Grossman:1997sk}. 
b) {The ratio of the  $K\to\pi\nu\bar\nu$ branching ratios as a
function of $\beta_X$ for $|X|=1.25,~1.5,~2.0$. The horizontal line is
again the GN bound.}  \label{fig:fig1}}

%
%

\end{figure}

The parameters $\abs{X}$ and $\theta_X$ {can be} extracted
from $\BR(\klpn)$ and $\BR(\kpn)$ without hadronic uncertainties,
while the function $X$ can be calculated in any extension of the SM
within perturbation theory. Of particular interest is the ratio
\be \label{KLKLSM}
\frac{\BR(\klpn)}{\BR(\klpn)_{\rm SM}}=\left|\frac{X} {X_{\rm
      SM}}\right|^2 \left[\frac{\sin (\beta-\theta_X)}{\sin
    \beta}\right]^2 \, . 
\ee
Bearing in mind that $\beta \approx 21.4^\circ$,
Eq.\ (\ref{KLKLSM}) shows that $\klpn$ is
a very sensitive function of the new phase $\theta_X$. The pattern of
the two $K\to \pi \nu \bar \nu$ branching ratios as a function of
$\theta_X$ is illustrated in {\Fig{fig:fig1}a}. We note that the ratio of
the two modes {shown in \Fig{fig:fig1}b}
depends very mildly on $\abs{X}$ and therefore provides
an excellent tool to extract the non-standard {CP-violating} phase
$\theta_X$.

An interesting and complementary window to $|\Delta S|=1$ SD
transitions is provided by the $K_L \to \pi^0 \ell^+\ell^-$ system
($\ell=\mu,e$). While the latter is theoretically not as clean as the
$K\to \pi \nu \bar \nu$ system, it is sensitive to different types of
SD operators. The $K_L \to \pi^0 \ell^+\ell^-$ decay amplitudes have
three main ingredients: i)~a clean direct-CP-violating (CPV) component
determined by SD dynamics; ii)~an indirect-CPV term due
to $K^0$--$\overline{K^0}$ mixing; iii)~a LD CP-conserving (CPC) component
due to two-photon intermediate states. Although generated by very
different dynamics, these three components are of comparable size and
can be computed (or indirectly determined) to good accuracy within
the SM \cite{Buchalla:2003sj, Isidori:2004rb}. In the presence of
non-vanishing NP contributions, the combined measurements of $K\to \pi
\nu \bar \nu$ and $K_L \to \pi^0 \ell^+\ell^-$ decays provide a unique
tool to distinguish among different NP models.

The following discussion concentrates on the $K\to \pi \nu \bar \nu$ and
$K_L \to \pi^0 \ell^+ \ell^-$ decays in the SM (\Sec{subsubsec:SM} and 
\Sec{subsubsec:SM2}) and its most popular 
extensions {(\Secsand{subsubsec:beyondSM}{subsubsec:KllbeyondSM})}. 
{In \Sec{subsubsec:end} we stress} the complementarity of $K$- and
$B$-physics as well as the interplay with the high-$p_T$ physics at
the LHC.
Recent theoretical updates on kaon decays are found 
in \cite{Haisch:2007pd,Tarantino:2007en,Smith:2007kx}.
 Experimental programs at CERN and J-PARC are described in 
 \Sec{subsubsec:k-cern} and  \Sec{subsubsec:k-jparc}, respectively.
The current experimental status is summarized in
Table \ref{rarekexp}.
\begin{table}[h]
\begin{center}
\begin{tabular}{cccc}
\hline
 $B(K^+\to\pi^+\nu\bar\nu)$ & $B(K_L\to\pi^0\nu\bar\nu)$ & 
 $B(K_L\to\pi^0e^+e^-)$ & $B(K_L\to\pi^0\mu^+\mu^-)$ \\
\hline
$(1.47^{+1.30}_{-0.89})\cdot 10^{-10}$ & $< 6.7\cdot 10^{-8}$ &
$< 2.8\cdot 10^{-10}$ & $< 3.8\cdot 10^{-10}$\\
\cite{Anisimovsky:2004hr,Adler:2001xv,Adler:1997am}   &
\cite{Ahn:2007cd} &
\cite{Alavi-Harati:2004} &
\cite{Alavi-Harati:2000} \\
\hline
\end{tabular}
\caption{\label{rarekexp}
Current experimental results or limits for rare $K$ decay branching 
fractions.}
\end{center}
\end{table}

\boldmath
\subsubsection{$\kpn$ and $\klpn$ in the SM}
\unboldmath
\label{subsubsec:SM}

After summation over the three lepton families the SM branching ratios
for the $\kpns$ decays can be written as
\begin{gather} \label{eq:BRSMKp}   
  \BR (\kpn)_{\rm SM} = \kappa_+ \left [ \left ( \f{\im
        \lambda_t}{\lambda^5} X_{\rm SM} \right )^2 + \left ( \f{\re
        \lambda_t}{\lambda^5} X_{\rm SM} + \f{\re \lambda_c}{\lambda}
      \left (P_c + \dPcu \right ) \right )^2 \right ] \, , \\
\label{eq:BRSMKL} \BR (\klpn)_{\rm SM} = \kappa_L \left ( \f{\im
    \lambda_t}{\lambda^5} X_{\rm SM} \right )^2 \, ,
\end{gather}
where $\lambda = |V_{us}|$, while $\kappa_+ = ( 5.26 \pm 0.06 ) \cdot
10^{-11} \hspace{0.5mm} (\lambda/0.225)^8$ and $\kappa_L = ( 2.29 \pm
0.03 ) \cdot 10^{-10} \hspace{0.5mm} (\lambda/0.225)^8$
\cite{Isidori:2006qy} include the leading IB corrections in relating
$\kpns$ to $K^+\to \pi^0 e^+ \nu$ \cite{Marciano:1996wy}.  The
dimension-six top quark contribution $X_{\rm SM} = 1.464 \pm 0.041$
\cite{Buras:2005gr, Buras:2006gb} accounts for around $63 \%$ and
almost $100 \%$ of the total rates. It is known to NLO
\cite{Misiak:1999yg, Buchalla:1998ba}, with a scale uncertainty of
about $1 \%$. In $\kpn$, dimension-six charm quark
corrections and subleading dimension-eight charm and LD up quark
effects, characterized by $P_c = 0.38 \pm 0.04$ \cite{Buras:2005gr,
  Buras:2006gb} and $\dPcu = 0.04 \pm 0.02$ \cite{Isidori:2005xm},
amount to a moderate $33 \%$ and a mere $4 \%$. Light quark
contributions are negligible in the case of the $\klpn$ decay 
{\cite{Buchalla:1998ux}}.

Taking into account all the indirect constraints from the latest
global unitarity triangle (UT) fit, the SM predictions for the two
$\kpns$ rates read 
\be \label{eq:BRKSM}
\BR (\kpn)_{\rm SM} = \left ( 8.4 \pm 1.0 \right ) \cdot
10^{-11} \, , \qquad \BR (\klpn)_{\rm SM} = \left ( 2.7
  \pm 0.4 \right ) \cdot 10^{-11} \, .
\ee
The quoted central value of $\kpn$ corresponds to $\mc = 1.3 \, {\rm
  GeV}$ and the given error breaks down as follows: residual scale
uncertainties ($13 \%$), $\mc$ ($22 \%$), CKM, $\alpha_s$, and $\mt$
($37 \%$), and matrix-elements from $K^+\to \pi^0 e^+ \nu$ and light
quark contributions ($28 \%$). The main source of uncertainty in
$\klpn$ is parametric ($74 \%$), while the impact of scales ($11 \%$)
and IB ($15 \%$) is subdominant. SM predictions for $\kpns$ with total
uncertainties at the level of $5 \%$ or below are thus possible
through a better knowledge of $\mc$, of the IB in the $K \to \pi$ form
factors, and/or by a lattice study \cite{Isidori:2005tv} of
higher-dimensional and LD contributions.

While the determination of $|V_{td}|$, $\sin 2 \beta$, and $\gamma$
from the $\kpns$ system is without doubt still of interest, with the
slow progress in measuring the relevant branching ratios and much
faster progress in the extraction of the angle $\gamma$ from the $B_s
\to DK$ system to be expected at the LHC, the role of the $\kpns$
system will shift towards the search for NP rather than the
determination of the CKM parameters.

In fact, determining the UT from tree-level dominated $K$- and
$B$-decays and thus independently of NP will allow to find the ``true"
values of the CKM parameters. Inserting these, hopefully accurate,
values in \Eqsand{eq:BRSMKp}{eq:BRSMKL} will allow to obtain very
precise SM predictions for the rates of both rare $K$-decays. A
comparison with future data on $\kpns$ may then give a clear signal of
potential NP contributions in a theoretical{ly} clean environment. Even
deviations by $20 \%$ from the SM expectations could be considered as
signals of NP, while such a conclusion cannot be drawn in most other
decays{,} in which the theoretical errors are at least $10 \%$.


\subsubsection{$K_{L}\rightarrow\pi^{0}\ell^{+}\ell^{-}$ in the SM}
\label{subsubsec:SM2}

As mentioned in the introduction, the $K_L \to \pi^0 \ell^+\ell^-$ amplitudes 
have three main components. The interesting direct-CPV 
component, proportional to
$\operatorname{Im} \lambda_t$, is generated by $Z^0$-, $\gamma$-penguins
and {box diagrams} and is {SD} dominated.
It is encoded by local dimension-six vector $Q_{7V}=(\bar{s}d)_{V}(\bar{\ell
}\ell)_{V}$ and axial-vector $Q_{7A}=(\bar{s}d)_{V}(\bar{\ell}\ell)_{A}$
operators, whose Wilson coefficients $y_{7V,7A}$ are known to
NLO \cite{Buras:1994qa}. The former produces the $\ell^{+}\ell^{-}$ pair in a
$1^{--}$ state, the latter both in $1^{++}$ and $0^{-+}$ states. As in the 
$K\rightarrow\pi\nu\overline{\nu}$ case, the corresponding hadronic matrix 
elements are obtained precisely from $K_{\ell3}$ decays \cite{Marciano:1996wy}.

The other two components are of electromagnetic origin and are dominated by 
{LD} dynamics. 
These contributions cannot be computed from first principles. {However,} they 
can be related to measurable quantities within Chiral Perturbation Theory
(CHPT).
The indirect {CPV} amplitude, $\mathcal{A}(K_{L}\approx
\varepsilon K_{1}\rightarrow\pi^{0}\gamma^{\ast}\rightarrow\pi^{0}\ell^{+}%
\ell^{-})$ is determined \cite{D'Ambrosio:1998yj} 
{--- up to a sign ambiguity ---} by the
measurements of $\BR(K_{S} \rightarrow\pi^{0}
\ell^{+}\ell^{-})$.
In this case the $\ell^{+}\ell^{-}$
pair is produced in a $1^{--}$ state and interferes with the SD 
contribution of $Q_{7V}$. As discussed {in \cite{Buchalla:2003sj,Friot:2004yr}}, 
various theoretical arguments {point} toward a constructive interference.
Finally, the {CPC} contribution  
($K_{L}\rightarrow\pi^{0}\gamma^{\ast}\gamma^{\ast}\rightarrow
\pi^{0}\ell^{+}\ell^{-}$) produces the $\ell^{+}\ell^{-}$ pair either 
in a helicity-suppressed $0^{++}$ state or in a phase-space suppressed $2^{++}$ state. 
Within CHPT, only the $0^{++}$ state is produced at LO through the finite 
two-loop process 
$K_{L}\rightarrow\pi^{0}P^{+}P^{-}\rightarrow\pi^{0}\gamma\gamma\rightarrow
\pi^{0}\ell^{+}\ell^{-}$ ($P=\pi,K$). Higher-order corrections are estimated
using $K_{L}\rightarrow\pi^{0}\gamma\gamma$ experimental data for both the
$0^{++}$ and $2^{++}$ contributions~\cite{Buchalla:2003sj,Isidori:2004rb}.

\begin{table}[t]
\begin{center}
\begin{tabular}{ccccc}
\hline
& $C_{dir}^{\ell}$ & $C_{int}^{\ell}$ & $C_{mix}^{\ell}$ & $C_{\gamma\gamma
}^{\ell}$\\\hline
$\ell=e$ & $\left(  4.62\pm0.24\right)  \;\left(  w_{7V}^{2}+w_{7A}%
^{2}\right)  $ & $\left(  11.3\pm0.3\right)  \;w_{7V}$ & $14.5\pm0.5,$ &
$\approx0$\\
$\ell=\mu$ & $\left(  1.09\pm0.05\right)  \left(  w_{7V}^{2}+2.32w_{7A}%
^{2}\right)  $ & $\left(  2.63\pm0.06\right)  \;w_{7V}$ & $3.36\pm0.20$ &
$5.2\pm1.6$\\
\hline
\end{tabular}
\caption{\label{Kpill1}
Numerical coefficients for the evaluation of 
$\mathcal{B}(K_{L}\rightarrow\pi^{0}\ell^{+}\ell^{-})$
 {as given in} Eq.~(\ref{eq:BllSM}).} 
\end{center}
\end{table}

Altogether, the branching ratios can be expressed as
\cite{Buchalla:2003sj, Isidori:2004rb}:
\be
\label{eq:BllSM}
\mathcal{B}(K_{L}\rightarrow\pi^{0}\ell^{+}\ell^{-})
=(C_{dir}^{\ell}\pm C_{int}^{\ell}\left|  a_{S}\right|  +C_{mix}^{\ell}\left|
a_{S}\right|  ^{2}+C_{\gamma\gamma}^{\ell})\cdot10^{-12}\,,
\ee
where the $C_i$  are reported in Table~\ref{Kpill1}, 
$w_{7A,7V}=\operatorname{Im}\left(  \lambda_{t}y_{7A,7V}\right)/\operatorname{Im}\lambda_{t}$, and 
 $\left|  a_{S}\right|  =1.2\pm0.2$ is fixed from 
$\mathcal{B}^{\exp}\left(  K_{S}\rightarrow\pi^{0}\ell^{+}\ell^{-}\right)$
\cite{Batley:2003mu,Batley:2004wg}. 
Using  the SM values of $y_{7A,7V}$ \cite{Buras:1994qa}, 
the predicted rates are
\be
\mathcal{B}_{\rm SM}^{e^{+}e^{-}}=3.54_{-0.85}^{+0.98}\;\left(
1.56_{-0.49}^{+0.62}\right)  \cdot10^{-11}~, \qquad 
\mathcal{B}_{\rm SM}^{\mu^{+}\mu^{-}}=1.41_{-0.26}^{+0.28}
\left(  0.95_{-0.21}^{+0.22}\right) \cdot10^{-11}~,
\ee
for constructive (destructive) interference. Currently, the
theory error ({see} Fig.~\ref{FigNPKpll}a) is dominated by 
the uncertainty on $\left| a_{S}\right|$. 
Better measurements of $\mathcal{B}(K_{S}\rightarrow\pi
^{0}\ell^{+}\ell^{-})$ would thus be very welcome. 
Also, better measurements of
$K_{L}\rightarrow\pi^{0}\gamma\gamma$ would help in reducing the error on the
$0^{++}$ and $2^{++}$ contributions. Alternatively, they can be partially 
cut away through energy cuts or Dalitz plot
analyses \cite{Buchalla:2003sj,Isidori:2004rb,Mescia:2006jd}.
As shown in Fig.~\ref{FigNPKpll}a, the irreducible theoretical errors 
on these modes can be pushed  below the $10\%$ level, allowing 
very significant tests of flavour physics. 

The integrated forward-backward (or lepton-energy) asymmetry (see
references {in \cite{Mescia:2006jd}}), generated by the
interference between {CPC and CPV} amplitudes, cannot be
reliably estimated at present for $\ell=e$ because of the poor
theoretical control on the $2^{++}$ contribution. {In the
  case of $A_{FB}^{\mu}$ the situation is better since the $2^{++}$
  part is negligible. One has $A_{FB}^{\mu}\approx20\%$ $\,(-12\%)$ for
  constructive (destructive) interference}. Interestingly, though the
error is large, $A_{FB}^{\mu}$ can be used to fix the sign of $a_{S}$.

Let us close with a short comment on $K_{L}\rightarrow\mu^{+}\mu^{-}$. 
Here the SD part is {CPC} and has recently been evaluated 
at {NNLO} \cite{Gorbahn:2006bm}. 
The much larger LD contribution proceeds via two photons. 
While its absorptive part is fixed from
$K_{L}\rightarrow\gamma\gamma$, its dispersive part is difficult to estimate,
requiring unknown counter\-terms in CHPT \cite{Isidori:2003ts}.
Moreover, in this case the two-photon LD amplitude interferes with the SD one
(they both produce a lepton pair in a $0^{-+}$ state).
This interference, which
depends on the sign of $\mathcal{A}(K_{L}\rightarrow\gamma\gamma)$, is
presumably constructive \cite{Gerard:2005yk} and better measurements of
$K_{S}\rightarrow\pi^{0}\gamma\gamma$ or $K^{+}\rightarrow\pi^{+}\gamma\gamma$
could settle this sign. However, even with the help of this information
it is difficult to reduce the theoretical error below $\sim 50\%$ 
of the SD contribution.


\boldmath
\subsubsection{$\kpn$ and $\klpn$ beyond the SM}
\unboldmath
\label{subsubsec:beyondSM}

\subparagraph{Minimal Flavour Violation}

In models with Minimal Flavour Violation (MFV) 
\cite{Buras:2000dm,D'Ambrosio:2002ex} both decays are, like
in the SM, governed by a single real function $X$ that can take a
different value than in the SM due to new particle exchange in the
relevant $Z^0$-penguin and box diagrams (see \Fig{fig:fig1}a). 
Restricting first our
discussion to the so-called constrained MFV (CMFV) (see
\cite{Blanke:2006ig}), in which strong correlations between $K$- and
$B$-decays exist, one finds that the branching ratios for $\kpn$ and
$\klpn$ cannot be much larger than their SM values given in
Eq.~(\ref{eq:BRKSM}). The $95 \%$ probability bounds read
\cite{Bobeth:2005ck}
\be 
\BR(\kpn)_{\rm CMFV}\leq11.9 \cdot 10^{-11} \, , \qquad \BR(\klpn)_{\rm CMFV}\leq 4.6
\cdot 10^{-11} \, .
\ee 
Explicit calculations in a model with one {Universal Extra Dimension (UED)}
\cite{Buras:2002ej} and in the Littlest Higgs model without {$T$-parity}
\cite{Buras:2006wk} give explicit examples of this scenario with the
branching ratios within $20 \%$ of the SM expectations. The latest
detailed analysis of $K\to \pi \nu \bar \nu$ in the Minimal
Supersymmetric SM (MSSM) with MFV can be found in
\cite{Isidori:2006qy}.

Probably the most interesting property of this class of models is a
theoretically clean determination of the angle $\beta$ of the standard
UT, which utilizes both branching ratios and is
independent of the value of $X$ \cite{Buchalla:1994tr,Buras:2001af}. Consequently, this determination is
universal within the class of MFV models and any departure of the
resulting value of $\beta$ from the corresponding one measured in
$B$-decays would signal non-MFV interactions.

\subparagraph{ Littlest Higgs Model with {$T$-parity }}

\begin{figure}[t]
\vspace*{0.3truecm}
\begin{minipage}{0.45 \textwidth}
%
\begin{center}
  \includegraphics[width=8.25cm]{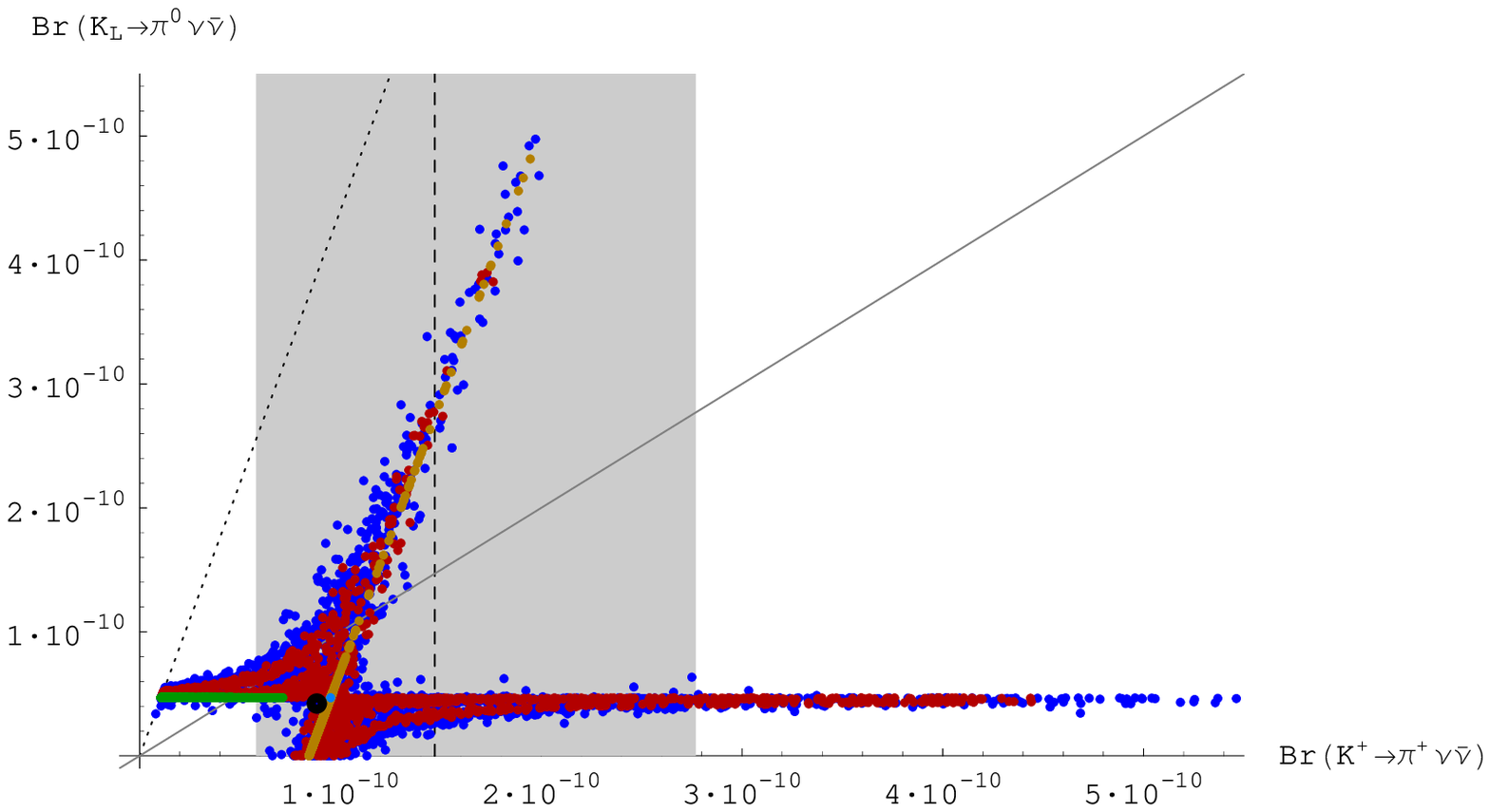}
\end{center}
\end{minipage}
\hspace{0.5cm}
\begin{minipage}{0.45 \textwidth}
%
\begin{center}
  \includegraphics[width=8.25cm]{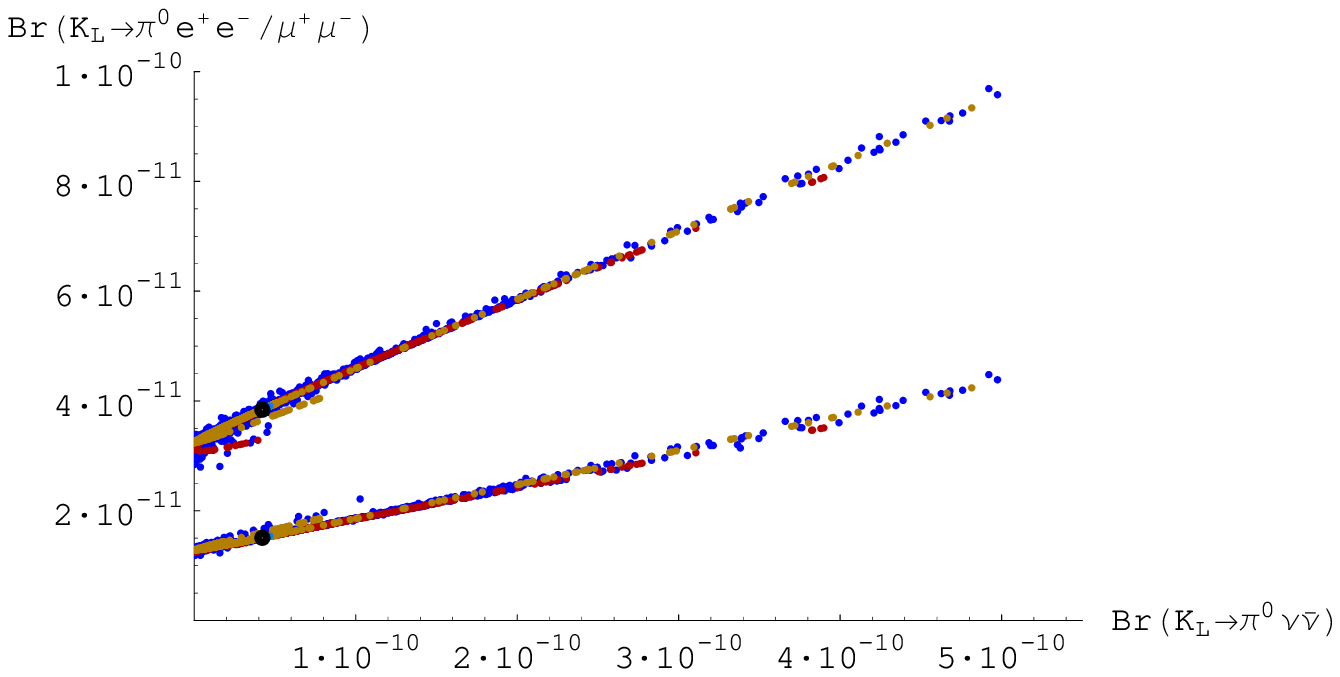}
\end{center}
\end{minipage}
\caption{a) $\BR(\klpn)$ vs. $\BR(\kpn)$ in the LHT model \cite{Blanke:2006eb}. The shaded
  area represents the experimental $1\sigma$-range for $\BR(\kpn)$
  . The GN bound  is displayed by the
  dotted line, while the solid line separates the two areas where
  $\BR(\klpn)$ is larger or smaller than $\BR(\kpn)$.
  {
  b) $\BR(K_L\to \pi^0 e^+e^-)$ (upper curve) and  $\BR(K_L \to \pi^0
  \mu^+\mu^-)$ (lower curve) as functions of $\BR(\klpn)$
  in the LHT model \cite{Blanke:2006eb}.} \label{fig:fig2}}
\begin{picture}(0,0)(0,0)
\put(0,218){a)} 
\put(235,215){b)} 
\end{picture}
\end{figure}

The structure of $K\to \pi \nu \bar \nu$ decays in the Littlest Higgs
model with {$T$-parity} (LHT) differs notably from the one found in MFV
models due to the presence of mirror quarks and leptons that interact
with the light fermions through the exchange of heavy charged
($W_H^{\pm}$) and neutral ($Z_H^0$, $A_H^0$) gauge bosons. The mixing
matrix $V_{Hd}$ that governs these interactions can differ from
$V_{\rm CKM}$, which implies the presence of non-MFV
interactions. Instead of a single real function $X$ that is universal
within the $K$-, $B_d$- and $B_s$-systems in MFV models, one now has
three functions
\be 
X_K=\abs{X_K} e^{i\theta_K} \, , \qquad X_d=\abs{X_d} e^{i\theta_d} \,
, \qquad X_s=\abs{X_s} e^{i\theta_s} \, ,
\ee 
that due to the presence of mirror fermions can have different phases
and magnitudes. 
{
Moreover, it is important to note that mirror fermion contributions 
are enhanced by a CKM factor $1/\lambda_t^{(i)}$ with $i= K, d, s$ for 
the $K$-, $B_d$- and $B_s$-systems respectively.
As  $\lambda_t^{(K)}\simeq 4\cdot 10^{-4}$, whereas
$\lambda_t^{(d)}\simeq 1\cdot 10^{-2}$ and $\lambda_t^{(s)}\simeq
4\cdot 10^{-2}$, the deviation from the SM prediction in the
$K$-system is found to be by more than an order of magnitude larger than in the
$B_d$-system, and even by two orders of magnitude larger than in the
$B_s$-system.}
This possibility can have a major impact on the $K\to
\pi \nu \bar \nu$ system, since the correlations between $K$- and
$B$-decays are partially lost and the presence of a large phase
$\theta_K$ can change the pattern of these decays from the one
observed in MFV. A detailed analysis \cite{Blanke:2006eb} shows that
both branching ratios can depart significantly from their SM values,
and can be as high as $5.0 \cdot 10^{-10}$. As shown in
{\Fig{fig:fig2}a}, there are two branches of allowed values with strong
correlations between both branching ratios within a given branch. In
the lower branch only $\BR(\kpn)$ can differ substantially from the SM
expectations reaching values well above the present central
experimental value. In the second branch $\BR(\klpn)$ and $\BR(\kpn)$
can be as high as $5.0 \cdot 10^{-10}$ and $2.3 \cdot 10^{-10}$,
respectively.  Moreover, $\BR(\klpn)$ can be larger than $\BR(\kpn)$
which is excluded within MFV models. Other features distinguishing
this model from MFV are thoroughly discussed in \cite{Blanke:2006eb}.

\subparagraph{Supersymmetry}

Within the MSSM with $R$-parity conservation, sizable non-standard contributions 
to $K \to \pi \nu\overline{\nu}$ decays can be generated 
if the soft-breaking terms have a non-MFV structure. 
The leading amplitudes giving rise to large effects are induced by:
i) chargino/up-squark 
loops~\cite{Nir:1997tf,Buras:1997ij,Colangelo:1998pm,Buras:1999da}
ii) charged Higgs/top quark loops~\cite{Isidori:2006jh}.
In the {first} case, large effects are generated if the left-right mixing
($A$ term) of the {up-squarks} has a non-MFV structure \cite{D'Ambrosio:2002ex}.
In the second case, deviations from the SM are induced by non-MFV terms 
in the right-right down sector, provided the ratio of the two Higgs {vacuum expectation values} 
($\tan \beta = v_u/v_d$) is large ($\tan \beta \sim 30-50$).

The effective Hamiltonian encoding SD contributions in the 
general MSSM has the following structure:
\be
\label{Ht}
{\mathcal H}_{\rm eff}^{({\rm SD})} \propto 
\sum_{l=e,\mu,\tau} V^{\ast}_{ts}V_{td} 
\left[X_L (\bar s_L \gamma^\mu d_L)(\bar\nu_{l L} \gamma_\mu \nu_{l L}) + 
X_R (\bar s_R \gamma^\mu d_R)(\bar\nu_{l L} \gamma_\mu \nu_{l L})\right]~,
\ee
where the SM case is recovered for ${X_R=0}$ and $X_L=X_{\rm SM}$. 
In general, both $X_R$ and  $X_L$ are {non-vanishing}, 
and the misalignment between quark and squark flavour 
structures implies that they are both complex quantities. 
Since the $K\to\pi$ matrix elements of {$(\bar s_L \gamma^\mu d_L)$ and
$(\bar s_R \gamma^\mu  d_R)$} are equal, the combination $X_L+X_R$
allows us to describe all the {SD} 
contributions to $K \to \pi \nu\overline{\nu}$ decays.
More precisely, we can simply use the SM expressions 
for the branching ratios in {Eqs.~(\ref{eq:BRSMKp}) to (\ref{eq:BRSMKL})}
with the following replacement 
\be
X_{\rm SM} \to  X_{\rm SM} + X^{\rm SUSY}_L + X^{\rm SUSY}_R~. 
\ee

\begin{figure}[t]
\begin{center}
\includegraphics[width=15.8cm]{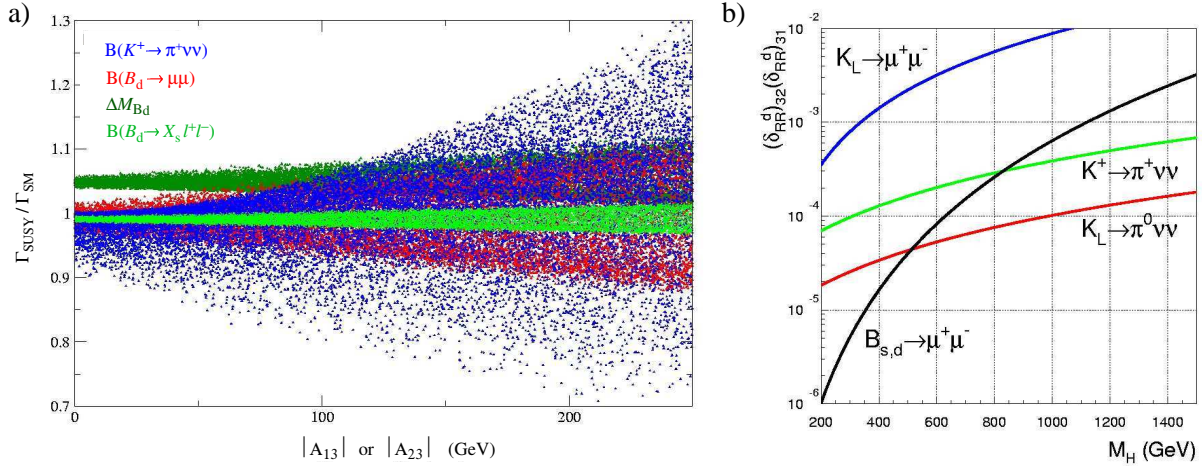}
\end{center}
\caption{\label{fig_SUSY} Supersymmetric contributions to 
$K \to \pi \nu\overline{\nu}$. {a)  
Dependence} of various FCNC 
observables (normalized to their SM value) on the up-type trilinear terms 
$A_{13}$ and $A_{23}$, for $A_{ij}\leq \lambda A_0$ and 
$\tan\beta =2$--$4$ (other key parameters in GeV: 
$\mu = 500 \pm 10$, $M_2 = 300 \pm 10$, $M_{\tilde u_R} = 600 \pm 20$, 
$M_{\tilde q_L} = 800 \pm 20$,  $A_0=1000$) \cite{Isidori:2006qy}.
{b) 
Sensitivity} to $(\delta^{d}_{RR})_{23}(\delta^{d}_{RR})_{31}$
of various rare {$K$- and $B$-decays} as a function 
of $M_{H^+}$,  setting $\tan\beta\!=\!50$, $\mu\!<\!0$
and assuming almost degenerate superpartners
(the bounds from the two $K\rightarrow \pi\nu\bar{\nu}$ modes are 
obtained assuming a $10\%$ measurement of their 
branching ratios while the $B_{s,d} \rightarrow \mu^+\mu^-$ bounds 
refer to the present experimental limits \cite{Isidori:2006jh}). }
\begin{picture}(0,0)(0,0)
\put(0,300){a)} 
\put(270,300){b)} 
\end{picture}
\end{figure}

In the limit of almost degenerate superpartners, the leading chargino/up-squarks
contribution is~\cite{Colangelo:1998pm}:
\be
X^{\chi^{\pm}}_L
\approx \frac{1}{96}
\left[\frac{(\delta^{u}_{LR})_{23} (\delta^{u}_{RL})_{31}}{\lambda_t}\right] 
~=~ 
\frac{1}{96 \lambda_t}~
\frac{(\tilde M^{2}_{u})_{2_L 3_R} (\tilde M^{2}_{u})_{3_R 1_L}
}{(\tilde M^{2}_{u})_{LL}
(\tilde M^{2}_{u})_{RR}}
\,.
\label{eq:XL_eff}
\ee
As pointed out {in \cite{Colangelo:1998pm}}, a remarkable feature of 
the above result is that no extra $\mathcal O(M_{W}/M_{\rm SUSY})$ suppression 
and no explicit CKM suppression is present (as it happens in
the chargino/up-squark contributions to other processes). 
{Furthermore}, the $(\delta^{u}_{LR})$-type mass insertions 
are not strongly constrained by other {$B$- and $K$-observables}.
This implies that large departures from the SM 
expectations in  $K \to \pi \nu\overline{\nu}$ decays are allowed,
as confirmed by the complete analyses {in \cite{Buras:2004qb,Isidori:2006qy}}. 
As illustrated {in Fig.~\ref{fig_SUSY}a},  
$K \to \pi \nu\overline{\nu}$ are the best observables 
to determine/constrain from experimental data the size of the off-diagonal 
 $(\delta^{u}_{LR})$ mass insertions or, equivalently, the
up-type trilinear terms $A_{i3}$ 
[$(\tilde M^{2}_{u})_{i_L 3_R} \approx m_t A_{i3}$].
Their measurement is therefore extremely interesting 
also in the LHC era. 

In the large $\tan\beta$ limit, 
the charged Higgs/{top quark} exchange leads to \cite{Isidori:2006jh}:
\beqa
X^{H^{\pm}}_R \approx
\left[\left(\frac{m_{s}m_{d}\,t^{2}_{\beta}}{2 M_W^2}\right)+
\frac{(\delta^d_{RR})_{31} (\delta^d_{RR})_{32}}{\lambda_t}
\left(\frac{m^{2}_{b}\,t^{2}_{\beta}}{2 M_W^2}\right)
\frac{\epsilon^{2}_{RR}t^{2}_{\beta}}{(1+\epsilon_{i}t_{\beta})^4}
\right]f_H(y_{tH})\,,
\label{eq:XR_eff}
\eeqa
where $y_{tH}=m_t^2/M_H^2$, $f_H (x)=x/4(1-x)+x\log x/4(x-1)^2$
and $\epsilon_{i,{RR}} t_{\beta} = \mathcal O(1)$ for 
$t_{\beta}=\tan\beta\sim 50$.
The first term of Eq.~(\ref{eq:XR_eff}) arises from MFV effects and
its potential $\tan\beta$ enhancement is more than compensated
by the smallness of $m_{d,s}$.
The second term on the r.h.s.~of Eq.~(\ref{eq:XR_eff}),
which would appear only at the three-loop level in a {standard} loop
expansion can be largely enhanced by the $\tan^4\beta$ factor
and does not contain any suppression due to light quark masses.
Similarly to the double mass-insertion mechanism of Eq.~(\ref{eq:XL_eff}),
also in this case the potentially leading effect is the one generated
when two off-diagonal squark mixing terms replace the two CKM
factors $V_{ts}$ and $V_{td}$.

The coupling of the $(\bar s_R \gamma^\mu d_R) (\bar\nu_L \gamma_\mu \nu_L)$
effective FCNC operator, generated by charged-Higgs/{top quark} loops
is phenomenologically relevant only at large $\tan\beta$ and with non-MFV 
right-right soft-breaking terms: a specific but well-motivated scenario
within grand-unified theories (see {e.g. \cite{Moroi:2000tk,Chang:2002mq}}).
These non-standard effects do not vanish in the limit of heavy squarks and 
gauginos, and have a slow decoupling with respect to the charged-Higgs boson mass.
As shown {in \cite{Isidori:2006jh}} 
the $B$-physics constraints still allow 
a large room of non-standard effects in $K \to \pi \nu\overline{\nu}$
even for flavour-mixing terms of CKM size {(see Fig.~\ref{fig_SUSY}b)}.

\subsubsection{$K_{L}\rightarrow\pi^{0}\ell^{+}\ell^{-}$ beyond the SM}
\label{subsubsec:KllbeyondSM}

Within the SM $K_{L}\rightarrow\pi^{0}e^{+}e^{-}$ and $K_{L}\rightarrow\pi^{0}\mu^{+}\mu^{-}$  
decays have a very similar dynamics, but for the different lepton masses.
This makes them an ideal probe of NP effects when taken in 
combination~\cite{Isidori:2004rb,Mescia:2006jd}. Moreover, $K_{L}\rightarrow\pi^{0}\mu
^{+}\mu^{-}$ is sensitive to Higgs-induced helicity-suppressed operators, to
which $K\rightarrow\pi\nu\bar{\nu}$ (and $K_{L}\rightarrow\pi^{0}e^{+}e^{-}$)
are blind.

\begin{figure}[t]
\begin{center}
\includegraphics[width=16.0cm]{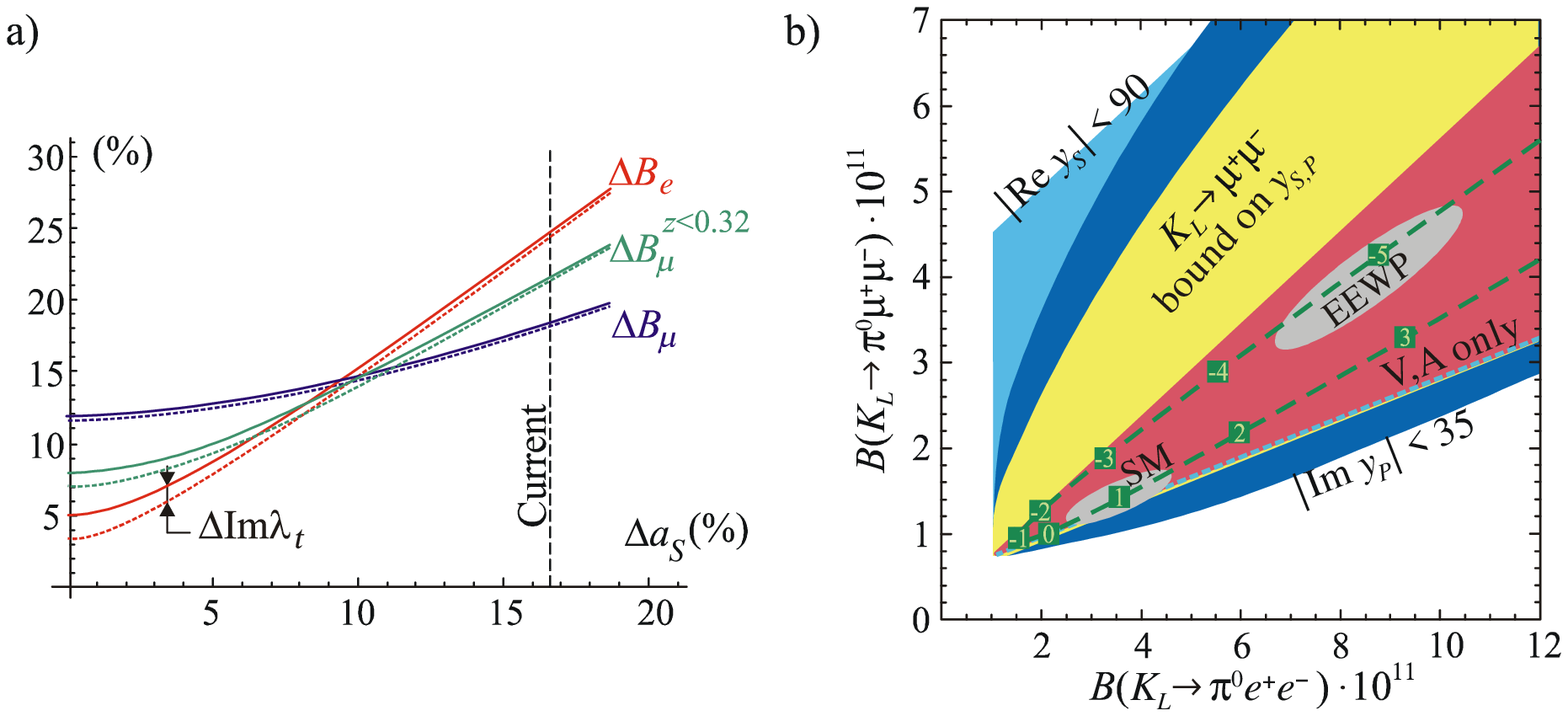}
\end{center}
\caption{a) Theory error as a function of the error on $\vert$$a_{S}$$\vert$.
b) $\BR(K_{L}\rightarrow\pi^{0}\mu^{+}\mu^{-})  $ against
$\BR(K_{L}\rightarrow\pi^{0}e^{+}e^{-})  $ for various NP
scenarios\cite{Mescia:2006jd}.
The red sector is allowed for 
the Wilson coefficients $y_{7A}$ and $y_{7V}$,
exclusively, to take arbitrary values; 
the green broken line with squares corresponds to
a common rescaling of the two coefficients.
{The LHT result of \cite{Blanke:2006eb} lies between EEWP and V,A only.}
Light blue (dark blue) corresponds to arbitrary 
$y_{7A, 7V}$ together with $|$Re$y_S$$|<90$ ($|$Im$y_P$$|<35$), respectively,
while the yellow region corresponds to 
$y_{7A, 7V, S, P}$ arbitrary but compatible with 
the $\BR(K_{L}\rightarrow\mu^+ \mu^-$) measurement, 
where $y_S$ and $y_S$ are the coefficients for 
scalar and pseudoscalar operators.
}%
\label{FigNPKpll}%
\end{figure}

\subparagraph{NP with SM operators}

In many scenarios, such as {enhanced electroweak penguins (EEWP)} ~\cite{Buras:2004ub}, 
{the MSSM} at moderate $\tan\beta$~\cite{Cho:1996we}, 
{Little Higgs models}
{(LHT)}~\cite{Blanke:2006eb}, UED~\cite{Buras:2002ej},
and {leptoquark models}~\cite{Davidson:1993qk}, NP only modifies the
strength of the SM operators, without introducing new structures. In general,
these models induce larger effects for $K_{L}\rightarrow\pi^{0}\nu\bar{\nu}$
than for $K_{L}\rightarrow\pi^{0}\ell^{+}\ell^{-}$. 
{Still, the latter modes should not be disregarded as they 
offer the possibility to disentangle effects in the vector and axial-vector currents.}
Indeed, $Q_{7A}$ produces the final
lepton pair also in a helicity-suppressed $0^{-+}$ state, hence contributes
differently to $K_{L}\rightarrow\pi^{0}e^{+}e^{-}$ and $K_{L}\rightarrow
\pi^{0}\mu^{+}\mu^{-}$, while the $Q_{7V}$ contributions are identical for
both modes (up to phase-space corrections, and assuming lepton flavour
universality) \cite{Isidori:2004rb}.

As a consequence, the area spanned in the $\BR(K_{L}%
\rightarrow\pi^{0}e^{+}e^{-})  -\BR(K_{L}\rightarrow
\pi^{0}\mu^{+}\mu^{-})  $ plane for arbitrary $w_{7A,7V}$ is
non-trivial, see {Fig.~\ref{FigNPKpll}b}. Taking all errors into account, this
translates into the bounds $0.1+0.24\,\mathcal{B}^{ee}\leq\mathcal{B}^{\mu\mu
}\leq0.6+0.58\,\mathcal{B}^{ee}$ with $\mathcal{B}^{\ell\ell}=\BR(
K_{L}\rightarrow\pi^{0}\ell^{+}\ell^{-})  \cdot10^{11}$ \cite{Mescia:2006jd}.

Usually, in specific models, there are correlations between the
effects of NP on $Q_{7V}$ and $Q_{7A}$ operators. In the MSSM at
moderate $\tan\beta$, the dominant effect is due to chargino
contributions to {$Z^0$- and
  $\gamma$-penguins}~\cite{Nir:1997tf,Buras:1997ij,Colangelo:1998pm,Buras:1999da}
sensitive to the double up-squark mass insertions.  Since {
  $Z^0$- and $\gamma$-penguins} are correlated, so are $Q_{7V}$ and
$Q_{7A}$ and only a subregion of the red area can be reached.
{This is true whether or not there are new CP-phases.
{Interestingly, in the LHT model~\cite{Blanke:2006eb},
the contributions to $w_{7V}$ cancel each
other to a large extent, leading to a quasi one-to-one correspondence, see
 \Fig{fig:fig2}b. This constitutes a powerful test of the model.}
In the case of MFV,} the {overall effect} is found to be
always {smaller} than for
$K_{L}\rightarrow\pi^{0}\nu\bar{\nu}$, with a maximum enhancement
w.r.t. the SM of about $10\%$~\cite{Isidori:2006qy}.
Finally, the contribution of the {dipole} operator $(\bar{s}\sigma^{\mu\nu}%
d)F_{\mu\nu}$ can be absorbed into $w_{7V}$~\cite{Buras:1999da}
and NP contributions of this type cannot be singled out.

\subparagraph{NP with New Operators}

{NP} could of course also induce new operators. A systematic analysis
of the impact of all possible dimension-six semileptonic operators on
$K_{L}\rightarrow\pi^{0}\ell^{+}\ell^{-}$ can be found in \cite{Mescia:2006jd}. 
Here we concentrate on the most interesting case of (pseudo-)scalar operators
$Q_{S}=(\bar{s}d)(\bar{\ell}\ell)$ and $Q_{P}=(\bar{s}d)(\bar{\ell}\gamma
_{5}\ell)$, inducing a {CPC~(CPV)} contribution. These
operators are enhanced in the MSSM at large $\tan\beta$ where they originate
from neutral Higgs exchanges and are sensitive to down-squark mass
insertions~\cite{Isidori:2002qe}. Being helicity-suppressed, they affect only
the muon mode and can lead to a clear signal outside the red region in
{Fig.~\ref{FigNPKpll}b}. Of course, in the MSSM, the $(\bar{s}\gamma_{5}%
d)(\bar{\ell}\ell)$ and $(\bar{s}\gamma_{5}d)(\bar{\ell}\gamma_{5}\ell)$
operators, contributing to $K_{L}\rightarrow\ell^{+}\ell^{-}$, are also
generated. Interestingly, the current $\BR(K_{L}\rightarrow
\mu^{+}\mu^{-}) ^{\exp}$ still leaves open the large yellow region in
{Fig.~\ref{FigNPKpll}b}, when combined with general $Q_{7V,7A}$ operators.

Finally, note that tree-level leptoquark exchange~\cite{Davidson:1993qk} 
or sneutrino exchange in SUSY without 
$R$-parity \cite{Barbier:1998fe,Grossman:2003rw,Deandrea:2004ae,Deshpande:2004xc}
can also induce {(pseudo-)scalar} operators, but without helicity-suppression. 
However, to evade the strong constraint from $\BR(K_{L}\rightarrow e^{+}%
e^{-})  ^{\exp}=(9_{-4}^{+6})\cdot10^{-12}$, one would need to invoke a
large breaking of lepton-flavour universality to have a visible effect in
$K_{L}\rightarrow\pi^{0}\mu^{+}\mu^{-}$.


\subsubsection{Conclusions on the theoretical prospects}
\label{subsubsec:end}

{Rare $K$-decays are excellent probes of New Physics. Firstly, their exceptional 
cleanness allows to access very high energy scales. As stressed recently in 
\cite{Isidori:2006qy,Blanke:2006eb,Bryman:2005xp,Grinstein:2006cg},
NP could be seen in rare $K$-decays without significant signals in 
$B_{d,s}$-decays and, in specific scenarios, even without new particles within the LHC reach.
Secondly, if LHC finds NP, its energy scale will be fixed. Then, the
combined measurements of the four rare $K$-modes would help in discriminating
among NP models. For instance, we have seen that specific correlations 
exist in MFV or LHT, which can be used as powerful tests (see \Fig{fig:fig2}). Further, 
in all cases, the information extracted from the four modes is essential to
establish the NP flavour structure in the $s\to d$ sector, as illustrated in 
the MSSM at both moderate (see \Fig{fig_SUSY}a) and large $\tan\beta$ 
(see Figs.\ \ref{fig_SUSY}b and \ref{FigNPKpll}b). 
Rare $K$-decays are thus an integral part, along with $B$-physics and collider 
observables, of the grand project of reconstructing the NP model from data.
Experimentally, together with these very rare modes, improving bounds on
forbidden decays (e.g.\ $K\to \pi e \mu$) can be interesting. 
Also, rare $K$-decays would benefit from experimental progress 
in (less rare) radiative $K$-decays like $K_S\to\pi^0\ell^+\ell^-$ (see \Fig{FigNPKpll}a). 
For all these reasons, it is very important to pursue ambitious $K$-physics programs 
in the era of the LHC.}





\subsubsection{Program at CERN}
\label{subsubsec:k-cern}

\begin{figure}[ht]
\begin{center}
\includegraphics[width=12cm]{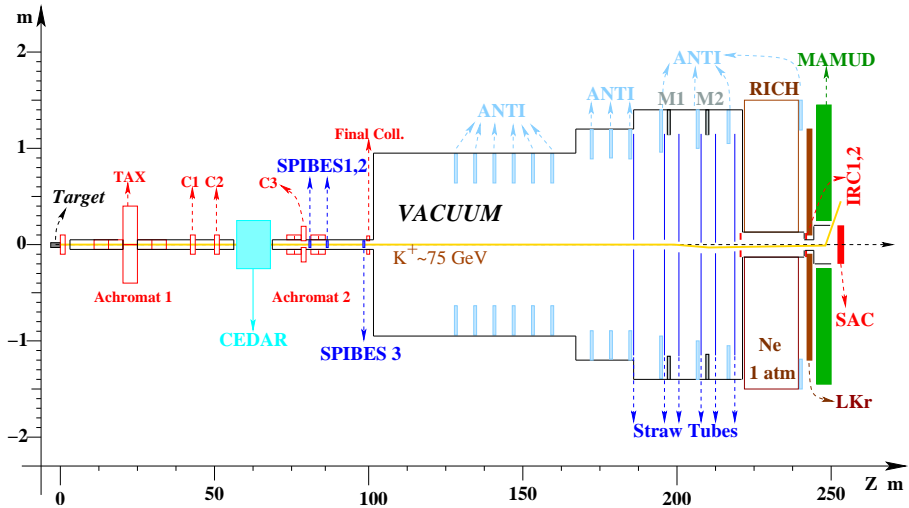}
\caption{Layout of the NA62 (NA48/3) experiment.}
\label{fig:p326det}
\end{center}
\end{figure}

The proposed experiment NA62 (formerly NA48/3) at CERN-SPS \cite{p326} aims 
to collect about $80$ $K^+\rightarrow\pi^+\nu\bar{\nu}$ events with an 
excellent signal over background ratio in two years of running, 
allowing for a $10\%$ measurement of the branching ratio of the $K^+\rightarrow\pi^+\nu\bar{\nu}$ decay. The data taking should start in 2010. 
NA62 will replace the NA48 apparatus at CERN and will make use of the existing beam line. The layout of the experiment is sketched in figure \ref{fig:p326det}.

The experiment proposes to exploit a kaon decay in flight technique to achieve $10\%$ of signal acceptance. An intense $400\GeVc$ proton beam, extracted from the SPS, produces a 
secondary charged beam by impinging on a Be target. A $100$ m long beam line selects a $75\GeVc$ momentum beam with a $1\%$ RMS momentum band. This beam
covers a $16\cm^2$ area, has an average rate of about $800$ MHz and is composed by $6\%$ of $K^+$ and $94\%$ of $\pi^+$, $e^+$ and protons. A differential Cerenkov counter (CEDAR) 
placed along the beam line ensures a positive kaon identification. The beam enters in a $80$ m long decay region evacuated at a level of $10^{-6}$ mbar, enough to avoid sizeable 
background from the interaction of the particles with the residual gas. The kaon decay rate in the decay region is about $6$ MHz$\,$: it provides about $10^{13}$ $K^{+}$ decays 
in two years of data taking, assuming 100 days as running time at $60\%$ of efficiency, which is a very realistic estimate based on the decennial NA48 experience at the SPS.

\begin{figure}[ht]
\begin{center}
\includegraphics[width=7cm]{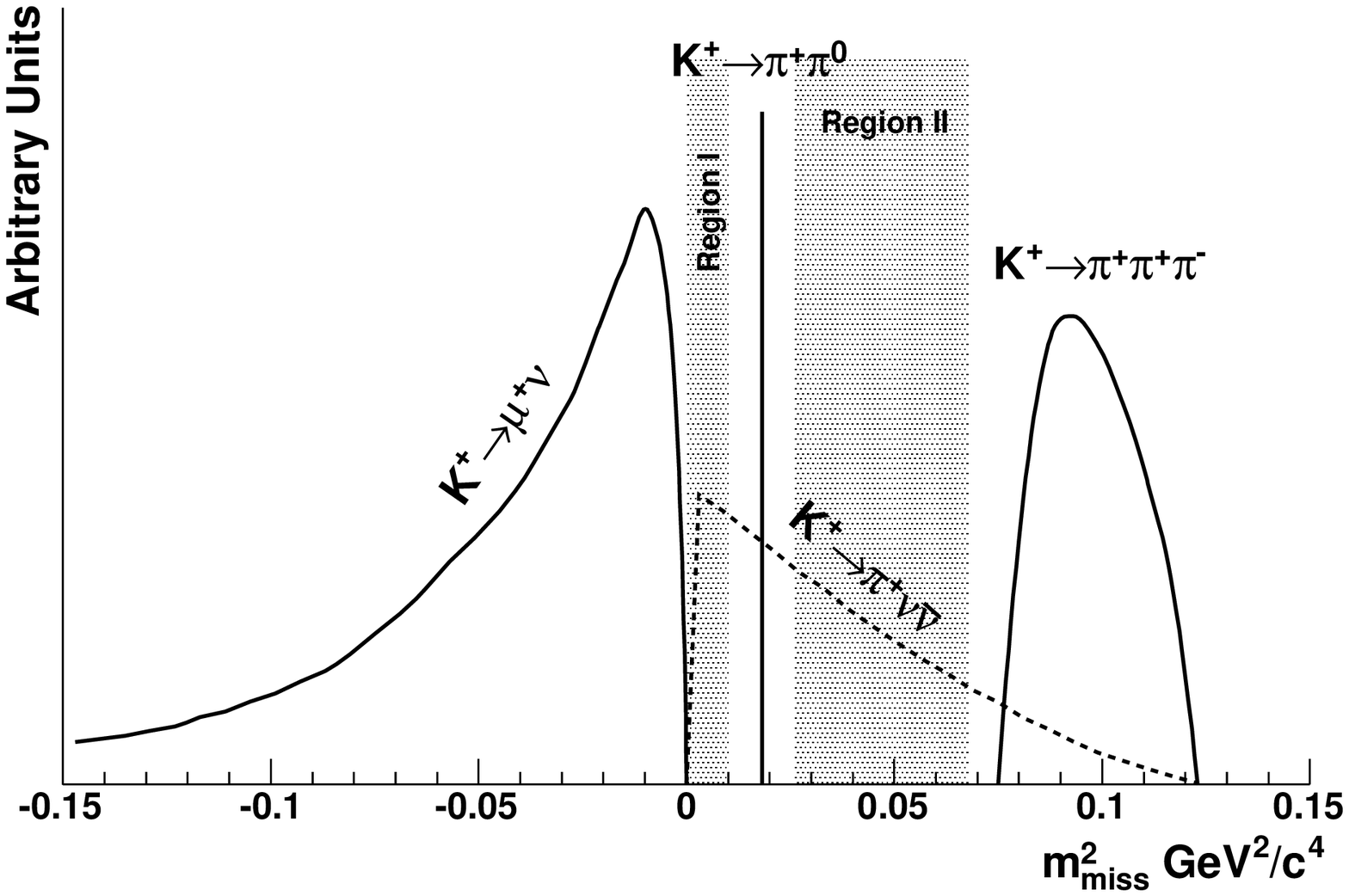}
\includegraphics[width=7cm]{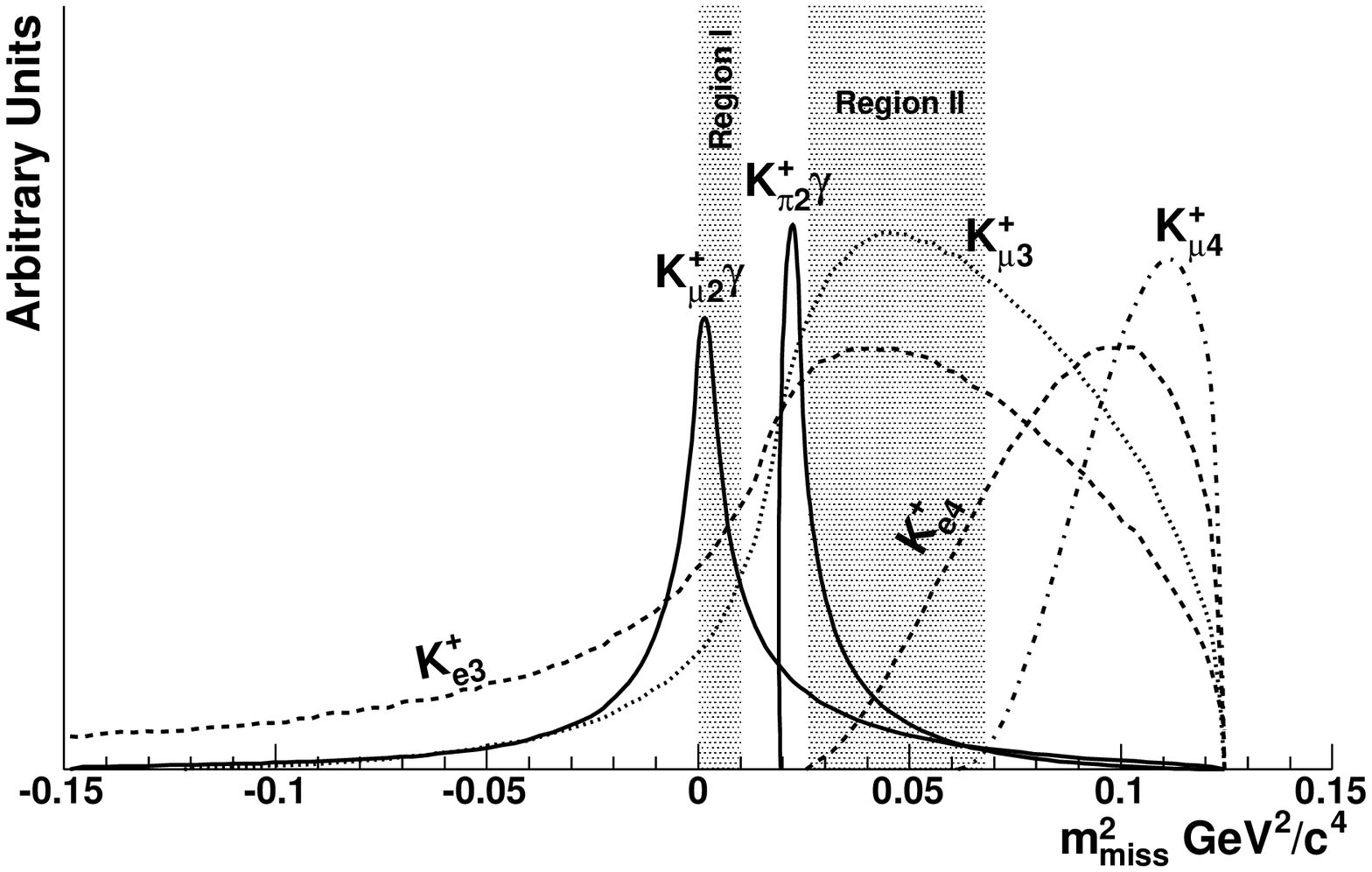}
\caption{Squared missing mass for Kaon decays. The squared missing mass is defined as the square of the difference between the 4-momentum of the kaon and of the decayed track
in the hypothesis that it is a pion.}
\label{fig:miss}
\end{center}
\end{figure}

The experimental signature of a $K^+\rightarrow\pi^+\nu\bar{\nu}$ is one reconstructed positive track in the downstream detector. The squared missing mass allows a kinematical
separation between the signal and about $90\%$ of the total background (see figure \ref{fig:miss}). The precise kinematical reconstruction of the event requires a
performing tracking system for the beam particles and the charged decay products of the kaons.

The beam tracker consists of three Si pixels stations (SPIBES) having a surface of $36\times48\mm^2$. The charged particle rate on each station is about $60$ MHz$\,\mathrm{cm^{-2}}$ 
on average. The stations are made up by $300\times300\mum^2$ pixels, $300\mum$ thick and containing the sensor and the chip bump-bonded on it. At least 200 ps time resolution 
per station is required to provide a suitable tag of the kaon track. A mistagging of the kaon, in fact, may be a source of background because it spoils the resolution of the 
reconstructed squared missing mass.

Six straw chambers, 0.5\% radiation length thick, placed in the same vacuum of the decay region form the downstream spectrometer. Two magnets provide a redundant 
measurement of the particle momentum, useful to keep the non gaussian tails of the reconstruction under control. The central hole of each station, which lets
the undecayed beam pass through, must be displaced in the bending plane of the magnets according to the path of the $75\GeVc$ positive beam. This configuration allows the tracker 
to be used as a veto for negative particles up to $60\GeVc$, needed for the rejection of backgrounds like $K^+\rightarrow\pi^+\pi^- e^+\nu$. A reduced size prototype will be built and tested in 2007. 

A system of $\gamma$ vetoes, a $\mu$ veto and a RICH complement the tracking 
system to guarantee a 10$^{13}$ level of background rejection.

A 18 m long RICH located after the spectrometer and filled with Ne at atmospheric pressure is the core of the e$^+/\pi/\mu$ separation. A $11\cm$ radius
beam pipe crosses the RICH and two tilted mirrors at the end reflect the Cerenkov light toward an array of about 2000 phototubes placed in the focal plane. Simulations showed that
enough photoelectrons can be collected per track to achieve a better than $3\sigma$ $\pi/\mu$ separation between $15$ and $35\GeVc$. The RICH provides also the timing of the 
downstream track with a $100$ ps time resolution. The construction and test of a full length prototype is planned for 2007. 

A combination of calorimeters covering up to 50 mrad serves to identify the photons. Ring-shaped calorimeters, most of them laying in the high vacuum of the decay region, 
cover the angular region between 10 and 50 mrad. Tests on prototypes built using lead scintillator tiles and scintillating fibers are scheduled for 2007 at a tagged $\gamma$ 
facility at LNF. The existing NA48 liquid krypton calorimeter (LKr) \cite{Unal:2000eu} is intended to be used as a veto for $\gamma$ down to 1 mrad. Data taken by NA48/2 in 2004 and a 
test run performed in 2006 using a tagged $\gamma$ beam at CERN show that the LKr matches our requests in terms of efficiency. A program of consolidation and update of 
the readout electronics of the LKr is under way. Small calorimeters around the beam pipe and behind the muon veto cover the low angle region. 

Six meters of alternated plates of iron and extruded scintillators form a hadronic sampling calorimeter (MAMUD), able to provide a $10^{5}$ $\mu$ rejection. An aperture 
in the center lets the beam pass through and a magnetic field inside deflects the beam out of the acceptance of the last $\gamma$ veto.

Simulations of the whole apparatus based on \GEANTthree and \GEANTfour showed that 10\% signal acceptance are safely achievable. The use of the RICH constrains the accepted 
pion track within the $(15,35)\GeV/c$ momentum range. The higher cut is an important loss of signal acceptance, but assures that events like $K^+\rightarrow\pi^+\pi^0$ deposit 
at least $40\GeV$ of electromagnetic energy, making their rejection easier. The simulations indicate that a 10\% background level is nearly achievable.

The overall experimental design requires a sophisticated technology for which an intense R\&D program is started. Actually we propose an experiment able to 
reach a sensitivity of $10^{-12}$
per event, employing existing infrastructure and detectors at CERN.




\subsubsection{Program at J-PARC}
\label{subsubsec:k-jparc}

\begin{figure}[ht]
\begin{center}
\includegraphics[width=10cm]{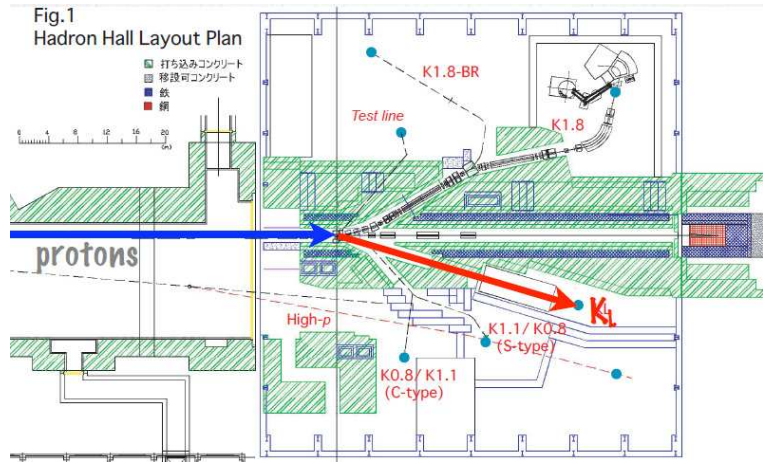}
\caption{A plan for the layout of NP Hall at J-PARC.}
\label{fig:jparcNPhall}
\end{center}
\end{figure}

The Japan Proton Accelerator Research Complex (J-PARC) \cite{JPARC_web}
is a new facility being constructed in the Tokai area of Japan
as a joint project of High Energy Accelerator Research Organization (KEK)
and Japan Atomic Energy Agency. 
Slow-extracted proton beam, which is of 30GeV and 
whose intensity is $2\times 10^{14}$ protons per 0.7-sec spill 
every 3.3 sec at the Phase-1, is transported to the experimental area
called NP Hall (figure \ref{fig:jparcNPhall}). 
The proton beam hits the target and produces a variety of 
secondary particles, including low-energy $K^+$'s and $K_L$'s.

The first PAC meeting 
for Nuclear and Particle Physics Experiments at J-PARC 
was held in the early summer of 2006 \cite{JPARC_PAC}. 
Concerning kaon physics, 
two proposals:
``Measurement of T-violating Transverse Muon Polarization in 
  $K^+\to\pi^0\mu^+ \nu$ Decays'' 
and 
``Proposal for $K_L\to \pi^0\nu\bar{\nu}$ Experiment at J-Parc''
received scientific approval.
The latter proposal on the $K_L\to \pi^0\nu\bar{\nu}$ decay 
is discussed in this section; 
the former one is discussed in the ``Charged Lepton CP/T'' section of WG3. 

The branching ratio for $K_L\to \pi^0\nu\bar{\nu}$ is predicted to be 
$(2.7\pm 0.4)\times 10^{-11}$ in the Standard Model, while
the experimental upper limit, $6.7\times 10^{-8}$
at the 90\% confidence level,
is currently set by the E391a Collaboration at the KEK 12-GeV PS
using the data collected during the second period of data taking 
\cite{Ahn:2007cd}.
E391a was the first dedicated experiment
for $K_L\to \pi^0\nu\bar{\nu}$ 
and aimed to be a pilot experiment.
The new proposal at J-PARC \cite{JPARC_Proposal} 
is to measure the branching ratio 
with an uncertainty less than 10\% and 
takes a step-by-step approach to achieve this goal. 

\begin{figure}[ht]
\begin{center}
\includegraphics[width=5.0cm,angle=90]{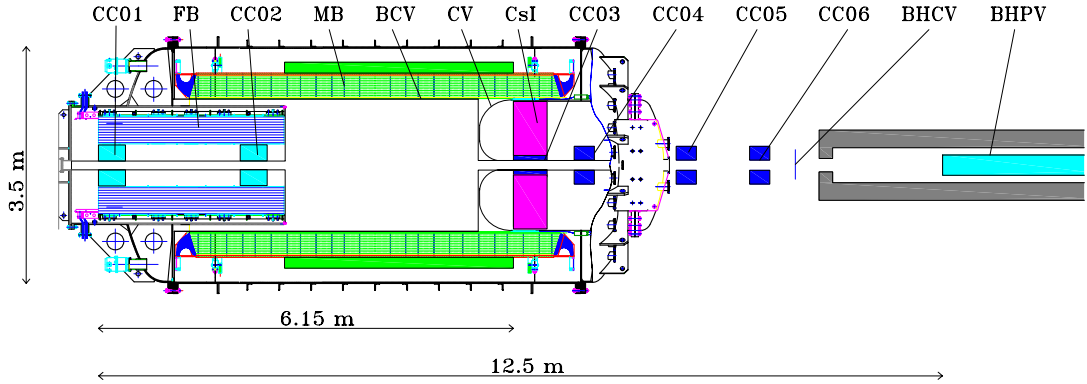}
\caption{Schematic view of the detector setup for the 
         E14 experimemt at J-PARC.}
\label{fig:jparcKLdet}
\end{center}
\end{figure}

\begin{figure}[ht]
\begin{center}
\includegraphics[width=7.5cm]{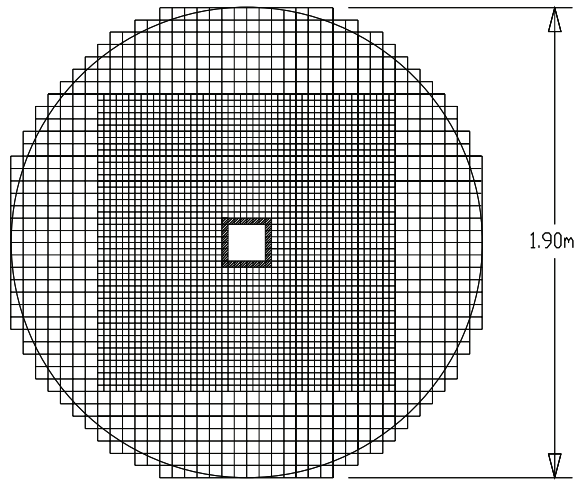}
\caption{Layout of the calorimeter for the J-PARC $K_L$ experiment
         with the KTeV CsI crystals.}
\label{fig:jparcCsI}
\end{center}
\end{figure}

The common T1 target on the A-line and the beamline 
with a 16-degree extraction angle, as shown in 
figure \ref{fig:jparcNPhall},
will be used in the first stage of the experiment (E14).
Survey of a new neutral beamline 
in the first year of J-PARC commissioning and operation
is essential to 
understands the beam-related issues at J-PARC.
The E14 experiment will be performed by the date 
of ``5 years of LHC'' ($\sim$ 2012/2013);
the goal is
to make the first observation of the decay.
In the current simulation, 
3.5 Standard Model events with $1.8\times 10^{21}$ protons on target
in total are expected with the S/N ratio of 1.4. 
The beamline elements and the detector of E391a will be re-used
by imposing necessary modifications.
A schematic view of the detector setup is shown in 
figure \ref{fig:jparcKLdet}. 
In particular, the undoped CsI crystals in the calorimeter 
for measuring the two photons from $\pi^0$ in $K_L\to \pi^0\nu\bar{\nu}$
will be replaced with the smaller-size and longer crystals used 
in the Fermilab KTeV experiment (figure \ref{fig:jparcCsI}); 
discussions on the loan of the crystals are in progress.
The technique of waveform digitization
will be used on the outputs of the counters in the detector
to distinguish pile-up signals from legitimate two-photon signals
under the expected high-rate conditions.
A new extra photon detection system to reduce
the $K_L\to \pi^0\pi^0$ background will cover 
the regions in or around the neutral beam.

After the E14 experiment 
establishes the experimental techniques to achieve the physics goal,
the beamline and the detector will be upgraded
for the next stage.
More than 100 Standard Model events 
(equivalent to a single event sensitivity of less than $3\times 10^{-13}$)
with a S/N ratio of 4.8 will be accumulated
by the era of a ``super B-factory'' ($\sim$ 2020).



\clearpage

%

\newpage 
\renewcommand{\arraystretch}{1.1}

\subsection{Charm physics}\label{sec:charm}

%
%
%
%

\subsubsection{Case for continuing charm studies in a nutshell}




While nobody can doubt the seminal role that charm studies played for the evolution and acceptance 
of the Standard Model (SM), conventional wisdom is less enthused  about their future. 
Yet on closer examination a 
strong case emerges in two respects, both of which are based on the 
weak phenomenology predicted by the SM for charm: 
\begin{itemize}
\item 
to gain new insights into and make progress in establishing theoretical 
control over QCD's nonperturbative dynamics, which will also calibrate our 
theoretical tools for $B$ studies; 
\item 
to use charm transitions as a novel window into New Physics (NP). 
\end{itemize}
Lessons from the first item will have an obvious impact on the tasks listed under the second one. They might actually be of great value even beyond QCD, if the New Physics anticipated 
for the TeV scale is of the strongly interacting variety. 

Detailed analyses of leptonic and semileptonic decays of charm hadrons provide a challenging testbed 
for validating lattice QCD, which is the only known framework with the promise for a truly 
quantitative treatment of charm hadrons that can be improved {\em systematically}. 

While significant `profit' can be `guaranteed' for the first item, the situation is 
less clear 
concerning the 
second one, the search for New Physics. While it had to be expected that no sign of New Physics 
would show up at the present level of experimental sensitivity, no clear-cut benchmark has been set 
at which level New Physics could emerge with even odds. In that sense one is dealing with 
hypothesis-generating rather than probing research. It will be essential to harness the statistical 
power of the LHC for high quality charm studies.

Yet the situation is much more promising than it seems at first glance. New Physics scenarios in general 
induce flavour changing neutral currents (FCNC) that {\em a priori} have little reason to be as much suppressed as in the SM. More specifically they could be substantially stronger for up-type than for down-type quarks; 
this can happen in particular in models which have to reduce strangeness changing neutral currents 
below phenomenologically acceptable levels by some alignment mechanism. 

In such scenarios charm plays a unique role among the up-type quarks $u$, $c$ and $t$; for only 
charm allows the full range of probes for New Physics in general and flavour-changing 
neutral currents in particular: 
(i) Since top quarks do not hadronise \cite{Bigi:1986jk}, there can be no $T^0- \bar T^0$ oscillations. 
More generally, hadronisation, while hard to bring under theoretical control, enhances the 
observability of \cp~violation. 
(ii)  
As far as $u$ quarks are concerned, $\pi^0$, $\eta$ and $\eta ^{\prime}$ decay 
electromagnetically, not weakly. They are their own antiparticles and thus cannot oscillate. 
\cp~asymmetries are mostly ruled out by \cpt~invariance. 

Our basic contention can then be formulated as follows: {\em Charm transitions provide a unique portal 
for a novel access to flavour dynamics with the experimental situation being a priori quite 
favourable (apart from the absence of Cabibbo suppression). Yet even that handicap can be overcome 
by statistics. }

The truly committed reader can find more nourishment for her/his curiosity in several recent reviews 
\cite{Burdman:2001tf,Bianco:2003vb,Burdman:2003rs}. 

These points alluded to above will be addressed in somewhat more detail in the following sections. 

\subsubsection{Charm Mixing} 
\label{sec:mixing}

Prior observations of mixing in all down-type quark mixing systems puts charm physics 
in a unique position in the modern investigations of flavour physics
as the system { {where the first evidence for the phenomena has emerged
only recently (just before the publication of this document). Results of these studies
are addressed after a short phenomenological introduction.}} 

The Standard Model contributions to charm mixing are suppressed to 
$\tan^2\theta_c\approx 5\%$ because $\DZ$ decays are Cabibbo favoured.
The GIM cancellation could further suppress mixing through off-shell
intermediate states to $10^{-2}-10^{-6}$. Standard Model predictions for
charm mixing rates span several orders of 
magnitude\cite{Petrov:2003un,Burdman:2003rs}. 
Fortunately, CP violation in mixing is ${\cal O}(10^{-6})$ in the SM so
CP violation involving $\DZ\DZB$ oscillations is a reliable probe of 
New Physics.

Charm physics studies are 
complementary 
to the corresponding programs in bottom or strange systems due to the 
fact that $\DZ\DZB$ mixing is influenced by the dynamical effects of {\it down-type 
particles}. 

Effective $\Delta C = 2$ interactions generate contributions to the effective operators 
that change a $\DZ$ state into a $\DZB$ state, leading to the mass eigenstates
\begin{equation} \label{definition1}
| D_{^1_2} \rangle =
p | \DZ \rangle \pm q | \DZB \rangle, ~~ R_m^2 = \left|\frac{q}{p}\right|^2,
\end{equation}
where the complex parameters $p$ and $q$ are obtained from diagonalising
the $\DZ\!-\!\DZB$ mass matrix with $|p|^2 + |q|^2 = 1$. If CP-violation
in mixing is neglected, $p$ becomes equal to $q$, so $| D_{^1_2} \rangle$
become CP eigenstates, $CP | D_{\pm} \rangle = \pm | D_{\pm} \rangle$.

The time evolution of a $D^0$ or $\bar{D}^0$ is conventionally described by an effective Hamiltonian which is non-Hermitian and allows the mesons to decay. We write
\begin{displaymath}
i\frac{\partial}{\partial t}\left[\begin{array}{c}|D^0(t)\rangle\\|\bar{D}^0(t)\rangle\end{array}\right]=
\left({\bf M}-\frac{i}{2}{\bf\Gamma}\right)\left[\begin{array}{c}|D^0(t)\rangle\\|\bar{D}^0(t)\rangle\end{array}\right]
\end{displaymath}
where ${\bf M}$ and ${\bf\Gamma}$ are $2\times2$ matrices. We invoke $CPT$ invariance so that $M_{11}=M_{22}\equiv M$ and $\Gamma_{11}=\Gamma_{22}\equiv\Gamma$. The eigenvalues of this Hamiltonian are
\begin{displaymath}
\lambda_{1,2}=M_{1,2}-\frac{i}{2}\Gamma_{1,2}\equiv
\left(M-\frac{i}{2}\Gamma\right)\pm\frac{q}{p}\left(M_{12}-\frac{i}{2}\Gamma_{12}\right)
\end{displaymath}
where $M_{1,2}$ are the masses of the $D_{1,2}$ and $\Gamma_{1,2}$ are their decay widths, and
\begin{displaymath}
\frac{q}{p}=
\sqrt{\frac{M_{12}^*-\frac{i}{2}\Gamma_{12}^*}{M_{12}-\frac{i}{2}\Gamma_{12}}}\,.
\end{displaymath}
The mass and width splittings between these eigenstates are given by
\begin{eqnarray} \label{definition}
x \equiv \frac{m_1-m_2}{\Gamma}, ~~
y \equiv \frac{\Gamma_1 - \Gamma_2}{2 \Gamma}, ~~ R_M=\frac{x^2+y^2}{2}.
\end{eqnarray}
These parameters are experimentally observable and can be studied using a 
variety of methods to be discussed below. SM and all reasonable models of NP 
predict $x,y \ll 1$~\cite{Petrov:2003un,Burdman:2003rs}, which influences the available 
strategies for those { {measurements.}} 



\subsubsection{Semileptonic decays}
The most natural way to search for charm mixing is to employ 
semileptonic decays. It is also not the most sensitive way, as it is only sensitive to 
$R_M$, a quadratic function of $x$ and $y$. Use of the $\DZ$ semileptonic decays for the mixing 
search involves the measurement of the time-dependent or time-integrated rate for the 
wrong-sign (WS) decays of $D$, where $c\to\overline{c}\to\overline{s}\ell^-\overline{\nu}$, 
relative to the right-sign (RS) decay rate, $c\to s\ell^+\nu$. 
Decays $\DZ\to K^{(\ast)-}\ell^+\overline{\nu}$ have been experimentally searched
for \cite{Abe:2005nq,Cawlfield:2005ze,Aubert:2004bn,Hosack:2002hr,Aitala:1996vz}. Although the time integrated rate is measured, several experiments
use the time dependence of $\DZ$ decays to increase the
sensitivity. Currently the best sensitivity is reached by the Belle experiment, 
$R_M=(0.20\pm 0.47\pm 0.14)\times 10^{-3}$ 
, using 253~fb$^{-1}$ of data in $e^\pm$ mode only. Projecting
to a possible 2~ab$^{-1}$ one can hope for a sensitivity of about $\pm
0.2\times 10^{-3}$, including also systematic uncertainty.  


\subsubsection{Nonleptonic decays to non-CP eigenstates}
{ {A decay mode providing one of the best sensitivities to the mixing
parameters is $\DZ\to K^+\pi^-$. }}
Time-dependent studies allow separation 
of the direct doubly-Cabibbo suppressed (DCS) $\DZ\to K^+\pi^-$ amplitude from 
the mixing contribution $\DZ \to \DZB \to K^+ \pi^-$\cite{Bigi:1986nr,Blaylock:1995ay},
\begin{eqnarray}\label{Kpi}
\Gamma[\DZ \to K^+ \pi^-]
=e^{-\Gamma t}|A_{K^-\pi^+}|^2 
~\left[
R_D+\sqrt{R_D}R_m(y'\cos\phi-x'\sin\phi)\Gamma t
+R_m^2 R_M^2 (\Gamma t)^2
\right],
\end{eqnarray}
where $R_D$ is the ratio of DCS and Cabibbo-favoured (CF) decay rates. 
Since $x$ and $y$ are small, the best constraint comes from the linear terms 
in $t$ that are also {\it linear} in $x$ and $y$.
A direct extraction of $x$ and $y$ from Eq.~(\ref{Kpi}) is not possible due 
to the unknown relative strong phase $\delta_{K\pi}$ of DCS and CF 
amplitudes, as $x'=x\cos\delta_{K\pi}+y\sin\delta_{K\pi}$, $y'=y\cos\delta_{K\pi}-x\sin\delta_{K\pi}$. 
This phase can be measured independently (see CLEO-c result in Section~\ref{sec:qc}).
The corresponding formula can 
also be written~\cite{Bergmann:2000id} for $\DZB$ decay with $x' \to -x'$ and 
$R_m \to R_m^{-1}$.

Experimentally, this method of $\DZ$ mixing search requires a good understanding of 
the detector decay time resolution to model correctly the measured distribution. 
Several experiments performed fits to disentangle the individual contributions in 
Eq.~(\ref{Kpi})~\cite{Zhang:2006dp,Abe:2004sn,Link:2004vk,Aubert:2003ae,Godang:1999yd,Aitala:1996fg,Barate:1998uy}. { { The most recent study by BaBar
collaboration~\cite{Aubert:2007wf} finds an evidence for non-zero values of the mixing
parameters. The preliminary 95\% C.L. contours of the measured values are
shown in Fig.~\ref{mixlimits}. In terms of 
single parameter errors 
to
be used for projections}} the most accurate is the measurement by Belle,
using 400~fb$^{-1}$ of data. Several fits to decay time distributions
are performed; assuming that the CP violation is negligible, the result is 
$x^{\prime 2}=(0.18\pm {0.21 \atop 0.23})\times 10^{-3}$, 
$y^\prime = (0.6\pm {4.0\atop 3.9})\times 10^{-3}$ and 
$R_D = (3.64\pm 0.17)\times 10^{-3}$, where the errors are statistical
only. 
Projections of the 95\% C.L. $(x^{\prime 2}, y^\prime)$ contour to
the axes yield confidence intervals of $x^{\prime 2}<0.72\times
10^{-3}$ and $y^\prime\in [-9.9, 6.8]\times 10^{-3}$. 
With a 2~ab$^{-1}$ data sample a statistical accuracy of $0.1\times 10^{-3}$ and 
$2\times 10^{-3}$ can be expected for $x^{\prime 2}$ and $y^\prime$,
respectively, similar to the current systematic uncertainties; a large
contribution to the latter is due to the background modelling, the
understanding of which might improve with a larger data sample as well. 


CDF has demonstrated the potential of experiments at hadron colliders
to make mixing-related measurements using hadronic decays
through the recent study of WS $\DZ \to K^+\pi^-$ events~\cite{Abulencia:2006sz}.
Using the distinctive $D^\ast \to \DZ \pi$ signature
and an integrated luminosity of $\rm 0.35\, fb^{-1}$ a sample of
around 2000 WS decays have been accumulated with a background to signal
level of order 1.  The ratio of WS to RS decays is found to 
be $4.05\pm 0.21 ({\rm stat}) \pm 0.11 ({\rm syst}) \times 10^{-3}$.
This ratio is equivalent to $R_D$ in the limit that 
$x'$ and $y'$ are zero, and CP violation is negligible.
Provided that the systematic uncertainties can continue to be 
kept under control, the full Tevatron dataset of several $\rm fb^{-1}$ 
will give a more precise result for $R_D$ than the 
B-factories, under the stated assumption.  More interesting
results are to be expected should it prove possible to 
perform a time-dependent measurement.

LHCb expects to collect very high statistics in all charged two-body
$\DZ$ decays through the inclusion of a dedicated $D^\ast \to \DZ(hh^\prime) \pi$
filter in the experiment's high level trigger~\cite{LHCBCHARM}.  
In one year of 
operation at nominal luminosity ($\rm 2 \, fb^{-1}$) 0.2 million 
WS and 50 million RS $K\pi$ events will be written to tape, 
where the triggered $D^\ast$ has originated from a $B$ decay.
A similar number of decays are expected where the $D^\ast$ is
produced in the primary event vertex.   

In a mixing analysis it is necessary
to measure the proper lifetime of the decaying $\DZ$.
LHCb's good vertexing allows the decay point of the
$\DZ$ to be well determined, and also the production point in the
case of $D^\ast$'s produced in the primary vertex.
For that sample where the $D^\ast$ arises from
a $B$ decay it is necessary to vertex the $\DZ$ direction
with other $B$ decay products in order to find
the production point, a procedure which entails a loss in efficiency.
Additional cuts are needed to enhance the purity of the WS signal,
and combat the most significant background source, where the wrong
`slow pion' is associated with a genuine $\DZ$.  This contamination
is dangerous for the reason that is the charge of the slow pion which tags
the initial flavour of the $\DZ$ meson.    After this selection,
46,500 WS decays are expected from $B$ events per $\rm 2 \, fb^{-1}$ , with a
background to signal ratio of around 2.5. 

These performance figures have been used as input to 
a `toy Monte Carlo' study to determine LHCb's sensitivity 
to the mixing parameters, including both the effects
of background and the estimated proper time resolution and acceptance.
The study was performed for event yields corresponding
to $10~\,{\rm fb^{-1}}$ of integrated luminosity, 
that is 5 years of operation
at nominal operation.  It was found that
with such a sample LHCb will  have a statistical sensitivity 
on $x^{\prime 2}$ and $y^\prime$ of $0.6 \times 10^{-4}$ and $0.9 \times 10^{-3}$ respectively.
Further work is needed to identify and combat the possible
sources of systematic uncertainty.


 
\subsubsection{Multi-body hadronic \mbox{\boldmath{$\DZ$}} decays}
In multi-body hadronic $\DZ$ decays
possible differences in the resonant structure between the CF 
and DCS decays must be taken into account, and, as discussed below, be exploited. 
The time integrated relative rates $R_{WS}=\Gamma(\DZ\to
K^+\pi^-(n\pi))/\Gamma(\DZ\to K^-\pi^+(n\pi))$, which assuming negligible
CP violation equal to $R_D+\sqrt{R_D}y^\prime +(x^{\prime 2}+y^{\prime
2})/2$, have been measured for $n\pi=\pi^0,\pi^+\pi^-$
\cite{Brandenburg:2001ze,Tian:2005ik,Aitala:1996sh,Dytman:2001rx}. For the latter mode Belle measures $R_{WS}(K\pi\pi\pi)=(0.320\pm
0.018\pm 0.013)\%$. Assuming a particular value of $x^\prime$ in
combination with the previous equation gives an allowed band in the $(R_D,
y^\prime)$ plane; however, one should note that the value of $x^\prime$ is decay
mode dependent.  
Studies with $\DZ\to K^\mp \pi^\pm \pi^-\pi^+$ events will also be possible
at LHCb, where plans are under consideration 
to extend the $D^\ast\to \DZ(h^+h^{\prime-}) \pi$ high level trigger stream 
to include charged 4-body $\DZ$ decays.  The foreseen event yields would
be similar to those anticipated for the $\DZ\to K^\mp\pi^\pm$ case.

The BaBar collaboration studied the time-dependence of the above multi-body decay
modes \cite{Aubert:2006kt}. Since the possible mixing contribution
followed by CF decay needs to be
distinguished from the DCS decays, the sensitivity of the measurement
is increased by selecting regions of phase space where the ratio of
the two is the largest. 
The preliminary value of $R_M$, which is not affected
by this selection, is found to be  
$R_M=(0.023\pm {0.018\atop 0.014} \pm 0.004)\%$ 
($R_M<0.054\%$ at 95\% C.L. using a Bayesian approach) 
in the $\DZ\to K^+\pi^-\pi^0$ mode, and without selecting a region of phase-space
$R_{WS}(K\pi\pi^0)=(0.214\pm
0.008\pm 0.008)\%$ is obtained. 
By combining the obtained
$\delta\log{\cal{L}(R_M)}$ curve with the one from the study of the
$\DZ\to K^+\pi-\pi^+\pi^-$ channel $R_M=(0.020\pm {0.011\atop
0.010})\%$ ($R_M<0.042\%$ at 95\% C.L. using a Bayesian approach) 
is obtained (stat. uncertainty only) . The combined data 
are compatible with the no-mixing hypothesis at the 2.1\% C.L.

\subsubsection{Time-dependent Dalitz-plot analysis}
Due to the strong variation of the interference effects over the
$\DZ\to K^+\pi^-(n\pi)$ phase-space a Dalitz analysis of these modes
can give further insight into the $\DZ$ mixing. Such an analysis 
has been performed for $\DZ\to K_S\pi^-\pi^+$
channel by CLEO collaboration \cite{Asner:2005sz}, { {and recently
results from Belle collaboration became available
\cite{Abe:2007rd}.}} 
Different intermediate states
contributing to $K_S\pi^-\pi^+$ (CP even or odd, like $K_Sf_0$ or
$K_S\rho^0$, or flavour eigenstates, like $K^\ast (892)^+\pi^-$), that
can be determined by inspection of the Dalitz plane, contribute
differently to the decay time distribution of $\DZ\to K_S\pi^-\pi^+$. 
A {{ simultaneous fit of the Dalitz and decay time distributions 
is used to determine the mixing parameters $x=(0.80\pm 0.29\pm
0.17)\%$ and $y=(0.33\pm 0.24 \pm 0.15)\%$.}} 
Important systematic error arises due to the uncertainty of
the model used for
the description of the Dalitz structure { {(around $\pm 0.15\%$ and 
$\pm 0.10\%$ on $x$ and $y$, respectively).}} 
Projecting the amount of data used in the analysis (540~fb$^{-1}$) to
the amount possibly available to the B-factories in the future
(2~ab$^{-1}$) the statistical precision on each parameter could be
improved to { {$\sim 0.15\%$.}} Hence the systematic error,
receiving contributions from the uncertainty of the $t$ distribution
modelling (similar as for the case of $\DZ\to
K^+\pi^-$ decays) as well as from the Dalitz model, will need to be
studied carefully.

%

\subsubsection{Nonleptonic decays to CP eigenstates} 
$\DZ$ mixing can be measured by comparing 
the lifetimes extracted from the analysis of $D$ decays into the CP-even and CP-odd 
final states. In practice, the lifetime measured in $D$ decays into CP-even final state $f_{CP}$,  
such as $K^+K^-, \pi^+\pi^-, \phi K_S$, etc., is compared to the one obtained from a 
measurement of decays to a non-CP eigenstate, such as $K^-\pi^+$. This analysis is also 
sensitive to a {\it linear} function of $y$ via
\begin{equation}
y_{CP} = \frac{\tau(D \to K^-\pi^+)}{\tau(D \to K^+K^-)}-1=
y \cos \phi - x \sin \phi \left[\frac{R_m^2-1}{2}\right],
\end{equation}
where $\phi$ is a CP-violating phase. In the limit of vanishing CP violation 
$y_{CP} = y$. This measurement requires precise determination of lifetimes. It
profits from some cancellation of the systematic uncertainties in
the ratio $\tau(K^-\pi^+)/\tau(f_{CP})$. To date $CP=+1$ final
states $K^+K^-$ and $\pi^+\pi^-$ have been used~\cite{Aubert:2003pz,Abe:2003ys,Abe:2001ed,Csorna:2001ww,Link:2000cu,Aitala:1999dt,Staric:2007dt}. 

{ {In the course of preparation of this document the Belle collaboration
obtained new 
result on $y_{CP}$ using 540~fb$^{-1}$ of
data \cite{Staric:2007dt}. It
represents evidence for the $\DZ\DZB$ mixing, with $y_{CP}=1.31\pm
0.32 \pm 0.25\%$ differing from zero by 3.2 standard deviations. }}

With the currently available statistical samples at 
the B-factories, the statistical uncertainty of the measurements using the $D^{\ast +}$ 
tag is comparable to the systematic one. The latter arises mainly from an 
imperfect modelling of the { {$t$ distribution of the background 
(although the overall background level is small, and the systematic
uncertainty due to this source might decrease with increased data
sample), and from the possible
non-cancellation of systematic errors on individual lifetime
measurements. With the final B-factories' data set one can hope for a
total uncertainty on $y_{CP}$ of around $\pm 0.25\%$. To this, 
systematic error contributes $\pm 0.10\%$ if the sources expected
to scale with the luminosity are taken into account. }}


LHCb intends to make an important contribution to the
{{measurements of}} a
non-zero value of $y_{CP}$ through the high statistics available
from the $D^\ast$ trigger, and the excellent particle identification
capabilities of its RICH system.  A sample of $1.6 \times 10^6$ 
$\DZ \to K^+K^-$  events is expected from $B$ decays alone after 
all selection cuts.  The expected sensitivity to $y_{CP}$ from this source
with 5 years of data is $0.5 \times 10^{-3}$.

%

\subsubsection{Quantum-correlated final states}
\label{sec:qc}
The construction of tau-charm factories introduces 
new {\it time-independent} methods that are sensitive to a linear function of $y$. One 
can use the fact that heavy meson pairs produced in the decays of heavy quarkonium 
resonances have the useful property that the two mesons are in the CP-correlated 
states~\cite{Bigi:1989ah,Atwood:2002ak}. For instance, by tagging one of the mesons as a CP 
eigenstate, a lifetime difference may be determined by measuring the leptonic 
branching ratio of the other meson. The final states reachable by neutral charmed 
mesons are determined by a set of selection rules according to the initial virtual 
photon quantum numbers $J^{PC}=1^{--}$~\cite{Atwood:2002ak,Asner:2005wf}. Currently, 
the decay rates of several singly-tagged (only a single meson is fully reconstructed) 
and doubly-tagged (both mesons reconstructed) final states of the $\DZ{\DZB}$
pairs are measured at CLEO-c~\cite{Asner:2006md}, where the individual fractions
depend on the mixing parameters $y$ and $R_M$, $\DZ$ branching
fractions and phases between DCS and CF decays. Types of decays
considered include semileptonic decays 
and decays to flavour and CP
eigenstates. The above parameters are determined from a fit to the
efficiency-corrected yields using 281~pb$^{-1}$ of data, 
with the preliminary results most relevant to the $\DZ$ mixing 
$y=-0.058\pm 0.066$, $R_M=(1.7\pm 1.5)\times 10^{-3}$ and
$\cos{\delta_{K\pi}}=1.09\pm 0.66$. The systematic
uncertainties, expected to be of smaller size, are being evaluated. 
At CLEO-c the precision of results is expected to be reduced by
increasing the data sample by a factor of three, increasing the number of
CP eigenstate modes, and using constraints from other measurements of $\DZ$
branching fractions. 
The same method will be exploited by BES III, with an expected data
sample of 20~fb$^{-1}$. Statistical uncertainty could be reduced
to $\sigma(y)\sim 0.002$, $\sigma(R_M)\sim 0.2\times 10^{-3}$ and 
$\sigma(\cos{\delta_{K\pi}})\sim 0.02$. 

\begin{center}
\begin{figure}[t]
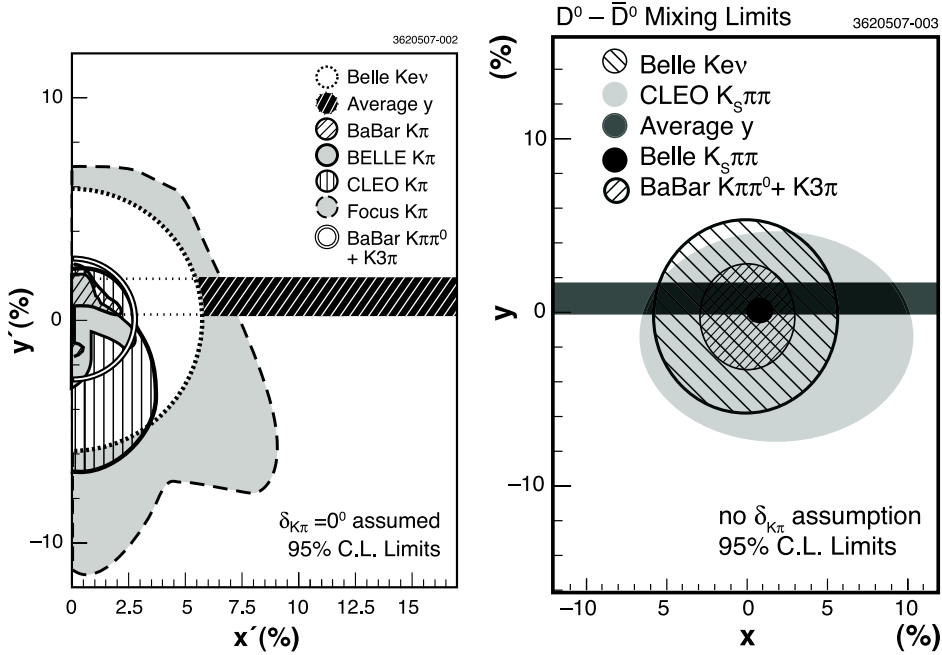

\hspace{1cm}
\includegraphics[width=6cm]{charm/fig_paper/3620507-002.epsi}\hspace{.4cm}\includegraphics[width=6cm]{charm/fig_paper/3620507-003.epsi}\hspace{1cm}
\caption{\it 
Allowed regions in the $x^\prime$ vs $ y^\prime$ plane (left) and $x$ vs $y$
for the measurements described in the text.
We assume $\delta_{K\pi}=0$ to place the $y$ results in $x^\prime$ vs $ y^\prime$.
A non-zero $\delta_{K\pi}$ would rotate the $\DZ \to CP$ eigenstates ($y$ results) 
confidence region clockwise about the origin by $\delta$.
}
\label{mixlimits}
\end{figure}
\end{center}

\subsubsection{Summary of Experimental $D$ Mixing Results}

The constraints in $x^\prime$ vs $y^\prime$ and $x$ vs $y$ are shown in Fig.~\ref{mixlimits}. 
Approximate uncertainties of the measured quantities, as expected from
the data samples assumed above, are shown in Table \ref{mixtab1}. 
\begin{table}
\begin{center}
\caption{Approximate expected precision ($\sigma$) on the measured
quantities using methods described in the text for the integrated
luminosity of 10~fb$^{-1}$ at LHCb, 2~ab$^{-1}$ at the 
B-factories at 10~GeV, and 20~fb$^{-1}$ at BESIII running at charm threshold.
The LHCb numbers do not include the effect
of systematic errors, but neglect the contribution of events from prompt charm
production.   Entries marked $`/'$ in the LHCb column are
where expected performance numbers are not yet available.
}
\begin{tabular}{|ccccc|}
\hline\hline
Mode & Observable & LHCb (10~fb$^{-1}$) &  B-factories  (2~ab$^{-1}$) &
$\psi(3770)$ (20~fb$^{-1}$) \\
$\DZ\to K^{(\ast)-}\ell^+\overline{\nu}$ & $R_M$ & $/$ & $0.2\times 10^{-3}$ & \\
$\DZ\to K^+\pi^-$ & $x^{\prime 2}$ & $0.6\times 10^{-4}$ & $1.5\times 10^{-4}$ & \\
                  & $y^{\prime}$   & $0.9\times 10^{-3}$ & $2.5\times 10^{-3}$ & \\
$\DZ\to K^+K^-$ & $y_{CP}$ & $0.5 \times 10^{-3} $  & $3\times 10^{-3}$ & \\
$\DZ\to K^0_S\pi^+\pi^-$ & $x$ & $/$ & $2\times 10^{-3}$ & \\
                        & $y$ & $/$ &  $2\times 10^{-3}$ & \\
$\psi(3770) \to \DZ\DZB$ & $x^2$  & & &  $3\times 10^{-4}$ \\
                         & $y$ & & & $4\times 10^{-3}$ \\
                & $\cos\delta$ & & & $0.05$ \\
\hline\hline
\end{tabular}
\label{mixtab1}
\end{center}
\end{table}
The errors shown include scaled statistical errors from the
most precise existing measurements and estimates of possible
systematic 
uncertainties. 

{ {As a simple illustration of the projected results, a $\chi^2$
minimization in terms of the mixing parameters $x$ and $y$, and
$\cos{\delta_{K\pi}}$ can be performed. For the unknown true values 
$x=5\times 10^{-3}$, $y=1\times 10^{-2}$ and 
$\delta_{K\pi}=0^\circ$, 
one finds the central 68\% C.L. intervals of $x\in [3,7]\times
10^{-3}$, $y\in
[0.85,1.15]\times 10^{-2}$ and $\delta_{K\pi}\in [-12^\circ,12^\circ]$.}} 
In some cases the p.d.f.'s for the estimated parameters are
significantly non-Gaussian.

\begin{center}
\begin{figure}[t]
\includegraphics[width=0.49\textwidth]{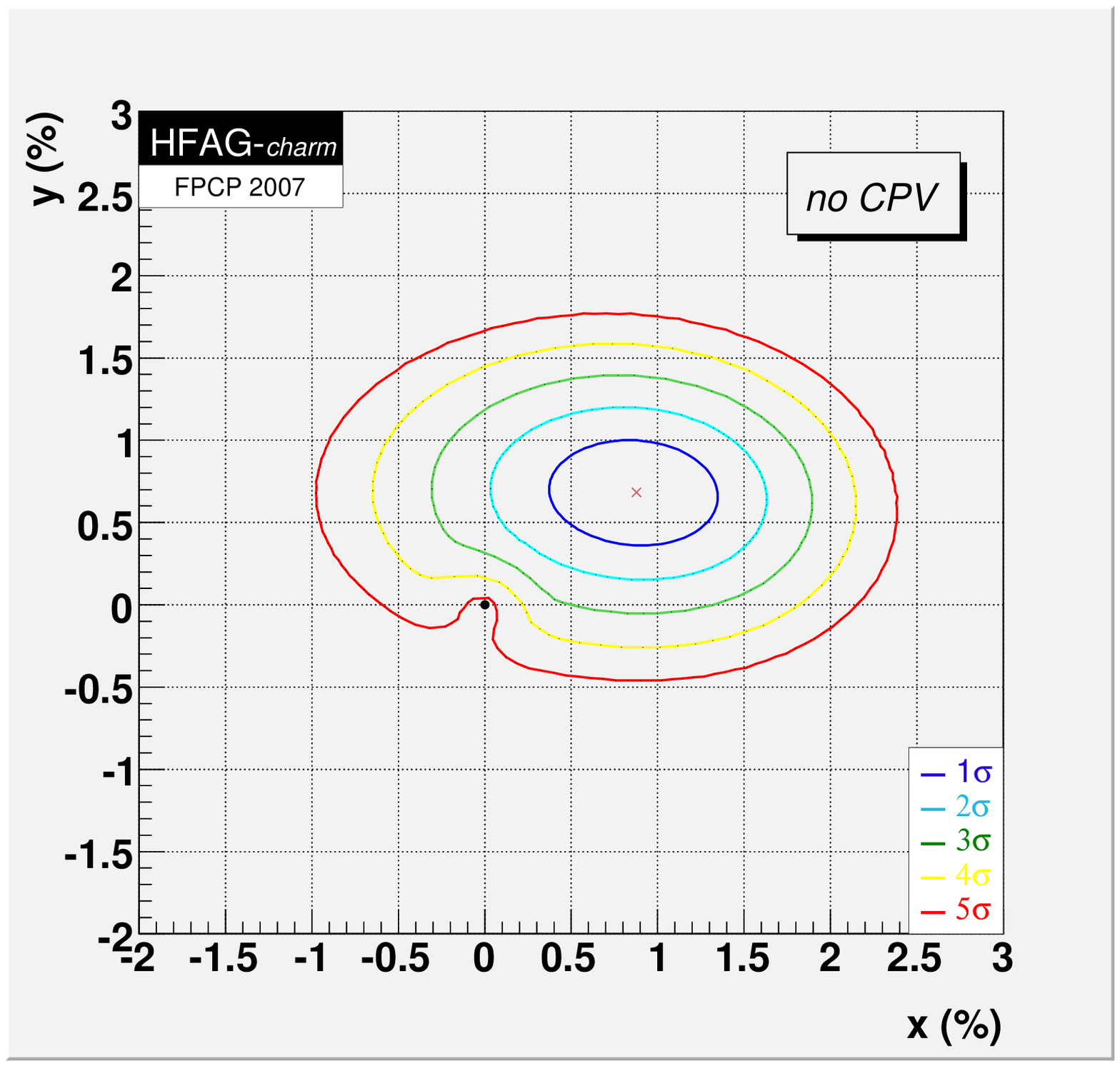}
\includegraphics[width=0.49\textwidth]{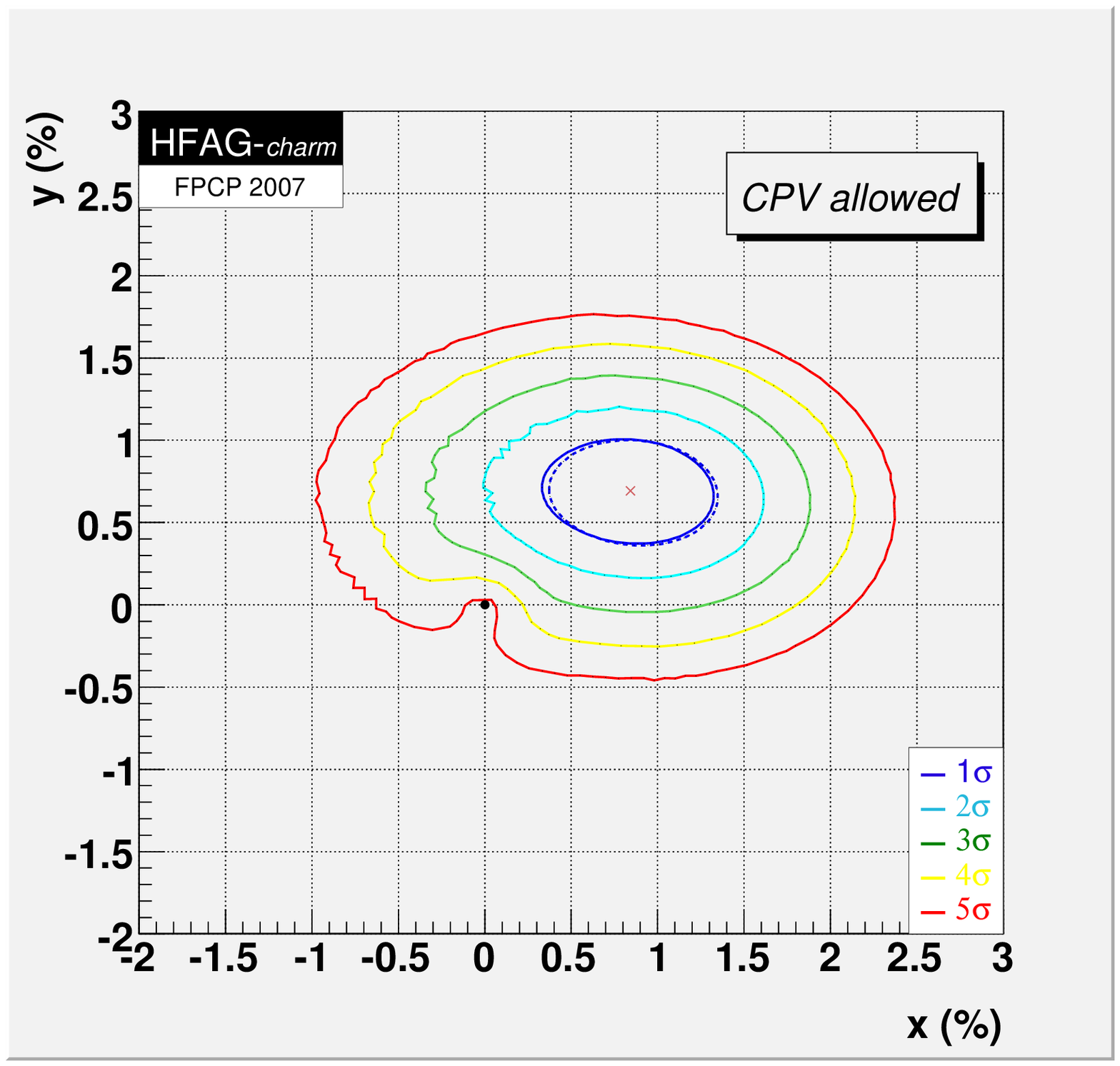}
\caption{\it 
All charm mixing measurements are combined by HFAG~\cite{Barberio:2007cr} to provide
constraints in the $x$ vs $y$ plane. Contours (1 through 5$\sigma$) of the allowed region are shown. The significance of the oscillation 
effect exceeds $5\sigma$.
}
\label{fig:mixave}
\end{figure}
\end{center}

{
The charm decays subgroup of the Heavy Flavour Averaging Group\cite{Barberio:2007cr} is
preparing world averages of all the charm measurements.
For charm mixing, the averages not only take into account correlations between meaurements but combine
the multidimensional likelihood functions associated with each measurement. A very preliminary average is
available\cite{Barberio:2007cr} giving 
$x = (8.7^{+3.0}_{-3.4})\times 10^{-3}$ 
and 
$y = (6.6^{+2.1}_{-2.0})\times 10^{-3}$. Allowing for CP violation the
very preliminary average is
$x = (8.4^{+3.2}_{-3.4})\times 10^{-3}$ 
and
$y = (6.9\pm 2.1)\times 10^{-3}$. 

The constraints in the $x$ vs $y$ plane are shown in Fig.~\ref{fig:mixave}. The significance of the oscillation 
effect exceeds $5\sigma$.}

\begin{table}
\caption{Approximate expected precision ($\sigma$) on the measured
quantities using methods described in the text for the integrated
luminosity of 100~fb$^{-1}$ at an upgraded LHCb, 75~ab$^{-1}$ at a Super 
B-factory at 10~GeV, and 200~fb$^{-1}$ at a Super B-factory running at charm threshold.
The upgraded LHCb numbers are merely the results from Table~\ref{mixtab1} scaled to the
new integrated luminosity.}
\begin{tabular}{|ccccc|}
\hline\hline
Mode & Observable & LHCb (100~fb$^{-1}$) &  Super B (75~ab$^{-1}$) &
$\psi(3770)$ (200~fb$^{-1}$) \\
$\DZ\to K^+\pi^-$ & $x^{\prime 2}$ & $2.0 \times 10^{-5}$  & $3\times 10^{-5}$ & \\
                  & $y^{\prime}$   & $2.8 \times 10^{-4}$  & $7\times 10^{-4}$ & \\
$\DZ\to K^+K^-$ & $y_{CP}$ & $1.5 \times 10^{-4}$ & $5\times 10^{-4}$ & \\
$\DZ\to K^0_S\pi^+\pi^-$ & $x$ & $/$ & $5\times 10^{-4}$ & \\
                        & $y$ & $/$  &  $5\times 10^{-4}$ & \\
$\psi(3770) \to \DZ\DZB$ & $x^2$  & & &  $<0.2\times 10^{-4}$ \\
                         & $y$ & & & $(1-2)\times 10^{-3}$ \\
                & $\cos\delta$ & & & $<0.05$ \\
\hline\hline
\end{tabular}
\label{mixtab2}
\end{table}

{ The interpretation of the new results in terms of New Physics is inconclusive. 
It is not yet clear whether the effect is caused by $x=0$ or $y=0$ or both, although the latter is favoured,
as shown in Table \ref{mixtab2}. Both 
an upgraded 
LHCb and a high luminosity Super B-factory will be able to observe both lifetime
and mass differences in the $\DZ$ system, if they lie in the range of Standard Model predictions.

A serious limitation in the interpretation of charm oscillations in terms of New Physics is the theoretical uncertainty on the Standard Model prediction. Nonetheless, if oscillations occurs at the level suggested by the recent
results, this will open the window to searches for $CP$ asymmetries that do provide unequivocal New Physics 
signals.}

\subsubsection{New Physics contributions to $D$ mixing}
As one can see from the previous discussion, mixing in the charm system is very small.
As it turns out, theoretical predictions of $x$ and $y$ in the Standard Model
are very uncertain, from a percent to orders of magnitude smaller~\cite{Petrov:2003un,Falk:2001hx}.
Thus, New Physics (NP) contributions are difficult to distinguish in the absence of large CP violation in mixing.

In order to see how NP might affect the mixing amplitude, it is instructive to
consider off-diagonal terms in the neutral D mass matrix,
\begin{eqnarray}\label{M12}
&& \left (M - \frac{i}{2}\, \Gamma\right)_{12} =
  \frac{1}{2 M_{\rm D}}\, \langle \DZB | 
{\cal H}_w^{\Delta C=-2} | \DZ \rangle \qquad\qquad
\\
&& + \frac{1}{2 M_{\rm D}}~\, \sum_n {\langle \DZB | {\cal H}_w^{\Delta
  C=-1} | n \rangle\, \langle n | {\cal H}_w^{\Delta C=-1} 
| \DZ \rangle \over M_{\rm D}-E_n+i\epsilon}\,
\nonumber
\end{eqnarray}
where ${\cal H}_w^{\Delta C=-1}$ is the effective $|\Delta C| = 1$ Hamiltonian. Since
all new physics particles are much heavier than the Standard Model ones, the most 
natural place for NP to affect mixing amplitudes is in the $|\Delta C|=2$ 
contribution, which 
corresponds to a local interaction at the charm quark mass scale.
%
\begin{center}
\begin{figure}[htb]
\hspace{4cm}\includegraphics[width=9cm]{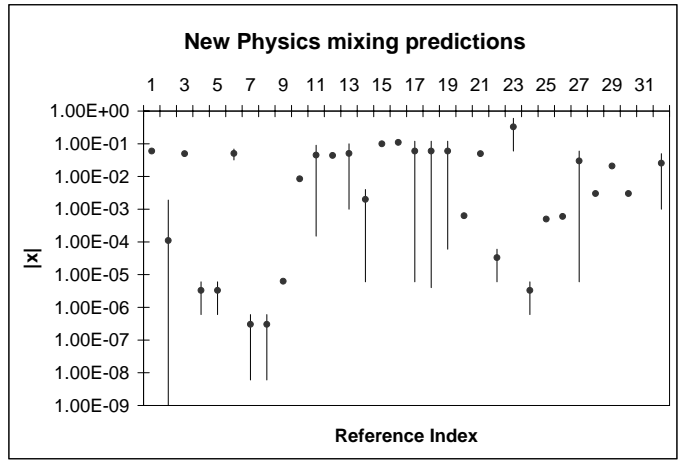}
\caption{\it NP predictions for $|x|$. Horizontal line references are 
tabulated in Table 5 of Ref.~\cite{Petrov:2003un}.}
\label{fig_NP}
\end{figure}
\end{center}
As can be seen from Fig.~(\ref{fig_NP}), predictions for $x$ vary by orders of magnitude for 
different models. It is interesting to note that some models {\it require} large signals 
in the charm system if mixing and FCNCs in the strange and beauty systems are to be small 
(e.g. the SUSY alignment model). 

The local $|\Delta C|=2$ interaction cannot, however, affect $\Delta \Gamma_{\rm D}$ because 
it does not have an absorptive part. Thus, naively, NP cannot affect the lifetime difference $y$.
This is, however, not quite correct. Consider a $\DZ$ decay amplitude which includes
a small NP contribution, $A[\DZ \to n]=A_n^{\rm (SM)} + A_n^{\rm (NP)}$.
Here, $A_n^{\rm (NP)}$ is assumed to be smaller than the current experimental
uncertainties on those decay rates. Then it is a good approximation to write $y$ as
\begin{eqnarray}\label{schematic}
y &\simeq& \sum_n \frac{\rho_n}{\Gamma_{\rm D}} 
A_n^{\rm (SM)} \bar A_n^{\rm (SM)}
+ 2\sum_n \frac{\rho_n}{\Gamma_{\rm D}}
A_n^{\rm (NP)} \bar A_n^{\rm (SM)} \ \ . 
\label{approx}
\end{eqnarray}
The SM contribution to $y$ is known to vanish in the limit of exact flavour $SU(3)$.
Moreover, the first order correction is also absent, so the SM contribution arises 
only as a {\it second} order effect. Thus, those NP contributions which do not vanish 
in the flavour $SU(3)$ limit must determine the lifetime difference there,
even if their contributions are tiny in the individual decay amplitudes~\cite{Golowich:2006gq}. 
A simple calculation reveals that NP contribution to $y$ can be as large as several percent
in R-parity-violating SUSY models or as small as $\sim 10^{-10}$ in the models with interactions 
mediated by charged Higgs particles~\cite{Golowich:2006gq}.
Assuming the projected precisions on $x$, $y$ and $cos(\delta_{K\pi})$ discussed below
 are achieved, a range of NP models can be ruled out. On the other hand, the
 uncertainty of SM predictions for the mixing parameters can in some
 scenarios (positive measurement, $y > x$) make the identification of NP
 contribution difficult. It is important to make a precise determination of
 individual parameters, using all the experimental methods mentioned (and
 possibly new ones) in order to pin down possible cracks in the SM.

\subsubsection{$D$ mixing impact on CKM angle $\gamma/\phi_3$}
Beside the importance of the mixing in the charm sector per-se, discussed above, the results of 
mentioned measurements can also have an impact on the determination of the Unitarity Triangle 
angle $\gamma/\phi_3$. Several proposed methods for measuring $\gamma/\phi_3$ use the interference 
between $B^-\to \DZ K^-$ and $B^-\to\DZB K^-$ which occurs when both $\DZ$ and $\DZB$ 
decay to the same final state \cite{Bigi:1988ym,Gronau:1990ra,Gronau:1991dp,Atwood:2000ck,Atwood:2003mj}.

The quantity 
sensitive to the angle $\gamma/\phi_3$ is the asymmetry 
$A_{DK}=[Br(B^-\to f_D K^-)-Br(B^+\to \overline{f}_D K^+)]/[Br(B^-\to f_D K^-)+Br(B^+\to \overline{f}_D K^+)]$, 
where $f_D$ denotes the common final state of $\DZ$ and $\DZB$. $A_{DK}$ can be expressed as 
\begin{equation}
A_{DK}={2r_Br_De^{-\epsilon}\sin{(\delta_B+\delta_D)}\sin{\gamma/\phi_3}\over 
r_B^2+r_D^2+2r_Br_De^{-\epsilon}\cos{(\delta_B+\delta_D)}\cos{\gamma/\phi_3}}~~,
\end{equation}
where $\delta_B$ is the difference of the strong phases in decays $B^-\to \DZ K^-$ and 
$B^-\to\DZB K^-$, $\delta_D$ is the difference of the strong phases for 
$\DZ\to f_D$ and $\DZB\to f_D$, $r_B$ is the ratio of amplitudes 
$|{\cal{A}}(B^-\to\DZB K^-)|/|{\cal{A}}(B^-\to \DZ K^-)|$ and $r_D$ is the ratio 
$|{\cal{A}}(\DZ\to f_D)|/|{\cal{A}}(\DZB\to f_D)|$. The dilution factor 
$e^{-\epsilon}$ arises if $x, y\ne 0$. 

In case of non-negligible $\DZ$ mixing the time integrated interference term between 
${\cal{A}}(\DZ\to f_D)$ and ${\cal{A}}(\DZB\to f_D)$ depends on $x$ and $y$, resulting in 
\cite{Grossman:2005rp}
\begin{equation}
\epsilon ={1\over 8}(x^2+y^2)\bigl({1\over r_D^2}+r_D^2\bigr)-
{1\over 4}(x^2\cos{2\delta_D}+y^2\sin{2\delta_D})~~.
\end{equation}

Using $f_D$ which is a CP eigenstate \cite{Gronau:1990ra,Gronau:1991dp} 
(the case where $f_D=K^0_S\pi^+\pi^-$ is dicussed in section~\ref{sec:Vub})
and neglecting CP violation in 
$\DZ$ decays the above expressions simplify due to $r_D=1$, $\delta_D=0$, and thus 
$\epsilon=y^2/4$. For $f=K^+K^-, \pi^+\pi^-$ the asymmetry $A_{DK}$ is measured to be 
$0.06\pm 0.14\pm 0.05$ using an integrated luminosity of 250~fb$^{-1}$ \cite{Abe:2006hc}. 
Projecting the result to 2~ab$^{-1}$ the expected statistical accuracy is $\pm 0.05$. 
An uncertainty on $y$ of 2\%, on the other hand, reflects in an error of 
$\sigma(A_{DK})\approx 5\times 10^{-5}$ using the above equations (conservatively assuming 
$r_B=0.25,~\sin{\delta_B}=\sin{\phi_3}=1$). It is thus save to conclude that 
neglecting the effect of $\DZ$ mixing in this method of $\gamma/\phi_3$ determination is 
appropriate. 

\begin{figure}[htb]
\includegraphics[width=7.25cm]{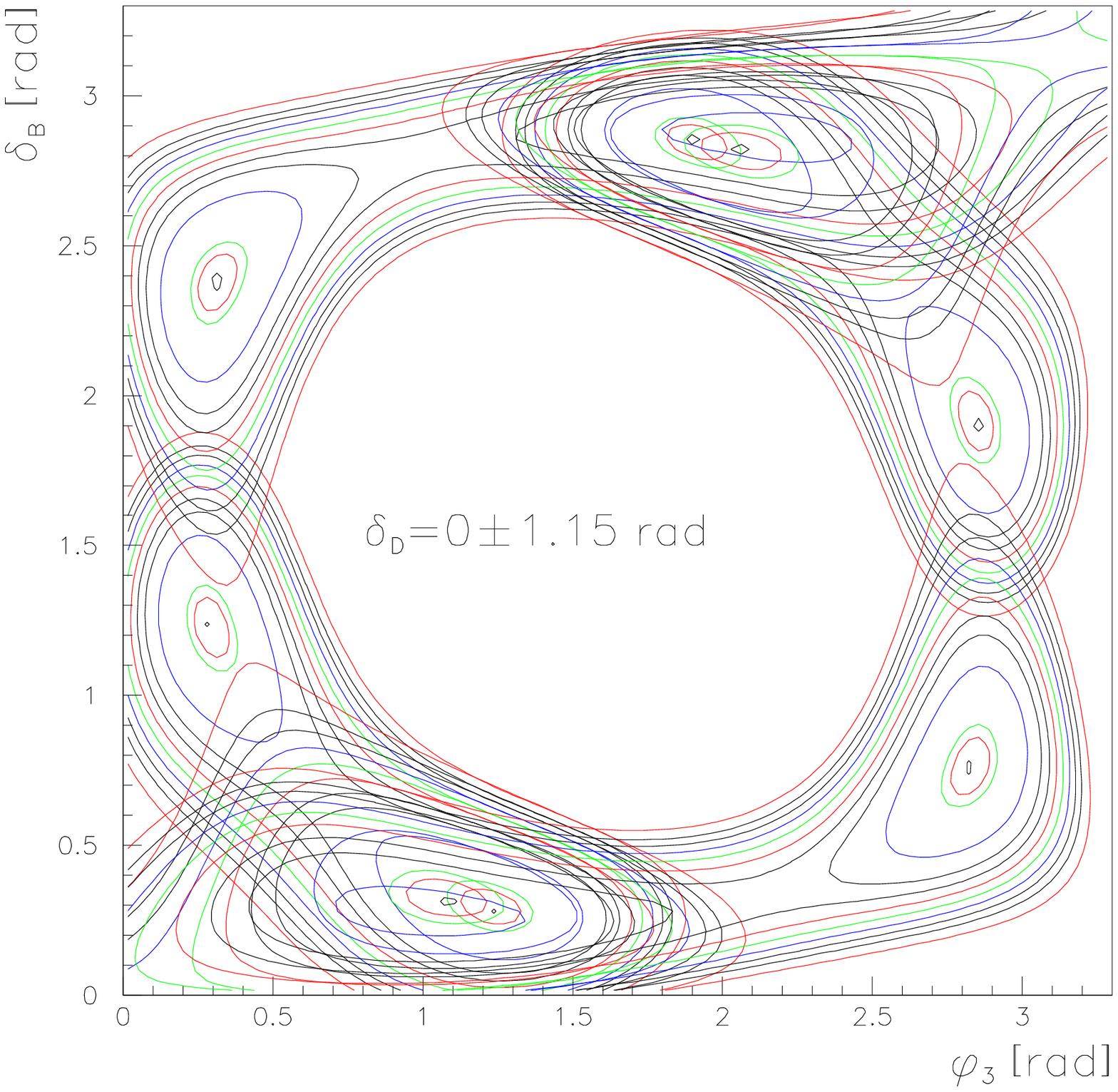}\hspace{.2cm}\includegraphics[width=7.25cm]{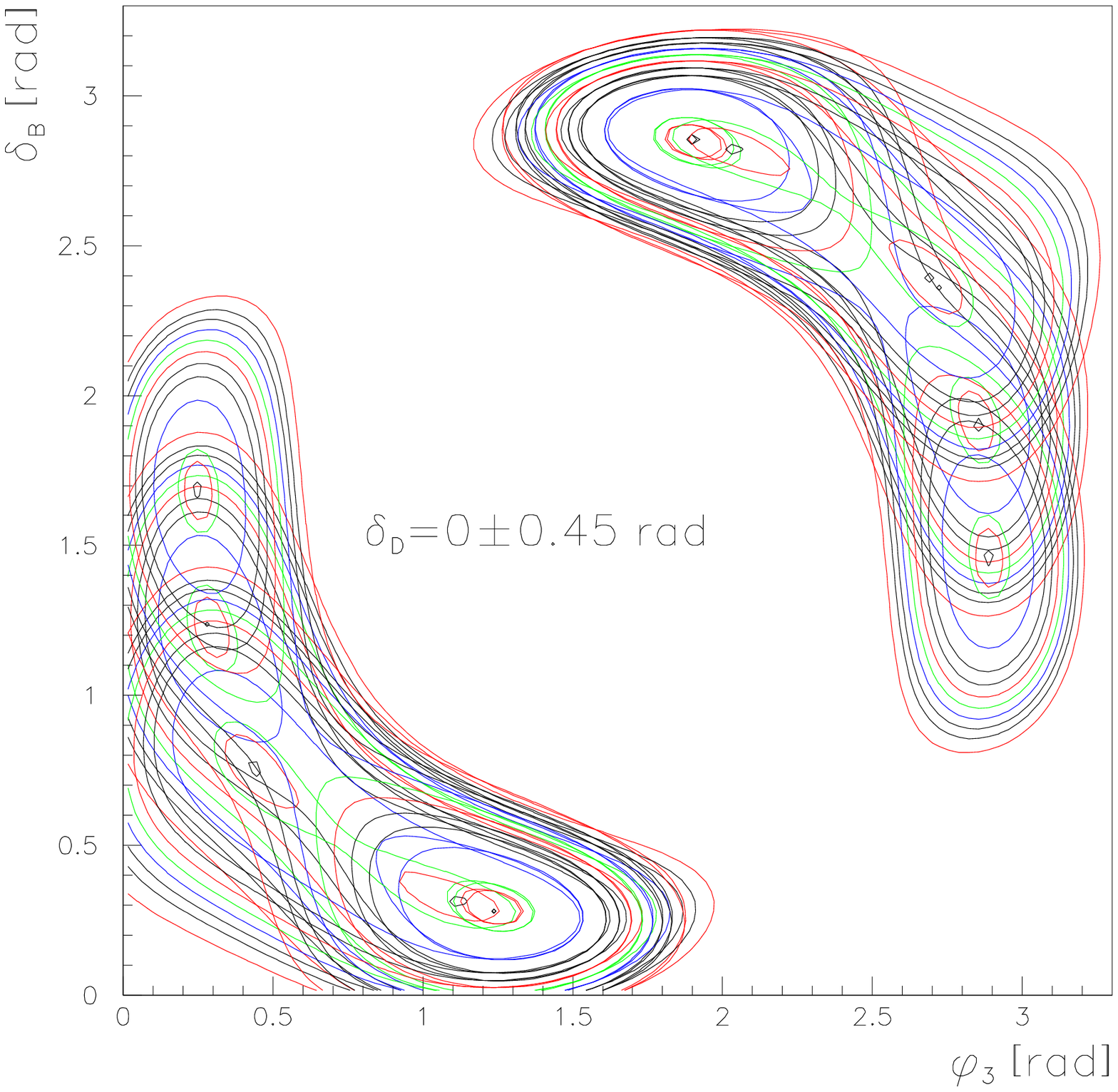}
\caption{\it 68\% C.L. contour for $\gamma/\phi_3$ and $\delta_B$ using the projected results of 
measurements described in the text. The strong phase difference $\delta_D$ between 
$\DZ\to K^+\pi^-/K^-\pi^+$ decays 
is assumed to have the values marked in the plots.} 
\label{fig_delD}
\end{figure}

Beside $f_D$ being a CP eigenstate, the final state can be chosen to arise from DCS 
decays \cite{Atwood:2000ck,Atwood:2003mj}. In this case the strong phase $\delta_D$ enters the 
expressions. To illustrate the effect of $\delta_D$ on extraction of the angle $\gamma/\phi_3$ 
one can envisage usage of two distinct final states, 
for example the above mentioned $f=K^+K^-, \pi^+\pi^-$ and $K^+\pi^-$ which can also be reached from either 
$\DZ$ or $\DZB$. For the former the same asymmetry $A_{DK}$ can be measured, while for the 
latter the ratio $R_{DK}=Br(B^-\to D_{sup}K^-)/Br(B^-\to D_{fav}K^-)$ is also sensitive to 
$\gamma/\phi_3$. Here, $D_{sup}$ denotes 
DCS decays $\DZ\to K^+\pi^-$ and $D_{fav}$ stands for $\DZ\to K^-\pi^+$. $R_{DK}$ depends on the unknown 
angles: 
\begin{equation}
R_{DK}=r_B^2+r_D^2+2r_Br_D\cos{(\delta_B+\delta_D)}\cos{\gamma/\phi_3}~~,  
\end{equation}
with $r_D=(6.2\pm 0.1)\times 10^{-2}$ \cite{Yao:2006px}. 
Assuming $r_B$ is known, measuring $A_{DK}$ and $R_{DK}$ constrains possible ranges for $\delta_B$ and $\gamma/\phi_3$. 
Knowledge of $\delta_D$ clearly helps in limiting the ($\gamma/\phi_3$, $\delta_B$) allowed region. We can use the projected 
result $A_{DK}=0.06\pm 0.05$ and the ratio $R_{DK}=(2.3\pm 1.5\pm 0.1)\times 10^{-2}$ as 
obtained using 250~fb$^{-1}$ of data \cite{Saigo:2004de}. Hence one can expect $R_{DK}=(2.3\pm 0.6)\times 10^{-2}$ 
with the final B-factories data set. The approximate two dimensional 68\% C.L. contour obtained by plotting 
the corresponding 
$\chi^2$ of the two projected measurements as a function of $\gamma/\phi_3$ and $\delta_B$ is shown in 
Fig.\ref{fig_delD}.  The left plot shows the allowed region for the current value of $\delta_D=(0\pm 1.15)$~rad 
\cite{Asner:2006md}. To show the effect of an improved knowledge of the $D$ meson decays strong phase the value 
$\delta_D=(0\pm 0.45)$~rad (see Table \ref{mixtab1}) is used in the right plot. The allowed region of the unknown 
angles is significantly reduced although it should be noted that the actual region strongly depends 
on the central 
values of $\delta_D$ as well as $r_B$ (for the latter the value 0.12 was used in the plots). 
%

\subsubsection{\cp~Violation with \& without Oscillations}



\label{sec:cpv}

\subsubsection{Theoretical overview}

Most factors favour or even call for dedicated searches for \cp~violation in charm transitions: 

$\oplus$ 
Since baryogenesis implies the existence of New Physics in \cp-violating dynamics, it would be unwise not to undertake dedicated searches for \cp~asymmetries in 
charm decays, where the `background' from known physics is between absent and small: 
for within the SM the effective weak phase is highly diluted, namely $\sim {\cal O}(\lambda ^4)$, and it can arise only in {\em singly-Cabibbo-suppressed} transitions, where one  
expects asymmetries to reach the ${\cal O}(0.1 \%)$ level; significantly larger values would signal New Physics.  
{\em Any} asymmetry in {\em Cabibbo-allowed or doubly-suppressed} channels requires the intervention of New Physics -- except for 
$D^{\pm}\to K_S\pi ^{\pm}$ \cite{Bianco:2003vb}, where the \cp~impurity in $K_S$ induces an asymmetry of $3.3\cdot 10^{-3}$. One should keep in mind that in going from Cabibbo-allowed to Cabibbo 
singly-  and doubly- suppressed channels, the SM rate is {\em suppressed} by factors of about 
twenty and four hundred, respectively: 
$$ 
\Gamma _{SM}( H_c \to [S=-1]) : \Gamma _{SM}( H_c \to [S= 0]) : \Gamma _{SM}( H_c \to [S= +1]) 
\simeq 
$$
\beq
1 : 1/20 : 1/400
\eeq
One would expect that this suppression will enhance the visibility of New Physics.  

$\oplus$ 
Strong phase shifts 
required for {\em direct} \cp~violation to emerge in partial widths are in general large as are the branching ratios into relevant modes;  while large final state interactions complicate the 
interpretation of an observed signal in terms of the microscopic parameters of the underlying dynamics, they enhance its observability.  

$\oplus$ 
Since the SM provides many amplitudes for charm decays, 
\cp~asymmetries can be linear in New Physics amplitudes thus increasing sensitivity to the 
latter.  

$\oplus$ 
Decays to final states of {\em more than} two pseudoscalar or one pseudoscalar and one vector meson contain 
more dynamical information than given by their  widths; their distributions as described by Dalitz plots 
or \ot~odd moments can exhibit \cp~asymmetries that might be considerably larger than those for the 
width. This will be explained in a bit more detail later on. 

$\oplus$ The distinctive channel $D^{\pm*} \to D \pi^{\pm}$ provides a powerful tag 
on the flavour identity of the neutral $D$ meson. 

$\ominus$ The `fly in the ointment' is that $D^0 - \bar D^0$ oscillations are on the slow side.  

$\oplus$ Nevertheless one should take on this challenge. For 
\cp~violation involving $D^0 - \bar D^0$ oscillations is a reliable probe of New Physics: the 
asymmetry is controlled by  
sin$\Delta m_Dt$ $\cdot$ Im$(q/p)\bar \rho (D\to f)$. Within the SM both factors are small, namely 
$\sim {\cal O}(10^{-3})$, making such an asymmetry unobservably tiny -- unless there is 
New Physics; for a recent New Physics model see \cite{Agashe:2004cp}.  
One should note 
that this observable is {\em linear} in $x_D$ rather than quadratic as for \cp~insensitive quantities 
like $D^0(t) \to l^-X$.  
$D^0 - \bar D^0$ oscillations, \cp~violation and New Physics might thus be discovered simultaneously in a transition. We will return to this point below. 

$\ominus$ Honesty compels us to concede there is no attractive, let alone compelling scenario of 
New Physics for charm transitions whose footprints should not be seen also in $B$ decays.  

$\oplus$ It is all too often overlooked that \cpt~invariance can provide nontrivial constraints on 
\cp~asymmetries. For it imposes equality not only on the masses and total widths of particles and antiparticles, but also on the widths for `disjoint' {\em sub}sets of channels. 
`Disjoint' subsets are the decays to final states that can{\em not} rescatter into each other. Examples are 
semileptonic vs. nonleptonic modes with the latter subdivided further into those with strangeness 
$S = -1,0.+1$. Observing a \cp~asymmetry in one channel one can then infer in which other channels 
the `compensating' asymmetries have to arise \cite{Bianco:2003vb}. 

\subsubsection{Direct \cp~violation in partial rates}

\cp~violation in $\Delta C =1$ dynamics can be searched for by comparing partial widths for 
\cp~conjugate channels. For an observable effect two conditions have to be satisfied simultaneously: 
a transition must receive contributions from two coherent amplitudes with (a) different weak 
and (b) different strong phases as well. While condition (a) is just the requirement of \cp~violation in the 
underlying dynamics, condition (b) is needed to make the relative weak phase observable. 
Since the decays of charm hadrons proceed in the nearby presence of many hadronic resonances 
inducing virulent final state interactions (FSI), requirement (b) is in general easily met; thus it provides 
no drawback for the {\em observability} of a \cp~asymmetry -- albeit it does for its 
{\em interpretation}. 

As already mentioned CKM dynamics does not support any \cp~violation in Cabibbo allowed and doubly suppressed channels due to the absence of a second weak amplitude; the only exception are 
modes containing a $K_S$ (or $K_L$) like $D^+ \to K_S \pi^+$ vs. $D^- \to K_S \pi^-$ which 
have to exhibit an asymmetry of $0.0032$ reflecting the \cp~impurity in the $K_S$ (or $K_L$) 
wave function. In once-Cabibbo-suppressed transitions one expects \cp~asymmetries, albeit highly 
diluted ones of order $\lambda ^4 \sim 10^{-3}$. 

While we have good information on the size of the weak phase, we do not know how to predict the 
size of the relevant matrix elements and strong phases in a reliable way. Even if a direct \cp~asymmetry 
larger than about $10^{-3}$ were observed in a Cabibbo-suppressed mode -- say even as large as $10^{-2}$ --, at present we could not 
claim such a signal to establish the intervention of New Physics. A judicious exercise in `theoretical 
engineering' could, however, solve our conundrum. 

\subsubsection{Theoretical Engineering}
\label{ENGIN}

\cp~asymmetries in integrated partial widths depend on hadronic matrix elements and (strong)  
phase shifts, neither of which can be predicted accurately. However the craft of theoretical 
engineering can be practised with profit here. One makes an ansatz for the general form of the matrix 
elements and phase shifts that are included in the description of $D\to PP, PV, VV$ etc. 
channels, where $P$ and $V$ denote pseudoscalar and vector mesons, and fits them to the measured branching ratios on the Cabibbo allowed, once and twice forbidden level. If one has sufficiently accurate and comprehensive data, one can use these fitted values of the hadronic parameters to predict \cp~asymmetries. Such analyses have been undertaken in the past \cite{Buccella:1994nf}, 
but the data base was not as broad and precise as one would like. 
{\em CLEO-c and BESIII measurements 
will certainly lift such studies to a new level of reliability.}

\subsubsection{\cp~violation in final state distributions}
\label{sec:cpvfs}
Once the final state in $D \to f$ is more complex than a pair of pseudoscalar mesons or a 
pseudoscalar plus a vector meson it contains more dynamical information than given by the 
modulus of its amplitude, since its kinematics are no longer trivial. 
 \cp~asymmetries in final state distributions can be substantially larger than in integrated partial 
widths. 

The simplest such case is given by decays into three pseudoscalar mesons, for which 
Dalitz plots analyses represent a very sensitive tool with the phase information they yield. They 
require large statistics; yet once those have been obtained, the return is very substantial. 
For the constraints one has on a Dalitz plot population provide us with powerful weapons to 
control systematic uncertainties. 

Such phenomenological advantages of having more complex final states apply also for four-body etc. final states. Measuring \ot~odd moments with 
\beq 
O_T \stackrel{\ot} \Longrightarrow - O_T 
\eeq
is an efficient way to make use of data with limited statistics. A simple example for a final state 
with four mesons $a$, $b$, $c$ and $d$ is given by 
$O_T = \langle \vec p_c \cdot (\vec p_a \times \vec p_b)\rangle $.

While FSI are not necessary for the emergence of such effects -- unlike the situation for 
partial width asymmetries --, they can fake a signal of \ot~violation with \ot~being an {\em anti}linear 
operator; 
yet that can be disentangled by comparing \ot~odd moments for \cp~conjugate modes \cite{Link:2005th}: 
\beq 
O_T(D\to f) \neq - O_T(\bar D \to \bar f) \; \; \; \Longrightarrow \; \; \; \cp~{\rm violation}
\eeq

A dramatic example for \cp~violation manifesting itself in a final state distribution much more 
dramatically than in a partial width has been found in $K_L$ decays. Consider the 
rare mode $K_L \to \pi^+\pi^- e^+e^-$ and define by $\phi$ the angle between the 
$\pi^+\pi^-$ and $e^+e^-$ planes. The differential width has the general form 
\beq 
\frac{d\Gamma}{d\phi}(K_L \to \pi^+\pi^- e^+e^-) = \Gamma_1 {\rm cos}^2 \phi + 
\Gamma_2 {\rm sin}^2 \phi + \Gamma_3 {\rm cos} \phi {\rm sin}\phi
\eeq
Upon integrating over $\phi$ the $\Gamma_3$ term drops out from the total width, which thus is 
given in terms of $\Gamma_{1,2}$ with $\Gamma_3$ representing a forward-backward asymmetry. 
\beq 
 \langle A \rangle \equiv 
 \frac{\int _0 ^{\pi/2}\frac{d\Gamma}{d\phi} -  \int _{\pi/2} ^{\pi}\frac{d\Gamma}{d\phi}}
{\int _0 ^{\pi}\frac{d\Gamma}{d\phi} } = 
\frac{2\Gamma_3}{\pi(\Gamma_1 + \Gamma_2)}
\eeq
Under \op~and \ot~one has cos$\phi$sin$\phi$ $\to$ $-$ cos$\phi$ sin$\phi$. 
Accordingly $\langle A \rangle $ and $\Gamma_3$ constitute a \ot~odd correlation, while $\Gamma_{1,2}$ are 
\ot~even. $\Gamma_3$ is driven by the \cp~impurity $\epsilon_K$ in the kaon wave function. 
$\langle A \rangle$ has been measured to be large in full agreement with theoretical 
predictions \cite{Sehgal:1992wm}: 
\beq
 \langle A \rangle = 0.138 \pm 0.022  \; . 
\eeq
One should note this observable is driven by $|\epsilon_K| \simeq 0.0023$. 

$D$ decays can be treated in an analogous way.  Consider the Cabibbo-suppressed channel 
\footnote{This mode can exhibit direct \cp~violation even within the SM.}
\beq 
\stackrel{(-)}D \to K \bar K \pi^+\pi^-
\eeq
and define $\phi$ to be the angle between the $K \bar K$ and $\pi^+\pi^-$ planes. Then 
one has 
\bea 
\frac{d\Gamma}{d\phi}(D \to K \bar K\pi^+\pi^-) &=& \Gamma_1 {\rm cos}^2 \phi + 
\Gamma_2 {\rm sin}^2 \phi + \Gamma_3 {\rm cos} \phi\, {\rm sin}\phi \\
\frac{d\Gamma}{d\phi}(\bar D \to K \bar K\pi^+\pi^-) &=& \bar \Gamma_1 {\rm cos}^2 \phi + 
\bar \Gamma_2 {\rm sin}^2 \phi + \bar \Gamma_3 {\rm cos} \phi\, {\rm sin}\phi 
\eea
As before the partial width for $D[\bar D] \to K\bar K \pi^+\pi^-$ is given by 
$\Gamma_{1,2} [\bar \Gamma_{1,2}]$; $\Gamma_1 \neq \bar \Gamma_1$ or 
$\Gamma_2 \neq \bar \Gamma_2$ represents direct \cp~violation in the partial width. 
$\Gamma_3 \& \bar \Gamma_3$ constitute \ot~odd correlations. By themselves they do not necessarily 
indicate \cp~violation, since they can be induced by strong final state interactions. However 
\beq 
\Gamma_3 \neq \bar \Gamma_3 \; \; \Longrightarrow \cp~{\rm violation!}
\eeq 
It is quite possible or even likely that a difference in $\Gamma_3$ vs. $\bar \Gamma_3$ 
is significantly larger than in $\Gamma_1$ vs. $\bar \Gamma_1$ or 
$\Gamma_2$ vs. $\bar \Gamma_2$. Furthermore one can expect that differences in detection 
efficiencies can be handled by comparing $\Gamma_3$ with $\Gamma_{1,2}$ and 
$\bar \Gamma_3$ with $\bar \Gamma_{1,2}$.

\subsubsection{\cp~asymmetries involving oscillations}

For final states that are common to $D^0$ and $\bar D^0$ decays one can search for 
\cp~violation manifesting itself with the help of $D^0-\bar D^0$ oscillations in qualitative -- 
though certainly not quantitative -- analogy to $B_d \to \psi K_S$. Such common states 
can be \cp~eigenstates -- like $D^0 \to K^+K^-/\pi^+\pi^-/K_S\eta^{(\prime)}$ --, but do not have to be: 
two very promising candidates are $D^0 \to K_S \pi^+\pi^-$, where one can bring the full Dalitz plot machinery to bear, and $D^0 \to K^+\pi^-$ vs. $\bar D^0 \to K^-\pi^+$, since its SM amplitude is 
doubly-Cabibbo-suppressed. Undertaking 
{\em time-dependent} Dalitz plot studies requires a higher initial overhead, yet in the long run this 
should pay handsome dividends exactly since Dalitz analyses can invoke many internal correlations 
that in turn serve to control systematic uncertainties. 

Searching for such effects with the required sensitivity (see below) will be quite challenging. 
Nevertheless one should take on this challenge. For 
\cp~violation involving $D^0 - \bar D^0$ oscillations is a reliable probe of New Physics: the 
asymmetry is controlled by  
sin$\Delta m_Dt$ $\cdot$ Im$(q/p)\bar \rho (D\to f)$. Within the SM both factors are small, namely 
$\sim {\cal O}(10^{-3})$, making such an asymmetry unobservably tiny -- unless there is 
New Physics; for a recent New Physics model see \cite{Agashe:2004cp}.  
One should note 
that this observable is {\em linear} in $x_D$ rather than quadratic as for \cp-insensitive quantities 
like $D^0(t) \to l^-X$.  
$D^0 - \bar D^0$ oscillations, \cp~violation and New Physics might thus be discovered simultaneously in a transition. 


\subsubsection{Experimental searches for \cp~violation}

Let the amplitude for $D^0$ to decay to a final state $f$ be written as
$$A_f\equiv\langle f|{\cal H}_{\rm int}|D^0\rangle$$
where $\cal{H}_{\rm int}$ is the interaction Hamiltonian responsible for $D^0\to f$. If \cp\ is conserved, that is if $\left[{\cal H}_{\rm int},CP\right]=0$, then we can clearly write
\begin{eqnarray}
A_f&=&\langle f|(CP)^\dagger(CP){\cal H}_{\rm int}|D^0\rangle\label{eq:ACP}\\
&=&\langle f|(CP)^\dagger{\cal H}_{\rm int}(CP)|D^0\rangle\nonumber\\
&=&-\langle\bar f|{\cal H}_{\rm int}|\bar D^0\rangle\equiv-\bar A_{\bar f}\nonumber
\end{eqnarray}
where $\bar f$ is the conjugate final state to $f$. Consequently, a measurement that shows $\Gamma(D^0\to f)\neq\Gamma(\bar D^0\to\bar f)$ is a demonstration that CP is violated in this decay.

Most CP violation results are from the FNAL fixed target experiments 
E791 and FOCUS, and the CLEO experiment 
and search for direct CP violation. 
The CP violation asymmetry is defined as 
\begin{equation}
A_{CP}\equiv \frac{\Gamma(D\to f) -\Gamma(\overline D \to \overline f)}{\Gamma(D\to f) +\Gamma(\overline D \to \overline f)}.
\end{equation}
A few results from CLEO, BaBar and Belle experiments consider CP violation in mixing. Typically, precisions of a few percent are obtained~\cite{Yao:2006px}.
No evidence for CP violation is observed consistent with Standard Model
expectations.

Certainly very large samples will be available from hadron colliders.
From an existing CDF measurement~\cite{Acosta:2004ts} it is possible to 
anticipate yields of over 0.5--1 million $D^0 \to K^+K^-$ events 
being available with the likely final Tevatron integrated luminosity of 5--10~$\rm fb^{-1}$.
This sample will have an intrinsic statistical precision of $\le 0.2\%$.
With the higher production cross-section and its dedicated $D^\ast$ trigger LHCb
will accumulate samples of up to 10 million tagged events in 
each year of nominal operation~\cite{LHCBCHARM}. The RICH system will ensure a low background,  
and these decays will be complemented by those selected in the
$D^0 \to \pi^+\pi^-$ mode.  In order to exploit these enormous statistics
it will be necessary to pay great attention to systematics biases.
Initial state asymmetries and detector asymmetries will be the main
concerns.

\subsubsubsection{Three-body decays}
Direct CP violation searches in analyses of charm decays to three-body final states are
more complicated than two-body decays. Three methods have been used to search
for CP asymmetries. (1) Integrate over phase space and construct $A_{CP}$ as in two-body decays; (2) Examine CP asymmetry in the quasi-two-body resonances; (3) Perform a full Dalitz-plot 
analysis for $D$ and $\overline D$ separately. 
The Dalitz-plot analysis procedure~\cite{pdgdalitz} 
allows increased sensitivity to CP violation by probing
decay amplitudes rather than the decay rate. E791\cite{Aitala:1996sh}, 
FOCUS\cite{Link:2000aw} and BABAR\cite{Aubert:2005gj} have analyzed 
$D^+\to K^+K^-\pi^+$ using method (1). E791 and BABAR have also analyzed 
$D^+\to K^-K^+\pi^+$ using method (2). 
FOCUS has a Dalitz-plot analysis in progress\cite{Malvezzi:2002xt}. The $D^+\to K^+K^-\pi^+$ 
Dalitz plot is well described by eight quasi-two-body decay channels. 
A signature of CP violation in 
charm Dalitz-plot analyses is different amplitudes and phases for $D$ and $\overline D$ 
samples. 
No evidence for CP violation is observed.

The decay $D^{*+} \to D^0\pi^+$ enables the discrimination between $D^0$ and 
$\DZB$.
The CLEO collaboration has searched for CP violation integrated across the Dalitz plot
in $D^0 \to K^\mp\pi^\pm\pi^0$\cite{Kopp:2000gv,Brandenburg:2001ze}, $K^0_S\pi^+\pi^-$\cite{Asner:2003uz} and $\pi^+\pi^-\pi^0$\cite{Cronin-Hennessy:2005sy} decays.
No evidence of CP violation has been observed.

CLEO has considered CP violation more generally in a simultaneous fit to the $D^0\to K^0_S\pi^+\pi^-$ and $\DZB \to K^0_S\pi^+\pi^-$ Dalitz plots. 
The possibility of interference between CP--conserving and CP--violating
amplitudes provides a more sensitive probe of CP violation. The constraints
on the square of the CP--violating amplitude obtained in the resonant submodes 
of $D^0 \to K^0_S \pi^+ \pi^-$ range from $3.5 \times 10^{-4}$ to $28.4 \times 10^{-4}$ at 95\%
confidence level\cite{Asner:2003uz}.
\subsubsubsection{Four-body decays}
FOCUS has searched for T-violation using the four-body decay modes $D^0\to K^+K^-\pi^+\pi^-$
\cite{Link:2005th}. As described in Section~\ref{sec:cpvfs},
a T-odd correlation can be formed with the momenta, $C_T\equiv (\vec{p}_{K^+}.(\vec{p}_{\pi^+}\times \vec{p}_{\pi^-}))$. Under time-reversal, $C_T \to -C_T$, however $C_T\ne 0$ does not
establish T-violation. Since time reversal is implemented by an anti-unitary operator, $C_T\ne 0$, can be induced by FSI\cite{Bigi:2000yz}. This ambiguity can be resolved by measuring 
$\overline C_T \equiv (\vec{p}_{K^+}.(\vec{p}_{\pi^+}\times \vec{p}_{\pi^-}))$ in 
$\overline D^0\to K^+K^-\pi^+\pi^-$; $C_T\ne \overline C_T$ establishes T violation. FOCUS
reports a preliminary asymmetry $A_T = 0.075\pm 0.064$ from a sample of $\sim 400$ decays. More restrictive constraints are anticipated from CLEO-c where
in 281 pb$^{-1}$ a sample of 2300 $D^\pm \to K^0_S K^\pm \pi^+\pi^-$ have been
accumulated.


\subsubsection{Experiments exploiting quantum correlations}

Most high-statistics measurements of $D^0$ decay employ ``flavour tagging'' through the sign of the slow pion in $D^*\to\pi_{\rm slow}D$. That is, if combined with a slow $\pi^+$ to make a $D^{*^+}$, the neutral $D$ meson is a $D^0$. Conversely, a slow $\pi^-$ implies a $\DZB$.

An entirely different way to tag flavour, and \cp, is to exploit quantum correlations in $D^0\bar D^0$ production in $e^+e^-$ annihilation~\cite{Bigi:1989ah,Atwood:2002ak,Asner:2005wf}. 

The production process
$e^+ e^- \rightarrow \psi(3770) \rightarrow \DZ\DZB$
produces an eigenstate of $CP+$, in the first step, since the $\psi$(3770) has
$J^{PC}$ equal to $1^{--}$. Now consider the case where both the $D^0$ and
the $\bar{D^0}$ decay into CP eigenstates. Then the decays
$\psi(3770) \rightarrow f^i_+ f^j_+ ~~or~~ f^i_- f^j_-$
are forbidden, where $f_+$ denotes a $CP+$ eigenstate and $f_-$ denotes a $CP-$
eigenstate. This is because
$CP(f^i_{\pm} ~f^j_{\pm}) = (-1)^\ell = -1$
for the $\ell = 1 ~\psi$(3770).
Hence, if a final state such as ($K^+K^-$)($\pi^+\pi^-$) is observed, one
immediately has evidence of CP violation. Moreover, all $CP+$ and $CP-$ 
eigenstates
can be summed over for this measurement. The expected sensitivity to direct CP
violation is $\sim 1\%$.
This measurement can also be performed at higher energies where the final
state $D^{*0} \bar{D^{*0}}$ is produced. When either $D^*$ decays into a
$\pi^0$ and a $D^0$, the situation is the same as above. When the decay is 
$D^{*0} \rightarrow \gamma D^0$ the CP parity is changed by a multiplicative
factor of -1 and all decays $f^i_+ f^j_-$ violate CP~\cite{Bigi:2000yz}. Additionally, CP
asymmetries in CP even initial states depend linearly on $x$ allowing sensitivity to
CP violation in mixing of $\sim 3\%$.

For $e^+e^-$ machines running at the $\psi(3770)$, the $D$ mesons are produced with very little momentum in the laboratory. Hence, their flight distance is virtually impossible to determine, and we instead measure time-integrated decay rates. From Ref.~\cite{Asner:2005wf}
\begin{equation}
\Gamma(j,k)=Q_M\left|A(j,k)\right|^2+R_M\left|B(j,k)\right|^2
\label{eq:RateGeneral}
\end{equation}
where
\begin{displaymath}
A(j,k)\equiv A_j\bar A_k-\bar A_jA_k
\end{displaymath}
is the ``unmixed'' contribution to the decay rate, and
\begin{displaymath}
B(j,k)\equiv\frac{p}{q}A_jA_k-\frac{q}{p}\bar A_j\bar A_k
\end{displaymath}
is the contribution from $\DZ\!-\!\DZB$ mixing. The integrations also yield the factors
\begin{eqnarray*}
Q_M&=&\frac{1}{2}\left[\frac{1}{1-y^2}+\frac{1}{1+x^2}\right]\approx1-\frac{x^2-y^2}{2}\\
R_M&=&\frac{1}{2}\left[\frac{1}{1-y^2}-\frac{1}{1+x^2}\right]\approx\frac{x^2+y^2}{2}
\end{eqnarray*}
Mixing does not occur if the eigenstates of the decay Hamiltonian have the same mass and width, i.e. $x=y=0$. In any case, we expect $R_M\ll Q_M\approx1$. Nevertheless, mixing would result in the second term of Eq.~\ref{eq:RateGeneral} and it is here that one obtains sensitivity to \cp\ violation through $q\neq p$. This will be exploited at CLEO-c, and eventually to a greater extent at BES~III.


\subsubsection{Benchmarks for future searches}

Since the primary goal is to establish the intervention of New Physics,   
one `merely' needs a sensitivity level above the reach of the SM; `merely' does not mean 
it can easily be achieved. As far as {\em direct} \cp~violation is concerned -- 
in partial widths as well as in final state distributions -- this means asymmetries down to the 
$10^{-3}$ or even $10^{-4}$ level in  Cabibbo-allowed channels and 1\% level or better 
in twice Cabibbo-suppressed modes; in Cabibbo-once-suppressed decays one wants 
to reach the $10^{-3}$ range although CKM dynamics can produce effects of that order 
because future advances might sharpen the SM predictions -- and one will get them 
along with the other channels. For  
{\em time dependent} asymmetries in $D^0 \to K_S\pi^+\pi^-$, $K^+K^-$, $\pi^+\pi^-$ etc. 
and in $D^0 \to K^+\pi^-$ 
one should strive for the  
${\cal O}(10^{-4})$ and ${\cal O}(10^{-3})$ levels, respectively.





Statisticswise these are not utopian goals considering the very large event
samples foreseen at LHCb.



When probing asymmetries below the $\sim1\%$ level one has to struggle against systematic uncertainties, in particular since detectors are made from matter. There are three powerful weapons in this struggle: 
(i) 
Resolving the time evolution of asymmetries that are controlled by $x_D$ and $y_D$, which requires 
excellent microvertex detectors; (ii)  Dalitz plot consistency checks;  
(iii) quantum statistics constraints on distributions, \ot~odd moments etc. \cite{Asner:2005wf,Bigi:1989ah}

\subsubsection{Rare Decays} 

Searches for rare-decay processes have played an important role
in the development of the SM.
Flavour changing neutral current (FCNC) processes have been studied
extensively for $K$ and $B$ mesons in both $\ko\!-\!\ok$ and $\bo\!-\ob$ mixing and in rare FCNC decays. The corresponding processes in the charm sector has 
recieved less attention and the experimental upper limits are currently above
SM predictions. 
Short-distance FCNC processes in charm decays are much more highly
suppressed by the GIM mechanism than the corresponding down-type quark decays 
because of the large top quark mass. 

Observation of $D^+$ FCNC
decays $D^+, D_s^+ \rightarrow \pi^+ l^+l^-$ and 
$K^+ l^+l^-$ could therefore provide an indication of New Physics or of
unexpectedly large rates for long-distance SM processes like
$D^+ \rightarrow \pi^+ V$, $V \rightarrow l^+l^-$, with a real
or virtual vector meson $V$. 
Detailed
description on rare charm decays can be found in references~\cite{Burdman:2001tf,Burdman:2003rs}.
The charm meson radiative decays are also very important to
understand final state interaction which may enhance the decay rates. 
In Ref.~\cite{Burdman:2001tf,Burdman:2003rs}, the decay rates of $D \rightarrow V \gamma$ ($V$ can
be $\phi$, $\omega$, $\rho$ and $K^*$ ) had
been estimated to be $10^{-5} - 10^{-6}$, which can be reached at BES-III and
the $B$-factories. 

\subsubsection{ Inclusive $c \to u$ transitions}

The $s \to d$ and $b \to s$ transitions  offer a possibility to investigate effects of 
New Physics in the down-type quark sector. The $c \to u$ transition, however, gives a chance to 
study effects of New Physics in the up-type quark sector. 
In the Standard Model the contribution coming from the penguin diagrams in 
$c\to u\gamma$ transition is
strongly GIM suppressed giving a 
branching ratio of order $10^{-18}$ \cite{Burdman:1995te}. 
The QCD-corrected
effective Lagrangian gives $BR(c\to u\gamma)\simeq3\times10^{-8}$
\cite{Greub:1996wn,Ho-Kim:1999bs}. 
A variety of models beyond the standard model were
investigated and it was found that the gluino exchange diagrams
\cite{Prelovsek:2000xy} within general minimal supersymmetric SM (MSSM) might lead to  the
enhancement
\begin{equation}
\rm\frac{BR(c\to u\gamma)_{{MSSM}}}{BR(c\to u\gamma)_{{SM}}} 
\simeq10^2.
\label{1}
\end{equation}
Within SM the $c\to ul^+l^-$ amplitude is given by the $\gamma$ and $Z$ penguin
diagrams and $W$ box diagram at one-loop electroweak order in the
standard model. It is dominated by the light quark contributions in
the loop.  
The leading order rate for the inclusive $c\to u l^+l^-$ calculated within 
 SM \cite{Fajfer:2001hj}
was found to be suppressed by QCD corrections in  \cite{Burdman:2001tf}. 
The inclusion of the renormalization group equations  for the Wilson coefficients 
gave an additional significant 
suppression \cite{Fajfer:2002gp} leading to the rates  
$\Gamma(c\to ue^+e^-)/\Gamma_{D^0}=2.4\times 10^{-10}$ and
$\Gamma(c\to u\mu^+\mu^-)/\Gamma_{D^0}=0.5\times 10^{-10}$.   
These transitions are largely driven by virtual photon at low dilepton mass $m_{ll}$.

The leading 
contribution to $c\to ul^+l^-$ in general MSSM with the conserved R parity 
comes from one-loop diagrams with 
gluino and squarks in the loop \cite{Prelovsek:2000xy,Fajfer:2001hj,Burdman:2001tf}. 
It proceeds via virtual photon  
and significantly enhances the $c\to ul^+l^-$ 
spectrum at small dilepton mass $m_{ll}$. 
The authors of \cite{Burdman:2001tf} have investigated supersymmetric 
(SUSY) extension of the SM with R parity breaking and they 
found that it can modify the rate. Using the most recent CLEO \cite{He:2005iz} 
results for the $D^+ \to \pi^+ e^+ e^-$ one can set the bound for the product of the 
relevant parameters entering 
the R parity violating $\tilde \lambda'_{22k} \tilde \lambda'_{21k} \simeq 0.001 $ 
(assuming that the 
mass of squark $M_{\tilde D_k} \simeq 100$ GeV). This bound give the rates 
$BR_R(c\to ue^+e^-) \simeq1.6 \times 10^{-8}$ and  
$BR_R(c\to u \mu^+\mu^-) \simeq1.8 \times 10^{-8}$.

Recently, the effects of 
Littlest Higgs models were investigated in rare $D$ decays \cite{Fajfer:2005ke} and it was 
found that there is a new tree
level coupling in which gives a $c \to u Z$ transition. However, that effect is insignificant  
due to  the parameters constrained by the present electroweak data  (see Ref.~[25] in 
\cite{Fajfer:2005ke}). A number of models of New Physics contain an extra up-type heavy quark 
\cite{Fajfer:2006yc} 
causing the appearance of the flavour changing neutral currents at tree level for 
the up-quark sector. 
The Lagrangian which describes this FCNC  
interaction  is given by 
\begin{equation}
{\cal L}_{NC} = \frac{g}{\cos \theta_W} Z_\mu (J_{W^3}^\mu - 
\sin^2 \theta_W J_{EM}^\mu ),
\label{e1}
\end{equation}
where $J_{EM}^\mu$ is the same electromagnetic current as in the SM, while $J_{W^3}^\mu$
is given by 
\begin{equation}
J_{W^3}^\mu = \frac{1}{2} \bar U_L^m \gamma^\mu \Omega U_L^m -  
\frac{1}{2} \bar D_L^m \gamma^\mu  D_L^m
\label{e2}
\end{equation}
with $L=\tfrac{1}{2}(1- \gamma_5)$ and mass eigenstates 
$U_L^m= (u_L,c_L,t_L,T_L)^T$, $D_L^m=(d_L,s_L,b_L)^T$.
The neutral current for the down-type quarks is the same as in 
the SM,  
 while the up sector has additional currents (see ref. \cite{Fajfer:2005ke}).
The unitarity conditions of the CKM matrix might constrain this coupling. However, the present 
bound on $\Delta 
m $ in $ D^0 - \bar  D^0$ transition 
limits the parameter describing the $c u Z$ vertex to be  $\Omega_{uc} \simeq 0.004$, 
giving the more strict limit on that parameter. The invariant dilepton mass distribution of 
the $c \to u l^+l^-$  distribution is only  moderately enhanced. \\

\subsubsection{ Exclusive rare $D$ decays}

The study of exclusive $D$ meson rare decay modes is very difficult due to the 
dominance of the long-distance effects \cite{Burdman:2001tf,Bianco:2003vb,Burdman:2003rs,Burdman:1995te,Greub:1996wn,Ho-Kim:1999bs,Prelovsek:2000xy,Fajfer:2001hj,Fajfer:2002gp,He:2005iz,Aubert:2006ak,Fajfer:2005ke,Fajfer:2006yc,Prelovsek:2000rj,Fajfer:1998dv,Fajfer:2000zx,Fajfer:1998rz} .
The $D \to V \gamma$ decay rates were calculated in Refs.  
\cite{Burdman:2003rs,Prelovsek:2000rj,Fajfer:1998dv,Burdman:1995te}. 
The long-distance
contribution is induced by the effective nonleptonic $|\Delta c|=1$
weak Lagrangian. 
In calculations of Ref.~\cite{Fajfer:1998dv} the long-distance effects 
were determined  using a heavy meson chiral
Lagrangian.  The factorization approximation has been used
for the calculation of weak transition elements.  The results of Ref.~\cite{Burdman:1995te} obtained within a 
different framework are in very good agreement with the results of Ref.~\cite{Fajfer:1998dv}.
In Table~\ref{tab:raretab1}
the branching ratios of $D\to V\gamma$ decays \cite{Fajfer:1998dv} are given.
The uncertainty is due to relative unknown phases of various
contributions.
\begin{table}[htb]
\begin{center}
\caption{Predicted branching ratios for $D\to V\gamma$ decays.}
\label{tab:raretab1}
\begin{tabular}{|l|l|}
\hline
$D\to V\gamma$ & $BR$ \\
\hline
$D^0 \to{\bar K}^{*0}\gamma$ &$[6-36]\times10^{-5}$ \\
$D_s^+\to\rho^+\gamma$ &$[20-80]\times10^{-5}$ \\
$D^0\to\rho^{0}\gamma$&$[0.1-1]\times10^{-5}$ \\
$D^0\to\omega\gamma$ &$[0.1-0.9]\times10^{-5}$  \\  
$D^0 \to \phi \gamma$ &$ [0.4 - 1.9 ]\times 10^{-5} $ \\
$D^+ \to \rho^+ \gamma$ &$ [0.4 -6.3]\times 10^{-5}$\\
$D_s^+ \to K^{*+ }\gamma$ &$[1.2 - 5.1]\times 10^{-5}$ \\
$D^+ \to K^{*+} \gamma$ &$ [0.3- 4.4]\times 10^{-6}$ \\
$D^0 \to K^{*0} \gamma$ &$ [0.3 - 2.0] \times 10^{-6}$   \\ 
\hline
\end{tabular}
\end{center}
\end{table}
Although the branching ratios are dominated by the long-distance
contributions, the size of the short-distance contribution can be
extracted from the difference of the decay widths
$\rm\Gamma(D^0\to\rho^{0}\gamma)$ and $\rm\Gamma(D^0\to\omega\gamma)$
\cite{Fajfer:2000zx}. Namely, the long-distance mechanism $ c\bar u\to d\bar
d\gamma$ screens the $ c\bar u\to u\bar u\gamma$ transition in $
D^0\to\rho^{0}\gamma$ and $D^0\to\omega\gamma$, the $\rho^{0}$ and
$\omega$ mesons being mixtures of $ u\bar u$ and $ d\bar d$.
Fortunately, the LD contributions are mostly cancelled in the ratio
\begin{align}
R&=\frac{BR({D}^0\to\rho^{0}\gamma)-BR({D}^0\to\omega\gamma)}
{BR({D}^0\to\omega\gamma)} \propto\operatorname{Re}\frac{A({D^0\to u\bar u\gamma})}
{A({D^0\to d\bar d\gamma})},
\label{2}
\end{align}
which is proportional to the SD amplitude $A({D^0\to u\bar
u\gamma})$ driven by $ c\to u\gamma$. This ratio is
$R_{\text{SM}}=(6\pm15)\%$ in Ref.\ \cite{Fajfer:2000zx}, and can be enhanced
up to ${\cal O}(1)$ in the MSSM. In addition to the $ c\to u\gamma$ 
searches in the charm meson decays,  in Ref. \cite{Fajfer:2001hj} it was 
suggested to search for
this transition in the decay $B_c\to B_u^*\gamma$, where
the long distance contribution is much smaller.

The inclusive $c \to u l^+ l^-$ process can be tested in the rare decays $D \to \mu^+ \mu^-$, 
$D \to P (V) l^+ l^-$ \cite{Burdman:2001tf,Fajfer:2001hj,Fajfer:2002gp,Burdman:2003rs}.
The branching ratio for the rare decay $D\to \mu^+ \mu^-$ is very small in the SM. 
The detailed treatment of this decay rate \cite{Burdman:2001tf} 
gives $Br(D \to \mu^+ \mu^-) \simeq 3\times 10^{-13}$ \cite{Burdman:2001tf}. This decay rate 
can be enhanced within a study which considers 
 SUSY with R-parity breaking effects \cite{Burdman:2001tf,Bianco:2003vb}. 
Using the bound $\tilde \lambda'_{22k} \tilde \lambda'_{21k} \simeq 0.001 $ 
one obtains the limit $Br(D \to \mu^+ \mu^-)_R\simeq 4\times 10^{-7}$,
a value which would be accessible at LHCb~\cite{LHCBCHARM}.
The $D \to P (V) l^+ l^-$ decays offer another possibility to study the $c \to u l^+ l^-$ transition in charm sector. 
The  $D^+\to \pi^+ l^+l^-$ and  $D^0\to \rho^0e^+e^-$ decay modes 
are simplest to be accessed by experiment \cite{Fajfer:2005ke}. 
The effects of SUSY with R parity violation were studied in \cite{Burdman:2001tf}.  
The recent 
experimental results of \cite{He:2005iz} restrict the R parity violating parameters  
found in \cite{Burdman:2001tf} more than one order of magnitude. 
 
\begin{figure}[t]
\includegraphics[width=7.25cm]{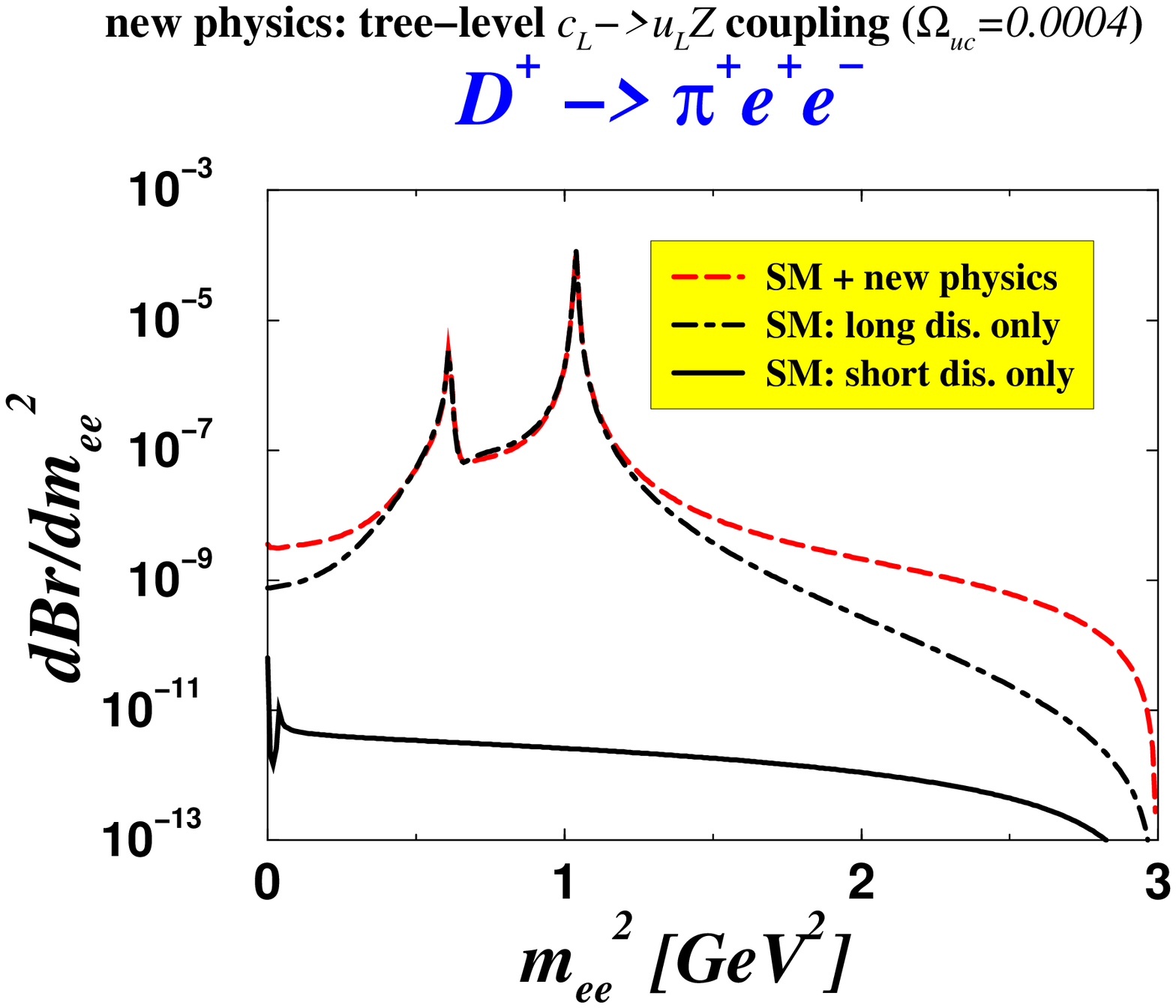}\hspace{.2cm}\includegraphics[width=7.25cm]{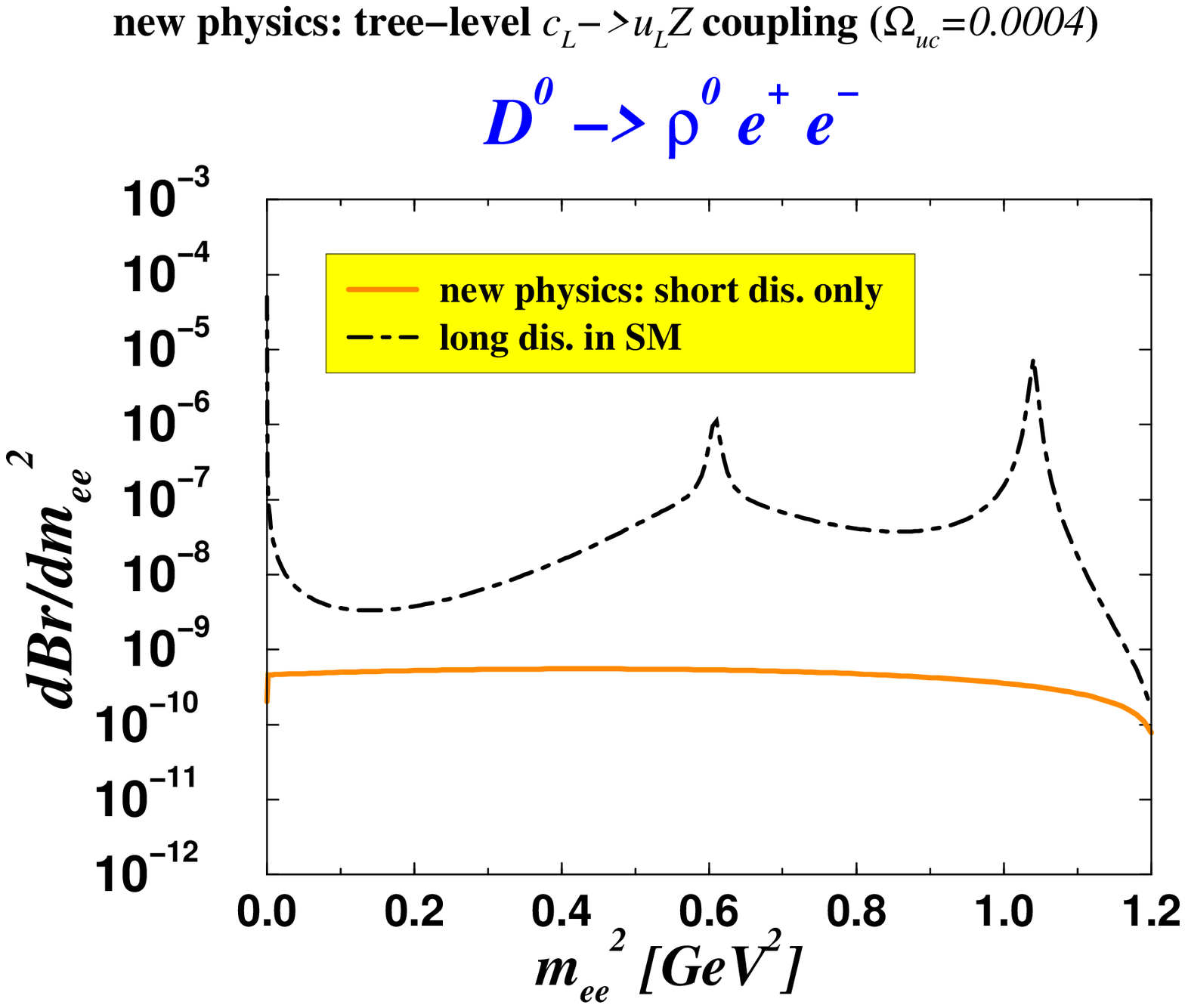}
\label{fig:brvsm2ll}
\caption{ \it (left) The dilepton mass distribution 
 $dBr/dm_{ee}^2$ for the decay 
$D^+\to \pi^+e^+e^-$   
as a function of the  dilepton mass square $m_{ee}^2=(p_++p_-)^2$.
(right) The figure shows the dilepton mass distribution for 
$D^0\to \rho^0e^+e^-$.}
\end{figure}

\begin{figure}[htb]
\begin{center}
\includegraphics[width=7.25cm]{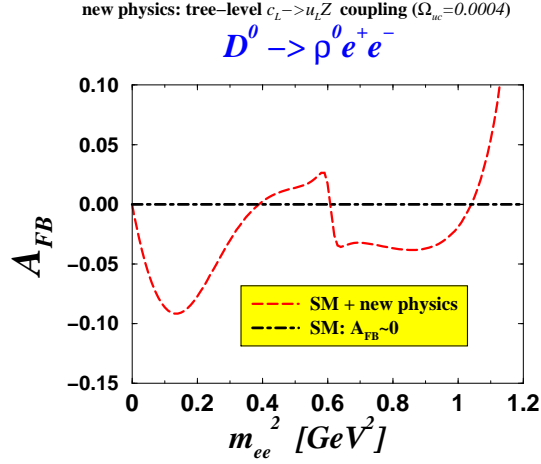}
\end{center}
\label{fig:Afbvsm2ll}
\caption{ \it  
 The figure  shows the forward-backward 
asymmetry for $D^0\to \rho^0e^+e^-$.}
\end{figure}

The most appropriate decay modes for the experimental searches of the 
New Physics coming from 
the FCNC tree level current are $D^+\to \pi^+ l^+l^-$ and  $D^0\to \rho^0e^+e^-$. 
The total rate for $D \to X l^+ l^-$ is dominated by the 
long-distance resonant contributions at dilepton mass 
$m_{ll}=m_\rho,~m_\omega,~m_\phi$ 
and even the largest contributions from New Physics are not expected to 
affect the total rate significantly \cite{Burdman:2001tf, Fajfer:2001hj}. 
New Physics could only modify the SM 
differential spectrum at low $m_{ll}$ below 
$\rho$ or the spectrum at high $m_{ll}$ above $\phi$. 
 In the case of $D\to\pi l^+l^-$ differential decay distribution there is a broad  
 region at high $m_{ll}$ (see Fig. \ref{fig:brvsm2ll}), 
 which presents an unique possibility to 
study the $c\to ul^+l^-$ transition \cite{Fajfer:2001hj,Fajfer:2005ke}.
\begin{table}[htb]
\begin{center}
\caption{\small Branching ratios for the decays probing the $c\to ul^+l^-$ transition}\label{tab:raretab2}
\begin{tabular}{|c|c|c|c|c|c|c|}
\hline
 {\bf Br} & \multicolumn{2}{c|}{short distance } & total rate $\simeq$ & experiment\\
 & \multicolumn{2}{c|}{contribution only } & long distance contr. & \\  

\hline 
 & SM & SM + NP &    &   \\
\hline
$D^+\to \pi^+ e^+e^-$ & $6\times 10^{-12}$ & $8 \times 10^{-9}$ & $1.9\times 10^{-6}$ & $<7.4\times 10^{-6}$\\
$D^+\to \pi^+ \mu^+\mu^-$ &  $6\times 10^{-12}$ & $8 \times 10^{-9}$ & 
$1.9\times 10^{-6}$ & $<8.8\times 10^{-6}$\\
\hline
$D^0\to \rho^0 e^+e^-$ & negligible &$5\times 10^{-10}$ & $1.6\times 10^{-7}$
&$<1.0\times 10^{-4}$\\
$D^0\to \rho^0 \mu^+\mu^-$ & negligible &$5\times 10^{-10}$ & $1.5\times 10^{-7}$ & $<2.2\times 10^{-5}$\\
\hline
\end{tabular}
\end{center}
\end{table} 
In Table~\ref{tab:raretab2} we present branching ratios for the $D^+\to \pi^+ e^+e^-$ and $D^0 \to \rho^0 l^+ l^-$ , giving the 
SM short-distance, long-distance contributions, as well as the effects of NP arising from the existence of one extra up-type quark. The total rates in Standard and New Physics 
models are completely dominated by the resonant 
long-distance contribution $D\to XV_0\to Xl^+l^-$ \cite{Burdman:2001tf,Fajfer:2005ke}. 
The SM short-distance contribution for $D^0\to \rho^0l^+l^-$ (see Fig.~\ref{fig:brvsm2ll}) is not shown since it is 
completely negligible in comparison to the long-distance contribution. 
The forward-backward asymmetry for  $D^0 \to \rho^0 l^+ l^-$ vanishes in 
SM, while it is reaching $0.05$ in a NP model with extra up-type quark as given in Fig.~\ref{fig:Afbvsm2ll}. 
Such an asymmetry is still small and it will be difficult to 
observed in present or planned experiments given that the rate itself is already 
small.


\subsubsection{Experimental Results}

There are a large number of FCNC charm decays including radiative, 
fully leptonic decays,
lepton flavour violating (LFV) and lepton number violating (LNV),  that have
been measured experimentally.

Belle has reported the observation of the decay $D^{0} \to \phi \gamma$ 
 This is the first observation of a flavor-changing radiative decay of a charmed meson. The Cabibbo- and colour-suppressed decays $D^0 \to \phi \pi^0$, 
$\phi \eta$ are also observed for the first time. The branching fractions are
${\cal B}(D^{0} \to \phi \gamma) 
= [ 2.60^{+0.70}_{-0.61}\!^{+0.15}_{-0.17}] \times 10^{-5}$ - somewhat higher than predicted in Table~\ref{tab:raretab1}, 
 ${\cal B}(D^{0} \to \phi \pi^{0}) 
= [ 8.01 \pm 0.26  \pm 0.47] \times 10^{-4}$, and 
${\cal B}(D^{0} \to \phi \eta) = [ 1.48 \pm 0.47  \pm 0.09 ] \times 10^{-4}$.

Recently, CLEO-c reported the branching
fraction of the resonant decay ${\cal BR}(D^+ \rightarrow \pi^+
\phi \rightarrow \pi^+ e^+e^-) = (2.8 \pm 1.9 \pm 0.2 ) \times 10^{-6}$\cite{He:2005iz}.
The lepton-number-violating (LNV) or lepton-flavour-violating (LFV) decays
$D^+ \rightarrow \pi^- l^+l^+$, $K^- l^+
l^+$ and $\pi^+ \mu^+ e^-$ are forbidden in the SM. Past searches have set upper
limits for the dielectron and dimuon decay modes~\cite{Yao:2006px}. 

The BABAR collaboration has recently reported on FCNC decays of the form
$D^+/D^+_s/\Lambda_c^+ \to \pi^+/K^+/p^+ \ell^+\ell^{\prime -}$, 
where the two leptons, $\ell^+$ and
$\ell^{\prime -}$, can each be either an electron or a muon. Upper limits are
set at the 90\% confidence level between $4\times 10^{-6}$ and 
$40\times 10^{-6}$ on the SM and LFV processes\cite{Aubert:2006ak}.

In Table~\ref{tab:rare}, the current
limits and expected sensitivities at BES-III are
summarized for $D^+$ and $D^0$, respectively.

\begin{table}[htbp]
 \centering
\caption{\it Current and projected 90\%-CL upper limits on
rare $D^+$ and $D^0$ decay modes at BES-III with 20 fb$^{-1}$ data at $\psi(3770)$
peak. 
}
\begin{tabular}{|llll|llll|}
\hline
                & Reference     & Best Upper  &  BES-III           &                                     & Reference     & Best Upper  &  BES-III  \\	      
Mode            & Experiment              & limits ($10^{-6}$)      & ($\times 10^{-6}$) & Mode            & Experiment              & limits ($10^{-6}$) & ($\times 10^{-6}$)       \\
\hline											
\multicolumn{4}{|c|}{$D^+$} & \multicolumn{4}{c|}{$D^0$} \\
$ \pi^+ e^+e^-$     & CLEO-c~\cite{He:2005iz} & 7.4 & 0.03 &                               $\gamma \gamma $  & CLEO~\cite{Coan:2002te} & 28 & 0.05  \\	      
$ \pi^+ \mu^+\mu^-$ & FOCUS~\cite{Link:2003qp}  & 8.8 & 0.03 &                             $\mu^+\mu^-$ & D0~\cite{Korn:2003pt}  & 2.4 & 0.03  \\		      
$ \pi^+ \mu^\pm e^\mp$ & BABAR~\cite{Aubert:2006ak}  & 5.9/10.8 & 0.03 &                              $\mu^+ e^-$ & E791~\cite{Aitala:1999db}  & 8.1 & 0.03  \\      	            
$ \pi^- e^+ e^+$ & CLEO-c~\cite{He:2005iz}  & 3.6 & 0.03 &                                 $e^+ e^-$ & E791~\cite{Aitala:1999db}  & 6.2 & 0.03  \\      	            
$ \pi^- \mu^+\mu^+$ & FOCUS~\cite{Link:2003qp}  & 4.8 & 0.03 &                             $\pi^0 \mu^+\mu^-$ & E653~\cite{Kodama:1995ia}  & 180 & 0.05  \\      
$ \pi^- \mu^+ e^+$ & E791~\cite{Aitala:1999db}  & 50 & 0.03 &                              $\pi^0 \mu^+ e^+$ & CLEO~\cite{Freyberger:1996it}  & 86 & 0.05  \\          
$ K^+ e^+e^- $ & CLEO-c~\cite{He:2005iz}  & 6.2 & 0.03 &                                   $\pi^0 e^+e^- $ & CLEO~\cite{Freyberger:1996it}  & 45 & 0.05  \\          
$ K^+ \mu^+\mu^- $ & FOCUS~\cite{Link:2003qp}  & 9.2 & 0.03 &                              $K_S \mu^+\mu^- $ & E653~\cite{Kodama:1995ia}  & 260 & 0.1  \\             
$ K^+ \mu^\pm e^\mp $ & BABAR~\cite{Aubert:2006ak}  & 5.9/5.7 & 0.03 &                               $K_S \mu^+ e^- $ & CLEO~\cite{Freyberger:1996it}  & 100 & 0.1  \\         
$ K^- e^+ e^+ $ & CLEO-c~\cite{He:2005iz}  & 4.5 & 0.03 &                                  $K_S e^+ e^- $ & CLEO~\cite{Freyberger:1996it}  & 110 & 0.1  \\	      
$ K^- \mu^+ \mu^+ $ & FOCUS~\cite{Link:2003qp}  & 13 & 0.03 &                              $\eta \mu^+ \mu^- $ & CLEO~\cite{Freyberger:1996it}  & 530 & 0.1  \\  
$ K^- \mu^+ e^+ $ & E687~\cite{Frabetti:1997wp}  & 130 & 0.03 &                            $\eta \mu^+ e^- $ & CLEO~\cite{Freyberger:1996it}  & 100 & 0.1  \\    
& & & &   										$\eta e^+ e^- $ & CLEO~\cite{Freyberger:1996it}  & 110 & 0.1  \\     \hline
\end{tabular} 
\label{tab:rare}
\end{table}

\subsubsection{Precision CKM Physics}


\def\calB{{\cal B}}

Precision measurements of the CKM matrix continue to be of great interest,
despite impressive strides in determining its parameters
\cite{Charles:2004jd,Bona:2005eu,Bona:2005vz,Bona:2006ah,Bona:2006sa,Bona:2007vi,Bona:2007qt}.  We first give an overview of ways in which studies of
charm can help this effort.  More details on some aspects are given in
subsequent subsections.

In section \ref{sec:dir} we discuss direct measurements of the CKM elements
governing $c \to d$ and $c \to s$ transitions.  We then turn in section
\ref{sec:ind} to ways in which charm can be of help in determining the
remaining elements.  An elementary constraint on new physics is discussed in
section \ref{sec:new}, while section \ref{sec:sum} summarizes.

\subsubsection{Direct determinations \label{sec:dir}}


\subsubsubsection{$V_{ud}$, $V_{us}$, and unitarity}

The parameter $V_{us} = \lambda$ is measured (with some recent contributions
playing a key role) to be $0.2257 \pm 0.0021$ \cite{Yao:2006px}.  To sufficient
accuracy, one then expects $V_{ud} = \sqrt{1 - |V_{us}|^2} = 0.9742 \pm
0.0005$, since $|V_{ub}| \simeq 0.004$ and hence its square can be neglected in
the unitarity relation $|V_{ud}|^2 + |V_{us}|^2 + |V_{ub}|^2 = 1$.  The
experimental value for $V_{ud}$, based primarily upon comparing beta-decays of
certain nuclei to muon decays, is $V_{ud} = 0.97377 \pm 0.00027$, so unitarity
is adequately satisfied for the first row.

\subsubsubsection{$V_{cd}$}

For the first column, one expects $|V_{ud}|^2 + |V_{cd}|^2 + |V_{td}|^2 = 1$.
With the value of $V_{ud}$ quoted above and $|V_{td}| \simeq 0.008$, one 
then expects $|V_{cd}| = 0.227 \pm 0.001$.  This is to be compared with the
value $0.230 \pm 0.011$ obtained from neutrino interactions \cite{Yao:2006px} and
$0.213 \pm 0.008 \pm 0.021$ from charm semileptonic decays
\cite{Artuso:2005jw}.  The first error is experimental and the second is
associated with uncertainty in the form factor.  Measurements of the branching fractions for $D\to\pi\ell\nu$ decay are improving somewhat (Sec.~\ref{sec:charmsl}) so the precision of $|V_{cd}|$ from this source will improve.   
However, from the current uncertainties in $\calB(D\to\pi\ell\nu)$
it is clear that one will not be able to match the precision of the unitarity
test for the first row of the CKM matrix anytime soon.  {\it Given} CKM unitarity, which
says to sufficient accuracy that we should expect the value of $|V_{cd}|$
mentioned above, one can use it to constrain form factors in semileptonic charm
decays and compare them with lattice QCD calculations.

\subsubsubsection{$V_{cs}$}

A similar philosophy applies to the CKM element $V_{cs}$.  Unitarity applied
to the second column of the CKM matrix implies $|V_{cs}| = \sqrt{1 - |V_{us}|^2
- |V_{ts}|^2}$.  Taking the experimental value of $V_{us}$ mentioned above and
the unitarity-based estimate $V_{ts} \simeq - V_{cb}$, we estimate $|V_{cs}|
= 0.9733 \pm 0.0006$.  This precision will not be matched by experiment soon.
The best measurements come from semileptonic charm decays and yield $|V_{cs}| =
0.957 \pm 0.017 \pm 0.093$, with the second error coming from uncertainty in
the form factor.  Again, assuming unitarity one will be able to subject
lattice gauge theory predictions to important tests.

\subsubsection{Indirect tests \label{sec:ind}}

\subsubsubsection{$V_{ub}$}
\label{sec:Vub}
The primary difficulty in measuring the matrix element $V_{ub}$ is that it must
be extracted from $b$ semileptonic decays which proceed to charm all but 2\% of
the time.  Inclusive methods must rely on kinematic separation techniques, the
oldest of which is the study of leptons with energies beyond the endpoint for
$b \to c \ell \nu$.  Exclusive decays such as $B \to \pi \ell \nu$ and $B \to
\rho \ell \nu$ do not share this problem, but one must understand the
corresponding form factors.  Tests of form factors in {\it charm} decays  
predicted by lattice gauge theories can help validate predictions for $B$
decays.

The phase of $V^*_{ub}$ ($\gamma$ or $\phi_3$ in the standard
parametrisations) can be measured in several ways with the help of information
from charm decays.  These help, for example, in using decays such as $B \to
D_{\rm CP} K$ decays to learn $\gamma$.  For $D$ modes such as $K_S \pi^+
\pi^-$, $\pi^+ \pi^- \pi^0$, $K^+ K^- \pi^0$, and $K_S K^\pm \pi^\mp$, Dalitz
plots yield information on CP-eigenstate and flavour-eigenstate modes and their
relative phases \cite{Asner:2003gh}.

The interference of $b \to c \bar u s$ (real) and $b \to u \bar c s ~(\sim \!
e^{-i \gamma})$ subprocesses in $B^- \to D^0 K^-$ and $B^- \to \overline{D}^0
K^-$, respectively, is sensitive to the weak phase $\gamma$.  This interference
may be probed by studying common decay products of $D^0$ and $\overline{D}^0$
into neutral $D$ CP eigenstates or into doubly-Cabibbo-suppressed modes
\cite{Bigi:1988ym,Gronau:1990ra,Gronau:1991dp,Atwood:2000ck,Atwood:2003mj}.

As one example, the decays $B^\pm \to K^\pm (K^{*+} K^-)_D$ and $B^\pm \to
K^\pm (K^{*-} K^+)_D$ provide information on $\gamma$ if the relative (strong)
phase between $D^0 \to K^{*+} K^-$ and $D^0 \to K^{*-} K^+$ is known
\cite{Grossman:2002aq}.  One can learn this relative phase from the study of
$D^0 \to K^+ K^- \pi^0$ since both final states occur and interfere with one
another where $K^{*+}$ and $K^{*-}$ bands cross on the Dalitz
plot \cite{Rosner:2003yk}.  This method was used recently by the CLEO
Collaboration \cite{Cawlfield:2006hm} to show that this interference was
predominantly destructive in the overlap region.

As another example, one can determine $\gamma$ using $B^\pm \to D K^\pm$
followed by $D \to K_S \pi^+ \pi^-$, $K_S K^+ K^-,~K_S \pi^+ \pi^- \pi^0$
\cite{Bondar:2002,Giri:2003ty}.  Recent high-statistics studies have been performed by
BaBar \cite{Aubert:2005iz} and Belle \cite{Poluektov:2006ia}.
The precision of these measurements will 
eventually be limited by the understanding of the $D \to K^0_S\pi^+\pi^-$ 
Dalitz plot. K-matrix descriptions of the $\pi\pi$ S-wave may yield improved
models of charm Dalitz plots and these models will be tested using the $CP$ 
tagged sample of charm decays at \hbox{CLEO-c} and later at BES-III. The model 
uncertainty, which is currently $\pm 10^\circ$, may be reduced to a few degrees.

Model independent methods\cite{Bondar:2005ki,Bondar:2006} use $CP$ tagged $K^0_S\pi^+\pi^-$ and $D\bar D \to
(K^0_S\pi^+\pi^-)^2$ to control the Dalitz plot model uncertainty. Analyses 
underway at CLEO-c are expected to control this systematic uncertainty on $\gamma/\phi_3$ to a few degrees.

\subsubsubsection{$V_{cb}$}

The semileptonic decays of $B$ mesons to $D$ or $D^*$ mesons are one source
of information about the element $V_{cb}$, but one must understand form
factors satisfactorily.  Lattice gauge theories make predictions for such
form factors; the validation of lattice form factor predictions in charm
decays again is a key ingredient in establishing credibility of the $B \to
D^{(*)}$ form factor predictions.  Moreover, under some circumstances it is
helpful to have precise information about $D$ branching ratios to specific
final states, which detailed charm studies can provide.

\subsubsubsection{$V_{td}$ {\rm and} $|V_{td}/V_{ts}|$}

The mixing of $\bo$ and $\ob$ is governed primarily by the CKM product
$|V^*_{tb}V_{td}|$.  If unitarity is assumed, $|V_{tb}|$ is very close to 1,
so the dominant CKM source of uncertainty is $|V_{td}|$.  However, the
matrix element of the short-distance operator inducing the $b \bar d \to
d \bar b$ transition contains an unknown factor $f_B^2 B_B$, where $f_B$ is
the $B$ {\it meson decay constant}, while $B_B = {\cal O}(1)$ is known as the
``bag constant'' or ``vacuum saturation factor'' and expresses the degree to
which the vacuum intermediate state dominates the transition.  The
corresponding mixing of strange $B$'s and their antiparticles is governed
by $|V^*_{tb}V_{ts}|$ and $f_{B_s}^2 B_{B_s}$.  

Lattice gauge theories predict not only $f_B$ and $f_{B_s}$ (as well as the
constants $B_B$ and $B_{B_s}$), but also the decay constants $f_D$ and
$f_{D_s}$ for charmed mesons.  Thus, the study of charmed meson decay
constants (Sec.~\ref{sec:leptonic}) and their ratios, and
comparison with lattice predictions, can shed indirect light on quantities
of interest in determining the CKM matrix elements $V_{td}$ and $V_{ts}$.

To give one example of the role charm measurements can play, it is expected on
rather general grounds \cite{Grinstein:1993ys} 
that $f_{B_s}/f_B$ and $f_{D_s}/f_D$ are
equal to within a few percent.  Now, the ratio $f_{B_s}/f_B$ is a key
ingredient in the extraction of $|V_{td}/V_{ts}|$ from measurements of
$\BZ$--$\BZB$ and $\BZS$--$\BZBS$ mixing.  The determination of Ref.\
\cite{Abulencia:2006mq} 
utilized an estimate $(f_{B_s} \sqrt{B_{B_s}}/f_B \sqrt{B_B}) =
1.21^{+0.047}_{-0.035}$ from the lattice \cite{Okamoto:2005zg}.  With a sufficiently
good measurement of $f_{D_s}/f_D$ and the theoretical input (again, from the
lattice) that $B_{B_s}/B_B \simeq 1$, one could check the lattice prediction
or simply substitute an experimental measurement for it.

\subsubsection{New physics constraint \label{sec:new}}

To see how great an impact even modest improvements in testing CKM unitarity
in the charm sector would have, we consider a model in which a fourth
family $(t',b')$ of quarks is added to the usual three, with neutrinos
heavy enough to evade the constraint $N_\nu = 3$ due to invisible $Z$ decays.
Unitarity relations involving the first two rows and columns of the expanded
$4 \times 4$ CKM matrix allow us to calculate the following 90\% c.l. upper
limits using the best-measured quantities mentioned above:
\bea
|V_{ub'}| & = & \sqrt{1 - |V_{ud}|^2 - |V_{us}|^2 - |V_{ub}|^2} \le 0.05~~,\\
|V_{cb'}| & = & \sqrt{1 - |V_{cd}|^2 - |V_{cs}|^2 - |V_{cb}|^2} \le 0.5~~,\\
|V_{t'd}| & = & \sqrt{1 - |V_{ud}|^2 - |V_{cd}|^2 - |V_{td}|^2} \le 0.07~~,\\
|V_{t's}| & = & \sqrt{1 - |V_{us}|^2 - |V_{cs}|^2 - |V_{ts}|^2} \le 0.5~~.\\
\eea
The poor quality of the bounds on $|V_{cb'}|$ and $|V_{t's}|$ is largely due
to the 10\% error on $|V_{cs}|$ which translates to errors of 0.18 on
$|V_{cb}|^2$ and $|V_{td}|^2$ and 90\% c.l. upper limits on them of about
1/4.  Thus improved measurements of $V_{cs}$ could have a great impact on
closing a rather gaping window for new physics or even revealing it.

\subsubsection{Summary of overview \label{sec:sum}}

The above examples show that charmed particle studies have a large role to play
in precision CKM physics, affecting nearly all the elements of the CKM matrix.
In turn, precision CKM physics is important as a clue to the very origin of
quark masses, since the CKM matrix arises from the same physics which generates
those masses.


\subsubsubsection{Leptonic Decays} 


\label{sec:leptonic}

\begin{figure}[ht]
\begin{center}
\includegraphics[width=6cm]{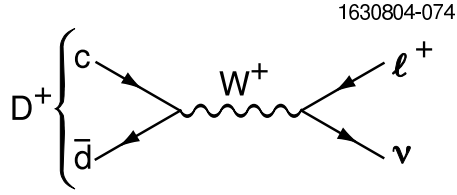}
\caption{The decay diagram for $D^+\to \ell^+\nu$.} \label{Dptomunu}
\end{center}
\end{figure}

Purely leptonic decays of charm mesons are of prime importance for
checks of theoretical QCD calculations and searches for New Physics.
Extraction of precise CKM information from neutral $B$ mixing
requires precision knowledge of the ratio of decay constants for
$B_s$ and $B^0$ \cite{Buchalla:1995vs}. While QCD calculations provide
this estimate, the uncertainties are large and the methods need to
checked by seeing if they can reproduce charm measurements. Leptonic
decays proceed in the Standard Model by annihilation of the
charm quark and spectator antiquark into a virtual $W^+$, that
transforms to a lepton-antineutrino pair as shown for the $D^+$
meson in Fig.~\ref{Dptomunu}.

In the SM the decay width is given by \cite{Silverman:1988gc}:
\begin{equation}
\Gamma(D^+\to \ell^+\nu) = {{G_F^2}\over
8\pi}f_{D^+}^2m_{\ell}^2M_{D^+} \left(1-{m_{\ell}^2\over
M_{D^+}^2}\right)^2 \left|V_{cd}\right|^2~~~, \label{eq:equ_rate}
\end{equation}
where $M_{D^+}$ is the $D^+$ mass, $m_{\ell}$ is the mass of the
final state lepton, $|V_{cd}|$ is a CKM matrix element assumed to be
equal to $|V_{us}|$, and $G_F$ is the Fermi coupling constant. (The
same formula applies to $D_s^+\to\ell^+\nu$ decays with the
replacement of $D_s^+$ mass and $|V_{cs}|$.)

New Physics can affect the expected widths; any undiscovered charged
bosons would interfere with the SM $W^+$. These effects may be
difficult to ascertain, since they would simply change the value of
the $f_i$'s. The ratio $f_{D_s^+}/f_{D^+}$ is much better predicted
in the SM than the values individually, so deviations see here could
point to beyond the SM charged bosons. For example, Akeroyd predicts
that the presence of a charged Higgs boson would suppress this ratio
significantly \cite{Akeroyd:2003jb}.

We can also measure the ratio of decay rates to different leptons,
and the predictions then are fixed only by well-known masses. For
example, for $\tau^+\nu$ to $\mu^+\nu$:

\begin{equation}
R\equiv \frac{\Gamma(D^+\to \tau^+\nu)}{\Gamma(D^+\to \mu^+\nu)}=
{{m_{\tau^+}^2 \left(1-{m_{\tau^+}^2\over
M_{D^+}^2}\right)^2}\over{m_{\mu^+}^2 \left(1-{m_{\mu^+}^2\over
M_{D^+}^2}\right)^2}}~~. \label{eq:rat}
\end{equation}

Any deviation from this formula would be a manifestation of physics
beyond the Standard Model. This could occur if any other charged
intermediate boson existed that coupled to leptons differently than
mass-squared. Then the couplings would be different for muons and
$\tau$'s. This would be a manifest violation of lepton universality,
which has identical couplings of the muon, the tau, and the electron
to the gauge bosons ($\gamma,~Z^0$ and $W^{\pm}$)
\cite{Hewett:1995aw}. (We note that in some models of supersymmetry
the charged Higgs boson couples as mass-squared to the leptons and
therefore its presence would not cause a deviation from
Eq.~\ref{eq:rat} \cite{Hou:1992sy}.)

\begin{figure}[htb]
\centerline{
\hfill\includegraphics[width=6.25cm]{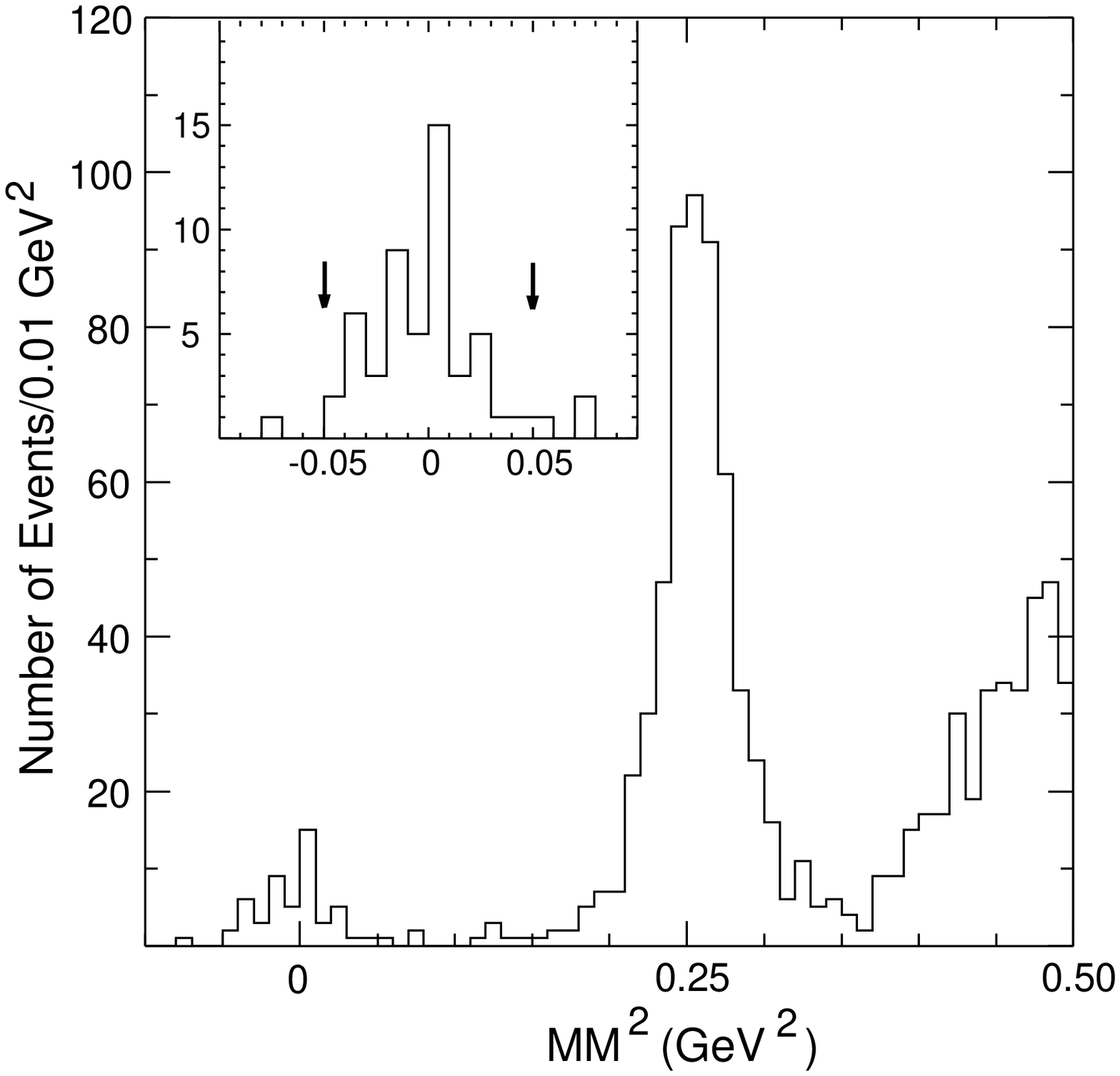}\hfill\includegraphics[width=6.25cm]{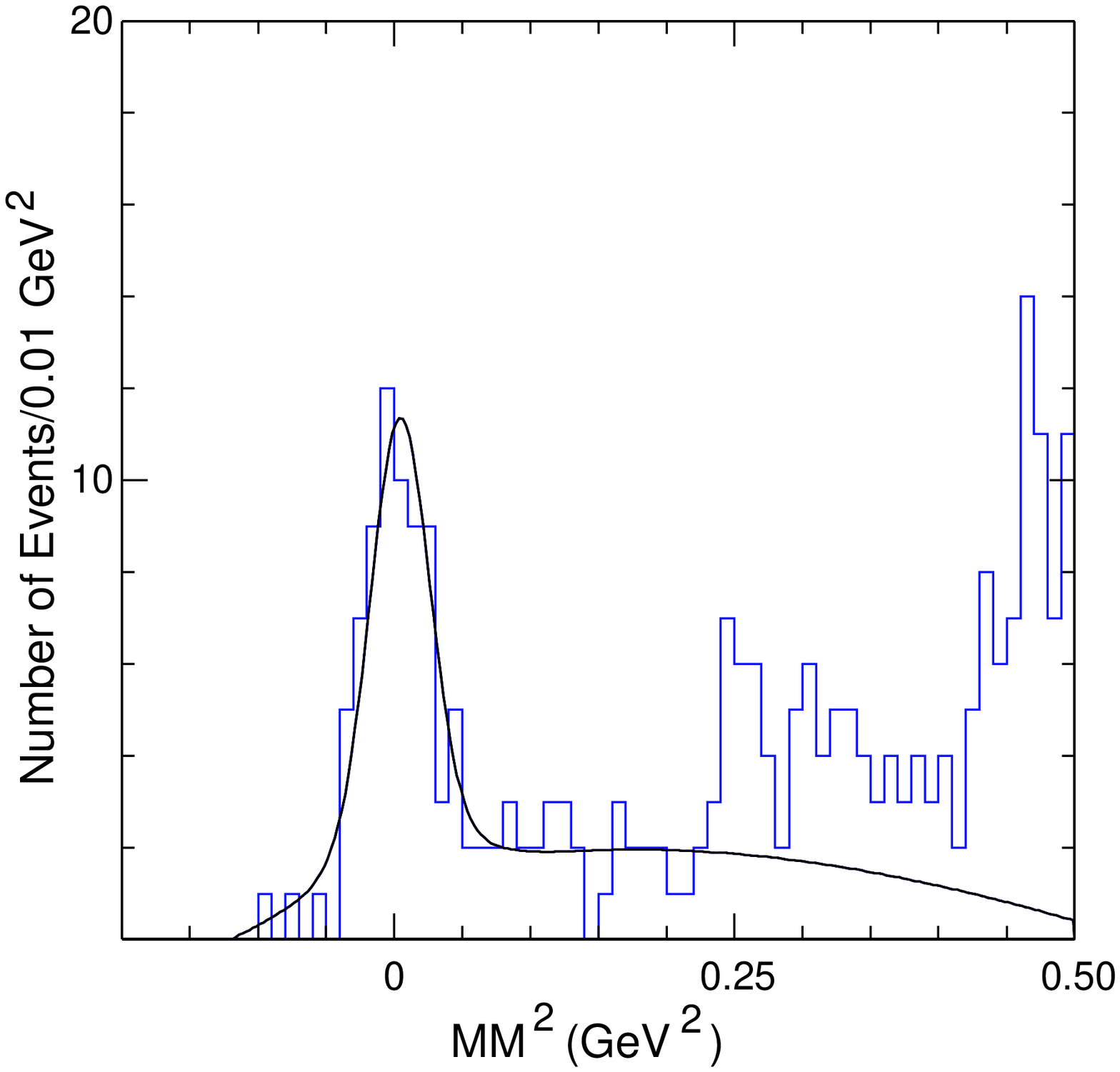}\hfill
} \caption{CLEO-c missing mass-squared distributions. (left) Using
$D^-$ tags and one additional opposite sign charged track depositing
$<$ 300 MeV (consistent with a muon) in the calorimeter and no extra
energetic clusters. The insert shows the signal region for
$D^+\to\mu^+\nu$ enlarged; the defined signal region is shown
between the two arrows. (right) Using $D_s^-$ tags but allowing any
energy deposit in the calorimeter (consistent with muon or pion).
The curve is the predicted shape for the sum $D_s^+\to \mu^+\nu$ +
$D_s^+\to \tau^+\nu$, $\tau^+\to \pi^+\nu$ normalized to the data
for MM$^2<0.2$ GeV$^2$.} \label{mm2}
\end{figure}

The CLEO-c collaboration has published a result for $f_{D^+}$
\cite{Artuso:2005ym,Bonvicini:2004gv}. Several results have been obtained for $f_{D_s^+}$,
the most precise being a preliminary result from CLEO-c. To measure
$f_{D^+}$ CLEO-c uses a ``double-tag" method, possible because at an
$e^+e^-$ centre-of-mass energy of 3770 GeV, the location of the
$\psi''$ resonance, $D^+D^-$ final states are produced without any
extra particles. Here one $D^-$ is fully reconstructed and then
there are enough kinematic constraints (energy and momentum) to
search for $D^+\to\mu^+\nu$ by constructing the missing mass-squared
(MM$^2$) opposite the $D^-$ and the muon, which should peak at the
essentially zero neutrino mass-squared. Explicitly
\begin{equation}
{\rm MM}^2=\left(E_{\rm
beam}-E_{\mu^+}\right)^2-\left(-\textit{\textbf{p}}_{D^-}
-\textit{\textbf{p}}_{\mu^+}\right)^2, \label{eq:MMsq}
\end{equation}
where $\textit{\textbf{p}}_{D^-}$ is the three-momentum of the fully
reconstructed $D^-$. The CLEO-c MM$^2$ distribution is shown in
Fig.~\ref{mm2}.  The peak near zero contains 50 signal events of
which 2.8 are estimated background.

The resulting rate is
\begin{equation}
{\cal{B}}(D^+\to\mu^+\nu)=(4.40\pm 0.66^{+0.09}_{-0.12})\times
10^{-4}~.
\end{equation}
The decay constant $f_{D^+}$ is then obtained from
Eq.~(\ref{eq:equ_rate}) using 1.040$\pm$0.007 ps as the $D^+$
lifetime \cite{Yao:2006px}, and $|V_{cd}|$ = 0.2238$\pm$0.0029, giving
\begin{equation}
f_{D^+}=(222.6\pm 16.7^{+2.8}_{-3.4})~{\rm MeV}~.
\end{equation}

CLEO-c also sets limits on ${\cal{B}}(D^+\to e^+\nu_{e})<2.4\times
10^{-5},$ \cite{Artuso:2005ym,Bonvicini:2004gv} and ${\cal{B}}(D^+\to\tau^+\nu)$
branching ratio to $<2.1\times 10^{-3}$ at 90\% C.L.
\cite{Rubin:2006nt}. These limits are consistent with SM
expectations.

Before turning to theoretical prediction of $f_{D^+}$, we discuss
the current status of $D_s^+\to\mu^+\nu$. Results here have been
obtained by several experiments \cite{Yao:2006px}. However, these results
have been subject to sizeable systematic errors, the largest of which
usually is the uncertainty on ${\cal{B}}(D_s^+\to\phi\pi^+)$, that
is important because the measurements are usually normalized by
taking the ratio of the observed number of $\ell^+\nu$ events to
$\phi\pi^+$ events.

CLEO-c eliminates this uncertainty by making absolute measurements
directly. Data are obtained near 4.170 GeV. Here the cross-section
for $D_s^{*\pm}D_s^{\mp}$ is $\sim$1 nb. Both $\mu^+\nu$ and
$\tau^+\nu$ decays are examined, with two different decay modes of
the $\tau^+$ used, $\pi^+\bar{\nu}$ and $e^+\nu\bar{\nu}$. The
MM$^2$ distribution for the sum of $D_s^+\to \mu^+\nu$ + $D_s^+\to
\tau^+\nu$, $\tau^+\to \pi^+\nu$ is shown on the right side of
Fig.~\ref{mm2}. Analysing these samples separately, they find the
ratio $R$ from Eq.~\ref{eq:rat} is consistent with the SM
expectation of 9.72. Combining both gives a measurement using
Eq.~\ref{eq:equ_rate} of $ f_{D_s}=282\pm 16 \pm 7 {~\rm MeV}.$
 CLEO-c also uses the $D_s^+\to \tau^+\nu$,
 $\tau\to e^+\nu\bar{\nu}$ to find
 $f_{D_s}=278\pm 17 \pm 12 {~\rm MeV}.$ Combining the two results gives
  \begin{equation}
 f_{D_s}=280.1\pm 11.6 \pm 6.0 {~\rm MeV}.
 \end{equation}
 Using only the  $D_s^+\to \tau^+\nu$,
 $\tau\to e^+\nu\bar{\nu}$ and the $D_s^+\to \mu^+\nu$, CLEO-c
 finds
\begin{equation}
R = \frac{\Gamma(D_s^+\to \tau^+\nu)}{\Gamma(D_s^+\to \mu^+\nu)}=
9.9\pm 1.7 \pm 0.7~, \label{eq:tntomu2}
\end{equation}
again consistent with the SM expectation. Furthermore CLEO-c also
sets limits on ${\cal{B}}(D_s^+\to e^+\nu_{e})<3.1\times 10^{-4}.$

 The
branching fractions, modes and derived values of $f_{D_s^+}$ from
all measurements are listed in Table~\ref{tab:fDs}.
\begin{table}[htb]
\begin{center}

\caption{Measurements of $f_{D_s^+}$ Results have been updated for
new values of the $D_s$ lifetime. ALEPH uses both measurements to
derive a value for the decay constant.\label{tab:fDs}}
\begin{tabular}{llccc}\hline\hline
\textbf{Exp.} & \textbf{Mode}  & \boldmath{${\cal{B}}$}& \boldmath{${\cal{B}}_{\phi\pi}(\%)$} & \boldmath{$f_{D_s^+}$} \textbf{(MeV)} \\
\hline CLEO-c & $\mu^+\nu$ & $(6.57\pm 0.90\pm 0.34)\cdot 10^{-3}$ & & $281\pm 19\pm 7$\\
CLEO-c & $\tau^+\nu$, $\tau\to\pi\nu$ & $(7.1\pm 1.4\pm 0.3)\cdot 10^{-2}$&& $296\pm 29 \pm 7 $ \\
CLEO-c & $\tau^+\nu$,  $\tau\to e\nu\nu$ & $(6.29\pm 0.78\pm 0.52)\cdot 10^{-2}$&& $278\pm 17 \pm 12 $ \\
CLEO-c & combined & -& & $280.1\pm 11.6 \pm 6.0$  \\
CLEO \cite{Chadha:1997zh}& $\mu^+\nu$ &$(6.2\pm 0.8\pm 1.3 \pm 1.6)\cdot
10^{-3}$&
3.6$\pm$0.9&$273\pm19\pm27\pm33$\\
BEATRICE \cite{Alexandrov:2000ns} & $\mu^+\nu$ &$(8.3\pm 2.3\pm 0.6 \pm
2.1)\cdot 10^{-3}$& 3.6$\pm$0.9&$315\pm43\pm12 \pm39$\\
ALEPH \cite{Heister:2002fp}& $\mu^+\nu$ &$(6.8\pm 1.1\pm 1.8)\cdot 10^{-3}$ & 3.6$\pm$0.9& $285\pm 19\pm 40$ \\
ALEPH \cite{Heister:2002fp}& $\tau^+\nu$ &$(5.8\pm 0.8\pm 1.8)\cdot 10^{-2}$ & &  \\
OPAL \cite{Abbiendi:2001nb} & $\tau^+\nu$ & $(7.0\pm 2.1 \pm 2.0)\cdot 10^{-3}$ & & $286\pm 44\pm 41$  \\
L3 \cite{Acciarri:1996bv} &$\tau^+\nu$ & $(7.4\pm 2.8 \pm 1.6\pm 1.8)\cdot 10^{-3}$ & & $302\pm 57\pm 32 \pm 37$  \\
BaBar \cite{Aubert:2006sd} & $\mu^+\nu$& $(6.7\pm 0.8\pm 0.3 \pm
0.7)\cdot 10^{-3}$ & 4.7$\pm$0.5 &
 $283\pm 17 \pm 7 \pm 14$\\
\\\hline\hline
\end{tabular}
\end{center}
\end{table}
Most measurements of $D_s^+\to\ell^+\nu$ are normalized with respect
to ${\cal{B}} (D_s^+\to\phi\pi^+)$. These measurements are difficult
to average because of the uncertainty in this scale, and we do not
attempt this here. We can extract a value for ratio using the CLEO-c
measurements only, since the scale error is absent
\begin{equation}
f_{D_s^+}/f_{D^+}=1.26\pm 0.11 \pm 0.03~.
\end{equation}

Theoretical calculations of $f_{D_s^+}$, $f_{D^+}$ and the ratio
$\frac{f_{D_s^+}}{f_{D^+}}$ are listed in Table~\ref{tab:Models}.
While the CLEO-c decay constant results are slightly higher than
most theoretical expectations, the ratio is quite consistent with
Lattice-Gauge theory and most other models. Furthermore, no
deviations from SM expectations are found in the ratio of decay
rates for various lepton species.

\begin{table}[h]
\begin{center}

\caption{Theoretical predictions of $f_{D^+}$ and
$f_{D_S^+}/f_{D^+}$. QL indicates quenched lattice calculations.}
\label{tab:Models}
\begin{tabular}{p{6cm}lccl} \hline\hline
    \textbf{Model} &\boldmath{$f_{D_s^+}$} \textbf{(MeV)} & \boldmath{ $f_{D^+}$} \textbf{(MeV)}
        &  ~~~~~\boldmath{$f_{D_s^+}/f_{D^+}$}           \\\hline
Lattice ($n_f$=2+1)  \cite{Aubin:2005ar} &
 $249 \pm 3 \pm 16 $&$201\pm 3 \pm 17 $&$1.24\pm 0.01\pm 0.07$ \\
QL (Taiwan) \cite{Chiu:2005ue} &
$266\pm 10 \pm 18$ &$235 \pm 8\pm 14 $&$1.13\pm 0.03\pm 0.05$ \\
QL (UKQCD) \cite{Lellouch:2000tw}&$236\pm 8^{+17}_{-14}$ & $210\pm 10^{+17}_{-16}$ & $1.13\pm 0.02^{+0.04}_{-0.02}$\\
QL \cite{Becirevic:1998ua} & $231\pm 12^{+6}_{-1}$&$211\pm 14^{+2}_{-12}$
&
$1.10\pm 0.02$\\
QCD Sum Rules \cite{Bordes:2005wi} & $205\pm 22$ & $177\pm 21$ & $1.16\pm
0.01\pm 0.03$\\
QCD Sum Rules \cite{Narison:2002hk} & $235\pm 24$&$203\pm 20$ & $1.15\pm 0.04$ \\
Quark Model \cite{Ebert:2006hj}&268 &$234$  & 1.15 \\
Quark Model \cite{Cvetic:2004qg}&248$\pm$27 &$230\pm$25  & 1.08$\pm$0.01 \\
Potential Model \cite{Wang:2004xs,Salcedo:2003yb} & 241& 238  & 1.01 \\
Isospin Splittings \cite{Amundson:1992rw} & & $262\pm 29$ & \\
\hline\hline
\end{tabular}
\end{center}
\end{table}



\subsubsubsection{Semileptonic Decays} 


\label{sec:charmsl}

\begin{figure}[hbt]
\begin{center}
\includegraphics[width=3.in]{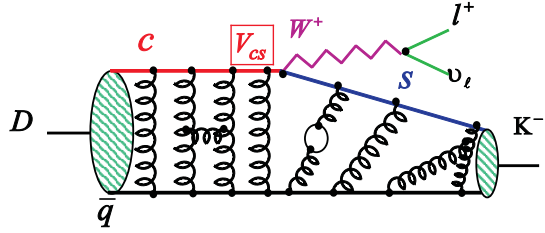} \caption{Feynman diagram for the
 semileptonic decay of charmed
mesons. The QCD non-perturbative effects are described by $q^2$
dependent form factors.}\label{sl-dia}
\end{center}
\end{figure}

The study of semileptonic charm decays has several important
ramifications. Figure~\ref{sl-dia} shows the Feynman diagram
describing these decays. It shows that the matrix element describing
these decays can be expressed as the product of a leptonic current,
unaffected by strong interactions, and a hadronic current, where the
non-perturbative QCD effects are generally modelled with form
factors. Theoretical predictions for these form factors have been
derived in the framework of quark models, QCD sum rules, and lattice
QCD.  Thus the study of inclusive and exclusive semileptonic decay
branching fractions and form factors provides the experimental
constraints needed to assess whether theoretical calculations are
reliable and feature well understood errors.

 On the other hand, once computational techniques
developed to predict relevant form factors demonstrate that they can
achieve reliable results with well understood errors, these data
allow precise determinations of the CKM matrix elements $V_{cs}$ and
$V_{cd}$. Moreover a combination of charm and beauty semileptonic
decay studies can be used to to determine $V_{ub}$.

\subsubsection{Branching Fractions}
We are now progressing towards a complete precision determination of
the absolute inclusive and exclusive charm semileptonic branching
fractions.   Inclusive semileptonic widths can provide some
information on weak annihilation diagrams \cite{Bianco:2003vb}.
Finally, better knowledge of the inclusive positron spectra can be
used to improved modelling of the ``cascade'' decays $b\rightarrow
c\rightarrow s e^+ \nu _e$ and thus it affects the precision of
several measurements of $b$ decays.

CLEO-c uses the two tagging modes with lowest background
($\bar{D}^0\rightarrow K^+\pi^-$ and $D^-\rightarrow K^+\pi^-\pi^-$)
to measure the inclusive $D^0$ and $D^+$ semileptonic branching
fractions \cite{Adam:2006nu}. The kinematic constraints available
through the use of $D$ tagged samples from data taken at the
$\psi(3770)$ provide a unique tool to select a pure sample of
electrons/positrons coming from $D$ semileptonic decays. They obtain
${\cal B}(D^0 \rightarrow X \ell \nu_e)=(6.46 \pm 0.17 \pm 0.13)\%$
and ${\cal B}(D^+ \rightarrow X \ell \nu_e)=(16.13 \pm 0.20 \pm
0.33)\%$. The inclusive branching fractions can be translated into
inclusive semileptonic widths $\Gamma_{D^+}$ and $\Gamma _{D^0}$,
using the well known $D$ lifetimes \cite{Yao:2006px}. These widths are
expected to be equal, modulo isospin violations, and indeed the
measured ratio $\Gamma^{sl}_{D^+}/\Gamma^{sl} _{D^0} = 0.985 \pm
0.028\pm 0.015$: thus isospin violations are limited to be below
$\sim$ 3\%.

BES-II \cite{Ablikim:2004ku,Ablikim:2004ej} and CLEO-c \cite{Coan:2005iu,Huang:2005iv} have recently
published data on exclusive semileptonic branching fractions. BES-II
results are based on 33 pb$^{-1}$; the CLEO-c published data are
based on the first 57 pb$^{-1}$, 
preliminary results
included in this report are based on 281 pb$^{-1}$.

The variable $U\equiv E_{miss} -|c\vec{p}_{miss}|$, where $E_{miss}$
and $\vec{p}_{miss}$ represent the missing energy and momentum of
the $D$ meson decaying semileptonically, is used to select signal
events. This variable is a non Lorentz invariant version of $MM^2$.
Table \ref{tab:brsemil} summarizes the recent data, as well as the
averages reported in the PDG 2006 \cite{Yao:2006px}.

A comparison between the inclusive branching fractions of the $D^+$
and $D^0$ mesons with the sum of the measured exclusive branching
fractions determines whether there are unobserved semileptonic decay
modes.The corresponding sums of exclusive branching fractions are:
$\Sigma _i {\cal B}(D^0\rightarrow X_i \ell \nu_e) =6.1\pm 0.2\pm
0.2$ and
 $\Sigma _i {\cal B}(D^+\rightarrow X_i \ell \nu_e)= 15.1\pm 0.50 \pm 0.50$: the measured
 exclusive modes are consistent with saturating the
inclusive widths, although there is some room left for higher
multiplicity modes. In particular, CLEO-c also provides the first
 evidence for $D^0\rightarrow K^-\pi^+\pi^- e^+\nu _e$ \cite{gao:ichep06}.
They study the $MM^2$, inferred from the missing energy and momentum
in the event and they obtain the preliminary branching fractions:
\begin{eqnarray}
{\cal B}(D^0\rightarrow K^-\pi^+\pi^- e^+\nu _e
=(2.9^{+1.5}_{-1.1}\pm 0.5)\times 10^{-4}\\
{\cal B}(D^0\rightarrow K_1(1270) e^+\nu _e) \times {\cal
B}(K_1(1270)\rightarrow K^-\pi^+\pi^-)=(2.2 ^{+1.4}_{-1.0}\pm
0.2)\times 10^{-4}
\end{eqnarray}
This branching fraction is about at the level predicted by Isgur and
Scora \cite{Scora:1995ty}, and is consistent with the expectation that
charm semileptonic decays are dominated by the pseudoscalar and
vector lowest mass resonances.

Finally, $D$ semileptonic decays are a tool to explore light quark
spectroscopy. For example, a few years ago the FOCUS collaboration
reported some evidence for an s-wave interference effect in the
decay amplitude of $D^+\rightarrow K^{\star 0} \mu^+ \nu _{\mu}$
\cite{Link:2002ev}. This observation can shed some light on our
understanding of the elusive scalar meson $\kappa$. This observation
has been recently confirmed by CLEO-c in the channel $D^+\rightarrow
K^{\star 0} e^+ \nu _{e}$ \cite{Shepherd:2006jm}. This study will acquire soon
a broader scope when CLEO-c will pursue similar analyses in the
$D_s$ system.

\begin{table}[htb]
\begin{center}
\caption{CLEO-c branching fractions and new world averages.}
\begin{tabular}{lcc|lcc}
\hline
 $D^+$ Mode &   Recent Data $\mathcal{B}$ (\%)   &   PDG
 2006 &   $D^0$ Mode &   Recent Data $\mathcal{B}$ (\%)   &   PDG 2006  \\
 \hline
$ \bar{K}^0 e^+ \nu_e  $ & $ 8.86\pm 0.17 \pm 0.20 $  & $8.7 \pm
0.5$ & $K^- e^+ \nu_e  $ &  3.58$\pm$0.05$\pm$0.05    &   3.47$\pm$0.13   \\
$\pi^0 e^+ \nu_e  $& $ 0.397 \pm  0.027 \pm 0.028 $  &  $0.44
\pm 0.06$ & $\pi^- e^+ \nu_e  $&  0.309$\pm$0.012$\pm$0.006  &   0.262$\pm$0.026\\
$\eta e^+ \nu_e$ & $ 0.129 \pm  0.019 \pm 0.07 $  & ~~ & ${K}^{*-} e^+ \nu_e  $ &2.16$\pm$0.15$\pm$0.08  &  2.16$\pm$0.16 \\
 $\bar{K}^{*0} e^+ \nu_e$ & $5.56 \pm 0.27 \pm 0.23 $  &  $5.61 \pm 0.31$ & $ \rho^- e^+ \nu_e  $& 0.156$\pm$0.016$\pm$0.009& 0.194$\pm$0.41 \\
$ \rho^0 e^+ \nu_e  $ & $ 0.232 \pm 0.020 \pm 0.012 $ &  $0.22 \pm 0.04$ & & & \\
$\omega e^+ \nu_e  $    &  $ 0.149\pm 0.027 \pm 0.005 $ & $0.16^{+0.07}_{-0.06} $ & & &  \\  \hline
\end{tabular}
\label{tab:brsemil}
\end{center}
\end{table}

\subsubsection{Form factors for $D\rightarrow K(\pi) \ell \nu$ and 
$D \to K^* (\rho)\ell \nu$  }

 Recently, non-quenched lattice $QCD$ calculations for
$D\rightarrow K\ell \bar{\nu}$ and $D\rightarrow \pi \ell \nu$ have
been reported \cite{Aubin:2004ej}. The chiral extrapolation is
performed at fixed $E =\vec{v}\cdot\vec{p}_P$, where $E$ is the
energy of the light meson in the centre-of-mass $D$ frame, $\vec{v}$
is the unit 4-velocity of the $D$ meson, and $\vec{p}_P$ is the
4-momentum of the light hadron $P$ ($K$ or $\pi$). The results are
presented in terms of a parametrisation originally proposed by
Becirevic and Kaidalov ($BK$) \cite{Becirevic:1999kt}:
\begin{equation}\label{eq:BK}
f_+(q^2) = \frac{F}{(1-\tilde{q}^2)(1-\alpha\tilde{q}^2)};\
 f_0(q^2) = \frac{F}{1-\tilde{q}^2/\beta},
\end{equation}
where $q^2$ is the 4-momentum of the electron-$\nu$ pair,
$\tilde{q}^2=q^2/m_{D_x^{*}}^2$, and $F=f_+(0)$, $\alpha$ and
$\beta$ are fit parameters. This formalism models the effects of
higher mass resonances other than the dominant spectroscopic pole
($D^{\star +}_S$ for the $K\ell \nu$ final state and $D^{\star +}$
for $\pi\ell \nu$ \cite{Fajfer:2005mk}).

Table \ref{fit-data1} shows the fit results obtained from FOCUS
\cite{Link:2004dh}, CLEO III \cite{Huang:2004fr}, Belle \cite{Abe:2005sh},
and BaBar \cite{Aubert:2006mc} compared to the lattice QCD predictions
\cite{Aubin:2004ej}. In addition, all these experiments perform a
single pole fit, traditionally used because of the conventional
ansatz of several quark models \cite{Stone:1992ft}, and the $BK$
parametrisation discussed before. In Table \ref{fit-data2} we include
preliminary results of fits obtained with the simple pole model by
CLEO-c. All of these experiments obtain very good fits also with
simple pole form factors; however the simple pole fit does not yield
the expected spectroscopic mass. This may hint that other higher
order resonances are contributing to the form factors \cite{Fajfer:2005mk}.
 It has been argued \cite{Hill:2005ju} that even
the $BK$ parametrisation is too simple and that a three parameter
form factor is more appropriate. This issue can be resolved  by
larger data samples, with better sensitivity to the curvature of the
form factor near the high recoil region.

\begin{table}[hbt]
\begin{center}
\caption{Measured shape parameter $\alpha$ compared to lattice QCD
predictions.}\label{fit-data1}
\begin{tabular}{lll}\hline
~~& {$\alpha(D^0\rightarrow K\ell \nu)$}& $\alpha(D^0\rightarrow
\pi\ell \nu)$\\\hline
Lattice QCD \cite{Aubin:2004ej} & $0.5\pm 0.04\pm 0.07$ & $0.44\pm 0.04\pm 0.07$\\
FOCUS \cite{Link:2004dh} & $0.28\pm 0.08\pm 0.07$\\
CLEOIII \cite{Huang:2004fr} & $0.36\pm 0.10^{+0.03}_{-0.07}$&
$0.37^{+0.20}_{-0.31}\pm 0.15$\\
Belle \cite{Abe:2005sh}& $0.40\pm 0.12\pm 0.09 $& $0.03\pm 0.27 \pm
0.13$\\
BaBar \cite{Aubert:2006mc}& $0.43\pm 0.03\pm 0.04 $& ~~ \\
\hline
\end{tabular}
\end{center}
\end{table}

\begin{table}[hbt]
\begin{center}
\caption{Measured shape parameter $\alpha$ compared to lattice QCD
predictions.}\label{fit-data2}
\begin{tabular}{lll}\hline
~~& {$M_{\text{pole}}(D^0\rightarrow K\ell \nu)$} (GeV)&
$M_{\text{pole}}(D^0\rightarrow \pi\ell \nu)$ (GeV)\\
\hline
FOCUS \cite{Link:2004dh} &  $1.93\pm 0.05\pm 0.03$ & $1.91 ^{+0.30}_{-0.15}\pm 0.07$\\
CLEOIII \cite{Huang:2004fr} &  $1.89\pm 0.05 ^{+0.04}_{-0.03}$ & $1.86^{+0.10+-.07}_{-0.06-0.03}$\\
Belle \cite{Abe:2005sh}& $1.88\pm 0.06\pm0.03$& $2.01\pm 0.13\pm 0.04$ \\
BaBar \cite{Aubert:2006mc}& $1.854\pm 0.016\pm 0.020$& ~~ \\
CLEO-c\cite{gao:ichep06} & $1.96\pm 0.03\pm 0.01$ & $1.95\pm 0.04\pm
0.02$ \\\hline
\end{tabular}
\end{center}
\end{table}
In experimental studies of $D \to K^* (\rho)\ell \nu$ usually single pole parametrisation of form factors was used. Following Becirevic- Kaidalov approach 
 in Ref \cite{Fajfer:2005ug,Fajfer:2006av}
new parametrisation of 
relevant form factors was given by
\begin{equation}
\begin{array}{rclrcl}
A_1(q^2) &=& \frac{A_1(0)}{1-b' x} & A_2(q^2) &=& \frac{A_2(0)}{(1-b'
x)(1-b'' x)}\\
A_0(q^2) &=& \frac{A_0(0)}{(1-y)(1-a' y)} & V(q^2) &=&
\frac{A_1(0)}{\xi(1-x)(1-a x)}
\end{array}\nonumber
\end{equation}

This parametrisation takes into account all known scaling properties of the decay to light vector semileptonic transition.
The study of nonparametric determination of helicity amplitudes in the semileptonic 
$D \to K^* (\rho)\ell \nu$ decays will shed more light on the corresponding decays in B physics.

\subsubsection{Lattice QCD Checks}
By combining the information of the measured leptonic and
semileptonic widths, a ratio
$R_{sl}=\sqrt{\frac{\Gamma(D^+\rightarrow \mu^+\nu
_{\mu})}{\Gamma(D\rightarrow \pi e\nu _{e})}}$, independent of
$|V_{cd}|$, can be evaluated: this is a pure check of the theory. We
assume isospin symmetry, and thus $\Gamma(D\rightarrow \pi e^+\nu
_e)=\Gamma(D^0\rightarrow \pi^- e^+\nu _e)=2\Gamma(D^+\rightarrow
\pi^0 e^+\nu _e)$. For the theoretical inputs, we use the recent
unquenched lattice QCD calculations in three flavours
\cite{Aubin:2005ar}, as they reflect the state of the
art of the theory and have been evaluated in a consistent manner.
The theory ratio is
$R^{th}_{sl}=\sqrt{\frac{\Gamma^{th}(D^+\rightarrow \mu^+\nu
_{\mu})}{\Gamma^{th}(D\rightarrow \pi e\nu _{e})}}=0.212\pm 0.028$.
The quoted error is evaluated through a careful study of the theory
statistical and systematic uncertainties, assuming Gaussian errors.
The corresponding experimental $R^{exp}_{sl}$ is calculated using
the CLEO-c $f_D$ and isospin averaged $\Gamma (D\rightarrow \pi
e^+\nu_e)$: $ R^{exp}_{sl}=\sqrt{ \frac
{\Gamma^{exp}(D^+\rightarrow\mu^+\nu)} {\Gamma^{exp}(D\rightarrow\pi
e \nu _e)}}= 0.237\pm 0.019.$ The theory and data are in good
agreement, though the errors need to be reduced both in theory and
experiment to validate the theory at the needed level of precision
($\sim1-3\%$).

\subsubsubsection{Hadronic Decays} 

\label{NLDEC}

While the dynamical issues are considerably more complex in nonleptonic than in semileptonic 
decays -- both a blessing and a curse --, the available theoretical tools are more limited. 
For inclusive rates like lifetimes one can turn to expansions in powers of $1/m_c$ to obtain at least 
a semi-quantitative description. 
For 
exclusive modes we have `Old Faithful', namely quark models, but also QCD sum rules and chiral dynamics with the latter two (in contrast to the first one) firmly rooted in QCD. Lattice QCD, usually 
perceived as panacea, faces much more daunting challenges in dealing with nonleptonic charm transitions than for 
semileptonic modes due to the central role played by strong final state interactions. Yet comprehensive measurements can teach us valuable lessons that can enlighten us about light flavour spectroscopy and also serve as cross checks on $B$ studies. 
Below we list some core examples for such lessons. 

\subsubsubsection{Lifetime ratios}

Heavy quark theory (HQT) allows to describe inclusive decays of charm hadrons through an expansion 
in powers of $1/m_c$ implemented by the OPE. With the charm quark mass $m_c$ exceeding ordinary 
hadronic scales merely by a moderate amount the expansion parameter is not much smaller than unity. 
In the description of fully integrated widths like lifetimes the leading nonperturbative contributions arise in order $1/m_c^2$ rather than $1/m_c$, which might be their saving grace. Indeed the resulting 
theoretical description of the lifetime ratios for the seven weakly decaying $C=1$ charm hadrons has been remarkably successful \cite{Bianco:2003vb}. 
Note that these seven charm lifetimes vary by a factor 
of 15, while the four singly-beautiful hadrons differ by less than 30\%.  The $B_c$ meson is shorter lived by a 
factor of three than the other four beauty hadrons -- not surprisingly, since it represents a glorified 
charm decay. 

The same framework allows one to predict also the lifetimes of the $C=2$ double-heavy baryons 
$\Xi_{cc}$, $\Omega_{cc}$ and even the $C=3$ $\Omega _{ccc}$ \cite{Bianco:2003vb}: 
\beq 
\tau (\Xi_{cc}^{++}) \sim 0.35 \; {\rm ps},\;  \tau (\Xi_{cc}^{+}) \sim 0.07 \; {\rm ps}, \; 
\tau (\Omega_{cc}^{+}) \sim 0.1 \; {\rm ps},\; \tau (\Omega_{ccc}^{++}) \sim 0.14 \; {\rm ps}
\eeq 
The SELEX collaboration has found tantalizing evidence for $\Xi_{cc}^{+,++}$ baryons all 
decaying with ultrashort lifetimes below $0.03$ ps. This feature cannot be accommodated in 
HQT. {\em If} confirmed, one would have to view the apparent successes of the HQT description 
of the $C=1$ lifetimes as mere coincidences.   

\subsubsubsection{Absolute branching ratios}

Precision absolute branching fraction measurements are difficult 
due to normalisation and systematic effects.  
Only one {\it golden mode} is needed to anchor the rest for each state.   
A desire to use all-charged final states necessitates use 
of some three-body modes where proper modeling of the Dalitz structure 
is needed to ensure an accurate efficiency simulation.  
These results serve not only to normalize charm physics, 
but also much $B$ physics due to dominance of $b \to c$ decays.  
For example, charm branching fractions affect $B \to D^* \ell \nu$, 
used to extract $V_{cb}$.  

Near-threshold $D\bar{D}$ pairs from $\psi(3770)$ decays and 
$D_s^{*\pm} D_s^\mp$ produced at 4170 MeV from CLEO-c 
now provide the best precision.  
Systematics are controlled and normalization provided with tagging: 
studying one $D$ vs. a fully-reconstructed {\it tag} $\bar{D}$.  
Precision on the golden modes $D^0 \to K^-\pi^+$ and 
$D^+ \to K^-\pi^+\pi^+$ results are limited by uncertainties 
of about 1\% per track \cite{He:2005bs} from tracking-finding 
and particle-identification efficiencies.  
Further studies \cite{Adam:2006me} are reducing these 
to less than 0.5\% per track.  
Current statistical precision for $D_s^+ \to K^+K^-\pi^+$ 
decays \cite{Adam:2006me} is 5\%; 
final CLEO-c accuracy should be about 3\%, limited by statistics.  
Producing a useful new result for the popular $D_s^+ \to \phi\pi^+$ mode 
is complicated by several factors: 
a non-resonant contribution under the $\phi$, Breit-Wigner tails 
of the $\phi$, treatment of nearby resonances like the $f(980)$, 
and lack of detail in existing publications.  The merit of such studies goes beyond 
determining the branching ratio for $D_s^+ \to \phi\pi^+$ and learning about hadronic 
resonances (see below). Their greatest impact might come in precision analyses of 
$B_d \to \phi K_S$ and its \cp~asymmetries.

\subsubsubsection{Dalitz plot studies \& light flavour spectroscopy}

Dalitz plot studies represent powerful analysis tools that are deservedly experiencing a renaissance 
in heavy flavour decays. Constructing a satisfactory description of the Dalitz plot populations allows one to 
extract the maximal amount of information from the data in a self-consistent way. One has to keep in mind, though, that a priori different parametrisations can be chosen; one has to make a judicious 
choice based on theoretical considerations. 
Along with better theoretical descriptions of the decay rate, 
improved treatments of background and efficiency may also be needed.  

One important application concerns the spectroscopy of light flavour hadrons, i.e. those made up from 
$u$, $d$ and $s$ quarks. Modes like $D_{(s)} \to 3 \pi, \, 3 K,  \, K\pi \pi, \, K \bar K\pi$ offer more than  
a treasure trove of additional data: since the final state evolves from a well defined initial one, we know 
some quantum numbers of the overall system. Finding evidence for, say, a $\pi\pi$ resonance like 
the $\sigma$ in Cabibbo favoured $D$ and Cabibbo suppressed $D_s$ modes with parameters 
consistent with what is inferred from low-energy $\pi\pi$ scattering would constitute a powerful validation for the $\sigma$ being a bona fide resonance. 

Such lessons possess considerable intrinsic value. The latter is greatly amplified, since these 
insights will turn out to be of great help in understanding $B$ decays into the analogous final states, when searching for \cp~asymmetries there. 

\subsubsubsection{QCD Sum Rules}

More than twenty years ago a pioneering analysis of $D$ and $D_s$ decays into two-body final states 
of the $PP$ and $PV$ type was performed by Blok and Shifman through a novel application of 
QCD sum rules. Those are -- unlike quark models -- genuinely based on the QCD. Their drawback,  
as for most applications of QCD sum rules, is that one has to allow for an irreducible 
theoretical uncertainty of about 20\%; furthermore they are very labour intensive. 
The authors of Ref.\cite{Blok:1986hm} assumed $SU(3)_{fl}$ symmetry to make their analysis manageable -- clearly 
a source of significant theoretical uncertainty. It would be marvellous, if some courageous minds would take up the challenge of updating and extending this study.

\subsubsubsection{On theoretical engineering}

Even without reliable predictions for exclusive nonleptonic widths, it makes a lot of sense 
to measure as many as precisely as possible on the Cabibbo allowed, once and twice suppressed 
levels. It can provide vital input into searches for direct \cp~violation in charm decays. 

\cp~asymmetries in integrated partial widths depend on hadronic matrix elements and (strong)  
phase shifts, neither of which can be predicted accurately. However the craft of theoretical 
engineering can be practised with profit here. One makes an ansatz for the general form of the matrix 
elements and phase shifts that are included in the description of $D\to PP, PV, VV$ etc. 
channels, where $P$ and $V$ denote pseudoscalar and vector mesons, and fits them to the measured branching ratios on the Cabibbo allowed, once and twice forbidden level. If one has sufficiently accurate and comprehensive data, one can use these fitted values of the hadronic parameters to predict \cp~asymmetries. Such analyses have been undertaken in the past \cite{Buccella:1994nf}
and more recently by \cite{Cheng:1994zx,Rosner:1999xd,Chiang:2001av,Chiang:2002mr,Wu:2004ht,Wu:2004bz}, 
but the data base was not as broad and precise as one would like. 
{ CLEO-c and BESIII measurements 
will certainly lift such studies to a new level of reliability.}  

Similar information can be obtained in a more subtle and model independent way using 
quantum entanglement in \cite{Bigi:1989ah}
\beq 
e^+e^- \to \psi (3770) \to D^0 \bar D^0 
\eeq 
and observing the subsequent decay of the neutral $D$ mesons into final states like 
$f(D) = K^- \pi^+, K^+\pi^-$, $K^+K^-, \pi^+\pi^-$. Since the $D^0\bar D^0$ pair forms a coherent system, 
one can extract the strong phases reliably. This procedure is described in detail in Subsection~\ref{sec:mixing}.

\subsubsubsection{Time dependent Dalitz studies}

Tracking three-body channels like $D^0 \to K \bar K \pi, K^0_S \pi \pi$ through time-dependent 
Dalitz plot studies is a very powerful way to look for New Physics through \cp~asymmetries 
involving $D^0 - \bar D^0$ oscillations, as described in more detail in 
Sections~\ref{sec:mixing} and \ref{sec:cpv}.



\subsubsection{Summary on Ongoing and Future Charm Studies}
Even accepting for the moment that the SM can provide a complete description of all charm transitions detailed and comprehensive measurements of the latter will continue to teach us important and quite possible even novel lessons on QCD. Those lessons are of considerable intellectual value and would  
also prepare us, if the anticipated New Physics driving the electroweak phase transition were of the strongly interacting variety. 

Yet most definitely those lessons will sharpen both our experimental and theoretical tools for 
studying $B$ decays and thus will be essential in saturating the discovery potential for New Physics 
there. Analyses of (semi)leptonic charm decays will yield powerful validation challenges to LQCD that 
if passed successfully will be of great benefit to extractions of $|V_{ub}|$ in particular. Careful 
studies of three-body final states in charm decays will yield useful constraints in analyses of the corresponding $B$ modes and their \cp~asymmetries. The relevant measurements can be made at 
the Tau-Charm, the $B$ and Super-flavour factories. Yet there is one area in {\em this} context, where hadronic experiments and in particular LHCb can make important contributions, namely in the search for and observation of doubly-heavy charm baryons of the $[ccq]$ type and their lifetimes. 

The study of charm dynamics was crucial in establishing the SM paradigm. Even so it is conceivable 
that another revolution might originate there in particular by observing non-SM type \cp~violation with and without oscillations. For on one hand the SM predicts practically zero results (except for direct 
\cp~violation in Cabibbo suppressed channels), and on the other hand flavour changing neutral currents might well be 
considerably less suppressed for up- than for down-type quarks. Charm is the only up-type 
quark that allows the full range of searches for \cp~violation. Modes like 
$D^0 \to K^+K^-$, $K^+\pi^-$ have the potential to exhibit (time dependent) \cp~asymmetries 
that -- if observed -- would establish the presence of New Physics. Likewise for asymmetries in 
final state distributions like Dalitz plots or for \ot~odd moments. Again especially 
LHCb appears well positioned to bring the statistical muscle of the LHC to bear on analyzing these transitions.

%


\newpage \section{Prospects for future facilities}
\label{sec:superb}


There are several new facilities 
for flavour physics discussed in the community among which
the Super Flavour Factories (SFF) and the upgrade of the LHCb experiment 
are the most important ones for $B$ physics. These are analysed in this 
chapter (for future kaon and charm physics facilities see also 
Sections~\ref{sec:kaon} and \ref{sec:charm}).

The physics case of a Super Flavour Factory is worked out 
in Section~\ref{sec:superff}. All opportunities of such a facility 
in $B$, charm and $\tau$ lepton physics are discussed. 
Then  the two existing proposals for such a machine,
namely Super$B$ and SuperKEKB, are presented in  Section~\ref{sec:super-b}
and Section~\ref{sec:super-kekb}, respectively. 
Finally, the physics, detector and accelerator issues of a 
possible future upgrade  of the LHCb experiment are discussed 
in Section~\ref{sec:lhcb-upgrade}.


\subsection{On the physics case of a Super Flavour Factory}
\label{sec:superff}













We summarize the physics case of a high-luminosity $e^+e^-$ flavour 
factory  collecting an integrated luminosity of $50-75 \ {\rm ab}^{-1}$.  
Many New Physics sensitive measurements involving $B$ and $D$ mesons 
and $\tau$ leptons, unique to a Super Flavour Factory, 
can be performed with excellent sensitivity to new particles with 
masses up to  $\sim 100$ (or even $\sim 1000$) TeV.
Flavour- and $\CP$-violating couplings of new particles 
that may be discovered at the LHC
can be measured in most scenarios, 
even in unfavourable cases assuming minimal flavour violation.  
Together with the LHC, a Super Flavour Factory, 
following either the SuperKEKB or the Super$B$ proposal, 
could be soon starting the 
project of reconstructing  the New Physics Lagrangian.


\subsubsection{Introduction}

Many open fundamental questions of particle physics are related to flavour: How
many families are there? What is their origin? How are neutrino and quark masses
and mixing angles generated? Do there exist new sources of flavour and $\CP$
violation beyond those we already know? What is the relation between the flavour
structure in the lepton and quark sectors? Future flavour experiments will
attempt to address these questions providing the exciting possibility to learn
something about physics at energy scales much higher than those reachable by
current experiments.

The Standard Model (SM) of elementary particles has been very successful in
explaining a wide variety of existing experimental data. It accounts for a range
of phenomena from low-energy physics (less than a GeV), such as kaon decays, to
high-energy (a few hundred GeV) processes involving real weak gauge bosons ($W$
and $Z$) and top quarks. There is, therefore, little doubt that the SM is the
theory to describe physics below the energy scale of several hundred GeV, namely
all that has been explored so far.

In spite of the tremendous success of the SM, it is fair to say that the flavour
sector of the SM is much less understood than its gauge sector, reflecting our
lack of answers to the questions mentioned above. Masses and mixing of the
quarks and leptons, which have a significant but unexplained hierarchy pattern,
enter as free parameters to be determined experimentally. In fact, while
symmetries shape the gauge sector, no principle governs the flavour structure of
the SM Lagrangian. Yukawa interactions provide a phenomenological description of
the flavour processes which, while successful so far, leaves most fundamental
questions unanswered. Hence the need to go beyond the SM.

Indeed the search for evidence of physics beyond the SM is the main goal of
particle physics in the next decades. The LHC at CERN will start soon looking
for the Higgs boson, the last missing building block of the SM. At the same time
it will intensively search for New Physics (NP), for which there are solid
theoretical motivations related to the quantum stabilization of the Fermi scale
to expect an appearance at energies around $1$ TeV.

However, pushing the high-energy frontier, {i.e.} increasing the available
centre-of-mass energy in order to produce and observe new particles, is not the
only way to look for NP. New particles could reveal themselves through their
virtual effects in processes involving only  standard particles 
as has been the case several  times in the history of particle physics. 
For these kind of searches
the production thresholds are not an issue.
Since quantum effects become typically smaller as the mass of the virtual
particles increases, the name of the game is rather high precision. 
As a matter of fact, high-precision measurements 
probe NP energy scales
inaccessible at present and next-generation colliders at the energy frontier.

Flavour physics is the best candidate as a tool for NP searches 
through quantum effects for several reasons. 
Flavour Changing Neutral Currents (FCNC), neutral
meson-antimeson mixing and $\CP$ violation occur at the loop level in the SM and
therefore are potentially subject to ${\cal O}(1)$ NP virtual corrections. In
addition, quark flavour violation in the SM is governed by the weak interaction
and suppressed by the small quark mixing angles. Both these features are not
necessarily shared by NP which, in such cases, could produce very large effects.
Indeed, the inclusion in the SM of generic NP flavour-violating terms with
natural ${\cal O}(1)$ couplings is known to violate present experimental
constraints unless the NP scale is pushed up to $10$--$100$ TeV depending on the
flavour sector. This difference between the NP scale emerging from flavour
physics and the one suggested by Higgs physics could be a problem for model
builders (the so-called flavour problem), but it clearly indicates that flavour
physics has the potential to push the explored NP scale in the $100$ TeV region.
On the other hand, if the NP scale is indeed close to $1$ TeV, the flavour
structure of NP must  be highly non-trivial and the experimental determination
of the flavour-violating couplings is particularly interesting.

Let us elaborate on this latter option. Any new-physics model, established at
the TeV scale to solve the gauge hierarchy problem, includes new flavoured
particles and new flavour- and $\CP$-violating parameters. Therefore, such a
model must provide a solution also to the flavour and $\CP$ problems, namely how
new flavour changing neutral currents and $\CP$-violating phenomena are
suppressed. This may be related to other interesting questions. For instance, in
supersymmetry the flavour problem is directly linked to the crucial issue of
supersymmetry breaking. Similar problems also occur in models of
extra-dimensions (flavour properties of Kaluza-Klein states), Technicolour models
(flavour couplings of Techni-fermions), little-Higgs models (flavour couplings
of new gauge bosons and fermions) and multi-Higgs models ($\CP$-violating Higgs
couplings). Once NP is found at the TeV scale, precision measurements of
flavour- and $\CP$-violating observables would 
shed light on the detailed structure of the underlying model.

On quite general grounds, 
quantum effects in flavour processes explore a parameter space including the
NP scale and the NP flavour- and $\CP$-violating couplings. In specific models
these are related to fundamental parameters such as masses and couplings of new
particles. In particular, NP effects tend to disappear at large NP scales as 
well as for small couplings. Therefore a crucial question is: could NP be
flavour-blind, thus making searches for it  with flavour physics unfeasible?
Fortunately, the concept of Minimal Flavour Violation (MFV) provides a negative
answer: even if NP does  not contain new sources of flavour and $\CP$ violation,
the flavour-violating couplings present in the SM are enough to produce a new
phenomenology that  makes flavour processes sensitive to the presence of new
particles. In other words, MFV puts a lower bound on the flavour effects
generated by NP appearing at a given mass scale, a sort of ``worst case''
scenario for the flavour-violating couplings extremely useful to exclude NP
flavour-blindness and assess the ``minimum'' performance of flavour physics in
searching for NP, always keeping in mind that larger effects are quite possible
and easily produced in many scenarios beyond MFV.

In the light of the above considerations, a Super Flavour Factory (SFF), 
following the recent proposals for SuperKEKB
(see Section~\ref{sec:super-kekb} and ref.~\cite{Hashimoto:2004sm}) 
and Super$B$ 
(see Section~\ref{sec:super-b} and ref.~\cite{Bona:2007qt}),
has one mission: 
to search for new physics in the flavour sector 
exploiting a huge leap in integrated luminosity
and the wide range of observables that it can measure.
However this goal can be pursued in different ways depending on whether
evidence of NP has been found at the time a SFF starts taking data.

In either scenario, a SFF can search for evidence of NP irrespective of the
values of the new particle masses and of the unknown flavour-violating
couplings. A large number of measurements could provide evidence for NP 
at a SFF. 
A first set is given by measurements of observables which are predicted 
by the SM with small uncertainty, including those which are 
vanishingly small (the so-called null tests). 
Among them are the flavour-violating $\tau$ decays,
direct $\CP$ asymmetries in $B\to X_{s+d}\gamma$, 
in $\tau$ decays and in some non-leptonic $D$ decays, 
$\CP$ violation in neutral charm meson mixing,
the dilepton invariant mass at which the
forward-backward asymmetry of $B\to X_s\ell^+\ell^-$ vanishes, 
and lepton universality violating $B$ and $\tau$ decays. 
Any deviation, as small as a SFF could measure, 
from its SM value of any observable in this
set could be ascribed to NP with 
essentially no uncertainty. 
A second set of NP-sensitive observables, 
including very interesting decays such as 
$b\to s$ penguin-dominated non-leptonic $B$ decays, $B\to \tau\nu$, 
$B \to D^{(*)}\tau\nu$, $B\to K^*\gamma$, $B\to \rho\gamma$, and many others, 
require more accurate determinations of SM contributions 
and improved control of the hadronic uncertainties with respect 
to what we can do today in order to match the experimental precision 
achievable at a SFF and to allow for an unambiguous
identification of a NP signal. 
The error on the SM can be reduced using the improved determination of 
the Cabibbo-Kobayashi-Maskawa (CKM) matrix provided by a SFF itself.
This can be achieved using generalized CKM fits which allow for
a $1\%$ determination of the CKM parameters using tree-level 
and $\Delta F=2$ processes even in the presence of generic NP contributions. 
As far as hadronic uncertainties are concerned, 
the extrapolation of our present knowledge
and techniques shows that it is possible to reach the required accuracy by the
time a SFF will be running using improved lattice QCD results obtained with
next-generation computers~\cite{Bona:2007qt} 
and/or bounding the theoretical uncertainties with
data-driven methods exploiting the huge SFF data sample.

As we already noted, the NP search at a  SFF could reveal the virtual effect of
particles with masses of hundreds of TeV and in some cases, notably $\Delta F=2$
processes, even thousands of TeV depending on the values of the
flavour-violating couplings. 
Therefore this search is worth doing 
irrespective of whether NP has already been found or not. 
If new particles are discovered at the energy frontier, 
a SFF could enlarge the spectrum 
providing evidence of heavier states not accessible otherwise; 
if not, quantum effects measurable at a SFF
could be the only option to look for NP for a long time.

If the LHC finds NP at the TeV scale --
in particular if the findings include one (or more) new flavoured particle(s) --
then a SFF could measure its flavour- and $\CP$-violating couplings. 
Indeed all terms of the NP Lagrangian non-diagonal in
the flavour space are barely  accessible at the LHC. A SFF would be needed to
accomplish the task of reconstructing them. It seems able to do that even in
the unfavourable cases provided by most MFV models. Indeed, for the purpose of
inferring the NP Lagrangian from experiments, the LHC and SFF physics programmes are
complementary. 

Finally, 
it must be emphasised that while a Super Flavour Factory 
will perform detailed studies of beauty, charm and tau lepton physics,
the results will be highly complementary to those on several 
important observables related to $B_s$ meson oscillations, 
kaon and muon decays that will be measured elsewhere.
Most benchmark charm measurements, 
in particular interesting NP-related measurements such as
$\CP$ violation in charm mixing,
  will still  be statistics-limited after the 
  CLEO$c$, BESIII and $B$ Factory  projects are completed, 
  and can only be pursued to their ultimate precision at a SFF.
  Operation at the $\Upsilon(5{\rm S})$ resonance
  provides the possibility of exploiting the clean $e^+e^-$ environment
  to measure $B^0_s$ decays with neutral particles in the final state,
  which will  complement the channels that can be measured at LHCb.
  A SFF has sensitivity for $\tau$ physics that is far superior 
  to any other existing or proposed experiment,
  and the physics reach can be extended even further by
  the possibility to operate with polarized beams.
  It is particularly noteworthy that 
  the combined information on $\mu$ and $\tau$ flavour violating decays 
  that will be provided by MEG~\cite{Ritt:2006cg} together with a SFF
  can shed light on the mechanism responsible for lepton flavour violation.

\subsubsection{Experimental Sensitivities}

A Super Flavour Factory (SFF) with integrated luminosity 
of $50$--$75 \ {\rm ab}^{-1}$
can perform a wide range of important measurements and dramatically improve 
upon the results from the current generation of $B$ Factories.
Many of these measurements cannot be made in a hadronic environment,
and are unique to a SFF. 
The experimental sensitivities of a SFF  can be 
schematically classified in two categories:

\begin{itemize}
\item {\it Searching for New Physics:} \\
  Many of the measurements that can be made at a SFF 
  are highly sensitive to NP  effects,
  and those with precise SM  predictions are potential 
  discovery channels.
  As an example: the mixing-induced $\CP$ asymmetry parameter
  for $B^0 \to \phi K^0$ decays 
  can be measured to a precision of $0.02$,
  as can equivalent parameters for numerous hadronic decay channels
  dominated by the $b \to s$ penguin transition.
  These constitute very stringent tests of any NP scenario
  which introduces new $\CP$ violation sources, beyond the Standard Model.
     The presence of new sources of $\CP$ violation 
    in $D^0$--$\bar{D}^0$ mixing, where the SM background is negligible,
    can be tested to similar precision.
    New physics that appears in the $D^0$ sector (involving up-type quarks) 
    may be different or complementary to that in the $B^0_d$ or $B^0_s$ 
sectors.
  Direct $\CP$ asymmetries can be measured to the 
  fraction of a percent level in $b \to s \gamma$ decays,
  using both inclusive and exclusive channels,
  and $b \to s \ell^+\ell^-$ can be equally thoroughly explored.
  Equally precise searches for direct $\CP$ violation in 
    charm or $\tau$ decays provide additional NP sensitivity,
    since the SM background is largely absent.
  At the same time, a SFF
  can access channels that are sensitive to NP
  even when there are no new sources of $\CP$ violation,
  such as the photon polarization in $b \to s\gamma$,
  and the branching fractions of $B^+ \to \ell^+ \nu_\ell$,
  the latter being sensitive probes of NP in MFV scenarios
  with large $\tan \beta$. 
  Furthermore, rare FCNC decays of the $\tau$ lepton are particularly
  interesting since lepton flavour violation sources involving the third 
  generation are naturally the largest.  
  Any of these measurements constitutes clear motivation for a SFF.
\item {\it Future metrology of the CKM matrix:} \\
  There are several measurements that are unaffected by NP
  in many likely scenarios, and which allow the extraction
  of the CKM parameters even in the presence of such NP effects.
  Among these, the angle $\gamma$ can be measured with a precision
  of $1$--$2{^\circ}$,
  where the precision is limited only by statistics,
  not by systematics or by theoretical errors.
  By contrast,
  the determination of the elements $|V_{ub}|$ and $|V_{cb}|$
  will be limited by theory,
  but the large data sample of a SFF  will allow
  many of the theoretical errors to be much improved.
  With anticipated improvements in lattice QCD calculations,
  the precision on $|V_{ub}|$ and $|V_{cb}|$ can be driven
  down to the percent level.
  These measurements could allow
  tests of the consistency of the Standard Model at a few per mille level and provide the
  NP phenomenological analyses with a determination of the CKM
  matrix at the percent level.
\end{itemize}

In Table~\ref{tab:superb} we give indicative estimates of the precision 
on some of the most important observables that can be achieved by a 
SFF with integrated luminosity of $50$--$75 \ {\rm ab}^{-1}$.
Here we have not attempted to comment on  the whole range of measurements 
that can be performed by such a machine,
but instead focus on channels with the greatest phenomenological impact.
For more details, including a wide range of additional measurements, 
we guide the reader to the reports~\cite{Hashimoto:2004sm,Bona:2007qt,Akeroyd:2004mj,Hewett:2004tv,superKEKB}, where also all original references are 
given.

\begin{table}[!ht]
  \begin{center}
    \caption{Expected sensitivity that can be achieved 
      on some of the most important observables,
      by a SFF with integrated luminosity of 
      $50$--$75 \ {\rm ab}^{-1}$. The range of values given
allow for possible variation in the total integrated luminosity, in the
accelerator and detector design, and in limiting systematic effects.
      For further details, refer to~\cite{Bona:2007qt,superKEKB}.
    }
    \label{tab:superb}    
    \begin{tabular}{l@{\hspace{15mm}}c}
      \hline \hline
      Observable                      & Super Flavour Factory  sensitivity \\
      \hline
      $\sin(2\beta)$ ($J/\psi\,K^0$)        &  0.005--0.012      \\
      $\gamma$ ($B \to D^{(*)}K^{(*)}$)       &  $1$--$2^\circ$  \\
      $\alpha$ ($B \to \pi\pi, \rho\rho, \rho\pi$)  &  $1$--$2^\circ$    \\
      $\left| V_{ub} \right|$ (exclusive)   &  $3$--$5\%$   \\
      $\left| V_{ub} \right|$ (inclusive)   &  $2$--$6\%$   \\
      \hline
      $\bar{\rho}$                          & $1.7$--$3.4\%$  \\
      $\bar{\eta}$                          & $0.7$--$1.7\%$  \\
      \hline
      $S(\phi K^0)$                         &  0.02--0.03        \\
      $S(\eta^\prime K^0)$                  &  0.01--0.02        \\
      $S(\KS\KS\KS)$                        &  0.02--0.04        \\
      \hline
      $\phi_D$                              & $1$--$3^\circ$      \\
      \hline
      $\BR(B \to \tau \nu)$                 &  $3$--$4\%$         \\
      $\BR(B \to \mu \nu)$                  &  $5$--$6\%$         \\
      $\BR(B \to D \tau \nu)$               &  $2$--$2.5\%$       \\
      \hline
      $\BR(B \to \rho \gamma)/\BR(B \to K^* \gamma)$ &  $3$--$4\%$ \\
      $A_{CP}(b \to s \gamma)$              &  $0.004$--$0.005$   \\
       $A_{CP}(b \to (s+d) \gamma)$          &  $0.01$             \\
      $S(\KS\pi^0\gamma)$                   &  $0.02$--$0.03$     \\
      $S(\rho^0\gamma)$                     &  $0.08$--$0.12$        \\
      $A^{\rm FB}(B \to X_s \ell^+ \ell^-) \ s_0$ &  $4$--$6\%$              \\
      $\BR(B \to K \nu \bar{\nu})$     &  $16$--$20\%$             \\
      \hline
      $\BR(\tau \to \mu\gamma)$         & $2$--$8 \times 10^{-9}$  \\
      $\BR(\tau \to \mu\mu\mu)$     & $0.2$--$1 \times 10^{-9}$  \\
      $\BR(\tau \to \mu\eta)$           & $0.4$--$4 \times 10^{-9}$  \\
      \hline \hline
    \end{tabular}
  \end{center}
\end{table}

{\it The most important measurements within the CKM metrology are} 
the angles of the Unitarity Triangle, 
the angle $\beta$ (also known as $\phi_1$), 
measured using mixing-induced $\CP$ violation in $B^0 \to J/\psi\,K^0$, 
the angle $\alpha$ ($\phi_2$),
measured using rates and asymmetries in 
$B \to \pi\pi$~\footnote{Notice that this method for extracting $\alpha$ is insensitive to NP in QCD penguins. However it could be affected by isospin-breaking NP contributions.}, $\rho\pi$ and $\rho\rho$, 
and the angle $\gamma$ ($\phi_3$),
measured using rates and asymmetries in $B \to D^{(*)}K^{(*)}$ decays,
using final states accessible to both $D^0$ and $\bar{D}^0$. 
Moreover, a SFF will improve our knowledge of the lengths  
of the sides of the Unitarity Triangle.
In particular, the CKM matrix element $\left| V_{ub} \right|$
will be precisely measured through both inclusive and exclusive
semileptonic $b \to u$ decays.

{\it Among the measurements sensitive for New Physics,} 
there are the mixing-induced $\CP$ violation parameters 
in charmless hadronic $B$ decays dominated by the $b \to s$ penguin transition,
$S(\phi K^0)$, $S(\eta^\prime K^0)$ and $S(K^0_S K^0_S K^0_S)$.
Within the Standard Model these give the same value of $\sin(2\beta)$
that is determined in $B^0 \to J/\psi\,K^0$ decays,
up to a level of theoretical uncertainty that is 
estimated to be $\sim 2$--$5\%$ within factorization.
(The theoretical error in these and other modes, such as $B\to K_S \pi^0$, 
can be also bounded with data-driven methods~\cite{Silvestrini:2007yf}. 
Presently these give larger uncertainties but will become more precise
as more data is available.)
Many extensions of the Standard Model result in deviations 
from this prediction. 
Another distinctive probe of new sources of $\CP$ violation is $\phi_D$, 
the $\CP$ violating phase in neutral $D$ meson mixing,
which is negligible in the SM and can be precisely measured using,
for example, $D \to \KS\pi^+\pi^-$ decays.
Furthermore, branching fractions for leptonic and semileptonic $B$ decays 
are sensitive to charged Higgs exchange.
In particular these modes are sensitive to new physics,
even in the unfavourable minimal flavour violation scenario,
with a large ratio of the Higgs vacuum expectation values, $\tan \beta$. 
Measurements of rare radiative and electroweak penguin processes
are well-known to be particularly  sensitive to new physics:
The ratio of branching fractions
$\BR(B \to \rho \gamma)/\BR(B \to K^* \gamma)$
depends on the ratio of CKM matrix parameters
$\left| V_{td} / V_{ts} \right|$,
with additional input from lattice QCD.
Within the Standard Model this result must be consistent with 
constraints from the Unitarity Triangle fits.
The inclusive $\CP$ asymmetries $A_{CP}(b \to s \gamma)$
or $A_{CP}(b \to (s+d)  \gamma)$
are predicted in the Standard Model to be small or exactly zero respectively
with well understood theoretical uncertainties.
The mixing-induced $\CP$ asymmetry in radiative $b \to s$ transitions,
measured for example through $S(K^0_S \pi^0\gamma)$,
is sensitive to the emitted photon polarization.
Within the SM the photon is strongly polarized,
and the mixing-induced asymmetry small,
but new right-handed currents can break this prediction
even without the introduction of any new $\CP$ violating phase.
Similarly, $S(\rho^0\gamma)$ probes radiative $b \to d$ transitions.
The dilepton invariant mass squared $s$ 
at which the forward-backward asymmetry 
in the distribution of $B \to X_s \ell^+ \ell^-$ decays is zero
(denoted $A^{\rm FB}(B \to X_s \ell^+ \ell^-) \ s_0$),
for which the theoretical uncertainty 
of the Standard Model prediction is small,
is sensitive to NP in electroweak penguin operators; 
finally, the branching fraction for the rare electroweak penguin decay 
$B \to K \nu \bar{\nu}$ is an important probe for NP  
even if this appears only well above the electroweak scale. 
A SFF also allows for the 
measurement of branching ratios of lepton flavour violating $\tau$ decays, 
such as  $\tau \to \mu\gamma$, $\tau \to \mu\mu\mu$ and $\tau \to \mu\eta$.
Within the Standard Model, these are negligibly small,
but many models of new physics create observable lepton flavour
violation signatures.

For some of the entries of Table~\ref{tab:superb}  
some additional comments are in order: 
\begin{itemize}
\item  
  With such large data samples as will be accumulated by a SFF,
  the uncertainty on several measurements 
  will be dominated by systematic errors.
  Estimating the ultimate precision therefore requires some knowledge
  of how these systematic uncertainties can be improved.
  One such important channel is the mixing-induced $\CP$ asymmetry
  in $B^0 \to J/\psi\,K^0$, which measures $\sin(2\beta)$ in the SM.
  The systematic uncertainties in the current $B$ Factory analyses
  are around $1$--$2\%$, 
  coming mainly from uncertainties in the vertex detector alignment 
  and beam spot position.
  Another example is direct $\CP$ asymmetry,
  both in exclusive and inclusive modes.
  Measurements with precision better than $1\%$ require knowledge
  of detector asymmetries at the same level.
  Reduction of these errors will be highly challenging,
  but there is some hope that improvement 
  by a factor of about two may be possible. 
\item 
  The precision that can be achieved on $\left| V_{ub} \right|$
  depends on improvements in the theoretical treatment.
  The most notable effect is for the exclusive channels,
  where reduction of the error on form factors calculated in 
  lattice QCD  is extremely important.
\item
    The sensitivities for some measurements depend on hadronic parameters
    that are not yet well known.
    For example, for $\phi_D$ to be measured at least one of the 
    $D^0$--$\bar{D}^0$ mixing parameters $x_D$ and $y_D$ must be nonzero.
    The first evidence for charm mixing has recently been reported~\cite{Staric:2007dt,Aubert:2007wf},
    but large ranges for the obtained parameters are still allowed.
    Our estimate of the sensitivity is obtained by extrapolating 
    results from the $D\to K_S \pi^+\pi^-$ time-dependent analysis~\cite{Abe:2007rd},
    which currently appears to be the single most sensitive channel,
    although better constraints can certainly be obtained by combining
    information from multiple decays modes.
\item 
  The specific details of the accelerator and detector configuration
  are important considerations for some measurements.
  For studies of mixing-induced $\CP$ asymmetry
  that obtain the $B$ decay vertex position from a reconstructed $\KS$ meson
  (such as $B^0\to\KS\KS\KS$ and $B^0\to\KS\pi^0\gamma$)
  the geometry of the vertex detector plays an important role --
  better precision is achieved for a larger vertex detector.
  Similarly, several channels with missing energy
  (such as $B\to\tau\nu_\tau$, $B \to D\tau\nu_\tau$ and $B\to K\nu\bar{\nu}$)
  make full use of the constraints available in 
  $\Upsilon(4{\rm S}) \to B\bar{B}$ decays
  by fully reconstructing one $B$ meson to know the kinematics of the other.
  Such measurements are dependent on the background condition 
  and the hermeticity of the detector.  
  Indeed, it is obvious that the sensitivity for all measurements 
  depends strongly on the detector performance,
  and improvements in, {\it e.g.}, vertexing and particle identification
  capability will be of great benefit to separate signal from background.
\item 
  The sensitivity to very rare processes,
  such as the lepton flavour violating decay $\tau \to \mu\gamma$
  depends strongly on how effectively the background may be reduced
  and on other possible improvements to the analysis techniques used.
\end{itemize}

\begin{figure}[!ht]
  \begin{center}
    \includegraphics[width=9cm]{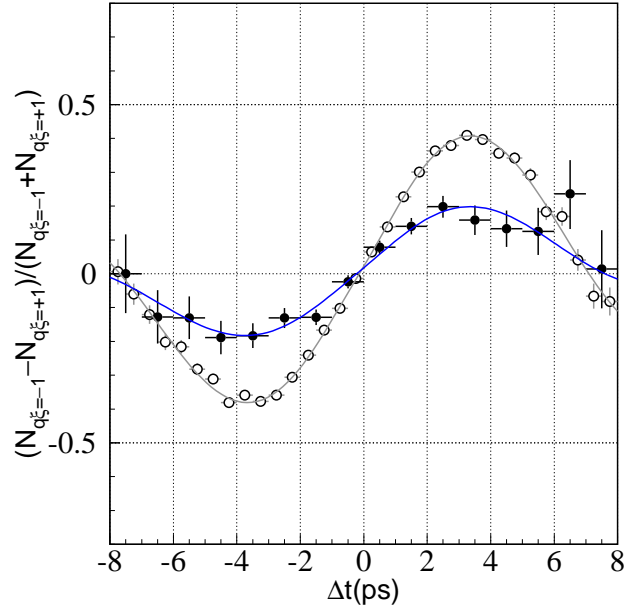}
    \caption{
      Simulation of new physics effects in $B^0 \to \phi K^0$,
      as could be observed by a SFF.  
      The open circles show simulated $B^0 \to J/\psi\,K^0$ events,
      the filled circles show simulated $B^0 \to \phi K^0$ events.
      Both have curves showing fit results superimposed.
      From~\cite{superKEKB}.
    }
    \label{fig:phiKs_data}
  \end{center}
\end{figure}

{\it The sensitivities  of these measurements to New Physics effects} 
may be shown  by a few examples:
In Figure~\ref{fig:phiKs_data} we show a simulation of the 
time-dependent asymmetry in $B^0 \to \phi K^0$,
compared to that for $B^0 \to J/\psi\,K^0$.
The events are generated using the current central values of the measurements.
With the precision of a SFF and the present central values, the difference between the two data sets is larger than the theoretical expectation, showing evidence of NP contributions.

\begin{figure}[!ht]
  \begin{center}
    \includegraphics[width=9cm]{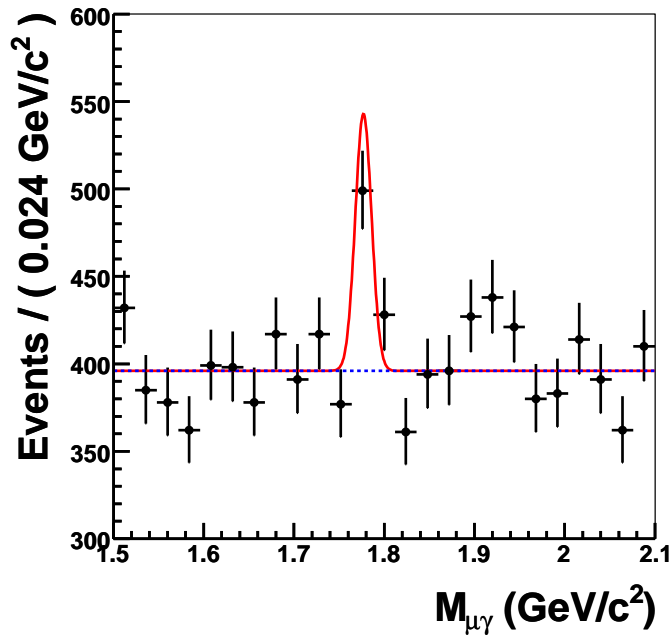}
    \caption{
      Monte Carlo simulation of the appearance of $\tau \to \mu \gamma$
      at a SFF.
      A clear peak in the $\mu\gamma$ invariant mass distribution 
      is visible above the background.
      The branching fraction used in the simulation is
      $\BR(\tau \to \mu \gamma) = 10^{-8}$,
      an order of magnitude below the current upper limit.
      With $75 \ {\rm ab}^{-1}$ of data the significance
      of such a decay is expected to exceed $5\sigma$.
    }
    \label{fig:tauLFV_data}
  \end{center}
\end{figure}

In Figure~\ref{fig:tauLFV_data} we show how lepton flavour violation
in the decay $\tau \to \mu \gamma$ may be discovered at a SFF.
The simulation corresponds to a branching fraction of 
$\BR(\tau \to \mu \gamma) = 10^{-8}$,
which is within the range predicted by many new physics models.
The signal is clearly observable, 
and well within the reach of a SFF.
The simulation includes the effects of irreducible background 
from initial state radiation photons,
though improvements in the detector and in the analysis 
may lead to better control of this limitation.
Other lepton flavour violating decay modes, 
such as $\tau \to \mu\mu\mu$ do not suffer from this background,
and have correspondingly cleaner experimental signatures.

The differences between the SFF physics programme and those of the 
current $B$ factories are striking.
At a SFF measurements of known rare processes such as 
$b\to s \gamma$ or $\CP$ violation in hadronic $b\to s$ penguin transitions
such as $B^0 \to \phi \KS$ will be advanced to unprecedented precision.
Channels which are just being observed in the existing data,
such as $B^0 \to \rho^0\gamma$, $B^+ \to \tau^+\nu_\tau$ 
and $B \to D^{(*)}\tau\nu$ will become precision measurements at a SFF.
Furthermore, detailed studies of decay distributions and asymmetries
that cannot be performed with the present statistics,
will enable the sensitivity to NP to be significantly improved.
Another salient example lies in $D^0$--$\bar{D}^0$ oscillations:
the current evidence for charm mixing, which cannot be interpreted
in terms of New Physics, opens the door for precise measurements of 
the $\CP$ violating phase in charm mixing, which is known to 
be zero in the Standard Model with negligible uncertainty.

In addition, these measurements will be accompanied by 
dramatic discoveries of new modes and processes. 
These will include decays such as $B \to K \nu \bar{\nu}$, 
which is the signature of the theoretically clean quark level process 
$b \to s \nu \bar{\nu}$. 
The high statistics and 
clean environment of a SFF allow for the accompanying $B$ meson
to be fully reconstructed in a hadronic decay mode, which then
in turn allows a one-charged prong rare decay to be isolated. 
Another example is $B^+ \to \pi^+ \ell^+ \ell^-$, 
the most accessible $b\to d \ell^+\ell^-$ process. 
These decays are the next level beyond $b\to s \ell^+\ell^-$ decays, 
which were first observed in the $B$ Factory era. 
Such significant advances will result in a strong phenomenological impact
of the Super Flavour Factory physics programme.

{\it Comparison with LHCb:} 
Since a SFF will take data in the LHC era,
it is reasonable to ask how the physics reach
compares with the $B$ physics potential of the LHC experiments,
most notably LHCb.
By 2014, the LHCb experiment is expected to have accumulated
$10 \ {\rm fb}^{-1}$ of data
from $pp$ collisions at a luminosity of
$\sim 2 \times 10^{32} \ {\rm cm}^{-2} {\rm s}^{-1}$.
In the following we assume the most recent estimates of LHCb 
sensitivity with that data set~\cite{ref:schneider}.
Note that LHCb is planning an upgrade where they would run
at 10 times the initial design luminosity and record a data sample of 
about $100 \ {\rm fb}^{-1}$, see Section~\ref{sec:lhcb-upgrade} and~\cite{ref:wilkinson}.

The most striking outcome of any comparison between SFF and LHCb
is that the strengths of the two experiments are largely complementary.
For example, the large boost of the $B$ hadrons produced at LHCb 
allows studies of the oscillations and mixing-induced CP violation 
of $B_s$ mesons
while many of the measurements that constitute the primary
physics motivation for a SFF cannot be performed
in the hadronic environment, including rare decay modes with missing energy
such as $B^+ \to \ell^+\nu_\ell$ and $B^+ \to K^+\nu\bar{\nu}$. 
Measurements of the CKM matrix elements $|V_{ub}|$ and $|V_{cb}|$
and inclusive analyses of processes such as $b \to s\gamma$
also benefit greatly from the SFF environment.
At LHCb the reconstruction efficiencies are reduced 
for channels containing several neutral particles and 
for studies where 
the $B$ decay vertex must be determined from a $K^0_S$ meson. 
Consequently, a SFF has unique potential to measure the photon polarization
via mixing-induced $\CP$ violation in $B^0 \to K^0_S \pi^0 \gamma$.
Similarly, a SFF is well placed to study possible NP effects in
hadronic $b \to s$ penguin decays as it 
can measure precisely the $\CP$ asymmetries in many $B^0_d$ decay modes 
including $\phi K^0$, $\eta^\prime K^0$, $K^0_S K^0_S K^0_S$ or 
$K^0_S \pi^0$. While LHCb will have limited capability for these channels,
it can achieve complementary measurements
using decay modes such as $B^0_s \to \phi \gamma$ and $B^0_s \to \phi\phi$
for radiative and hadronic $b \to s$ transitions respectively.

Where there is overlap,
the strength of the SFF programme in its ability to use multiple 
approaches to reach the objective becomes apparent.
For example, LHCb will be able to measure
$\alpha$ to about $5^\circ$ precision using $B \to \rho\pi$,
but would not be able to access the full information in the
$\pi\pi$ and $\rho\rho$ channels, which is necessary to drive
the uncertainty down to the $1$--$2^\circ$ level of a SFF.
Similarly, LHCb can certainly measure $\sin(2\beta)$
through mixing-induced $\CP$ violation in $B^0 \to J/\psi K^0_S$ decay
to high accuracy (about 0.01),
but will have less sensitivity to make the complementary measurements
({\it e.g.}, in $J/\psi\,\pi^0$ and $Dh^0$)
that help to ensure that the theoretical uncertainty is under control.
LHCb plans to measure the angle $\gamma$ with a precision
of $2$--$3^\circ$.
A SFF is likely to be able to improve this precision to about $1^\circ$. 
LHCb can make a precise measurement of the zero of the
forward-backward asymmetry in $B^0 \to K^{*0}\mu^+\mu^-$,
but a SFF  can also measure the inclusive channel $b \to s \ell^+\ell^-$,
which is theoretically a significantly cleaner observable~\cite{Ghinculov:2003qd}. 

The broad program of a SFF thus provides a very comprehensive set
of measurements, extending what will already have been achieved
by LHCb at that time.  This will be of great importance for the study of
flavour physics in the LHC era and beyond.


\subsubsection{Phenomenological Impact}
\begin{figure}[tb!]
  \begin{center}
    \includegraphics[width=0.45\textwidth]{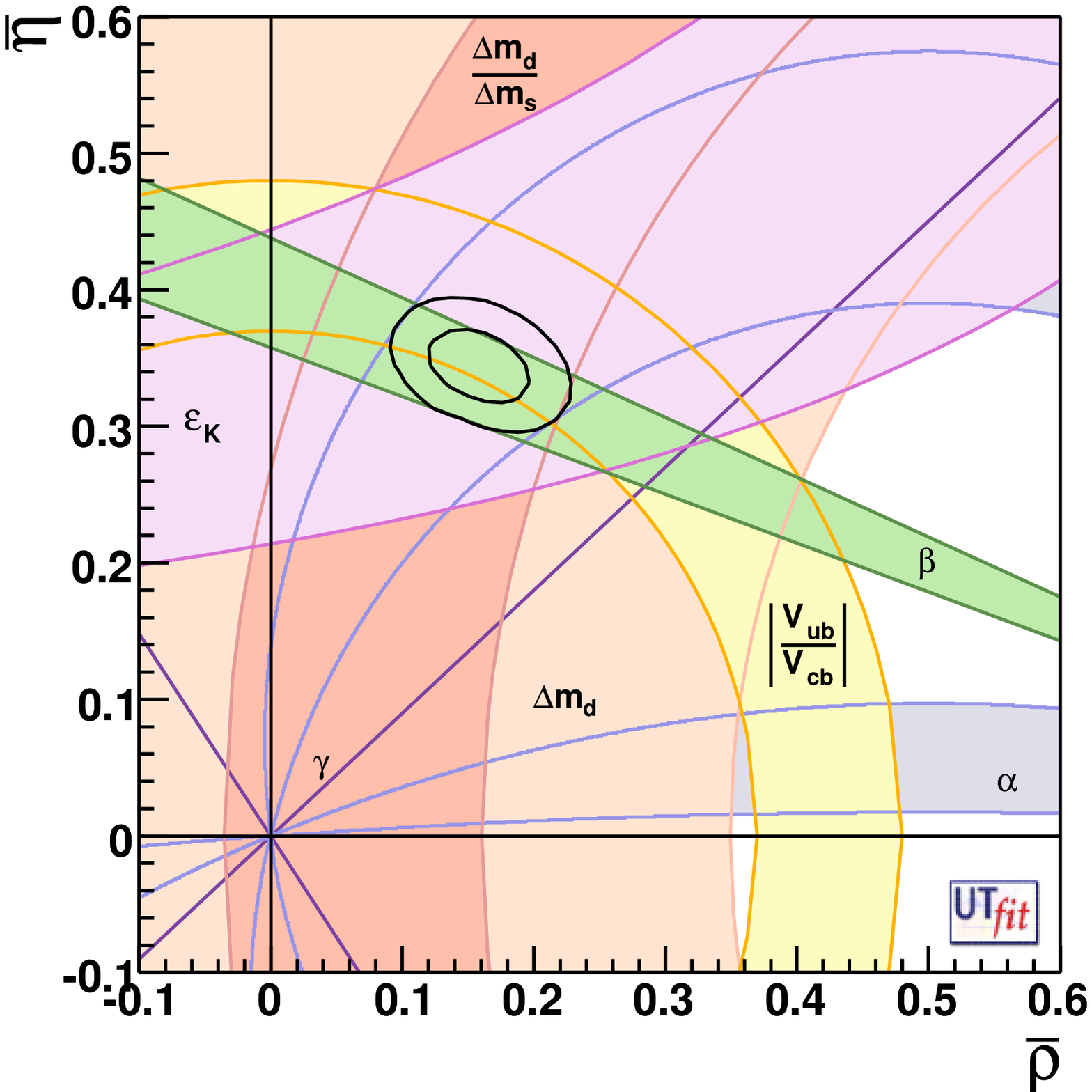}
    \includegraphics[width=0.45\textwidth]{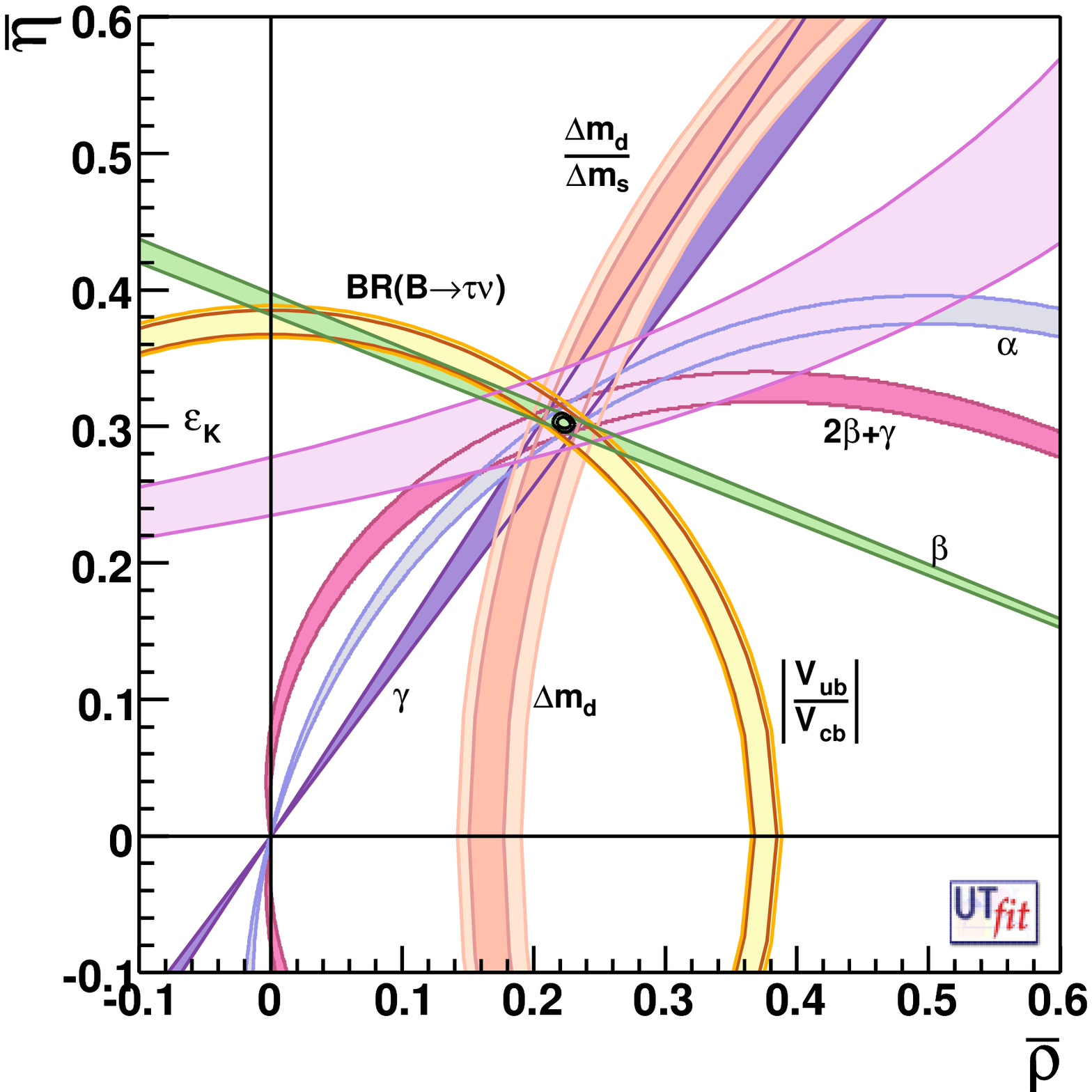}
    \caption{
      Regions corresponding to $95\%$ probability for the CKM
      parameters $\bar\rho$
      and $\bar\eta$ selected by different constraints, assuming
      present central values with present errors (left) or with
      errors expected at a  SFF tuning central values to have
      compatible constraints (right).
    }
    \label{fig:ckm}
  \end{center}
\end{figure}
The power of a SFF to observe NP and to determine the 
CKM parameters precisely is manifold.
In the following, we present a few highlights  of  the phenomenological
impact (for more detailed analyses see~\cite{Hashimoto:2004sm,Bona:2007qt,Akeroyd:2004mj,Hewett:2004tv,superKEKB}).

\begin{figure}[htb!]
  \begin{center}
    \includegraphics[width=0.45\textwidth]{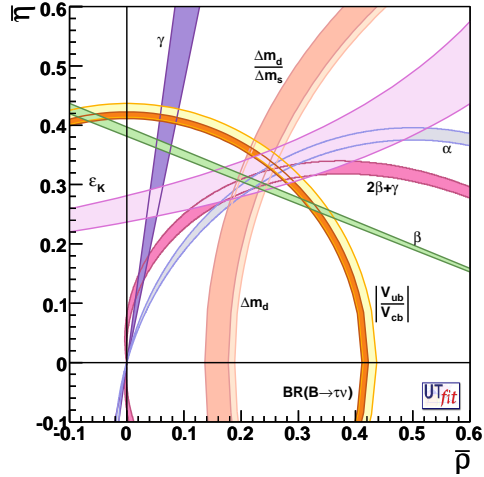}
    \caption{Region corresponding to 95\% probability for the CKM parameter
   $\rhobar$ and $\etabar$ selected by the different constraints, assuming
   todays central values with the precision of a SFF. Note for example
that the band corresponding to the $\gamma$  measurement does not pass through
the intersection of other constraints.}\label{fig:smfit}
  \end{center}
\end{figure}

{\it Precise Determination of CKM Parameters in the SM:} Most of 
the measurements described in the previous section can be used to select a 
region in the $\rhobar$--$\etabar$ plane as shown in Figure~\ref{fig:ckm}.
The corresponding numerical results are given in Table~\ref{tab:smfit}.
The results indicate that a precision of a fraction of a percent can be 
reached, significantly improving the current situation, and providing 
a generic test of the presence of NP at that level of precision. 
Note that in the right plot of  Figure~\ref{fig:ckm} -  
where the  expected precision  offered by a SFF is used -  
the validity of the SM is assumed, so the compatibility of all constraints 
is put in by hand.
In contrast, in Figure~\ref{fig:smfit} we assume that all results take
the central values of their current world averages
with the expected precision of a SFF.
In this case, the hints of discrepancies present in today's data
 have evolved into fully fledged NP discoveries.

\begin{table}[ht]
  \caption{
    Uncertainties of the CKM parameters obtained from the Standard Model fit
    using the experimental and theoretical information available today (left)
    and at the time of a SFF (right). The precision corresponds to the plots 
    in  Figures~\ref{fig:ckm} and  \ref{fig:smfit}.}
  \begin{center}
    \begin{tabular}{lll}
      \hline\hline
      Parameter               &   SM Fit today        &  SM Fit at a SFF \\
      \hline
      $\overline {\rho}$      &   $0.163\pm 0.028$    &  $\pm 0.0028$      \\
      $\overline {\eta}$      &   $0.344\pm 0.016$    &  $\pm 0.0024$      \\
      $\alpha$ ($^{\circ}$)   &   $92.7\pm4.2$        &  $\pm 0.45$        \\
      $\beta$ ($^{\circ}$)    &   $22.2\pm0.9$        &  $\pm 0.17$        \\
      $\gamma$ ($^{\circ}$)   &   $64.6\pm4.2$        &  $\pm 0.38$        \\
      \hline
    \end{tabular}
  \end{center}
  \label{tab:smfit}
\end{table}
Of course,  many of the measurements used for the SM  determination of
$\rhobar$--$\etabar$ can be affected by the presence of NP.
Thus, unambiguous NP searches require a determination of $\rhobar$  and $\etabar$
in the presence of arbitrary NP contributions, which can be done using $\Delta F=2$ processes.

{\it New Physics in Models with Minimal Flavour Violation:}
The basic assumption of Minimal Flavour Violation (MFV)~\cite{Gabrielli:1994ff,Buras:2000dm,D'Ambrosio:2002ex} is that 
NP does not introduce new sources of flavour and $\CP$ violation.
Hence the only flavour-violating couplings are the SM Yukawa couplings.
One can assume that the top Yukawa coupling is dominant
in the simplest case with one Higgs doublet and - with some exceptions - also
in the case with two Higgs doublets with small $\tan \beta$; this means
that all NP effects amount to a real contribution added to the SM
loop function generated by virtual top exchange. In particular, in the $\Delta B = 2$ amplitude,
MFV NP may be parameterized as $$S_0(x_t) \to S_0(x_t) + \delta S_0$$
where the function $S_0(x_t)$ represents the top contribution
in the box diagrams and $\delta S_0$ is the NP contribution.
Therefore, in this class of MFV models, the NP contribution to 
all $\Delta F = 2$ processes is universal,
and the effective Hamiltonian retains the SM  structure.

\begin{figure}[t]
  \begin{center}
    \includegraphics[width=0.45\textwidth]{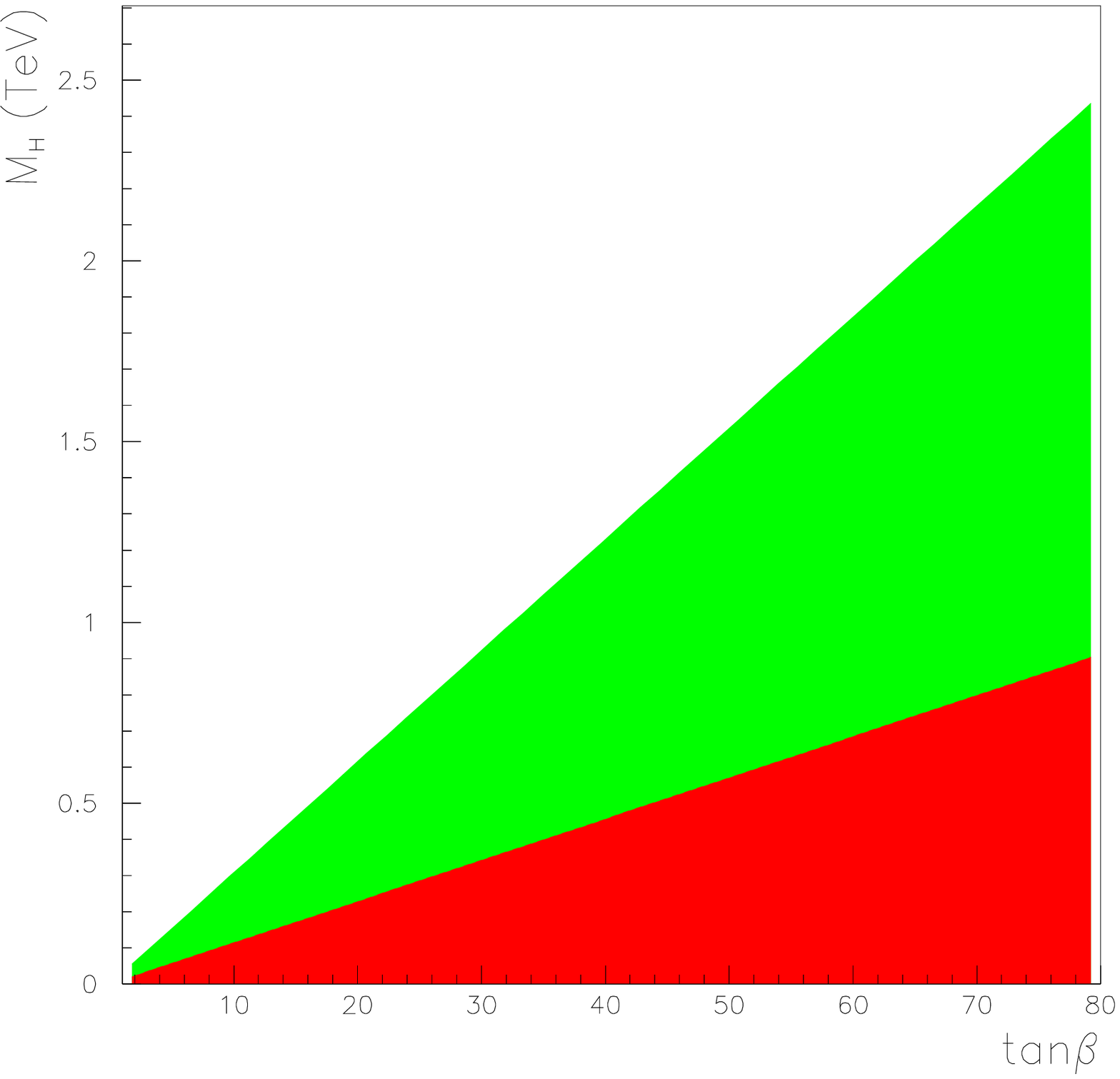} 
    \includegraphics[width=0.45\textwidth]{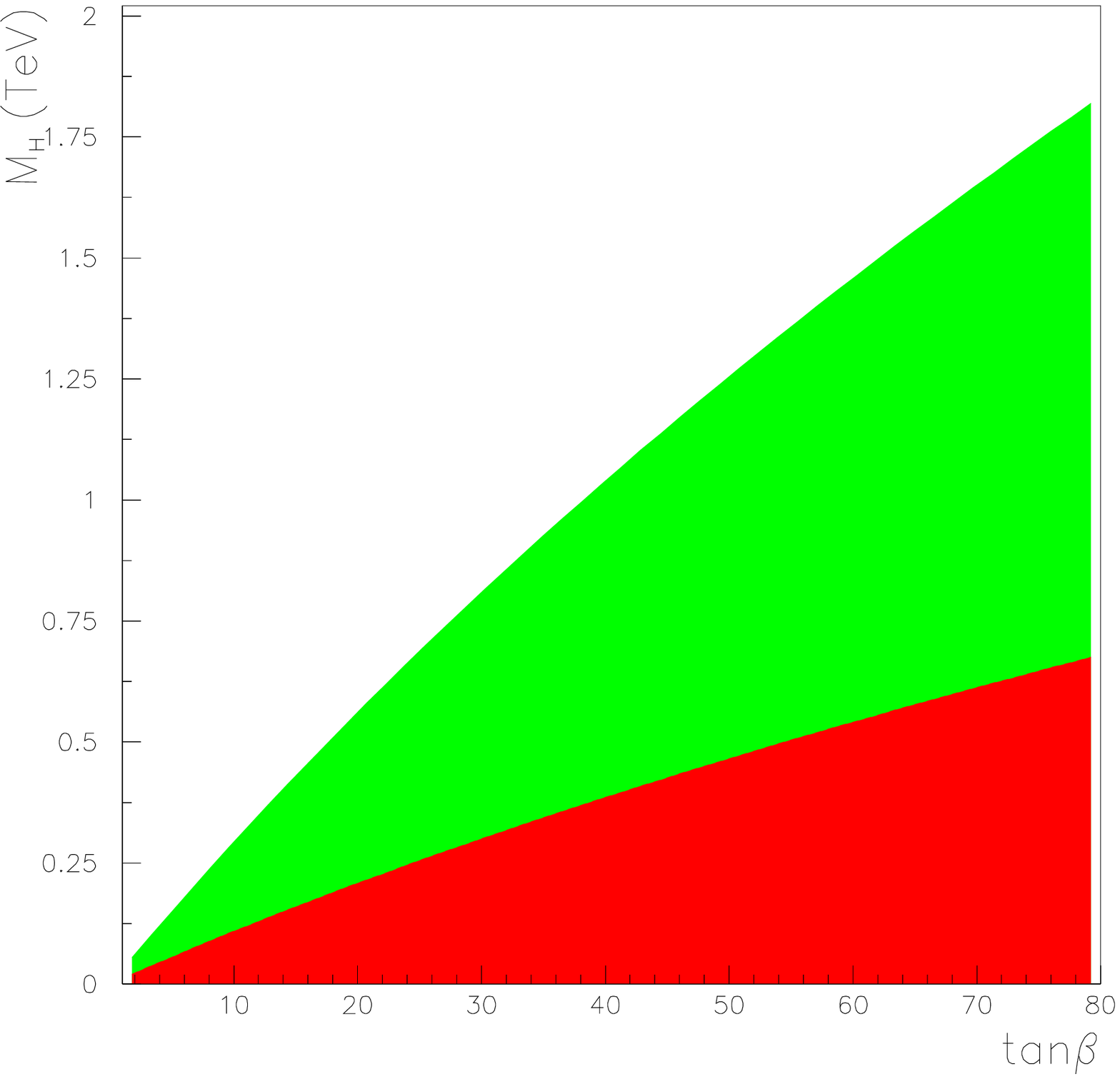} 
    \caption{
      Exclusion regions at 95\% probability 
      in the $M_{H^\pm}$--$\tan \beta$ plane for the 2HDM-II (left) 
      and the MSSM (right) obtained
      assuming the Standard Model value of ${\cal B}(B\to\ell\nu)$ measured 
      with $2 \ {\rm ab}^{-1}$ (dark (red) area) and
      $75 \ {\rm ab}^{-1}$ (dark (red) + light (green) area). 
      In the MSSM case, we have used
      $\epsilon_0 \sim 10^{-2}$~\cite{Isidori:2007zm}.
    }
    \label{fig:btaunu}
  \end{center}
\end{figure}

Following Ref.~\cite{D'Ambrosio:2002ex}, this value can be converted into a NP scale using
\begin{equation}
  \delta S_0 =  4 a \left( \frac{\Lambda_0}{\Lambda}\right)^2\,,
\end{equation}
where $\Lambda_0=Y_t \sin^2 \theta_W M_W/\alpha \approx 2.4 \ {\rm TeV}$
is the SM  scale, $Y_t$ is the top Yukawa coupling,
$\Lambda$ is the NP  scale and $a$ is an unknown (but real)
Wilson coefficient of ${\cal O}(1)$.

The UT analysis can constrain the value of the NP  parameter $\delta S_0$ together
with $\rhobar$ and $\etabar$.
In the absence of a NP  signal, $\delta S_0$ is distributed around zero.
From this distribution, we can obtain a lower bound on the NP scale $\Lambda$.

For a one-Higgs-doublet model (1HDM) or a two-Higgs-doublet model (2HDM) in the low 
$\tan \beta$ regime, the combination of measurements at a SFF 
and the improved lattice 
results give
\begin{equation}
  \Lambda > 14 \ {\rm TeV}~@~95\% \ {\rm CL}
\end{equation}

\begin{figure}[t]
  \begin{center}
    \includegraphics[width=0.45\textwidth]{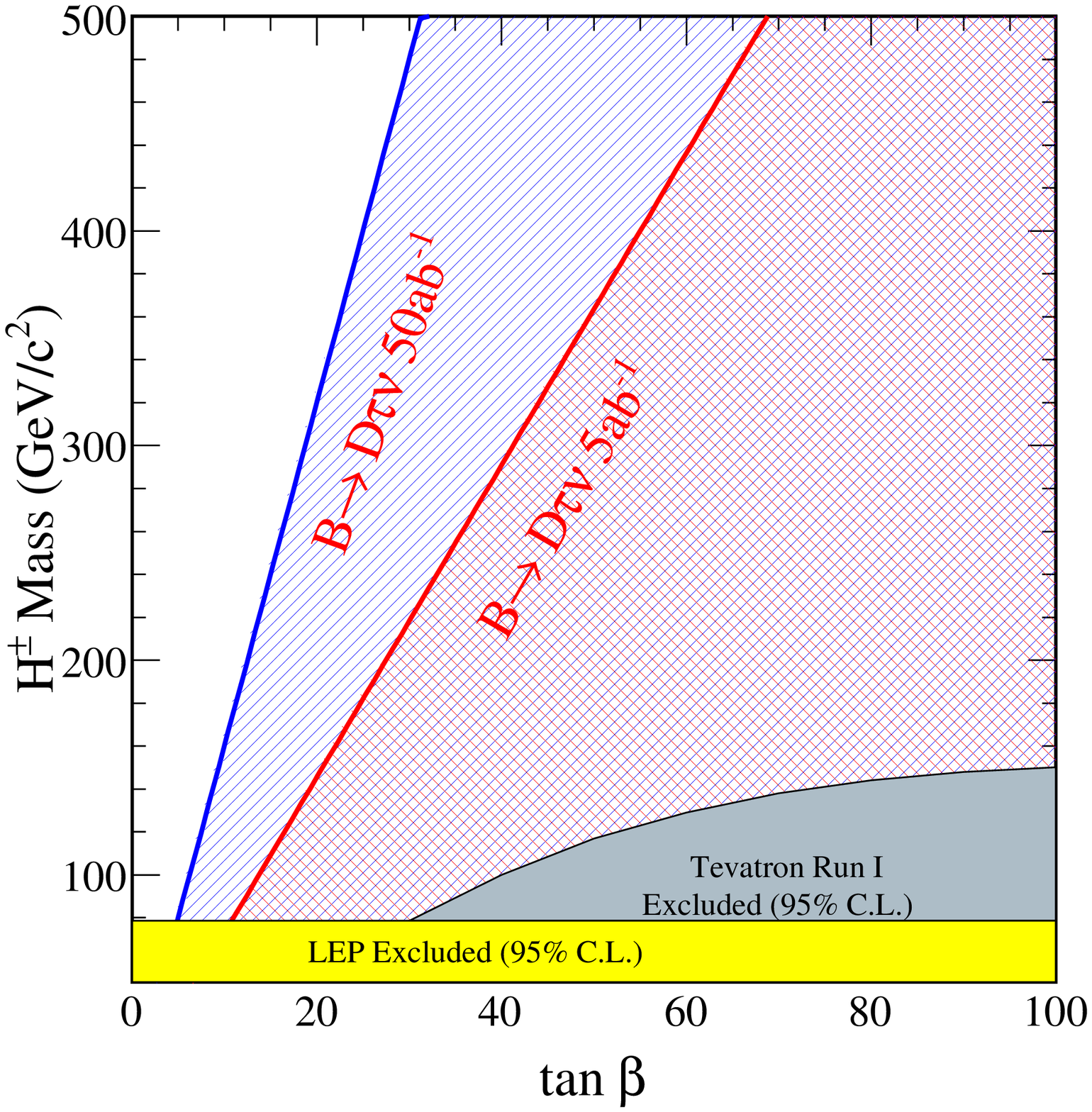}
    \caption{
      Exclusion region 
      in the $M_{H^\pm}$--$\tan \beta$ plane 
      assuming the SM value of ${\cal B}(B\to D\ell\nu)$ measured with
      $5 \ {\rm ab}^{-1}$ and with $50 \ {\rm ab}^{-1}$.}
    \label{fig:btaunu2}
  \end{center}
\end{figure}

These bounds are a factor of three larger than those available 
today~\cite{Bona:2005eu}.
This means that even in the ``worst case'' scenario,
{\it i.e.},~in models with MFV at small $\tan \beta$,
the sensitivity of flavour-violating processes to NP 
is strong enough to allow for the study of the flavour-violating
couplings of new particles with masses up to $600$ GeV.
This conversion to a NP scale in the MFV case deserves 
further explanation.
Consider that the SM reference scale
corresponds to virtual $W$-exchange in the loops. 
As MFV has the same flavour violating couplings as the SM, the
MFV-NP scale is simply translated to a new virtual particle
mass as $\Lambda/\Lambda_0\times  M_W$.
It must be noted, however, that as soon as one considers large $\tan \beta$,
or relaxes the MFV assumption in this kind of analysis,
the NP scale is raised by at least a factor of three,
covering the whole range of masses accessible at the LHC.
In fact the RGE-enhanced contribution of the scalar operators
(absent or subleading in the small $\tan \beta$ MFV case)
typically sets bounds an order of magnitude stronger than those
on the SM current-current operator,
correspondingly increasing the lower bound on the NP scale.
This is the case, for instance, in the Next-to-Minimal Flavour Models (NMFV)  
discussed in Ref.~\cite{Agashe:2005hk} as described in 
the analysis of Ref.~\cite{Bona:2007vi}.

The large $\tan \beta$ scenario offers additional opportunities
to reveal NP by enhancing flavour-violating couplings in
$\Delta B = 1$ processes with virtual Higgs exchange.
This can be the case in decays such as $B\to\ell\nu$ or $B\to D\tau\nu$
whose branching ratios are strongly affected by a charged Higgs
for large values of $\tan\beta$.
In Figure~\ref{fig:btaunu} we show the region excluded in the
$M_{H^\pm}$--$\tan \beta$ plane by the measurement of ${\cal B}(B\to\ell\nu)$
with the precision expected at the end of the current $B$~Factories
and at a SFF, assuming the central value given by the SM.
It is apparent that a SFF  pushes the lower bound on $M_{H^\pm}$,
corresponding, for example, to $\tan \beta \sim 50$ from the hundreds 
of GeV region up to about 2 TeV, both in the 2HDM-II and in the MSSM.
Another interesting possibility is 
to test lepton flavour universality by measuring the ratio
$R_B^{\mu/\tau} = {\cal B}(B\to\mu\nu)/{\cal B}(B\to\tau\nu)$,
which could have a ${\cal O}(10\%)$ deviation from its SM value
at large $\tan \beta$~\cite{Isidori:2006pk,Masiero:2005wr},
whereas the relative error on the individual branching fraction measurements
at a SFF  is expected to be $5\%$ or less. 
In Figure \ref{fig:btaunu2} we show the region excluded in the
$M_{H^\pm}$--$\tan \beta$ plane by the measurement of ${\cal B}(B\to D\ell\nu)$
at a SFF, assuming the central value given by the SM.

{\it MSSM with Generic Squark Mass Matrices:}
\begin{figure}[t]
  \begin{center}
    \begin{tabular}{cc}
      \includegraphics[width=0.46\textwidth]{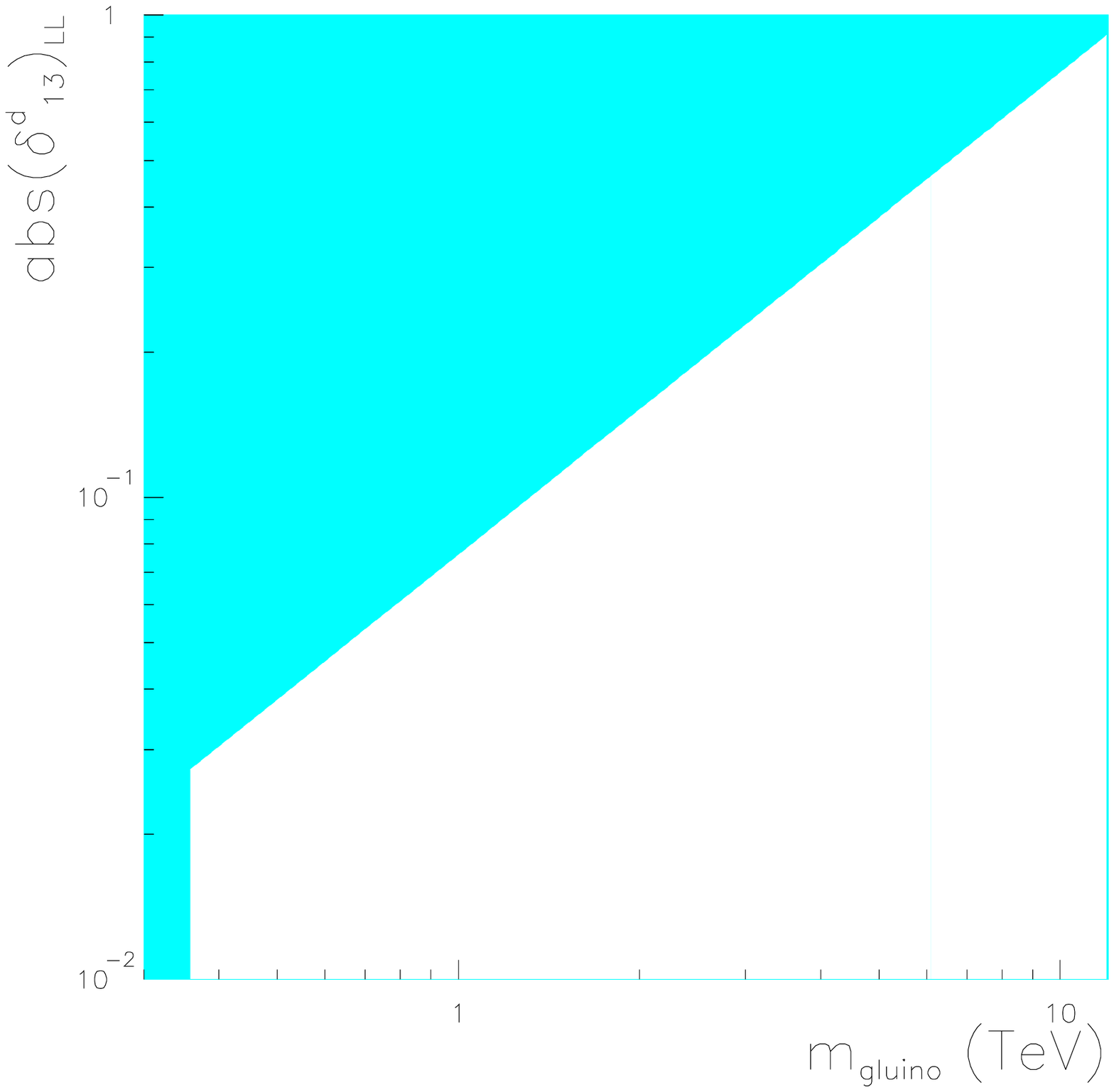} &
      \includegraphics[width=0.46\textwidth]{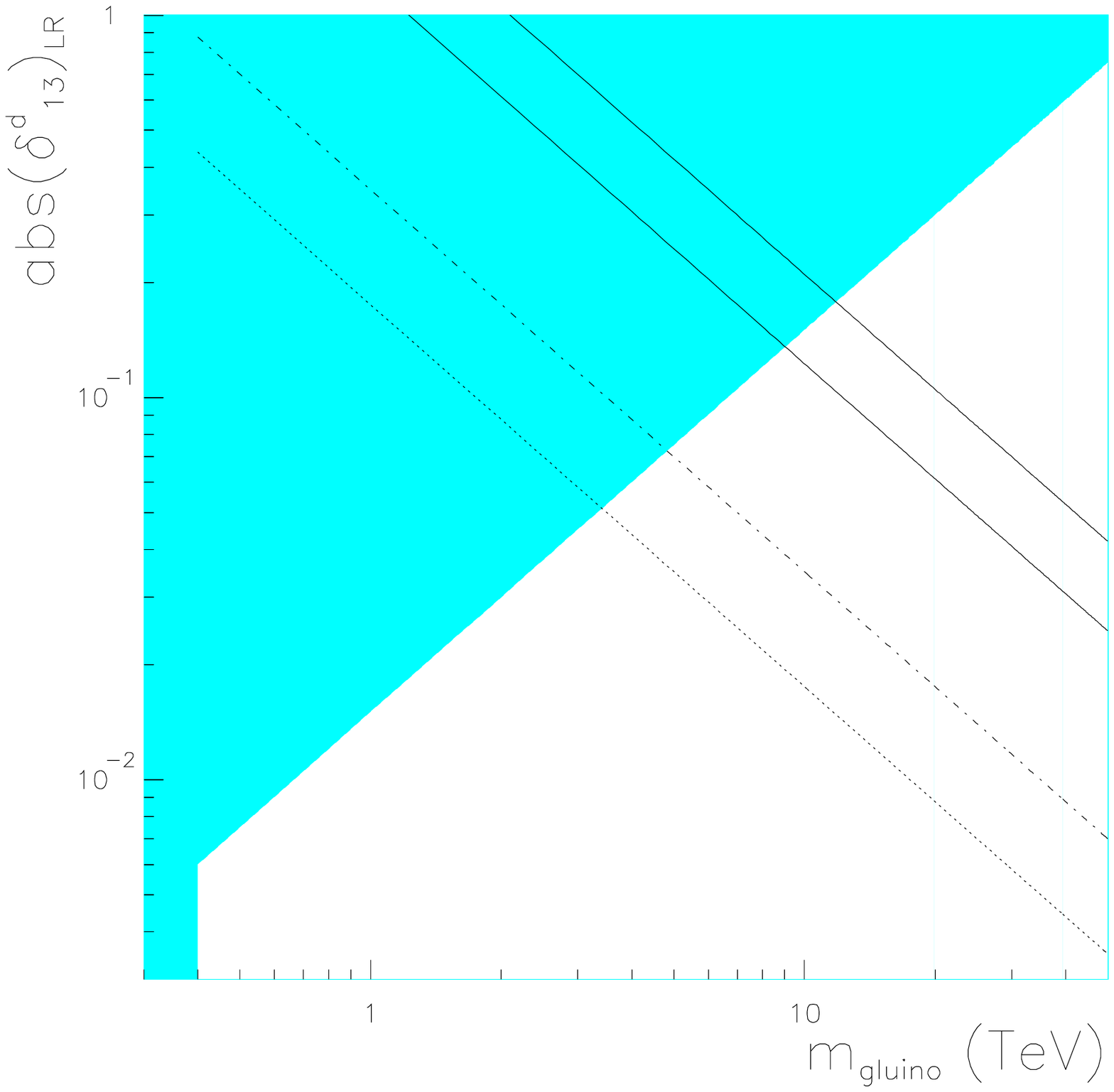} \\
      \includegraphics[width=0.46\textwidth]{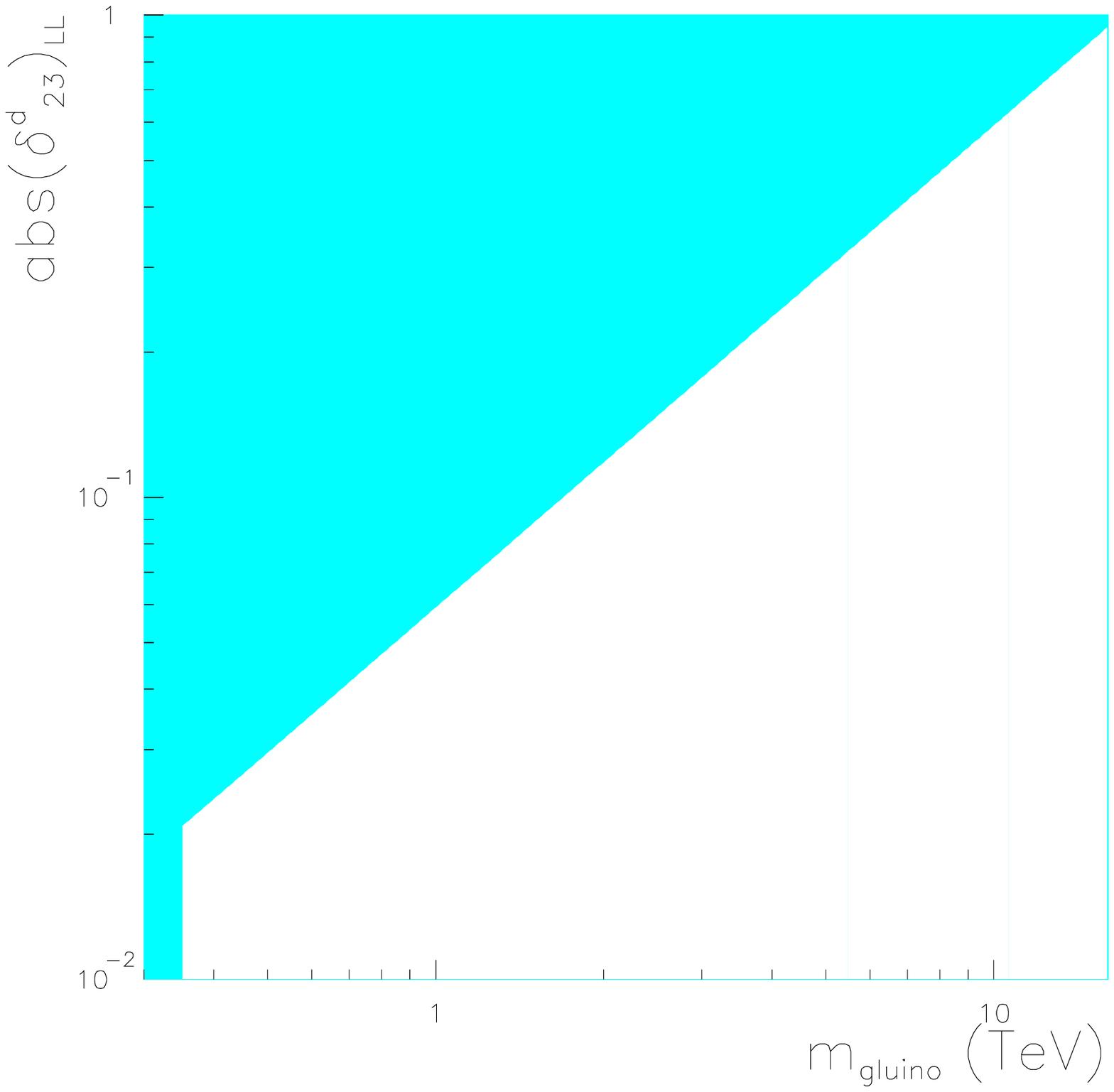} &
      \includegraphics[width=0.46\textwidth]{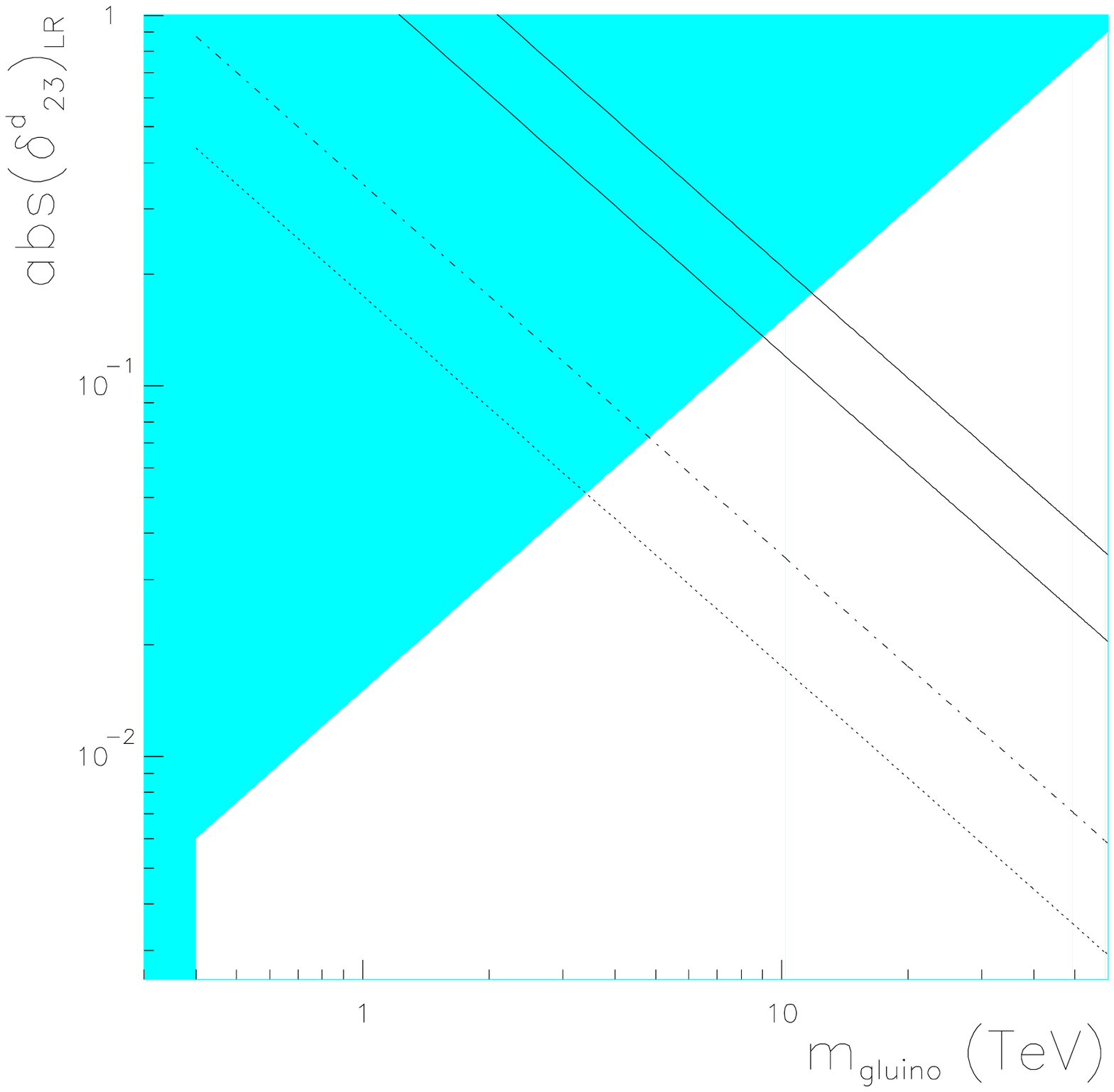}
    \end{tabular}
    \caption{
      Sensitivity region of SFF in the
      $m_{\tilde g}$--$\vert(\delta^d_{ij})_{AB}\vert$ plane.
      The region is obtained by requiring that the reconstructed
      MI is $3\sigma$ away from zero.
      The cases of $(\delta^d_{13})_{LL}$ (upper left),
      $(\delta^d_{13})_{LR}$ (upper right),
      $(\delta^d_{23})_{LL}$ (lower left) and
      $(\delta^d_{23})_{LR}$ (lower right) are shown.
      For LR MIs the theoretical upper bound (allowed parameter region is below these lines)
      discussed in the text is also shown for $\tan \beta= 5, 10, 35, 60$
      (dashed, dotted, dot-dashed, solid line respectively).
    }
    \label{fig:MIvsMg}
  \end{center}
\end{figure}
There is also an impressive  impact of a SFF  on
the parameters of the MSSM with generic squark mass matrices
parameterized using the mass insertion (MI) approximation~\cite{Hall:1985dx}.
In this framework, the NP  flavour-violating couplings are the complex MIs.
For simplicity, we consider only the dominant gluino contribution.
The relevant parameters are therefore the gluino mass $m_{\tilde g}$,
the average squark mass $m_{\tilde q}$ and the MIs $(\delta^{d}_{ij})_{AB}$,
where $i,j=1,2,3$ are the generation indices and
$A,B=L,R$ are the labels referring to the helicity of the SUSY partner quarks.
For example, the parameters relevant to $b \to s$ transitions are
the two SUSY masses and the four MIs $(\delta^{d}_{23})_{LL,LR,RL,RR}$.
In order to simplify the analysis, we consider the
contribution of one MI at a time.
This is justified to some extent by the hierarchy of
the present bounds on the MIs.
In addition, barring accidental cancellations,
the contributions from two or more MIs would produce larger NP effects
and therefore make the detection of NP  easier,
while simultaneously making the phenomenological analysis more 
involved~\cite{Borzumati:1999qt,Besmer:2001cj}.
\begin{figure}[t]
  \begin{center}
    \begin{tabular}{cc}
     \multicolumn{2}{c}{
      \includegraphics[width=0.65\textwidth]{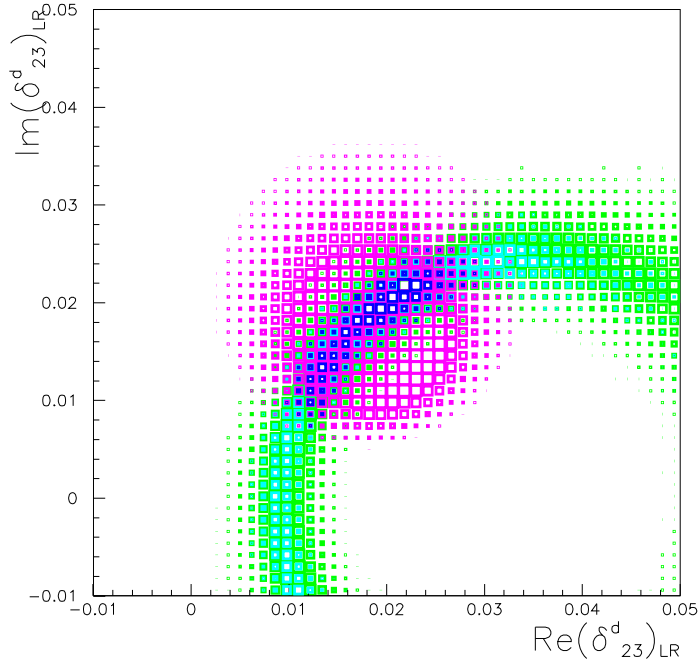}} \\
    \end{tabular}
    \caption{
      Density plot of the region in the
      $\Re(\delta^d_{23})_{LR}$--$\Im(\delta^d_{23})_{LR}$ for
      $m_{\tilde q}=m_{\tilde g}=1 \ \mathrm{TeV}$ generated
      using SFF measurements. Different colours correspond to different
      constraints: $\BR(B\to X_s\gamma)$ (green),
      $\BR(B\to X_s\ell^+\ell^-)$ (cyan), $A_{CP}(B\to X_s\gamma)$ (magenta),
      all together (blue). Central values of constraints corresponds to
      assuming $(\delta^d_{13})_{LL}=0.028 e^{i\pi/4}$.
    }
    \label{fig:MI23LR}
  \end{center}
\end{figure}
The analysis presented here  is based on results and techniques
developed in Refs.~\cite{Gabbiani:1996hi,Becirevic:2001jj,Ciuchini:2002uv}.
The aim of this analysis is twofold.
On the one hand, we want to show the bounds on the MSSM parameter space
as they would appear at a SFF.
For this purpose, we first simulate the signals produced by the MSSM
for a given value of one MI.
We then check how well we are able to determine this value
using the constraints coming from a SFF.
In particular, we examine  the ranges of masses
and MIs for which clear NP  evidence,
given by a non-vanishing value of the extracted MI, can be obtained.
In Figure~\ref{fig:MIvsMg} we show for some of the different MIs,
the observation region in the plane $m_{\tilde g}$--$\vert\delta^d\vert$
obtained by requiring that the absolute value of the reconstructed MI
is more than $3\sigma$ away from zero.
For simplicity we have taken $m_{\tilde q}\sim m_{\tilde g}$.
From these plots, one can see that  a SFF  could detect NP  effects
caused by SUSY masses up to $10$--$15$ TeV
corresponding to $(\delta^d_{13,23})_{LL}\sim 1$.
Even larger scales could be reached by $LR$ MIs.
However overly large $LR$ MIs are known to produce
charge- and colour-breaking minima in the MSSM potential~\cite{Casas:1996de},
which can be avoided by imposing the bounds
shown in the $LR$ plots of Figure~\ref{fig:MIvsMg}.
These bounds decrease as $1/m_{\tilde q}$ and
increase linearly with $\tan \beta$.
Taking them into account,
we can see that still $LR$ MIs are sensitive to gluino masses
up to $5$--$10$ TeV for $\tan \beta$ between 5 and 60.
The plots of Figure~\ref{fig:MIvsMg} show the values of the MI
that can be reconstructed if SUSY masses are below $1$ TeV.
In the cases considered we find
$(\delta^{d}_{13})_{LL}=2$--$ 5\times 10^{-2}$,
$(\delta^{d}_{13})_{LR}=2$--$15\times 10^{-3}$,
$(\delta^{d}_{23})_{LL}=2$--$ 5\times 10^{-1}$ and
$(\delta^{d}_{23})_{LR}=5$--$10\times 10^{-3}$.
These value are typically one order of magnitude smaller than
the present upper bounds on the MIs~\cite{silvckm}.

Figure~\ref{fig:MI23LR} shows a
simulation of how well the the mass insertions (MIs), related to the
off-diagonal entries of the squark mass matrices, could be reconstructed 
at a  SFF. Figure \ref{fig:MI23LR} displays 
the allowed region in the plane
$\Re(\delta^d_{ij})_{AB}$--$\Im(\delta^d_{ij})_{AB}$
with a value of $(\delta^d_{ij})_{AB}$
allowed from the present upper bound,
$m_{\tilde g}=1$ TeV and using the SFF measurements as constraints.
The relevant constraints come from
$\BR(b\to s\gamma)$, $A_{\CP}(b\to s\gamma)$, $\BR(b\to s\ell^+\ell^-)$,
$A_{\CP}(b\to s\ell^+\ell^-)$, $\Delta m_{B_s}$ and $A^s_{\rm SL}$.
It is apparent the key role of $A_{\CP}(b\to s\gamma)$
together with the branching ratios of $b\to s\gamma$ and $b\to s\ell^+\ell^-$.
The zero of the forward-backward asymmetry
in $b\to s\ell^+\ell^-$, missing in the present analysis,
is expected to give an additional strong constraint,
further improving the already excellent extraction of
$(\delta^d_{23})_{LR}$ shown in Figure~\ref{fig:MI23LR}.

{\it Lepton Flavour Violation in $\tau$ Decays:}   
The search for Flavour Changing Neutral Current (FCNC) transitions
of charged leptons is one of the most promising directions
to search for physics beyond the SM. In the last 
few years neutrino physics has provided unambiguous indications 
about the non-conservation of lepton flavour, 
we therefore expect this phenomenon to occur also in the charged lepton sector.
FCNC transitions of charged leptons could occur well beyond any 
realistic experimental resolution if the light neutrino mass matrix ($m_{\nu}$)
were the only source of Lepton Flavour Violation (LFV).
However, in many realistic extensions of the SM this is not the case. 
In particular, the overall size of $m_{\nu}$
is naturally explained by a strong suppression associated to the
breaking of the total Lepton Number (LN), 
which is not directly related to the size of  LFV interactions. 

Rare FCNC decays of the $\tau$ lepton are particularly interesting
since the LFV sources involving the third generation are naturally
the largest. In particular, searches of  $\tau \to \mu \gamma$
at the $10^{-8}$ level or below are extremely interesting even
taking into account the present stringent bounds on
$\mu \to e \gamma$.  We illustrate this with one example where the
comparison of possible bounds on (or evidences for)
$\tau \to \mu \gamma$, $\mu \to e \gamma$ and other LFV rare decays 
provides a unique tool to identify the nature of the NP  model.

\begin{figure}
  \centering
  \includegraphics[angle=-90, width=0.70\textwidth]{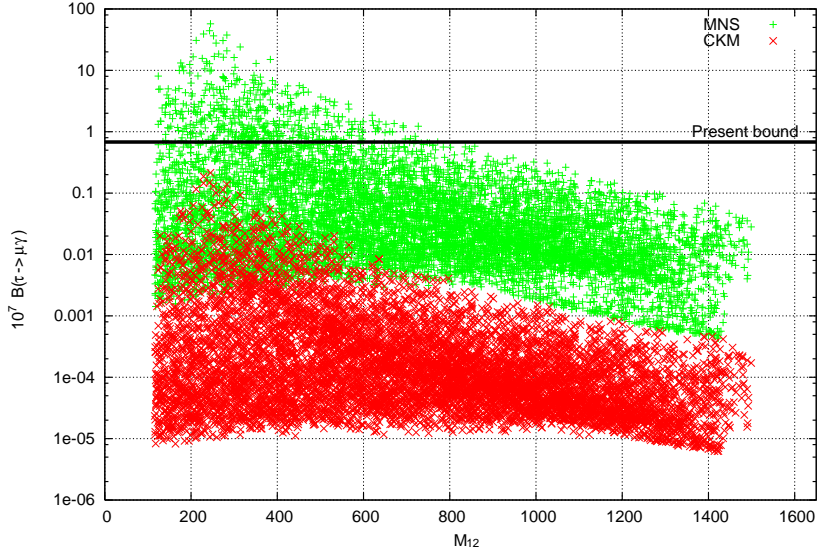}
  \caption{
    $\BR(\tau\to \mu\gamma)$ in units of $10^{-7}$ {\it vs.}
    the high energy universal gaugino mass ($M_{1/2}$)
    within a $SO(10)$ framework~\cite{Calibbi:2006nq}.
    The plot is obtained by scanning the LHC accessible parameter space
    $m_0\leq 5 \ {\rm TeV}$ for $\tan \beta = 40$.
    Green or light (red or dark) points correspond to the
    scenario where LFV is governed by the PMNS (CKM) mixing
    matrix.
    The thick horizontal line denotes the present experimental sensitivity.
The expected  SFF sensitivity is $2\times 10^{-9}$.
  }
  \label{masieroLFV06}
\end{figure}

In Figure~\ref{masieroLFV06}, we show the prediction
for $\BR(\tau\to\mu\gamma)$ within a SUSY SO(10) framework
for the accessible LHC SUSY parameter space $M_{1/2}\leq 1.5 \ {\rm TeV}$,
$m_0 \leq 5 \ {\rm TeV}$ and $\tan \beta = 40$~\cite{Calibbi:2006nq}.
Note that the measurement of $\BR(\tau \to \mu\gamma)$ at a SFF
can distinguish the scenario where LFV is governed by neutrino mixing matrix $U_{\rm PMNS}$ from the scenario where LFV is governed by the quark mixing matrix $V_{\rm CKM}$.

{\it Little Higgs Models:} 
These models address the tension 
between the naturalness of the electroweak scale and the precision electroweak 
measurements showing no evidence for new physics up to $5-10$ TeV.
The Littlest Higgs model~\cite{Arkani-Hamed:2002qy}
is based on a $SU(5)/SO(5)$ non-linear sigma model. 
It is strongly 
constrained by the electroweak precision data due to tree-level 
contributions of the new particles.

\begin{table}
  \caption{Upper bounds on some LFV decay branching ratios in the LHT model 
    with a new physics scale $f = 500 \ {\rm GeV}$, 
    after imposing constraints on 
    $\mu^-\to e^-\gamma$, $\mu^-\to e^-e^+e^-$, $\tau^- \to \mu^- \pi^0$ 
    and $\tau^- \to e^- \pi^0$. 
    \label{tab:bounds}
  }
  \begin{center}
    \begin{tabular}{|c|c|c|c|}
      \hline
      Decay & Upper bound \\
      \hline\hline
      $\tau^- \to e^-\gamma$       & $1   \cdot 10^{- 8}$ \\
      $\tau^- \to \mu^-\gamma$     & $2   \cdot 10^{- 8}$ \\
      $\tau^- \to e^-e^+e^-$       & $2   \cdot 10^{- 8}$ \\
      $\tau^- \to \mu^-\mu^+\mu^-$ & $3   \cdot 10^{- 8}$ \\
      \hline
    \end{tabular}
  \end{center}
\end{table}

Implementing an additional discrete symmetry, 
so-called T-parity~\cite{Cheng:2003ju}, 
constrains the new particles to contribute at the loop-level only 
and allows for a NP scale around $500 \ {\rm GeV}$. 
It also calls for additional (mirror) 
fermions providing an interesting flavour phenomenology.

The high sensitivity for $\tau$ decays serves as an important tool 
to test the littlest Higgs model with T-parity (LHT), in particular 
to distinguish it from the MSSM~\cite{Blanke:2007db}.
Upper bounds on some lepton flavour violating decay branching ratios
are given in Table~\ref{tab:bounds}.

\begin{table}
  {\renewcommand{\arraystretch}{1.5}
    \caption{
      Comparison of various ratios of branching ratios in the LHT model 
      and in the MSSM without and with significant Higgs contributions.
      \label{tab:ratios}
    }
    \begin{center}
      \begin{tabular}{|c|c|c|c|}
        \hline
        Ratio & \hspace{.8cm} LHT \hspace{.8cm} & MSSM (dipole) & MSSM (Higgs) \\
        \hline\hline
        $\frac{\BR(\mu^-\to e^-e^+e^-)}{\BR(\mu^-\to e^-\gamma)}$  & 
        0.4 -- 2.5 & $\sim6\cdot10^{-3}$ & $\sim6\cdot10^{-3}$ \\
        $\frac{\BR(\tau^-\to e^-e^+e^-)}{\BR(\tau^-\to e^-\gamma)}$ & 
        0.4 -- 2.3 & $\sim1\cdot10^{-2}$ & $\sim1\cdot10^{-2}$ \\
        $\frac{\BR(\tau^-\to \mu^-\mu^+\mu^-)}{\BR(\tau^-\to \mu^-\gamma)}$ &
        0.4 -- 2.3 & $\sim2\cdot10^{-3}$ & $\sim1\cdot10^{-1}$ \\
        \hline 
      \end{tabular}
    \end{center}
    \renewcommand{\arraystretch}{1.0}
  }
\end{table}

By comparison with Table~\ref{tab:superb}, 
these are seen to be well within the reach of a SFF.
However, the large LFV branching ratios are not a specific 
feature of the LHT but a general property  of many new physics models
including the MSSM. 
Nevertheless, as Table~\ref{tab:ratios} clearly shows,  
specific correlations are very suitable to distinguish 
between the LHT and the MSSM. 
The different ratios are a consequence of the fact that in the MSSM 
the dipole operator plays the crucial role in those observables while 
in the LHT the $Z^0$ penguin and the box diagram contributions 
are dominant. 
The pattern is still valid when there is a significant
Higgs contribution in the MSSM, as can be read off 
from Table~\ref{tab:ratios}.

{\it Comparison of different SUSY Breaking Scenarios:} 
In SUSY models the squark and slepton mass matrices
are determined by various SUSY breaking parameters, 
and hence a SFF has the potential to study SUSY breaking scenarios 
through quark and lepton flavour signals. 
This will be particularly important when SUSY particles are found at the LHC,
because flavour off-diagonal terms in these mass matrices 
could carry information on the origin of SUSY breaking and
interactions at high energy scales such as the GUT and the seesaw
neutrino scales. 
Combined with the SUSY mass spectrum obtained at 
energy frontier experiments, it may be possible
to clarify the whole structure of SUSY breaking.
In order to illustrate the potential of a SFF to explore the SUSY
breaking sector, three SUSY models are considered and various 
flavour signals are compared. 
These are $(i)$ the minimal supergravity model (mSUGRA), 
$(ii)$ a SU(5) SUSY GUT model with right-handed neutrinos, 
$(iii)$ the MSSM with U(2) flavour symmetry~\cite{Goto:2003iu}. 
In mSUGRA, the SUSY breaking terms are assumed to be 
flavour-blind at the GUT scale. 
The SU(5) SUSY GUT with right-handed neutrinos is a well-motivated 
SUSY model which can accommodate the gauge coupling unification 
and the seesaw mechanism for neutrino mass generation. There is 
interesting interplay between the quark and lepton sectors in 
this model. Since quarks and leptons are unified in the same GUT
multiplets, quark flavour mixing can be a source of flavour 
mixings in the slepton sector that  induce LFV in the charged 
lepton processes. Furthermore, the neutrino Yukawa coupling 
constants introduce new flavour mixings that are not related to
the CKM matrix. Due to the SU(5) GUT multiplet structure
sizable flavour mixing can occur in the right-handed
sdown sector as well as the left-handed slepton sector, and 
contributions to various LFV and 
quark FCNC processes become large. 
When we require that the neutrino Yukawa coupling 
constants only induce flavour mixing in the 2-3 generation, then the 
constraint from the $\mu \to e \gamma$ process is somewhat relaxed
(so-called non-degenerate case). 
Finally, in the MSSM with U(2) flavour symmetry, 
the first two generations of quarks and squarks are assigned as doublets with
respect to the same U(2) flavour group, whereas those in the third generation 
are singlets. 
Therefore this model explains the suppression  of the FCNC processes 
between the first two generations, but it still provides sizable 
contributions for $b \to s$ transition processes. 

\begin{figure}[t]
  \begin{center}
    \hspace{-0.7cm}\includegraphics[width=0.52\textwidth]{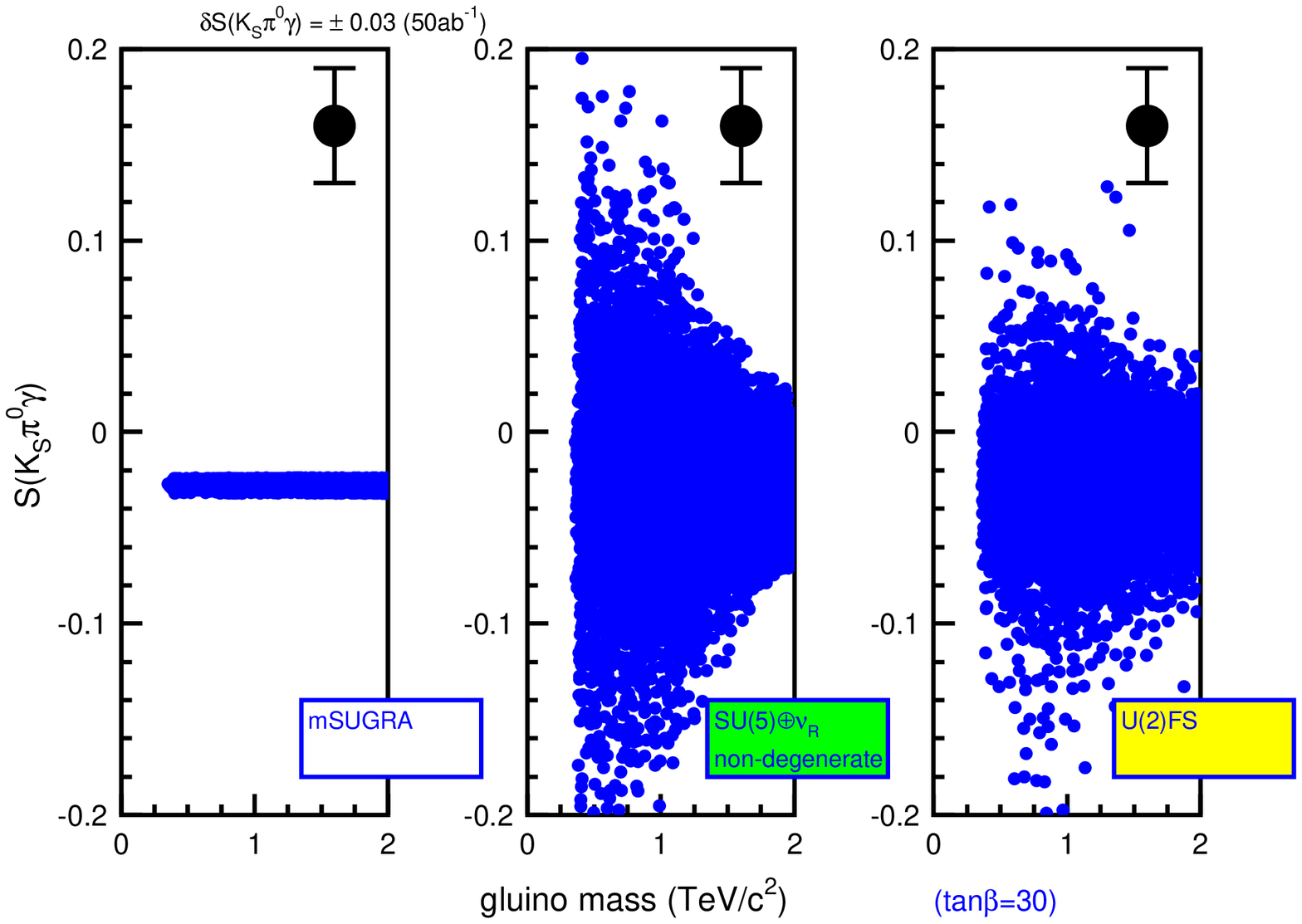}
    \hspace{-0.5cm}\includegraphics[width=0.52\textwidth]{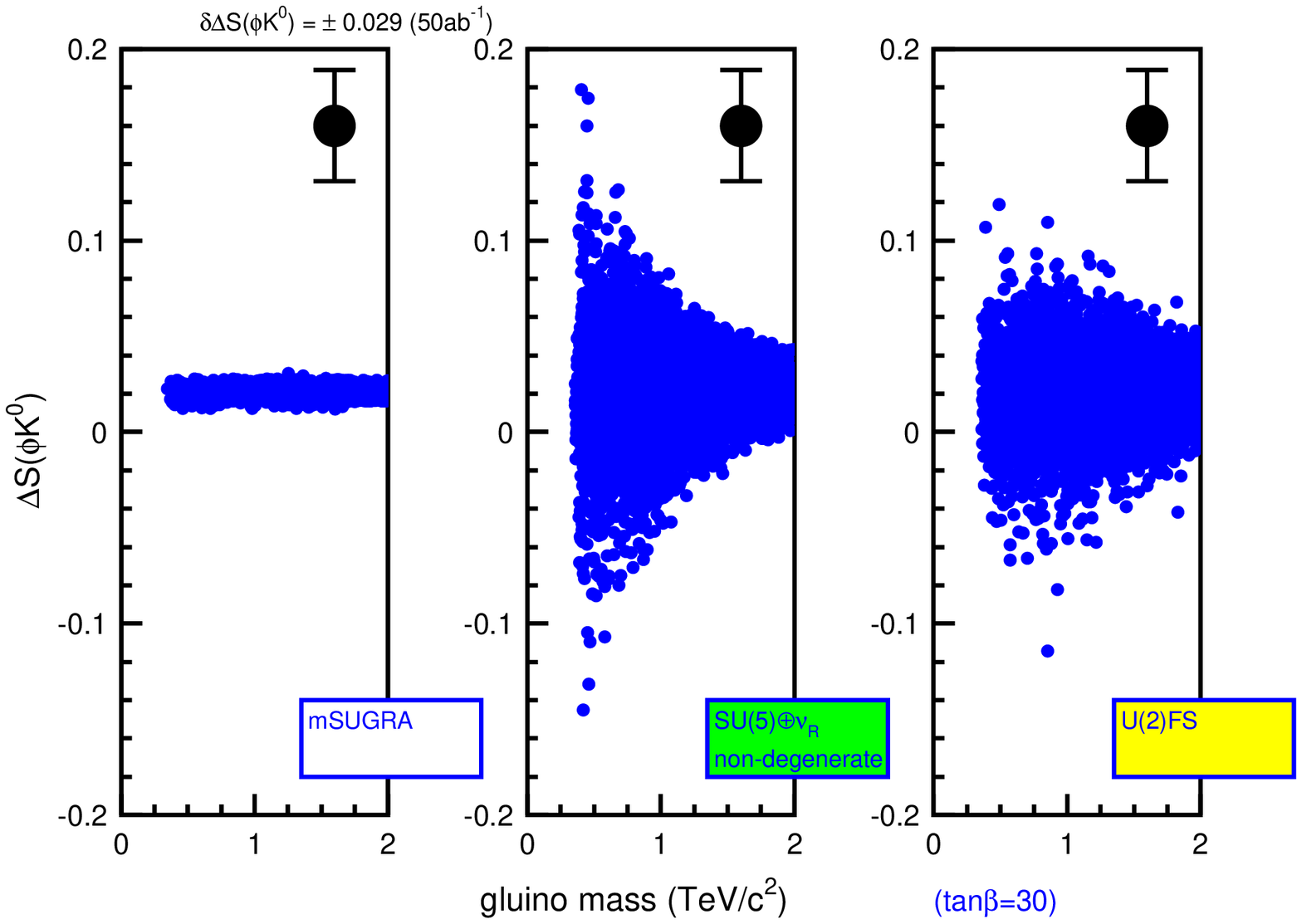}
    \caption{
      Time-dependent asymmetry of $B \to \KS \pi^0 \gamma$
      and the difference between the time-dependent asymmetries of 
      $B \to \phi \KS$ and $B \to J/\psi\,\KS$ modes 
      for three SUSY breaking scenarios:
      mSUGRA(left), SU(5) SUSY GUT with right-handed neutrinos 
      in non-degenerate case (middle), 
      and MSSM with U(2) flavour symmetry (right).  
      The expected SFF sensitivities are also shown.
    }
    \label{Okada1}
  \end{center}
\end{figure}

Flavour signals in the $b \to s$ sector are shown in Figure~\ref{Okada1}
for these three SUSY breaking scenarios.
Scatter plots of
the time-dependent asymmetry of $B \to \KS \pi^0 \gamma$
and the difference between the time-dependent asymmetries of 
$B \to \phi \KS$ and $B \to J/\psi\,\KS$ modes are presented
as a function of the gluino mass. Various phenomenological constraints 
such as ${\cal B}(b \to s \gamma)$, the rate of $B_s$ mixing, 
and neutron and atomic electic dipole moments 
are taken into account as well as SUSY and Higgs particle 
search limits from LEP and TEVATRON experiments. For the SUSY
GUT case, the branching ratios of muon and tau LFV processes 
are also calculated and used to limit the allowed parameter 
space. Sizable deviations can be seen for SU(5) SUSY GUT and
U(2) flavour symmetry cases even if the gluino mass is 1 TeV.
The deviation is large enough to be identified at SFF. On the 
other hand, the deviations are much smaller for the mSUGRA case. 

\begin{figure}[t]
  \begin{center}
    \includegraphics[width=0.55\textwidth]{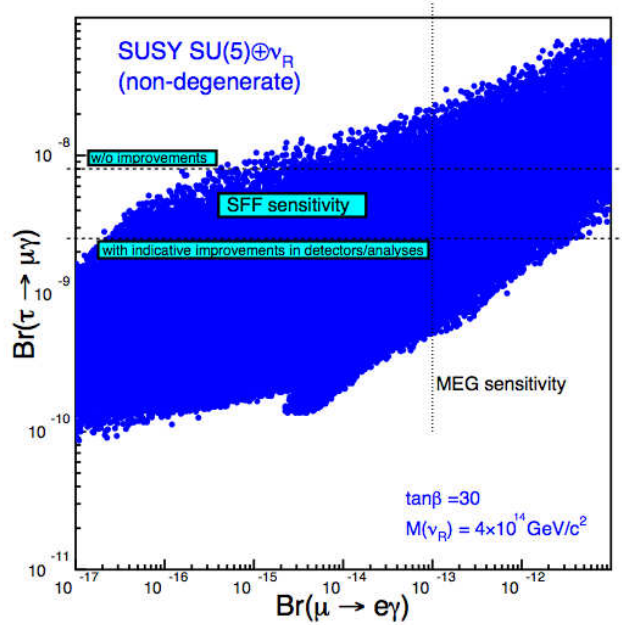}\\
    \caption{
      Correlation between ${\cal B}(\tau \to \mu \gamma)$ and 
      ${\cal B}(\mu \to e \gamma)$  for  SU(5) SUSY GUT 
      with right-handed neutrinos in non-degenerate case. 
      Expected search limits at the SFF for  ${\cal B}(\tau \to \mu \gamma)$
      and for  ${\cal B}(\mu \to e \gamma)$ from MEG are also shown.
    }
    \label{Okada2}
  \end{center}
\end{figure}

The correlation between 
${\cal B}(\tau \to \mu \gamma)$ and ${\cal B}(\mu \to e \gamma)$
is shown in Figure~\ref{Okada2} for the non-degenerate SU(5) SUSY GUT case. 
In this case, both processes can reach current upper bounds. 
It is thus possible that improvements in the $\mu \to e \gamma$ search 
at the MEG experiment and in the $\tau \to \mu \gamma$ search at a SFF 
lead to discoveries of muon and tau LFV processes, respectively.
Notice that the Majorana mass scale that  roughly corresponds to the heaviest 
Majorana neutrino mass is taken to be $M_R = 4 \times 10^{14} \ {\rm GeV}$
in these figures. When the Majorana mass scale is lower, flavour signals
become smaller because the size of the neutrino Yukawa coupling 
constant is proportional to $\sqrt{M_R}$ and LFV branching ratios
scale with $M_R^2$. This means that a SFF can cover some part of the
parameter space from $\tau \to \mu \gamma$ if the Majorana
scale is larger than $10^{13} \ {\rm GeV}$.   
The pattern of LFV signals also depends on the choice of SUSY 
breaking scenarios.
If we take the degenerate case of three heavy Majorana masses in a 
SU(5) SUSY GUT, 
${\cal B}(\mu \to e \gamma)$ can be close to the present experimental
bound while branching ratios of tau LFV processes are generally
less than $10^{-9}$.
The LFV branching ratios for both muon and tau LFV processes are negligible
for the mSUGRA case. In MSSM with U(2) flavour symmetry, LFV signals
depend on how the flavour symmetry is implemented in the lepton sector
so that there is a large model dependence.

\subsubsection{Summary}

In conclusion, the physics case of a Super Flavour Factory 
collecting an integrated luminosity of 
$50$--$75$ ab$^{-1}$ is well established.
Many NP sensitive measurements involving $B$ and $D$ mesons and $\tau$ leptons,
unique to a Super Flavour Factory, 
can be performed with excellent sensitivity to new particles with masses up to 
$\sim 100$ (or even $\sim 1000$) TeV. 
The possibility to operate at the $\Upsilon(5{\rm S})$ resonance
makes some measurements with $B_s$ mesons also accessible,
and options to run in the tau-charm threshold region and 
possibly with one or two polarized beams further broadens the physics reach.
Flavour- and $\CP$-violating couplings of new particles accessible at the 
LHC can be measured in most scenarios, even in the unfavourable cases
assuming minimal flavour violation.  
Together with the LHC, a Super Flavour Factory could be soon starting the 
project of reconstructing  the NP  Lagrangian. 
Admittedly, this daunting task would be difficult and take many years,
but it provides an exciting objective for accelerator-based 
particle physics in the next decade and beyond.



\newpage
\subsection{Super$B$ proposal}
\label{sec:super-b}


The  two asymmetric $B$ Factories, PEP-II~\cite{ref:PEP-II} and KEKB\cite{ref:KEKB}, and their companion detectors, \babar\cite{ref:babar} and Belle\cite{ref:belle}, have produced a wealth of flavour physics results, subjecting the quark and lepton sectors of the Standard Model to a series of stringent tests, all of which have been passed. With the much larger data sample that can be produced at a Super~$B$~Factory, qualitatively new studies will be possible, including searches for flavour-changing neutral currents, lepton-flavour violating processes, and new sources of \CP\ violation, at sensitivities that could reveal New Physics beyond the Standard Model. These studies will provide a uniquely important source of information about the details of the New Physics uncovered at hadron colliders in the coming decade~\cite{ref:disc}.

In light of this strong physics motivation, there has been a great deal of activity over the past six years aimed at designing an $e^+e^-$ $B$ Factory that can produce samples of $b$, $c$ and $\tau$ decays 50 to 100 times larger than will exist when the current $B$ Factory programs end. 

Upgrades of PEP-II~\cite{ref:SuperPEP-II} and
KEKB~\cite{ref:SuperKEKB} to Super $B$ Factories that accomplish this goal have been considered
at SLAC and at KEK. These machines are extrapolations of the existing $B$ Factories, with higher currents, more bunches, and smaller $\beta$ functions (1.5 to 3 mm). They also use a great deal of power (90 to 100 MW), and the high currents, approaching 10A, pose significant challenges for detectors. To minimize the substantial wallplug power, the SuperPEP-II design doubled the current RF frequency, to 958 MHz. In the case of SuperKEKB, a factor of two increase in luminosity is assumed for the use of crab crossing, 
which is currently being tested at KEKB, see Section~\ref{sec:super-kekb}.

SLAC has no current plans for an on-site accelerator-based
high energy physics program, so the SuperPEP-II proposal
is moribund.
The SuperKEKB proposal is considered as a future option of KEK. 
The problematic power consumption and background issues associated with the SLAC and KEK-based Super $B$ Factory designs have now, however, motivated a new approach to Super $B$~Factory design, using low emittance beams to produce a collider with a luminosity of $10^{36}$, but with reduced power consumption and lower backgrounds. This collider is called \hbox{Super$\!\!\;${\sl B}}.  Design parameters  of the exisiting colliders PEP-II and KEKB are compared with those of SuperPEP-II, SuperKEKB, and Super$\!\!\;${\sl B}\ in Table~\ref{table:comparison}.

\begin{table}[htb]
\caption{\label{table:comparison}
Comparison of $B$ Factory and Super $B$ Factory designs.}
\vspace*{2mm}
\centering
\setlength{\extrarowheight}{2pt}
\begin{tabular}{lccccc}
\hline
\hline
                   & PEP-II & KEKB & SuperPEP-II & SuperKEKB & Super$\!\!\;${\sl B} \\
                 \hline
$E_{LER} \ (\gev)$ & 3.1 & 3.5 & 3.5 & 3.5 & 4 \\
$E_{HER} \ (\gev)$ & 9 & 8 & 8 & 8 & 7 \\
$N_{part}$ $(\times 10^{10})$   & 8 & 5.8 & 10 & 12 & 6 \\
$I_{LER} \ (A)$ & 2.95 & 1.68 & 4.5 & 9.4 & 2.28 \\
$I_{HER} \ (A)$ & 1.75 & 1.29 & 2.5 & 4.1 & 1.3 \\
Wallplug power (MW) & 22.5 & 45 & $\sim$100& $\sim$90 & 17 \\
Crossing angle (mrad) & 0 & $\pm 15$ & 0 & 0 & $\pm 17$ \\
Bunch length $\sigma_z$ (mm) &  11 &  6 & 1.7 & 3  & 7 \\
$\sigma_y^*$ (nm) &6900 & 2000&700 & 367& 35\\
$\sigma_x^*$ ($\mu$m)&160 & 110&58 &42 & 5.7 \\
$\beta_y^*$ (mm) & 11 & 6 & 1.5 & 3 & 0.3 \\
Vertical beam-beam tune shift $\xi_y$ & 0.068 & 0.055 & 0.12 & 0.25 & 0.17\\
Luminosity (cm$^{-2}$s$^{-1}$) ($\times10^{34})$ &  \multicolumn{1}{p{0.5in}}{\centering 1.1} &  \multicolumn{1}{p{0.5in}}{\centering 1.6} & \multicolumn{1}{p{0.7in}}{\centering 70} &
  \multicolumn{1}{p{0.7in}}{\centering 80} &  \multicolumn{1}{p{0.5in}}{\centering 100} \\
\hline
\end{tabular}
\end{table}


The \hbox{Super$\!\!\;${\sl B}}\ Conceptual Design Report~\cite{ref:CDR} describes a nascent international effort to construct a very high luminosity asymmetric $e^+e^-$ Flavour Factory. The machine can use an existing tunnel or it could be built at a new site, such as the campus of the University of Rome ``Tor Vergata'', near the INFN National Laboratory of Frascati. The report was prepared by an international study group set up by the President of INFN at the end of 2005, with the charge of studying the physics motivation and the feasibility of constructing a Super Flavour Factory that would come into operation in the first half of the next decade with a peak luminosity in excess of \hbox{$10^{36}$ cm$^{-2}$ s$^{-1}$} at the \FourS resonance. 

The key idea in the \hbox{Super$\!\!\;${\sl B}}\  design is the use of low emittance beams produced in an accelerator lattice derived from the ILC Damping Ring Design, together with a new collision region, again with roots in the ILC final focus design, but with important new concepts developed in this design effort.  Remarkably, Super$\!\!\;${\sl B}\ produces this very large improvement in luminosity with circulating currents and wallplug power similar to those of the current $B$ Factories. There is clear synergy with ILC R\&D; design efforts have already influenced one another, and many aspects of the ILC Damping Rings and Final Focus would be operationally tested at Super$\!\!\;${\sl B}.  

There is quite a lot of siting flexibility in the Super$\!\!\;${\sl B}\ CDR design.  Since the required damping times are produced by wigglers in straight sections, the radius of the ring can be varied (within limits, of course) to accommodate other sites and/or to optimize cost. Smaller radius designs are also being explored, in which the bending magnets bear a greater burden in producing the needed damping.

Employing concepts developed for the ILC damping rings and final focus in the design of the Super$B$ collider, one can produce a two-order-of-magnitude increase in luminosity with beam currents that are comparable to those in the existing asymmetric $B$ Factories.  Background rates and radiation levels associated with the circulating currents are comparable to current values; luminosity-related backgrounds such as those due to radiative Bhabhas, increase substantially. With careful design of the interaction region, including appropriate local shielding, and straightforward revisions of detector components, upgraded detectors based on \babar\ or Belle are a good match to the machine environment:  in this discussion, we use \babar\ as a specific example.  Required detector upgrades include: reduction of the radius of the beam pipe, allowing a first measurement of track position closer to the vertex and improving the vertex resolution (this allows the energy asymmetry of the collider  to be reduced to 7 on 4 GeV); replacement of the drift chamber, as the current chamber will have exceeded its design lifetime; replacement of the endcap calorimeter, with faster crystals having a smaller Moli{\`e}re radius, since there is a large increase in Bhabha electrons in this region.

Super$\!\!\;${\sl B}\ has two additional features: the capability of running at center-of-mass energies in the $\tau$/charm threshold region, and longitudinal polarization of the electron (high energy) beam. The luminosity in the 4 GeV region will be an order of magnitude below that in the \FourS\ region, but even so, data-taking runs of only one month at each of the interesting energies ($\psi(3770)$, 4.03 GeV, $\tau$ threshold, {\it etc.}) would produce an order of magnitude more integrated luminosity than will exist at the conclusion of the BES-II program.  The polarization scheme is discussed in some detail in the Super$\!\!\;${\sl B} CDR~\cite{ref:CDR}. The electron beam can be polarized at a level of 85\%, making it possible to search for $T$ violation in $\tau$ production due to the presence of an electric dipole moment, or for \CP\ violation in $\tau$ decay, which is not expected in the Standard Model.

The Super$\!\!\;${\sl B}\ design has been undertaken subject to two important constraints: 1) the lattice is closely related to the ILC Damping Ring lattice, and 2) as many PEP-II  components as possible have been incorporated into the design. A large number of PEP-II components can, in fact, be reused: The majority of the HER and LER magnets, the magnet power supplies, the RF system, the digital feedback system, and many vacuum components. This will reduce the cost and engineering effort needed to bring the project to fruition.

The crabbed waist design employs a large ``Piwinski angle'' $\phi= \frac{\theta}{2}\frac{\sigma_z}{\sigma_x}$, where $\theta$ is the full geometric crossing angle of the beams at the interaction point. By producing the large Piwinski angle through the use of a large crossing angle and a very small horizontal beam size, and having $\beta_y$ comparable to the size of the beam overlap area, it is possible simultaneously to produce a very small beam spot, reduce the vertical tune shift and suppress vertical synchrobetatron resonances. However, new beam resonances then arise, which can be suppressed by using sextupoles in phase with the IP in the $x$ plane and with a $\pi/2$ phase difference in the $y$ plane. This is the crabbed waist transformation. These optical elements have an impact on the dynamic aperture of the lattice; studies carried out after the Super$\!\!\;${\sl B}\ CDR indicate that an adequate dynamic aperture can be achieved. The longer bunch length made possible by the new scheme has the further advantage of reducing the problems of higher order mode heating, coherent synchrotron radiation and high power consumption. Beam sizes and particle densities are, however, in a regime where Touschek scattering is an important determinant of beam lifetime.

The Super$B$ concept is a breakthrough in collider design. The invention of the ``crabbed waist'' final focus can, in fact, have impact even on the current generation of colliders. A test of the crabbed waist concept is planned to take place at Frascati in late 2007 or early 2008; a positive result of this test would be an important milestone as the Super$\!\!\;${\sl B}\ design progresses. The low emittance lattice, fundamental as well to the ILC damping ring design, allows high luminosity with modest power consumption and demands on the detector.

\begin{table}[t!]
\setlength{\extrarowheight}{2pt}
    \caption{Parameters of the Super$\!\!\;${\sl B}\ HER and LER rings compared with the ILC damping rings.}
    \label{tab:LatticeParam}
    \vspace*{2mm}
    \centering
\begin{tabular}{lccc}
\hline
\hline
& \multicolumn{1}{p{1.2in}}{\centering LER} &
 \multicolumn{1}{p{1.2in}}{\centering HER} &
 \multicolumn{1}{p{1.2in}}{\centering ILC DR} \\
\hline
   Energy (GeV)                            &       4       & 7 & 5 \\
   Luminosity (cm$^{-2}$s$^{-1}$) & \multicolumn{2}{p{2.6in}}{\centering{$1\times10^{36}$}} & - \\
   C (m)                                   &    \multicolumn{2}{p{2.6in}}{\centering{2249}}  &  6695 \\
   Crossing angle (mrad) &  \multicolumn{2}{p{2.6in}}{\centering{$2\times 17$}} & - \\
   Longitudinal polarization (\%) & 0 & 80 & 80 \\
   Wiggler field Bw (T)                                  &     1.00      & 0.83 & 1.67 \\
   $L_{\textrm{bend}}$ (m) (Arc/FF)        & 0.45/0.75/5.4 & 5.4/5.4 & 3/6/-  \\
   Number of Bends (Arc/FF)                &  120/120/16   & 120/16 & 126/- \\
   $U_0$ (MeV/turn)                        &     1.9       & 3.3 & 8.7 \\
   Wiggler length: $L_{\textrm{tot}}$(m) &     100 &       50 & 200 \\
   Damping time $\tau_{s},\ \tau_x$ (ms)                         &     16/32     & 16/32 & 12.9/25.7 \\
   $\sigma_z$ (mm)                              &    6      & 6 & 9\\
   $\epsilon_x$  (nm-rad)                       &    1.6      & 1.6 & 0.8 \\
   $\epsilon_y$ (pm-rad)                     &   4    & 4  &  2\\
   $\sigma_{E} (\%)$                            &   0.084 & 0.09 & 0.13 \\
   Momentum compaction                     &  $1.8 \times 10^{-4}$   & $3.1\times 10^{-4}$  & $4.2\times 10^{-4}$  \\
   Synchrotron tune $\nu_{s}$                               &     0.011     & 0.02 & 0.067 \\
   $V_{\textrm{RF}}$ (MV), $N_{\textrm{cavities}}$&     6, 8      & 18, 24  & 24, 18 \\
   $N_{part}$ $(\times 10^{10})$           &     6.16      & 3.52 &  2.0 \\
   $I_{\textrm{beam}}$ (A)                 &     2.3       & 1.3  &  0.4 \\
   $P_{\textrm{beam}}$ (MW)                &     4.4       & 4.3 & 3.5  \\
   $f_{rf}$ (MHz)                          &     \multicolumn{2}{p{2.6in}}{\centering{476}} & 650 \\
   $N_{\textrm{bunches}}$                  &     \multicolumn{2}{p{2.6in}}{\centering{1733}}  &  2625  \\
\hline
\end{tabular}
\end{table}

Since the circulating currents in \hbox{Super$\!\!\;${\sl B}}\ are comparable to those in the current $B$ Factories, an upgrade of one of the existing $B$ Factory detectors, \babar\ or Belle is an excellent match to the \hbox{Super$\!\!\;${\sl B}}\ machine environment. As an example, we will describe the changes envisioned in an upgrade of \babar, beginning with those components closest to the beamline. 

Developments in silicon sensors and materials technology make it possible to improve the resolution of the silicon vertex tracker (SVT) and to reduce the diameter of the beam pipe. This allows reduction of the energy asymmetry of \hbox{Super$\!\!\;${\sl B}}\ to 7 on 4 GeV, saving on power costs, and slightly improving solid angle coverage. The first layer of the SVT will initially be composed of striplets, with an upgrade to pixels in the highest luminosity regime. The main tracking chamber will still be a drift chamber, although with smaller cell size. The radiators of the DIRC particle identification system will be retained, but the readout system will be replaced with a version that occupies a smaller volume. The barrel CsI (Tl) electromagnetic calorimeter will also be retained, but the forward endcap will be replaced with LYSO (Ce) crystals, which are faster and more radiation-hard. A small backward region calorimeter will be added, mainly to serve as a veto in missing energy analyses. The superconducting coil and instrumented flux return (IFR) will be retained, with the flux return segmentation and thickness modified to improve muon identification efficiency. The instrumentation in the endcap regions of the IFR will be replaced with scintillator strips for higher rate capability.  The basic architecture of the trigger and data acquisition system will be retained, but components must be upgraded to provide a much-increased bandwidth.

Super$\!\!\;${\sl B}~\cite{ref:web} is an extremely promising approach to
producing the very high luminosity asymmetric $B$ Factory that is required to observe
and explore the contributions of physics beyond the Standard Model
to heavy quark and $\tau$ decays. Its physics capabilities are complementary to those of
an experiment such as LHC$b$ at a hadron machine~\cite{ref:LHCb} .
The $B$~Factories, building on more than thirty years of work in heavy flavour studies, have developed an extraordinarily vibrant and productive physics community. They have produced more than four hundred refereed publications on mixing-induced and direct \CP\ violation, improved the measurements of leptonic, semileptonic and hadronic decays and discovered a series of surprising charmonium states. The $B$~Factories have also been an excellent training ground for hundreds of graduate students and postdoctoral fellows. Super$\!\!\;${\sl B}\ will no doubt be similarly productive. The physics emphasis would, however, shift to constraining or elucidating physics beyond the Standard Model.

INFN has formed an International Review Committee to critically examine the Super$\!\!\;${\sl B}\ Conceptual Design Report and give advice as to further steps, including submission of the CDR to the CERN Strategy Group, requests for funding to the Italian government, and application for European Union funds.

Should the proposal process move forward, it is expected that the collider and detector projects will be realized as an international collaborative effort. Members of the Super$\!\!\;${\sl B}\ community will apply to their respective funding agencies for support, which will ultimately be recognized in Memoranda of Understanding. A cadre of accelerator experiments must be assembled to detail the design of Super$\!\!\;${\sl B}, while an international detector/physics collaboration is formed. The prospect of the reuse of substantial portions of PEP-II and \babar\ raises the prospect of a major in-kind contribution from the US DOE and/or other agencies that contributed to \babar\ construction; support of the project with other appropriate in-kind contributions is also conceivable. It is anticipated that the bulk of the US DOE contribution would be in kind, in the form of PEP-II components made available with the termination of the SLAC heavy flavour program. These include the HER and LER magnets, the RF and digital feedback systems, power supplies and vacuum components and the \babar\ detector as the basis for an upgraded Super$\!\!\;${\sl B}\ detector.

The \babar\ model of international collaboration, based on experience gained at CERN and other major laboratories in building and managing international collaborations over the past several decades is expected to serve as a model for the Super$\!\!\;${\sl B}\ effort~\cite{ref:web}. The funding agencies of the participating countries will have a role, together with the host agency and host laboratory, in the management of the enterprise, as well as a fiscal role through an International Finance Committee and various review committees.

\newpage
\subsection{Accelerator design of SuperKEKB}
\label{sec:super-kekb}


The design of SuperKEKB has been developed since 2002\cite{LoI}. The baseline design extends the same scheme as the present KEKB, as described below. The recently developed nano-beam scheme will be further studied as an option of SuperKEKB, while maintaing the baseline design for the time being. The possibility of an intermediate solution between these two schemes is not excluded {\it a priori}.

\subsubsection{Baseline Design of SuperKEKB}
SuperKEKB is a natural extension of present KEKB. The baseline parameters of SuperKEKB are listed in Table \ref{param}.The luminosity goal, $8\times 10^{35}~{\rm cm }^{-2}{\rm s}^{-1}$, is about 50 times higher than present KEKB. The gains of  the luminosity  will be achieved by higher currents($\times 3$ -$\times 6$), smaller $\beta^*_y$($\times 2$), and higher beam-beam parameter $\xi_y$($\times 4.5$). 

\begin{table}[htdp]
\caption{Parametes of SuperKEKB and present KEKB, for the low (LER) and high (HER) energy rings.} \label{param}
\begin{center}
\begin{tabular}{|l|c|c|c|}
\hline
 & SuperKEKB & KEKB & \\
  & LER / HER & LER / HER & \\
  \hline
  Flavor & e$^+$ / e$^-$ & e$^-$ / e$^+$ & \\ 
 Beam energy & 3.5 / 8 & 3.5 / 8 & GeV\\
 Beam current & 9.4 / 4.1 & 1.7 / 1.4 & A\\
 $\beta^*_y$ / $\beta^*_x $ & 3 / 200 & 6 / 600 & mm\\ 
Beam-beam $\xi_y$ & $\sim 0.25$ & 0.055 &\\
Number of bunches / beam & 5000 & 1400 & \\
Horizontal emittance $\varepsilon_x$ & 6 - 12 & 18 - 24 & nm\\
Bunch length $\sigma_z$ & 3 & 6 & mm\\
Peak luminosity ${\cal L}$ & 8 & 0.17 & $10^{35} {\rm cm }^{-2}{\rm s}^{-1}$ \\
Wall-plug power & $\sim 100$ & 45 & MW\\
\hline
\end{tabular}
\end{center}
\label{default}
\end{table}%

A higher stored current requires more rf sources and accelerating cavities. The baseline design adopts the same rf frequency, 509~MHz, as the present KEKB. The number of klystrons will be doubled and the number of cavities will be increased by 50\%. The total wall-plug power will be doubled. An option to adopt 1~GHz rf system to reduce the power is under consideration. The cavities will be modified for high current operation. The normal conducting accelerator with resonantly-coupled energy storage (ARES) cavity will have higher stored energy ratio of the storage cavity to the accelerating cavity. The superconducting cavity will have a new higher-order mode (HOM) absorber to dissipate 5 times more HOM power, $50$ kW per cavity. These designs of rf system and cavities have been basically done and prototyping is 
going on 
\cite{akai,kageyama,abe,takeuchi,abe_a}. 

To store the high current, it is necessary to replace all existing beam pipes in both rings. In the positron ring, beam pipes with antechamber and special surface treatment such as TiN coating are required to suppress the electron cloud. The antechambers are necessary to store such high currents to absorb the power of the synchrotron radiation in both rings. Also all vacuum components such as bellows and gate valves must be replaced with low-impedance and high-current capable version. The small $\beta^*_y$ requires shorter bunch length, which raises another reason to replace the beam pipes, otherwise the HOM loss and associated heating of the components will be crucial.  The designs of beam pipes, bellows, gate valves for SuperKEKB have been done and some prototypes were tested at present KEKB. There still remain a few R\&D issues in beam collimators and 
coherent synchrotron radiation 
\cite{suetsugu,suetsugu_a,suetsugu_b,suetsugu_c,kanazawa,flanagan}. 

SuperKEKB will switch the charges of the beams from present KEKB  to store positrons and electrons in the HER and the LER, respectively. The charge switch will relax the electron-cloud instability and reduce the amount of the positron production. For the charge switch, the injector linac will be upgraded with C-band system, whose prototype has already been built and tested successfully. Also new ideas such as single-crystal target for the positron production have been already utilized to increase the 
intensity of the positrons 
{\it et al})\cite{suwada,kamitani}. 

All existing magnets of KEKB will be reused in SuperKEKB, except the interaction region (IR), which must be renewed for smaller $\beta^*$. The final focusing superconducting quadrupole with compensation solenoid will be made stronger and their prototype has already been produced. Also the crossing angle will be increased from 22~mrad to 30~mrad. A local chromaticity correction system, which is currently installed in the LER, will be added in the HER. Another issue with the smaller $\beta^*$ is the aperture for the injected beam, especially for positrons. A new damping ring for positrons will be necessary in the injector linac to reduce the injection emittance and to increase the 
capture efficiency of the positrons 
\cite{ohuchi}. 

The boost in the beam-beam parameter $\xi_y$ assumes the success of ``crab crossing", which recovers an effective head-on collision under crossing angle by tilting each bunch by a half crossing angle. The crab cavities have been built and operated at KEKB since February 2007, basically showing the design performance in the voltage, Q-value, and phase stability, etc. The associated tilt of the beam and the effective head-on collision have been confirmed in various observations including streak cameras. The resulting beam-beam parameter reached 0.086, which is higher than the geometrical gain by about 15\%. Further study is necessary to realize higher beam-beam parameter ($> 0.1$) predicted by 
simulations for the present KEKB 
\cite{ohmi,ohmi_a,ohmi_b,ohmi_c,morita,crab}.

A number of beam instrumentations and controls will be upgraded at SuperKEKB, including beam position monitors, feedbacks, visible light and X-ray monitors, etc. Also utilities such as water 
cooling system will be reinforced 
\cite{tobiyama}.

The current estimate of the total cost of the upgrade for SuperKEKB is about 300 M\euro\  (1~\euro $\sim$ 150~Y), excluding the salaries for KEK employee in the accelerator group (about 90 FTE/year). If the upgrade of the rf system is deferred, the initial cost will be reduced to 200 M\euro.

One of the options to reduce the cost of the construction and electricity is to change the energy asymmetry from 8 GeV + 3.5 GeV to 7 GeV + 4 GeV. An early study has been done for the option resulting in a reduction by about 30 M\euro\ in the construction, and 12 MW in the electricity. Such a possibility will be investigated further.

This machine should have a flexibility to run at the charm threshold. The damping time and the emittance can be controlled by adding wigglers in the HER for that purpose. A polarized beam for the collision needs intensive study for implementation of spin rotators.

\subsubsection{Studies for Nano-beam Scheme at KEK}
 The crab waist scheme is one of the most innovative features of the 
 nano-beam Super$B$ design (Section~\ref{sec:super-b} and \cite{ref:CDR}).
Simulation by K. Ohmi has shown that the crab waist scheme can improve the luminosity of present KEKB as powerfully as crab crossing with crab cavities. Actually crab waist can be even better than crab crossing, as it only needs conventional sextupole magnets whose construction and operation will be much easier than the state-of-art crab cavities. Efforts have been made at KEK to make such a design of lattice to involve sextupole magnets at present KEKB (H.~Koiso, A.~Morita). A number of possibilities have been studied to locate the crab sextupoles, close or apart from the interaction point (IP), one pair or two pairs, which are necessary to cancel the unnecessary $x^3$ term at the IP.
 
 This study of lattice has realized that the dynamic aperture of the ring is drastically reduced by tuning on the crab sextupole magnets. These sextupoles are paired via $I$ or $-I$ transformation, and the IP is located within the pair. If the transformation between the pair is completely linear, the nonlinearity of the first sextupole is completely absorbed by the second. This kind of cancellation has been succesfully working in existing machines including KEKB. In the case of the crab waist, however, there is the IP in the middle of the pair, and the nonlinearities around the IP violates the cancellation of the nonlinear terms of the sextupoles. At least two kinds of nonlinearity, the fringe field of the final focusing quadrupoles and the kinematical terms in the drift space around the IP, has been known to be inevitable, and either one of them is enough to degrade the dynamic aperture by 50\%. As the fringe field and the kinematical terms are quite fundamental for the elements around the IP, it is not possible to remove them. The hope is to put several nonlinear magnets around the IP to cancel the nonlinearity at the IP. A. Morita has tried such possibility by introducing many octupole magnets, but not yet successful so far.
 
 The degradation of dynamic aperture by crab waist sextupoles will be also serious for future Super-B. Y. Ohnishi has studied the dynamic aperture for a Super-B lattice given by P. Raimondi. The stable horizontal amplitude with the crab-sextupoles were dropped by 70\% on the on-momentum particles, and even worse for off-momentum, synchrotron-oscillating particles. Again it has been known that the fringe field and the kinematical terms at the IP are the reason of the reduction of the dynamic aperture.

One of the questions on the nano-beam scheme is that no strong-strong simulation has been done. Because of the relatively long bunch length, such a simulation will take the computer power more than 100 times than that for usual schemes. Some preliminary efforts are going on by K. Ohmi for intermediate bunch length or with simplified models.

 Anyway the nano-beam scheme can be still attractive even without the crab waist, because it has a potential to achieve $10^{36}$~cm$^{-2}$s$^{-1}$ with smaller beam current. Therefore the KEKB team has decided to study the nano-beam scheme as an option of SuperKEKB, to make a flexible lattice and an IP design which is compatible both with the nano-beam and high-current schemes. Such a design study will identify fundamental and technical issues on the nano-beam scheme more specifically.

\newpage
%
%
%

\subsection{LHCb upgrade}
\label{sec:lhcb-upgrade}

\subsubsection{Introduction}

Flavour Physics has played a major role in the formulation
of the Standard Model (SM) of particle physics. 
As example is the observation of CP violation 
which, in the SM,  can be explained with three generations of quarks.
However despite its success, the SM is seen as an effective
low-energy theory 
because it cannot explain dark matter and the force hierarchy.
The search for evidence of 
new physics (NP) beyond the Standard Model  is the main goal
of particle physics over the next decade.

The Large Hadron Collider (LHC) at CERN will start operating in 2008 
and will start to look for the Higgs boson and for NP particles
which are expected in many models at the 1 TeV scale.  
However probing NP at the TeV scale is not restricted to 
direct searches at the high-energy frontier.

Flavour physics also has excellent potential to probe NP.
In the SM, flavour-changing neutral currents (FCNC)
are suppressed as these only occur through loop diagrams.
Hence these decays are very sensitive to NP contributions
which, in principle, could contribute with magnitude ${\cal O}(1)$ 
to these virtual quantum loops.
The NP flavour sector could also exhibit CP violation  and be very different 
from what is observed in the SM. In fact, the existing experimental limits
from the flavour physics point to either a suppression of the couplings
also for NP or an even higher NP mass scale. 

LHCb is a dedicated heavy-flavour physics experiment
designed to make precision measurements of CP violation and of rare decays 
of B hadrons at the LHC~\cite{ref:lhcbreopt}. 
LHCb will start taking data in 2008 and  
plans to record an integrated luminosity of $\sim 0.5\; \fbinv$ in the first physics run.
During the following five years LHCb expects to accumulate
a data sample of $\sim 10\; \fbinv$.
This will put LHCb in an excellent position to probe
new physics beyond the SM. The expected performance 
is summarised in Section~\ref{sec:lhcb-physics}.

During this first phase of LHC operations, particle physics will
reach a branch point. Either new physics 
beyond the Standard Model (SM)
will have been discovered at the general purpose detectors (ATLAS and CMS) 
and LHCb or new physics will be at a higher mass scale.
In both scenarios we will then almost certainly 
require a substantial increase in sensitivities
to flavour observables, either to study the flavour structure
of the newly discovered particles or to probe NP through 
loop processes at even higher mass scales.

The LHCb detector is optimised to operate at a luminosity of 
$2\ \rm to \  5 \times 10^{32}\,\rm cm^{-2} s^{-1}$,
which is a factor of 20 to 50 below the LHC design luminosity.
The LHC accelerator will reach its design luminosity
of $10^{34}\,\rm cm^{-2} s^{-1}$ after a few years of operation.
The LHC machine optics allows LHCb to focus the beams in order
to run at a luminosity of up to 50\% of the LHC luminosity.
To profit from the higher peak luminosities that are available 
at the LHC 
the LHCb experiment is proposing an upgrade to extend its 
physics programme.
The plan to operate the LHCb detector at 
ten times the design luminosity, i.e. at 
$2 \times 10^{33}\,\rm cm^{-2} s^{-1}$, 
is described in Section~\ref{sec:lhcb-highlumi}.
The LHCb upgrade would the allow the LHCb experiment 
to probe NP in the flavour sector at unprecedented sensitivities.

Initial studies of the physics reach of the proposed LHCb upgrade are discussed 
in Section~\ref{sec:lhcb-100fbinv}.
To profit from these higher luminosities the LHCb experiment 
requires an upgrade 
such that the detectors and triggers 
are able to cope with these larger luminosities.
This is described in Section~\ref{sec:superlhcb}.
A summary and conclusions are given in Section~\ref{sec:lhcb-conclusions}.

\subsubsection{LHCb Physics Programme - The First Five Years}
\label{sec:lhcb-physics}

The large cross section of $500\, \mu \rm b$ for $b \bar{b}$-quark production
in $p p$ collisions at 14 TeV centre-of-mass energy 
will allow the LHCb experiment to collect much larger
data samples of $B$ mesons than previously available. 
The expected performance for measurements with LHCb 
has been determined  by a full simulation~\cite{ref:schneider}. 
Many of these results have been described in detail 
in Section~\ref{sec:npbc} of this report.
We expect exciting results from the LHCb experiments over the next five years.
Here we summarise some of the anticipated highlights.

In the Standard Model flavour-changing neutral current (FCNC)
$b \to s$ transitions are suppressed as these only occur through
loop diagrams.
Of particular interest is the decay $B^0_s \to \mu^+\mu^-$ 
which is very rare. The SM branching ratio ${\cal B}(B^0_s \to \mu^+\mu^-)$ 
is calculated at   
$(3.86 \pm 0.15) \times 10^{-9}$~(Equ.~\ref{smunum})~\cite{Babu:1999hn}. 
New physics beyond the SM can enhance this branching ratio considerably.
For example, in the constrained minimal supersymmetric extension 
of the SM (CMSSM)~\cite{Ellis:2005sc}
the branching ratio increases as $\tan^6\beta$ where $\tan \beta$ 
is the ratio of the Higgs vacuum expectation values.
The current limits from CDF and D0 are about a factor 20 above the 
SM prediction. 
Using their good invariant mass resolution 
$\sigma (M_{\mu\mu}) \approx 20\; \rm MeV$ 
and low trigger threshold on the transverse momentum 
$p_T \geq 1 \; \rm GeV$, 
LHCb will to be able to probe the full CMSSM parameter space.
With 10~\fbinv\ of data LHCb expects 
to discover $B^0_s \to \mu^+ \mu^-$ with $5\sigma$ significance at
the SM level~\cite{Martinez:2007mi}.

Another major goal is to probe the weak phase $\phi_s$ 
of $B^0_s$ mixing.  
This is another excellent NP probe as the SM prediction
for $\phi_s$ is very small: 
$\phi_s = -2\lambda^2 \eta \approx -0.035$ 
where $\lambda$ and $\eta$ are Wolfenstein parameters 
of the CKM matrix~\cite{Bigi:1981qs}. 
Currently there are no strong constraints on $\phi_s$ available 
and large CP violation in $B^0_s$ mixing is 
allowed~\cite{Abazov:2007tx,Ligeti:2006pm,Grossman:2006ce,Lenz:2006hd,Ball:2006xx}. 
The LHCb experiment expects to collect 
131~k $B^0_s \to J/\psi \phi$ decays with a 2~\fbinv\ data sample.
The corresponding precision on $\phi_s$ is estimated 
to be $\sigma(\phi_s) \approx 0.023$~\cite{bib:LHCb-2006-047_LF}. 
A value of $\phi_s$ of ${\cal O}(0.1)$ or larger could be clearly observed by
LHCb. This would be a clear signal for Non-Minimal
Flavour Violation (NMFV) beyond the SM~\cite{D'Ambrosio:2002ex}.

LHCb will perform measurements of the CKM angle $\gamma$ using two interfering diagrams
in neutral and charged $B \to D K$ decays 
as well as $B^0_s \to D_s^{\mp} K^{\pm}$ decays.
The interference arises due to decays which are common to $D^0$ and $\bar D^0$ mesons
such as $D^0 (\bar D^0) \to K^0_S \pi^+\pi^-$ (Dalitz decay~\cite{Giri:2003ty}) and
$D^0 (\bar D^0) \to K^{\mp}\pi^{\pm}, K^+ K^-$ 
(ADS and GLW~\cite{Atwood:1996ci,Gronau:2002mu}), or through $B_s$ mixing.
The expected $\gamma$ sensitivities for 2~\fbinv\ of LHCb data are estimated
at $\sigma(\gamma) \sim 7^\circ - 15^\circ$.
When combining these measurements
LHCb expects to achieve a precision $\sigma(\gamma) \sim 2.5^{\circ}$ 
in a 10~\fbinv\ data sample~\cite{ref:schneider}.  
This will improve substantially 
the $\gamma$ measurements from the B-factories 
which currently have an uncertainty of about $30^\circ$~\cite{Barberio:2007cr}.

\subsubsection{LHCb Luminosity Upgrade}
\label{sec:lhcb-highlumi}

After the first five years of operation with the LHCb experiment, 
the LHC will hopefully provide answers to some of the 
open questions of particle physics and, 
very possible, produce a few new puzzles. 
To be able to make progress in determining the flavour structure of 
new physics beyond the SM or probing higher mass scales, 
it is very likely that the required precision for several 
flavour physics observables 
will need to be improved substantially.
It is also expected that the precision of many LHCb physics results 
will remain limited by the 
statistical error of the collected data. 
The following questions arise:
What is the scientific case for collecting even larger
data samples?
Is LHCb exploiting the full potential for $B$ physics
at hadron colliders?
Note that LHCb is the only dedicated heavy flavour experiment approved to run after 2010.
In the remainder of this report we will try to answer these questions.

The LHCb experiment has commenced  studying the feasibility of 
upgrading the detector such that it can operate at a luminosity  
${\cal L} \sim 2 \times 10^{33}\, \rm  cm^{-2}\,s^{-1}$, which is 
ten times larger than the design luminosity~\cite{ref:muheim}.
This upgrade would allow LHCb to collect a data
sample of about $100\; \fbinv$ during five years of running. 
This increased luminosity is achievable by decreasing 
the amplitude function $\beta^*$ at the LHCb interaction point.
The LHCb upgrade does not require the planned LHC luminosity upgrade (Super-LHC)
as the LHC design luminosity is $10^{34}\; \rm  cm^{-2}\,s^{-1}$,
although it  could operate at Super-LHC. 
Thus an upgrade of LHCb could be implemented as early as 2014.

As the number of interactions per beam crossing will increase to $n \sim 4$
this will require improvements to the LHCb sub-detectors and trigger.
A major component of the LHCb upgrade will be the addition of  
a first level detached vertex trigger which will use information
from the tracking detectors~\cite{ref:dijkstra:fpcp,ref:parkes:eps}. 
This trigger has the potential of 
increasing the trigger efficiencies for 
decays into hadronic final states by at least a factor of two. 
The implementation of this detached vertex trigger will require
large modifications to the detector read-out electronics which will be discussed in
Section~\ref{sec:superlhcb}.

\subsubsection{Physics with the LHCb Upgrade}
\label{sec:lhcb-100fbinv}

A 100~\fbinv\ data sample would allow to improve the sensitivity
of LHCb to unprecedented levels such that new physics beyond the SM
can be probed at the 1\% level. Here we present estimates for  
a few selected channels.
These are based on the following assumptions, 
which have yet to be demonstrated:
maintaining  trigger and reconstruction efficiencies 
at high luminosity running and,
making use of a detached vertex trigger
to double the  
trigger efficiency for hadronic modes.
Systematic errors are only treated in a very simple way.
Hence the quoted sensitivities have very large uncertainties 
and should be treated with caution. 
However, these estimates are extremely useful to motivate simulation studies
for validating these assumptions.
In addition, as soon as LHCb will start taking data, the simulations 
for low luminosity running can be verified with data.

New physics can be probed for by studying FCNC in hadronic $b \to s$ transitions.
One approach is to compare the time-dependent CP asymmetry 
in a hadronic penguin loop decay   
with a decay based on a tree diagram 
when both decays have the same weak phase. 
In hadronic FCNC transitions unknown massive particles could 
make a sizable contribution to the $b\to s$ penguin loop whereas 
tree decays are generally insensitive to NP.
The B-factories  measure  the CP asymmetry $\sin 2 \beta^{\rm eff}$ 
in the penguin decay \Bdtophiks.  
A value for $\sin 2 \beta^{\rm eff}$ which is different from
$\sin 2 \beta$ measured in \Bdtojpsiks\ would signal physics
beyond the SM. 
Within the current available precision, 
all $\sin 2 \beta^{\rm eff}$ measurements are in reasonable agreement
with the SM, but most central values are lower than expected. 
For example, we find for the decay \Bdtophiks\ that 
$\Delta S(\phi K^0_S) = \sin 2 \beta^{\rm eff} - \sin 2 \beta = 0.29 \pm 0.17$~\cite{ref:hfag}.

This approach can also be applied to $B^0_s$ mesons which will be exploited by LHCb.
Within the SM the weak mixing phase \phis\ is expected to be almost the same when comparing
the time-dependent CP asymmetry of the hadronic penguin decay \Bstophiphi\ with 
the tree decay \Bstojpsiphi. Due to a cancellation of the $B^0_s$ mixing and decay phase, 
the SM prediction for the sine-term, \phisphiphi, 
in the time-dependent asymmetry of \Bstophiphi\ 
is very close to zero~\cite{Raidal:2002ph}.
Thus any measurement of $\phisphiphi \neq 0$ would be a clear signal for 
new physics and definitively rule out 
Minimal Flavour Violation~\cite{D'Ambrosio:2002ex}.
From a full simulation, 
LHCb expects to collect 3100 \Bstophiphi\ events in 2 \fbinv\ of data 
with a background to signal ratio $B/S < 0.8$ at 90\% C.L~\cite{ref:LHCbphiphi}.
The $\phisphiphi$ sensitivity 
has been studied using a toy Monte Carlo,
taking resolutions and acceptance from the full simulation.
After about 5 years LHCb expects to have accumulated a data 
sample of 10~\fbinv\ and will measure $S(\phi\phi)$ 
with a precision of $\sigma(S(\phi\phi))=0.05$~\cite{ref:LHCbphiphi}.
This precision is expected to be statistically limited, 
systematic errors are likely much lower.

The LHCb upgrade will substantially improve the measurement of \phisphiphi,
since this is a hadronic decay mode which will benefit most from 
the first level detached vertex trigger.
Scaling the sensitivity up to a data sample of 100~\fbinv,   
we estimate a precision of $\sphisphiphi \sim 0.01$ to $0.02$\,rad. 
This sensitivity presents a exciting NP probe at the 
percent level which will arguably be (one of) the most precise time-dependent 
CP study in $b\to s$ transitions.

In a similar study LHCb investigated the $b\to s$ penguin 
decay $B^0_d \to \phi K^0_S $.
A yield of 920 events is expected in $2~\fbinv$ of integrated luminosity 
and the background to signal ratio is $0.3 < B/S <1.1$.
The sensitivity for the time-dependent 
CP violating asymmetry $\sin 2 \beta ^{\rm eff}$ 
is estimated to be  0.10 in a 10~\fbinv\ data sample~\cite{ref:LHCbphiks}.
This is a hadronic decay which will also profit  from a first level
detached vertex trigger.
With 100~\fbinv\ of integrated luminosity 
LHCb upgrade will allow to improve the $\sin 2 \beta ^{\rm eff}$ 
sensitivity for $B^0_d \to \phi K^0_S $ to $\sim 0.025$ to 0.035.
  
Using the  tree decay $B^0_s \to J/\psi \phi$
LHCb will also probe NP in the CP violation of $B^0_s$ mixing. 
With a 10~\fbinv\ data sample the weak phase $\phi_s$ 
will be determined with a precision of $0.01$~\cite{ref:schneider}. 
This corresponds to $\sim 3.5\sigma$ significance for the SM expectation
of $\phi_s$ for which 
the theoretical uncertainty is very precise (${\cal O}(0.1\%)$).
This precision is expected to be still statistically limited.
A significantly larger data-set would allow LHCb to search for NP in 
$B$ meson mixing at an unprecedented level. 
An upgrade of LHCb has the potential to measure the SM value of \phis\ 
with $\sim 10 \sigma$ significance ($\sphis \sim 0.003$) in \Bstojpsiphi\ decays.
To control systematic errors at this level will be very challenging.

In the SM, the angle $\gamma$ can be determined very precisely with
tree decays which are theoretically very clean.
When combining all $\gamma$ 
measurements in $B \to DK$ and $B^0_s \to D_s^{\mp} K^{\pm}$ (including systematics)
LHCb  will constrain the value of $\gamma$ to about 2.5\degrees.
However, it will not be possible to push below the 
desired $1^{\circ}$ precision.
Therefore, a very precise determination of $\gamma$ in
tree decays is an important objective of the LHCb upgrade physics programme.
The expected yields in 100~\fbinv\ of data are very large:
Examples are  620k $B^0_s \to D_s^{\mp} K^{\pm}$, 
500k $B \to D(K^0_S\pi^+ \pi^-) K$ and 
5600k $B \to D(K \pi) K$ events, respectively.
All these $\gamma$ modes will benefit greatly from an 
improved first-level trigger strategy that does not rely solely on high transverse 
momentum hadrons. 
Simple statistical extrapolations show that several individual modes 
will give a potential statistical uncertainty close to  $1^{\circ}$. 
Systematic uncertainties will clearly be very important. 
However, these uncertainties are largely uncorrelated amongst 
the modes and, in many cases, can be measured in control samples.
Therefore, a global determination to below $1^{\circ}$ of the tree level 
unitarity triangle will  be possible~\cite{ref:wilkinson}.
 This will act as a standard candle to be compared to all loop determinations 
of the unitarity triangle parameters.

The very rare decay $B^0_s \to \mu^+ \mu^-$ is key to many extensions beyond the SM.
With a 100~\fbinv\ data sample LHCb upgrade would be able to make 
a precision measurement of the branching ratio
${\cal B}( B^0_s \to \mu^+ \mu^-)$ to about $\sim 5 \%$ at the SM level.
This will allow LHCb upgrade to either measure precisely 
the flavour properties of new SUSY particles
discovered at the LHC or to put very stringent constraints on all
SUSY models in the large $\tan \beta$ regime~\cite{Ellis:2005sc}.

LHCb upgrade should also aim to observe the even rarer decay
$B^0_d \to \mu^+ \mu^-$ which has a SM branching ratio of 
$(1.06 \pm 0.04)\times 10^{-10}$ (Equ.~\ref{dmunum}).
The ratio
${\cal B}( B^0_d \to \mu^+ \mu^-)/{\cal B}( B^0_s \to \mu^+ \mu^-)$ 
is sensitive to new physics beyond the SM and will allow to distinguish
between different models. 
This search will be extremely challenging as it requires an excellent 
understanding of the detector to reduce the muon fake rate 
due to backgrounds from hadronic two body modes to an acceptable level.

LHCb will exploit the semileptonic decay $B\to K^{*0} \mu^+ \mu^-$ 
which is sensitive to new physics in the 
small $\tan \beta$ range.
Using a full simulation LHCb expects to collect 7200 $B\to K^{*0} \mu^+ \mu^-$ 
per 2~\fbinv~\cite{LHCb-2007-038}.
In addition to the forward-backward asymmetry, $A_{\rm FB}$,
these large data samples will allow LHCb to measure the 
differential decay rates in the di-muon mass squared, $q^2$, 
and the angular distributions, and probe NP 
through the transversity amplitude $A_{T}^{(2)}$ 
and the $K^{*0}$ longitudinal polarisation~\cite{Kruger:2005ep}.
In the theoretically favoured region of $1 < q^2 < 6\; \rm GeV^2/c^4$ 
the resolution in  $A_{T}^{(2)}$  is estimated at 0.16 with
10~\fbinv\ of integrated luminosity~\cite{LHCb-2007-057}.
While this data sample might provide a hint of NP, a ten-fold increase 
in statistics will allow to probe new physics 
at the few percent level and cover a large region of the MSSM parameter space.
With a 100~\fbinv\ data sample LHCb upgrade expects to collect 
360k $B\to K^{*0} \mu^+ \mu^-$ events.
The corresponding precision for $A_{T}^{(2)}$  is estimated to be 0.05 to 0.06. 

There are several other channels which have a large potential for probing NP 
with a 100~\fbinv\ data sample. An excellent example is $B^0_s \to \phi \gamma$ 
which is sensitive to the photon polarisation and
right-handed currents~\cite{Atwood:1997zr}.
Using a full simulation LHCb expects a yield of 11500 $B^0_s \to \phi \gamma$ events 
in 2~\fbinv\ of data with a background to signal ratio $<0.91$ at 90\% C.L.~\cite{lhcb-bkstgam}.
The sensitivity of this decay to NP arising in
right-handed currents is under study.
LHCb upgrade would also be able to search for NP by studying                                    
the decays $B_s \to \phi \mu^+\mu^-$ and                                                                
$B \to \pi (\rho) \mu^+\mu^-$.

The very large charm sample would allow LHCb upgrade to search for NP
in $D^0$ mixing and CP violation in charm decays.
The expected statistical sensitivity on the parameters
$x'^2$, $y'$ and $y_{CP}$ are 
$2 \times 10^{-5}$, $2.8 \times 10^{-4}$ and $1.5 \times 10^{-4}$,
respectively (Table~\ref{mixtab2}).
An LHCb upgrade could also probe 
lepton flavour violation in the decay mode  $\tau \to \mu^+\mu^-\mu^+$ 
with a an estimated sensitivity of $2.4 \times 10^{-9}$~\cite{ref:tau2mumumu}.

The Standard Model (SM) as well as SUSY or Extra Dimension models can be
augmented by additional gauge sectors
\cite{matt-hidden,matt-higgs,matt-super}. 
This is a very general
consequence of string theories \cite{Cvetic:2002qa,Arkani-Hamed:2005yv,Barger:2007ay}. 
These gauge sectors can
only be excited by high energy collisions. 
An example is the ``hidden valley'' sector.
The manifestations of many of these  models
could be new $v$-flavoured particles with a long lifetime~\cite{matt-hidden}. 
These can decay to a pair of $b$ and $\overline{b}$ quarks that produce jets
in the detector. An example is the Higgs decay process
$H \to \pi^0_v \pi^0_v$ followed by $\pi^0_v \to b \bar b$.  
LHCb is designed to detect $b$-flavored hadrons 
and thus in a good position to detect decays of long-lived new particles.
The LHCb vertex detector (VELO) is $\sim$1~m long making it possible
to measure these decays.
LHCb upgrade will increase the sensitivity to much lower production 
cross section for these processes.

In Table~\ref{tab:lhcb-upgrade} 
we present a summary of the expected sensitivities for selected key measurements,
discussed above and 
that could be performed with an upgrade of the LHCb experiment. 
These sensitivities will exceed the range for probing NP  from LHCb and 
B-factories considerably,
and they will also improve upon the precision of SM parameters.

\begin{table}[thb]
  \begin{center}
    \caption{Expected sensitivity for LHCb upgrade with an integrated luminosity
of 100 \fbinv. A factor two of improvement 
for the L0 hadron trigger and systematic error estimates are shown as a range.}
    \label{tab:lhcb-upgrade}    
    \begin{tabular}{l@{\hspace{15mm}}c}
      \hline \hline
      Observable                      & LHCb upgrade sensitivity \\
      \hline
      $S(B_s \to \phi\phi)$                    & $0.01 - 0.02$ \\
      $S(B_d \to \phi K^0_S)$                 &  $0.025 - 0.035$        \\
      $\phi_s$ ($J/\psi\phi$)                 &  $0.003$ \\
      \hline
      $\sin(2\beta)$ ($J/\psi\,K^0_S$)        &  $0.003 - 0.010$      \\
      $\gamma$ ($B \to D^{(*)}K^{(*)}$)       &  $< 1^\circ$  \\
      $\gamma$ ($B_s \to D_s K$)              &  $1 - 2^\circ$  \\
      \hline
      ${\cal B}(B_s \to \mu^+ \mu^-) $        &  $5 - 10\%$             \\
      ${\cal B}(B_d \to \mu^+ \mu^-) $        &  $3\sigma$            \\
      $A_T^{(2)} (B \to K^{*0} \mu^+ \mu^-) $ &  $0.05 - 0.06$             \\
      $A_{\rm FB}(B \to K^{*0}\mu^+ \mu^-) \ s_0$ &  $0.07\; \rm GeV^2$            \\
      \hline \hline
    \end{tabular}
  \end{center}
\end{table}

We now compare the physics potential of LHCb upgrade
collecting a 100~\fbinv data sample, with
that of a Super Flavour Factory (SFF), 
based on a $50\ \rm to \ 75\; \rm ab^{-1}$ 
data sample which is discussed in Section~\ref{sec:superff} of this report.

The strengths of the two proposals are surprisingly complementary.
For example the more benign environment of an $e^+e^-$ collider allows the 
SFF to make inclusive measurements of $b\to s \gamma$ and the
CKM matrix element $V_{ub}$ 
and of rare decays with missing energy such as $B^+ \to \ell^+\nu$.
However, LHCb upgrade is unique in its potential to exploit
the physics of $B^0_s$ mesons, especially in $B^0_s$ oscillations.
A key motivation for LHCb upgrade is 
the ability to probe new physics in  hadronic $b\to s$ penguin transitions by 
measuring the time-dependent CP asymmetry in 
the decay $B^0_s \to \phi\phi$ with a precision of 0.01 to 0.02. 
The SFF will make complementary measurements 
by studying the  time-dependent CP asymmetries of $b \to s$ transitions in 
several $B^0_d$ decays. 

LHCb upgrade will be able to measure CP violation in the interference of mixing
and decay in both $B^0_s$ and $B^0_d$ mesons. This will allow 
LHCb to probe NP simultaneously 
in FCNC  with $B^0_d \to J/\psi K^0_s$ and $B^0_s \to J/\psi \phi$  (tree)
and $B^0_d \to \phi K^0_s$ and $B^0_s \to \phi\phi$ (hadronic $b\to s$ penguin) 
to the unprecedented level of $\sim 1\%$.

The LHCb upgrade will probe NP contributions to right-handed currents by
measuring the time-dependent CP asymmetry in the decay $B^0_s \to \phi \gamma$.
The SFF will make complementary measurements and
exploit their better reconstruction efficiencies 
for decays with several neutral particles in the final state
to measure the photon polarisation 
of $B^0_d \to K^0_S \pi^0 \gamma$. 

In channels where both approaches are possible, 
the sensitivities are often comparable.
LHCb upgrade usually will have larger statistics, 
but systematic errors in the hadronic environment
will be more difficult to control.
Both, LHCb upgrade and SFF propose to 
measure $\sin 2 \beta$ to $0.01$ and the CKM angle  $\gamma$ with 
1\degrees\ precision.    

A SFF can measure the zero of the forward-backward asymmetry in 
the inclusive channel $b \to s \ell^+ \ell^-$, 
but LHCb upgrade will collect a substantially larger  
sample of 360k $B^0_d \to K^{*0} \mu^+ \mu^-$ decays compared
to 11k at a SFF. 
This  will enable LHCb to measure the asymmetry $A_T^{(2)}$
to $\sim 5\%$. 
Only LHCb upgrade will be able to measure 
the $B^0_s \to \mu^+ \mu^-$ branching ratio to $\sim 5\%.$
This will precisely determine the flavour structure of new particles 
discovered at the LHC 
or severely constrain the SUSY parameter space.

\subsubsection{LHCb Detector and Trigger Upgrade }
\label{sec:superlhcb}

We start out by presenting the limitations of the LHCb detector and trigger which
prevent LHCb from operating the detectors at higher luminosity.
At the design luminosity of $2\times 10^{32}\, \rm cm^{-2}\,s^{-1}$
the visible cross section is 63 mb which corresponds to about
10 MHz of bunch crossings with at least one visible interaction.
Note that  increasing the luminosity from 
$2\ \rm to \ 10 \times 10^{32}\, \rm cm^{-2}\,s^{-1}$ 
will only increase the number of interactions by a factor of two
since the number of bunch crossings with visible interactions increases 
from 10 to 26 MHz.

The LHCb experiment has a two level trigger system.
The Level-0 trigger (L0) is 
implemented in hardware and the Higher Level Trigger (HLT) is running on a 
large CPU farm. The L0 trigger operates at 40 MHz. 
The purpose of L0 is to reduce this rate 
to 1.1 MHz which is the maximum at which all LHCb detectors can be read-out by
the front-end electronics. 
The L0 trigger selects objects (hadron $h$, $e$, and $\gamma$) 
with high transverse energy, $E_T^{h,e,\gamma}$, 
in the electromagnetic and hadronic calorimeters 
and the two highest transverse momentum ($p_T^{\mu}$) muons in the muon system. 
At the nominal luminosity of $2\times 10^{32}\, \rm cm^{-2}\,s^{-1}$
the typical trigger thresholds are 
$E_T^h \geq 3.5\; \rm GeV$,
$E_T^{e,\gamma} \geq 2.5\; \rm GeV$ and 
$p_T^\mu \geq 1\; \rm GeV$.
Events with multiple interactions are vetoed.

Simulations show that the L0 muon trigger efficiency 
for reconstructible events at 
the design luminosity of $2\times 10^{32}\, \rm cm^{-2}\,s^{-1}$
is around 90\% and that the 
output rate raises almost linearly 
with luminosity up to $5\times 10^{32}\, \rm cm^{-2}\,s^{-1}$.
For larger luminosities the loss in efficiency is minor.
At the design luminosity 
the muon trigger uses about 15\% of the L0 bandwith.
However, the L0 hadron trigger has a lower performance.
The efficiencies of this trigger for hadronic decays
are only about 40\% at the design luminosity, 
whereas the L0 hadron trigger uses about $\sim 70\%$ of the L0 bandwith.
At higher peak luminosity the rate of visible $pp$ interaction
increases which requires an increase in the threshold
and the corresponding loss in efficiency 
results in an almost constant yield for the hadron trigger~\cite{ref:dijkstra:fpcp}.


This illustrates that the existing trigger
does not scale with luminosity, in particular the hadronic trigger
will not 
allow operating the LHCb experiment at ten times the design luminosity.
The total trigger efficiency including the HLT for hadronic $B$ decays
is expected to be 25 to 30\%~\cite{ref:schneider}. 
The goal of the LHCb upgrade should also
be to improve the hadron trigger efficiency by at least a factor two.

We have commenced initial 
studies which investigate how to upgrade the LHCb detector and triggers
such that the experiment can operate 
at luminosities ${\cal L} \sim 2\times 10^{33}\, \rm cm^{-2}\,s^{-1}$. 
These show that the only way to achieve this is to measure both the momentum and
the impact parameter of charged $B$ decay products simultaneously.
The present front-end architecture is not compatible with this requirement.
The vertex and tracking detectors are read-out at a maximum rate of 1.1~MHz,
thus this information is not available to the L0 trigger.

Hence the LHCb upgrade has opted for a front-end electronics which
will read-out all LHCb sub-detectors at the full bunch crossing rate
of 40~MHz of the LHC. Data will be transmitted over optical fibres 
to a off detector interface board which is read out by the DAQ. 
This has clear advantages as it would allow the implementation of
a L0 displaced vertex trigger in a CPU farm. 
In fact all trigger decisions would be 
software-based which allows flexibility.

A initial study for the 40~MHz trigger uses \Bstodsk\ decays simulated at 
a luminosity of $6\times 10^{32}\, \rm cm^{-2}\,s^{-1}$. 
Events with large numbers of interactions are employed to simulate 
larger effective luminosities up to $2\times 10^{33}\, \rm cm^{-2}\,s^{-1}$. 
Assuming enough CPU power to process an event rate of 5~MHz
we obtain  a trigger efficiency of 66\% for this channel. 
The requirements are a transverse energy $E_T > 3\; \rm GeV$ from the L0 hadron trigger 
which has an efficiency of 76\% for signal 
combined  with a matched track 
that has a transverse momentum $p_T > 2$ GeV/$c$ and an 
impact parameter $\delta > 50 \mu \rm m$. 
In this combined trigger the minimum bias rate does not depend 
strongly on the luminosity 
and the triggered event yield scales linearly with the luminosity. 
In addition, the total trigger efficiency
is 60\% larger when compared with the existing baseline.

However this approach requires a replacement 
of the front-end electronics for all sub-detectors, with the exception of the
muon chambers which are already read out at 40 MHz.  
Replacing the front-end electronics will require new sensors for several
sub-systems.
Besides the VELO silicon sensors, the silicon sensors of the tracking 
stations will need to be replaced.
The sensors close to the beam will suffer from a ten-fold increase in radiation 
and hence more radiation hard sensors will be required.
The RICH photon detectors have encapsulated front-end electronics and need
to be replaced entirely.

The vertex detector (VELO) silicon sensors
undergo radiation damage and it is expected that 
these will need to be replaced when 6 to 8 \fbinv\ of luminosity 
has been collected~\cite{ref:parkes}.
However the channel occupancy in the VELO is $\sim 1\%$ at design luminosity.
When increasing  the luminosity by a factor of ten to 
$2\times 10^{33}\, \rm cm^{-2}\,s^{-1}$ the occupancy only increases to 
$\sim 3\%$ and the corresponding efficiency loss is small.

A preliminary study of the performance of the 
electromagnetic calorimeter (ECAL) at high luminosity 
shows only a small
degradation for the selection efficiency 
of the decay $B^0_s \to \phi\gamma$.
It might be necessary to upgrade the inner section 
of ECAL to improve its granularity and energy resolution.
The increased radiation level of irradiation leads to a
degradation of the energy resolution and will require
that half the inner ECAL section will need to be replaced
after 3 years of operation at $2\times 10^{33}\rm cm^{-2}s^{-1}$.

R\&D efforts have started on technologies for radiation-hard vertex detectors 
that will be able to operate in the LHC radiation environments at LHCb upgrade 
luminosities.
The detector sensors will need to be able to operate at radiation doses of 
about $10^{15}\; 1\, \rm MeV \; equivalent\; neutrons/cm^2$. 
Initial studies of Czochralski and $n$-on-$p$ sensors
irradiated up to $4.5 \times 10^{14}$ 24 GeV protons/cm$^2$ 
are promising and show that the charge collection efficiencies saturate at 
acceptable bias voltages~\cite{ref:parkes}.
Pixel sensors are very radiation hard and R\&D on this technology
has started.

Two different vertex-detector geometries are envisaged. 
One is to shorten the strips, the other is to use pixels. 
Removing the RF foil that separates
the VELO sensors from the primary beam-pipe vacuum would reduce the radiation length before
the first measurement by 3\% and improve the proper time resolution of $B$ meson decays.

\subsubsection{Summary and Conclusions}
\label{sec:lhcb-conclusions}

The LHC will open a new window 
for discovering new physics (NP) beyond the Standard Model.
The LHCb experiment will probe NP with precision studies 
of flavour observables, 
whereas the general purpose detectors ATLAS and CMS aim to directly 
observe new particles. 
Both approaches are required to study the mass hierarchy 
and the couplings of the new physics.
LHCb will collect an integrated luminosity of about 10~\fbinv\ during its first 
five years. Very likely the LHC results will show   
that a significantly  better sensitivity will be required  
for both, the direct and indirect approaches.
Here we present a proposal to upgrade the LHCb detectors to be able
to operate at ten times the design luminosity, i.e. at 
 $2\times 10^{33}\, \rm cm^{-2}\,s^{-1}$, and to collect a data sample of 
100~\fbinv\ with an improved detector.
Initial sensitivities for physics with LHCb upgrade are presented.
These show that LHCb upgrade has the potential to probe new physics at 
unprecedented levels that is mainly complementary to the 
proposed Super Flavour Factory.
The upgraded LHCb experiment will include a first level detached vertex trigger for which
a new front-end architecture must be designed. A more radiation hard vertex detector
is required to cope with the increased radiation doses.


\newpage \section{Assessments}
\label{sec:asm}

In Sect.~\ref{sec:nps} we briefly introduced several NP scenarios and
discussed their impact on FCNC and CP violating processes. Then, in
Sect.~\ref{sec:npbc} we considered several benchmark channels that are
particularly sensitive to NP, discussing the present status and future
developments. The aim of this Section is to summarize the present
status of NP flavour scenarios, to identify possible patterns of NP
signals, and to describe the first attempts that have been made during
the workshop to connect constraints on NP (and possible NP signals) in
flavour and high-energy physics. The first two items are discussed in
Sect.~\ref{sec:patterns}, the last one is presented in
Sect.~\ref{sec:connections}. 

\subsection{New-physics patterns and correlations}
\label{sec:patterns}


The past decade has witnessed enormous progress in the field of
flavour physics: B-factories have studied flavour and CP violation in
$B_d - \bar B_d$ mixing and in an impressive number of $B$ decays; the
Tevatron has produced the first results on $B_s - \bar B_s$ mixing and
has studied several BRs and CP asymmetries in $B$ and $B_{s}$ decays;
very recently, B-factories have established the first evidence of $D
- \bar D$ mixing. This flourishing of experimental results has been
accompanied by several remarkable improvements on the theory side,
both in perturbative and non-perturbative computations. Let us just
mention the NNLO calculation of BR$(b \to s \gamma)$, the proof of
factorization in nonleptonic $B$ decays in the infinite mass limit and
the first unquenched results on $B$ physics from lattice QCD.

Thanks to these experimental and theoretical achievements, we now have
a rather precise idea of the flavour structure of viable NP extensions
of the SM. The general picture emerging from the generalized Unitarity
Triangle analysis performed in
ref.~\cite{Bona:2005eu,Bona:2006sa,Bona:2007vi} and from the very
recent data on $D - \bar D$
mixing~\cite{Aubert:2007wf,Staric:2007dt,Abe:2007rd,Ciuchini:2007cw}
is that no new sources of CP violation of $\mathcal{O}(1)$ are
observed in $B_d$, $K$ and $D$ mixing amplitudes. However, the
possibility of NP CP-violating effects in $B_s$ mixing is still
open. Concerning $\Delta F=1$ processes, the situation is quite
different. In particular, large NP contributions to $s \to d g$,
$b \to d g$ and $b \to s g$ transitions are not at all
excluded. Sizable NP effects in $s \to d Z$, $b \to d Z$ and $b \to s
Z$ vertices are also possible, although the available experimental
data excludes order-of-magnitude enhancements. Finally, FC Higgs
interactions generated by NP can still give large enhancements of
scalar vertices, although the upper bounds on $B_s \to
\mu^+ \mu^-$ are getting tighter and tighter. 

To summarize, we can say that, although the idea of minimal flavour
violation is phenomenologically
appealing~\cite{Gabrielli:1994ff,Misiak:1997ei,Ciuchini:1998xy,Buras:2000dm,D'Ambrosio:2002ex,Bobeth:2005ck,Blanke:2006ig},
an equally possible alternative is that NP is contributing more to
$\Delta F=1$ transitions than to $\Delta F=2$ ones. Within the class
of $\Delta F=1$ transitions, (chromo)-magnetic and scalar vertices are
peculiar since they require a chirality flip to take place, which
leads to a down-type quark mass suppression within the SM. On the
other hand, NP models can weaken this suppression if they contain
additional heavy fermions and/or additional sources of chiral
mixing. In this case, they can lead to spectacular enhancements for
the coefficients of (chromo)-magnetic and scalar
operators. Furthermore, if the relevant new particles are colored,
they can naturally give a strong enhancement of chromomagnetic
operators while magnetic operators might be only marginally
modified. The electric dipole moment of the neutron puts strong
constraints on new sources of CP violation in chirality-flipping
flavour-conserving operators involving light quarks, but this does not
necessarily imply the suppression of flavour-violating operators,
especially those involving $b$ quarks.  Therefore, assuming that NP is
sizable in several $\Delta F=1$ processes is perfectly legitimate
given the present information available on flavour physics.

Thus, we can identify at least three classes of viable
weakly-interacting NP extensions of the
SM:\footnote{Strongly-interacting NP most probably lies beyond the
reach of direct searches at the LHC and so will not be discussed
here~\cite{Bona:2007vi}.}
\begin{enumerate}
\item Models with exact MFV;
\item Models with small $(\mathcal{O}(10\%))$ departures from MFV;
\item Models with enhanced scalar or chromomagnetic $\Delta F=1$
vertices, and a suitable suppression of NP contributions to $\Delta
F=2$ processes.
\end{enumerate}

In models belonging to the third class, we expect sizable NP effects in
$B$ physics. From a theoretical point of view, a crucial
observation is the strong breaking of the SM $SU(3)^5$ flavour symmetry
by the top quark Yukawa coupling. This breaking necessarily propagates
in the NP sector, so that in general it is very difficult to suppress
NP contributions to CP violation in $b$ decays, and these NP
contributions could be naturally larger in $b \to s$ transitions than
in $b \to d$ ones. This is indeed the case in several flavour models
(see for example Ref.~\cite{Masiero:2001cc}).

Another interesting argument is the connection between quark and
lepton flavour violation in grand unified
models~\cite{Baek:2000sj,Harnik:2002vs,Hisano:2003bd,Huang:2003fv}. The
idea is very simple: the large flavour mixing present in the neutrino
sector, if mainly generated by Yukawa couplings, should be shared by
right-handed down-type quarks that sit in the same $SU(5)$ multiplet
with left-handed leptons. Once again, one expects in this case large
NP contributions to $b \to s$ transitions. 

\subsection{Correlations between FCNC processes}

On general grounds, it is difficult to establish correlations between
FCNC processes without specifying not only the NP flavour structure,
but also the details of the NP model. However, there is a notable
exception, given by models of Constrained Minimal Flavour Violation
(see Sect.~\ref{sec:nps} for the definition of this class of MFV
models). While correlating $\Delta F=1$ to $\Delta F=2$ processes is
not possible without specifying the details of the model, in the case
of CMFV there are several interesting correlations between FCNC
processes. In CMFV, all NP effects can be reabsorbed in a redefinition
of the top-mediated contribution to FCNC amplitudes. Thus, all
processes that involve the same top-mediated amplitude are exactly
correlated. This has interesting phenomenological consequences,
allowing for stringent tests of CMFV by looking at correlated observables
\cite{Buras:2000dm,Buras:2000xq,Buras:2001af,D'Ambrosio:2002ex,Bobeth:2005ck}.

It is enough to go from CMFV to MFV to destroy many of these
correlations: for example, in MFV models with two Higgs doublets at
large $\tan \beta$ it is in general not possible to connect $K$, $B$
and $B_s$ decays in a model-independent way. However, interesting
correlations remain present also at large $\tan \beta$. For example,
the enhancement of $B_s \to \mu^+ \mu^-$ corresponds in general to a depletion
of $\Delta m_s$~\cite{Buras:2002wq} (actually, both features might be
phenomenologically acceptable~\cite{Isidori:2006pk}).

Of course, within a specific model it is in general possible to
correlate $\Delta F=1$ and $\Delta F=2$ processes and to fully exploit
the constraining power of flavour physics. The most popular example is
given by the minimal supergravity models, where one can combine not
only all the information from flavour physics, but also the available
lower bounds on SUSY particles and the constraints from electroweak
physics, dark matter and cosmology
\cite{deBoer:1996vq,deBoer:1996hd,Cho:1999km,Cho:2001nf,Erler:1998ur,Altarelli:2001wx,Djouadi:2001yk,deBoer:2001nu,deBoer:2003xm,Belanger:2004ag,Ellis:2003si,Ellis:2004tc,Ellis:2005tu,Ellis:2006ix,Ellis:2007fu,Baltz:2004aw,Allanach:2005kz,Allanach:2006jc,Allanach:2006cc,Allanach:2007qk,deAustri:2006pe,Buchmueller:2007zk}. Interesting
correlations between FCNC processes are also present in the CMSSM if
one considers more general SUSY spectra than minimal
supergravity~\cite{Gabrielli:1994ff,Buras:2000qz}. 

Even allowing for new sources of flavour and CP violation to be
present, correlations remain present between the several flavour
observables generically affected by the same NP flavour violating
parameter. An interesting example is given by SUSY models with
enhanced chromomagnetic $b \to s$ vertices (see \textit{e.g.}
ref.~\cite{Ciuchini:2002uv}). 

\begin{figure}
\begin{center}
\includegraphics[width=15cm]{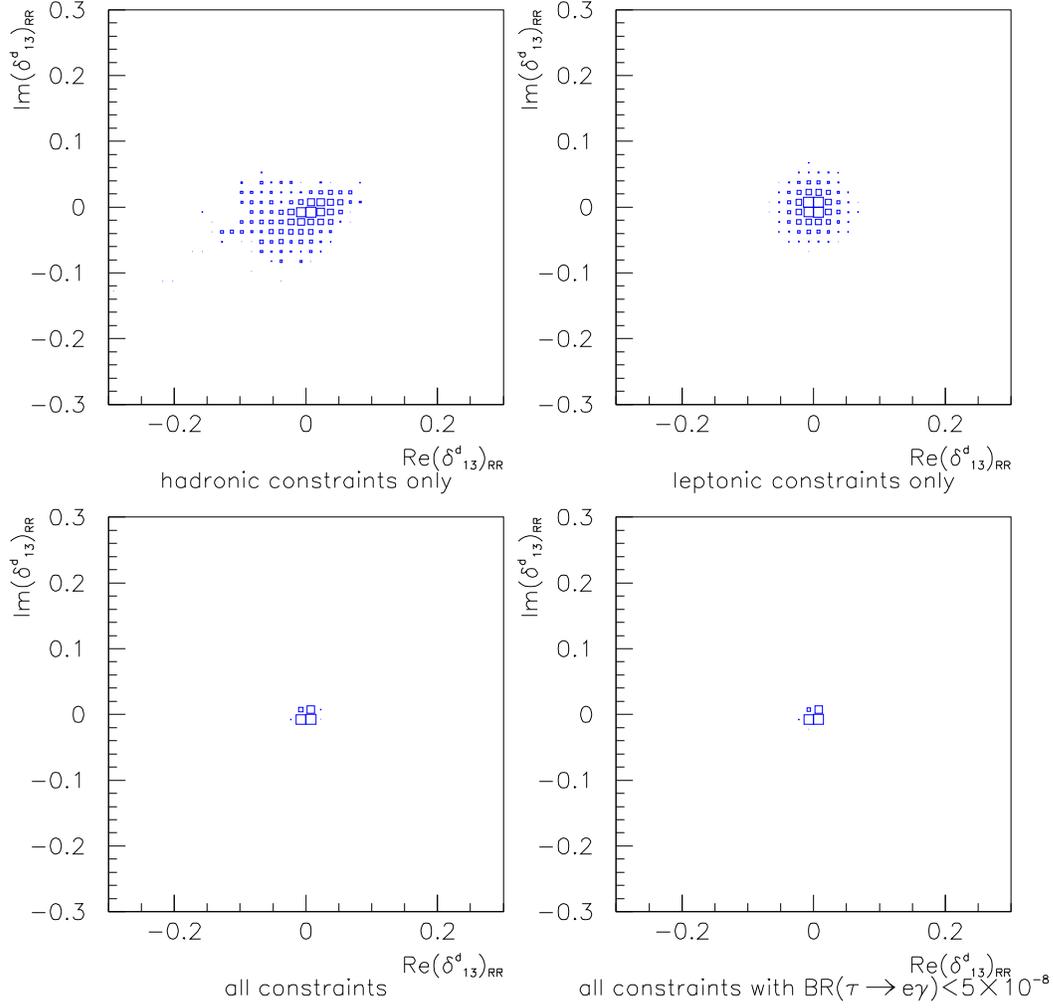}
\end{center}
\caption{Allowed region in the
  Re$\left(\delta^d_{13}\right)_\mathrm{RR}$-Im$\left(\delta^d_{13}\right)_\mathrm{RR}$
  plane using hadronic constraints only (upper left), leptonic
  constraints only (upper right), all constraints (lower left) and all
  constraints with improved leptonic bounds (lower right).} 
\label{fig:RR13}
\end{figure}

Another general class of NP models in which interesting correlations
between FCNC processes can be established is given by SUSY-GUTs. Grand
unification implies the equality of soft SUSY breaking terms at the
GUT scale. Thus, any new source of flavour and CP violation present in
squark masses must also be present in slepton masses, leading to a
correlation between squark and slepton FCNC
processes~\cite{Ciuchini:2003rg}. An extensive discussion of these
correlations has been carried out in ref.~\cite{Ciuchini:2007ha}. As
an example, we present in Fig.~\ref{fig:RR13} (from
ref.~\cite{Ciuchini:2007ha}) the constraints on $
\left( \delta^d_{13} \right)_\mathrm{RR}$ (defined in
Sec.~\ref{sec:genFCNC}) from hadronic constraints only (upper left), leptonic
constraints only (upper right), all constraints (lower left) and all
constraints with improved leptonic bounds (lower right). In this
interesting case, hadronic and
leptonic bounds have comparable strengths. Exploiting the GUT
correlation, it is possible to combine them to obtain a much tighter
constraint on $\left( \delta^d_{13} \right)_\mathrm{RR}$.

\newpage 
\newpage

\subsection{Connection to high-energy physics}
\label{sec:connections}


Recent low-energy data from flavour physics experiments showed
relatively good agreement with the SM prediction (taking into account
the theory uncertainties). This imposes strong constraints on any new
physics scenario. In view of the new results and the new bounds on
physics beyond the SM the demand for scenarios that could be used for
studies at ATLAS or CMS (or more generally for setting up the
infrastructure for future studies once ATLAS and CMS have collected
their first data) was issued. These scenarios should be in agreement
with all existing $B$~and $K$~physics data and possibly show
interesting signatures at the LHC experiments. 

In this respect the question which parameter choices are useful as a
benchmark scenario 
depends on the purpose of the actual investigation. If one is
interested, for instance, in setting exclusion limits on the SUSY
parameter space from the non-observation of SUSY signals at the
experiments performed up to now, it is useful to use a benchmark
scenario which gives rise to ``conservative'' exclusion bounds. An
example for a benchmark scenario of this kind is the
$\mh^{\mathrm{max}}$-scenario~\cite{Carena:2002qg,Carena:2005ek} used for the  
Higgs search at LEP~\cite{Schael:2006cr} and the
Tevatron~\cite{Abulencia:2005kq,Abazov:2005yr}. 
Another purpose for using benchmark scenarios is to study ``typical''
experimental signatures of e.g.\ SUSY models and to investigate the
experimental sensitivities and the achievable experimental precisions
for these cases. For this application it seems reasonable to choose
``typical'' parameters (a notion which is of course hard to define) of 
certain SUSY-breaking scenarios (see e.g.\ the ``Snowmass Points and
Slopes''~\cite{Allanach:2002nj}). In this context it can also be
useful to consider ``pathological'' regions of parameter space or
``worst-case'' scenarios. 

In the perspective of future improvements on $B$~and $K$~physics data,
it is also worth to consider the possibility of a {\em positive} signal 
of new physics selected by some low-energy observable. In this 
perspective, it is useful to consider benchmark scenarios with
well-defined low-energy signatures, such as the MFV scenario with large 
$\tan\beta$ discussed in Ref.~\cite{Isidori:2006pk},
or models with small flavour-breaking structures 
departing from the minimal structure of the constrained MSSM.
These cases are particularly useful to explore the capability of 
future flavour-physics measurements in constraining a limited 
set of the SUSY parameter space, both separately and in conjunction 
with future ATLAS/CMS data.

A related issue concerning the definition of appropriate scenarios is
whether a benchmark scenario chosen for investigating physics at 
ATLAS and CMS should be 
compatible with additional information from other experiments (beyond
$B$~and $K$~physics). This refers in
particular to constraints from cosmology or the measurement of
the anomalous magnetic moment of the muon, 
$(g-2)_\mu$~\cite{Bennett:2006fi}. On the one hand, applying 
constraints of this kind gives rise to ``more realistic'' benchmark
scenarios (see e.g. Ref.~\cite{Allanach:2002nj}).  
On the other hand, one relies in this way on further 
assumptions (and has to take account of experimental and theoretical
uncertainties related to these additional constraints), and it could
eventually turn out that one has narrowed down the range of
possibilities too much by applying these constraints. This applies in
particular if slight modifications of the model under investigation
have a minor impact on collider phenomenology but could
significantly alter the bounds from cosmology and low-energy
experiments. E.g.\ the presence of 
a small amount of R-parity violation in a SUSY model would
strongly affect the constraints from dark matter relic abundance
while leaving the phenomenology at high energy colliders essentially 
unchanged. 
Thus we restrict ourselves to scenarios which are compatible with
flavour physics, with existing lower bounds on new particles (e.g.\
the bound on the lightest MSSM Higgs
boson~\cite{Barate:2003sz,Schael:2006cr}) and with other electroweak
precision data, see Ref.~\cite{Heinemeyer:2004gx} and references therein. 

The general procedure of setting up new scenarios follows the steps:
\begin{enumerate}
\item
identify the models of interest;
\item
identify within these models the regions of the parameter space 
that are compatible with the existing constraints from flavour physics, 
electroweak precision physics and direct bounds; 
\item
identify specific sub-regions which could be selected by future improvements
on flavour physics; 
\item
study the most interesting points in view of their
high-energy phenomenology that can be explored at ATLAS and CMS;
\item
set up the infrastructure for the analysis of (possible) data that
will be collected at ATLAS and CMS to test the new high-energy results
against existing low-energy data.
\end{enumerate}
Concerning the first step, the model(s) which exhibited most interest during
the workshop are the
MSSM with (N)MFV. Consequently, in the following we concentrate on
this class of SUSY models.

Within the second and third step 
it is desirable to connect different codes (e.g.\ working in
the (N)MFV MSSM, see Section~\ref{sec:flavortools}) to each other. 
Especially interesting is the combination of codes that provide the evaluation
of (low-energy) flavour observables and others that deal with high-energy 
(high $p_T$) calculations for the same set of parameters.
This combination would allow to test the ((N)MFV MSSM) parameter space with
the results from flavour experiments as well as from high-energy experiments
such as ATLAS or CMS. 

A relatively simple approach for the combination of different codes is their
implementation as sub-routines, called by a ``master code'' (see
Sections~\ref{sec:mastertool2}, \ref{sec:mastertool}). 
This master codes takes care of the correct definition
of the input parameters for the various subroutines. 
Concerning the last step, 
the application and use of the master code would change once experimental data
showing a deviation from the SM predictions is available. This can come
either from the on-going flavour experiments, or latest (hopefully) from ATLAS
and CMS. If such a ``signal'' appears at the LHC, it has to be determined to
which model and to which parameters within a model it can correspond. Instead
of checking parameter points (to be investigated experimentally) for their
agreement with experimental data, now a scan over a chosen model could be
performed. Using the master code with its subroutines each scan point can be
tested against the ``signal'', and preferred parameter regions can be obtained
using a $\chi^2$ evaluation. It is obvious that the number of evaluated
observables has to be as large as possible, i.e.\ the number of subroutines
(implemented codes) should be as big as possible.


\subsubsection{The first approach:\\
prediction of $b$-physics observables from SUSY measurements}

The first approach was followed in collaboration with ATLAS. 

An LHC experiment will hopefully be able to measure
a significant number of SUSY parameters based on the 
direct measurement of SUSY decays. The experimental potential 
in  this field has been studied in detail for various benchmark points.
Based on these studies, a possible approach is to 
focus on specific models for which many SUSY parameters
can be measured at the LHC, and to try to answer the following
questions:
\begin{enumerate}
\item
How precisely can $b$-physics variables be predicted using 
measured SUSY parameters?
\item
Vice versa: can we use $b$-physics measurements to constrain 
badly measured SUSY parameters?
\item
Is the precision of the measurements on the two sides
adequate to rule out minimal flavour violation and/or to 
constrain flavour violation in the squark sector?
\end{enumerate}
We will show in the following the application of this approach,
especially of question~(1), to 
a point of the MSSM space 
which was adopted as a benchmark point
by the Supersymmetry Parameter Analysis (SPA) group
\cite{Aguilar-Saavedra:2005pw}. This model is defined in terms
of the parameters of the mSUGRA model
($m_0=70$ GeV, $m_{1/2}=250$ GeV, $A_0=-300$ GeV,
$\tan{\beta}=10$, $\mu>0$). This is a modification
of the point SPS1a,  essentially achieved
by lowering $m_0$ from 100 to 70~GeV,  originally defined in
Ref.~\cite{Allanach:2002nj} to take into account more recent
results on dark matter density.\par
The values 
of the sparticle masses at tree level, 
computed with the program ISASUSY 7.71 \cite{isa}, 
are given in Table~\ref{tab:masses}.
\begin{table}[htb]
\begin{center}
\vskip 0.2cm
\begin{tabular}{|c|c|c|c|}
\hline
Sparticle & mass [GeV] & Sparticle & mass [GeV] \\
\hline
\hline
$\lsp$ & 97.2 & $\tchi^0_2$ & 180.1 \\
$\tchi^0_3$ & 398.4 & $\tchi^0_4$ & 413.8 \\
$\tl_L$ & 189.4 & $\tl_R$ & 124.1 \\
$\ttau_1$ & 107.7 & $\ttau_2$ & 194.2 \\
$\ttop_1$ & 347.3 & $\ttop_2$ & 562.3 \\
$\tu_L$ & 533.3 & $\tg$ & 607.0 \\
$h$ & 116.8 & $A$ & 424.6 \\
\hline
\end{tabular}
\caption{\label{tab:masses}  {\it Masses of the sparticles
in the considered model as calculated at tree level
with ISAJET 7.71 \cite{isa}
}}
\end{center}
\end{table}
Constraints on the sparticles masses can be obtained 
from measurements of the kinematics of the SUSY
cascade decays Ref.~\cite{Bachacou:2000zb,atltdr,
Allanach:2000kt}.
This program has been carried out recently for the SPS1a
model point \cite{LHCLC}, assuming the performance
of the ATLAS detector.
The resulting constraints allow the measurement
of the masses of $\tchi^0_1$, $\tchi^0_2$, $\tchi^0_4$,
$\tg$, $\tq_L$, $\tq_R$, $\sbot_1$, $\sbot_2$ $\tl_R$ $\tl_L$, $\ttau_1$,
where $\tq_L$ and $\tq_R$ are the average of the masses
of the squarks of the first two generations.
All these masses should be measurable with an uncertainties
of a few percent, for  an integrated luminosity of 300~fb$^{-1}$.
The estimated uncertainties will be used as an input to this 
study. \par
For the stop sector a detailed study is 
available \cite{Hisano:2003qu}, always performed in the
framework of the ATLAS collaboration. 
\begin{figure}
\vspace{-3em}
\includegraphics[width=0.5\textwidth]{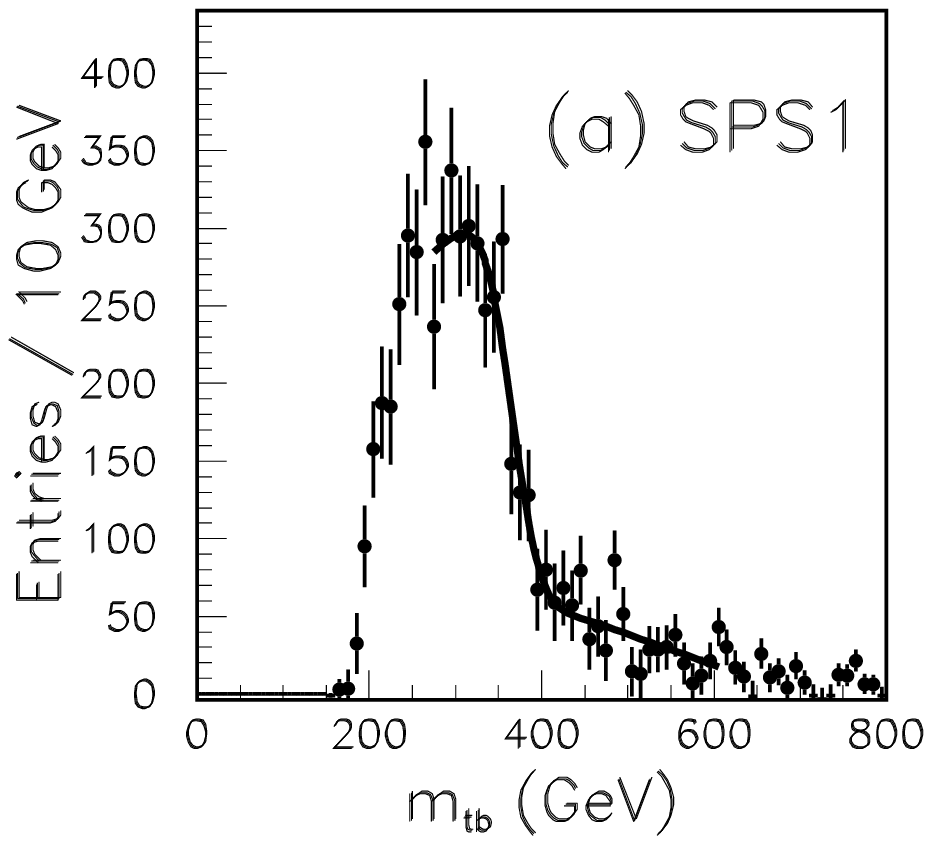}\\[-16em]
\mbox{}\hspace{0.5\textwidth}
\includegraphics[width=0.4\textwidth]{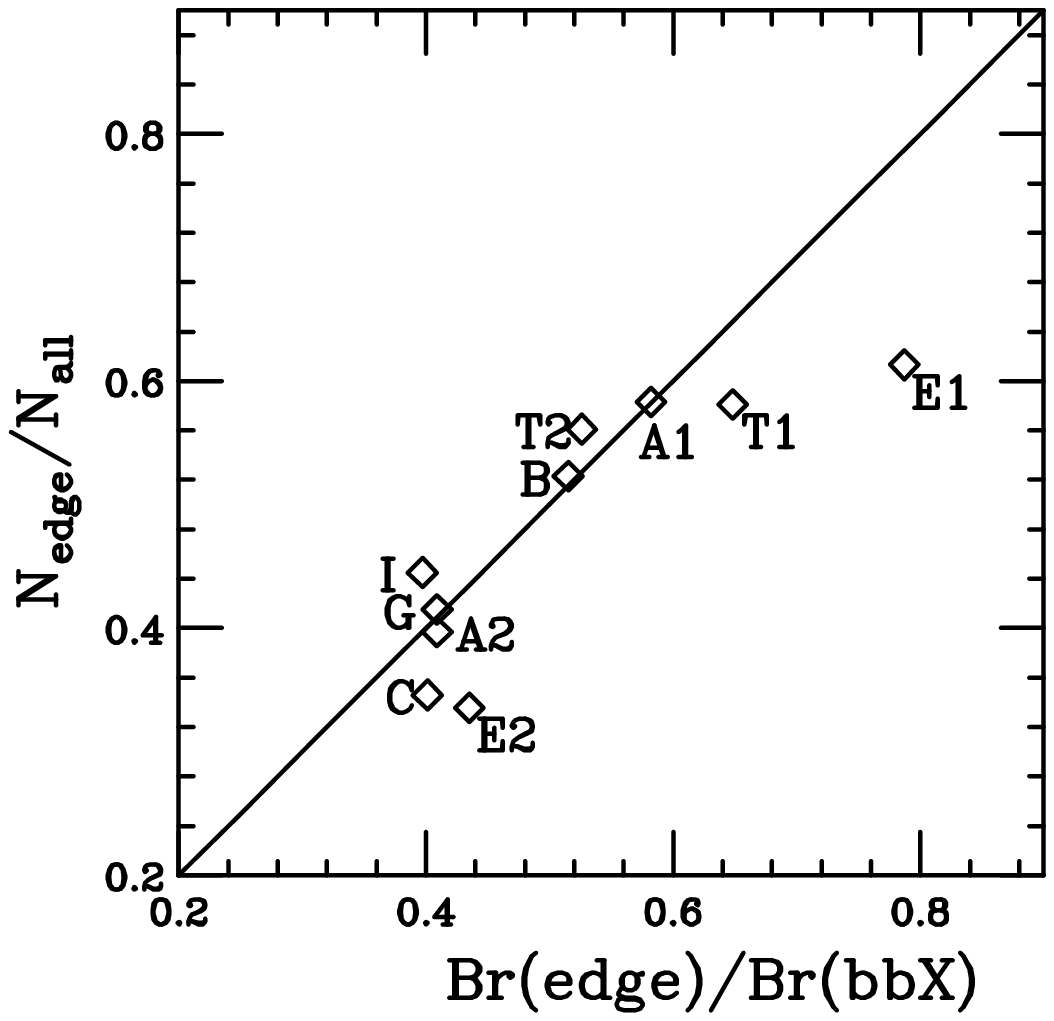}
\caption{\label{fig:stopmihoko} {\it Left: $m_{tb}$ distribution 
for model point SPS1a. Right: relationship between $N_{\rm edge}/N_{\rm all}$
and $\br({\rm edge})/\br(bbX)$ for different model points as described
in \cite{Hisano:2003qu}. Both figures from \cite{Hisano:2003qu}.
}}
\end{figure}
This analysis studies the $tb$ invariant mass distribution 
in SUSY events. This distribution, shown in 
the left panel of Fig.~\ref{fig:stopmihoko} shows the characteristic
kinematic edge which can be expressed as a function of the
masses. 
Two main  SUSY decay chains yield a $tb$ final state signature:
\begin{equation}
\mbox{$\tg\to\ttop_1 t\to tb\tchi^\pm_1$} 
\label{eq:dec1}
\end{equation}
and
\begin{equation}
\mbox{$\tg\to\sbot_1 b\to tb\tchi^\pm_1$}.
\label{eq:dec2}
\end{equation}
Therefore the position of the end-point in the $tb$ 
mass distribution ($M_{tb}^{\rm fit}$) will measure  the average 
of the edges for the two decays weighted by the relative BR, 
which yields a constraint on a number of MSSM parameters:
$$
M_{tb}^{\rm fit}=f(\mste,\msbe,\mgl,\mcha{1},\tst,\tsb)
$$
From the height of the observed kinematic distribution
one can also measure the ratio of
events in the $tb$ mass distribution to all SUSY events 
with a $b$ pair in the final state, $N_{\rm edge}/N_{\rm all}$.
This observable is well correlated, as shown in the
right panel of Fig.~\ref{fig:stopmihoko}, with the quantity
$\br({\rm edge})/\br(\tg\to bbX)$ where $\br({\rm edge})$ is the
sum of the BR's for the decays (\ref{eq:dec1}) and 
(\ref{eq:dec2}) above. 
Finally direct searches in the SUSY Higgs sector yield
additional constraints on the MSSM soft parameters.\par

The next step is the extraction of the soft SUSY-breaking parameters 
from the measured sparticle masses and branching ratios.
We use a Monte Carlo
technique relying on the generation of simulated experiments sampling
the probability density functions of the measured observables. 
We proceed in the following way:
\begin{enumerate}
\item
An `experiment' is defined as a set of measurements, each of which
is generated by picking a value from a Gaussian distribution with mean
given by the central value calculated from the input parameters 
of the considered model and width given by the estimated statistical+
systematic uncertainty of each measurement.
\item
For each experiment, we extract the constraints on the MSSM
model as we will describe in the following.
\end{enumerate}
We obtain as a result of this calculation a set of MSSM models, each
of which is the ``best'' estimate for a given Monte Carlo experiment 
of the model generating the observed measurement pattern. For each of
these models the $b$-physics observables can be calculated.

Three groups of soft SUSY-breaking parameters are relevant for the prediction 
of $b$-physics observables:
\begin{itemize}
\item
The parameters of the neutralino mixing matrix,
$M_1$, $M_2$, $\mu$, $\tan\beta$
\item
$\mA$, the mass of the pseudoscalar Higgs, defining (together with
$\tb$) the Higgs sector at tree level
\item
The masses and mixing angles of third generation squarks $\ttop$ and
$\sbot$
\end{itemize}
For the first two a detailed discussion is given in \cite{Nojiri:2005ph}
which we will briefly summarize here.\par
In the SPA point only the  mass of three neutralinos (1,2 and 4) 
can be measured. The three masses give a strong 
constraint on $M_1$, $M_2$, $\mu$, but have little 
sensitivity to  $\tb$. Therefore we use a 
fixed input value for  $\tb$, and 
we calculate  the values of
$M_1$, $M_2$, $\mu$ from numerical inversion of the neutralino 
mixing matrix. We will then study `a posteriori' the dependence on $\tb$.
The resultant uncertainty on $M_1$, $M_2$, $\mu$ is  $\sim$5-6 GeV, 
corresponding to the uncertainty on neutralino masses.  
By varying $\tb$ in the range $3<\tb<30$, the calculated values 
vary by less than $5$~GeV. \par

\begin{figure}[htb]
\begin{minipage}[c]{0.49\textwidth}
\dofig{0.96\textwidth}{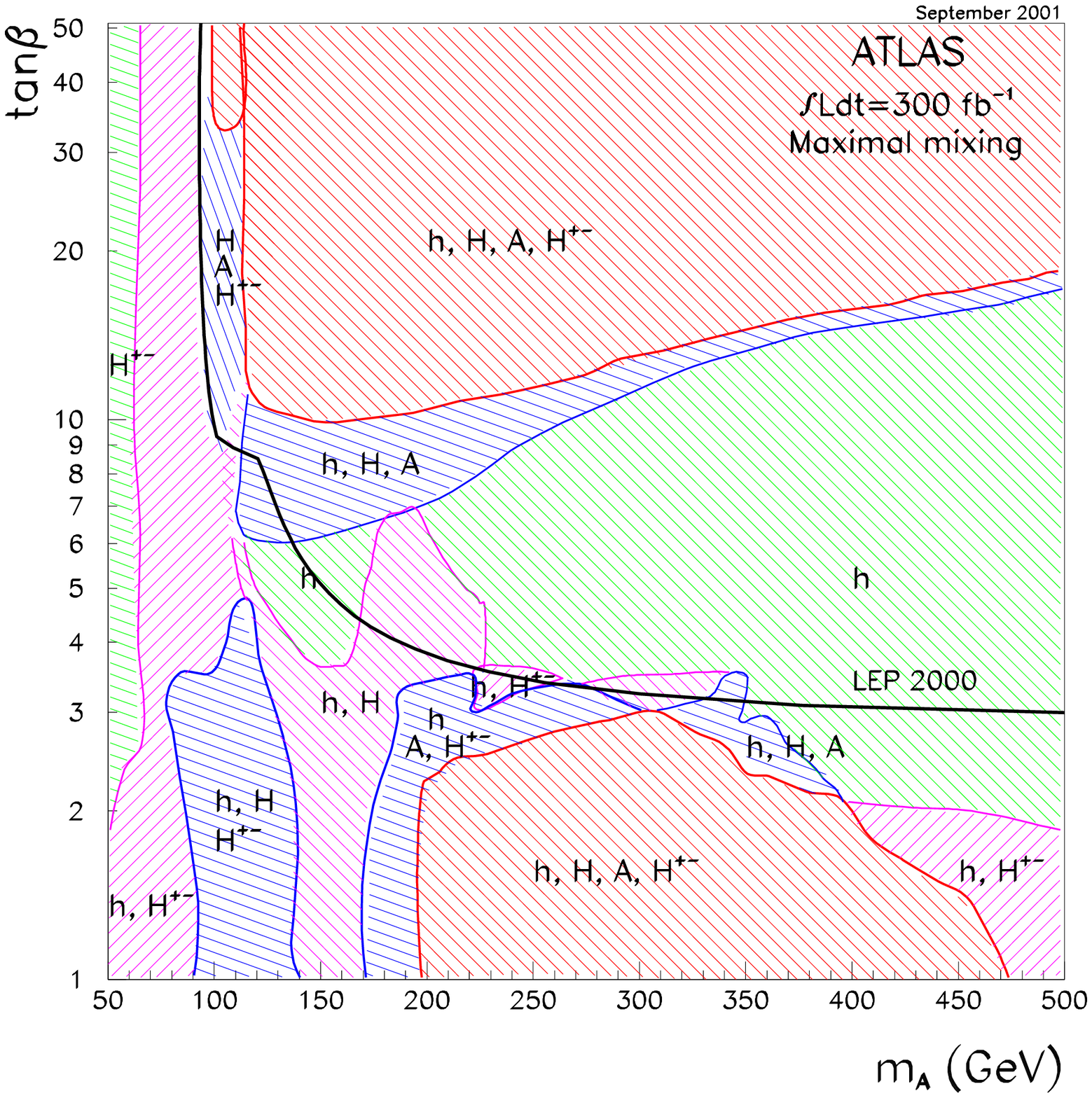}
\end{minipage}
\begin{minipage}[c]{0.49\textwidth}
\dofig{0.96\textwidth}{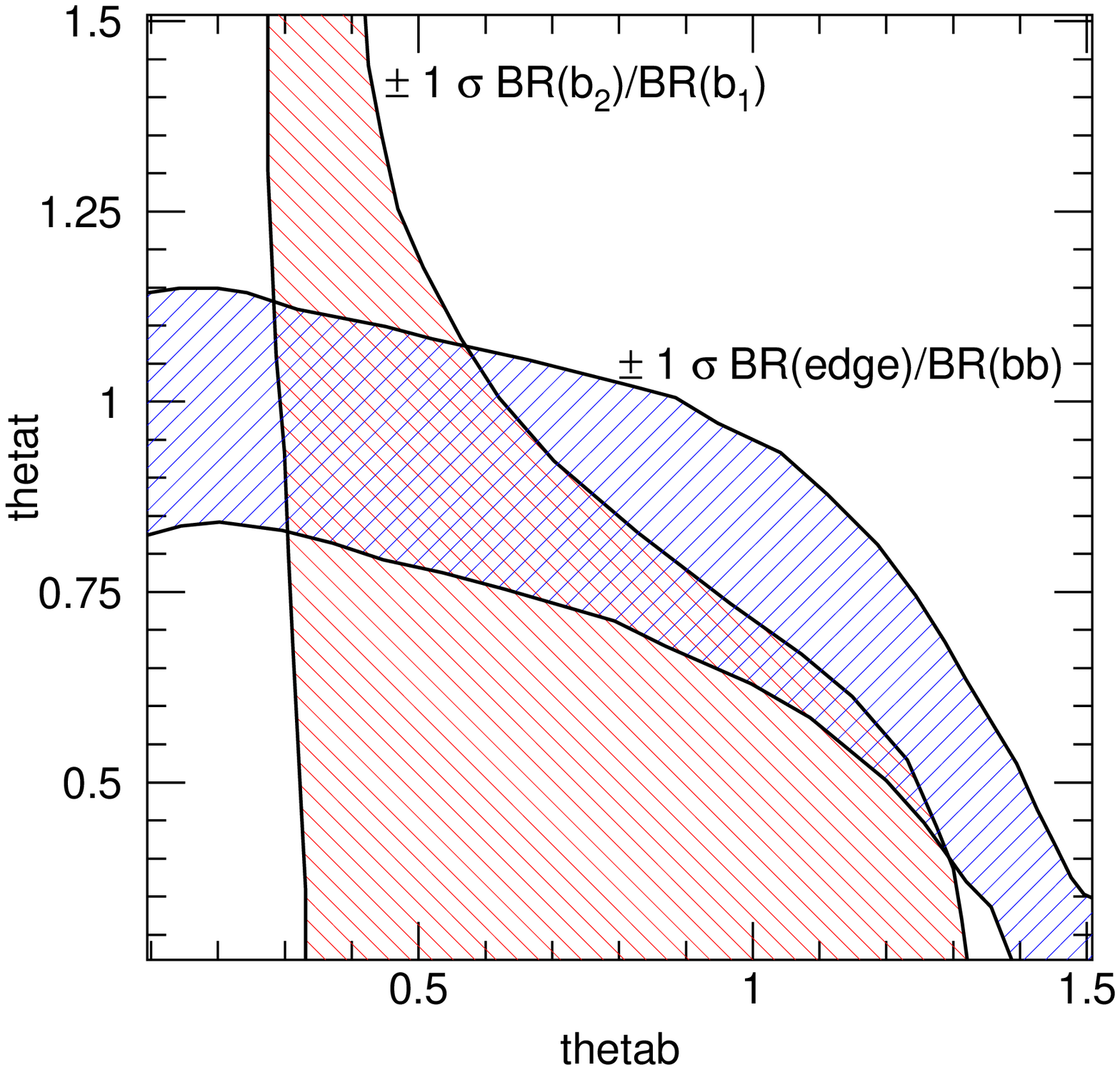}
\end{minipage}
\begin{minipage}[c]{0.45\textwidth}
\caption{\label{fig:mssmatlas} {\it
Reach of the ATLAS experiment in the \mbox{$\mA-\tb$}
plane for an integrated luminosity of 300~fb$^{-1}$. For each region
in the plane, the detectable Higgs bosons are marked.
}}
\end{minipage}
\begin{minipage}[c]{0.10\textwidth}
\mbox{}
\end{minipage}
\begin{minipage}[c]{0.44\textwidth}
\caption{\label{fig:thtthb} {\it
Allowed 1$\sigma$ bands on the $\tsb$-$\tst$ plane respectively 
for the measurement of $\br(\tb)$ (red downwards hatching) and of
$\br(\ttop)$ (blue upwards hatching). 
}}
\end{minipage}
\end{figure}

Information on $\tb$ and $\mA$ can in principle be extracted from the
study of the Higgs sector.
The ATLAS potential for discovery is shown in Fig.~\ref{fig:mssmatlas}, from 
\cite{atltdr}.  The light Higgs boson $h$ can be discovered 
over the whole parameter space, but the measurement of its
mass  only provides somewhat loose constraints, depending on 
the knowledge of the parameters of the stop sector.
Much stronger constraints would be provided by the measurement 
of the mass and production cross-section of one or more of the
heavy Higgs bosons. For the model under consideration, with $\tb=10$ and 
$\mA\sim$425~GeV, 
heavy Higgs bosons cannot be discovered at the LHC in their
SM decay modes.
Moreover, the heavy Higgs bosons can not be produced in chargino-neutralino
cascade decays because the decays are kinematically closed.
The only possibility would be
the detection of $A/H\to\tchi^0_2\tchi^0_2\to 4\ell\ell$.
Unfortunately the rate is very small, $\sim 40$ events/experiment 
for 300~fb$^{-1}$ before experimental cuts. 
A very detailed background study would 
be needed to assess the detectability of this signal.\par

We can now turn to the extraction of parameters of the stop-sbottom sector.
The sector is defined by 5 soft SUSY-breaking parameters:
$m(Q_3)$, the mass of the left-handed third generation 
doublet; $m(t_R)$ and  $m(b_R)$, the masses of the stop and
sbottom right-handed singlets; $A_t$ and  $A_b$, the stop
and sbottom trilinear couplings. More convenient mixing variables
would be $\tsb$ and $\tst$, the left-right sbottom and stop 
mixing angles.
For the considered point 5 measurements will be available
at the LHC: 
\begin{itemize}
\item
$\msbe$,~~~ $\msbz$,~~~~ 
\mbox{$\br(\tg\to b\sbot_2\to bb\tchi^0_2)/\br(\tg\to b\sbot_1\to bb\tchi^0_2)$} ($\br(\sbot)$) \cite{LHCLC}
\item
$M_{tb}^{\rm fit}$,~~~~ \mbox{$\br({\rm edge})/\br(\tg\to bbX)$} 
($\br(\ttop)$) \cite{Hisano:2003qu}
\end{itemize}
The assumed experimental errors on these variables are given 
in Table~\ref{tab:stop}.
\begin{table}[htb]
\begin{center}
\vskip 0.2cm
\begin{tabular}{|c|c|c|}
\hline
Variable & Value & Error  \\
\hline
\hline
$\mgl-\msbe$ &     128.7~GeV &    1.6~GeV    \\
$\mgl-\msbz$ &        86.9~GeV  &    2.5~GeV    \\
$\br(\sbot)$  &      0.70  &    0.05 \\
$\br(\ttop)$ &      0.21  &    0.08 \\ 
$M_{tb}$  &      411.3~GeV  &    5.4~GeV \\ 
\hline
\end{tabular}
\caption{\label{tab:stop}  {\it Assumed uncertainties
for the LHC measurements in stop-bottom sector.
The assumed statistics is 300~fb$^{-1}$. The only 
systematic error considered is the jet energy scale 
error on the mass/end point measurements.
}}
\end{center}
\end{table}

\begin{figure}[htb]
\dofigs{0.5\textwidth}{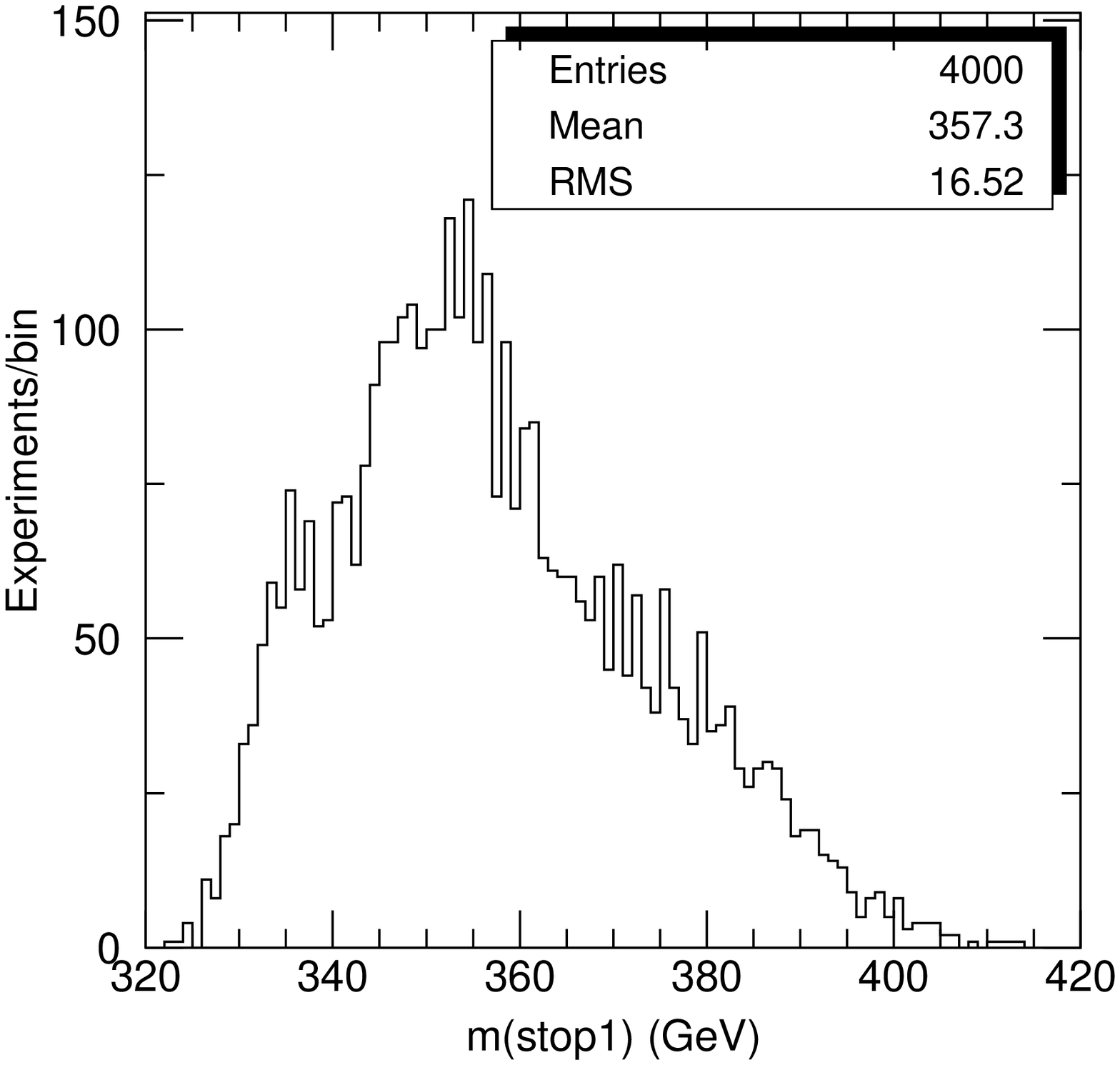}{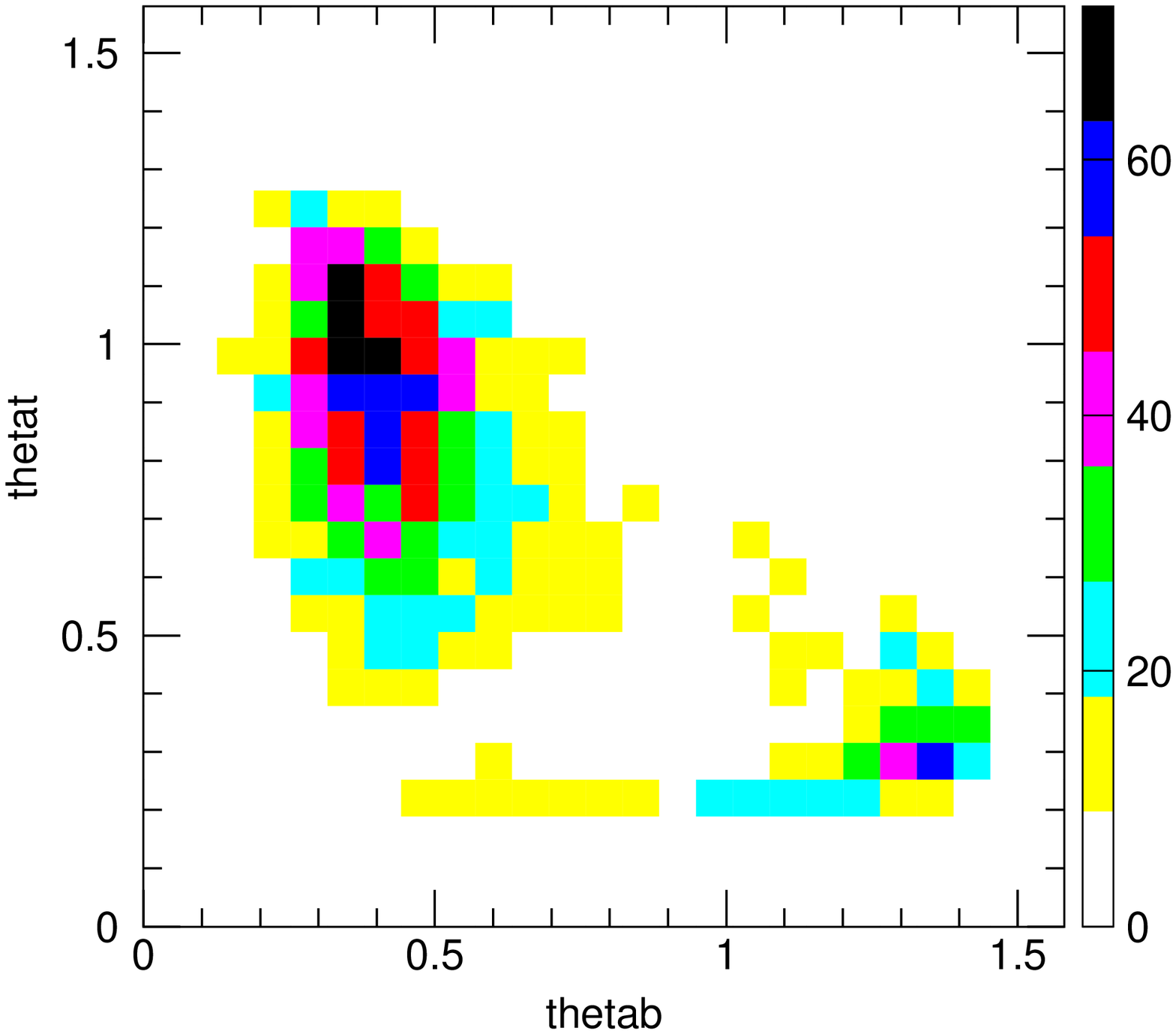}
\caption{\label{fig:var1} {\it
Left: distribution of the calculated $\ttop_1$ mass for an 
ensemble of Monte Carlo experiments at the LHC.
Right: distribution of the calculated $\tst$ versus $\tsb$ 
for an ensemble of Monte Carlo experiments. The assumed statistics is
300~fb$^{-1}$.
}}
\end{figure}

It is therefore possible to  
solve the available constraints for $\mste$, $\tsb$, $\tst$,
as discussed in \cite{mihokoweiglein}. 
In \cite{mihokoweiglein} the parameters of the gaugino matrix were
assumed to be measured with infinite precision at the ILC,  
and the errors on the parameters in the stop sector
were estimated by mapping the region in the \mbox{$\tst-\mste$}
plane compatible within the estimated errors with the 
nominal values of the five observables.\par
We incorporate the LHC uncertainties on the measurement
of $M_1$, $M_2$, $\mu$, and we use the technique of 
building Monte Carlo experiments described above.\par
The strategy is to scan the three-dimensional space $\mste$,
$\tsb$, $\tst$, and to find the point in space which 
reproduces the measured values of $M_{tb}$, $\br(\ttop)$, $\br(\sbot)$.
For fixed $\mste$, the measurement of the
position  in the \mbox{$\tsb$-$\tst$} plane is given 
by combining the crossing of the line corresponding to the
measured value of $\br(\sbot)$ with the line corresponding to the measured 
values of $\br(\sbot)$. We show in Fig.~\ref{fig:thtthb} respectively 
the band constrained by $\pm 1\sigma$ around the input values
of $\br(\sbot)$ and $\br(\ttop)$ when all the other MSSM parameters
are kept fixed.  Because of the rather loose 
constraints on $\br(\sbot)$, and the low statistics in the $\sbot_2$
peak, the region where the two bands cross, which roughly 
represents the allowed region in the plane, extends 
from the region around the input value 
($\tst=0.933$, $\tsb=0.42)$ with  a very low tail
towards the region of high $\tsb$ and low $\tst$.\\
The results of the scan are shown in Fig.~\ref{fig:var1}.
In the left plot we show the distribution of the measured 
$\mste$ values for the considered ensemble of MC experiments.
The RMS of the distribution is $\sim 17$~GeV, corresponding
to a $\sim 5$\% uncertainty on the light stop mass.  
The measured values in the $\tst$ versus $\tsb$ plane
are shown in the plot on the right of  Fig.~\ref{fig:var1}.
As expected from the discussion above, a significant number 
of experiments yield a high value of $\tsb$ and a low value
of $\tst$. \par

The conclusions on the MSSM parameter measurement for the SPA model point
under the assumption  of no FCNC effects from sfermion 
mixing matrices are thus:
\begin{itemize}
\item Neutralino/chargino mixing matrices fixed with $\sim 5\%$
if the value of $\tb$ is known.
\item
Slepton sector well constrained, including stau mixing angle
\item
Masses of first two generations squarks (L \& R) and of gluino 
measured at $\sim$5-10\% level
\item
Enough constraints to fix the 5 parameters of the stop/sbottom
sector. For fixed $\tb$ uncertainty of $\sim$5\% on stop mass, 
long tails in the measurement of $\tsb$ and $\tst$.
\item
Weak constraints on $\tb$ and $\mA$
\end{itemize}

We can now, based on the expected 
precision for the measurement of MSSM parameters 
estimate how precisely observables in the $b$-sector can 
be predicted.
We focus on two variables:
\begin{itemize}
\item
$\br(B_s\to \mu\mu)$
\item
$\br(B\to X_s\gamma)$
\end{itemize}
Two public programs 
micrOMEGAs 1.3.6 \cite{Belanger:2001fz} and ISARED \cite{isa}
allow the evaluation  of these two variables from 
an input set of MSSM parameters. Both programs work 
in the MFV framework, and are based on the most recent
NLO calculations. The results from  micrOMEGAs 1.3.6
were used for the present exercise.\par
The study is done in different steps. We first perform 
scans in the parameter space to evaluate the
sensitivity of the two observables to the key parameters.
Thereafter, based on the method of Monte Carlo experiments
described above, we evaluate  the expected value of $\br(B_s\to\mu\mu)$
and $\br(B\to X_s\gamma)$ for each Monte Carlo experiment.
The spread of the obtained distributions is taken as the experimental
uncertainty of the observables. Since $\mA$ and $\tb$
are badly constrained by the LHC measurements, this is done
keeping  $\mA$ and $\tb$ fixed.\par 

\begin{figure}[htb]
\dofigs{0.45\textwidth}{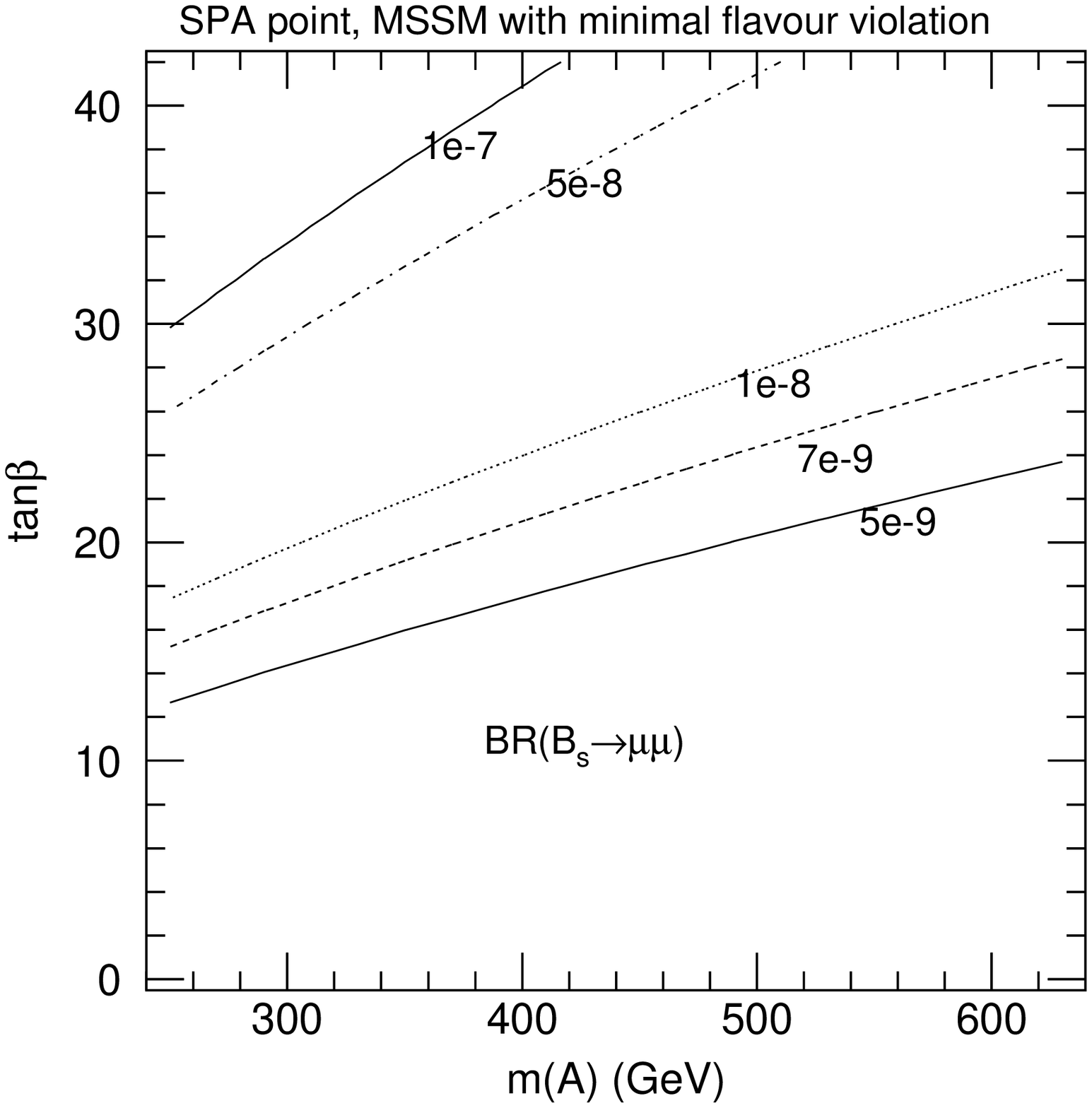}{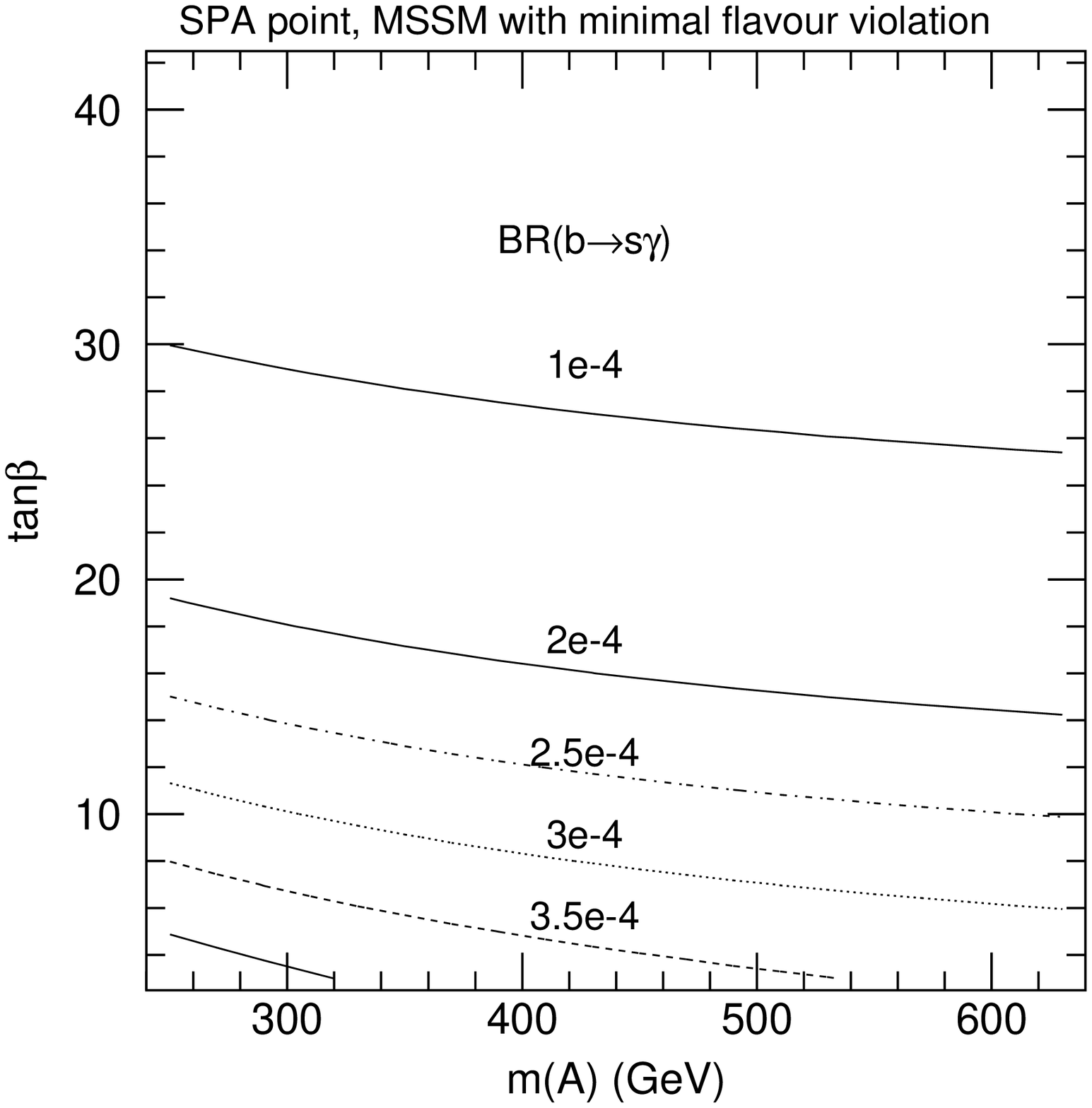}
\caption{\label{fig:pardep} {\it
Left: curves of equal value for $\br(B_s\to \mu\mu)$ in the
\mbox{$\mA-\tb$} plane. Right: curves of equal value 
for $\br(B\to X_s\gamma)$. The MSSM parameters
are as defined for the SPA point and the calculations are
performed using MicrOMEGAs.}}
\vspace{-1em}
\end{figure}

The dependence of $\br(B_s\to \mu\mu)$ on $\mA$, $\tb$ is 
shown in the left panel of Fig.~\ref{fig:pardep}.
Since $\br(B_s\to \mu\mu) \propto \tan^6\beta/\mA^4$,
this measurement has a
strong constraining power on $\tb$ if $\tb\simge 15$.
For lower values of $\tb$ $\sim$ the effect 
becomes too small and SUSY is indistinguishable from the SM.
The present limits from the Tevatron 
experiments only eliminate a small region of the parameter space with
small $\mA$ and large $\tb$.
The expected 90\% bound from ATLAS:   $6.6\times10^{-9}$ for 30~fb$^{-1}$
\cite{atlasbmumu} would allow us to 
exclude a region in $\mA-\tb$ similar to the one excluded by 
non-discovery of $H/A\to\tau\tau$.
For higher $\tb$ the measurement of a 
deviation from the SM would provide a 
nice cross-check with $\tb$ as measured 
from $H/A$ production.\par
The value of $\br(B\to X_s\gamma)$ in the $\mA-\tb$ plane 
is shown in the right panel of Fig.~\ref{fig:pardep}.
The present world average for $\br(B\to X_s\gamma)$~\cite{Barberio:2006bi}:
$$
(3.3\pm 0.4) \times 10^{-4}
$$
would select a narrow band in the $\mA-\tb$ plane, thus providing
essentially no bound on $\mA$ and 
a strong constraint on the allowed $\tb$ range, in the MFV 
hypothesis.\par

\begin{figure}[htb]
\dofigs{0.45\textwidth}{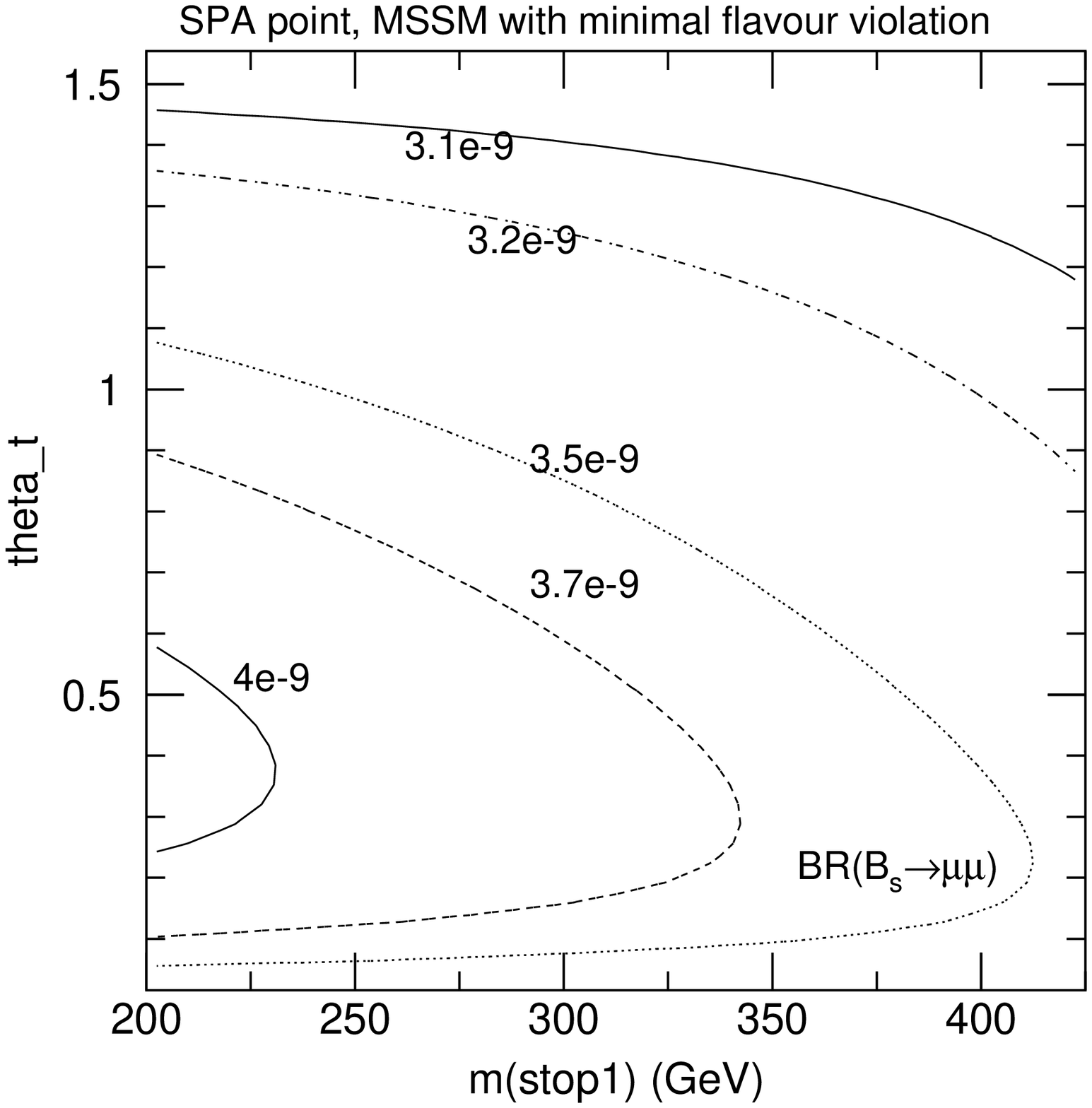}{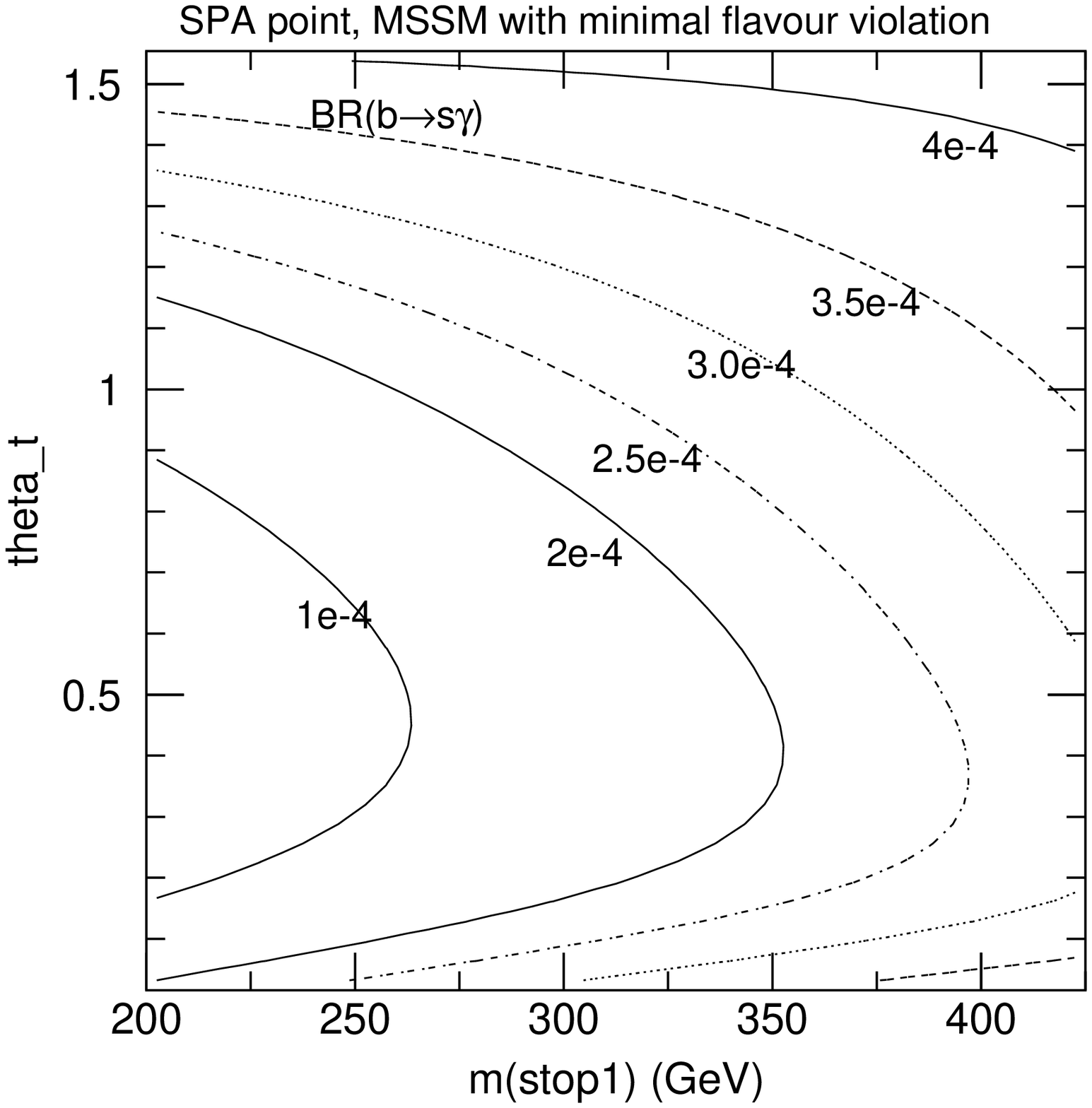}
\caption{\label{fig:pardep1} {\it
Left: curves of equal value for $\br(B_s\to \mu\mu)$ in the
\mbox{$\mste-\tst$} plane. Right: curves of equal value
for $\br(B\to X_s\gamma)$ in the
\mbox{$\mste-\tst$} plane. The MSSM parameters
are as defined for the SPA point and the calculations are
performed using MicrOMEGAs.}}
\end{figure}

\begin{figure}[htb]
\dofigs{0.45\textwidth}{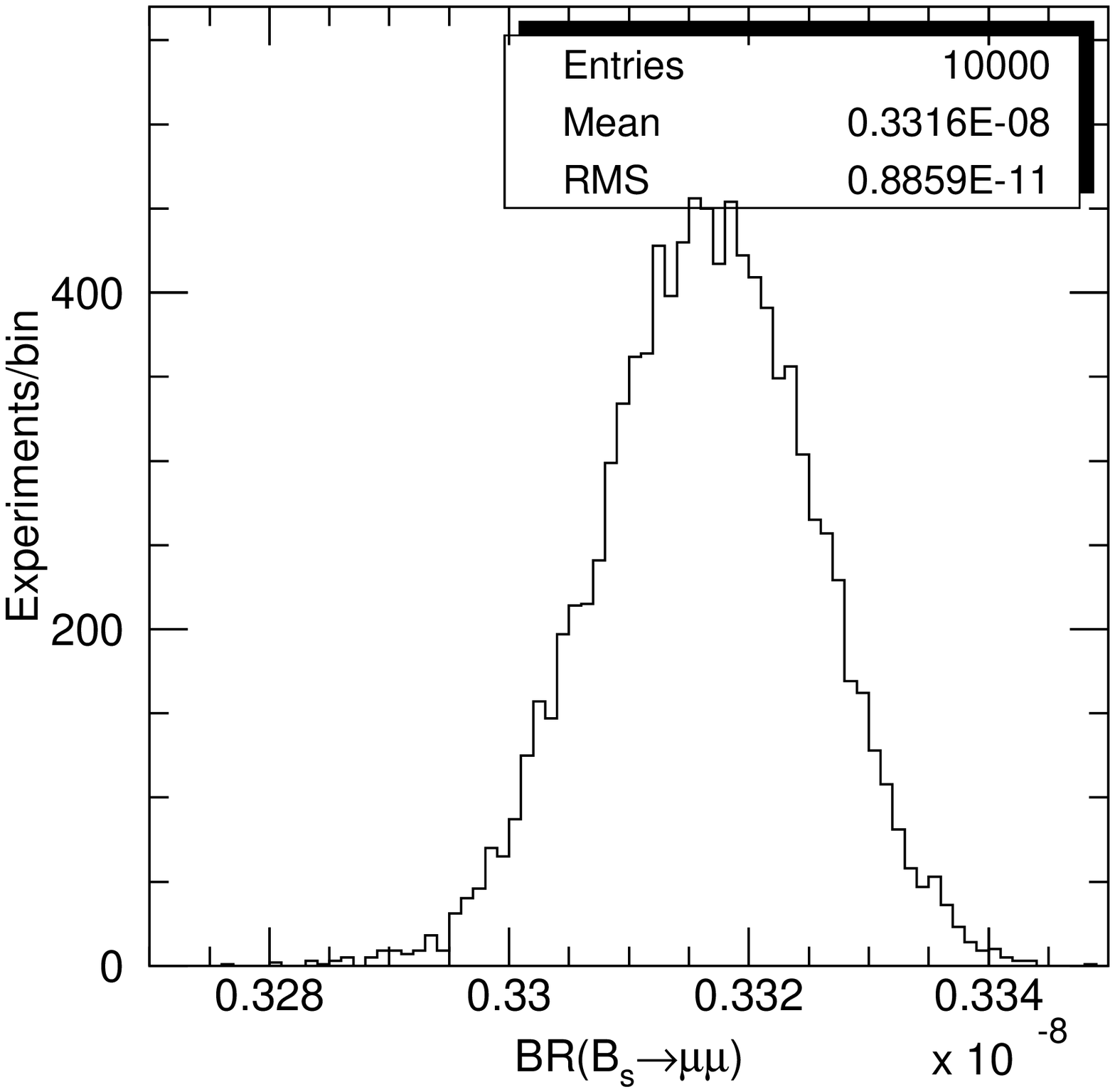}{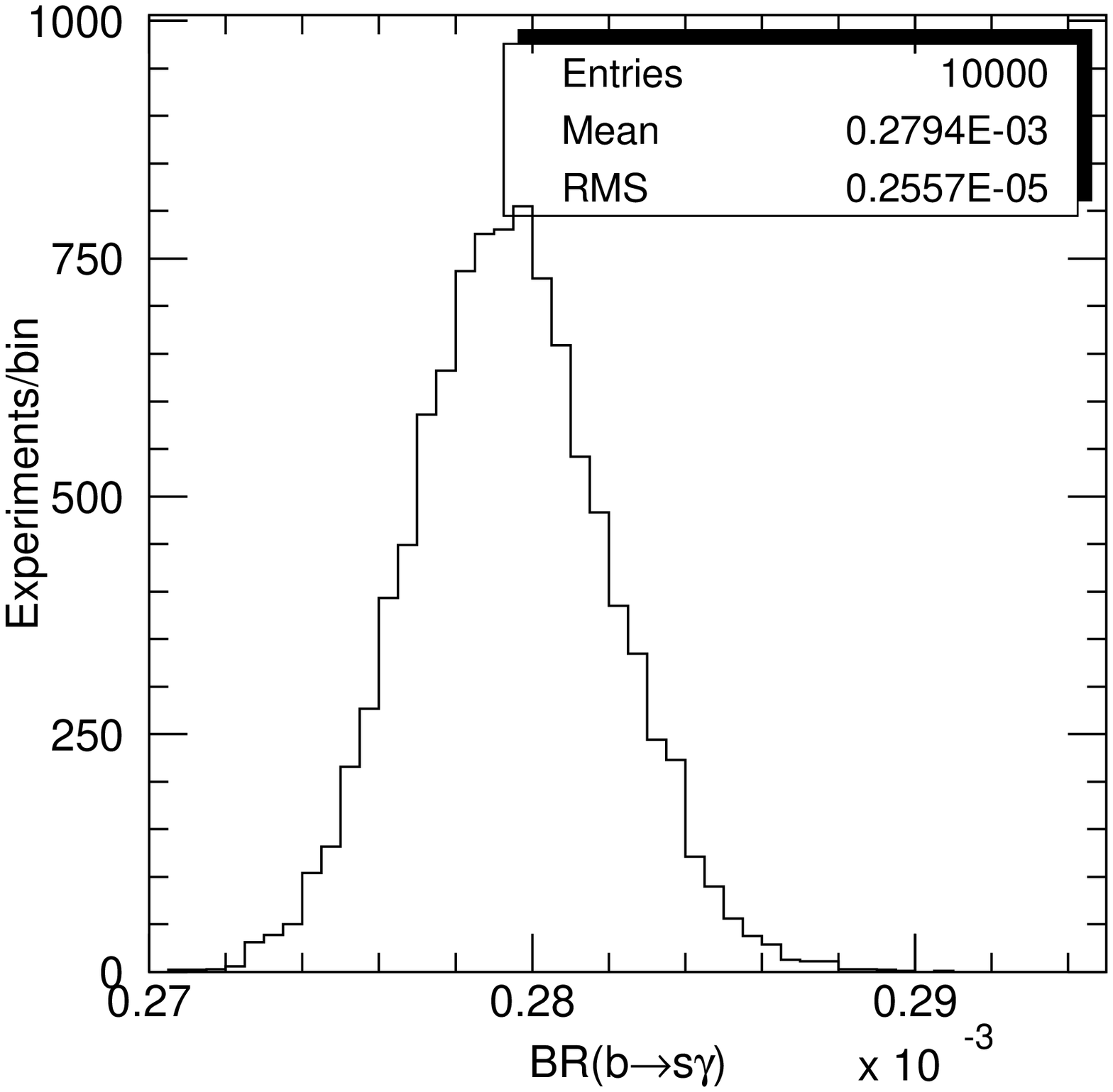}
\caption{\label{fig:meas1} {\it Distribution of the predictions 
$\br(B_s\to \mu\mu)$ (left) and $\br(B\to X_s\gamma)$
(right) for an ensemble of LHC experiments when $\mA$, $\tb$,
$\mste$, $\tst$, $\tsb$ are kept fixed at the nominal
values and all the remaining MSSM parameters are smeared
according to the expected measurement uncertainty 
}}
\vspace{-1em}
\end{figure}

We show in Fig.~\ref{fig:pardep1} the values of 
$\br(B_s\to \mu\mu)$ and $\br(B\to X_s\gamma)$ in the 
$\mste-\tst$ plane with the other parameters fixed (see
Fig.~\ref{fig:meas1} below for an analysis of the effect of their uncertainty).
The variation of $\br(B_s\to \mu\mu)$ over the 
considered space is moderate.
The present experimental error on the 
measurement of  $\br(B\to X_s\gamma)$ already  defines a very 
small slice in the $\mste-\tst$ plane.  
For fixed $\tst$ the  dependence on $\mste$ is not very strong.
We therefore conclude that a precise measurement of $\tst$ is
the key ingredient for the prediction of   $\br(B\to X_s\gamma)$
from the LHC SUSY data.\par

As a  next step we  verify that the experimental uncertainty 
on the two considered observables is indeed dominated by 
the measurement of $\mA$, $\tb$, $\mste$ and $\tst$.
To this effect we calculate $\br(B_s\to \mu\mu)$  
and $\br(B\to X_s\gamma)$ for all the Monte Carlo experiments,
letting all of the MSSM parameters fluctuate according to 
the experimental error, except the four parameters mentioned above.
The result  is shown in Fig.~\ref{fig:meas1}. In these conditions
the uncertainty is small, 
0.3\% on the prediction of  $\br(B_s\to \mu\mu)$
and 1\% for the prediction  of 
$\br(B\to X_s\gamma)$. These parametric uncertainties do not include 
the theoretical uncertainties in the calculation of the two 
observables.\par
\begin{figure}[htb]
\dofigs{0.45\textwidth}{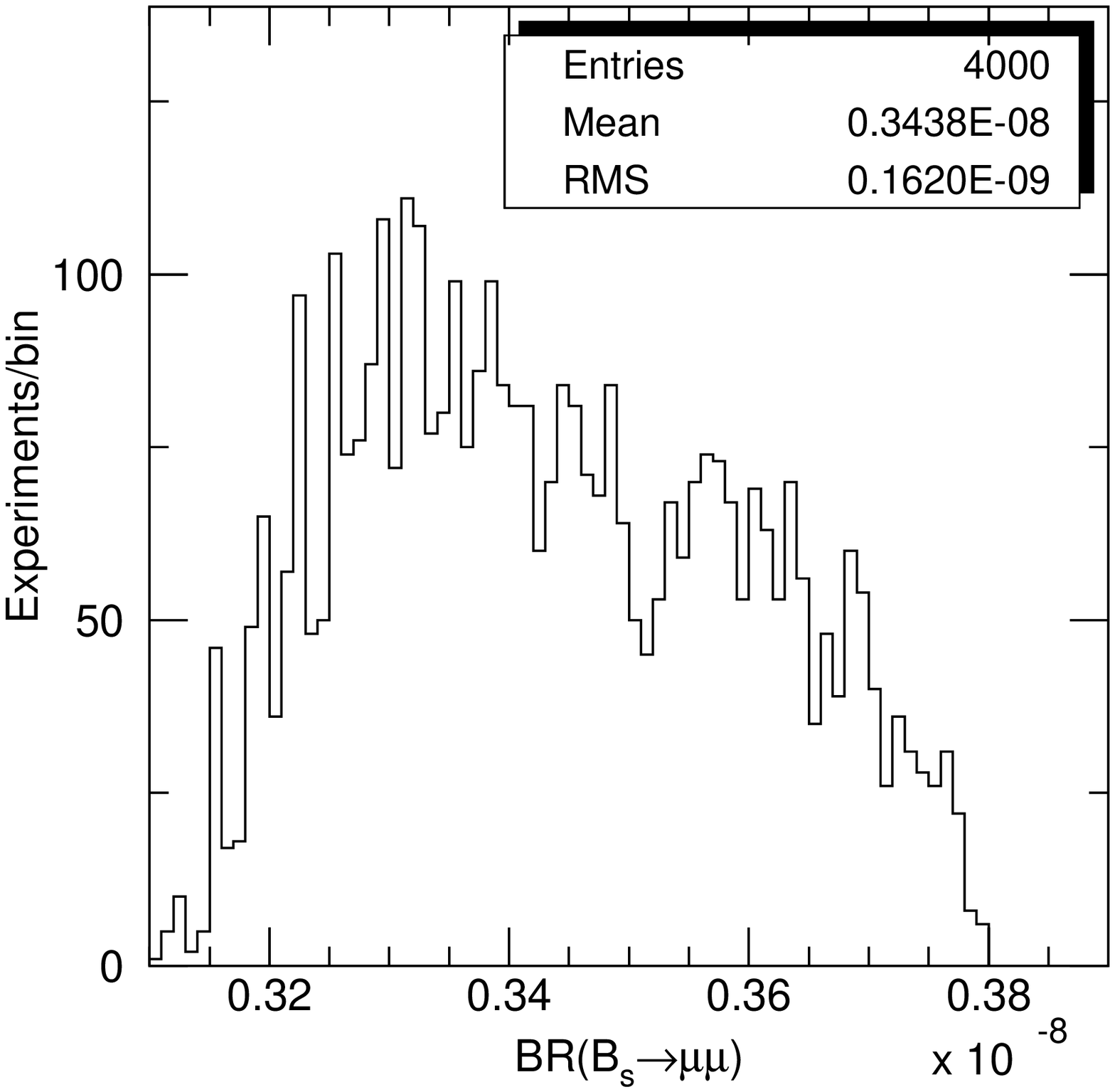}{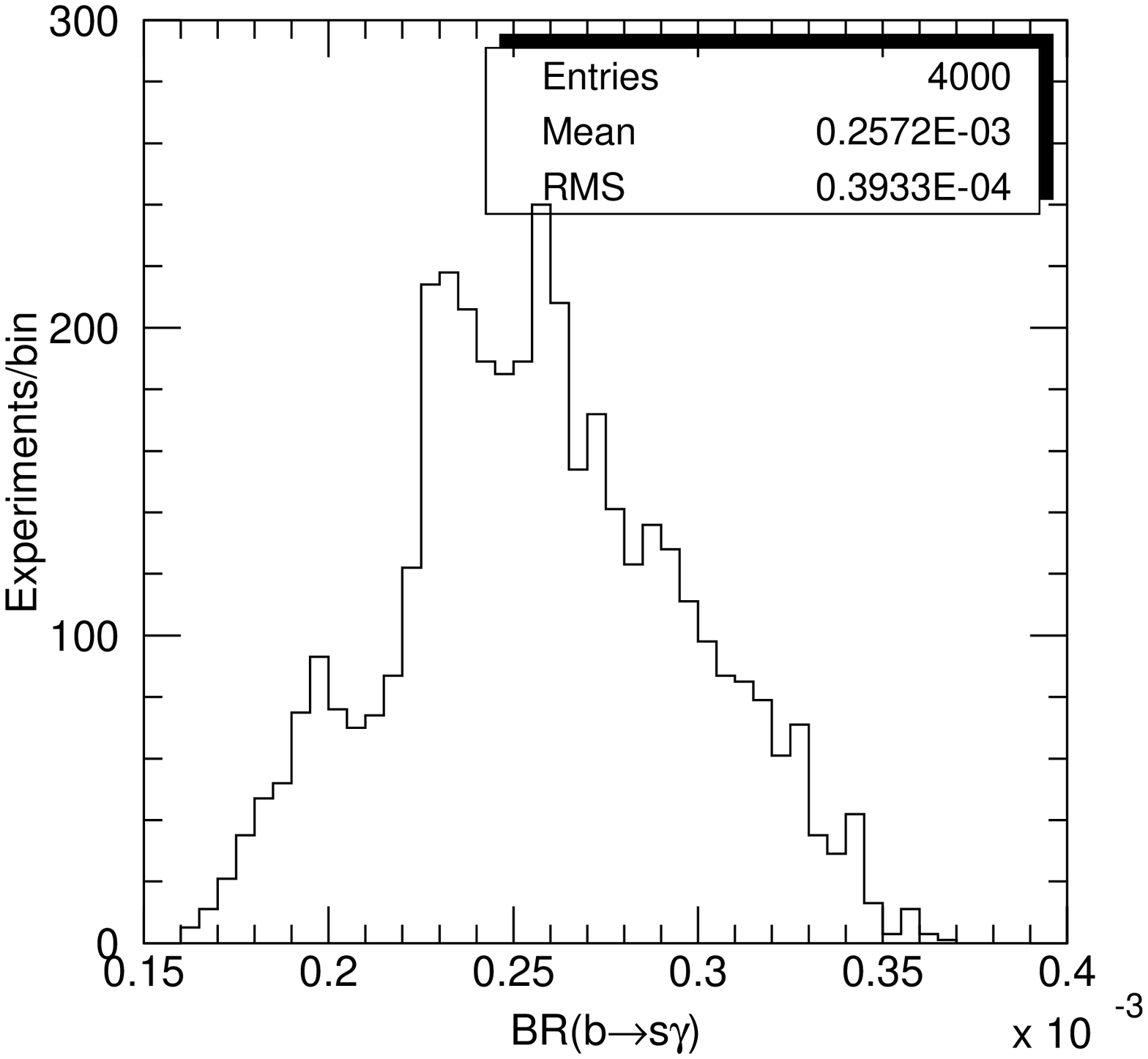}
\caption{\label{fig:full} {\it
Distribution of the predictions
$\br(B_s\to \mu\mu)$ (left) and $\br(B\to X_s\gamma)$
(right) for an ensemble of LHC experiments when $\mA$, $\tb$,
are kept fixed at the nominal
values and all the remaining MSSM parameters, including
the ones defining the stop sector are smeared
according to the expected measurement uncertainty
}}
\vspace{-1em}
\end{figure}

Finally, we can evaluate how precisely we can predict 
the $b$-physics observables, by varying 
all of the MSSM parameters, according to 
the expected measurement precision at the LHC for the SPA point,
except $\mA$ nd $\tb$, which are kept fixed.
The results are shown in Fig.~\ref{fig:full}.
We observe a $\sim$5\% uncertainty  on the prediction 
for $\br(B_s\to \mu\mu)$, 
and a $\sim$15\% uncertainty  on the prediction 
for $\br(B\to X_s\gamma)$.
For both observables one can roughly observe two populations, corresponding
to the regions in \mbox{$\tsb$-$\tst$} observed in 
Fig.~\ref{fig:var1}. The experiments in the tail of mismeasured
$\tst$ and $\tsb$ contribute respectively to the region 
of high values of $\br(B_s\to \mu\mu)$, and to the bump
for low values of $\br(B\to X_s\gamma)$.\par
We have thus shown that for the considered model
good enough measurements of MSSM parameters are possible
at the LHC to provide predictions for
$\br(B\to X_s\gamma)$, $\br(B_s\to \mu\mu)$ 
as a function of the two unconstrained variables: $\mA$ and $\tb$.

Once the LHC data are available, one can imagine different scenarios, e.g. 
\begin{itemize}
\item
$A/H\to\tau\tau$ is  observed and $\tb$ and 
$\mA$ measured. \\
At this point a consistency check would be possible among the $\tb$
constraints provided by the Higgs measurement and the one provided
by the $b$-physics observables calculated in the MFV scheme. 
A significant disagreement, once all the
experimental and statistical uncertainties are 
evaluated, would indicate the presence of flavour violation in the
squark sector.
\item
$\tb$ is not constrained by high-$p_T$ searches. \\
A signal for non-minimal flavour violation could still be provided
by the inconsistency of the $\tb$ regions constrained 
by respectively $m(h)$, $\br(B\to X_s\gamma)$,  
and $\br(B_s\to \mu\mu)$. In case of consistency the 
results could be taken as a measurement of the $\tb$ parameter. 
\end{itemize}
Relevant questions at this point are:
what are the precisions required on the MSSM and 
on the $b$-physics measurements and on the theoretical calculations 
to be able to claim a signal for flavour-changing terms in the 
squark mass matrices?\\
In case the measurements are consistent with MFV, what additional constraints
on the flavour violation sector can be extracted by combining MSSM studies
and $b$-physics measurements?\par
Various analyses  are available in the 
literature \cite{Ciuchini:2002uv}, \cite{Foster:2005wb}, 
based on assessing present allowed regions of
non-diagonal elements in the super-CKM matrix, parametrised in terms of
$(\delta^d_{23})_{AB}$, where AB can be $RR$, $LL$, $RL$, $LR$.
Bounds on $\delta$ are normally given for some special choice 
of soft SUSY-breaking parameters, e.g.\ $m(\tq)=\mgl=\mu=-A_u$ for different 
choices of $m(\tq)$. Additional variables are 
also considered such as $\Delta M_B$, 
$\br(B\to X_s\ell^+\ell^-)$, $A_{CP}(B\to X_s\gamma)$.\par
Based on the study presented here it would be interesting
to repeat these analyses but for the
parameters of a specific SUSY point, incorporating the expected 
experimental errors on the SUSY parameters. As a result of these studies, 
one could get guidance on which are the MSSM measurements  
crucial to discover flavour violation, thus pointing the way
for the investigation of SUSY models in high-$p_T$ physics.



\subsubsection{The second approach:\\
               SUSY measurements in $b$-physics favoured parameter spaces}
\label{sec:newscen2app}

A second, somewhat complementary, approach was followed in
collaboration with CMS physicists. 

\subsubsubsection{$b$-physics favoured parameter space}

The model under investigation is the MSSM, in the first step with MFV,
and possibly in a later stage also with NMFV. 
The compatibility with flavour physics was taken into account following
Ref.~\cite{Isidori:2006pk}, where the MSSM parameter space was analyzed
under the assumption of heavy scalar quarks and leptons, 
and  large $\tan\beta$.
The range of SUSY parameters has been restricted to the values
listed in Table~\ref{tab:CMSparameters}.
Here $\tb$ is the ratio of the two vacuum expectation values, $\MA$ denotes
the mass of the CP-odd Higgs boson, $\mu$ is the Higgs mixing
parameter, $M_{\tilde q, \tilde l}$ are
the diagonal soft SUSY-breaking parameters in the scalar quark and
scalar lepton sector, respectively. All the trilinear couplings are 
set to be equal to $\At$ (the tri-linear Higgs-stop
coupling), while $\Mgl$, $M_2$
and $M_1$ are the gluino mass and the soft SUSY-breaking parameters in
the chargino/neutralino sector. All parameters are assumed to be real. The upper
part of Table~\ref{tab:CMSparameters} are the more relevant parameters,
while the lower part has a smaller impact on the flavour physics
phenomenology. 

The ranges in Ref.~\cite{Isidori:2006pk} are generally compatible 
with the existing low-energy constrains. However, one expects to be able 
to select narrow sub-regions by more precise measurements 
of specific $B$-physics observables, such as $\br(B\to\tau\nu)$ or 
$\br(B_s\to\mu^+\mu^-)$. The ``best'' values denote specific 
points for which a more detailed investigation of the high-energy 
signatures at CMS has been performed.

\begin{table}[h]
\renewcommand{\arraystretch}{1.2}
\begin{center}
\begin{tabular}{|l|c|c|}
\cline{2-3} \multicolumn{1}{l|}{}
 & range & ``best'' value(s) \\ \hline\hline
$\tb$          & 30 -- 50       & 40 \\ \hline
$\MA$ [GeV]    & 300 -- 1000    & 300, 500, 800, 1000 \\ \hline
$\At$ [GeV]    & -2000 -- -1000 & -1000, -2000 \\ \hline
$\mu$ [GeV]    & 500 -- 1000    & 500, 1000 \\ \hline
$M_{\tilde q}$ [GeV] & $>$ 1000 & 1000, 2000 \\ \hline \hline
$M_{\tilde l}$ & 1/2 $M_{\tilde q}$ & \\ \hline
$M_{\gl}$      & $M_{\tilde q}$ & \\ \hline
$M_2$ [GeV]    &  & 300, 500 \\ \hline
$M_1$          & 1/2 $M_2$ & \\ \hline
\end{tabular}
\caption{Selected ranges and ``best values'' of the SUSY parameters
for the ``CMS analysis'' in the MFV MSSM (following Ref.~\cite{Isidori:2006pk}):
$\tb$ is the ratio of the two vacuum expectation values, $\MA$ denotes
the mass of the CP-odd Higgs boson, $\mu$ is the Higgs mixing
parameter, $M_{\tilde q, \tilde l}$ are
the diagonal soft SUSY-breaking parameters in the scalar quark and
scalar lepton sector, respectively; $\At$ is the tri-linear Higgs-stop
coupling, where all trilinear couplings are set equal; $\Mgl$, $M_2$
and $M_1$ are the gluino mass and the soft SUSY-breaking parameters in
the gaugino sector. All parameters are assumed to be real.} 
\label{tab:CMSparameters}
\end{center}
\renewcommand{\arraystretch}{1}
\end{table}


\subsubsubsection{Experimental analysis}


The strategy followed by CMS physicists is to apply an already understood
search analysis to the sample of MSSM points that are consistent
with flavour constraints as described above.  
The starting point is Ref.~\cite{CMSNOTE2006134}, in which CMS
studied the production and decay of SUSY particles via
inclusive final states including muons, high-$\PT$ jets, and large
missing transverse energy.  In that work, a fully simulated and
reconstructed low mass (LM1) Constrained MSSM (CMSSM) point was taken
as the benchmark for 
selection optimisation and study of systematic effects.  Even though the
study was performed within the context of CMSSM, the method is not
specific to the CMSSM framework and should apply equally well in other
contexts including, i.e.\ also in the general MSSM. 

The response of the CMS detector to incident particles was simulated
using a~GEANT4-based framework~\cite{Geant4}, known as the
Object-oriented Simulation for CMS Analysis and Reconstruction
(OSCAR)~\cite{PTDRV1}.  The inclusion of pile-up and the reconstruction
of analysis objects (muons, jets, etc) from hits in the detector was
performed by a software framework known as the Object-oriented
Reconstruction for CMS Analysis (ORCA)~\cite{PTDRV1}. In addition, a
standalone fast simulation, known as the CMS FAst MOnte Carlo Simulation
(FAMOS) framework~\cite{PTDRV1}, was used to facilitate simulations
involving CMSSM parameter scans.  The fast simulation FAMOS has been
shown to adequately represent the full CMS simulation
\cite{CMSNOTE2006134}.  In both the full and fast simulations, hits from
minimum bias events are superimposed on the main simulated event to
reproduce the pile-up conditions expected for a luminosity 
of~$2 \times 10^{33}{\rm cm^{-2}s^{-1}}$.   

Because the work presented in Ref.~\cite{CMSNOTE2006134} is an inclusive
study of signatures involving at least one muon accompanied by multiple
jets and large \mMET, several SM processes contribute as
sources of background and had to be taken into account. Accordingly, the
main backgrounds studied in Ref~\cite{CMSNOTE2006134} correspond to QCD
dijet (2.8 million events with $0<\!\hat{p}_{\rm T}\!<4{\rm TeV}/c$),
top (${t\bar{t}}$) production (3.3 million events), electroweak
single-boson production (4.4 million events with 
$0<\!\hat{p}_{\rm T}\!<\!4.4{\rm TeV}/c$) and electroweak dibosons
production (1.2 million events). All backgrounds used were fully
simulated and reconstructed.   

The method employed in Ref.~\cite{CMSNOTE2006134} is to search for an excess
in the number of selected events, compared with the number of events
predicted from the SM.  A Genetic Algorithm
(GARCON~\cite{Abdullin:2006nu}) was used for the optimisation of cuts to
select the LM1 CMSSM point and results in:  
$E_{\rm T}^{\rm miss} >130 \gev$, 
$E^{\rm j1}_{\rm T}\!>\!440 \gev$, 
$E^{\rm j2}_{\rm T}\!>\!440 \gev$, 
$|\eta^{\rm j1}|<1.9$, 
$|\eta^{\rm j2}|<1.5$,
$|\eta^{\rm j3}|<3$, 
$\cos \big[\Delta \phi({\rm j1,j2}) \big]\!<\!0.2$,
$-0.95 < \cos \big[ \Delta \phi(\MET,{\rm j1}) \big]\!<\!0.3$, 
$\cos\big[ \Delta \phi(\MET,{\rm j2}) \big]\!<\!0.85$.  
Assuming 10 fb$^{-1}$
of collected data, this set of cuts would expect to select a total of
2.5 background events from the SM and 311 signal events from
the CMSSM LM1 benchmark signal point \cite{CMSNOTE2006134}. 

In order to extend the work presented in \cite{CMSNOTE2006134} to the
context of the MSSM parameter space suggested by flavour
considerations as described above,
several points within the ranges of the MSSM parameters listed in
Table~\ref{tab:CMSparameters} were sampled and simulated using the CMS fast
simulation FAMOS.  (The Pythia parameters used to generate each MSSM
point may be found in Ref.~\cite{CMSNOTE2006134}.)   In the CMS
exercise, the same set of selection cuts presented above, is directly
applied (i.e.\ not re-optimised) to each simulated MSSM point.  Finally,
the number of selected events from each simulated MSSM point is tallied
and compared with the expected number of standard model background
events ($N_B = 2.5$). 


It has been shown that the analysis method also works for this ``new''
part of the MSSM parameter space. Clearly, an optimization could enhance
the analysis power. More detailed results will be presented elsewhere.


\subsubsection{The ``master code'': multi-parameter fit to electroweak
  and low-energy observables}
\label{sec:mastertool2}

A first attempt to develop a ``master code'' as described above (see
also Section~\ref{sec:mastertool}) has been started in the course of the
workshop in collaboration with physicists from CMS~\cite{mastertool}. 

Based on flavour physics computer code from \cite{Isidori:2006pk} and
the more high-energy observable oriented computer code 
{\tt FeynHiggs}~\cite{Heinemeyer:1998np,Degrassi:2002fi,Heinemeyer:2004by},
a first version of a ``master code'' has been developed. This ``master
code'' combines calculations from both low-energy and electroweak
observables in one common code. Great care has been taken to ensure that
both sets of calculations are steered with a consistent set of input
parameters. The current version of the ``master code'' is restricted to
applications in the MSSM parameter space assuming Minimal Flavour
Violation (MFV). 
Table~\ref{tab:constraints} shows the
observables which are currently considered in the ``master code''. 

However, in the future it is foreseen to significantly
extend the ``master code'' by including other calculations both for
different New Physics models as well as additional observables
(e.g. cosmology constraints), see \cite{Buchmueller:2007zk} for the
latest updates and developments. 
With the help of the ``master code'' it will eventually be possible to
test model points from the low-energy side (via flavour and electroweak
observables) and from the high-energy side (via the measurements of
ATLAS/CMS). Thus a model point can be tested with {\em all} existing
data.

\begin{table}[tbh!]
\renewcommand{\arraystretch}{1.2}
\begin{center}
\begin{tabular}{|c|c|c|c|} \hline
Observable & Source & Constraint & theo. error \\ \hline \hline
$R_{\br_{b \to s \gamma}} = \br_{b \to s \gamma}^{\rm SUSY}/
\br_{b \to s \gamma}^{\rm SM}$  &    \cite{Isidori:2006pk}   & $1.127
\pm 0.12$  & $0.1$  \\ \hline 
$R_{\Delta M_s} = \Delta M_s^{\rm SUSY}/ \Delta M_s^{\rm SM}$  &
\cite{Isidori:2006pk}   & $0.8 \pm 0.2$  & $0.1$  \\ \hline 
$\br_{b \to \mu \mu}$ &    \cite{Isidori:2006pk}   & $<8.0 \times
10^{-8}$  & $2 \times 10^{-9}$  \\ \hline 
$R_{\br_{b \to \tau \nu}} = \br_{b \to \tau \nu}^{\rm SUSY}/
\br_{b \to \tau \nu}^{\rm SM}$  &    \cite{Isidori:2006pk}   & $1.125
\pm 0.52$  & $0.1$  \\ \hline 
$\Delta a_{\mu} =  a_{\mu}^{\rm SUSY}-  a_{\mu}^{\rm SM}$  &   {\tt FeynHiggs}
& $(27.6 \pm 8.4) \times 10^{-10}$  & $2.0 \times 10^{-10}$  \\ \hline 
$M_W^{\rm SUSY}$ &   {\tt FeynHiggs}  & $80.398 \pm 0.025 $ GeV & $0.020$
GeV\\ \hline 
$\sin^2 \theta_W^{\rm SUSY}$ &   {\tt FeynHiggs}  & $0.23153 \pm 0.00016 $
& $0.00016$  \\ \hline 
$M_h^{light}$(SUSY) &   {\tt FeynHiggs}  & $>114.4$ GeV  & $3.0$ GeV  \\
\hline 
\end{tabular}
\caption{List of available constraints in the ``master code''. The shown
  values and errors represent the current best understanding of these
  constraints. Smaller errors for $\MW^{\rm SUSY}$ and 
$\sin^2 \theta_W^{\rm SUSY}$ are possible using a dedicated
code~\cite{Heinemeyer:2006px,Heinemeyer:2007bw}, 
which is, however, so far not included in
the ``master code'' (see, however, \cite{Buchmueller:2007zk}).}
\label{tab:constraints}
\end{center}
\end{table}

Using the ``master code'' as a foundation, an additional code layer
containing a $\chi^2$ fit~\cite{MINUIT_REFERENCE} has been added to
determine the consistency of a given set of MSSM parameters with the
constraints defined in Table~\ref{tab:constraints}. Other studies of
this kind using todays data can been performed in 
Refs.~\cite{Ellis:2004tc,Allanach:2005kz,deAustri:2006pe,Ellis:2006ix,Allanach:2006cc}. 
Studies using the anticipated data
from the LHC and the ILC are carried out and documented in 
Ref.~\cite{Bechtle:2004pc,Lafaye:2004cn}.

Using the ``master code'' we will present a few showcases for a global
$\chi^2$ fit using a {\em simplified} version of the MSSM.
The fit considers the following parameters: $\MA$ (the CP-odd Higgs
boson mass), $\tb$ (the ratio of the two vacuum expectation values),
$M_{\sq, \tilde l}$ (a common diagonal soft SUSY-breaking parameter
for squark and sleptons, respectively), $A$ (a common trilinear
Higgs-sfermion coupling), $\mu$ (the Higgs mixing parameter), 
$M_1$ and $M_2$ (the soft SUSY-breaking parameters in the
chargino/neutralino sector) and $\mgl = M_3$ (the gluino mass). All
parameters are assumed to be real.
Some further simplifying restrictions are applied: 
For the parameter $\mu$ we require $|\mu|>M_2$. 
This ad-hoc Ansatz is fully sufficient for our illustrative studies  
but in the future it will be replaced with a more sophisticated
treatment of the parameters and of the experimentally excluded phase
space regions  (e.g.\ sparticle mass limits, etc.)
In addition the Ansatz assumes $M_{\tilde l} = a_{\sq , \tilde l} \times
M_{\sq}$ as well as fixed values for $M_1,M_2,$ and $M_3$. The
initially assumed values of $a_{\sq , \tilde l}=0.5$, $M_2=200$ GeV,
$M_3=300$ GeV and $M_1=M_2/2$ are later varied within reasonable ranges to
evaluate the systematic impact of the assumption on the final results.   

The $\chi^2$ is defined as:
\begin{equation}
\chi^2 = \sum_i^{N_{const.}}\frac{(Const._i - Pred._i({\rm MSSM}))^2}
                                 {\Delta Const.^2 + \Delta Pred. ^2} 
\end{equation}
where $Const._i$ represents the measured values (constraints) and $Pred._i$
defines the MSSM parameter dependent predictions of a given constraint.
These predictions are obtained from the ``master code''. They depend
on SM parameters like $m_t$, $m_b$ and $\alpha_s$. Some of
these parameters still exhibit significant uncertainties which need to
be taken into account in the fit procedure.  
In a simple $\chi^2$ approach it is straightforward to include these
parametric uncertainties as fit parameters with penalty constraints. For
our study the uncertainty of the top quark mass was found to be by far
the dominating parametric uncertainty. 
The required minimization of the $\chi^2$ is carried out by the well
known and very reliable fit package Minuit~\cite{MINUIT_REFERENCE}.

In the following section we present some illustrative showcases that
utilize this global $\chi^2$ fit to extract quantitative
results. However,  
these studies are mainly meant to demonstrate the potential and
usefulness of ``external'' constraints for the interpretation of
forthcoming discoveries and for the corresponding model parameter
extraction.      

\subsubsubsection{Scan in the lightest Higgs-boson mass $\Mh$} 

One of the most important predictions of the MSSM is the existence of a
light neutral Higgs boson with 
$\Mh \le 135 \gev$~\cite{Heinemeyer:1998np,Degrassi:2002fi}. 
This upper limit together with the lower limit obtained at LEP, 
$\Mh^{\rm direct} \ge 114.4 \gev$~\cite{Barate:2003sz,Schael:2006cr}
\footnote{It is possible that the current lower limit could be even
  further improved before the LHC will 
  start data taking in 2008 by the currently running Tevatron
  experiments CDF and D0.} 
represent a tight constraint on the remaining
allowed parameter space of the MSSM. In the MSSM (with the
simplifications explained above), $\Mh$ depends mainly
on the average squark mass $M_{\sq}$, the Higgs mixing parameter
$\mu$, the tri-linear Higgs-squark coupling $A$, and $\tb$.
However, these parameters are also important for the predictions
of low-energy and electroweak observables in the MSSM. Therefore, a
global fit using the constraints listed in Table~\ref{tab:constraints}
not only allows a consistent extraction of the important MSSM parameters
but will also provide a prediction for the most probable light Higgs
boson mass $\Mh$ in the MSSM.  A convenient way to illustrate the
sensitivity of these parameters to $\Mh$ is a scan of the preferred
parameter space as a function of this variable. For this procedure the
global $\chi^2$ fit is performed repeatedly each time with a different
value for the $\Mh$ constraint. Therefore, the extracted set of MSSM
parameters for each individual fit correspond to the preferred parameter
space for a given value of $\Mh$.  While all $\Mh$ scan values below the
lower limit  of $\Mh^{\rm direct}>114.4 \gev$ are already excluded by
experiment, it is nevertheless interesting to see the results of the
$\Mh$ scan over the entire parameter space (i.e.\ also for $\Mh$ values
$\lsim 115 \gev$). For that reason the lower $\Mh$ limit from the
direct search at LEP has not been included in the $\chi^2$ fit.
    
\begin{figure}[ht]
\begin{center}
  \begin{minipage}[t]{.49\textwidth}
    \begin{center}  
      \epsfig{file=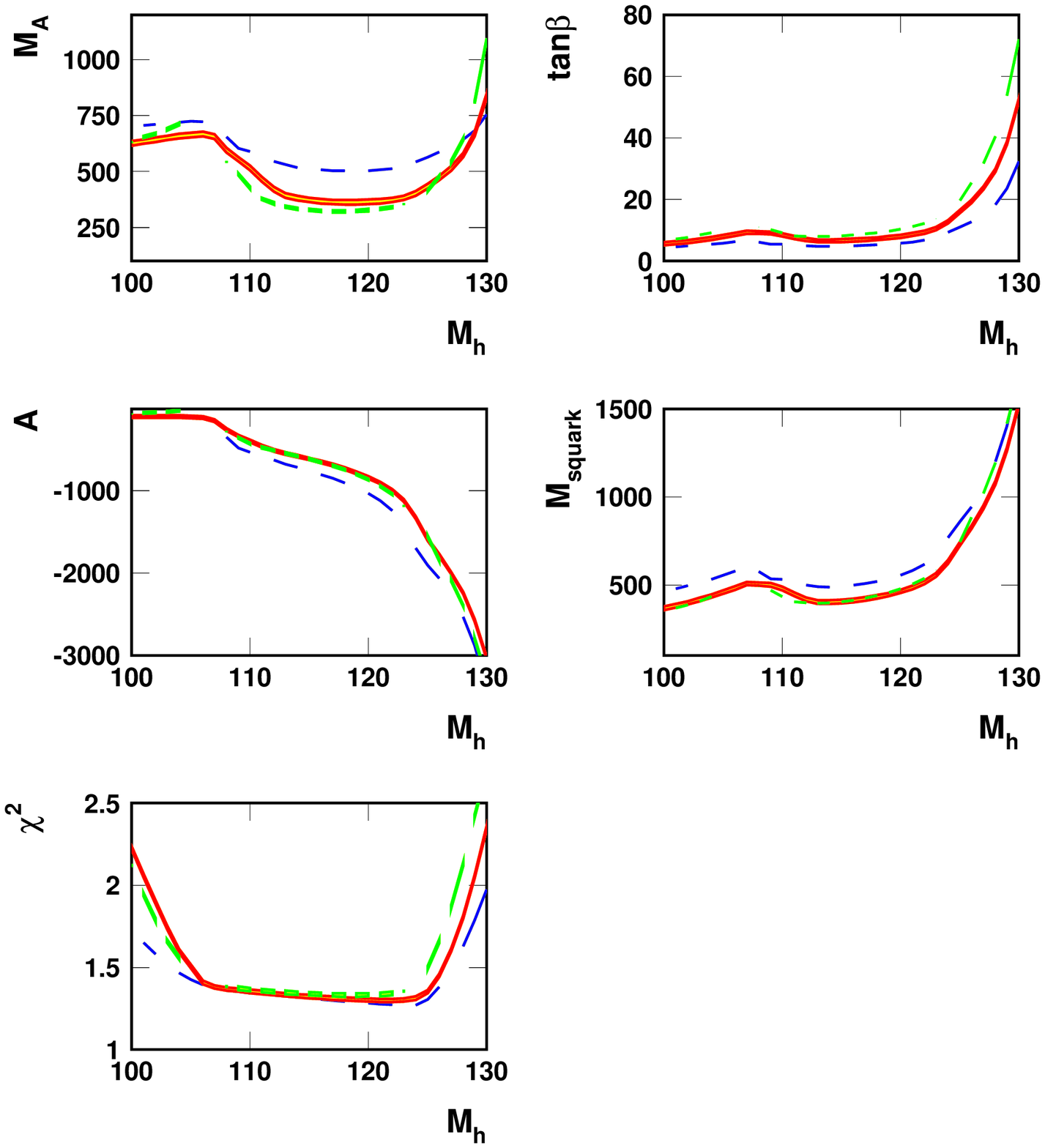, scale=0.4}
    \end{center}
  \end{minipage}
  \hfill
  \begin{minipage}[t]{.49\textwidth}
    \begin{center}  
      \epsfig{file=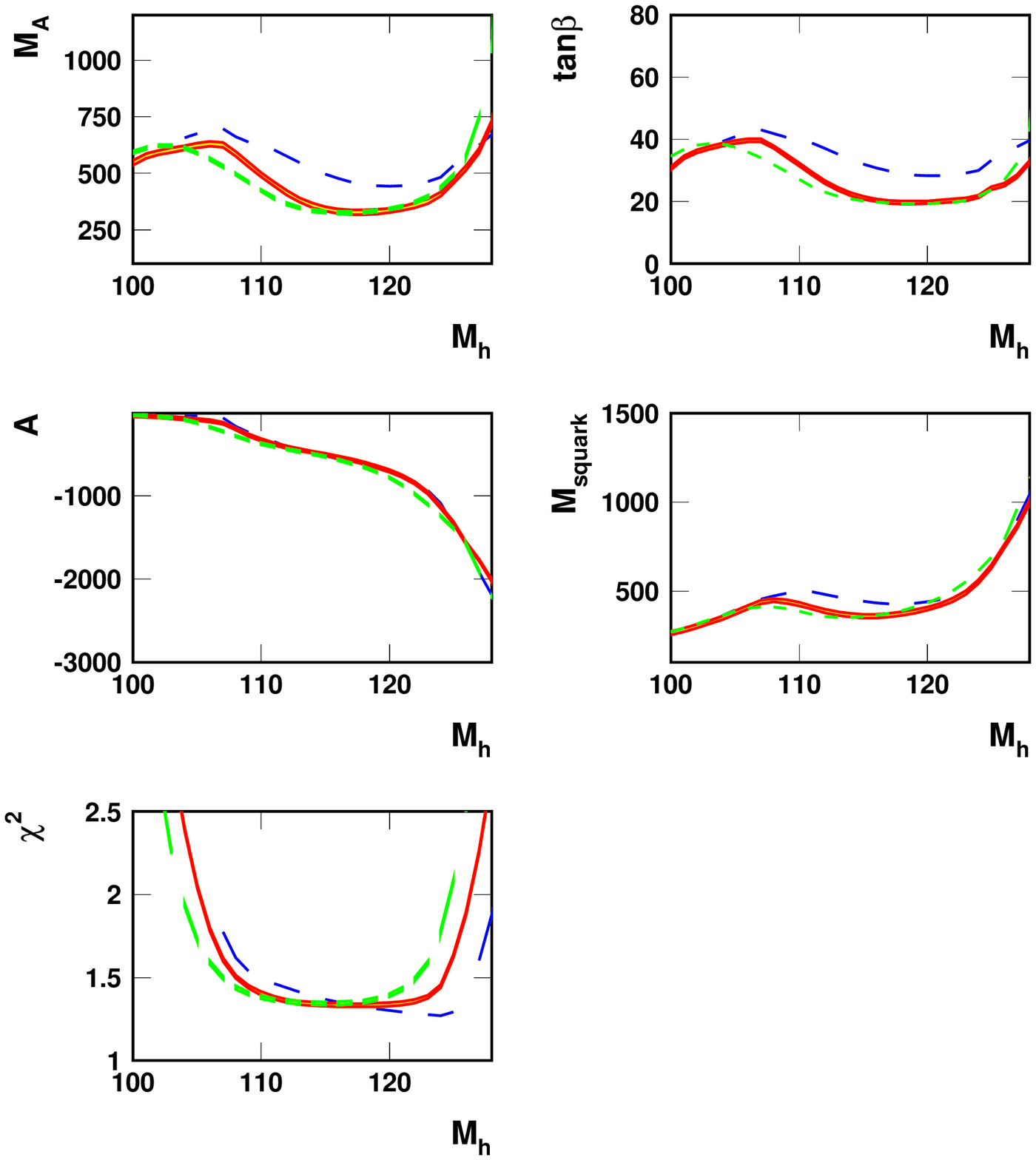, scale=0.4}
    \end{center}
  \end{minipage}
\caption{This figure shows the result of the extracted MSSM fit
  parameters and the corresponding $\chi^2$ distribution (lower right
  plot in each case) for the two scan scenarios: {\it today's $\Mh$
    scan} (left five plots) and {\it 2009-EW-LowE $\Mh$ scan} (right
  five plots). Each plot shows three scan results where the full-red
  curve corresponds to the default assumptions of $M_2=200 \gev$,
  $M_3=300 \gev$ and $a_{\sq , \tilde l}=0.5$. The blue-dashed line
  (large dash) changes $a_{\sq , \tilde l}=0.33$ with respect to
  the default setting, while the green-dashed line (small dash) modifies
  $M_2=300 \gev$, $M_3=500 \gev$ with respect to the default setting.} 
 \label{fig:scan_todayand2009}
\end{center}
\end{figure}

\subsubsubsection{$\Mh$ scan using today's constraint values and errors}

Fig.~\ref{fig:scan_todayand2009} shows the results of the $\Mh$ scan
using the constraint values listed in Table~\ref{tab:constraints}. Since
these values represent today's best knowledge of these observables, this
result provides a first estimate of how low-energy and electroweak
measurements constrain the MSSM parameter space. In the following we
will refer to this scan result as {\it today's $\Mh$ scan}.

It is important to note that the $\Mh\approx [110,125] \gev$ region seems
to be preferred by the $\chi^2$ scan. On the one hand, all $\Mh$ values
in this distinguished region of minimal $\chi^2$ are almost equally
likely.  On the other hand, values outside this window (i.e.\ $<110 \gev$
or $>125 \gev$) are clearly disfavoured by the low-energy and electroweak
constraints.  This is an interesting observation suggesting that today's
low-energy and electroweak data prefer a light MSSM Higgs boson with a
mass significantly higher than the most probable value for the 
SM Higgs boson. For comparison, the current preferred value from
the general electroweak fit is 
$\Mh^{\rm SM}\approx 80 \gev$~\cite{unknown:2005em,:2005di,MHworld}.

In order to qualitatively estimate the systematic impact of the assumed
parameter values ($M_2=200 \gev$, $M_3=300 \gev$ and 
$a_{\sq , \tilde l}=0.5$) on the scan results, 
a variation of the parameter values
within reasonable ranges has been carried out.
Fig.~\ref{fig:scan_todayand2009} shows the results of two of these
cross checks:  the blue-dashed line corresponds to the parameter setting
$M_2=200 \gev$, $M_3=300 \gev$ and $a_{\sq , \tilde l}=0.33$, while
the green-dashed line uses $M_2=300 \gev$, $M_3=500 \gev$ and 
$a_{\sq, \tilde l}=0.5$. The observed variation is rather small indicating
that the general conclusions are not strongly affected by the assumed
parameter setting of these quantities. In particular the preferred
minimal $\chi^2$ region of $\Mh$ remains almost unchanged.

The overall $\chi^2$ minimum of {\it today's $\Mh$ scan} is at
$\Mh\approx 123 \gev$ and the preferred values of the important MSSM
parameters are $\MA \approx 400 \gev$, $\tb \approx 10$, 
$A \approx -1000 \gev$, and $M_{\sq} \approx 500 \gev$.  These values are
qualitatively compatible with the range of ``allowed'' MSSM parameter
space reported in section~\ref{sec:newscen2app}. The 
fact that {\it today's $\Mh$ scan} prefers somewhat lower values for
$\tb$ and $M_{\sq}$ is mainly explained by the change in the
experimental Belle result of $R_{\br_{b \to \tau \nu}}$ from
$0.7\pm0.3$ to $1.125\pm0.52$~\cite{Ikado:2006un}. 
Using $0.7\pm0.3$ instead of the other
more recent (corrected) value yields $\tb \approx 30$, 
$M_{\sq}\approx 700 \gev$, and  $A\approx -1500 \gev$ but does not change the
general conclusion of the results (e.g. the preferred $\Mh$ range
remains the same).

\begin{figure}[ht]
  \begin{minipage}[t]{.49\textwidth}
    \begin{center}  
      \epsfig{file=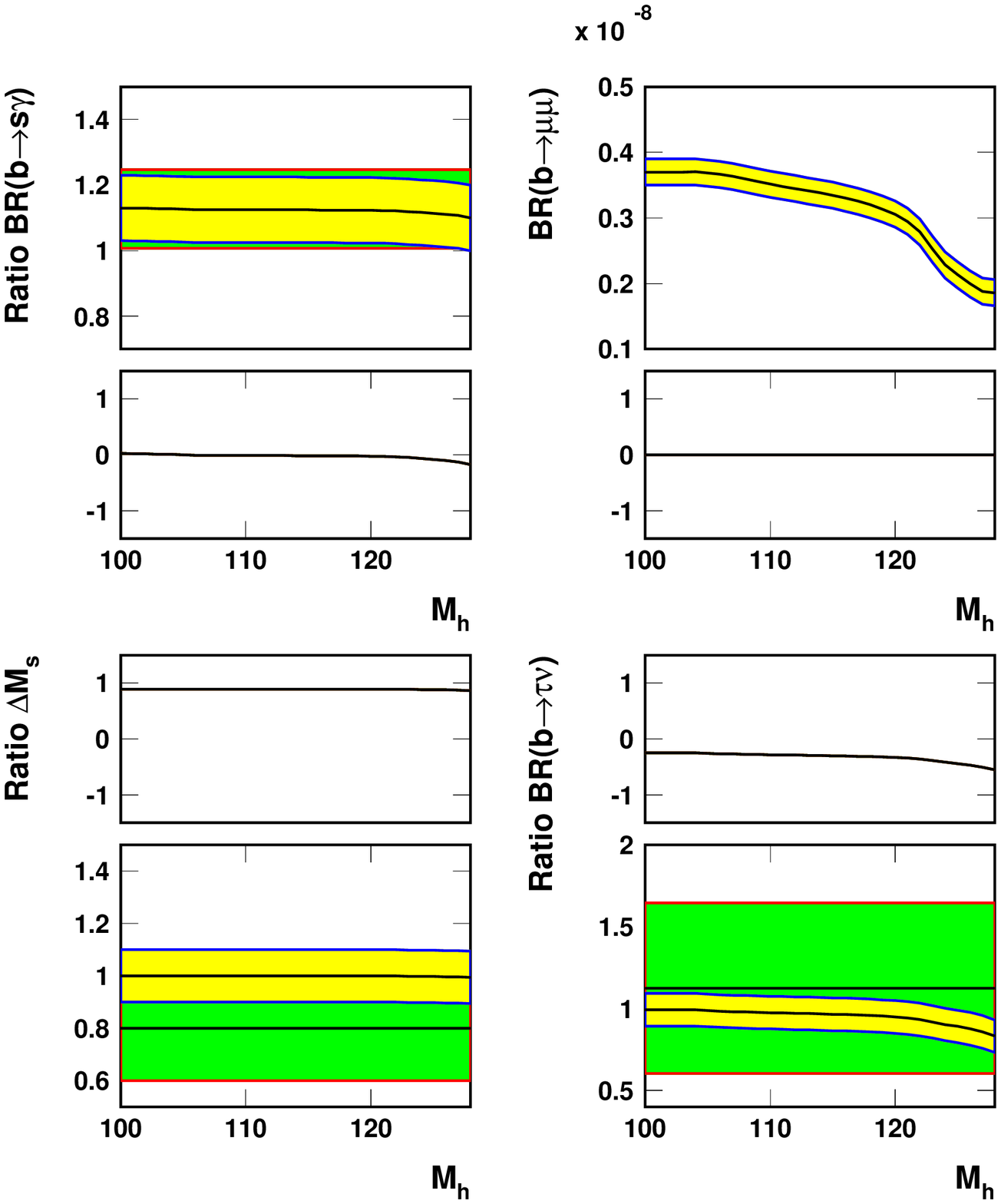, scale=0.4}
    \end{center}
  \end{minipage}
  \hfill
  \begin{minipage}[t]{.49\textwidth}
    \begin{center}  
      \epsfig{file=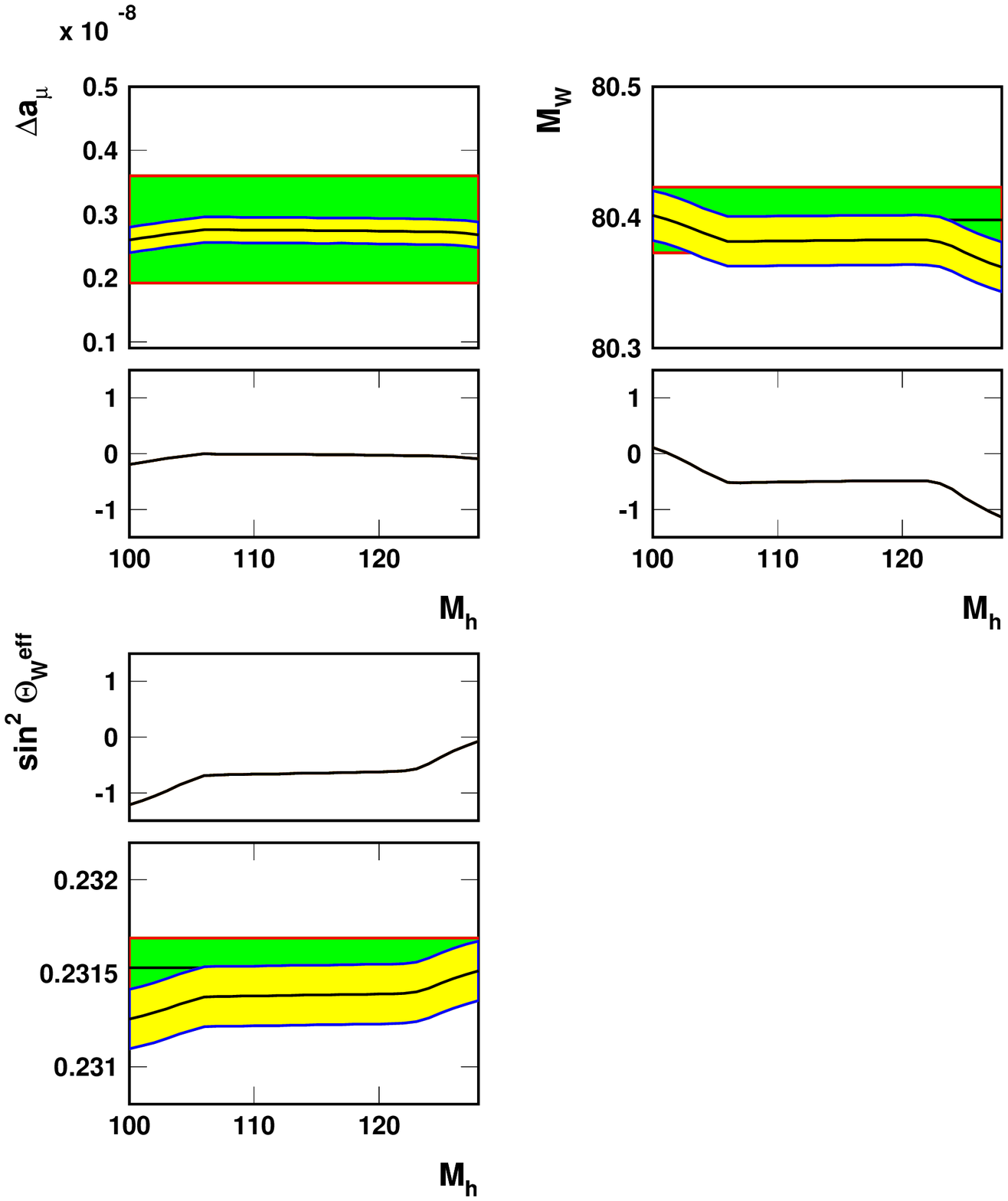, scale=0.4}
    \end{center}
  \end{minipage}
\caption{This figure shows a comparison of the predicted constraint
  values (yellow band) and their corresponding measurements (constant
  green band) obtained from {\it today's $\Mh$ scan}. All plots show a
  comparison of prediction versus measurement (plots with bands) as well
  as their corresponding pull contributions 
$\frac{Const._i - Pred._i({\rm MSSM})}
      {\sqrt{\Delta Const.^2 + \Delta Pred. ^2}}$ to the
  overall $\chi^2$.} 
 \label{fig:plot_scan_mh_varall_today}
\end{figure}

Fig.~\ref{fig:plot_scan_mh_varall_today} shows a comparison of the
predicted constraint values and their corresponding measurements
obtained from {\it today's $\Mh$ scan}. The measurements and their
errors are also listed in Table~\ref{tab:constraints}. In general, good
agreement between prediction and measurement is observed in the
preferred minimal $\chi^2$ region of $\Mh\approx [110,125] \gev$. The
fact that the $\chi^2$ scan prefers a prediction of $R_{\Delta M_s}$
very close to unity is explained by (1) the already rather tight limit
on $\br(B_s\to\mu^+\mu^-)<8\times10^{-8}$ and (2) the large value of
$R_{\br_{b \to \tau \nu}}$. Both constraints prefer low values of
$\tb$ and thus result in a prediction of 
$R_{\Delta M_s}\approx 1$.  
However, today's experimental value is still within one sigma
compatible with this prediction.  

\begin{table}[tbh!]
\renewcommand{\arraystretch}{1.4}
\begin{center}
\begin{tabular}{|c|c|c|} \hline
Observable & Constraint & theo. error \\ \hline \hline
$R_{\br_{b \to s \gamma}} $ & $1.127 \pm 0.1$  & $0.1$  \\ \hline
$R_{\Delta M_s} $  & $0.8 \pm 0.2$  & $0.1$  \\ \hline
$\br_{b \to \mu \mu}$ & $(3.5 \pm 0.35) \times 10^{-8}$  & $2
\times 10^{-9}$  \\ \hline 
$R_{\br_{b \to \tau \nu}} $  & $0.8 \pm 0.2$  & $0.1$  \\ \hline
$\Delta a_{\mu} $  & $(27.6
 \pm 8.4) \times 10^{-10}$  & $2.0 \times 10^{-10}$  \\ \hline
$M_W^{\rm SUSY}$ & $80.392 \pm 0.020  \gev$ & $0.020$  GeV\\ \hline
$\sin^2 \theta_W^{\rm SUSY}$ & $0.231

53 \pm 0.00016 $  & $0.00016$  \\ \hline
$\Mh^{\rm light}$(SUSY) & $>114.4 \gev$  & $ 3.0 \gev$  \\ \hline
\end{tabular}
\caption{Assumed constraint values and errors for the {\it 2009-EW-LowE}
  scenario.} 
\label{tab:constraints_2009}
\end{center}
\end{table}

Another interesting observation is the prediction of
$\br(B_s \to \mu^+\mu^-)$.  Although the constraint used for this
quantity allows values up to $\br(B_s \to \mu^+\mu^-) < 8 \times 10^{-8}$,
the scan predicts (in the interesting $\Mh$ region) an almost constant
value of $\br(B_s \to \mu^+\mu^-)\approx [3.0-4.0] \times 10^{-9}$. This
is an interesting observation because this value coincides well with the
standard model prediction of $\br(B_s \to \mu^+\mu^-)^{\rm SM}\approx 3.5
\times 10^{-9}$.  This might suggest that the current low-energy and
electroweak data prefer a value of $\br(B_s \to \mu^+\mu^-)$ close to its
SM prediction. It will be interesting to see whether the
soon forthcoming combined result of  $R_{\br_{b \to \tau \nu}}$ 
from BABAR and Belle will confirm this trend. If this is the case
spectacular effects from new (MSSM) physics  
contributions seem rather unlikely for $B_s \to \mu^+\mu^-$ . 

\begin{figure}[ht]
\begin{center}
\begin{minipage}[t]{.49\textwidth}
\begin{center}       
\epsfig{file=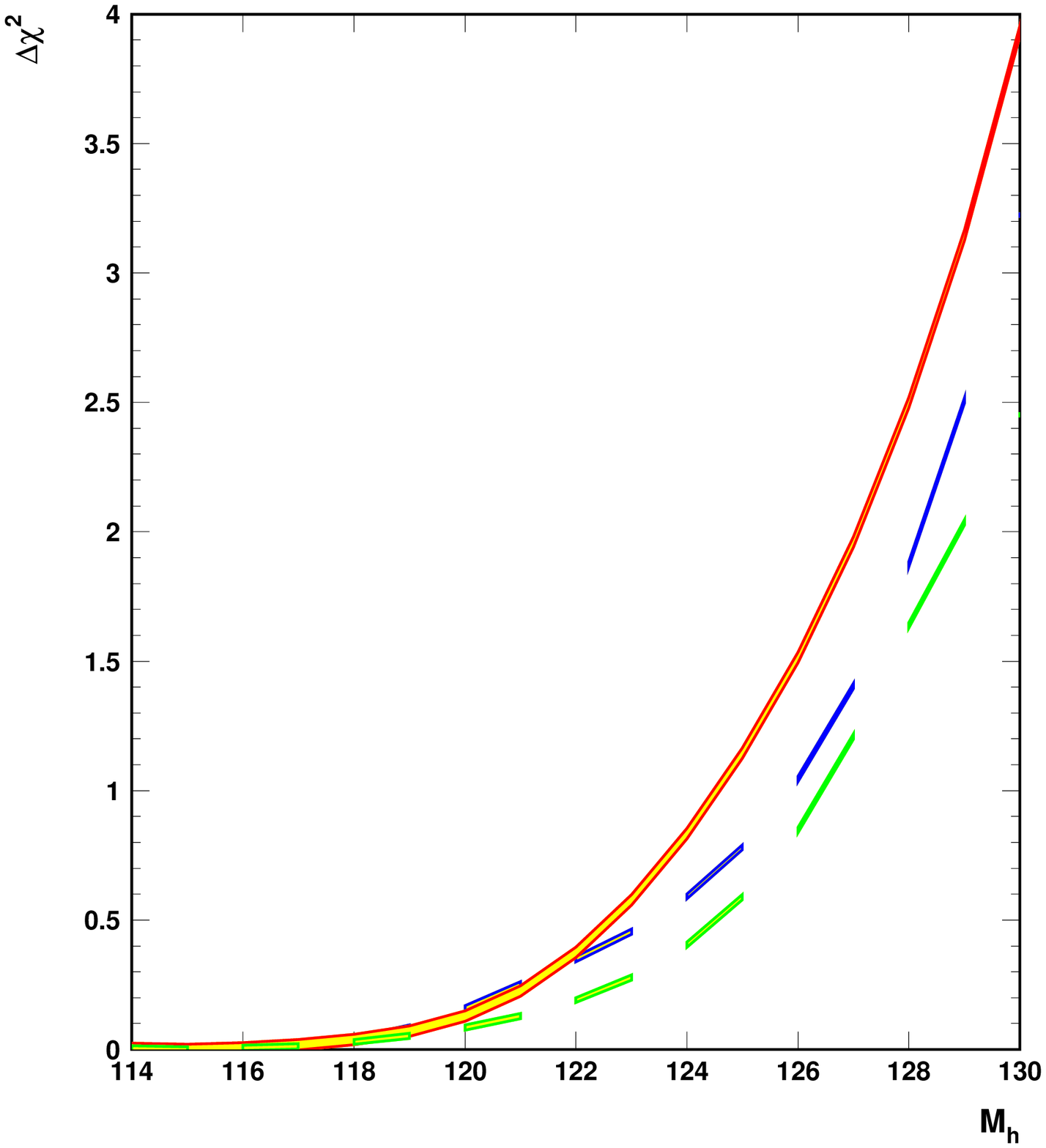, scale=0.35}     
\caption{$\Delta \chi^2$ distribution for scenario 
{\it LHC-$M_{\sq}$-$M_{A}$} (red curve),   
{\it LHC-$M_{\sq}$} (blue dashed curve), and {\it 2009-EW-LowE }
(green dashed curve). All curves are evaluated 
with an assumed error of  $\Delta \Mh=$ 3 GeV. }      
\label{fig:dchi_com}  
\end{center}
\end{minipage}  
\hfill
\begin{minipage}[t]{.49 \textwidth}
\begin{center}        
\epsfig{file=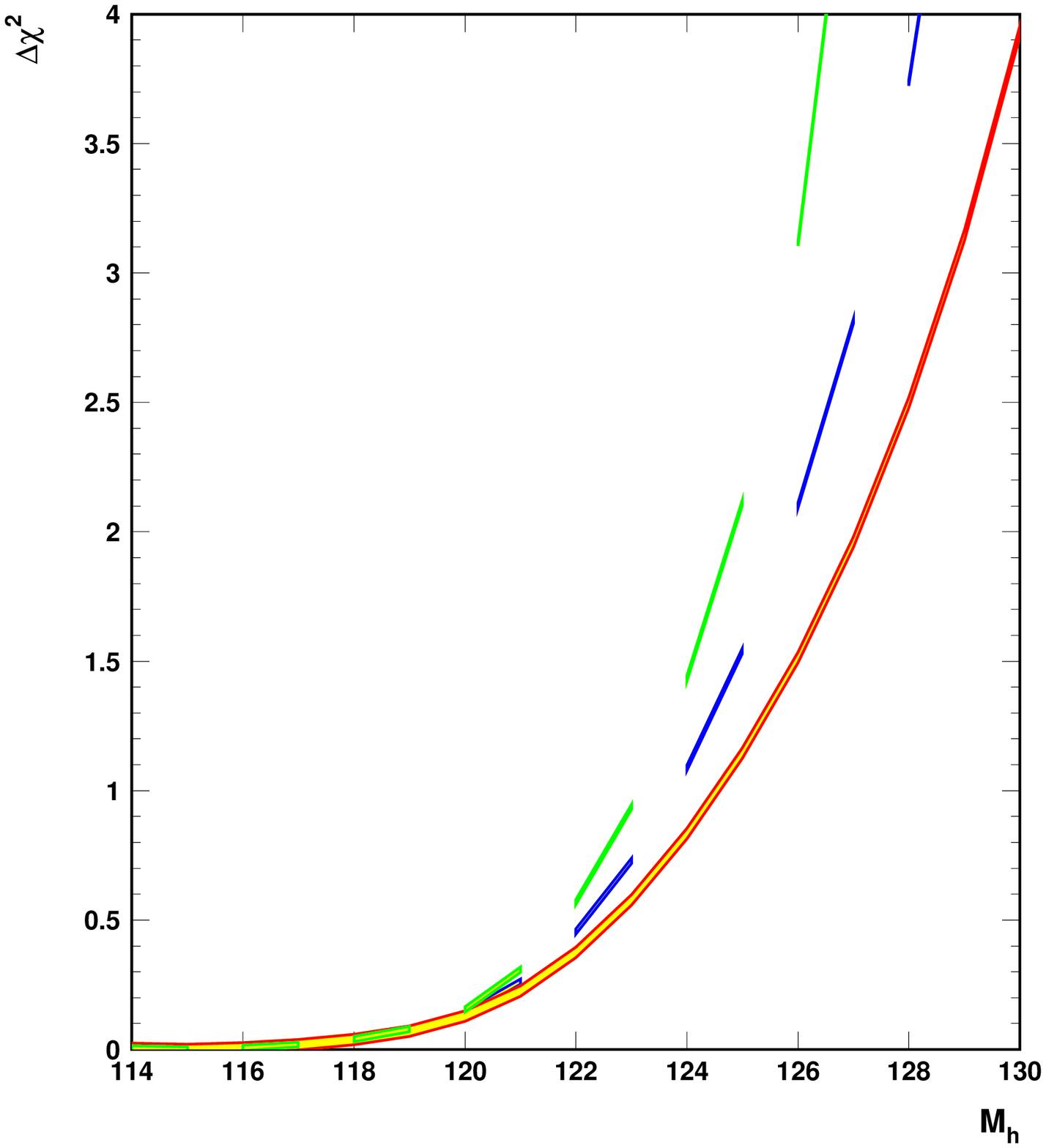, scale=0.35}      
\caption{$\Delta \chi^2$ distribution for scenario 
{\it LHC-$M_{\sq}$-$\MA$} testing the hypothesis 
that a discovered light higgs boson candidate with a mass error of:
$\Delta \Mh=$ 3 GeV (red curve),  
2 GeV (blue dashed curve), and 1 GeV (green dashed curve) is compatible
with the MSSM.  }       
\label{fig:dchi_error}    
\end{center}  
\end{minipage}  
\hfill
\end{center}
\end{figure}

\subsubsubsection{Interpretation of potential LHC discoveries}

The LHC will start collecting physics data in 2008. For that reason, the
first results are not expected before early 2009. In the meantime,
however, it is likely that most of the considered low-energy and
electroweak constraints will further improve. Therefore, in 2009 it will
be possible to even more strongly restrict the allowed MSSM parameter
space. Table~\ref{tab:constraints_2009} lists the assumed constraint
values that might be achieved by this time period. The assumed values
and errors are only chosen for illustrative purposes. The sole intention
of this study is to demonstrate the potential of low-energy and
electroweak data to constrain the parameter space of new physics and to
eventually provide guidance for the interpretation of potential new
physics discoveries at the LHC.  Fig.~\ref{fig:scan_todayand2009}
(five plots on the right) shows the results of the $\chi^2$ scan using
the constraints listed in Table~\ref{tab:constraints_2009}. In the
following, we refer to these results as {\it 2009-EW-LowE $\Mh$
  scan}. Similar to the results from the {\it today's $\Mh$ scan}, the
general results and conclusions of this study are largely unaffected by
the variation of the assumed values for $M_2$, $M_3$ and 
$a_{\sq,\tilde l}$. As shown in Fig.~\ref{fig:scan_todayand2009} the
$\chi^2$ preferred $\Mh$ region becomes even more pronounced. Hence, the
allowed MSSM parameters space is further reduced. In particular this
information will become very useful in the case of LHC discoveries and
their corresponding interpretation. In order to illustrate this property
we define a few hypothetical scenarios: 

\begin{figure}[ht]
\begin{center}
\begin{minipage}[t]{.49 \textwidth}
\begin{center}        
\epsfig{file=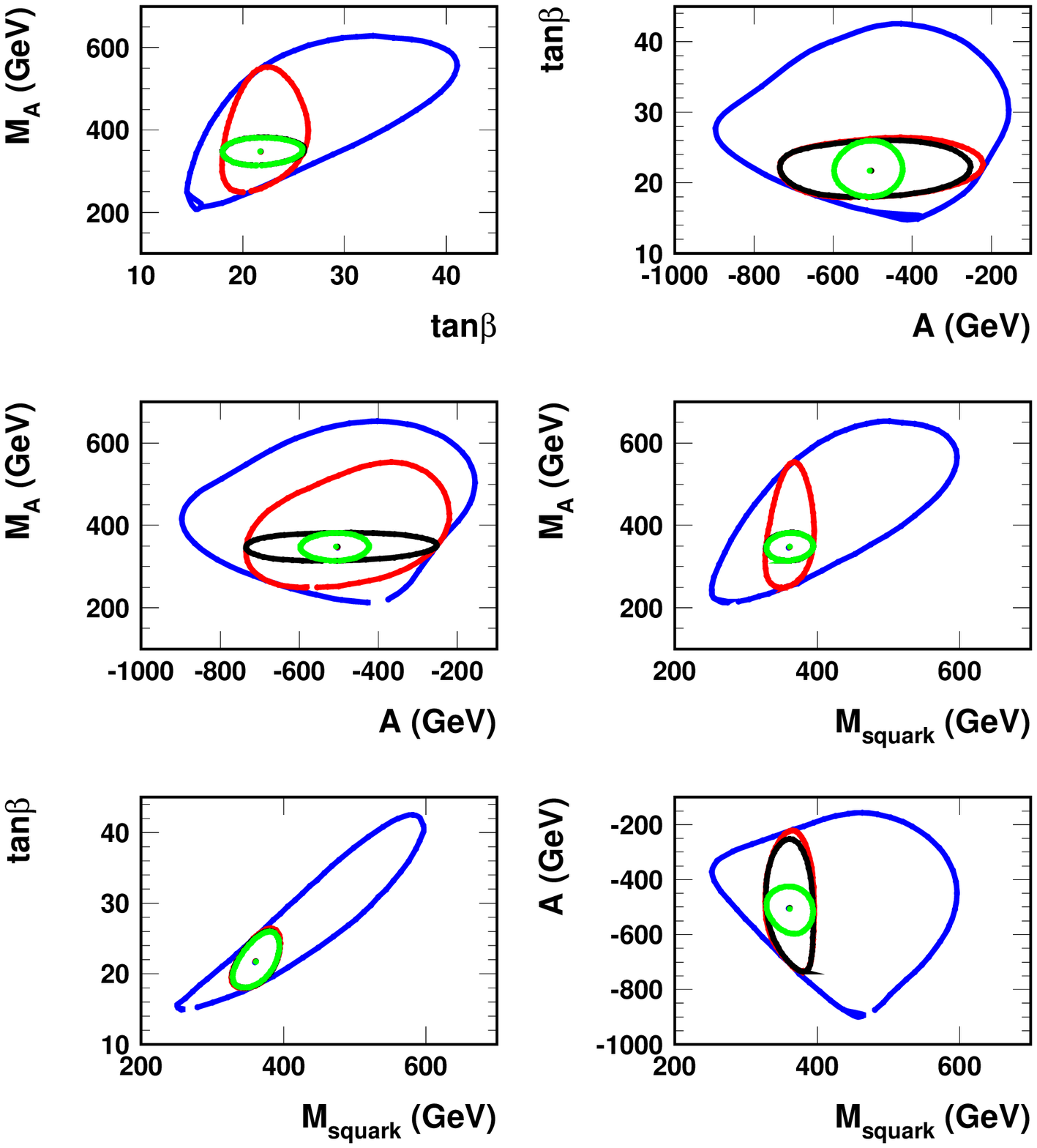, scale=0.4}      
\caption{This figure shows the $\Delta \chi^2=1$ contours of the four
  scenarios: {\it 2009-EW-LowE} (blue contour), {\it LHC-$M_{\sq}$}
  (red contour), {\it LHC-$M_{\sq}$-$\MA$} (black contour), and
  {\it LHC-$M_{\sq}$-$\MA$-$\Mh$} (green contour).  }       
\label{fig:contour}    
\end{center}  
\end{minipage}  
\hfill
\begin{minipage}[t]{.49\textwidth}
\begin{center}       
\epsfig{file=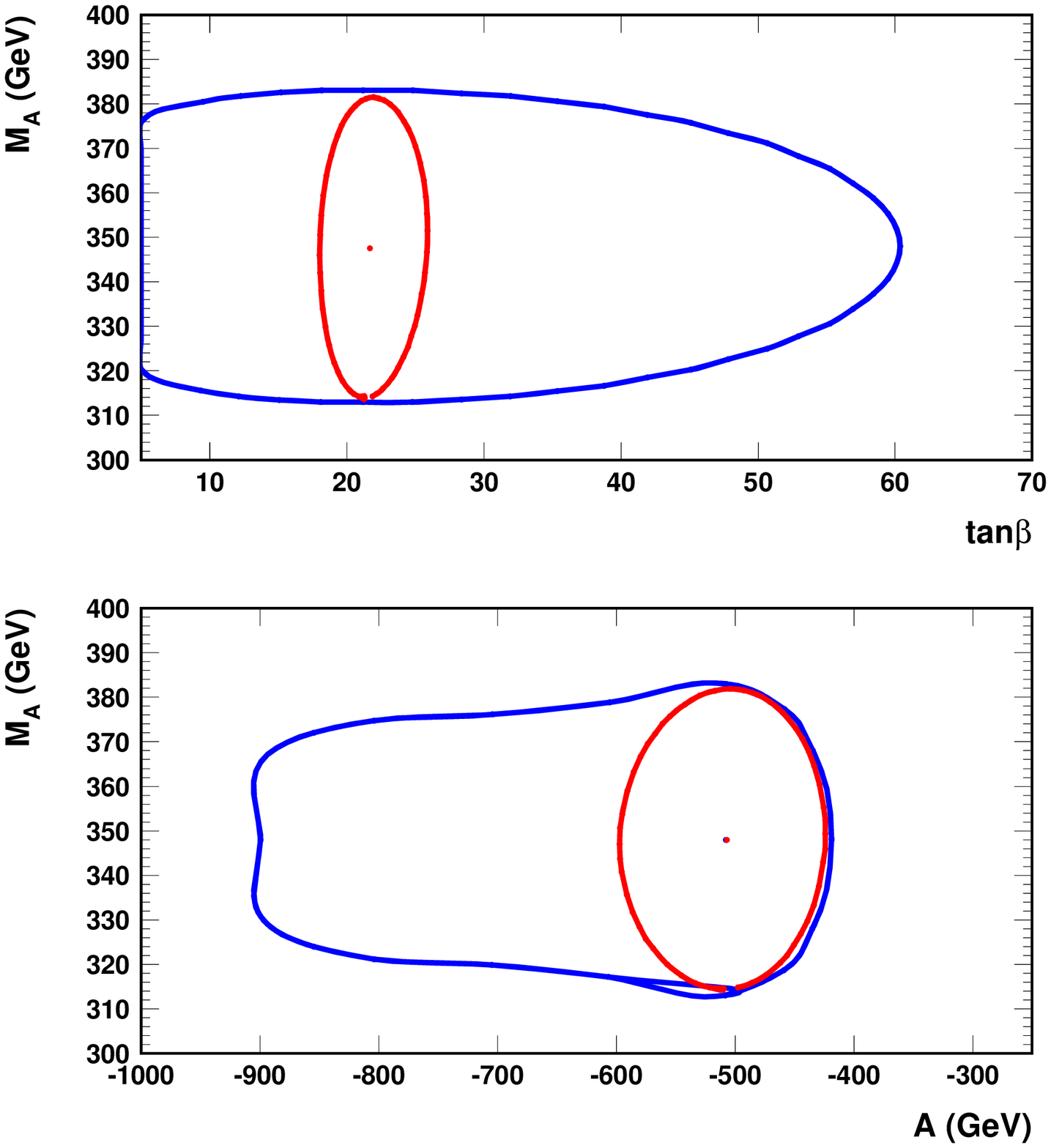, scale=0.4}     
\caption{The red contour corresponds to scenario 
{\it LHC-$M_{\sq}$-$\MA$-$\Mh$} that includes the low-energy and
electroweak constraints, while the blue contour makes the same
assumptions about the assumed LHC discoveries, but does not include any
external constraints.}       
\label{fig:tanbeta_a}    
\end{center}  
\end{minipage}  
\hfill
\end{center}
\end{figure}

\begin{figure}[ht]
\begin{center}
\hfill
\begin{minipage}[t]{.49\textwidth}       
\begin{center}  
\epsfig{file=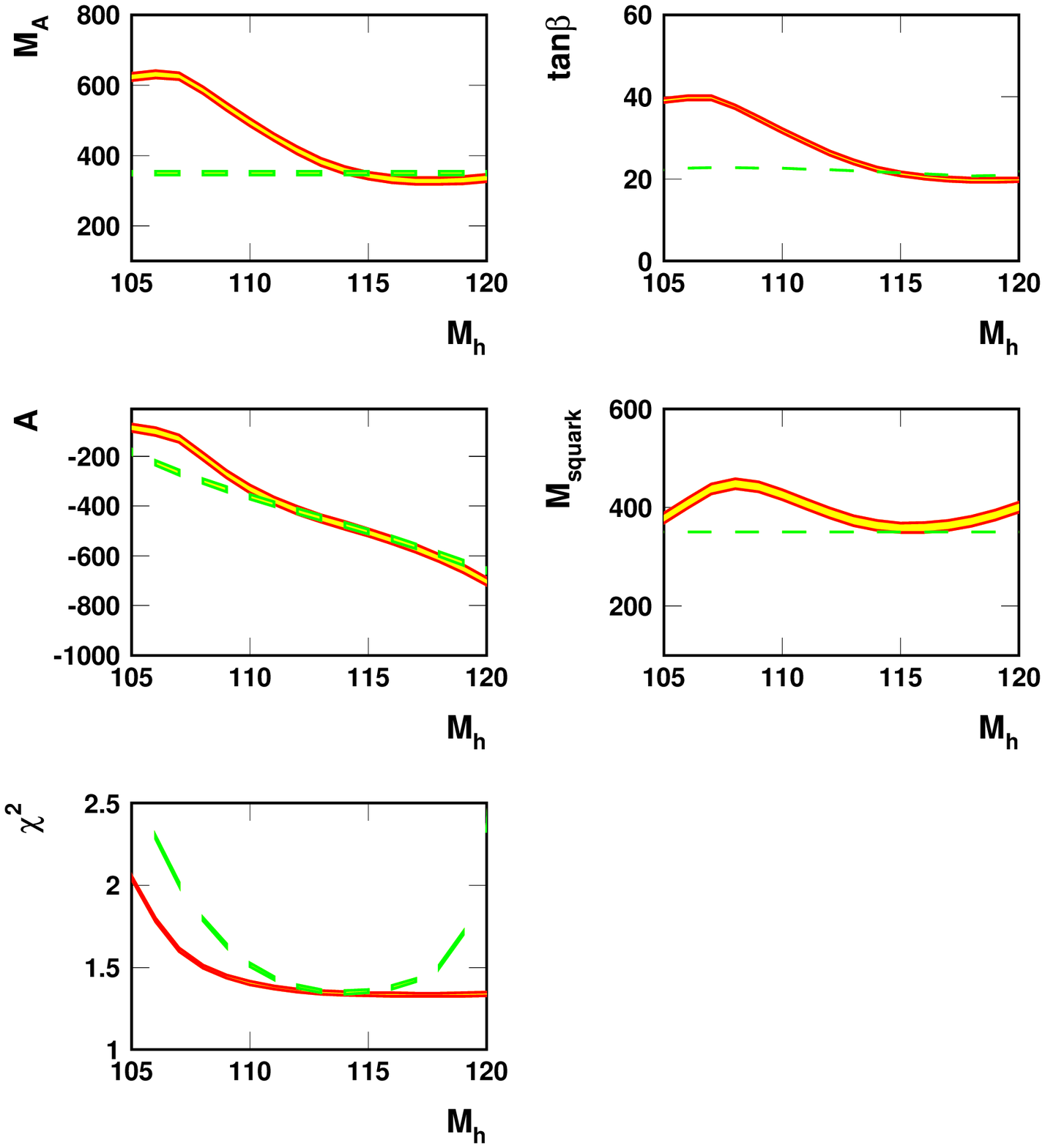,scale=0.4}
\caption{This figure shows the result of the extracted MSSM fit
  parameter and the corresponding  
$\chi^2$ distribution for the two scan scenarios: {\it 2009-EW-LowE
  $\Mh$ scan} (full-red curve)  
and {\it LHC-$M_{\sq}$-$\MA$-$\Mh$ scan}  (green-dashed curve).}
\label{fig:scan_2010}    
\end{center}
\end{minipage}
\hfill  
\begin{minipage}[t]{.49\textwidth}    
\begin{center}  
\epsfig{file=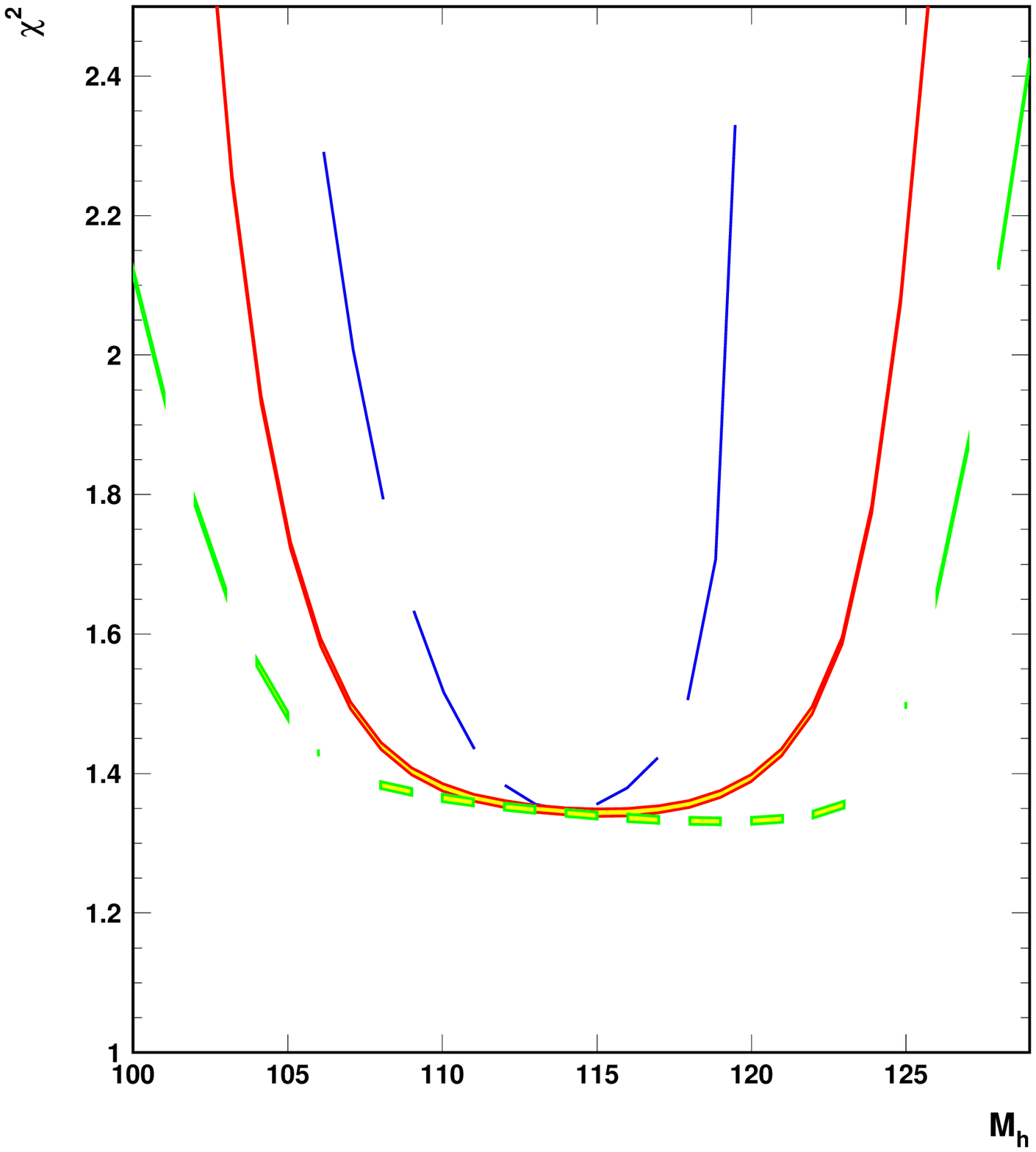, scale=0.4}      
\caption{$\chi^2$ distribution as a function of $\Mh$ for the three scenarios:
{\it 2009-EW-LowE $\Mh$ scan} (full-red curve),  
{\it today's $\Mh$ scan} (green-small-dashed curve), 
and {\it LHC-$M_{\sq}$-$\MA$-$\Mh$ scan} 
(blue-large-dashed curve).}    
\label{fig:chi2}   
\end{center}  
\end{minipage}
\hfill
\end{center}
\end{figure}

\begin{itemize}
\item {\it 2009-EW-LowE}:\\ 
This scenario includes only the observables listed in
Table~\ref{tab:constraints_2009}.  The overall $\chi^2$ minima for this
scenario is achieved for $\MA\approx 350 \gev$, $\tb \approx 22$,
$\mu\approx 5 \gev$, $A\approx -450 \gev$, and $M_{\sq} \approx 350 $
GeV. The corresponding prediction of the light MSSM Higgs boson mass is
$\Mh\approx 115 \gev$. 
\item {\it LHC-$M_{\sq}$}:\\  
This scenario includes {\it 2009-EW-LowE} and additionally assumes that
the relevant squark mass\footnote{For example this could be achieved by
  a determination of the stop mass. In particular this mass is important
  for the determination of the lightest Higgs boson mass $\Mh$ in the
  MSSM.} $M_{\sq}$ is known at the level of 10\%. To be consistent
with {\it 2009-EW-LowE}, we therefore define: $M_{\sq}=350\pm35 \gev$. 
\item {\it LHC-$M_{\sq}$-$\MA$}:\\  
This scenario includes  {\it LHC-$M_{\sq}$} and additionally
assumes that the mass of $M_{A{H^{\pm}}}$ is known to 10\%. To be
consistent with {\it 2009-EW-LowE}, we therefore define:
$\MA=355\pm35 \gev$.  
\item {\it LHC-$M_{\sq}$-$\MA$-$\Mh$}:\\  
This scenario includes {\it LHC-$M_{\sq}$-$\MA$} and additionally
assumes that the mass of $M_{h}$ is measured with a 3 GeV error. To be
consistent with {\it 2009-EW-LowE}, we therefore define: $M_{h}=115\pm3$
GeV. 
\end{itemize}

Fig.~\ref{fig:scan_2010} shows the results of the $\Mh$ scan for the
scenario {\it 2009-EW-LowE} and the scenario 
{\it  LHC-$M_{\sq}$-$\MA$}. As expected, the $\chi^2$ allowed region
of $\Mh$ is 
reduced to a small window by including the additional information of
$\MA=355\pm35 \gev$ and $M_{\sq}=350\pm35 \gev$.  This information
can, for example, be utilized to test the consistency of a discovered
light Higgs boson candidate with: 
\begin{itemize} 
\item [] a) other discoveries of MSSM particle candidates (in our case
  squark and heavy Higgs candidates), 
\item [] b) low-energy and electroweak constraints.
\end{itemize}
Assuming that a light Higgs boson candidate has been observed and that
its mass is measured with an error of $\Delta \Mh=\pm 3 \gev$,
Fig.~\ref{fig:dchi_error} shows the $\Delta \chi^2$ distributions for
the scenario {\it 2009-EW-LowE} (green small-dashed line), 
{\it LHC-$M_{\sq}$} (blue large-ashed line) and,  
{\it LHC-$M_{\sq}$-$\MA$} (red full line). 

As defined above, all scenarios correspond to one MSSM parameter set
that has a $\chi^2$ minimum for $\Mh\approx 115 \gev$.  The $\Delta
\chi^2$, and therefore also the exclusion limits, are defined with
respect to this MSSM parameter set. For the most constraining scenario,
all masses above $\approx 130 \gev$ are excluded at 95\%~CL. Therefore,
in this hypothetical case $\Mh$ must be below $130 \gev$ in order to be
compatible with the other observed LHC discoveries as well as with the
low-energy and electroweak constraints. A discovery of a lightest Higgs
boson with a mass above $130 \gev$ would rule out the MSSM at 95\%~CL.
It is clear that the exclusion limit depends on the assumed error
for $\Mh$. For scenario {\it LHC-$M_{\sq}$-$\MA$},
Fig.~\ref{fig:dchi_com} compares the results for $\Delta \Mh=\pm 3$,
$\Delta \Mh=\pm 2$, and  $\Delta \Mh=\pm 1$. With an assumed error of 2
GeV, the 95\%~CL exclusion limit would be around $\Mh\approx 128 \gev$,
while for a 1 GeV error it would be as stringent as $\Mh\approx 126 \gev$.

Therefore, together with the discoveries of a stop candidate and a heavy
Higgs candidate, the consistency of a measured light Higgs candidate
within the MSSM hypothesis can be tested. It should be noted that
without the use of low-energy and electroweak constraints, this
consistency test would be much weaker. For example the three LHC
discoveries alone will not significantly constrain the important MSSM
parameters $\tb$ and $A$. This feature is clearly demonstrated in
Fig.~\ref{fig:tanbeta_a}. Without the inclusion of the low-energy and
  electroweak constraints, the parameters $\tb$ and $A$ are much
  less determined.  Thus, the overall sensitivity of the consistency test
  is significantly worse. 

Another way to illustrate the potential of external constraints for the
interpretation of new physics discoveries and the eventual extraction of
the model parameters is shown in Fig.~\ref{fig:contour} which displays
the $\Delta \chi^2=1$ contours of the four different scenarios for
various parameter combinations. Although {\it 2009-EW-LowE} (blue
contour) only utilizes indirect constraints (i.e.\ no direct measurement
of new physics quantities) the MSSM parameter space is already rather
restricted.  Adding $M_{\sq}=350\pm35 \gev$ (red contour) in
particular helps to further constrain $\tb$ and to some extent
also $\MA$, while measuring also the heavy (black contour) and also
the light Higgs boson mass (green contour) will restrict the allowed
range for $A$ rather significantly. Also here the use of the external
low-energy and electroweak constraints is essential to determine the
important MSSM parameters $\tb$ and $A$.  



\subsubsubsection{Outlook}

In order to fully exploit this interesting potential, it will be
important to extend the ``master code'' by adding additional
calculations such as extra low-energy observables, as well as,
potentially, constraints from cosmology data (see \cite{Buchmueller:2007zk}). 
This will eventually yield
an important tool for the comprehensive interpretation of future new
physics discoveries.


-

\newpage \subsection{Discrimination between new physics scenarios}
\label{sec:discrimination}

At present, the SM gives a fully consistent description of all
experimental data in the flavour sector, apart from a few, not yet
statistically significant deviations. This means that flavour physics
can at present only rule out models that produce too large deviations
from the SM; in practice, this means giving an upper bound on new sources of
flavour and CP violation for a fixed NP scale, or giving a lower bound
on the NP scale for fixed values of the NP flavour parameters. As
discussed in Sec.~\ref{sec:patterns}, this gives us hints on the
flavour structure of NP models with new particles up to the TeV
range. However, to fully exploit the constraining power of flavour
physics, additional (external) information on the spectrum of new
particles must be provided. First examples of the combination of flavour and
high-$p_T$ information have been presented in
Sec.~\ref{sec:connections}, and there is increasing activity in this
direction.

\clearpage
\section*{Acknowledgements}

%
%
%
%
%
This work was supported in part by the National Science Foundation, the 
Natural Sciences and Engineering Council of Canada and 
the DFG cluster of excellence 
Origin and Structure of the Universe (www.universe-cluster.de), Germany. 
J.L.R. was supported by 
the United States Department of Energy through Grant No.\ DE FG02 90ER40560. 
A.A.P.~was supported in part by the U.S.\ National Science Foundation CAREER 
Award PHY--0547794 and by the U.S.\ Department of Energy under 
Contract DE-FG02-96ER41005.

The authors of this report thank all further participants in this
workshop series
for their presentations during the five workshop meetings,
for contributing to the discussions, and for useful comments. 
%

%
%

\end{document}